\documentclass[PhD]{dukethesis2018}
\usepackage{amsmath}
\usepackage{amssymb}
\usepackage{amsthm}
\usepackage{array}
\usepackage{microtype}
\usepackage{booktabs}
\usepackage{multirow}
\usepackage{bbm}
\usepackage{mathtools}
\usepackage[dvipsnames]{xcolor}
\usepackage{cite}
\usepackage{tabularx}
\usepackage{enumitem}
\usepackage{cprotect}
\usepackage{listings}
\usepackage{tablefootnote}
\usepackage{xifthen}
\usepackage[pdftex, pdfusetitle, plainpages=false,
				bookmarks, bookmarksnumbered,
				colorlinks, linkcolor=black, citecolor=black,
	         filecolor=black, urlcolor=black]
				{hyperref}
\newcommand{\fig}{\begin{figure}[tp]\centering}
\newcommand{\figsec}{\begin{figure}[htp!]\centering}

\newcommand{\efig}{\end{figure}}
\newcommand{\tab}{\begin{table}[tp]\centering}
\newcommand{\tabsec}{\begin{table}[htp!]\centering}
\newcommand{\etab}{\end{table}}

\newcommand{\eq}{\begin{equation}}
\newcommand{\eeq}{\end{equation}}
\newcommand{\eqa}{\begin{eqnarray}}
\newcommand{\eeqa}{\end{eqnarray}}
\newcommand{\etal}{\nobreak\mbox{\it et al.}}
\newcommand{\vs}{\nobreak\mbox{\it vs.}}
\newcommand{\etc}{\nobreak\mbox{\it etc.}}
\newcommand{\ie}{\nobreak\mbox{\it i.e.}}
\newcommand{\eg}{\nobreak\mbox{\it e.g.}}
\newcommand{\abinitio}{\nobreak\mbox{\it ab-initio}}
\newcommand{\Abinitio}{\nobreak\mbox{\it Ab-initio}}
\newcommand{\abinitiospace}{\nobreak\mbox{\it ab initio}}

\newcommand{\tabvspace}{\vspace{3mm}}

\newcommand{\boldsection}[1]{\noindent{\bf #1}}
\newcommand{\mycaption}[2][]{
  \ifthenelse{\equal{#1}{}}
  {\caption{#2}}
  {\caption[#1]{\textbf{#1} #2}}
}

\newcommand{\singlespacing}{
  \let\CS=\small\renewcommand{\baselinestretch}{1.0}\CS}
\newcommand{\doublespacing}{
  \let\CS=\small\renewcommand{\baselinestretch}{1.6}\CS}

\DeclareMathOperator*{\argmin}{arg\,min}
\DeclareMathOperator*{\avg}{avg}
\DeclareMathOperator*{\std}{std}

\DeclareMathOperator*{\fraction}{frac}
\definecolor{pranab_green}{rgb}{0.31,0.53,0.10}
\definecolor{pranab_red}{rgb}{0.85,0.23,0.11}

\newcommand{\onlinecite}[1]{\nocite{#1}\citenum{#1}}
\newcommand{\onlinecitesq}[1]{\cite{#1}}

\def\BZ{{\small BZ}}
\def\DOS{{\small DOS}}

\lstdefinelanguage{POSCAR}{
  alsoletter=0123456789,
  alsodigit={.-}
}

\def\AFLOW{{\small AFLOW}}
\def\QUANTUMESPRESSO{\textsc{Quantum {\small ESPRESSO}}}
\def\QE{\QUANTUMESPRESSO}
\def\SIESTA{{\small SIESTA}}
\def\GAUSSIAN{{\small GAUSSIAN}}
\def\FHIAIMS{{\small FHI-AIMS}}
\def\ABINIT{{\small ABINIT}}

\def\OQMD{{\small OQMD}}
\def\NOMAD{{NOMAD}}
\def\VASP{{\small VASP}}
\def\AIIDA{{\small AiiDA}}

\def\AEL{{\small AEL}}
\def\AGL{{\small AGL}}
\def\APL{{\small APL}}
\def\AFLOWPOCC{{\small AFLOW-POCC}}
\def\POCC{{\small POCC}}

\def\AFLOWorg{{\sf \AFLOW.org}}
\def\AFLOWHULLtitle{AFLOW-CHULL}
\def\AFLOWHULL{{\small \AFLOWHULLtitle}}
\def\AFLOWXTALMATCH{{\small AFLOW-XTAL-MATCH}}

\def\AFLUX{{\small AFLUX}}
\def\LUX{{\small LUX}}
\def\AFLOWSYM{{\small AFLOW-SYM}}
\def\QHAAPL{{\small QHA-APL}}
\def\AAPL{{\small AAPL}}
\def\AFLOWML{{\small AFLOW-ML}}

\def\ML{{\small ML}}
\def\EOS{{\small EOS}}
\def\VRH{{\small VRH}}
\def\RESTAPI{{\small REST-API}}
\def\GIBBS{{\small GIBBS}}
\def\RMSrD{{\small RMSrD}}
\def\ICSD{{\small ICSD}}
\def\QMSPR{{\small QMSPR}}
\def\QSAR{{\small QSAR}}
\def\GBDT{{\small GBDT}}
\def\PLMF{{\small PLMF}}
\def\GW{{\small GW}}
\def\DFT{{\small DFT}}
\def\IFC{{\small IFC}}
\def\RMSE{{\small RMSE}}
\def\MAE{{\small MAE}}
\def\AUC{{\small AUC}}
\def\ROC{{\small ROC}}
\def\CCR{{\small CCR}}
\def\MOF{{\small MOF}}
\def\PDF{{\small PDF}}
\def\URL{{\small URL}}
\def\PAW{{\small PAW}}
\def\PBE{{\small PBE}}
\def\GGA{{\small GGA}}
\def\LDA{{\small LDA}}
\def\ASM{{\small ASM}}
\def\QHULL{{\small Qhull}}
\def\UNIX{{\small UNIX}}
\def\JSON{{\small JSON}}
\def\KPPRA{{\small KPPRA}}
\def\API{{\small API}}
\def\TDDFT{{\small TDDFT}}

\def\POSCAR{{\small POSCAR}}
\def\PARTCAR{{\small PARTCAR}}

\def\LIBONE{{\small LIB1}}
\def\LIBTWO{{\small LIB2}}
\def\LIBTHREE{{\small LIB3}}
\def\AURL{{\small AURL}}
\def\AUID{{\small AUID}}

\def\WmK{{\small (W\,m$^{-1}$K$^{-1}$)}}
\def\K{{\small (K)}}
\def\GPa{{\small (GPa)}}

\def\acoustic{{\mathrm{a}}}

\def\EXP{{\mathrm{exp}}}

\def\sDebye{{\substack{\scalebox{0.6}{D}}}}
\def\sD{{\substack{\scalebox{0.6}{D}}}}
\def\sDFT{{\substack{\scalebox{0.6}{DFT}}}}
\def\sAGL{{\substack{\scalebox{0.6}{AGL}}}}
\def\sAEL{{\substack{\scalebox{0.6}{AEL}}}}
\def\sMP{{\substack{\scalebox{0.6}{MP}}}}
\def\sBM{{\substack{\scalebox{0.6}{BM}}}}
\def\sBCN{{\substack{\scalebox{0.6}{BCN}}}}
\def\sVIN{{\substack{\scalebox{0.6}{Vinet}}}}
\def\sVRH{{\substack{\scalebox{0.6}{VRH}}}}
\def\sVoigt{{\substack{\scalebox{0.6}{Voigt}}}}
\def\sReuss{{\substack{\scalebox{0.6}{Reuss}}}}
\def\sStatic{{\substack{\scalebox{0.6}{Static}}}}

\def\svib{{\substack{\scalebox{0.6}{vib}}}}
\def\sopt{{\substack{\scalebox{0.6}{opt}}}}
\def\sL{{\substack{\scalebox{0.6}{L}}}}
\def\sT{{\substack{\scalebox{0.6}{T}}}}
\def\sB{{\substack{\scalebox{0.6}{B}}}}
\def\sS{{\substack{\scalebox{0.6}{S}}}}
\def\sV{{\substack{\scalebox{0.6}{V}}}}
\def\sGGA{{\substack{\scalebox{0.6}{GGA}}}}
\def\sLDA{{\substack{\scalebox{0.6}{LDA}}}}
\def\sXC{{\substack{\scalebox{0.6}{XC}}}}
\def\sBG{{\substack{\scalebox{0.6}{BG}}}}
\def\sP{{\substack{\scalebox{0.6}{P}}}}
\def\sp{{\substack{\scalebox{0.6}{p}}}}
\def\sa{{\substack{\scalebox{0.6}{a}}}}
\def\sbond{{\substack{\scalebox{0.6}{bond}}}}
\def\svapor{{\substack{\scalebox{0.6}{vapor}}}}
\def\satom{{\substack{\scalebox{0.6}{atom}}}}
\def\sfusion{{\substack{\scalebox{0.6}{fusion}}}}
\def\smolar{{\substack{\scalebox{0.6}{molar}}}}
\def\scov{{\substack{\scalebox{0.6}{cov}}}}
\def\seff{{\substack{\scalebox{0.6}{eff}}}}

\DeclareFontEncoding{LGR}{}{}
\DeclareTextSymbol{\~}{LGR}{126}

\def\citeAFLOW{\cite{curtarolo:art49,curtarolo:art53,curtarolo:art57,aflowBZ,curtarolo:art63,aflowPAPER,curtarolo:art85,curtarolo:art110,monsterPGM,aflowANRL,aflowPI}}
\def\citeAFLOWLIB{\cite{aflowlibPAPER,aflowAPI,curtarolo:art104,aflux}}

\def\citeVASP{\cite{vasp_prb1996}}
\def\citeICSD{\cite{ICSD,ICSD3}}

\newcommand\PLMFelectronicTotal{26,674}
\newcommand\PLMFinsulatorTotal{13,812}
\newcommand\PLMFmetalTotal{12,862}
\newcommand\PLMFthermoTrainingTotal{2,829}
\newcommand\PLMFthermoTestTotal{770}

\def\AFLOWVERSION{3.1.207}
\def\CHULLCountPromisingTernaries{17}
\def\CHULLCountBinaryHulls{493}
\def\CHULLCountTernaryHulls{873}

\DeclareFixedFont{\ttb}{T1}{txtt}{bx}{n}{8}
\DeclareFixedFont{\ttm}{T1}{txtt}{m}{n}{8}

\definecolor{tmrwBlue}{rgb}{0.259,0.443,0.68.2}
\definecolor{tmrwRed}{rgb}{0.784,0.157,0.161}
\definecolor{tmrwGreen}{rgb}{0.443,0.549,0}
\definecolor{tmrwPurple}{rgb}{0.537,0.349,0.659}
\definecolor{tmrwAqua}{rgb}{0.243,0.6,0.624}
\definecolor{tmrwYellow}{rgb}{0.918,0.718,0}
\definecolor{tmrwOrange}{rgb}{0.871,0.576,0.373}
\definecolor{tmrwComment}{rgb}{0.557,0.565,0.549}

\newcommand\pythonstyle{\lstset{
language=Python,
basicstyle=\ttm,
otherkeywords={__init__, self},
keywordstyle=\ttb\color{tmrwBlue},
emph={and,break,class,continue,def,yield,del,elif,else,
except,exec,finally,for,from,global,if,import,as,
lambda,not,or,pass,print,raise,return,try,while,assert,with},
emphstyle=\ttb\color{tmrwPurple},
emph={[2]},
emphstyle=[2]\ttb\color{tmrwBlue},
emph={[3] in },
emphstyle=[3]\ttb\color{tmrwAqua},
emph={[4]object,type,list,set,len,dict,tuple,str,repr,int,float},
emphstyle=[4]\ttb\color{tmrwYellow},
emph={[5]aflow_hull, pprint, CHull, json, subprocess, os, aflow_command, get_hull, get_distance_to_hull, get_stability_criterion, get_hull_energy, loads, path, realpath, join},
emphstyle=[5]\ttm\color{tmrwBlue},
emph={[6]True, False, None},
emphstyle=[6]\ttm\color{tmrwOrange},
emph={[7]self},
emphstyle=[7]\ttm\color{tmrwRed},
stringstyle=\color{tmrwGreen},
morecomment=[s]{"""}{"""},
commentstyle=\color{tmrwComment}\ttm,
literate=
 {-}{{{-}}}1
 {<}{{{<}}}1,
frame=tb,
breaklines=true,
postbreak=\mbox{\textcolor{tmrwRed}{$\hookrightarrow$}\space},
showstringspaces=false
}}

\lstnewenvironment{python}[1][]
{
  \pythonstyle
  \lstset{#1}
}
{}

\newcommand\pythoninline[1]{{\pythonstyle\lstinline!#1!}}

\lstdefinelanguage{mylang}{
  basicstyle=\ttfamily,
  alsoletter=0123456789,
  alsodigit={.-}
}
\setlist[itemize]{noitemsep, topsep=0pt}
\setlist[itemize,1]{label=--,leftmargin=1em}
\newlist{myitemize}{itemize}{3}
\setlist[myitemize,1]{label=\textbullet,leftmargin=1em}
\setlist[myitemize,2]{label=--,leftmargin=1em}
\setlist[myitemize,3]{label=$\diamond$,leftmargin=1em}
\setlist[myitemize]{noitemsep, topsep=0pt}

\newcolumntype{L}[1]{>{\raggedright\arraybackslash}p{#1} }
\newcolumntype{C}[1]{>{\centering  \arraybackslash}p{#1} }
\newcolumntype{R}[1]{>{\raggedleft \arraybackslash}p{#1} }
\author{Corey Oses}
\advisor{Stefano Curtarolo}
\member{Laurens E. Howle}
\member{Cormac H. Toher}
\member{Donald W. Brenner}
\department{Mechanical Engineering and Materials Science}
\title{Machine Learning, Phase Stability, and Disorder with the Automatic Flow Framework for Materials Discovery}
\begin{document}
\hypersetup{pageanchor=false}
\maketitle
\hypersetup{pageanchor=true}
\makeabstract
\Copyright
\abstract

Traditional materials discovery approaches --- relying primarily on laborious experiments ---
have controlled the pace of technology.
Instead, computational approaches offer an accelerated path:
high-throughput exploration and characterization of virtual structures.
These ventures, performed by automated \abinitio\ frameworks, have rapidly expanded
the volume of programmatically-accessible data, cultivating opportunities for data-driven approaches.
Herein, a collection of robust characterization methods are presented, implemented within
the Automatic Flow Framework for Materials Discovery (\AFLOW), that
leverages materials data for
the prediction of phase diagrams and properties of disordered materials.
These methods directly address the issue of materials synthesizability, bridging
the gap between simulation and experiment.
Powering these predictions is the \AFLOWorg\ repository for inorganic crystals, the largest and
most comprehensive database of its kind, containing more than 2 million compounds
with about 100 different properties computed for each.
As calculated with standardized parameter sets,
the wealth of data also presents a favorable learning environment.
Machine learning algorithms are employed for property prediction,
descriptor development, design rule discovery, and the identification of candidate functional materials.
When combined with physical models and intelligently formulated descriptors,
the data becomes a powerful tool, facilitating the discovery of new materials for applications ranging from
high-temperature superconductors to thermoelectrics.
These methods have been validated by the synthesis of two new permanent magnets introduced herein --- the first
discovered by computational approaches.
\clearpage
\section*{Dedication}

This thesis is dedicated to my parents, Sonia and Oscar, and my brother John Paul.
Their love and support are everything to me, and this milestone is as much a celebration
of accomplishments as it is a testament to the strength of family.
\clearpage
\acknowledgements

These past five years have been an incredible privilege for, if nothing else,
the extraordinary individuals with whom I met and worked.
Day in and day out, my group members challenge, encourage, and support
me, and I am thankful for their companionship throughout this journey:
David Hicks,
Demet Usanmaz,
Eric Gossett,
Pinku Nath,
Marco Esters,
Rico Friedrich,
Pranab Sarker,
Denise Ford,
Pauline Colinet,
Camilo Calderon,
Carlo De Santo,
Geena Gomez,
Harvey Shi,
Allison Stelling,
Yoav Lederer,
Luis Agapito,
and
Manuela Damian.

I am fortunate to have so many mentors and guides who, in various
capacities, have opened my eyes to new fields and fueled my scientific curiosity:
Cormac Toher,
Frisco Rose,
Olexandr Isayev,
Kesong Yang,
Stefano Sanvito,
Amir Natan,
Michael Mehl,
Patrick McGuire,
Jes\'{u}s Carrete,
Natalio Mingo,
Matthias Scheffler,
Claudia Draxl,
Valentin Stanev,
Ichiro Takeuchi,
and
Ohad Levy.

I am incredibly appreciative of my professors, teachers,
and advisors from Cornell University and Bloomfield High School
who helped me construct the vision in which this milestone is achievable:
Sara Xayarath Hern\'{a}ndez,
Joel Brock,
Ernest Fontes,
Kenneth Card,
Daniel Di Domenico,
Marian Connolly,
Brian Miller,
Lou Cappello,
and
Manuela Gonnella.

Above all, I am especially grateful to my PhD Advisor, Stefano Curtarolo,
for the opportunity to discover my passion in this field.
I have been blessed with many in my life who believe in me, but few as
emphatically as Stefano.
``\textit{Non ducor, duco}''.

Finally, I acknowledge support from the National Science Foundation Graduate Research Fellowship under Grant No. DGF1106401.

\clearpage
\tableofcontents
\listoffigures
\listoftables
\chapters
\chapter{Introduction}

\begin{center}
``\textit{Nihil est in intellectu quod non sit prius in sensu}''\footnote{``Nothing is in the intellect that was not first in the senses.''} \\
---  Thomas Aquinas's \textit{Quaestiones Disputatae de Veritate}, \\ quaestio 2, articulus 3, argumentum 19.
\end{center}

Materials discovery drives technological innovation, spanning
the stones and simple metals that forged the first tools to the semiconductors that power today's computers.
Historically, these advancements follow from intuition and
serendipity~\cite{curtarolo:art94,curtarolo:art124,MGI,Norman_RPP_2016,Eberhart_NMat_2004}.
As such, major breakthroughs --- which are few and far between --- are seldom predictable.
Fortunately, ``\textit{big data}'' is powering a paradigm shift:
materials informatics.
Integration of data-centric approaches in an otherwise \textit{a posteriori} field promises to
bridge the widening gap between observation and understanding,
accelerating the pace of technology.
More importantly, data-driven modeling --- offering predictions grounded in empirical evidence ---
may finally break with tradition, enabling control over discovery and
achieving rational materials design.

Wielding data to accelerate innovation is not a new idea,
since it constitutes standard practice in biology~\cite{Reichhardt_Nature_1999,Luscombe_MIM_2001}
and chemistry~\cite{Brown_Chemoinformatics_1998,Gasteiger_Chemoinformatics_2003}.
Yet its adoption in materials science has been slow, as it was first introduced in the early 2000's~\cite{curtarolo:art13}.
This delay can be attributed to the ongoing development of standard \abinitio\
packages~\cite{kresse_vasp,VASP4_2,vasp_cms1996,vasp_prb1996,quantum_espresso_2009,gonze:abinit,Blum_CPC2009_AIM},
particularly to better address calculation of the exchange correlation energy~\cite{PBE,Perdew_SCAN_PRL_2015}.
Nevertheless, the impact of density functional theory (\DFT) on computational materials science cannot be understated~\cite{nmatHT},
offering a reasonable compromise between cost and accuracy~\cite{Haas_PRB_2009}.
The success of these implementations has stimulated the rapid development
of automated frameworks and corresponding data repositories,
including \AFLOW\ (\underline{A}utomatic \underline{Flow} for Materials Discovery)~\cite{aflowPAPER,curtarolo:art110,curtarolo:art85,curtarolo:art63,aflowBZ,curtarolo:art57,curtarolo:art53,curtarolo:art49,monsterPGM,aflowANRL,aflowPI},
Novel Materials Discovery Laboratory~\cite{nomad},
Materials Project~\cite{APL_Mater_Jain2013},
Open Quantum Materials Database~\cite{Saal_JOM_2013},
Computational Materials Repository~\cite{cmr_repository},
and Automated Interactive Infrastructure and Database for Computational Science~\cite{Pizzi_AiiDA_2016}.
These house an abundance of materials data.
For instance, the \AFLOW\ framework, described in Section~\ref{sec:aflow_chp},
has characterized more than 2 million compounds, each by about
100 different properties accessible via the \AFLOWorg\ online database~\cite{aflowlibPAPER,aflowAPI,curtarolo:art104,aflux}.
Investigations employing this data have not only led to advancements in modeling
electronics~\cite{nmatTI,curtarolo:art94,curtarolo:art124,ceder:nature_1998},
thermoelectrics~\cite{curtarolo:art96,curtarolo:art114,curtarolo:art115,curtarolo:art119,curtarolo:art120,curtarolo:art125,curtarolo:art129},
superalloys~\cite{curtarolo:art113},
and metallic glasses~\cite{curtarolo:art112},
but also to the synthesis of two new magnets --- the first
discovered by computational approaches~\cite{curtarolo:art109}.

Further advancements are contingent on continued development and expansion of these materials repositories.
New entries are generated both by
\textbf{i.} calculating the properties of previously observed compounds
from sources such as the Inorganic Crystal Structure Database~\cite{ICSD} (\ICSD),
and
\textbf{ii.} decorating structure prototypes~\cite{aflowANRL,curtarolo:art130}.
Considering all possible crystals of different arrangements and decorations~\cite{curtarolo:art124,Walsh_NChem_2015},
the analysis of existing structures --- a small subset --- is a critical first-step in determining fruitful directions for exploration.
For example, Section~\ref{sec:art130} presents a general overview of the structure types appearing in an important
class of the solid compounds, \ie, binary and ternary compounds of the 6A column oxides, sulfides, and selenides.
It contains an in-depth statistical analysis of these compounds, including the prevalence of various structure types,
their symmetry properties, compositions, stoichiometries and unit cell sizes.
Results reveal that these compound families include preferred stoichiometries and structure
types that may reflect both their specific chemistry and research bias in the
available empirical data.
Detection of non-overlapping gaps and missing stoichiometries in such
populations will guide subsequent studies: structures are avoided in the event that they are chemically
unfavorable, or targeted to complement existing measurements.

With materials of interest identified, accurate computation of their properties demands
a set of reliable calculation parameters/thresholds~\cite{curtarolo:art104}.
These inputs need to be understood by researchers, and should be reported by the originators
to ensure reproducibility and enable collaborative database expansion.
As described in Section~\ref{sec:art104}, the \AFLOW\ Standard defines these parameters
for high-throughput electronic structure calculations of crystals --- the basis for all \AFLOW\ characterizations.
Standard values are established for reciprocal space grid density,
plane wave basis set kinetic energy cut-off, exchange-correlation
functionals, pseudopotentials, \DFT$+U$ parameters, and convergence criteria.

Exploration of more complex properties~\cite{curtarolo:art96,curtarolo:art115} and materials~\cite{curtarolo:art110,curtarolo:art112}
typically warrants advanced (and expensive) characterization techniques~\cite{Hedin_GW_1965,GW,ScUJ,Malashevich_GW_TiO_PRB2014,Patrick_GW_TiO2_JPCM2012}.
Fortunately, state-of-the-art workflows~\cite{curtarolo:art96,curtarolo:art110,curtarolo:art115} and
careful descriptor development~\cite{curtarolo:art112} have
enabled experimentally-validated modeling within a \DFT\ framework.
For instance, a thorough description of thermomechanical properties
requires difficult and time-consuming experiments.
This limits the availability of data:
one of the main obstacles for
the development of effective accelerated materials design strategies.
Section~\ref{sec:art115} introduces an automated, integrated workflow with robust error-correction
within the \AFLOW\ framework {that combines} the newly devised
``Automatic Elasticity Library'' with the previously implemented \GIBBS\ method~\cite{curtarolo:art96}.
The former extracts the mechanical properties from several automatic self-consistent stress-strain calculations,
while the latter employs those mechanical properties to evaluate the thermodynamics within the Debye model.
{The} thermomechanical {workflow} is benchmarked against a set of
74 experimentally characterized systems to pinpoint a
robust computational methodology for the evaluation of bulk and shear moduli,
Poisson ratios, Debye temperatures, Gr{\"u}neisen parameters, and thermal conductivities of a wide variety of materials.
The effect of different choices of equations of state {and exchange-correlation functionals}
is examined and the optimum combination of properties for the
Leibfried-Schl{\"o}mann prediction of thermal conductivity is identified,
leading to improved agreement with experimental results compared to the \GIBBS-only approach.
The \AEL-\AGL\ framework has been applied to the \AFLOWorg\ data repositories to compute the thermomechanical properties
of over 5,000 unique materials.

Similar to thermomechanical characterizations,
descriptions of thermodynamic stability and
structural/chemical disorder are also resolved through an analysis of aggregate sets of \abinitio\ calculations.
\textit{A priori} prediction of phase stability
requires
knowledge of all energetically-competing structures at formation conditions.
Large materials repositories
offer a path to prediction through the construction of
\abinitio\ phase diagrams, \ie, the convex hull
at a given temperature/pressure.
However, limited access to relevant data and software infrastructure has
rendered thermodynamic characterizations largely peripheral,
despite their continued success in dictating synthesizability.
In Section~\ref{sec:art146}, a new module is presented for autonomous thermodynamic stability analysis
implemented within \AFLOW.
Powered by the \AFLUX\ Search-\API, \AFLOWHULL\ leverages data of more than
2 million compounds characterized in the \AFLOWorg\ repository,
and can be employed locally from any \UNIX-like computer.
This module integrates a range of functionality:
the identification of stable phases and equivalent structures, phase coexistence,
measures for robust stability, and determination of decomposition reactions.
As a proof-of-concept, thermodynamic characterizations have been performed
for more than 1,300 binary and ternary systems, enabling the identification of several
candidate phases for synthesis based on their relative stability criterion --- including
\CHULLCountPromisingTernaries\ promising $C15_{b}$-type structures and two half-Heuslers.
In addition to a full report included herein, an interactive online web application
has been developed, showcasing the results of the analysis, and is
located at {\sf aflow.org/aflow-chull}.

The convex hull construction has fueled the generation of
novel descriptors for glass forming ability~\cite{curtarolo:art112} and,
more generally, modeling structurally disordered systems.
Statistical methods are employed to address chemically disordered
structures, where system-wide properties are resolved through an analysis of
representative ordered supercells~\cite{curtarolo:art109}.
Incorporating the effects of disorder is a necessary, albeit difficult, step in materials modeling.
Not only is disorder intrinsic to all materials,
but it also offers a route to enhanced and even otherwise inaccessible functionality,
as demonstrated by its ubiquity in technological applications.
Prominent examples include glasses~\cite{kelton1991crystal,kelton2010nucleation,kelton1998new},
superalloys~\cite{Donachie_ASM_2002},
fuel cells~\cite{Xie_ACatB_2015},
high-temperature superconductors~\cite{Bednorz_ZPBCM_1986,Maeno_Nature_1994},
and low thermal conductivity thermoelectrics~\cite{Winter_JACerS_2007}.

Predicting material properties of chemically disordered systems remains a
formidable challenge in rational materials design.
A proper analysis of such systems by means of a supercell approach requires
consideration of all possible superstructures, which can be a time-consuming process.
On the contrary, the use of quasirandom-approximants, while
computational effective, implicitly bias the analysis toward disordered states with the lowest site correlations.
In Section~\ref{sec:art110}, a novel framework is proposed
to investigate stoichiometrically driven trends of disordered systems
(\ie, having partial occupation and/or disorder in the atomic sites).
At the heart of the approach is the identification and analysis of unique supercells of a
virtually equivalent stoichiometry to the disordered material.
Boltzmann statistics are employed to resolve system-wide properties at a high-throughput level.
To maximize efficiency and accessibility, this method has been integrated within \AFLOW.
As proof of concept, the approach is applied to three systems of interest,
a zinc chalcogenide (ZnS$_{1-x}$Se$_x$),
a wide-gap oxide semiconductor (Mg$_{x}$Zn$_{1-x}$O),
and an iron alloy (Fe$_{1-x}$Cu$_{x}$)
at various stoichiometries.
These systems exhibit properties that are highly tunable as a function of composition,
characterized by optical bowing and linear ferromagnetic behavior.
Not only are these qualities
predicted, but additional insight into underlying physical mechanisms is revealed.

The aforementioned frameworks --- offering characterizations of thermomechanical
and thermodynamic properties, as well as resolving features of disordered systems ---
have both benefited from and stimulated the development of the \AFLOWorg\ repository.
The combination of plentiful and diverse materials data~\cite{aflowlibPAPER,aflowAPI,curtarolo:art104,aflux}
and its programmatic accessibility~\cite{aflowAPI,aflux} also
justify the application of data-mining techniques.
These methods can
resolve subtle trends and correlations among materials and their
properties~\cite{curtarolo:art94,Ghiringhelli_PRL_2015,curtarolo:art124,curtarolo:art129,curtarolo:art135},
as well as motivate the formulation of novel property descriptors~\cite{curtarolo:art112,curtarolo:art139}.
In fact, materials data generated by automated frameworks are
conducive to such approaches,
where strict standardizations of calculation parameters~\cite{curtarolo:art104} not only ensure
reproducibility, but also a minimum accuracy threshold.
Errors from approximations or choice in parameters can therefore be treated as systematic,
which are easily identified and rectified by \underline{m}achine \underline{l}earning (\ML) algorithms.
Models have been generated for predicting electronic~\cite{curtarolo:art124},
thermomechanical~\cite{curtarolo:art124,deJong_SR_2016} and vibrational~\cite{curtarolo:art120,curtarolo:art129} properties,
as well as the thermodynamic stability of both ordered~\cite{Ghiringhelli_PRL_2015}
and disordered~\cite{Ward_ML_GFA_NPGCompMat_2016} phases.
In Section~\ref{sec:art124}, data from the \AFLOW\ repository for \abinitio\ calculations
is combined with Quantitative Materials Structure-Property Relationship models to predict important properties:
metal/insulator classification, band gap energy, bulk/shear moduli, Debye temperature, and heat capacities.
The prediction's accuracy compares well with the quality of the training data for virtually any
stoichiometric inorganic crystalline material, reciprocating the available thermomechanical experimental data.
The universality of the approach is attributed to the construction of the descriptors: Property-Labeled Materials Fragments.
The representations require only minimal structural input allowing straightforward implementations of simple heuristic design rules.

\ML\ approaches are expected to become indispensable in two specific scenarios, prediction
of complex properties and screening of large sets of materials.
For example, feature-importance analyses have informed on the interactions that elicit high-temperature
superconductivity~\cite{curtarolo:art94,curtarolo:art137}, an
elusive phenomenon in which the driving mechanisms are still contested.
Superconductivity has been the focus of enormous research efforts since its discovery more than a century ago.
Yet, some features
remain poorly understood; mainly the connection
between superconductivity and chemical/structural properties of materials.
To bridge the gap, several machine learning schemes are developed in Sections~\ref{sec:art094} and \ref{sec:art137}
to model the critical temperatures $\left(T_{\mathrm{c}}\right)$ of
known superconductors available via the SuperCon database.
As expected, these analyses suggest distinct mechanisms are responsible for driving superconductivity in
different classes of materials.
However, they also hint at very complex physical interactions.
Fortunately, \ML\ algorithms like random forests~\cite{randomforests} are
capable of extracting very complicated functional relationships.
In the case of predicting $T_{\mathrm{c}}$, these ``black-box'' models are
quite valuable as
few alternative practical modeling schemes exist.

In Section~\ref{sec:art094}, novel analytical approaches are introduced based on structural and electronic materials fingerprints
and applied to predict the $T_{\mathrm{c}}$
of known superconductors.
The framework is employed to \textbf{i.} query large databases of materials using similarity concepts,
\textbf{ii.} map the connectivity of materials space (\ie, as materials cartograms)
for rapidly identifying regions with unique organizations/properties,
and \textbf{iii.} develop predictive Quantitative Materials Structure-Property Relationship models for guiding materials design.
The materials fingerprinting and cartography approaches are
effective computational tools to analyze,
visualize, model, and design new materials.

Superconductors are revisited in a much more in-depth study presented in Section~\ref{sec:art137},
leveraging the full set of the $12,000+$ materials in the SuperCon database.
Materials are first divided into two classes based on their $T_{\mathrm{c}}$ values,
above and below $10$~K,
and a classification model predicting this label is trained.
The model uses coarse-grained features based only on the chemical compositions.
It shows strong predictive power, with out-of-sample accuracy of about $92\%$.
Separate regression models
are developed to predict the values of $T_{\mathrm{c}}$ for cuprate, iron-based, and low-$T_{c}$ compounds.
These models also demonstrate good performance,
with learned predictors offering
insights into the mechanisms behind superconductivity in different families of materials.
To improve the accuracy and interpretability of these models,
new features are incorporated using materials data
from the \AFLOWorg\ repository.
To find potential new superconductors, the classification and regression models are combined into a single integrated pipeline
and employed to search the entire Inorganic Crystallographic Structure Database (\ICSD).
More than 30 non-cuprate and non-iron-based oxides are selected as candidate materials.

Beyond superconductors, \ML\ models are created to predict properties of thermoelectrics (Section~\ref{sec:art120})
and permanent magnets (Section~\ref{sec:art109}).
Thermoelectric materials generate an electric voltage when subjected to a temperature gradient, or
conversely create
a temperature gradient when a voltage is applied~\cite{snyder_complex_2008, nolas_thermoelectrics:_2001}.
With no moving parts and their resulting scalability, thermoelectrics
have potential applications in power generation for spacecraft,
energy recovery from waste heat in automotive and industrial facilities~\cite{bell_cooling_2008, disalvo99},
and spot cooling for nanoelectronics using the Peltier cooling effect~\cite{bell_cooling_2008, disalvo99}.
However, most of the available thermoelectric materials have low efficiency, only converting a few percent
of the available thermal energy into electricity.
Therefore, a major goal of thermoelectrics research is to develop new materials that have
higher thermoelectric efficiency as determined by a figure of merit~\cite{snyder_complex_2008, nolas_thermoelectrics:_2001}.
The metric is dependent on quantities such as the Seebeck coefficient and electrical/thermal conductivities.
One promising path to optimizing the figure of method is to
minimize the lattice thermal conductivity.

In Section~\ref{sec:art120}, the thermal conductivity $\left(\kappa\right)$ is analyzed for
semiconducting oxides and fluorides with cubic perovskite structures.
Using finite-temperature phonon calculations and \ML\ methods,
the mechanical stability of about 400 structures is resolved at 0~K, 300~K, and 1000~K.
Of these, 92 compounds are determined to be mechanically stable at high temperatures
--- including 36 not mentioned in the literature so far --- for which $\kappa$ is calculated.
Several trends are revealed, including
\textbf{i.} $\kappa$ generally being smaller in fluorides than in oxides,
largely due to the lower ionic charge,
and \textbf{ii.} $\kappa$ decreasing more
slowly than the usual $T^{-1}$ behavior for most cubic perovskites.
Analyses expose the simple structural descriptors that correlate with $|\kappa|$.
This set is also screened for materials exhibiting negative thermal expansion.
The study highlights a general strategy coupling force constants calculations with an iterative \ML\ scheme
to accelerate the discovery of mechanically stable compounds at high temperatures.

The role of \ML\ models in predicting magnetic properties is of particular significance,
as their \textit{a priori} predictions were validated with the discovery of two new magnets.
Magnetic materials underpin modern technologies, ranging from data storage to energy conversion and contactless sensing.
However, the development of a new high-performance magnet is a long and often unpredictable process, and only
about two dozen feature in mainstream applications.
In Section~\ref{sec:art109}, a systematic pathway is described to the discovery of novel
magnetic materials.
Based on an extensive electronic structure library of Heusler alloys containing 236,115 compounds,
alloys displaying magnetic order are selected, and it is determined
whether they can be fabricated at thermodynamic equilibrium.
Specifically, a full stability analysis is carried out for intermetallic Heusler alloys made only of transition metals.
Among the possible 36,540 candidates, 248 are found to be thermodynamically stable but only 20 are magnetic.
The magnetic ordering temperature, $T_\mathrm{C}$, has then been estimated by a regression
calibrated on the experimental $T_\mathrm{C}$ of about 60 known compounds.
As a final validation, the synthesis is attempted for a few of the predicted compounds,
and two new magnets are produced.
One, Co$_2$MnTi, displays a remarkably high $T_\mathrm{C}$ in perfect agreement with
the predictions, while the other, Mn$_2$PtPd, is an antiferromagnet.
This work paves the way for large-scale design of novel magnetic materials at unprecedented speed.

Overall, data-driven approaches have extended materials modeling capabilities within a \DFT\ framework.
Descriptors for thermodynamic stability and formation/features of disordered materials
are accessible through analyses of ensembles of ordered structures,
stimulating the development of large materials repositories.
To match the growth of these databases, insight-extraction must also be automated.
\ML\ methods are employed to reveal structure-property relationships and expose similarities among materials.
Ultimately, the power in \ML\ lies in the speed of its predictions, which out-paces
\DFT\ calculations by orders of magnitude~\cite{Isayev_ChemSci_2017}.
Efforts to explore the full materials space through brute-force \DFT\
calculations are impractical;
studies conservatively enumerate the size
of possible hypothetical structures to be as large as 10$^{100}$~\cite{Walsh_NChem_2015}.
Given that the number of currently characterized materials pales in comparison
to the true potential diversity, methods --- like those presented here ---
to filter/screen the most interesting candidate materials
will play an integral role in future materials discovery workflows.
\clearpage
\chapter{The Automatic Flow Framework for Materials Discovery}
Materials informatics requires large repositories of materials data to identify trends in and correlations between materials properties,
as well as for training machine learning models.
Such patterns lead to the formulation of descriptors that guide rational materials design.
Generating large databases of computational materials properties requires robust, integrated, automated frameworks~\cite{nmatHT}.
Built-in error correction and standardized parameter sets enable the production and analysis of data without direct intervention from human researchers.
Current examples of such frameworks include
\AFLOW\ (\underline{A}utomatic \underline{{\small FLOW}})~\cite{aflowPAPER, aflowBZ, aflowlibPAPER, aflowAPI, curtarolo:art104, aflowlib.org, aflow_fleet_chapter, aflowPI, paoflow},
Materials Project~\cite{materialsproject.org, APL_Mater_Jain2013, CMS_Ong2012b, Mathew_Atomate_CMS_2017},
\OQMD\ (\underline{O}pen \underline{Q}uantum \underline{M}aterials \underline{D}atabase)~\cite{Saal_JOM_2013, Kirklin_AdEM_2013, Kirklin_ActaMat_2016},
the Computational Materials Repository~\cite{cmr_repository} and its associated scripting interface {\small ASE} (\underline{A}tomic \underline{S}imulation \underline{E}nvironment)~\cite{ase},
\AIIDA\ (\underline{A}utomated \underline{I}nteractive \underline{I}nfrastructure and \underline{Da}tabase for Computational Science)~\cite{aiida.net, Pizzi_AiiDA_2016, Mounet_AiiDA2D_NNano_2018},
and the Open Materials Database at \verb|httk.openmaterialsdb.se| with its associated \underline{H}igh-\underline{T}hroughput \underline{T}ool\underline{k}it ({\small HTTK}).
Other computational materials science resources include the aggregated repository maintained by the \underline{No}vel \underline{Ma}terials \underline{D}iscovery (\NOMAD) Laboratory~\cite{nomad},
the Materials Mine database available at \verb|www.materials-mine.com|,
and the \underline{T}heoretical \underline{C}rystallography \underline{O}pen \underline{D}atabase ({\small TCOD})~\cite{Merkys_TCOD_2017}.
For this data to be consumable by automated machine learning algorithms,
it must be organized in programmatically accessible repositories~\cite{aflowlibPAPER, aflowAPI, aflowlib.org, materialsproject.org, APL_Mater_Jain2013, Saal_JOM_2013, nomad}.
These frameworks also contain modules that combine and analyze data from various calculations to predict complex thermomechanical phenomena, such as lattice thermal conductivity and mechanical stability.

Computational strategies have already had success in predicting materials for
applications including photovoltaics~\cite{YuZunger2012_PRL},
water-splitters~\cite{CastelliJacobsen2012_EnEnvSci},
carbon capture and gas storage~\cite{LinSmit2012_NMAT_carbon_capture, Alapati_JPCC_2012},
nuclear detection and scintillators~\cite{Derenzo:2011io, Ortiz09, aflowSCINT, curtarolo:art46},
topological insulators~\cite{nmatTI, Lin_NatMat_HalfHeuslers_2010},
piezoelectrics~\cite{Armiento_PRB_2011, Vanderbilt_Piezoelectrics_PRL2012},
thermoelectric materials~\cite{curtarolo:art68, madsen2006, aflowKAPPA, curtarolo:art85},
catalysis~\cite{Norskov09},
and battery cathode materials~\cite{Hautier-JMC2011, Hautier-ChemMater2011, Mueller-ChemMater2011}.
More recently, computational materials data has been combined with machine learning approaches
to predict electronic and thermomechanical properties~\cite{curtarolo:art124, deJong_SR_2016},
and to identify superconducting materials~\cite{curtarolo:art94}.
Descriptors are also being constructed to describe the formation of disordered
materials, and have recently been used to predict the glass forming
ability of binary alloy systems~\cite{curtarolo:art112}.
These successes demonstrate that
accelerated materials design can be achieved by combining structured data sets generated
using autonomous computational methods with intelligently formulated descriptors and machine learning.

\section{Automated computational materials \texorpdfstring{design \\ frameworks}{design frameworks}}
\label{sec:aflow_chp}

Rapid generation of materials data relies on automated frameworks such as
\AFLOW~\cite{aflowPAPER, aflowBZ, aflowlibPAPER, aflowAPI, curtarolo:art104},
Materials Project's \verb|pymatgen|~\cite{CMS_Ong2012b} and \verb|atomate|~\cite{Mathew_Atomate_CMS_2017},
\OQMD~\cite{Saal_JOM_2013, Kirklin_AdEM_2013, Kirklin_ActaMat_2016},
{\small ASE}~\cite{ase}, and \AIIDA~\cite{Pizzi_AiiDA_2016}.
The general automated workflow is illustrated in Figure~\ref{fig:aflow_chp:materials_design_workflow}.
These frameworks begin by creating the input files required by the electronic structure
codes that perform the quantum-mechanics level calculations, where the initial geometry is
generated by decorating structural prototypes (Figure~\ref{fig:aflow_chp:materials_design_workflow}(a, b)).
They execute and monitor these calculations, reading any error messages written to the
output files and diagnosing calculation failures.
Depending on the nature of the errors, these frameworks are equipped with a catalog of prescribed solutions ---
enabling them to adjust the appropriate parameters and restart the calculations (Figure~\ref{fig:aflow_chp:materials_design_workflow}(c)).
At the end of a successful calculation, the frameworks parse the output files to extract the relevant materials data
such as total energy, electronic band gap, and relaxed cell volume.
Finally, the calculated properties are organized and formatted for entry into machine-accessible, searchable and sortable databases.

\fig
\includegraphics[width=1.00\linewidth]{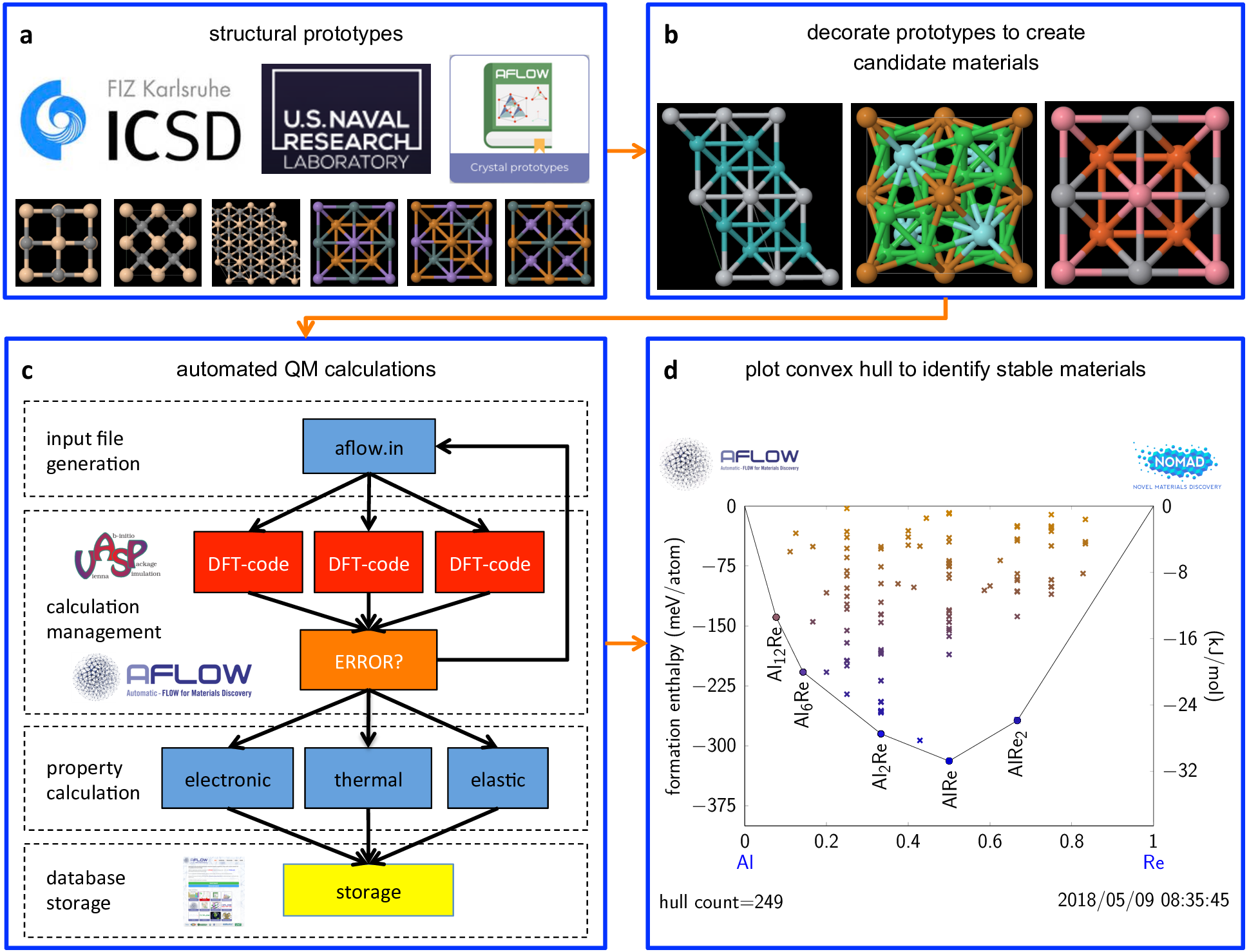}
\mycaption[Computational materials data generation workflow.]
{({\bf a}) Crystallographic prototypes are extracted from databases such as the
\ICSD\ or the NRL crystal structure library, or generated by enumeration algorithms.
The illustrated examples are for the rocksalt, zincblende, wurtzite, Heusler, anti-Heusler and
half-Heusler structures.
({\bf b}) New candidate materials are generated by decorating the atomic sites with different elements.
({\bf c}) Automated \DFT\ calculations are used to optimize the geometric structure and calculate energetic, electronic,
thermal, and elastic properties.
Calculations are monitored to detect errors.
The input parameters are adjusted to compensate for the problem and the calculation is re-run.
Results are formatted and added to an online data repository to facilitate programmatic access.
({\bf d}) Calculated data is used to plot the convex hull phase diagrams for each alloy system to identify stable compounds.}
\label{fig:aflow_chp:materials_design_workflow}
\efig

In addition to running and managing the quantum-mechanics level calculations, the frameworks also
maintain a broad selection of post-processing libraries for extracting additional properties,
such as calculating x-ray diffraction (XRD) spectra from relaxed atomic coordinates, and the
formation enthalpies for the convex hull analysis to identify stable compounds (Figure~\ref{fig:aflow_chp:materials_design_workflow}(d)).
Results from calculations of distorted structures can be combined to calculate
thermal and elastic properties~\cite{aflowPAPER, curtarolo:art96, curtarolo:art100, curtarolo:art115},
and results from different compositions and structural phases can be amalgamated to generate thermodynamic phase diagrams.

\subsection{Generating and using databases for materials discovery}

A major aim of high-throughput computational materials science is to identify new, thermodynamically stable compounds.
This requires the generation of new materials structures, which have not been previously reported in the literature,
to populate the databases. The accuracy of analyses involving sets of structures, such as that used to determine thermodynamic stability,
is contingent on sufficient exploration of the full range of possibilities. Therefore, autonomous materials design frameworks
such as \AFLOW\ use crystallographic prototypes to generate new materials entries consistently and reproducibly.

Crystallographic prototypes are the basic building blocks used to generate the wide range of materials entries involved in
computational materials discovery.
These prototypes are based on \textbf{i.} structures commonly observed in nature~\cite{ICSD, navy_crystal_prototypes, aflowANRL},
such as the rocksalt, zincblende, wurtzite or Heusler structures illustrated in Figure~\ref{fig:aflow_chp:materials_design_workflow}(b),
as well as \textbf{ii.} hypothetical structures, such as those enumerated by the methods described in References~\onlinecite{enum1, enum2}.
The \AFLOW\ Library of Crystallographic Prototypes~\cite{aflowANRL} is also available online at \url{aflow.org/CrystalDatabase/}, where
users can choose from hundreds of crystal prototypes with adjustable parameters, and which can be decorated to generate new input
structures for materials science calculations.

New materials are then generated by decorating the various atomic sites in the crystallographic prototype with different elements.
These decorated prototypes serve as the structural input for \abinitio\ calculations.
A full relaxation of the geometries and energy determination follows, from which phase diagrams for stability analyses can be constructed.
The resulting materials data are then stored in an online data repository for future consideration.

The phase diagram of a given alloy system can be approximated by considering the low-temperature limit in
which the behavior of the system is dictated by the ground state~\cite{monster, monsterPGM}.
In compositional space, the lower-half convex hull defines the minimum energy surface and the
ground-state configurations of the system.
All non-ground-state stoichiometries are unstable, with the decomposition described by the
hull facet directly below it.
In the case of a binary system, the facet is a tie-line as illustrated in Figure~\ref{fig:aflow_chp:convex_hulls}(a).
The energy gained from this decomposition is geometrically represented by the (vertical-)distance of the
compound from the facet and quantifies the excitation energy involved in forming this compound.
While the minimum energy surface changes at finite temperature (favoring disordered structures),
the $T=0$~K excitation energy serves as a reasonable descriptor for relative thermodynamic
stability~\cite{curtarolo:art113}.
This analysis generates valuable information such as ground-state structures,
excitation energies, and phase coexistence for storage in the
online data repository.
This stability data can be visualized and displayed by online modules,
such as those developed by \AFLOW~\cite{curtarolo:art113}, the Materials Project~\cite{Ong_ChemMat_2008},
and the \OQMD~\cite{Akbarzadeh2007, Kirklin_AdEM_2013}.
An example visualization from \AFLOW\ is shown in Figure~\ref{fig:aflow_chp:convex_hulls}(b).

\fig
\includegraphics[width=0.8\linewidth]{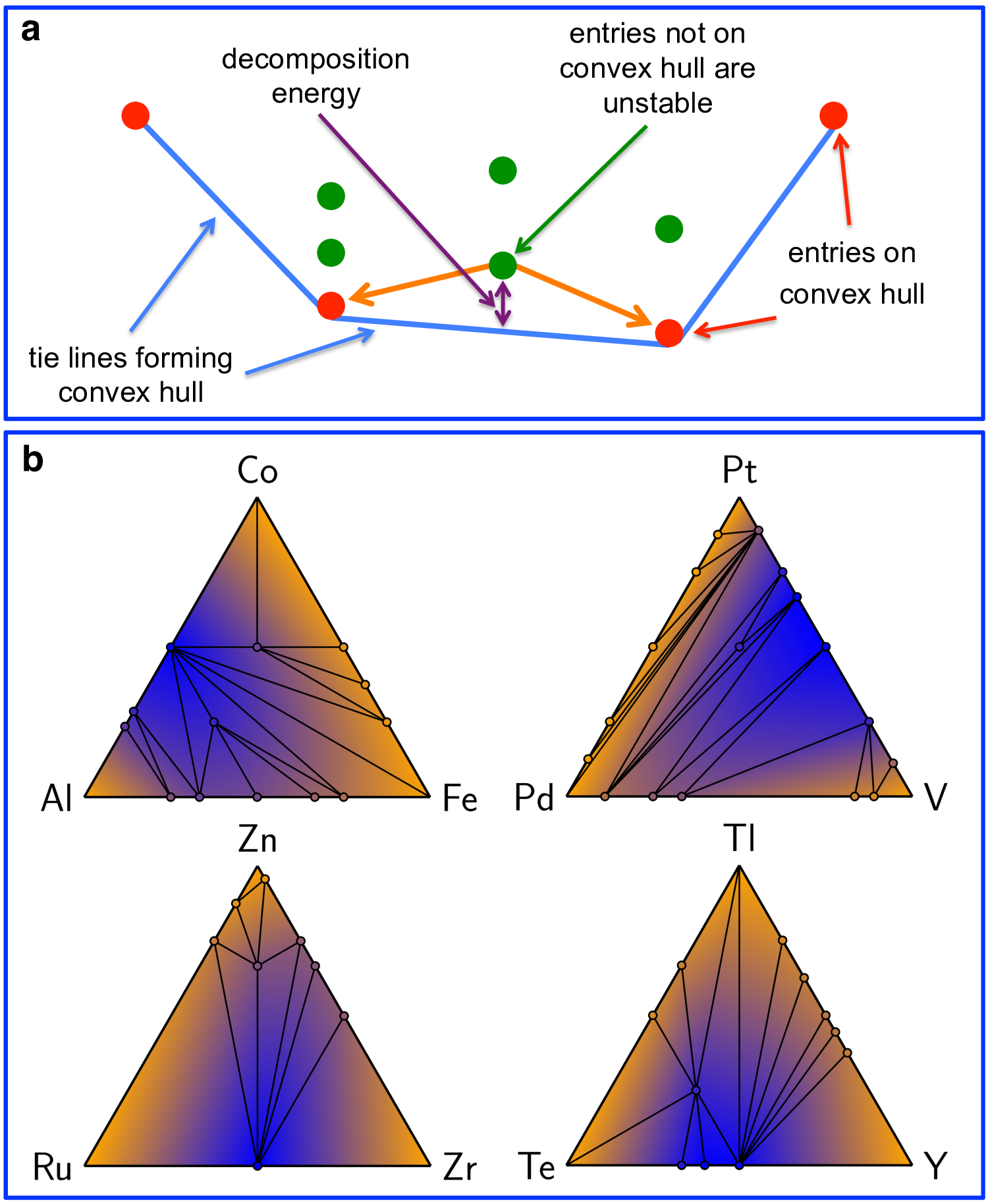}
\mycaption[Convex hull phase diagrams for multicomponent alloys systems.]
{({\bf a}) Schematic illustrating construction of convex hull for a general
binary alloy system $A_{x}B_{1-x}$. Ground state structures are depicted as red points, with the minimum energy
surface outlined with blue lines. The minimum energy surface is formed by
connecting the lowest energy structures with tie lines which form a convex hull.
Unstable structures are shown in green, with the decomposition reaction indicated
by orange arrows, and the decomposition energy indicated in purple.
({\bf b}) Example ternary convex hulls as generated by \AFLOW.}
\label{fig:aflow_chp:convex_hulls}
\efig

Convex hull phase diagrams have been used to discover new thermodynamically
stable compounds in a wide range of alloy systems, including hafnium~\cite{curtarolo:art49, curtarolo:art51},
rhodium~\cite{curtarolo:art53}, rhenium~\cite{curtarolo:art63}, ruthenium~\cite{curtarolo:art67}, and technetium ~\cite{curtarolo:art70}
with various transition metals, as well as the Co-Pt system~\cite{curtarolo:art66}. Magnesium alloy systems such as the lightweight
Li-Mg system~\cite{curtarolo:art55} and 34 other Mg-based systems~\cite{curtarolo:art54} have also been investigated.
This approach has also been used to calculate the solubility of elements in titanium alloys~\cite{curtarolo:art47}, to study the effect of hydrogen
on phase separation in iron-vanadium~\cite{curtarolo:art74}, and to find new superhard tungsten nitride compounds~\cite{curtarolo:art90}.
The data has been employed to generate structure maps for hcp metals~\cite{curtarolo:art57},
as well as to search for new stable compounds with the Pt$_8$Ti phase~\cite{curtarolo:art56},
and with the $L1_1$ and $L1_3$ crystal structures~\cite{curtarolo:art71}.
Note that even if a structure does not lie on the ground state convex hull, this does not rule out its existence.
It may be synthesizable under specific temperature and pressure conditions, and then be metastable under ambient
conditions.

\subsection{Standardized protocols for automated data generation}

Standard calculation protocols and parameters sets~\cite{curtarolo:art104} are essential to
the identification of trends and correlations among materials properties.
The workhorse method for calculating quantum-mechanically resolved materials properties
is \underline{d}ensity \underline{f}unctional \underline{t}heory (\DFT).
\DFT\ is based on the Hohenberg-Kohn theorem~\cite{Hohenberg_PR_1964}, which proves that for a ground state system,
the potential energy is a unique functional of the density: $V (\mathbf{r}) = V(\rho(\mathbf{r}))$.
This allows for the charge density $\rho(\mathbf{r})$ to be used as the central variable for the calculations
rather than the many-body wave function $\Psi(\mathbf{r}_{1}, \mathbf{r}_{2}, ..., \mathbf{r}_{N})$,
dramatically reducing the number of degrees of freedom in the calculation.

The Kohn-Sham equations~\cite{DFT} map the $n$ coupled equations for the system of $n$ interacting particles
onto a system of $n$ independent equations for $n$ non-interacting particles:
\begin{equation}
\label{eq:aflow_chp:kohnshameqns}
\left[ -\frac{\hbar^2}{2m} \nabla^2 + V_s (\mathbf{r}) \right] \phi_i (\mathbf{r}) = \varepsilon_i \phi_i(\mathbf{r}),
\end{equation}
where $\phi_i(\mathbf{r})$ are the non-interacting Kohn-Sham eigenfunctions and $\varepsilon_i$ are their eigenenergies.
$V_s (\mathbf{r})$ is the Kohn-Sham potential:
\begin{equation}
\label{eq:aflow_chp:kohnshampotential}
  V_s (\mathbf{r}) = V(\mathbf{r}) + \int e^2 \frac{\rho_s (\mathbf{r}^{\prime})}{|\mathbf{r} - \mathbf{r}^{\prime}|}
  d^3 \mathbf{r}^{\prime} + V_\sXC\left[\rho_s(\mathbf{r})\right],
\end{equation}
where $V(\mathbf{r})$ is the external potential
(which includes influences of the nuclei, applied fields, and the core electrons when pseudopotentials are used),
the second term is the direct Coulomb potential, and $V_\sXC\left[\rho_s(\mathbf{r})\right]$ is the exchange-correlation term.

\fig
\includegraphics[width=1.00\linewidth]{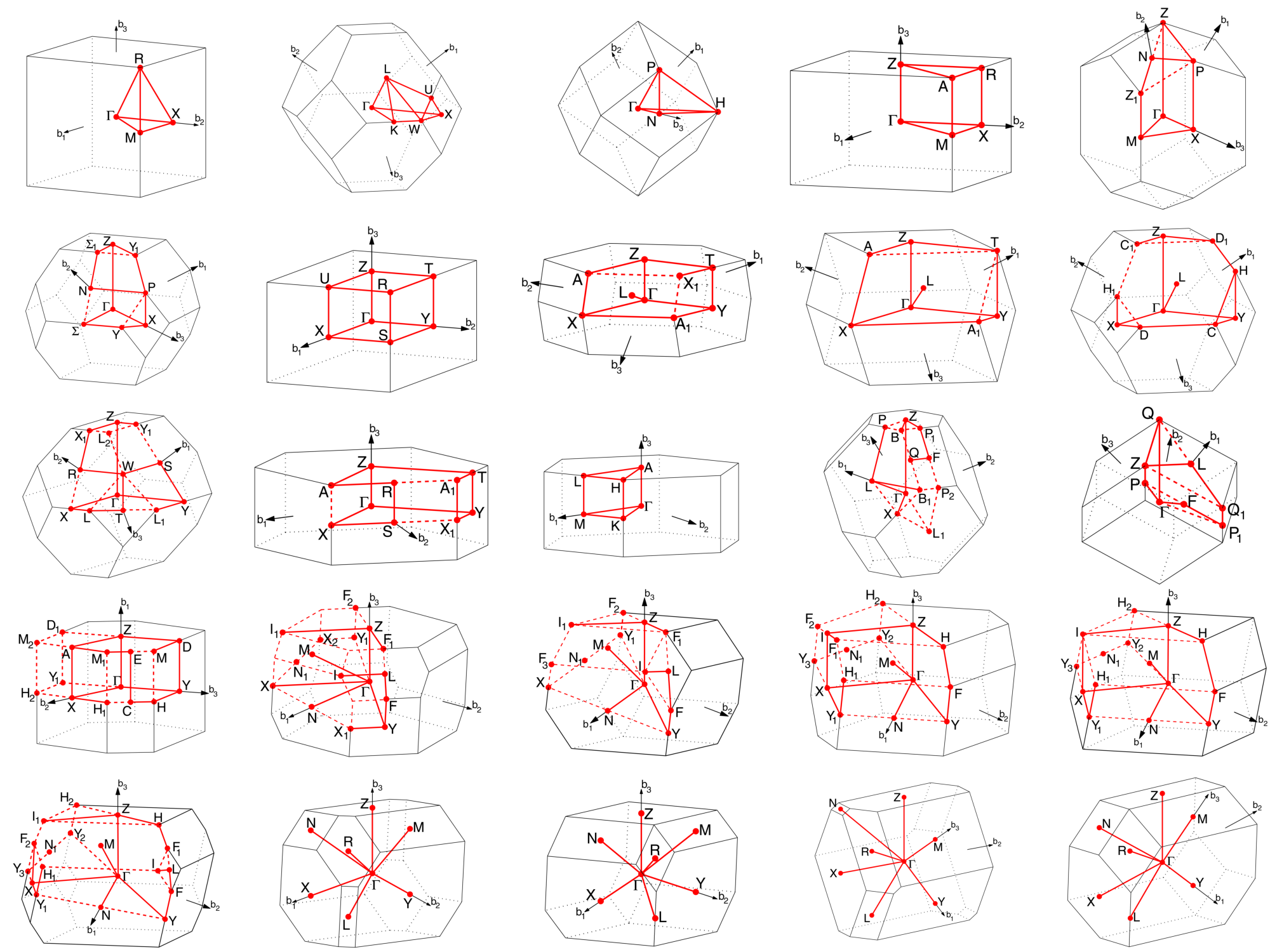}
\mycaption{Standardized paths in reciprocal space for calculation of the electronic band
structures for the 25 different lattice types~\cite{aflowBZ}.}
\label{fig:aflow_chp:band_structure_paths}
\efig

\fig
\includegraphics[width=1.00\linewidth]{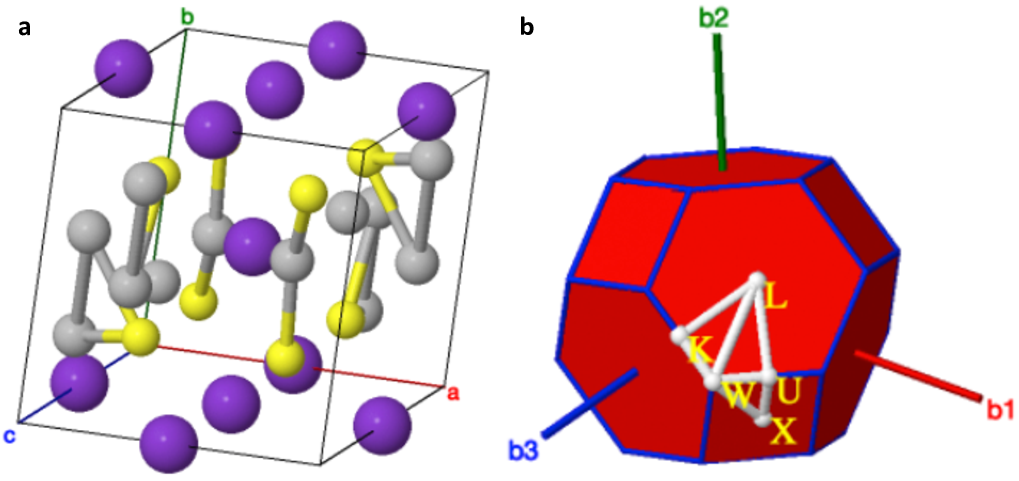}
\mycaption[Side-by-side visualization of the crystal structure and Brillouin Zone using Jmol~\cite{Jmol_Hanson,Jmol}.]
{(\textbf{a}) The structure highlighted is Ag$_{3}$KS$_{2}$ (\ICSD\ \#73581): \url{http://aflow.org/material.php?id=Ag6K2S4_ICSD_73581}.
(\textbf{b}) The \AFLOW\ Standard path of high-symmetry \textbf{k}-points is illustrated in the Brillouin Zone~\cite{aflowBZ}.}
\label{fig:aflow_fleet:jmol_bz}
\efig

The mapping onto a system of $n$ non-interacting particles comes at the cost of introducing the
exchange-correlation potential $V_\sXC\left[\rho_s(\mathbf{r})\right]$, the exact form of which is unknown and must be approximated.
The simplest approximation is the \underline{l}ocal \underline{d}ensity \underline{a}pproximation (\LDA)~\cite{Perdew_prb_1981},
in which the magnitude of the exchange-correlation energy at a particular point in space is
assumed to be proportional to the magnitude of the density at that point in space.
Despite its simplicity, \LDA\ produces realistic results for atomic structure, elastic and vibrational properties
for a wide range of systems. However, it tends to overestimate the binding energies of materials, even
putting crystal bulk phases in the wrong energetic order~\cite{Zupan_LDAperformance_PRB_1998}.
Beyond \LDA\ is the \underline{G}eneralized \underline{G}radient \underline{A}pproximation (\GGA), in which the exchange correlation term
is a functional of the charge density and its gradient at each point in space.
There are several forms of \GGA\, including those developed by Perdew, Burke and Ernzerhof (\PBE~\cite{PBE}), or by Lee, Yang and Parr ({\small LYP}~\cite{LYP_1988}).
A more recent development is the meta-\GGA\ \underline{S}trongly \underline{C}onstrained and \underline{A}ppropriately \underline{N}ormed ({\small SCAN})
functional~\cite{Perdew_SCAN_PRL_2015}, which satisfies all 17 known exact constraints on meta-\GGA\ functionals.

The major limitations of \LDA\ and \GGA\ include their inability to adequately describe systems with strongly correlated or localized electrons,
due to the local and semilocal nature of the functionals.
Treatments include the Hubbard $U$ corrections~\cite{LiechDFTU, Dudarev_dftu}, self-interaction corrections~\cite{Perdew_prb_1981}
and hybrid functionals such as Becke's 3-parameter modification of {\small LYP} ({\small B3LYP}~\cite{B3LYP_1993}), and that of Heyd, Scuseria and Ernzerhof ({\small Heyd2003}~\cite{Heyd2003}).

Within the context of \abinitio\ structure prediction calculations, \GGA-\PBE\ is the usual standard since it tends to produce
accurate geometries and lattice constants~\cite{monster}.
For accounting for strong correlation effects, the \DFT$+U$ method~\cite{LiechDFTU, Dudarev_dftu}
is often favored in large-scale automated database generation due to its low computational overhead.
However, the traditional \DFT$+U$ procedure requires the addition of an empirical factor to the potential~\cite{LiechDFTU, Dudarev_dftu}.
Recently, methods have been implemented to calculate the $U$ parameter self-consistently from first-principles, such as the ACBN0 functional~\cite{curtarolo:art93}.

\DFT\ also suffers from an inadequate description of excited/unoccupied states, as the theory
is fundamentally based on the ground state.
Extensions for describing excited states include time-dependent \DFT\ (\TDDFT)~\cite{Hedin_GW_1965} and the GW correction~\cite{GW}.
However, these methods are typically much more expensive than standard \DFT, and are not generally considered for large scale database generation.

\fig
\includegraphics[width=1.00\linewidth]{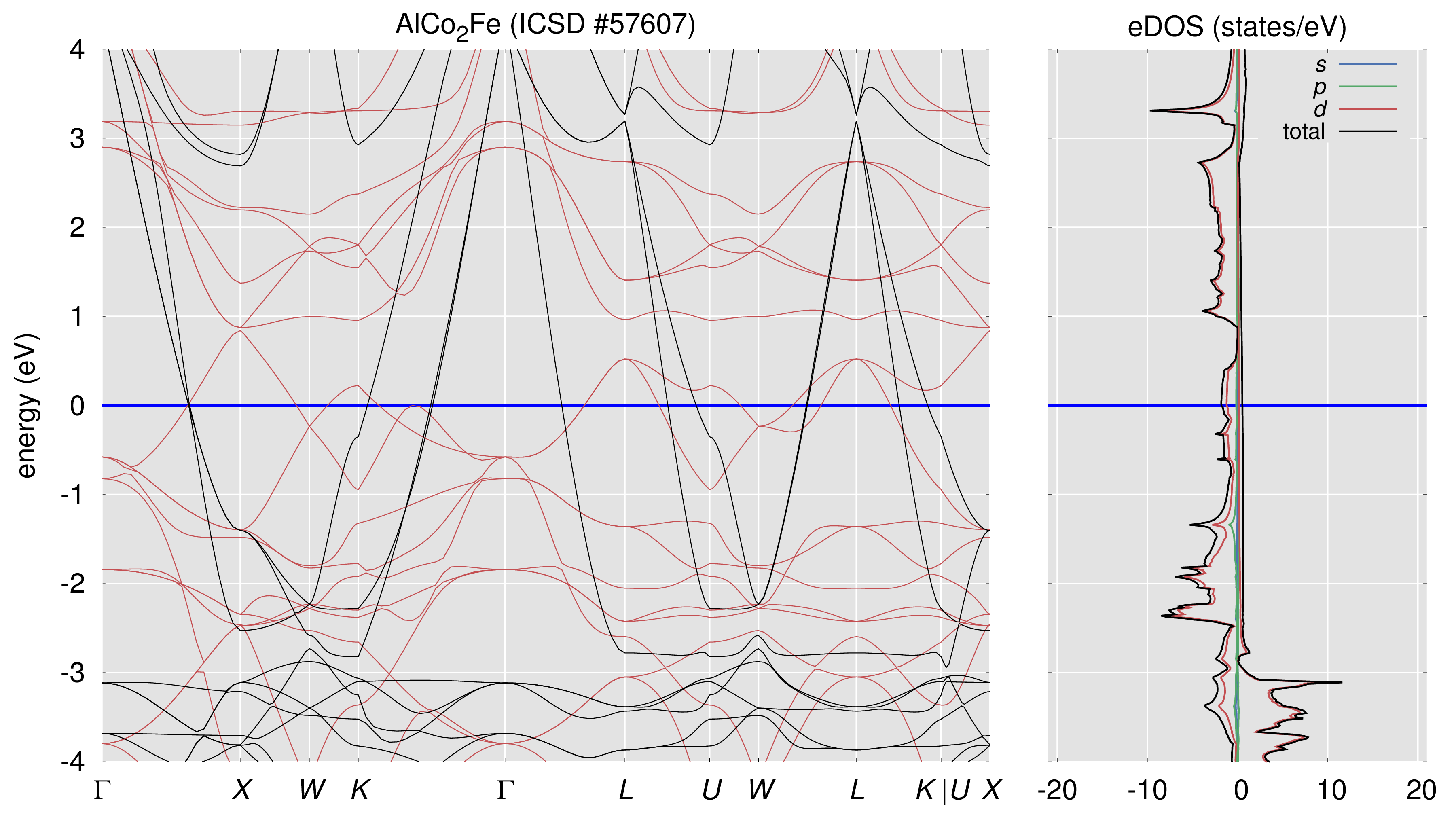}
\mycaption[Example band structure and density of states images automatically generated and
served through the \AFLOWorg\ data repository.]
{The structure highlighted is AlCo$_{2}$Fe (\ICSD\ \#57607): \url{http://aflow.org/material.php?id=Al1Co2Fe1_ICSD_57607}.
The results of the spin-polarized calculation are differentiated by: color on the band structure plot
(black/red for majority/minority spin), and sign on the density of states plot (positive/negative for majority/minority spin).
The band structure is calculated following the \AFLOW\ Standard path of high-symmetry \textbf{k}-points~\cite{aflowBZ}.}
\label{fig:aflow_fleet:bs_plot}
\efig

At the technical implementation level, there are
many \DFT\ software packages available, including
\VASP~\cite{kresse_vasp, vasp_prb1996, vasp_cms1996, kresse_vasp_paw},
\QE~\cite{qe, Giannozzi:2017io}, \ABINIT~\cite{gonze:abinit, abinit_2009},
\FHIAIMS~\cite{Blum_CPC2009_AIM}, \SIESTA~\cite{Soler2002SIESTA} and \GAUSSIAN~\cite{Gaussian_2009}.
These codes are generally distinguished by the choice of basis set.
There are two principle types of basis sets: plane waves, which take the form $\psi (\mathbf{r}) = \sum e^{i \mathbf{k}\cdot\mathbf{r}}$,
and local orbitals, formed by a sum over functions $\phi_a (\mathbf{r})$ localized at particular points in space, such as
gaussians or numerical atomic orbitals~\cite{Hehre_self_consistent_molecular_orbit_JCP1969}.
Plane wave based packages include \VASP, \QE\ and \ABINIT, and are generally better suited to periodic systems such as bulk inorganic materials.
Local orbital based packages include \FHIAIMS, \SIESTA\ and \GAUSSIAN, and are generally better suited to non-periodic systems such as organic molecules.
In the field of automated computational materials science, plane wave codes such as \VASP\ are generally preferred:
it is straightforward to automatically and systematically generate well-converged basis sets
since there is only a single parameter to adjust, namely the cut-off energy determining the number
of plane waves in the basis set.
Local orbital basis sets tend to have far more independently adjustable degrees of freedom,
such as the number of basis orbitals per atomic orbital as well as their respective cut-off radii,
making the automated generation of reliable basis sets more difficult.
Therefore, a typical standardized protocol for automated materials science calculations~\cite{curtarolo:art104} relies on
the \VASP\ software package with a basis set cut-off energy higher than that recommended by the \VASP\ potential files,
in combination with the \PBE\ formulation of \GGA.

Finally, it is necessary to automate the generation of the \textbf{k}-point grid and pathways in
reciprocal space used for the calculation of forces, energies and the electronic band structure.
In general, \DFT\ codes use standardized methods such as the Monkhorst-Pack scheme~\cite{MonkhorstPack} to generate reciprocal lattice \textbf{k}-point grids,
although optimized grids have been calculated for different lattice types and are available online~\cite{Wisesa_Kgrids_PRB_2016}.
Optimizing \textbf{k}-point grid density is a computationally expensive process that is difficult to automate,
so instead standardized grid densities based on the concept of
``\underline{$k$}-\underline{p}oints \underline{p}er \underline{r}eciprocal \underline{a}tom'' (\KPPRA) are used.
The \KPPRA\ value is chosen to be sufficiently large to ensure convergence for all systems.
Typical recommended values used for \KPPRA\ range from 6,000 to 10,000~\cite{curtarolo:art104},
so that a material with two atoms in the calculation cell will have a \textbf{k}-point mesh of at least 3,000 to 5,000 points.
Standardized directions in reciprocal space have also been defined for the calculation of the
band structure as illustrated in Figure~\ref{fig:aflow_chp:band_structure_paths}~\cite{aflowBZ} and Figure~\ref{fig:aflow_fleet:jmol_bz}.
These paths are optimized to include all of the high-symmetry points of the lattice.
A standard band structure plot as generated by \AFLOW\ is illustrated in Figure~\ref{fig:aflow_fleet:bs_plot}.

\subsection{Integrated calculation of materials properties}
\label{subsec:aflow_chp:thermomechanical}

Automated frameworks such as \AFLOW\ combine the computational analysis of properties including symmetry, electronic structure,
elasticity, and thermal behavior into integrated workflows.
Crystal symmetry information is used to find the primitive cell to reduce the size of \DFT\ calculations,
to determine the appropriate paths in reciprocal space for electronic band structure calculations (see Figure~\ref{fig:aflow_chp:band_structure_paths}~\cite{aflowBZ}),
and to determine the set of inequivalent distortions for phonon and elasticity calculations.
Thermal and elastic properties of materials are important for predicting the thermodynamic and mechanical stability
of structural phases~\cite{Greaves_Poisson_NMat_2011, Poirier_Earth_Interior_2000, Mouhat_Elastic_PRB_2014, curtarolo:art106}
and assessing their importance for a variety of applications.
Elastic properties such as the shear and bulk moduli are important for predicting the hardness
of materials~\cite{Chen_hardness_Intermetallics_2011, Teter_Hardness_MRS_1998},
and thus their resistance to wear and distortion.
Elasticity tensors can be used to predict the properties of composite
materials~\cite{Hashin_Multiphase_JMPS_1963, Zohdi_Polycrystalline_IJNME_2001}.
They are also important in geophysics for modeling the propagation of seismic waves
in order to investigate the mineral composition of geological
formations~\cite{Poirier_Earth_Interior_2000, Anderson_Elastic_RGP_1968, Karki_Elastic_RGP_2001}.
The lattice thermal conductivity $\left(\kappa_\sL\right)$ is a crucial
design parameter in a wide range of important
technologies, such as the development of new thermoelectric
materials~\cite{zebarjadi_perspectives_2012,aflowKAPPA,Garrity_thermoelectrics_PRB_2016},
heat sink materials for thermal management in electronic devices~\cite{Yeh_2002},
and rewritable phase-change memories~\cite{Wright_tnano_2011}.
High thermal conductivity materials, which typically have a zincblende or diamond-like structure, are essential
in microelectronic and nanoelectronic devices for achieving
efficient heat removal~\cite{Watari_MRS_2001}, and have
been intensively studied for the past few decades~\cite{Slack_1987}.
Low thermal conductivity materials constitute
the basis of a new generation of thermoelectric materials and thermal
barrier coatings~\cite{Snyder_jmatchem_2011}.

The calculation of thermal and elastic properties offer an excellent example of the power of
integrated computational materials design frameworks.
With a single input file, these frameworks can automatically set-up and run calculations of
different distorted cells, and combine the resulting energies and
forces to calculate thermal and mechanical properties.

\subsubsection{Autonomous symmetry analysis}

Critical to any analysis of crystals is the accurate determination of the symmetry profile.
For example, symmetry serves to
\textbf{i.} validate the forms of the elastic constants
and compliance tensors, where the crystal symmetry dictates equivalence or absence
of specific tensor elements~\cite{nye_symmetry, curtarolo:art100, Mouhat_Elastic_PRB_2014}, and
\textbf{ii.} reduce the number of \abinitio\ calculations needed for phonon
calculations, where, in the case of the finite-displacement method, equivalent
atoms and distortion directions are identified through factor group and site symmetry
analyses~\cite{Maradudin1971}.

Autonomous workflows for elasticity and vibrational characterizations
therefore require a correspondingly robust symmetry analysis.
Unfortunately, standard symmetry packages~\cite{stokes_findsym,Stokes_FROZSL_Ferroelectrics_1995,platon_2003,spglib},
catering to different objectives, depend on tolerance-tuning to
overcome numerical instabilities and atypical data --- emanating from
finite temperature measurements and uncertainty in experimentally reported observations.
These tolerances are responsible for validating mappings and identifying isometries,
such as the $n$-fold operator depicted in Figure~\ref{fig:aflow_chp:sym}(a).
Some standard packages define separate tolerances for space, angle~\cite{spglib},
and even operation type~\cite{stokes_findsym,Stokes_FROZSL_Ferroelectrics_1995,platon_2003}
(\eg, rotation \vs\ inversion).
Each parameter introduces a factorial expansion of unique inputs, which can result in
distinct symmetry profiles as illustrated in Figure~\ref{fig:aflow_chp:sym}(b).
By varying the spatial tolerance $\epsilon$, four different space groups can be observed
for AgBr (\ICSD\ \#56551\footnote{{h}ttp://www.aflow.org/material.php?id=56551}), if one is found at all.
Gaps in the range, where no consistent symmetry profile can be resolved, are
particularly problematic in automated frameworks, triggering critical failures in subsequent analyses.

\fig
\includegraphics[width=1.00\linewidth]{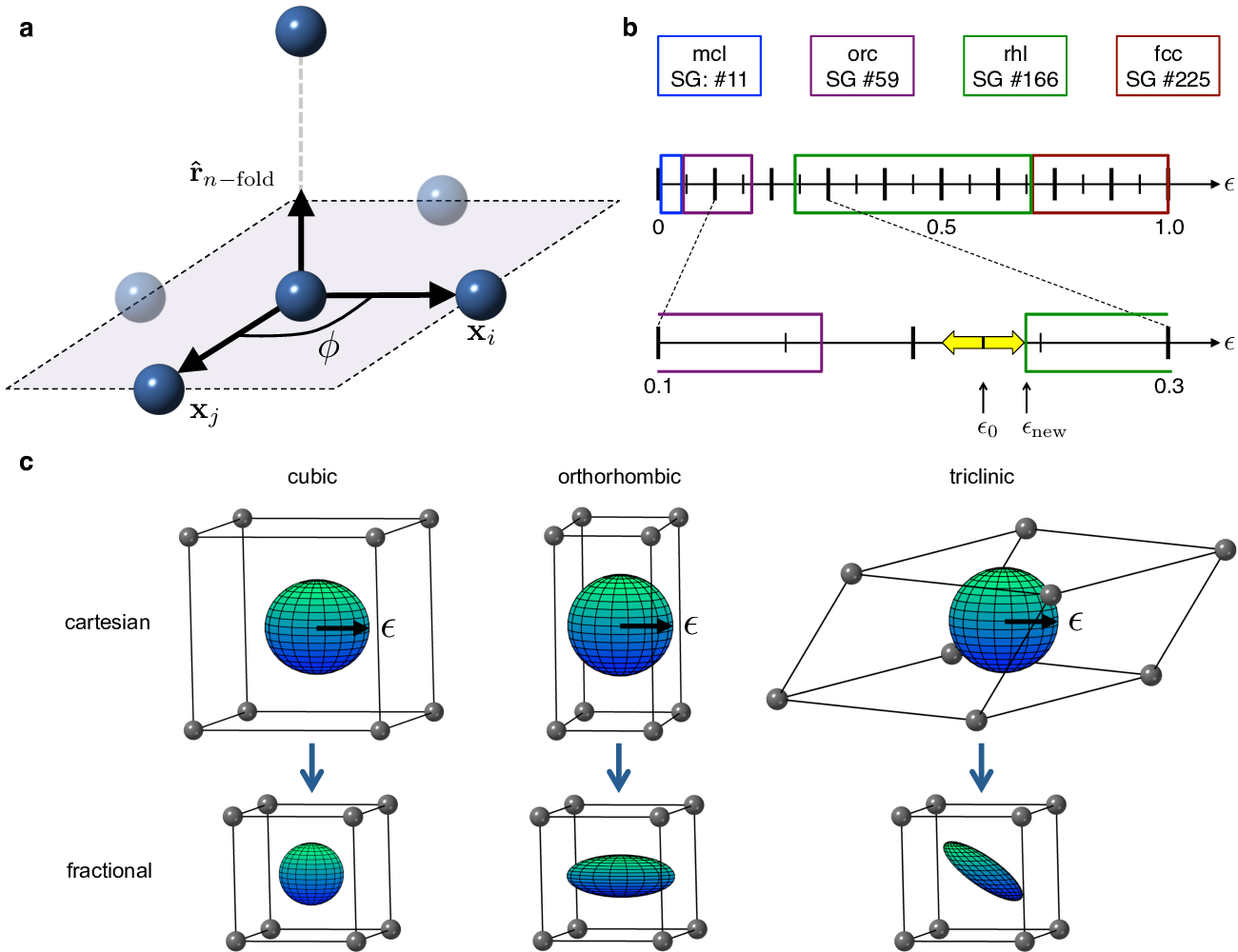}
\mycaption[Challenges in autonomous symmetry analysis.]
{(\textbf{a}) An illustration of a general $n$-fold symmetry operation.
(\textbf{b}) Possible space group determinations with mapping tolerance $\epsilon$ for AgBr (\ICSD\ \#56551).
(\textbf{c}) Warping of mapping tolerance sphere with a transformation from cartesian to fractional basis.}
\label{fig:aflow_chp:sym}
\efig

Cell shape can also complicate mapping determinations.
Anisotropies in the cell, such as skewness of lattice vectors, translate
to distortions of fractional and reciprocal spaces.
A uniform tolerance sphere in cartesian space, inside which points are considered mapped,
generally warps to a sheared spheroid, as depicted in Figure~\ref{fig:aflow_chp:sym}(c).
Hence, distances in these spaces are direction-dependent, compromising the integrity
of rapid minimum-image determinations~\cite{hloucha_minimumimage_1998} and generally warranting
prohibitively expensive algorithms~\cite{curtarolo:art135}.
Such failures can result in incommensurate symmetry profiles, where the real space
lattice profile (\eg, bcc) does not match that of the reciprocal space (fcc).

The new \AFLOWSYM\ module~\cite{curtarolo:art135} within \AFLOW\ offers careful treatment of tolerances, with extensive
validation schemes, to mitigate the aforementioned challenges.
Although a user-defined tolerance input is still available, \AFLOW\ defaults to one of two pre-defined
tolerances, namely \texttt{tight} (standard) and \texttt{loose}.
Should any discrepancies occur, these defaults are the starting values of a large tolerance scan,
as shown in Figure~\ref{fig:aflow_chp:sym}(b).
A number of validation schemes have been incorporated to catch such discrepancies.
These checks are consistent with crystallographic group theory principles, validating operation
types and cardinalities~\cite{tables_crystallography}.
From considerations of different extreme cell shapes, a heuristic threshold has been defined
to classify scenarios where mapping failures are likely to occur --- based on skewness and mapping tolerance.
When benchmarked against standard packages for over 54,000 structures in the Inorganic Crystal Structure Database,
\AFLOWSYM\ consistently resolves
the symmetry characterization most compatible with experimental observations~\cite{curtarolo:art135}.

Along with accuracy, \AFLOWSYM\ delivers a wealth of symmetry properties and representations
to satisfy injection into any analysis or workflow.
The full set of operators --- including that of the point-, factor-, crystallographic point-, space groups,
and site symmetries --- are provided in matrix, axis-angle, matrix generator, and quaternion representations in
both cartesian and fractional coordinates.
A span of characterizations, organized by degree of symmetry-breaking, are available, including
those of the lattice, superlattice, crystal, and crystal-spin.
Space group and Wyckoff positions are also resolved.
The full dataset is made available in both plain-text and \JSON\ formats.

\fig
\includegraphics[width=1.00\linewidth]{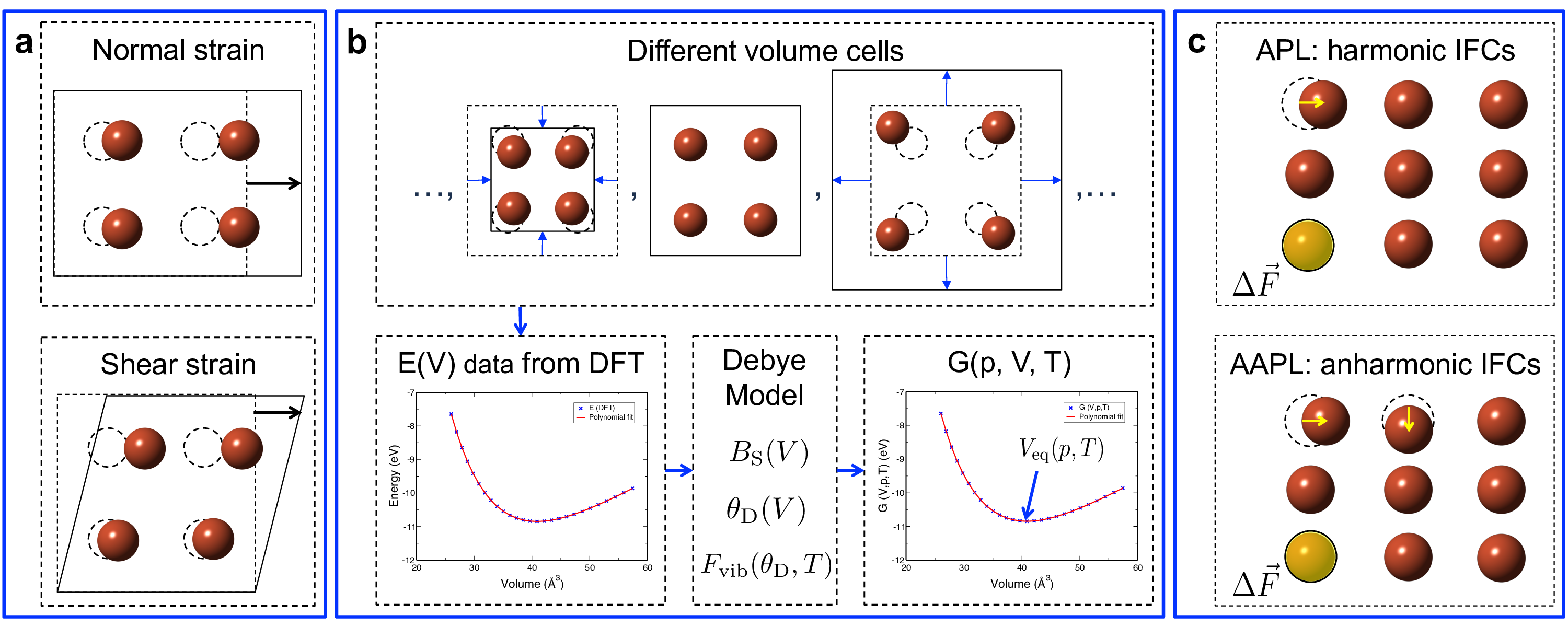}
\mycaption[Calculation of thermomechanical properties.]
{({\bf a}) \AEL\ applies a set of independent normal and shear strains to the crystal structure to obtain
the elastic constants.
({\bf b}) \AGL\ applies a set of isotropic strains to the unit cell to obtain
energy \vs\ volume data, which is fitted by a polynomial in order to
calculate the bulk modulus as a function of volume, $B_{\mathrm S} (V)$.
$B_{\mathrm S} (V)$ is then used to calculate the Debye temperature as a function of volume and thus
the vibrational free energy as a function of temperature.
The Gibbs free energy as a function of volume is then minimized for each pressure and temperature
point to obtain the equilibrium volume and other thermomechanical properties.
({\bf c}) \APL\ obtains the harmonic \underline{i}nteratomic \underline{f}orce \underline{c}onstants (\IFC{}s) from supercell calculations where inequivalent
atoms are displaced in inequivalent directions, and then the changes in the forces on the other atoms are calculated.
The \IFC{}s are then used to construct the dynamical matrix, which is diagonalized to obtain the phonon eigenmodes.
\AAPL\ calculates three-phonon scattering effects by performing supercell calculations where pairs of inequivalent atoms are displaced in inequivalent directions, and
 the changes in the forces on the other atoms in the supercell are calculated to obtain the third-order anharmonic \IFC{}s.}
\label{fig:aflow_chp:thermomechanical}
\efig

\subsubsection{Harmonic phonons}

Thermal properties can be obtained by directly
calculating the phonon dispersion from the dynamical matrix of \IFC{}s.
The approach is implemented within the {\small \underline{A}FLOW} \underline{P}honon \underline{L}ibrary
(\APL)~\cite{aflowPAPER}.
The \IFC{}s are determined from a set of supercell calculations in which the atoms are
displaced from their equilibrium positions~\cite{Maradudin1971} as shown in Figure~\ref{fig:aflow_chp:thermomechanical}(c).

The \IFC{}s derive from a Taylor expansion of the potential energy, $V$, of the crystal
about the atoms' equilibrium positions:
\begin{multline}
  V=\left.V\right|_{\mathbf{r}(i,t)=0,\forall i}+
  \sum_{i,\alpha}\left.\frac{\partial V}{\partial r(i,t)^{\alpha}}\right|_{\mathbf{r}(i,t)=0,\forall i} r(i,t)^{\alpha} \\+
  \frac{1}{2}\sum_{\substack{i,\alpha,\\ j,\beta}}\left.\frac{\partial^2V}
{\partial r(i,t)^{\alpha}\partial r(j,t)^{\beta}}\right|_{\mathbf{r}(i,t)=0,\forall i}
  r(i,t)^{\alpha}r(j,t)^{\beta}+
\ldots,
\label{eq:aflow_chp:PE_harmonic}
\end{multline}
where $r(i,t)^{\alpha}$ is the $\alpha$-cartesian component ($\alpha=x,y,z$) of the time-dependent atomic displacement
$\mathbf{r}(t)$ of the $i^{\mathrm{th}}$ atom about its equilibrium position,
$\left.V\right|_{\mathbf{r}(i,t)=0,\forall i}$ is the potential energy of the crystal in its equilibrium configuration,
$\left.\partial V/\partial r(i,t)^{\alpha}\right|_{\mathbf{r}(i,t)=0,\forall i}$
is the negative of the force acting in the $\alpha$ direction on atom $i$ in the equilibrium configuration
(zero by definition), and
$\left.\partial^2V/\partial r(i,t)^{\alpha}\partial r(j,t)^{\beta}\right|_{\mathbf{r}(i,t)=0,\forall i}$
constitute the \IFC{}s $\phi(i,j)_{\alpha,\beta}$.
To first approximation, $\phi(i,j)_{\alpha,\beta}$ is the negative of the force exerted
in the $\alpha$ direction on atom $i$ when atom
$j$ is displaced in the $\beta$ direction with all other atoms maintaining their equilibrium positions,
as shown in Figure~\ref{fig:aflow_chp:thermomechanical}(c).
All higher order terms are neglected in the harmonic approximation.

Correspondingly, the equations of motion of the lattice are as follows:
\begin{equation}
  M(i)\ddot{r}(i,t)^{\alpha}=
-\sum_{j,\beta}\phi(i,j)_{\alpha,\beta}
r(j,t)^{\beta}\quad\forall i,\alpha,
\label{eq:aflow_chp:eom_harmonic}
\end{equation}
and can be solved by a plane wave solution of the form
\begin{equation}
r(i,t)^{\alpha}=\frac{v(i)^{\alpha}}{\sqrt{M(i)}} e^{\mathrm{i}\left(\mathbf{q}\cdot\mathbf{R}_{l} - \omega t \right)},
\end{equation}
where $v(i)^{\alpha}$ form the phonon eigenvectors (polarization vector),
$M(i)$ is the mass of the $i^{\mathrm{th}}$ atom,
$\mathbf{q}$ is the wave vector,
$\mathbf{R}_{l}$ is the position of lattice point $l$,
and $\omega$ form the phonon eigenvalues (frequencies).
The approach is nearly identical to that taken for electrons in a periodic potential (Bloch waves)~\cite{ashcroft_mermin}.
Plugging this solution into the equations of motion (Equation~\ref{eq:aflow_chp:eom_harmonic}) yields the following set of linear equations:
\begin{equation}
\omega^{2}v(i)^{\alpha}=
  \sum_{j,\beta}D_{i,j}^{\alpha,\beta}(\mathbf{q})
  v(j)^{\beta}\quad\forall i,\alpha,
\end{equation}
where the dynamical matrix $D_{i,j}^{\alpha,\beta}(\mathbf{q})$ is defined as
\begin{equation}
D_{i,j}^{\alpha,\beta}(\mathbf{q})=
  \sum_{l}\frac{\phi(i,j)_{\alpha,\beta}}{\sqrt{M(i)M(j)}} e^{-\mathrm{i}\mathbf{q}\cdot\left(\mathbf{R}_{l}-\mathbf{R}_{0}\right)}.
\end{equation}
The problem can be equivalently represented by a standard eigenvalue equation:
\begin{equation}
\omega^{2}
\begin{bmatrix}
                                            \\
\mathbf{v} \\
 ~
\end{bmatrix}
=
\begin{bmatrix}
 &                                                                                       & \\
  & \mathbf{D}(\mathbf{q}) & \\
 &                                                                                       &
\end{bmatrix}
\begin{bmatrix}
                                            \\
\mathbf{v} \\
 ~
\end{bmatrix},
\label{eq:aflow_chp:dyn_eigen}
\end{equation}
where
the dynamical matrix and phonon eigenvectors have dimensions $\left(3 n_{\mathrm{a}} \times 3 n_{\mathrm{a}}\right)$
and $\left(3 n_{\mathrm{a}} \times 1 \right)$, respectively, and $n_{\mathrm{a}}$ is the number of atoms in the cell.
Hence, Equation~\ref{eq:aflow_chp:dyn_eigen} has $\lambda=3 n_{\mathrm{a}}$ solutions/modes referred to as branches.
In practice, Equation~\ref{eq:aflow_chp:dyn_eigen} is solved for discrete sets of $\mathbf{q}$-points to compute
the phonon density of states (grid over all possible $\mathbf{q}$) and dispersion
(along the high-symmetry paths of the lattice~\cite{aflowBZ}).
Thus, the phonon eigenvalues and eigenvectors are appropriately denoted $\omega_{\lambda}(\mathbf{q})$ and
$\mathbf{v}_{\lambda}(\mathbf{q})$, respectively.

Similar to the electronic Hamiltonian, the dynamical matrix is Hermitian, \ie,
$\mathbf{D}(\mathbf{q})=\mathbf{D}^{*}(\mathbf{q})$.
Thus $\omega_{\lambda}^{2}(\mathbf{q})$ must also be real, so $\omega_{\lambda}(\mathbf{q})$ can either be real or purely imaginary.
However, a purely imaginary frequency corresponds to vibrational motion of the lattice that increases exponentially in time.
Therefore, imaginary frequencies, or those corresponding to soft modes, indicate the structure is dynamically unstable.
In the case of a symmetric, high-temperature phase, soft modes suggest there exists a lower symmetry structure
stable at $T=0$~K.
Temperature effects on phonon frequencies can be modeled with
\begin{equation}
\widetilde{\omega}_{\lambda}^{2}(\mathbf{q},T)=\omega_{\lambda}^{2}(\mathbf{q},T=0)+\eta T^2,
\end{equation}
where $\eta$ is positive in general.
The two structures, the symmetric and the stable, differ by the distortion
corresponding to this ``frozen'' (non-vibrating) mode.
Upon heating, the temperature term increases until the frequency reaches zero, and a phase transition occurs from
the stable structure to the symmetric~\cite{Dove_LatDynam_1993}.

In practice, soft modes~\cite{Parlinski_Phonon_Software} may indicate:
\textbf{i.} the structure is dynamically unstable at $T$,
\textbf{ii.} the symmetry of the structure is lower than that considered, perhaps due to magnetism,
\textbf{iii.} strong electronic correlations, or
\textbf{iv.} long range interactions play a significant role, and a larger supercell should be considered.

With the phonon density of states computed, the following thermal properties can be calculated:
the internal vibrational energy
\begin{equation}
  U_\svib(\mathbf{x},T)=\int_{0}^{\infty} \left( \frac{1}{2} + \frac{1}{e^{\left(\beta \hbar \omega\right)}-1} \right) \hbar \omega g(\mathbf{x};\omega) d\omega,
\end{equation}
the vibrational component of the free energy $F_\svib(\mathbf{x}; T)$
\begin{equation}
F_\svib(\mathbf{x}; T) \!=\!\! \int_0^{\infty} \!\!\left[\frac{\hbar \omega}{2} \!+\!
\frac{1}{\beta} \ \mathrm{log}\!\left(1\!-\!{\mathrm e}^{- \beta \hbar \omega }\right)\!\right]\!g(\mathbf{x}; \omega) d\omega,
\label{eq:aflow_chp:fvib}
\end{equation}
the vibration entropy
\begin{equation}
S_\svib(\mathbf{x},T)=\frac{U_\svib(\mathbf{x},T)-F_\svib(\mathbf{x}; T)}{T},
\end{equation}
and the isochoric specific heat
\begin{equation}
C_{\sV, \svib}(\mathbf{x},T)=\int_{0}^{\infty} \frac{  k_\sB \left(\beta \hbar \omega \right)^2 g(\mathbf{x};\omega)}{\left(1-e^{-\left(\beta \hbar \omega\right)}\right) \left(e^{\left(\beta \hbar \omega\right)}-1\right) } d\omega.
\end{equation}

\subsubsection{Quasi-harmonic phonons}

The harmonic approximation does not describe phonon-phonon scattering, and so cannot be used to
calculate properties such as thermal conductivity or thermal expansion.
To obtain these properties, either the quasi-harmonic approximation can be used,
or a full calculation of the higher order anharmonic \IFC{}s can be performed.
The quasi-harmonic approximation is the less computationally demanding of these two methods,
and compares harmonic calculations of phonon properties at different volumes to predict anharmonic properties.
The different volume calculations can be in the form of harmonic phonon calculations as described
above~\cite{curtarolo:art114, curtarolo:art119},
or simple static primitive cell calculations~\cite{Blanco_CPC_GIBBS_2004, curtarolo:art96}.
The \underline{Q}uasi-\underline{H}armonic \underline{A}pproximation
is implemented within \APL\ and referred to as \QHAAPL~\cite{curtarolo:art96}.
In the case of the quasi-harmonic phonon calculations, the anharmonicity of the system is described by
the mode-resolved Gr{\"u}neisen parameters, which are given by the change in the phonon frequencies as a function of volume
\begin{equation}
\label{eq:aflow_chp:gamma_micro}
\gamma_{\lambda}(\mathbf{q}) = - \frac{V}{\omega_{\lambda}(\mathbf{q})} \frac{\partial \omega_{\lambda}(\mathbf{q})}{\partial V},
\end{equation}
where $\gamma_{\lambda}(\mathbf{q})$ is the parameter for the wave vector $\mathbf{q}$ and the $\lambda^{\rm{th}}$ mode of the phonon dispersion.
The average of the $\gamma_{\lambda}(\mathbf{q})$ values, weighted by the specific heat capacity of each mode $C_{\sV,\lambda}(\mathbf{q})$, gives the average
Gr{\"u}neisen parameter:
\begin{equation}
\label{eq:aflow_chp:gamma_ave}
\gamma = \frac{\sum_{\lambda,\mathbf{q}} \gamma_{\lambda}(\mathbf{q})  C_{\sV,\lambda}(\mathbf{q})}{C_\sV}.
\end{equation}
The specific heat capacity, Debye temperature and Gr{\"u}neisen parameter can then be combined to
calculate other properties such as the specific heat capacity at constant pressure $C_{\rm p}$,
the thermal coefficient of expansion $\alpha$, and the lattice thermal
conductivity $\kappa_\sL$~\cite{curtarolo:art119}, using similar expressions to those described in Section~\ref{sec:art115}.

\subsubsection{Anharmonic phonons}

The full calculation of the anharmonic \IFC{}s requires performing supercell calculations in which pairs of
inequivalent atoms are displaced in all pairs of
inequivalent directions~\cite{Broido2007, Wu_PRB_2012, ward_ab_2009, ward_intrinsic_2010, Zhang_JACS_2012, Li_PRB_2012, Lindsay_PRL_2013, Lindsay_PRB_2013, Li_ShengBTE_CPC_2014, curtarolo:art125}
as illustrated in Figure~\ref{fig:aflow_chp:thermomechanical}(c).
The third order anharmonic \IFC{}s can then be obtained by calculating the change in the forces on
all of the other atoms due to these displacements.
This method has been implemented in the form of a fully
automated integrated workflow in the \AFLOW\ framework,
where it is referred to as the {\small \underline{A}FLOW} \underline{A}nharmonic
\underline{P}honon \underline{L}ibrary (\AAPL)~\cite{curtarolo:art125}.
This approach can provide very accurate results for the lattice thermal conductivity when combined
with accurate electronic structure methods
\cite{curtarolo:art125},
but quickly becomes very expensive for systems with multiple
inequivalent atoms or low symmetry.
Therefore, simpler methods such as the quasi-harmonic Debye model
tend to be used for initial rapid screening~\cite{curtarolo:art96, curtarolo:art115}, while
the more accurate and expensive methods are used for characterizing systems
that are promising candidates for specific engineering applications.

\subsection{Online data repositories}

Rendering the massive quantities of data generated using automated \abinitio\ frameworks available
for other researchers requires going beyond the conventional methods
for the dissemination of scientific results in the form of journal articles.
Instead, this data is typically made available in online data repositories, which can usually be accessed both
manually via interactive web portals, and programmatically via an \underline{a}pplication \underline{p}rogramming \underline{i}nterface (\API).

\subsubsection{Computational materials data web portals}

Most computational data repositories include an interactive web portal front end that enables manual data access.
These web portals usually include online applications to facilitate data retrieval and analysis.
The front page of the \AFLOW\ data repository is displayed in Figure~\ref{fig:aflow_chp:aflow_web_apps}(a).
The main features include a search bar where information such as
\ICSD\ reference number, {\small \underline{A}FLOW} \underline{u}nique \underline{id}entifier (\AUID) or the chemical formula can be entered
in order to retrieve specific materials entries.

\fig
\includegraphics[width=1.00\linewidth]{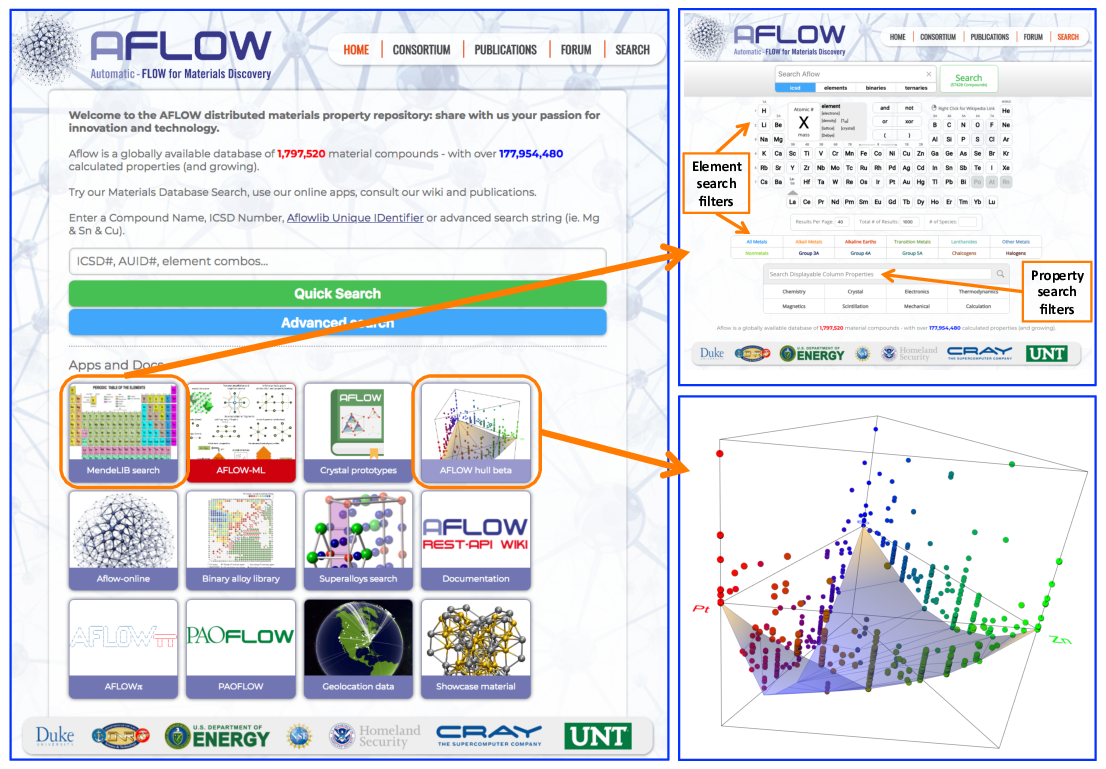}
\mycaption[\AFLOW\ web applications.]
{({\bf a}) Front page of the \AFLOW\ online data repository, highlighting the link to
({\bf b}) the \AFLOW\ advanced search application, which facilitates complex search
queries including filtering by chemical composition and materials properties and
({\bf c}) the \AFLOW\ interactive convex hull generator, showing the 3D hull for the Pt-Sc-Zn ternary alloy system.}
\label{fig:aflow_chp:aflow_web_apps}
\efig

Below are buttons linking to several different online applications such as the advanced search functionality,
convex hull phase diagram generators, machine learning applications~\cite{curtarolo:art124, curtarolo:art129, curtarolo:art136} and \AFLOW-online data analysis tools.
The link to the advanced search application is highlighted by the orange square, and the application page is shown in Figure~\ref{fig:aflow_chp:aflow_web_apps}(b).
The advanced search application allows users to search for materials that contain (or exclude) specific elements or groups of elements,
and also to filter and sort the results by properties such as electronic band structure energy gap (under the ``Electronics'' properties filter group)
and bulk modulus (under the ``Mechanical'' properties filter group).
This allows users to identify candidate materials with suitable materials properties for specific applications.

Another example online application available on the \AFLOW\ web portal is the convex hull phase diagram generator.
This application can be accessed by clicking on the button highlighted by the orange square in
Figure~\ref{fig:aflow_chp:aflow_web_apps}(a), which will bring up a periodic table allowing users to
select two or three elements for which they want to generate a convex hull.
The application will then access the formation enthalpies and stoichiometries of the materials entries in the
relevant alloy systems, and use this data to generate a two or three dimensional convex hull phase diagram
as depicted in Figure~\ref{fig:aflow_chp:aflow_web_apps}(c).
This application is fully interactive, allowing users to adjust the energy axis scale,
rotate the diagram to view from different directions, and select specific points to obtain more information on the
corresponding entries.

\subsubsection{Programmatically accessible online repositories of computed materials properties}

In order to use materials data in machine learning algorithms, it should be stored in a structured online database
and made programmatically accessible via a \underline{re}presentational \underline{s}tate \underline{t}ransfer \API\ (\RESTAPI).
Examples of online repositories of materials data include \AFLOW~\cite{aflowlibPAPER, aflowAPI},
Materials Project~\cite{materialsproject.org}, and \OQMD~\cite{Saal_JOM_2013}.
There are also repositories that aggregate results from multiple sources such as
\NOMAD~\cite{nomad} and Citrine~\cite{citrine_database}.

\RESTAPI{}s facilitate programmatic access to data repositories.
Typical databases such as \AFLOW\ are organized in layers,
with the top layer corresponding to a project or catalog (\eg, binary alloys),
the next layer corresponding to data sets (\eg, all of the entries for a particular alloy system),
and then the bottom layer corresponding to specific materials entries, as illustrated in Figure~\ref{fig:aflow_chp:aflow_restapi_layers_aurl}(a).

\fig
\includegraphics[width=1.00\linewidth]{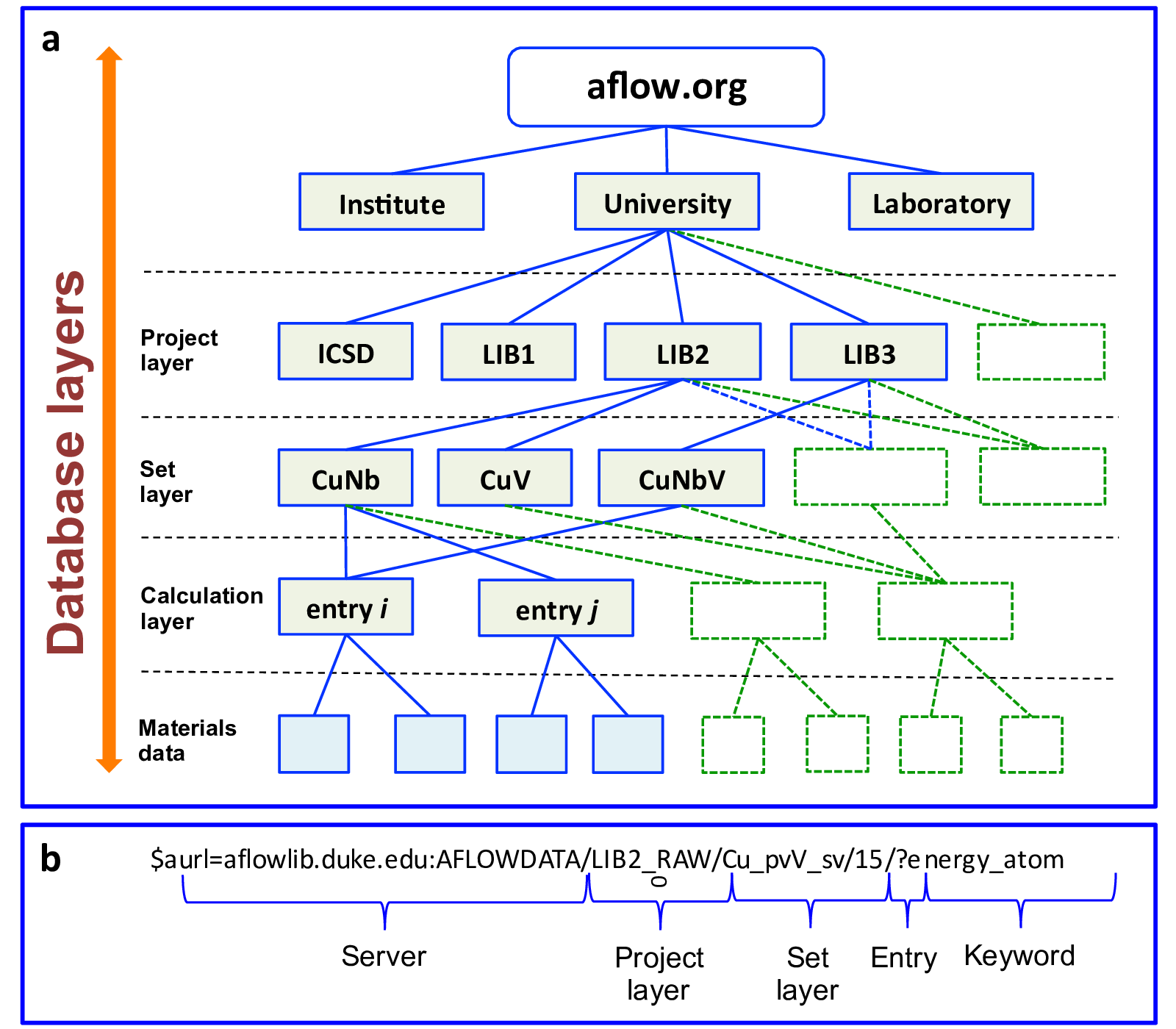}
\mycaption[\AFLOW\ \RESTAPI\ structure.]
{\small (\textbf{a}) The \AFLOW\ database is organized as a multilayered system.
(\textbf{b}) Example of an \AURL\ which enables direct programmatic access to specific materials entry properties in the \AFLOW\ database.}
\label{fig:aflow_chp:aflow_restapi_layers_aurl}
\efig

In the case of the \AFLOW\ database, there are currently four different ``projects'', namely the
``ICSD'', ``LIB1'', ``LIB2'' and ``LIB3'' projects; along with three
more under construction: ``LIB4'', ``LIB5'' and ``LIB6''.
The ``ICSD'' project contains calculated data for previously observed compounds~\cite{ICSD},
whereas the other three projects contain calculated data for single elements, binary alloys,
and ternary alloys respectively, and are constructed by decorating prototype
structures with combinations of different elements.
Within ``LIB2'' and ``LIB3'', there are many different data sets, each corresponding to a specific
binary or ternary alloy system.
Each entry in the set corresponds to a specific prototype structure and stoichiometry.
The materials properties values for each of these entries are encoded via keywords,
and the data can be accessed via \URL{}s constructed from the different layer names and the appropriate keywords.
In the case of the \AFLOW\ database, the location of each layer and entry is
identified by an {\small \underline{A}FLOW} \underline{u}niform \underline{r}esource \underline{l}ocator (\AURL)~\cite{aflowAPI},
which can be converted to a \URL\ providing the absolute path to a particular layer, entry or property.
The \AURL\ takes the form \url{server:AFLOWDATA/project/set/entry/?keywords},
for example \url{aflowlib.duke.edu:AFLOWDATA/LIB2_RAW/Cu_pvV_sv/15/?energy_atom},
where \url{aflowlib.duke.edu} is the web address of the physical server where the data is located,
\url{LIB2_RAW} is the binary alloy project layer, \url{Cu_pvV_sv} is
the set containing the binary alloy system Cu-V, \url{15} is a specific entry with the composition
Cu$_3$V in a tetragonal lattice, and \url{energy_atom} is the keyword corresponding
to the property of energy per atom in units of eV, as shown in Figure~\ref{fig:aflow_chp:aflow_restapi_layers_aurl}(b).
Each \AURL\ can be converted to a web \URL\ by changing the ``\url{:}'' after the server name to a ``\url{/}'',
so that the \AURL\ in Figure~\ref{fig:aflow_chp:aflow_restapi_layers_aurl}(b) would become the
\URL\ \url{aflowlib.duke.edu/AFLOWDATA/LIB2_RAW/Cu_pvV_sv/15/?energy_atom}.
This \URL, if queried via a web browser or using a UNIX utility such as \texttt{wget},
returns the energy per atom in eV for entry \url{15} of the Cu-V binary alloy system.

In addition to the \AURL, each entry in the \AFLOW\ database is also associated with an
\AUID~\cite{aflowAPI},
which is a unique hexadecimal (base 16) number constructed from a checksum of the \AFLOW\ output file for that entry.
Since the \AUID\ for a particular entry can always be reconstructed by applying the checksum
procedure to the output file, it serves as a permanent, unique specifier for each calculation,
irrespective of the current physical location of where the data are stored.
This enables the retrieval of the results for a particular calculation from different servers,
allowing for the construction of a truly distributed database that is robust
against the failure or relocation of the physical hardware. Actual database versions can be
identified from the version of \AFLOW\ used to parse the calculation output files and
postprocess the results to generate the database entry. This information can be retrieved using the
keyword \url{aflowlib_version}.

The search and sort functions of the front-end portals can be combined with the programmatic
data access functionality of the \RESTAPI\ through the implementation of a Search-\API.
The \AFLUX\ Search-\API\ uses the \LUX\ language to enable the embedding of logical
operators within \URL\ query strings~\cite{aflux}.
For example, the energy per atom of every entry in the \AFLOW\ repository containing the element Cu or V,
but not the element Ti, with an electronic band gap between 2 and 5~eV, can be retrieved using the command:
\url{aflowlib.duke.edu/search/API/species((Cu:V),(!Ti)),Egap(2*,*5),energy_atom}.
In this \AFLUX\ search query, the comma ``\verb|,|'' represents the logical {\small AND} operation, the colon ``\verb|:|'' the logical {\small OR} operation,
the exclamation mark ``\verb|!|'' the logical {\small NOT} operation, and the asterisk ``\verb|*|''  is the ``loose'' operation that defines a range of values to search within.
Note that by default \AFLUX\  returns only the first 64 entries matching the search query.
The number and set of entries can be controlled by appending the \verb|paging| directive to the end of the search query as follows:
\url{aflowlib.duke.edu/search/API/species((Cu:V),(!Ti)),Egap(2*,*5),energy_atom,paging(0)},
where calling the  \verb|paging| directive with the argument ``0'' instructs \AFLUX\ to return all of the matching entries
(note that this could potentially be a large amount of data, depending on the search query).
The \AFLUX\ Search-\API\ allows users to construct and retrieve customized data sets, which they can feed into materials
informatics machine learning packages to identify trends and correlations for use in rational materials design.

The use of \API{}s to provide programmatic access is being extended beyond materials data retrieval,
to enable the remote use of pre-trained machine learning algorithms.
The \AFLOWML\ \API~\cite{curtarolo:art136} facilitates access to the two machine learning models
that are also available online at \url{aflow.org/aflow-ml}~\cite{curtarolo:art124, curtarolo:art129}.
The \API\ allows users to submit structural data for the material of interest using a utility such as \verb|cURL|,
and then returns the results of the model's predictions in \JSON\ format.
The programmatic access to machine learning predictions enables the incorporation of machine learning into
materials design workflows, allowing for rapid pre-screening to automatically select
promising candidates for further investigation.

\clearpage
\section{The Structure and Composition Statistics of 6A Binary and Ternary Crystalline Materials}
\label{sec:art130}

This study follows from a collaborative effort described in Reference~\cite{curtarolo:art130}.

\subsection{Introduction}
The creation of novel materials with optimal properties for diverse applications requires a fundamental
understanding of the factors that govern the formation of crystalline
solids from various mixtures of elements.
Compounds of the non-metallic elements of column 6A, oxygen, sulfur and selenium, are of particular interest.
They serve in a large variety of applications
in diverse fields of technology, \eg, chemistry, catalysis, optics,
gas sensors, electronics, thermoelectrics, piezoelectrics,
topological insulators, spintronics and more~\cite{eranna2004oxide,fortunato2012oxide,tsipis2008electrode,jiang1998new,panda2009review,shi2017,lorenz20162016,ruhle2012all}.
Given the very large number of possibilities, many of the alloy systems of these elements have not
been fully investigated, some of them even not at all.

In recent years, high-throughput computational techniques based on \abinitio\ calculations
have emerged as a potential route to bridge these experimental gaps and
gain understanding of the governing principles of compound formation~\cite{nmatHT}.
This led to the creation of large databases of computational materials
data~\cite{aflowPAPER,CMS_Ong2012b}.
Yet, these computational approaches are practically limited by the number and size of structures
that can be thoroughly analyzed, and fundamental issues that limit
the applicability of standard semi-local \DFT\ for non-metallic compounds.
The sought-after governing principles are thus still largely unknown.

Nevertheless, the considerable body of experimental data that is already available,
although by no means complete, is a useful basis for large-scale data analysis.
This experimental data is usually presented in
compendiums that lack statistical analysis.
Presenting this data in a structured manner may be conducive for gaining insights
into the essential factors that determine structure formation, and may help to provide
material scientists with the necessary foundation for rational
materials design.

Analyses recently carried out for the intermetallic binaries~\cite{dshemuchadse2014some}
and ternaries~\cite{dshemuchadse2015more} have uncovered interesting Bravais lattices distributions and an unexpected large prevalence of unique structure types.
Here we extend the analysis and discuss trends, as well as special phenomena, across
binary and ternary compounds of the 6A non-metals.
This analysis reveals the following
interesting observations:
\begin{itemize}[leftmargin=*]

  \item Considerable overlap exists between the sulfides and selenides:
   about a third of the total number of structure types are shared among
  both compound families.
  In contrast, the overlap between the oxides and the other two families is rather small.

  \item The prevalence of different compound stoichiometries in the sulfide
	and selenide families is very similar to each other
	but different from that of the oxides. Some stoichiometries
	are abundant in the oxides but are {\it almost
  absent} in the sulfides or  selenides, and vice versa.

  \item The number of ternary oxide stoichiometries, $A_{x}B_{y}$O$_{z}$, decreases when the product of
    binary oxide stoichiometries, of participating elements, increases. This behavior can be explained by general thermodynamic arguments and is discussed in the text.

  \item Overall, oxide compounds tend to have richer oxygen content than the sulfur and selenium content in their corresponding compounds.

  \item Across all three compound families, most structure types are represented
    by only one compound.

  \item High symmetry lattices, \eg, the orthorhombic face centered,
    orthorhombic body centered and cubic lattices
	  are relatively rare among these compounds.
    This reflects the spatial arrangement of the compound forming orbitals of the 6A non-metals,
    whose chemistry does not favor these structures.

\end{itemize}

In the analysis presented here, we adopt the ordering of the elements by Mendeleev numbers as
defined by Pettifor~\cite{pettifor:1984,pettifor:1986},
and complement it by investigating the crystallographic properties of
the experimentally reported compounds.
Pettifor maps constructed for these compound families exhibit similar separation between different structure types as the
classical Pettifor maps for binary structure types~\cite{pettifor:1984,pettifor:1986}.
For some stoichiometries, the structure types show similar patterns in
the maps of the three compound families, suggesting
that similar atoms tend to form these stoichiometries with all three elements.
Such similarity of patterns is more common between
the sulfides and selenides than between either of them and the oxides.

These findings suggest a few possible guiding principles for directed searches of new compounds.
Element substitution could be used to examine favorable candidates within the
imperfect overlaps of the structure distributions, especially between the sulfides and selenides.
Moreover, the missing stoichiometries and structure symmetries mean that data-driven approaches, \eg,
machine learning, must use training sets not limited to one compound family, even in studies  directed at that specific set of compounds.
This hurdle may be avoided by augmenting the known structures with those of the other families.
In addition, identified gaps in the Mendeleev maps suggest potential new compounds,
both within each family or by correlations of similar structure maps across the different families.

\subsection{Data methodology}

\tab
\mycaption{Data extraction numerical summary.}
\tabvspace
\begin{tabular}{l|r|r|r}
 & compounds & unique compounds & structure types \\
\hline
total                            & 88,373    & 50,294& 13,324   \\
\hline
unary              & 1752    & 499   & 197\\
\hline
binary & 27,487 & 10,122 & 1,962 \\
binary oxides  & 3,256 & 844 & 538 \\
binary sulfides & 1,685 & 495 & 270 \\
binary selenides & 1,050 & 332 & 168 \\
\hline
ternary  & 37,907 & 23,398&  4,409\\
ternary oxides  & 10,350 & 5,435 & 2,079 \\
ternary sulfides & 3,190 & 2,041 & 784 \\
ternary selenides & 1,786 & 1,256 & 521\\
\hline
quaternary & 15,138 & 11,050 & 3,855 \\
5 atoms         & 4,638 & 3,899 & 2,053 \\
6 atoms & 1,219 & 1,101 & 682 \\
7 atoms & 212	& 201 &	154 \\
8 atoms          & 20  & 20   & 12\\
\end{tabular}
\label{tab:art130:ICSD_DATA}
\etab

The \ICSD~\cite{ICSD_database} includes approximately 169,800 entries (as of August 2016).
For this study we exclude all entries with partial or random occupation and those that do not have full structure data.
The remaining set of structures has been filtered using the \AFLOW\ software~\citeAFLOW,
which uses an error checking protocol to ensure the integrity of each entry.
\AFLOW\ generates each structure by appropriately propagating the Wyckoff positions of the specified spacegroup.
Those structures that produce inconsistencies, \eg, overlapping atoms or a different stoichiometry
than the structure label are ignored.
If atoms are detected to be too close ($\leq 0.6$\AA), alternative standard ITC
(International Table of Crystallography)~\cite{tables_crystallography} settings of the spacegroup are attempted.
These settings define different choices for the cell's unique axes, possibly
causing atoms to overlap if not reported correctly.
Overall, these considerations reduce the full set of \ICSD\ entries to a
much smaller set of 88,373 ``true'' compounds.
These entries are contained in \AFLOW\ Database~\citeAFLOWLIB.
They include the results of the \AFLOW\ generated full symmetry analysis for each structure, \ie, Bravais lattice,
space group and point group classifications, and Pearson symbol
(the method and tolerances used for this analysis follow the \AFLOW\ standard~\cite{curtarolo:art104}).
For the analysis presented here we identify all the binary and ternary compounds included in this set,
27,487 binary entries and 37,907 ternary entries.
From these, we extract all the entries that contain oxygen, sulfur or selenium as one of the components.
Of the binaries, we find 3,256 oxides, 1,685 sulfides and 1,050 selenides.
10,530 oxides, 3,190 sulfides and 1,786 selenides are found among the ternaries.
Duplicate entries representing different experimental reports of the same compound,
\ie, the same elements, stoichiometry, space group and Pearson designation, are then eliminated
to obtain a list in which every reported compound is represented by its most recent corresponding entry in the \ICSD.
This reduces our list of binaries to 844 oxides, 495 sulfides and 332 selenides, and
the list of ternaries to 5,435 oxides, 2,041 sulfides and 1,256
selenides.
These results are summarized in Table~\ref{tab:art130:ICSD_DATA}.
Throughout the rest of the study, we will refer to these sets of
binary and ternary compounds. We choose not to discuss multi-component structures with four or more elements since their
relative scarcity in the database most probably indicates incomplete
experimental data rather than fundamental issues of their chemistry.
It is also instructive to check the effect of element abundance on the number of compounds.
The abundance of oxygen in the earth's crust is $\sim47\%$ by weight,
around 1000 times more than that of sulfur ($\sim697$~ppm) which is around $5,000$
more abundant than selenium ($120$~ppb)~\cite{wedepohl1995composition}.
Comparison with the number of elements (O/S/Se) binary compounds, 844/495/332,
or ternary compounds, 5,435/2,041/1,256, makes it clear that while a rough correlation
exists between the elements' abundance and the number of their known compounds,
it is by no means a simple proportion.

\fig
\includegraphics[width=0.6\linewidth]{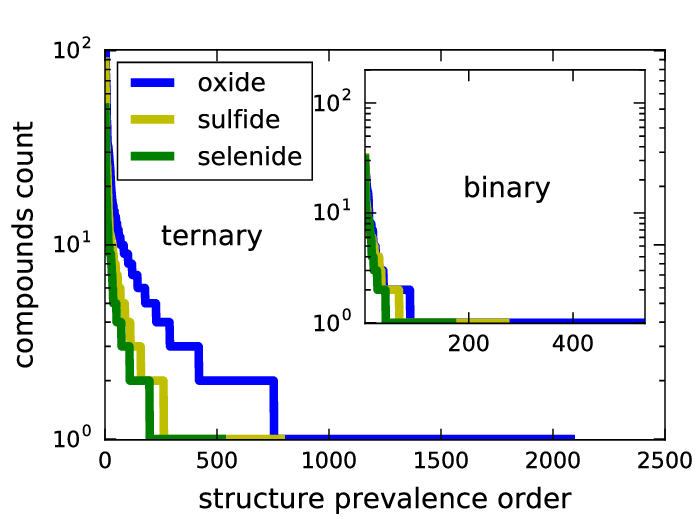}
\mycaption[Distributions of the compounds among structure types for binary (inset) and ternary compounds.]
{Oxides are shown in blue, sulfides in yellow and selenides in green.
The binary distributions differ mostly by the length of their single-compound prototypes tails,
while the ternary distribution of the oxides deviates significantly from those of the sulfides and selenides.}
\label{fig:art130:prototypes_distribution_curves_log}
\efig

In the next stage, we identify unique structure types.
Structure types are distinguished by stoichiometry, space group, and Pearson designation, without consideration
of the specific elemental composition.
This implicit definition of structure type is common in the literature~\cite{Villars2013, PaulingFile},
and we use it throughout the study as
providing a good balance of clarity and simplicity.
However, it should be noted that there are a few rare cases of complex structures where a given
structure type under this definition includes a few sub-types (see Figure~\ref{fig:art130:structure_types_comparison}).
Examples exist of more complex definitions of structure types, formulated to define similarities
between inorganic crystals structures~\cite{lima1990nomenclature}.

The binary structure type lists contain 538 oxides, 270 sulfides and 168 selenides.
The ternary lists contain 2,079 oxides, 784 sulfides and 521 selenides.
This means that 64\% of the binary oxides, 55\% of the sulfides and 51\% of the selenides are distinct structure types.
The corresponding ratios for the ternaries are 38\% of the oxides, 38\% of the sulfides and 41\% of the selenides.
All the other entries in the compound lists represent compounds of the same
structure types populated by different elements.
Differently put, this means that there are on average about 1.6 compounds per structure type in the binary oxides,
$1.8$ in the binary sulfides and $2$ in the binary selenides.
Among the ternaries, the corresponding numbers are 2.6 compounds per structure type in the oxides,
$2.6$ in the sulfides and $2.4$ in the selenides.
These numbers may be compared to the intermetallics, where there are
20,829 compounds of which 2,166, about 10\%,
are unique structure types~\cite{dshemuchadse2014some}.
There are about seven compounds per structure types in the binary intermetallics and about nine in the ternaries.
The number for binary intermetallics is considerably larger than for ternary oxides, sulfides or selenides. Together with the higher proportion of unique structure types in the latter, this reflects the limits on
materials chemistry imposed by the presence of one of those 6A elements.

\fig
\includegraphics[width=1.0\linewidth]{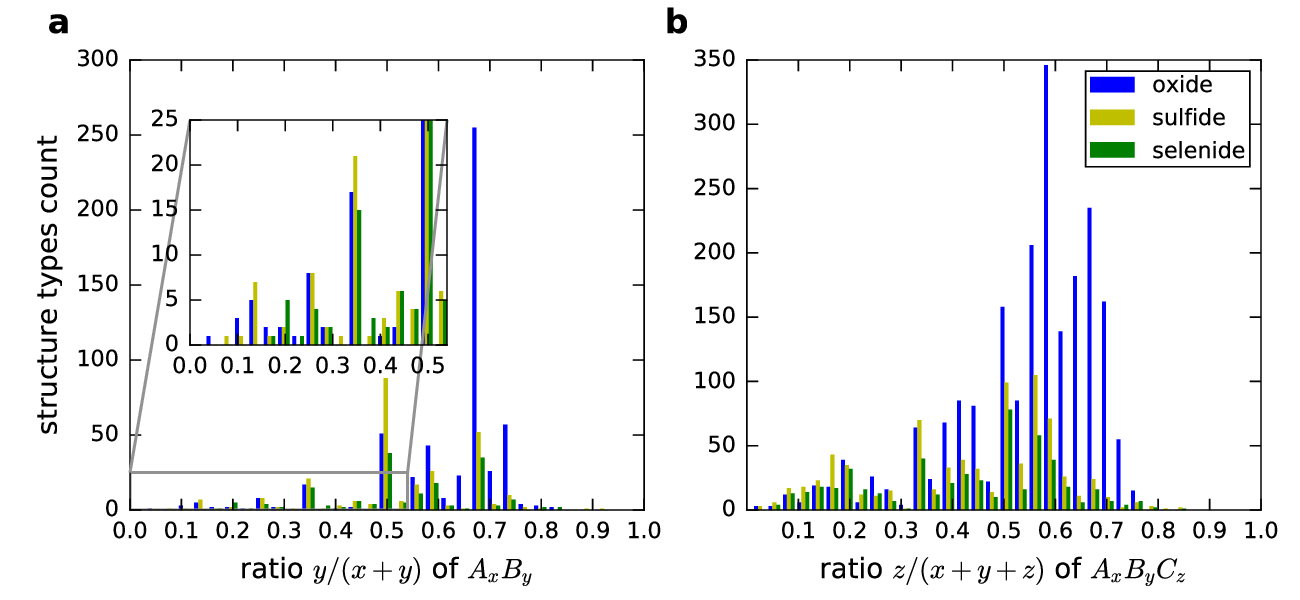}
\mycaption[Distributions of structure types among (\textbf{a}) binary and (\textbf{b}) ternary stoichiometries.]
{Oxides are shown in blue, sulfides in yellow and selenides in green.
The distributions of the selenides and sulfides are quite similar while those of the
oxides deviate significantly, as detailed in the text.}
\label{fig:art130:stoi_hist}
\efig

It should be noted that this structure selection procedure produces lists that partially overlap,
\ie, certain structure types may appear in more than one list,
since there might be oxide structure types that are also represented among
the sulfide or selenide structures, and vice versa.
11\% of the binary oxide structure types also appear in the binary sulfides list and 8\% are represented
in the binary selenides list.
33\% of the binary sulfide are also represented in the selenides list.
The total number of binary oxides, sulfides and selenides structure types is 976, which is reduced by 16\%,
to 818 structure types, by removing all overlaps.
The corresponding overlap ratios for the ternaries are 10\% for the oxides and sulfides,
6\% for the oxides and selenides and 31\% for the sulfides and selenides.
The total number of entries in the ternary oxides, sulfides, and selenides structure type lists is 3,384,
which is reduced to 2,797 structure types by removing all overlaps, a 17\% reduction.
Therefore, the overlaps between these three compound families are similar for the binaries and ternaries.
In both, the overlap between the oxides and the other two families is rather small,
whereas the overlap between the sulfides and selenides represents about a third of the total number of structure types.

The sequence of Mendeleev numbers includes 103 elements, from hydrogen to lawrencium
with numbers 1-6 assigned to the noble gases, 2-16 to the alkali metals and alkaline earths,
17-48 to the rare earths and actinides, 49-92 to the metals and metalloids and 93-103 to the non-metals.
Of these, noble gases are not present in compounds and artificial elements
(metals heavier than uranium) have very few known compounds.
We are thus left with 86 elements, of which the above compounds are composed.
That means there are about ten times more binary oxides than
element-oxygen combinations, about six times more sulfides than element-sulfur
combinations and four times more selenides than element-selenium combinations.
Oxides are much more common than sulfides and selenides.
The corresponding numbers for the ternaries are much lower.
There are about 1.6 times more ternary oxides than two-element-oxygen ternary possible systems,
about 0.6 times less ternary sulfides and about 0.4 times less ternary selenides than the corresponding two-element combinations.
The ternaries are relatively quite rare, more so as we progress from oxides to sulfides and then to selenides.
A similar analysis of the intermetallic binaries in Reference~\onlinecite{dshemuchadse2014some} shows that of the 20,829 intermetallics,
277 are unaries (about three times more than possible metal elements), 6,441 are binaries
(about two times more than possible metal binary systems),
and 13,026 are ternaries (6.5 times less than possible metal ternary systems).
This means that unary metal structures are less common among the metallic
elements than the oxide, sulfide and binary selenide compounds among their corresponding binary systems.
This seems to reflect simply the larger space of stoichiometries available to binaries over unaries.
However, on the contrary, the intermetallic binary compounds are more common among the metallic binary
systems than the oxide, sulfide and ternary selenide compounds among their corresponding ternary systems.
This discrepancy again reflects either the chemical constraints imposed by the presence of a 6A non-metal on the
formation of a stable ternary structure, or simply gaps in the
experimental data since many ternary systems have not been thoroughly investigated.

\fig
\includegraphics[width=0.55\linewidth]{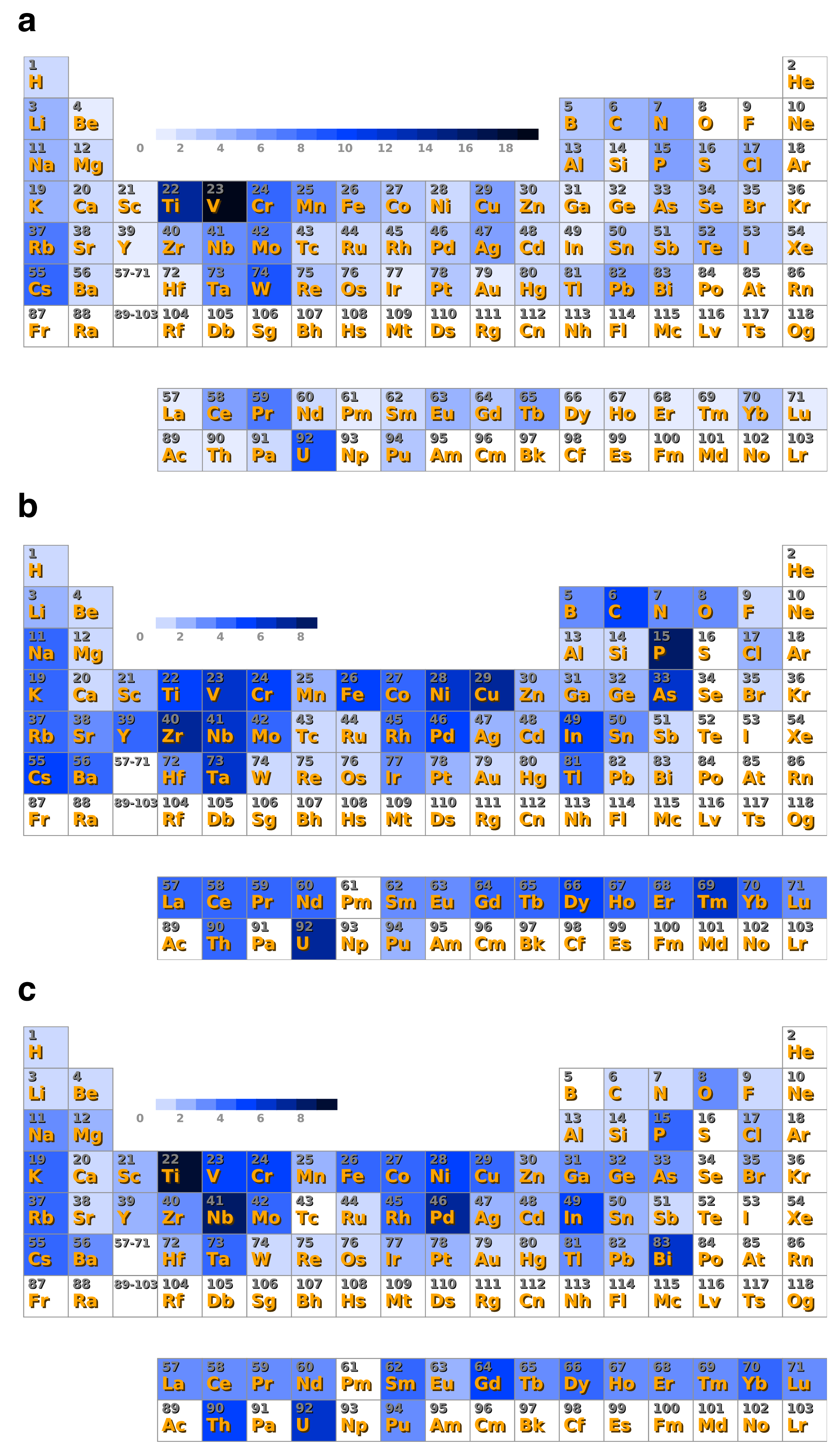}
\mycaption[Composition distributions of binary (\textbf{a}) oxide, (\textbf{b}) sulfide, and (\textbf{c}) selenide stoichiometries.]
{The count indicates the number of different stoichiometries that include the respective element.
The colors go from no stoichiometries (white) to the maximal number of stoichiometries
(dark blue) which is different for each element, 19/8/9 for O/S/Se.
Islands of high prevalence appear for the 4B and 5B transition metals and
the heavy alkalis in all three compound families.
Additional, smaller islands appear in the sulfides and selenides for the 8 and
1B transition metals and the 3A and 5A semi-metals.}
\label{fig:art130:stoi_periodic_oxide}
\efig

\subsection{Results and discussion}
\boldsection{Structure types.}
The distribution of the binary and ternary compounds among the corresponding structure types is shown
in Figure~\ref{fig:art130:prototypes_distribution_curves_log}.
Detailed data for the most common structure types is presented in Tables~\ref{tab:art130:oxide_binary_data}-\ref{tab:art130:selenide_ternary_data}.

About 84\% of the binary oxide structure types represent a single
compound, characterizing the tail end of the binary oxide distribution.
They include about 53\% of the binary oxide compounds.
The most common structure type represents 29 compounds,
3.4\% of the oxide compounds list.
Among the binary sulfides, 76\% of the structure types represent a single compound.
They include 41\% of the binary sulfide compounds.
The most common structure type represents 32 compounds, 6.5\% of the
sulfide compounds list.
Among the binary selenides, 76\% of the structure types represent a single compound.
They include 39\% of the binary selenide compounds.
The most common structure type represents 31 compounds, 9.3\% of the selenide compounds list.

In all three binary lists the most common structure type is rock salt (NaCl).
The binary oxide structure type distribution has a much longer tail than the sulfides and selenides,
\ie, more oxide compounds have unique structure types.
The most common structure type in these three distributions represents
a similar number of compounds but a smaller proportion of the corresponding compounds in the oxides.
The middle regions of the distributions are very similar
(inset Figure~\ref{fig:art130:prototypes_distribution_curves_log}).
This means that the much larger number of binary oxide compounds, compared to the sulfides and selenides,
is expressed at the margin of the distribution, in the long tail of unique compounds.

This discrepancy between the three binary distributions is much less
apparent among the ternary compounds.
64\% of the ternary oxide structure types represent a single compound.
They include 24\% of the ternary oxide compounds.
The two most common structure types, pyrochlore and perovskite, represent 116 and 115 compounds,
respectively, about 2\% each of the entire compounds list.
Among the ternary sulfides, 70\% of the structure types represent a single compound.
They include 34\% of the ternary sulfide compounds.
The most common structure type, delafossite, represents 65 compounds,
4\% of the entire compounds list.
Among the ternary selenides, 62\% of the structure types represent a single compound.
They include 26\% of the ternary selenide compounds.
The most common structure type, again delafossite, represents 51 compounds, 4\% of the ternary sulfides.

In contrast to the binaries, the larger count of ternary oxides, compared to the sulfides and selenides,
is expressed by a thicker middle region of the structure type distribution,
whereas the margins have a similar weight in the distributions of the three compound families.

\boldsection{Binary stoichiometries.}
The structure types stoichiometry distribution for the binary oxide, sulfide and selenide compounds is shown in
Figure~\ref{fig:art130:stoi_hist}(a).
We define the binaries as $A_xB_y$, where $B$ is O, S or Se, and the number of structure types is shown as a function of $y/(y+x)$.
A very clear peak is found for the oxides at the stoichiometry 1:2,
$A$O$_2$, while both the sulfides and selenides have a major peak at 1:1, $A$S and $A$Se, respectively.

For $y/(y+x)<0.5$, there are more gaps in the plot (missing stoichiometries) for the oxides compared
to the sulfides and selenides, while for $y/(y+x)>0.6$ there are more gaps in the sulfides and selenides,
this behavior is shown in detail in Tables~\ref{tab:art130:Prevalence_of_Elements_in_Binaries_Stoichiometries}-\ref{tab:art130:Prevalence_of_Elements_in_Binaries_Stoichiometries_3}.
An important practical conclusion is that augmenting the binary oxide structure types with
those of sulfides and selenides will produce a more extensive coverage of possible stoichiometries.

Another interesting property is the number of stoichiometries
for each of the elements in the periodic table.
The prevalence of binary oxide stoichiometries per element is shown in Figure~\ref{fig:art130:stoi_periodic_oxide}(a).
A few interesting trends are evident --- the first row of transition metals shows a peak near vanadium (19 stoichiometries)
and titanium (14 stoichiometries).
Hafnium, which is in the same column of titanium has only a single stoichiometry --- HfO$_{2}$.
Both the beginning and end of the $d$-elements exhibit a small amount of stoichiometries --- scandium
with only one and zinc with only two.
The two most abundant elements, silicon and oxygen, form only a single stoichiometry in the
\ICSD\ --- SiO$_{2}$, with 185 {\it different} structure types.
Another interesting trend is evident for the alkali metals, where rubidium and cesium have more
stoichiometries --- perhaps related to the participation of $d$-electrons in the chemical bonds.

Figures~\ref{fig:art130:stoi_periodic_oxide}(b) and (c)
show the binary stoichiometries prevalence per element for sulfur and selenium respectively.
Similar trends are exhibited --- there are two ``islands'' of large number of stoichiometries
in the transition metals: one around vanadium and titanium and the other near nickel and copper.
Evidently, prime candidates for new compounds should be searched among structures in the vicinity of
these high density islands, especially for elements that exhibit a considerably higher density in one family.

\boldsection{Ternary stoichiometries.}
Similar to the binaries, the ternary stoichiometries are designated
$A_xB_yC_z$, where $C$ is O, S or Se.
The distributions of the ternaries are, as might be expected, more complex,
with maxima at $z/(x+y+z)=0.6$ for the oxides, $z/(y+x+z)=0.55$ for the sulfides and $z/(y+x+z)=0.5$ for the selenides.
The major peaks still appear at integer and half integer values, but with more minor peaks at intermediate values.
This behavior is shown in Figure~\ref{fig:art130:stoi_hist}(b).
The ternary selenide and sulfides distributions are again nearly identical, and there are
almost no compounds with ratios larger than $0.75$ in the oxides or larger than $0.66$ in the sulfides and selenides.
However, there are few sulfide and selenide compounds around 0.8 and 0.85 but no oxides.

\tab
\mycaption[Ternary stoichiometry data: $A_xB_yC_z$.]{``$C$-rich'' refers to stoichiometries where $z>x+y$.}
\tabvspace
\begin{tabular}{ l|r|r|r  }
& oxygen & sulfur & selenium \\
\hline
Number of stoichiometries    & 585   & 282 & 206   \\
$C$-rich stoichiometries ratio & 0.85    & 0.67   & 0.66 \\
$C$-rich compound ratio  & 0.92 & 0.77 & 0.73 \\
\end{tabular}
\label{tab:art130:ternary_stoi}
\etab

\fig
\includegraphics[width=1.0\linewidth]{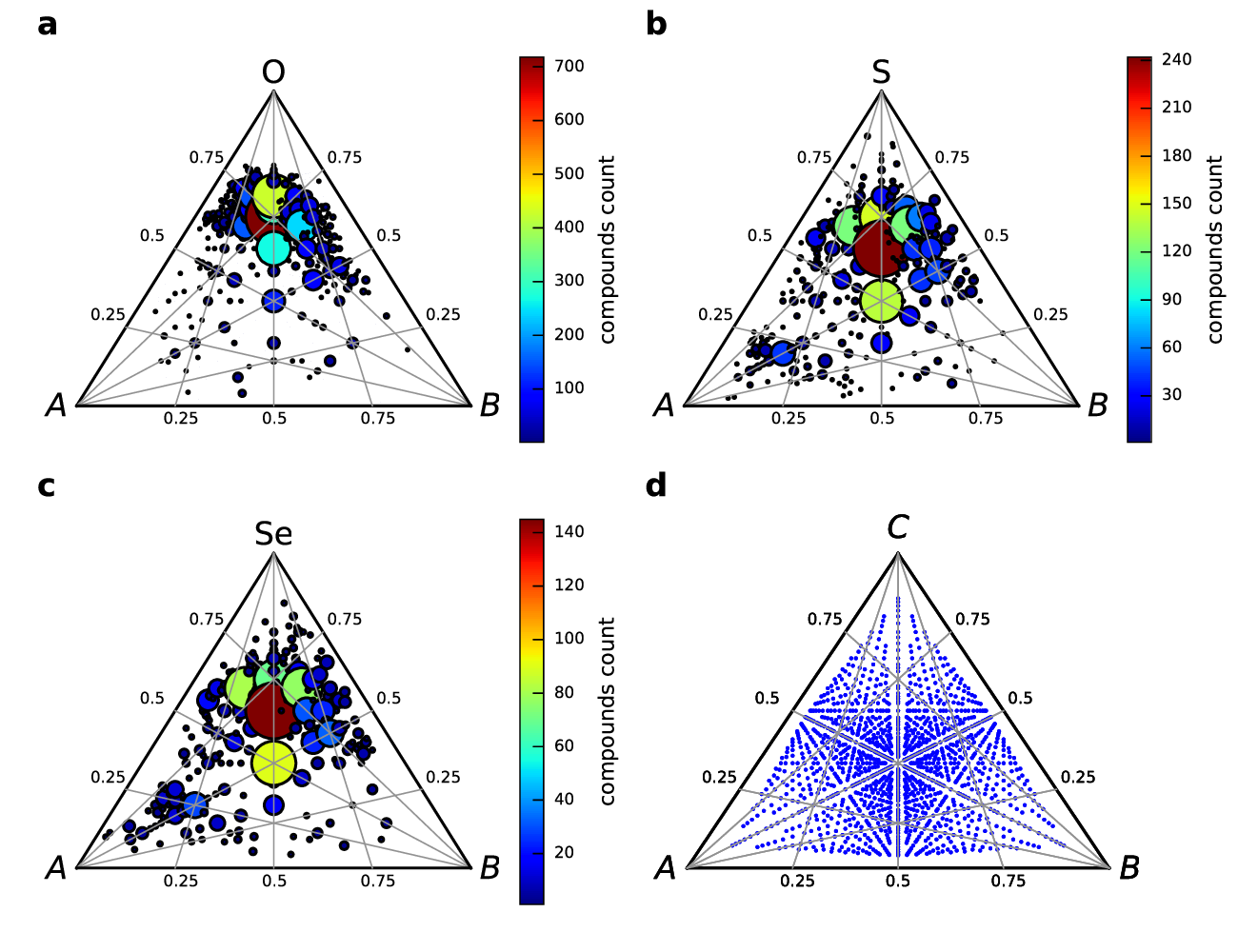}
\mycaption[Prevalence of stoichiometries among ternary compounds.]
{Panels include
(\textbf{a}) oxide,
(\textbf{b}) sulfide,
(\textbf{c}) selenide compounds,
and, for reference,
(\textbf{d}) all the possible stoichiometries with up to 12 atoms
of each component per unit cell.
In each figure, the smaller circles are normalized to the biggest one, which denotes the highest prevalence, \ie,
718 for oxides, 242 for sulfides, and 145 for selenides,
in addition a heat map color scheme is used where blue means low prevalence and red means the highest prevalence for each element.
The $x$ and $y$ axes denote the atomic fractions in the ternaries $A_xB_yC_z$, where $C$ is O, S or Se, respectively.
$A$ and $B$ are ordered by Mendeleev number where $M_A>M_B$.}
\label{fig:art130:triangles}
\efig

Another perspective of ternary stoichiometries is demonstrated in Figure~\ref{fig:art130:triangles}
which shows the abundance of the most common stoichiometries.
The biggest circle in each diagram denotes the prevalence of the most common stoichiometry
(number of unique compounds for this stoichiometry),
which is 718 ($x=1$, $y=1$, $z=3$) for oxides, 242 ($x=1$, $y=1$, $z=2$) for sulfides, and 145 ($x=1$, $y=1$, $z=2$) for selenides.
The smaller circles in each plot are normalized to the corresponding highest prevalence.

These diagrams highlight the similarities as well as important differences between the three families of compounds.
In all three cases, the most common stoichiometries appear on the symmetry axis of the diagram, \ie,
at equal concentrations of the $A$ and $B$ components, or very close to it.
For the oxides, they are concentrated near 0.5-0.6 fraction of oxygen, representing the
$A_1B_1$O$_2$ and $A_1B_1$O$_3$ stoichiometries, respectively,
and form a very dense cluster with many similar reported stoichiometries of lower prevalence.
Outside this cluster, the occurrence of reported compositions drops sharply, and other regions
of the diagram are very sparsely populated, in particular near the vertices of the $B$ and O components.

The sulfide and selenide diagrams also exhibit prominent clusters on the $AB$ symmetry axes,
but they appear at a lower S or Se concentration of about 0.5, \ie, $A_1B_1C_2$ stoichiometry.
They are considerably more spread out and include a significant contribution at the $ABC$ stoichiometry.
In both sulfides and selenides, an additional minor cluster appears closer to the $A$ vertex (Figure~\ref{fig:art130:triangles}).
A few members of this cluster are ternary oxides, reflecting the high electronegativity and high Mendeleev number (101) of oxygen.
The $B$ and $C$ vertex regions are still sparsely populated, but less so than in the oxides case.
Overall, the sulfide and selenide diagrams are very similar to each other and different from that of the oxides.
They are more spread out, less $AB$ symmetric than the oxide diagram and less tilted towards rich $C$-component concentration.
This discrepancy may reflect some uniqueness of oxygen chemistry compared to sulfur and selenium,
or rather simply reflect the oxygen rich environment in which naturally formed compounds are created in the atmosphere.
The number of stoichiometries and the differences in the $C$-component concentration are summarized in Table~\ref{tab:art130:ternary_stoi}.

\tab
\mycaption{Distribution of the oxide, sulfide and selenide compounds and structure types among the 14 Bravais lattices.}
\tabvspace
\resizebox{\linewidth}{!}{
\begin{tabular}{l|r|r|r|r|r|r|r|r|r|r|r|r|r|r|r|r|r|r}
& \multicolumn{3}{R{3cm}|}{binary compounds}
& \multicolumn{3}{R{3cm}|}{binary structure types}
& \multicolumn{3}{R{3cm}|}{binary compounds per structure type}
& \multicolumn{3}{R{3cm}|}{ternary compounds}
& \multicolumn{3}{R{3cm}|}{ternary structure types}
& \multicolumn{3}{R{3cm}}{ternary compounds per structure type}\\ \hline
& O & S & Se & O & S & Se & O & S & Se & O & S & Se & O & S & Se & O & S & Se \\ \hline
aP & 51 & 13 & 5 & 39 & 12 & 5 & 1.3 & 1.1 & 1 & 378 & 79 & 60 & 219 & 56 & 39 & 1.7 & 1.4 & 1.5 \\
mP & 82 & 54 & 31 & 62 & 36 & 20 & 1.3 & 1.5 & 1.6 & 918 & 318 & 198 & 363 & 166 & 109 & 2.5 & 1.9 & 1.8 \\
mS & 88 & 31 & 22 & 58 & 21 & 15 & 1.5 & 1.5 & 1.5 & 672 & 251 & 170 & 292 & 117 & 77 & 2.3 & 2.1 & 2.2 \\
oP & 123 & 82 & 48 & 81 & 37 & 30 & 1.5 & 2.2 & 1.6 & 950 & 481 & 266 & 373 & 139 & 105 & 2.5 & 3.5 & 2.5 \\
oS & 39 & 24 & 11 & 36 & 19 & 9 & 1.1 & 1.3 & 1.2 & 334 & 84 & 60 & 133 & 40 & 25 & 2.5 & 2.1 & 2.4 \\
oF & 11 & 7 & 11 & 10 & 6 & 4 & 1.1 & 1.2 & 2.8 & 51 & 32 & 23 & 28 & 14 & 8 & 1.8 & 2.3 & 2.9 \\
oI & 22 & 5 & 2 & 20 & 4 & 2 & 1.1 & 1.25 & 1 & 89 & 36 & 27 & 39 & 15 & 12 & 2.3 & 2.4 & 2.25 \\
tI & 41 & 20 & 10 & 31 & 17 & 8 & 1.3 & 1.2 & 1.25 & 418 & 80 & 72 & 101 & 34 & 23 & 4.1 & 2.4 & 3.1 \\
tP & 78 & 27 & 28 & 48 & 13 & 16 & 1.6 & 2.1 & 1.75 & 239 & 73 & 52 & 107 & 39 & 26 & 2.2 & 1.9 & 2.0 \\
hP & 94 & 87 & 66 & 62 & 50 & 32 & 1.5 & 1.7 & 2.1 & 435 & 224 & 103 & 198 & 75 & 41 & 2.2 & 3.0 & 2.5 \\
hR & 40 & 44 & 20 & 30 & 33 & 15 & 1.3 & 1.3 & 1.3 & 420 & 230 & 133 & 123 & 49 & 33 & 3.4 & 4.7 & 4.0 \\
cP & 42 & 22 & 20 & 21 & 6 & 4 & 2.0 & 3.7 & 5.0 & 187 & 58 & 43 & 45 & 18 & 13 & 4.2 & 3.2 & 3.3 \\
cF & 75 & 65 & 48 & 19 & 10 & 6 & 3.9 & 6.5 & 8.0 & 251 & 80 & 43 & 27 & 17 & 7 & 9.3 & 4.7 & 3.9 \\
cI & 58 & 14 & 10 & 21 & 6 & 2 & 2.8 & 2.3 & 5.0 & 92 & 15 & 6 & 30 & 5 & 3 & 3.1 & 3.0 & 2.0 \\
\end{tabular} }
\label{table:bravais_lattice_distribution}
\etab

Another interesting observation is that while some stoichiometries are abundant in the oxides
they are almost absent in the sulfides or the selenides. For example,
there are 299 compounds with the $A_2B_2$O$_7$ stoichiometry (ignoring
order between $M_A$ and $M_B$), but only two $A_2B_2$S$_7$ compounds
and no $A_2B_2$Se$_7$ compounds. Also, there are 71 $A_1B_3$O$_9$
compounds but no $A_1B_3$S$_9$ and $A_1B_3$Se$_9$ compounds. On the
other hand, there are no $A_4B_{11}X_{22}$ oxides, but 20 sulfides and 8 selenides.
If we require that $M_A>M_B$, there are no oxides of the $A_3B_2X_2$
stoichiometry, but 25 sulfides and 7 selenides.

\fig
\includegraphics[width=0.55\linewidth]{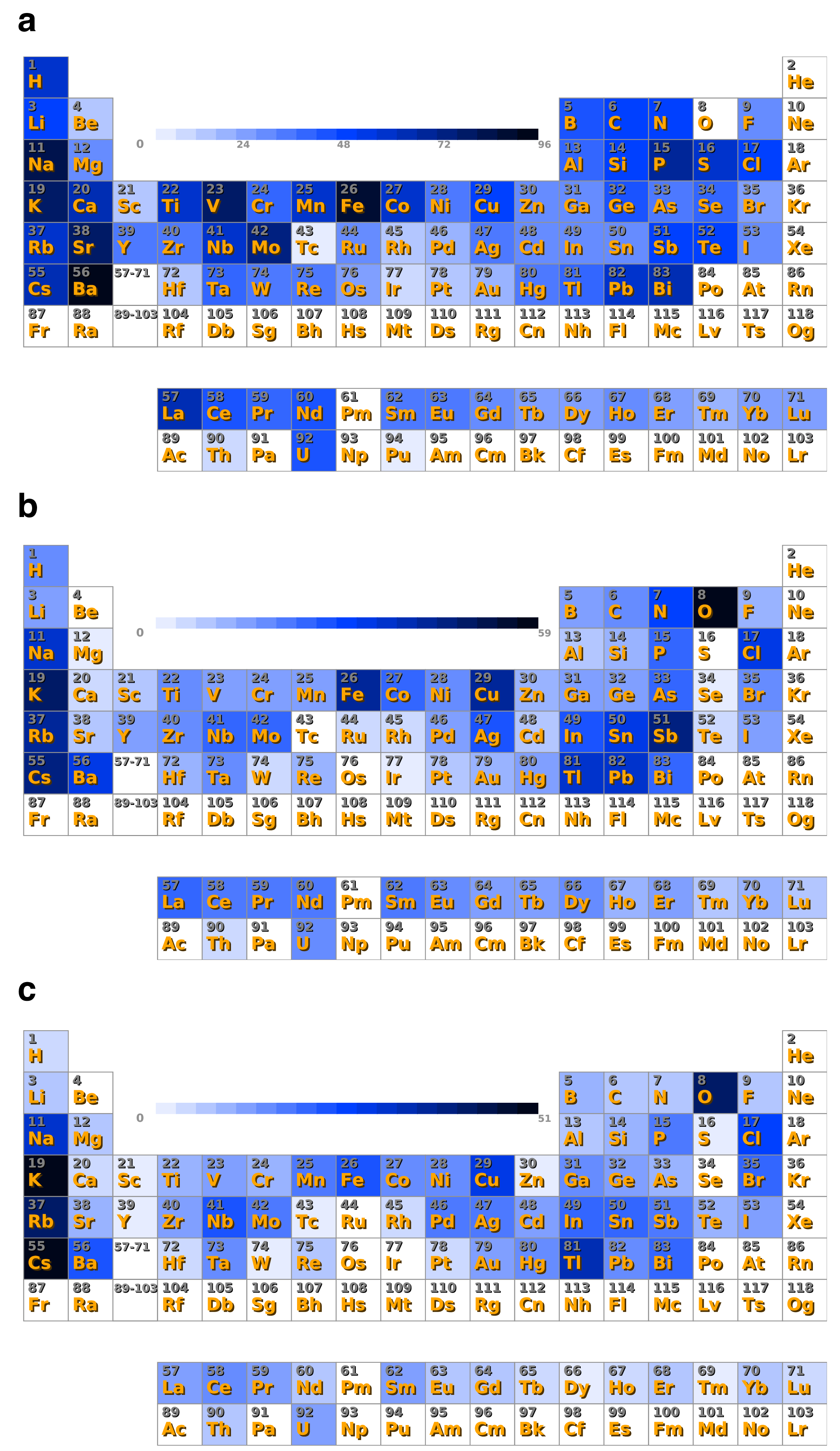}
\mycaption[Composition distributions of ternary (\textbf{a}) oxide, (\textbf{b}) sulfide, and (\textbf{c}) selenide stoichiometries.]
{The count indicates the number of different stoichiometries that include the respective element.
The colors go from no stoichiometries (white) to the maximal number of stoichiometries (dark blue)
which is different for each element, 96/59/51 for O/S/Se.
High prevalence appears for the alkali metals in all three compound families.
An additional island in the transition metals is much more pronounced in the oxides.
The sulfides and selenides distributions are nearly identical, and show high prevalence of oxygen containing ternaries.}
\label{fig:art130:tern_stoi_periodic_oxide}
\efig

Again, an important conclusion is that there are many missing stoichiometries,
Figure~\ref{fig:art130:triangles}(d) shows all the possible stoichiometries for $A_xB_yC_z$ for $x,y,z \le 12$,
clearly showing rich concentration in the middle, which is not the case for oxides, and also to a
lesser degree to sulfides and selenides.

We can repeat the analysis of the binary stoichiometries and ask how many stoichiometries
per element are there for the ternaries.
This is shown in Figure~\ref{fig:art130:tern_stoi_periodic_oxide}.
Here, also, the similarity of sulfides and selenides is clear.
In addition, while there are similarities between the distributions of binary stoichiometries
per element to the ternary distributions, there are also obvious differences.
One might guess that there should be a correlation between the binary and ternary distributions.
This is examined in Figure~\ref{fig:art130:bintern1a}(a).

It is evident that the correlation between ternary and binary number of stoichiometries is not strong
but the minimal number of ternary stoichiometries tends to grow with the number of binary stoichiometries.
We check this further in Figure~\ref{fig:art130:bintern1a}(b), by comparing the number of ternary stoichiometries of
$A_xB_y$O$_z$ to the product of stoichiometry numbers of $A_x$O$_y$ and $B_x$O$_y$.
The general trend obtained is an inverse correlation, \ie, as the product of the numbers of binary
stoichiometries increases, the number of ternaries decreases. This trend can be explained by the following argument: when the two binaries are rich with stable compounds, the ternaries need to compete with more possibilities of
binary phases, which makes the formation of a stable ternary more difficult.
In Figure~\ref{fig:art130:bintern1a}(b), this trend is highlighted for vanadium,
the element with the most binary stoichiometries, but this pattern repeats itself for most elements.
We analyze this behavior for the sulfides and selenides in Section~\ref{subsec:art130:prev_stoich_supp}, similar trends are found but
they are less pronounced due to a smaller number of known compounds.

\fig
\includegraphics[width=1.0\linewidth]{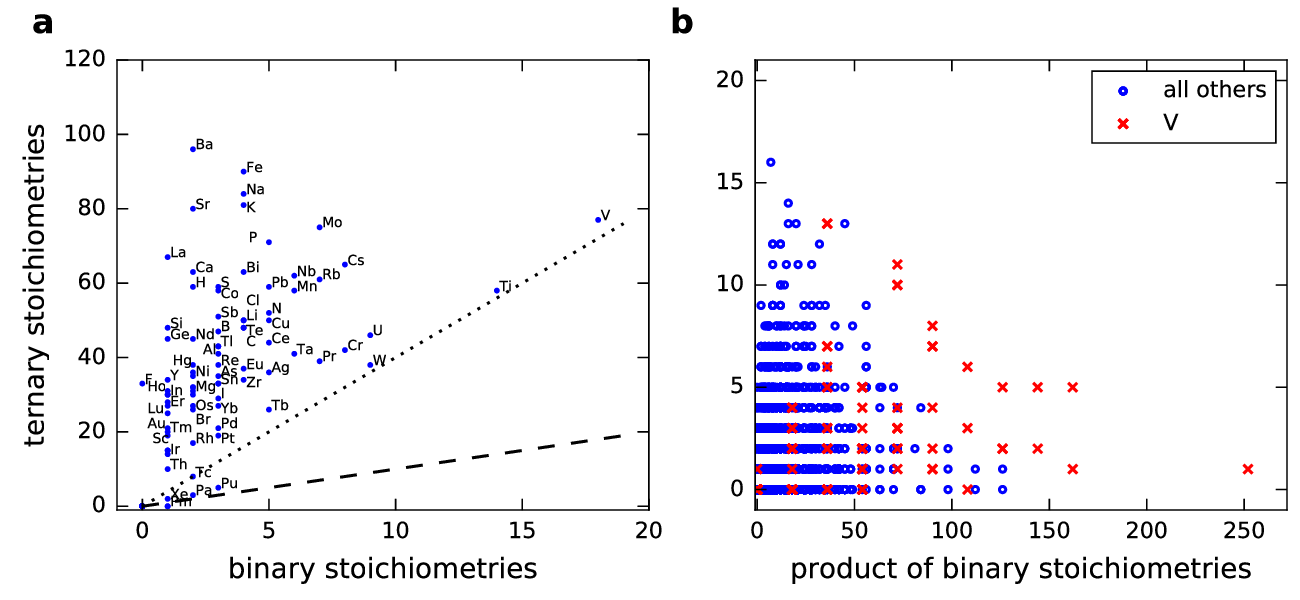}
\mycaption[Analysis of ternary \vs\ binary stoichiometry counts for oxides.]
{(\textbf{a}) The number of ternary oxide stoichiometries per element as a function
of the count of its binary stoichiometries.
The dashed line marks perfect similarity $(y=x)$, and the dotted line marks the ratio $y=4x$.
(\textbf{b}) The number of ternary oxide stoichiometries as a function of the
product of the numbers of the binary stoichiometries of participating elements.
The data for vanadium is shown with red crosses, all the rest is shown with blue circles.}
\label{fig:art130:bintern1a}
\efig

\boldsection{Composition and Mendeleev maps.}
The occurrence of each element in the binary and ternary compound lists has been
counted and tabulated.
The results are described in Figure~\ref{fig:art130:mendeleev_distribution_all_in_one}.
For the binary oxides a very prominent peak appears at $M=85$, the
Mendeleev number of silicon.
It represents the 185 different silicon oxide
structures types reported in the \ICSD\ database for just a {\it single} stoichiometry, SiO$_2$.
Smaller peaks appear for $M=51$ (titanium, 42 structure types, 14 stoichiometries,
leading stoichiometry is
TiO$_2$ with 14 structure types), $M=54$ (vanadium, 42 structure types,
18 stoichiometries, leading stoichiometry is VO$_2$ with 10 structure types),
$M=56$ (tungsten, 24 structure types, 9 stoichiometries, leading stoichiometry is WO$_3$ with 13 structure types),
and $M=45$
(uranium, 22 structure types, 9 stoichiometries, leading stoichiometries are UO$_2$ and U$_3$O$_8$ with 6 structure types each).
Unlike the silicon peak which is composed of a single stoichiometry,
the other leading peaks evidently include multiple stoichiometries, reflecting the different chemistry of those elements.
These differences also carry over into the ternary oxide compounds involving those elements.
For example, the stoichiometry distribution of silicon ternary oxides is more tilted towards
the silicon poor compounds compared to the corresponding distributions of vanadium and titanium ternary oxides,
as is shown in Figure~\ref{fig:art130:specific_triangle_stoichiometries}.

The distribution of the sulfides is generally much lower than
that of the oxides, due to the much smaller total number of known binaries, but is also more uniformly structured.
It has one major peak
for $M=76$ (zinc, 40 structure types, 2 stoichiometries, leading stoichiometry is ZnS with 39 structure types),
and quite a few smaller ones such as $M=51$ (titanium, 16 structure types, 5 stoichiometries, leading stoichiometry is TiS$_2$ with 9 structure types),
$M=61$ (iron, 18 structure types, 5 stoichiometries, leading stoichiometry is FeS with 6 structure types),
$M=67$ (nickel, 16 structure types, 6 stoichiometries, leading stoichiometry is NiS$_2$ with 8 structure types),
$M=90$ (phosphorus, 13 structure types, 8 stoichiometries, of which
P$_2$S$_7$, P$_4$S$_9$, P$_4$S$_6$, P$_4$S$_5$ and P$_4$S$_3$ have 2 structure types each).
The $M$~=~8--33 region also exhibits a minor concentration of
participating elements.
The selenides distribution is yet smaller than that of the sulfides, and
even more uniform.
Several peaks appear, $M=51$ (titanium, 13 structure types, 9 stoichiometries, leading stoichiometry is
TiSe with 3 structure types),
$M=52$ (niobium, 15 structure types, 8 stoichiometries, leading stoichiometry is
NbSe$_2$ with 8 structure types),
$M=53$ (tantalum, 15 structure types, 4 stoichiometries, leading stoichiometry is
TaSe$_2$ with 10 structure types) and
$M=79$ (indium, 14 structure types, 5 stoichiometries, leading stoichiometry is In$_2$Se$_3$ with 6 structure types).
All distributions cover most of the elements except two obvious gaps, one at $M<9$,
which includes the noble gases and the two heaviest alkali metals, cesium and francium, and another
at $34\leq M\leq 42$ which represents the heavy actinides. Another gap appears in the sulfide and selenide distributions at $91\leq M\leq 97$,
which reflects the rarity of polonium and astatine compounds and shows that the elements of the 6A column,
except oxygen, do not coexist, in the known compounds, with each other
or with the heavier halogen iodine.

The element occurrence distributions for the ternary oxides, sulfides and
selenides exhibit greater similarity than the
corresponding binary distributions. The most apparent difference, however, is the
most common component, which is sulfur, $M=90$, in the
oxides, but oxygen itself, $M=101$, in the sulfides and selenides. The
sulfide and selenide distributions are almost the same, except for
generally lower numbers in the selenides (due to the smaller total
number of compounds) and an apparent lower participation of the
lanthanides $M$~=~17--35.

Mendeleev maps for the ternaries are shown in
Figures~\ref{fig:art130:mendeleev_bigger_x_upper_all_in_one}-\ref{fig:art130:mendeleev_sulfur_selenium_prototypes}.
Figure~\ref{fig:art130:mendeleev_bigger_x_upper_all_in_one} shows the cumulated maps for all
stoichiometries reported for the respective ternary family.
They reflect the same major gaps as the binary distributions.
The maps show that most of the reported compositions are represented by one or two compounds
with just a few hot-spots that include up to 20 compounds in the oxides and
10 compounds in the sulfides and selenides.
The oxides map is obviously denser, reflecting the much richer, currently known, chemistry of the oxides compared
to the other two elements.
The chemistry becomes more constrained as we proceed down the periodic table column from
oxygen to sulfur and then to selenium.

Next, we examine maps of specific stoichiometries.
Maps of a few notable oxide stoichiometries and their
leading structure types are shown in Figure~\ref{fig:art130:mendeleev_oxide_prototypes}.
These maps reflect the dominant features of the
full ternary oxides map (Figure~\ref{fig:art130:mendeleev_bigger_x_upper_all_in_one}),
but with significant new additional gaps of absent compounds. These gaps are naturally
wider for less prevalent stoichiometries, \ie, the map of the most
prevalent stoichiometry,  $A_1B_1$O$_3$, is denser than the three
other maps in Figure~\ref{fig:art130:mendeleev_oxide_prototypes}.
Different structure types in all stoichiometries tend to accumulate at
well defined regions of the map. The separation between them
is not perfect, but is similar to that exhibited by the classical Pettifor maps for
binary structure types~\cite{pettifor:1984,pettifor:1986}.
A similar picture is obtained for the sulfide and selenide structure types, although more sparse
(Figure~\ref{fig:art130:mendeleev_sulfur_selenium_prototypes}).
It is interesting to note that the maps of, \eg,
$A_1B_2C_4$ ($C=$ O, S, Se), show similar patterns in the map for oxides
(Figure~\ref{fig:art130:mendeleev_oxide_prototypes}) and sulfides/selenides
(Figure~\ref{fig:art130:mendeleev_sulfur_selenium_prototypes}) ---
suggesting that similar elements tend to form this stoichiometry.
In the same manner, the 2:1:1 stoichiometry shows very similar patterns in oxides, sulfides and selenides (see also Figure~\ref{fig:art130:mend_211_stoichiometries}).

\boldsection{Symmetries.}
The distribution of the compounds and structure types among the 14 Bravais lattices
is presented in Table~\ref{table:bravais_lattice_distribution} and
Figure~\ref{fig:art130:bravais_combined}.
It is interesting to note that in all six cases (binary and ternary oxides, sulfides and selenides)
the distribution is double peaked, with the majority of the compounds belonging to the
monoclinic and orthorhombic primitive lattices,
and a smaller local maximum at the hexagonal and tetragonal lattices.
All distributions exhibit a local minimum for the orthorhombic face and body centered lattices.
The high symmetry cubic lattices are also relatively rare.
This reflects the complex spatial arrangement of the compound forming electrons of oxygen, sulfur and selenium,
which does not favor the high symmetry cubic structures or the
densely packed face and body centered orthorhombic structures.

Figure~\ref{fig:art130:symmetry_distribution_of_structures}
shows a more detailed distribution of the compounds among the different space groups.
The binary compounds show a distinct seesaw structure, with a few
local peaks near the highest symmetry groups of each crystal system.
The corresponding ternary distributions have three sharp peaks in the triclinic,
monoclinic and orthorhombic systems, and much smaller peaks in the hexagonal and cubic groups.
It is interesting to note that the three compound families, exhibit distributions of very similar structure.
The oxide distributions are the densest, simply due to the existence of more oxide compounds in the database,
and become sparser in the sulfide and selenide cases.
The compounds of all these families are distributed among a rather limited number of space groups,
with most space groups represented by just a single compound or not at all.

\boldsection{Unit cell size.}
The distributions of unit cell sizes (\ie, the number of atoms per unit cell) for the six compound families we discuss
are shown in Figure~\ref{fig:art130:number_of_atoms_distribution}.
All of these distributions have strong dense peaks at small cell sizes and decay sharply at sizes above a few tens of atoms.
However, the details of the distributions differ quite significantly from group to group.
Among the binaries, the oxides exhibit the highest and widest peak with
its maximum of 102 binary oxide compounds located at 12 atoms per cell.
90\% of the binary oxides have less than 108 atoms in the unit cell and 50\% of them have less than 24 atoms.
The sulfides distribution has a lower and narrower peak of 70 compounds at 8 atoms.
The distribution of the selenides has a still lower peak of 60 compounds at 8 atoms.
The fact that oxygen has a peak at 12 atoms in the unit cell and not at 8 as the sulfides and selenides,
is related to the fact that binary oxides prefer the $A$O$_2$ stoichiometry over $A$O,
where as both sulfides and selenides prefer the 1:1 stoichiometry over 1:2.
This is probably related to the different chemistry of oxygen \vs\ sulfur and selenium.
Additional computational analysis would be required to fully understand the effect of the
different chemistry on the stoichiometry and number of atoms.
Detailed data for these dense parts of the distributions is tabulated in Tables~\ref{tab:art130:Number_of_atoms_in_Binaries_unit_cells}-\ref{tab:art130:Number_of_atoms_in_Binaries_unit_cells_3}.
The oxides distribution exhibits the longest tail of the binaries, with the largest
binary oxide unit cell including 576 atoms.
The largest binary sulfide and selenide unit cells include 376 and 160 atoms, respectively.

The distributions of the ternary compounds have higher, wider peaks and longer tails than
their binary counterparts.
The relative differences between the oxide, sulfide and selenide distributions
remain similar to the distributions of the binaries.
The ternary oxides exhibit a high and wide peak.
Its maximum of 465 compounds is located at 24 atoms per cell, and
90\% of the compounds have less than 92
atoms in the unit cell and 50\% of the compounds have less than 32 atoms.
As in the binary case, the distribution of the ternary sulfides has a lower and narrower peak than the oxides,
where the maximum of 190 compounds at 28 atoms and 90\% of the compounds have less than 72 atoms in the unit cell.
The distribution of the selenides has a still lower and narrower peak, where the corresponding numbers are
130 compounds at 28 atoms and 90\% of the compounds having less than 28 atoms in the unit cell.
Detailed data for these dense parts of the distributions is shown in Tables~\ref{tab:art130:Number_of_atoms_in_ternary_unit_cells}-\ref{tab:art130:Number_of_atoms_in_ternary_unit_cells_3}.
The ternary oxides distribution exhibits the longest tail of
the three types, with the largest ternary oxide unit cell having 1,080 atoms.
The largest ternary sulfide and selenide unit cells have 736 and 756
atoms, respectively.

\fig
\includegraphics[width=1.0\linewidth]{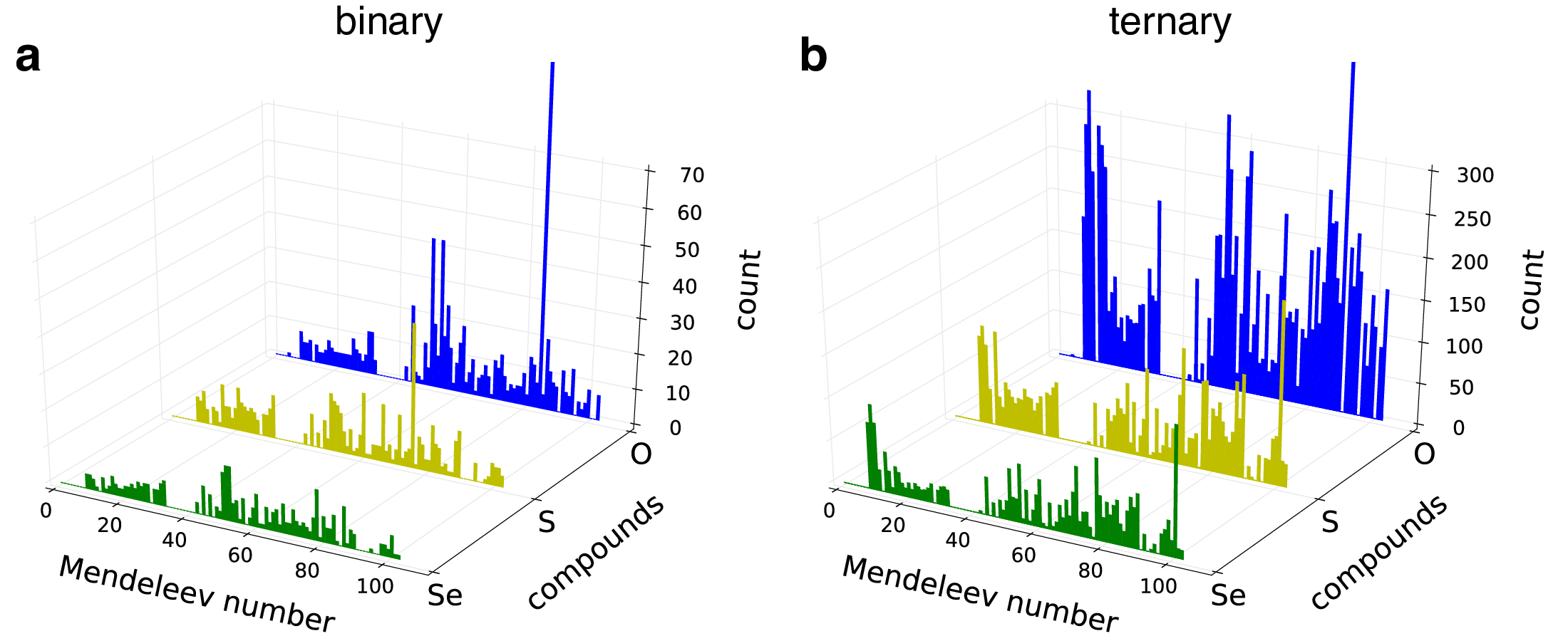}
\mycaption[Distributions of (\textbf{a}) binary and (\textbf{b}) ternary compounds among the elements.]
{The binary oxides exhibit a structures distribution with two prominent peaks. The distributions
of the binary sulfides and selenides are less structured and more similar to each other.
The distributions of the ternary compounds have higher, wider peaks than
their binary counterparts. The relative differences between the oxide, sulfide and selenide
distributions remain similar to the distributions of the binaries.}
\label{fig:art130:mendeleev_distribution_all_in_one}
\efig

\fig
\includegraphics[width=0.375\linewidth]{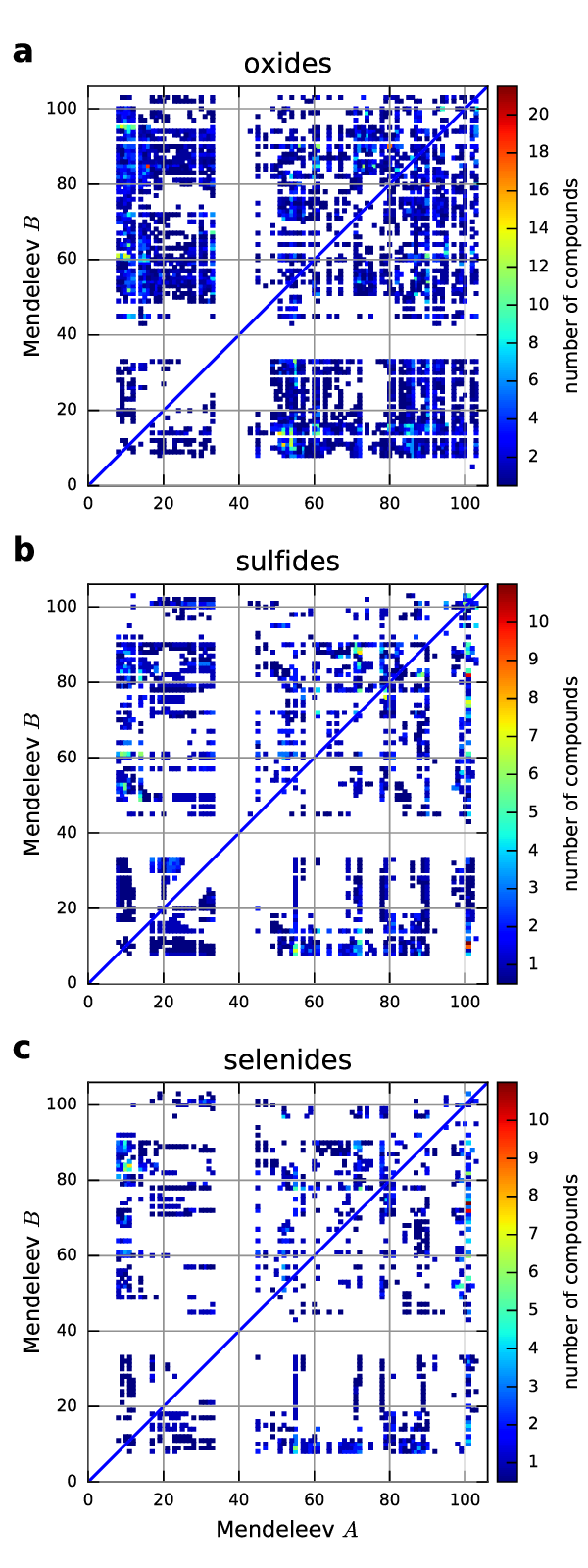}
\mycaption[Mendeleev maps of ternary (\textbf{a}) oxide $A_xB_y$O$_z$, (\textbf{b}) sulfide $A_xB_y$S$_z$
and (\textbf{c}) selenide $A_xB_y$Se$_z$ compounds.]
{It is assumed that $x\geq y$ with the $x$-axis indicating $M_A$
and the $y$-axis $M_B$.
If the stoichiometry is such that $x=y$, the compound is counted as $0.5 A_xB_y$O$_z + 0.5 B_xA_y$O$_z$.
A color scheme is used to represent the compound count for each composition, blue means the minimal number (one)
and green means the maximal number which is different for each element.}
\label{fig:art130:mendeleev_bigger_x_upper_all_in_one}
\efig

\fig
\includegraphics[width=1.0\linewidth]{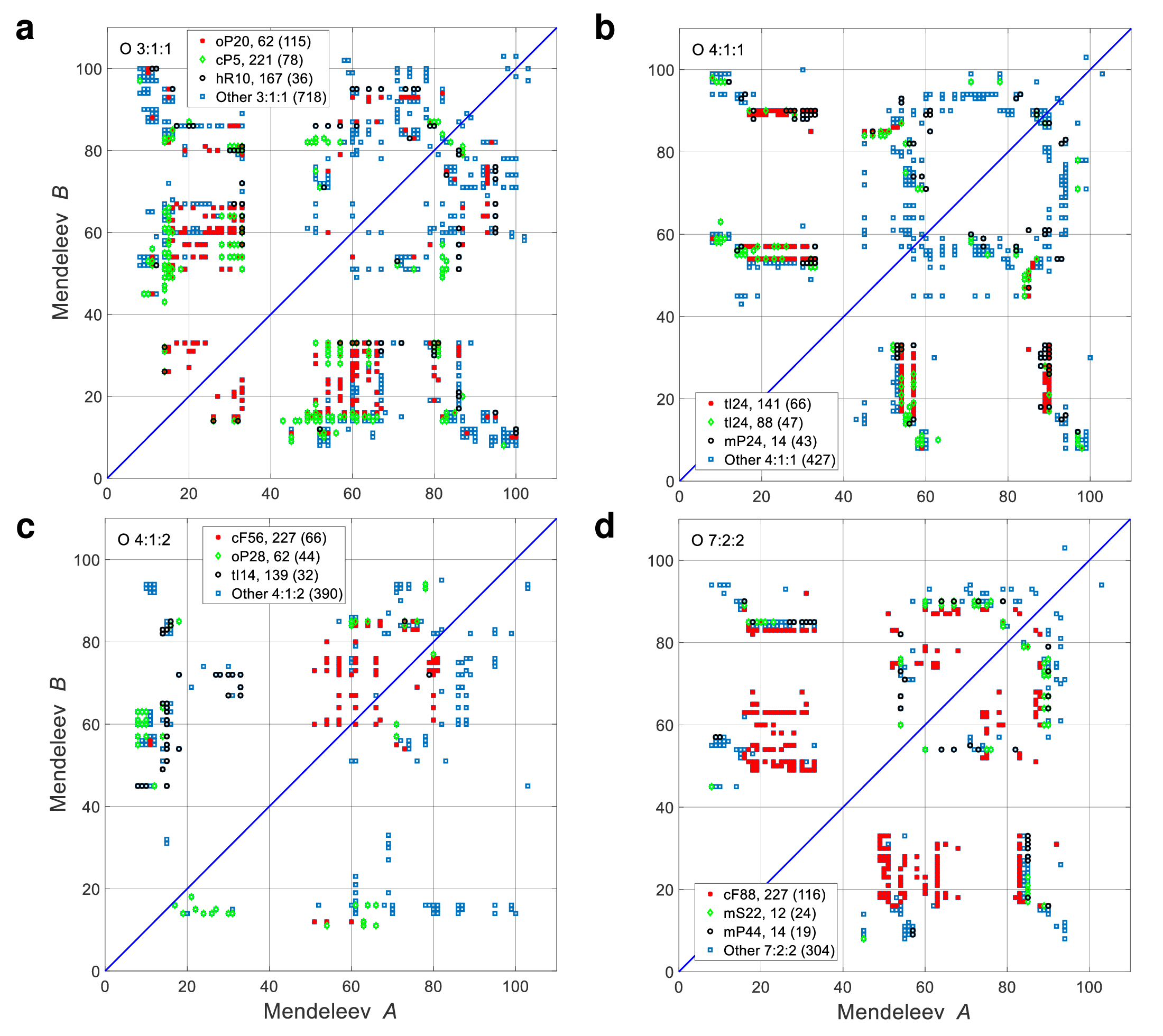}
\mycaption[Mendeleev maps of the three leading structure types in each of the four leading stoichiometries in ternary oxides.]
{(\textbf{a}) $A_1B_1$O$_3$,
(\textbf{b}) $A_1B_1$O$_4$,
(\textbf{c}) $A_1B_2$O$_4$, and
(\textbf{d}) $A_2B_2$O$_7$.
The legend box appears at a region with no data points.
The number in parenthesis is the number of compounds for this structure type, for ``Other'',
it refers to the total number of compounds with this stoichiometry.}
\label{fig:art130:mendeleev_oxide_prototypes}
\efig

\fig
\includegraphics[width=1.0\linewidth]{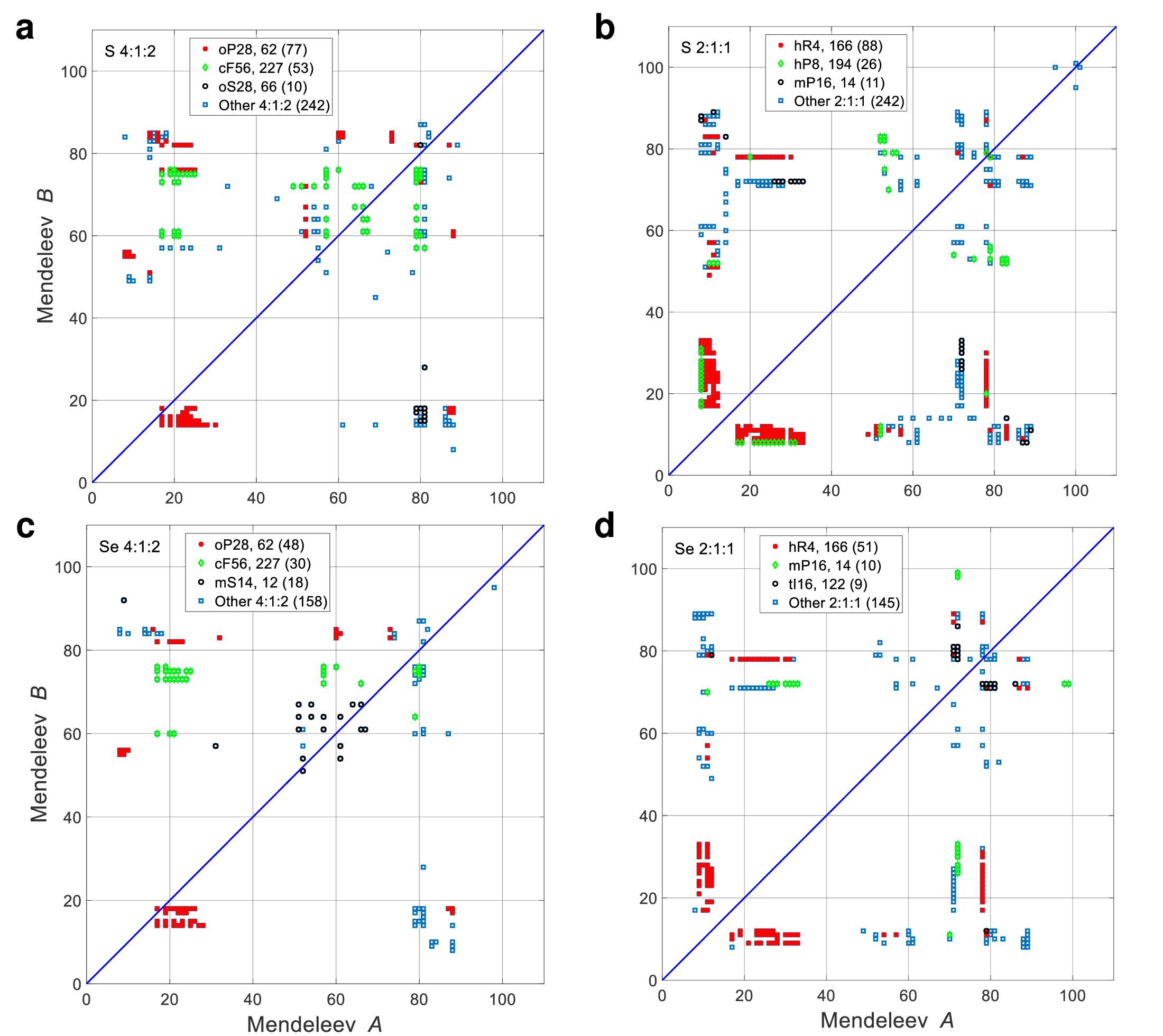}
\mycaption[Mendeleev maps of the three leading structure types in each of the two leading stoichiometries in sulfur and selenium ternaries.]
{(\textbf{a}) $A_1B_2$S$_4$,
(\textbf{b}) $A_1B_1$S$_2$,
(\textbf{c}) $A_1B_2$Se$_4$, and
(\textbf{d}) $A_1B_1$Se$_2$.
The number in parenthesis is the number of compounds for this structure type, for ``Other'',
it refers to the total number of compounds with this stoichiometry.}
\label{fig:art130:mendeleev_sulfur_selenium_prototypes}
\efig

\fig
\includegraphics[width=1.0\linewidth]{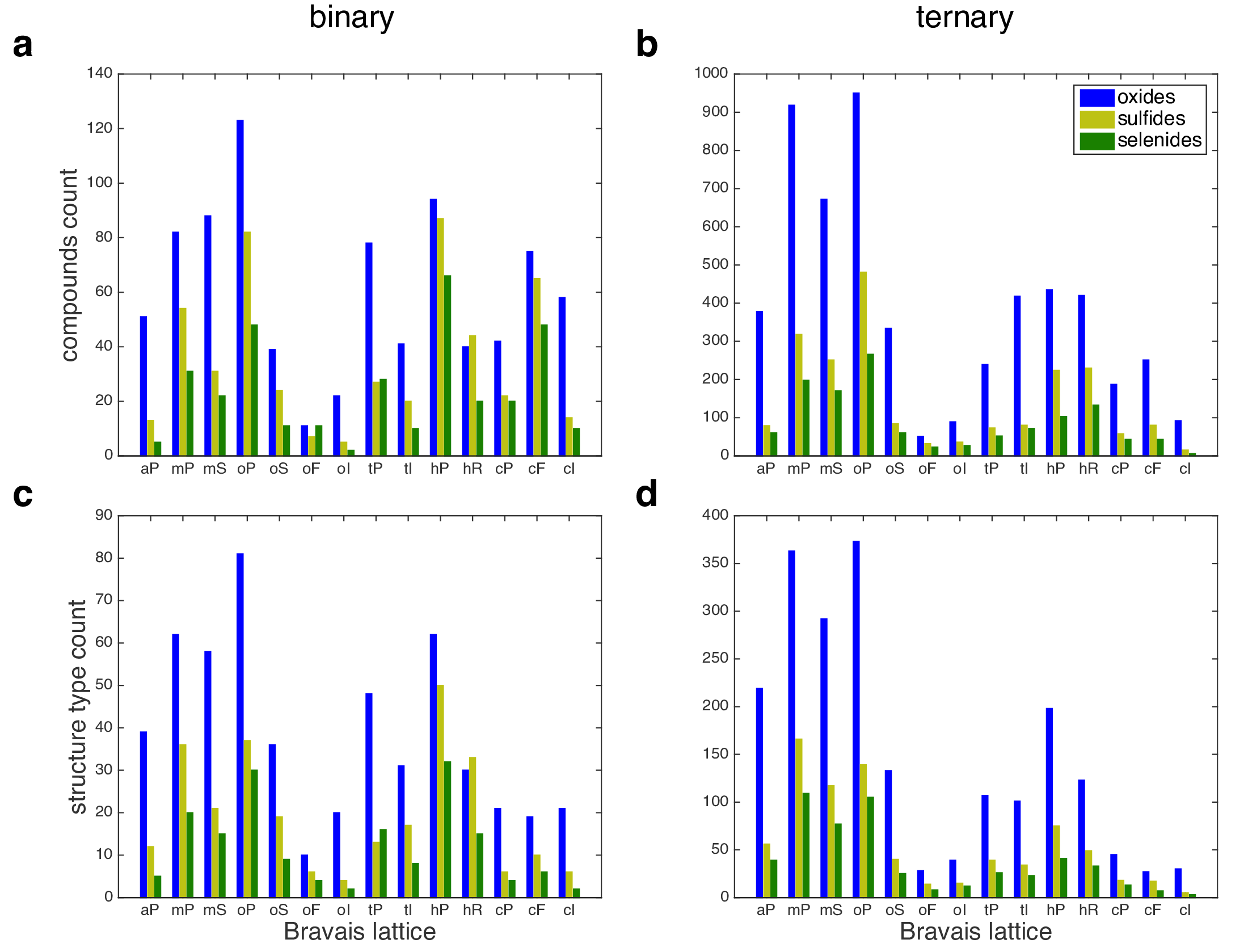}
\mycaption[Number of compounds (\textbf{a} and \textbf{b}) and
structure types (\textbf{c} and \textbf{d}) for each Bravais lattice.]
{Binaries are on the left (\textbf{a} and \textbf{c}) and
ternaries on the right (\textbf{b} and \textbf{d}).
Oxides are shown in blue, sulfides in light green and selenides in darker green. All six
distributions (binary and ternary oxides, sulfides and selenides) are double peaked with a
local minimum for the orthorhombic face and body centered lattices. The high symmetry
cubic lattices are also relatively rare. This reflects the complex spatial arrangement of
the compound forming electrons of the 6A elements, which does not favor the
high symmetry of these
structures.}
\label{fig:art130:bravais_combined}
\efig

\fig
\includegraphics[width=1.0\linewidth]{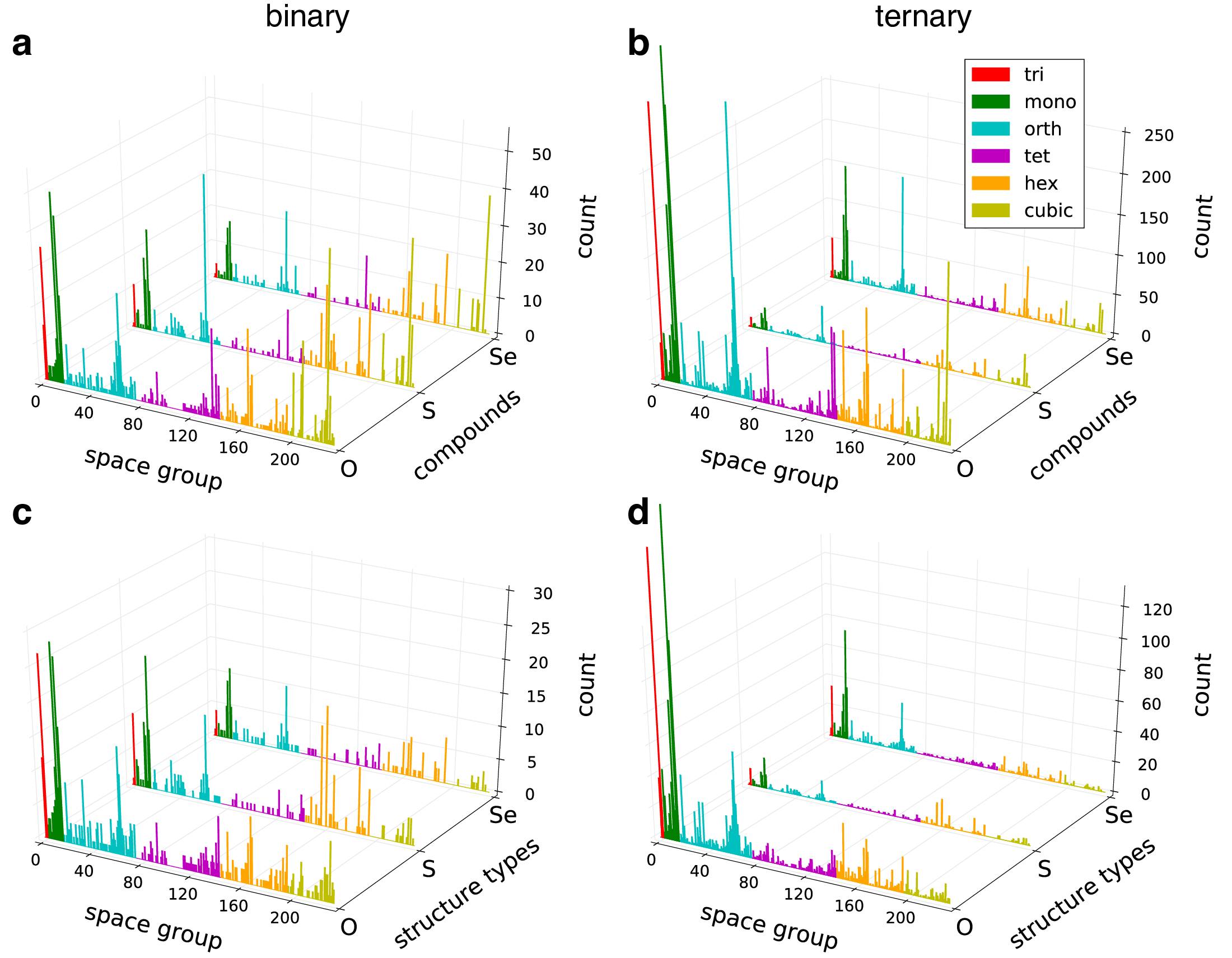}
\mycaption[Distributions of compounds (\textbf{a} and \textbf{b}) and structure types (\textbf{c} and \textbf{d}) among the 230 space groups.]
{Binaries are on the left (\textbf{a} and \textbf{c}) and
ternaries on the right (\textbf{b} and \textbf{d}).
Compounds are depicted on the top (\textbf{a} and \textbf{b})
and structure types on the bottom (\textbf{c} and \textbf{d}).}
\label{fig:art130:symmetry_distribution_of_structures}
\efig

It should be noted that large unit cells, within the tails of all distributions, tend to have very few representatives,
with just one compound with a given unit cell size in most cases.
Notable exceptions are local peaks near $80$ atoms per unit cell in the binary
distributions and near 200 atoms per unit cell in the ternary distributions.
The oxide distributions exhibit additional peaks, near 300 atoms per unit cell for the
binaries and near 600 atoms per unit cell for the ternaries.
These minor peaks may indicate preferable arrangements of cluster-based structures.

\fig
\includegraphics[width=0.9\linewidth]{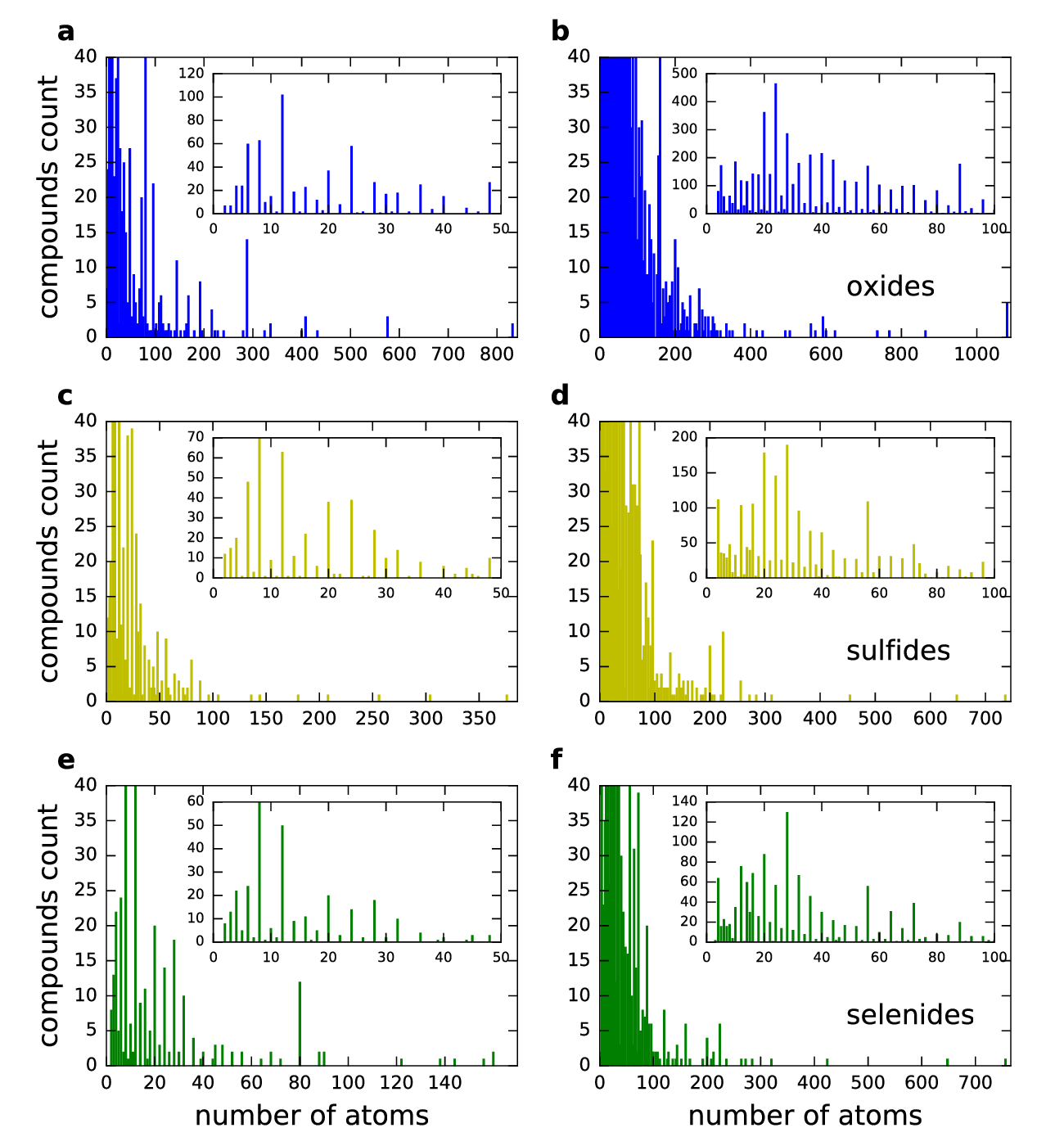}
\mycaption[Unit cell size distributions for oxides, sulfides, and selenides.]
{Binaries are on the left (\textbf{a}, \textbf{c} and \textbf{e}) and
ternaries on the right (\textbf{b}, \textbf{d} and \textbf{f}).
Oxides are at the top (\textbf{a} and \textbf{b}),
sulfides in the middle (\textbf{c} and \textbf{d}) and
selenides at the bottom (\textbf{e} and \textbf{f}).
The insets show the compounds with up to 50 atoms per unit cell in each case.
All distributions exhibit long tails of rare very large unit cells which extend much further in the oxides.
The dense cores of the distributions reflects the higher prevalence of oxides and are very similar for the sulfides and selenides.}
\label{fig:art130:number_of_atoms_distribution}
\efig

\subsection{Structure sub-types}
The definition of structure type by the combination of stoichiometry,
Pearson symbol and symmetry is common in the literature, but it is not
necessarily unique. A given structure, according to this definition,
can contain few sub-types.
As an example, the structure types
($A_1B_1$O$_3$:oP20:62),
($A_1B_1$O$_3$:hR10:167) and ($A_1B_1$O$_3$:cP5:221)  contain 115, 36, and
78 unique compounds, respectively, of mostly perovskites. However, the
oP20 also includes the aragonite structure, the MgSeO$_{3}$
structure, and others.
The hR10 contains also calcite-like
structures. The cP5 group, which has a more strict symmetry, contains
only perovskites.
These three structure types belong to a common parent class, the high symmetry
cP5, with two different types of symmetry breaking.
The different sub-types within each structure type may be discerned by
examining relations between structural descriptors, \eg, the volume
as a function of nearest neighbor distance cubed, as shown in Figure~\ref{fig:art130:structure_types_comparison}.

\fig
\includegraphics[width=1.0\linewidth]{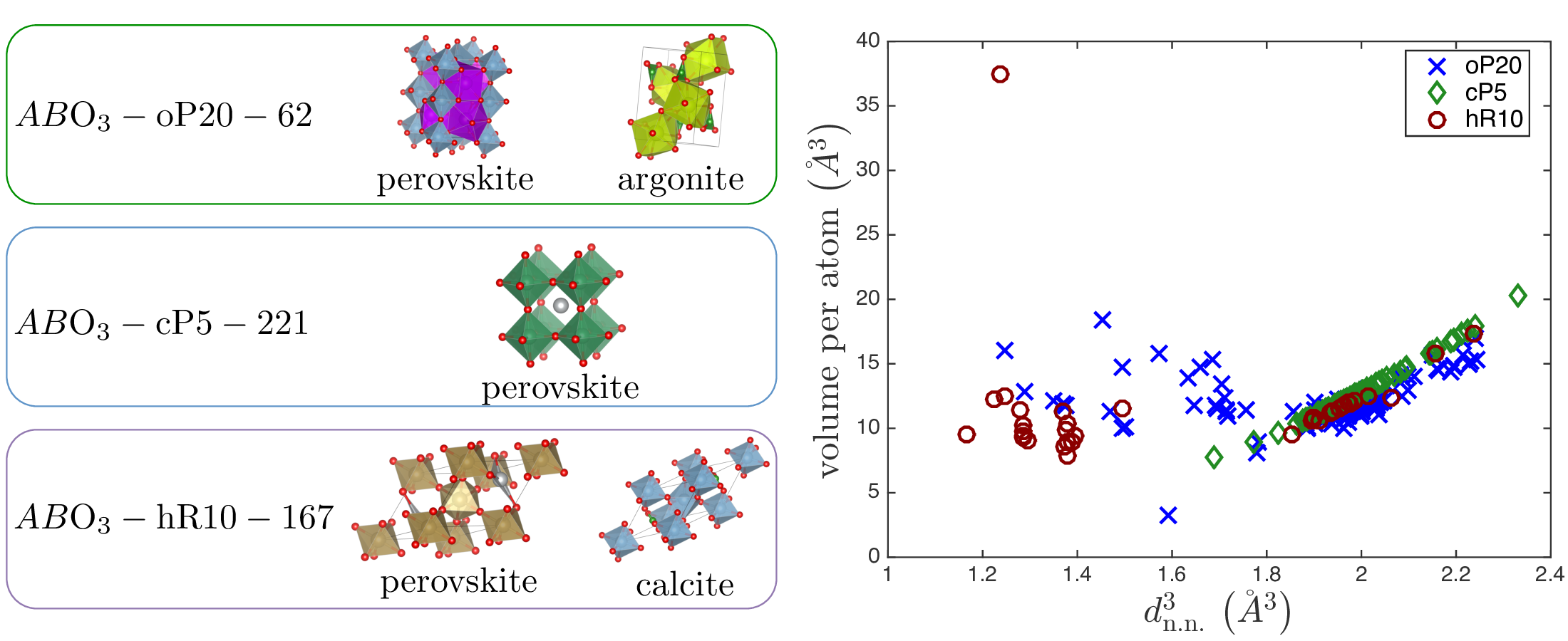}
\mycaption{Comparison of different structure types according to volume per atom \vs\
shortest nearest-neighbor distance cubed $\left(d^{3}_{\mathrm{n.n.}}\right)$.}
\label{fig:art130:structure_types_comparison}
\efig

It can be easily seen that the ($A_1B_1$O$_3$:cP5:221) group follows a
perfect linear relation, as is expected from a uniform structure type.
However, both the ($A_1B_1$O$_3$:oP20:62) and the ($A_1B_1$O$_3$:hR10:167) types
include points that are close to the ($A_1B_1$O$_3$:cP5:221) line but also clusters of
points that deviate from it.

Those points represent non-perovskite
structures, including those that are close to aragonite and calcite.

\subsection{Ternary stoichiometry triangles}

\fig
\includegraphics[width=1.0\linewidth]{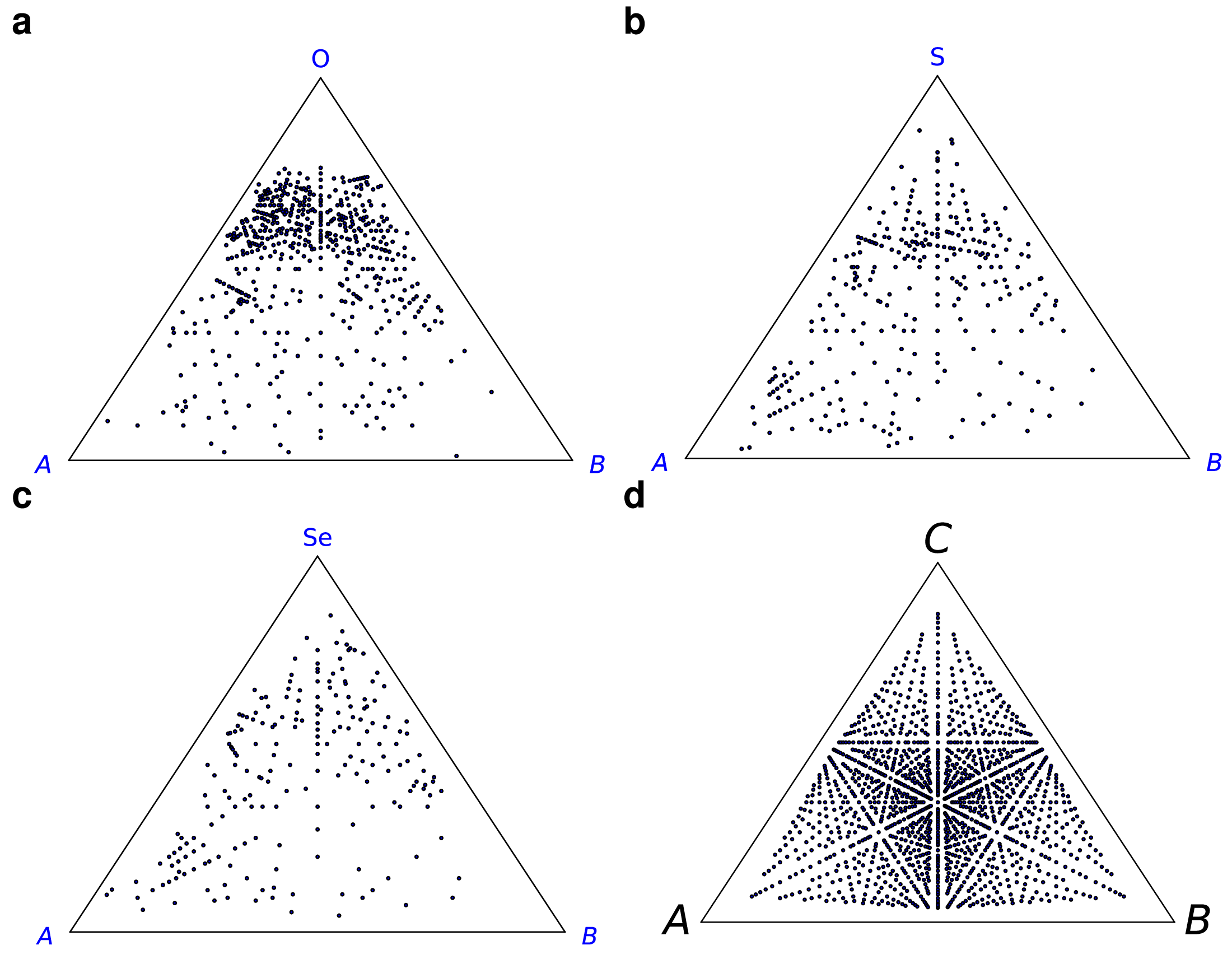}
\mycaption{Comparison of ternary stoichiometries for (\textbf{a}) oxygen, (\textbf{b}) sulfur and (\textbf{c}) selenium compounds.
All stoichiometries of $A_xB_yC_z$, $x,y,z \leq 12$ are shown in (\textbf{d}).}
\label{fig:triangle_stoichiometries}
\efig

Figure~\ref{fig:art130:stoi_periodic_oxide} shows the prevalence of different ternary
stoichiometries in a triangle shape.
The points inside the triangle are defined by
the intersection of lines that connect the vertex points with the
corresponding binary stoichiometry on the opposing edge.
For example, the stoichiometry $A_{u}B_{v}$O$_{w}$ is represented by the
intersection of three lines, one from the O vertex to the point
$u/(u+v)$ on the $AB$ edge,
another from the $A$ vertex to the point $v/(v+w)$ on the $B$O edge, and
the third from the $B$ vertex to the point $u/(u+w)$ on the $A$O edge.
The different stoichiometries in Figure~\ref{fig:art130:stoi_periodic_oxide} are denoted by
circles that vary in size according to the number of compounds for
each stoichiometry.
Figure~\ref{fig:triangle_stoichiometries}(a-c) shows the same data but without reference to prevalence,
showing just the stoichiometries locations.
Figure~\ref{fig:triangle_stoichiometries}(d) shows, for comparison, the locations of all possible stoichiometries up to
12 atoms per species ($A_xB_yC_z$ where $x,y,z \leq 12$). The differences in the distributions of the
reported compositions of the three compound families are clearly apparent.

\subsection{Prevalence of structure types among the oxide, sulfide and binary and ternary selenide compounds}

Numerical data for the leading 40 structure types of the oxides,
sulfides and selenides are shown in Tables
\ref{tab:art130:oxide_binary_data} through
\ref{tab:art130:selenide_binary_data} for the binaries, and in Tables
\ref{tab:art130:oxide_ternary_data} through
\ref{tab:art130:selenide_ternary_data} for the ternaries.

\tab
\mycaption{Prevalence of the 40 most common structure types among the binary oxide compounds.}
\tabvspace
\resizebox{\linewidth}{!}{
\begin{tabular}{L{2.5cm}|L{2.5cm}|L{2.5cm}|L{5cm}|R{3.25cm}}
stoichiometry & Pearson & symmetry  & name & compounds count \\
\hline
1:1& cF8 & 225&NaCl&29\\
3:2& cI80 & 206&Bixbyite-Mn$_{2}$O$_{3}$&24\\
2:1& tP6 & 136&KrF$_{2}$&22\\
3:2& hP5 & 164&La$_{2}$O$_{3}$&19\\
2:1& cF12 & 225&Fluorite-CaF$_{2}$&16\\
3:2& mS30 & 12&Sm$_{2}$O$_{3}$&16\\
3:2& cI80 & 199&Sm$_{2}$O$_{3}$ (c180)&15\\
2:1& cP12 & 205&CO$_{2}$ (cP12)&13\\
2:1& mP12 & 14&Baddeleyite-ZrO$_{2}$ (mP12)&8\\
2:1& oP12 & 60&&8\\
2:1& oP12 & 62&&8\\
2:1& oP6 & 58&&8\\
3:2& hR10 & 167&&8\\
1:1& hP4 & 186&&7\\
2:1& oP24 & 61&&7\\
2:1& tI6 & 139&&7\\
1:1& cF8 & 216&&5\\
1:2& cF12 & 225&&5\\
1:2& cP6 & 224&&5\\
3:2& mP20 & 14&&5\\
3:2& oP20 & 60&&5\\
2:1& aP24 & 1&&4\\
2:1& tP12 & 92&&4\\
3:1& mP16 & 14&&4\\
3:2& mS20 & 12&&4\\
3:2& oP20 & 62&&4\\
5:2& mS28 & 15&&4\\
1:1& tP4 & 129&&3\\
2:1& hP9 & 152&&3\\
2:1& mS6 & 12&&3\\
3:1& cP4 & 221&&3\\
3:2& cF80 & 227&&3\\
3:2& hP5 & 150&&3\\
3:2& oS20 & 63&&3\\
7:4& aP22 & 2&&3\\
12:7& hR19 & 148&&3\\
1:1& cP2 & 221&&2\\
1:1& hP4 & 194&&2\\
1:1& mS4 & 12&&2\\
1:1& mS8 & 15&&2\\
\end{tabular}}
\label{tab:art130:oxide_binary_data}
\etab

\clearpage

\tab
\mycaption{Prevalence of the 40 most common structure types among the binary sulfide compounds.}
\tabvspace
\resizebox{\linewidth}{!}{
\begin{tabular}{L{2.5cm}|L{2.5cm}|L{2.5cm}|L{5cm}|R{3.25cm}}
stoichiometry & Pearson & symmetry  & name & compounds count \\
\hline
1:1& cF8 & 225&NaCl&32\\
3:2& oP20 & 62&Sb$_{2}$S$_{3}$&20\\
2:1& tP6 & 129&PbClF/Cu$_{2}$Sb&13\\
2:1& cP12 & 205&Pyrite-Fe$_{2}$S$_{2}$ (cP12)&12\\
2:1& cF24 & 227&Laves(Cub)-Cu$_{2}$Mg&11\\
1:1& hP4 & 194&Nickeline-NiAs&8\\
4:3& cI28 & 220&Th$_{3}$P$_{4}$&8\\
1:1& cF8 & 216&Sphalerite-ZnS (cF8)&7\\
2:1& hP3 & 164&CdI$_{2}$&7\\
2:1& mP12 & 14&CeSe$_{2}$&7\\
1:1& cP2 & 221&&6\\
1:1& hP4 & 186&&6\\
1:2& cF12 & 225&&6\\
2:1& hP6 & 194&&6\\
7:5& mS24 & 12&&6\\
1:1& oP8 & 62&&5\\
2:1& oP6 & 58&&5\\
3:2& hR10 & 167&&5\\
3:2& mP30 & 11&&5\\
1:1& hP2 & 187&&4\\
1:2& hP6 & 194&&4\\
1:2& oP12 & 62&&4\\
2:1& hR3 & 160&&4\\
2:1& oP12 & 62&&4\\
2:1& oP24 & 62&&4\\
3:1& mP8 & 11&&4\\
4:3& cF56 & 227&&4\\
5:2& oP28 & 19&&4\\
1:1& oS8 & 63&&3\\
2:2& hP12 & 189&&3\\
3:2& hP30 & 185&&3\\
3:2& oS20 & 36&&3\\
1:1& hP16 & 186&&2\\
1:1& hP8 & 194&&2\\
1:1& hR2 & 160&&2\\
1:1& hR4 & 166&&2\\
1:1& hR6 & 160&&2\\
1:1& mP8 & 14&&2\\
1:1& mS8 & 5&&2\\
1:1& oS8 & 39&&2\\
\end{tabular}}
\label{tab:art130:sulfide_binary_data}
\etab

\clearpage

\tab
\mycaption{Prevalence of the 40 most common structure types among the binary selenide compounds.}
\tabvspace
\resizebox{\linewidth}{!}{
\begin{tabular}{L{2.5cm}|L{2.5cm}|L{2.5cm}|L{5cm}|R{3.25cm}}
stoichiometry & Pearson & symmetry  & name & compounds count \\
\hline
1:1& cF8 & 225&NaCl&31\\
2:1& tP6 & 129&PbClF/Cu$_{2}$Sb&13\\
2:1& cP12 & 205&Pyrite-FeS$_{2}$ (cP12)&11\\
1:1& hP4 & 186&Wurtzite-ZnS(2H)&9\\
1:1& hP4 & 194&Nickeline-NiAs&9\\
2:1& hP3 & 164&CdI$_{2}$&9\\
3:2& oP20 & 62&Sb$_{2}$S$_{3}$&9\\
4:3& cI28 & 220&Th$_{3}$P$_{4}$&9\\
1:1& cF8 & 216&Sphalerite-ZnS (cF8)&8\\
3:2& oF80 & 70&Sc$_{2}$S$_{3}$&8\\
1:1& cP2 & 221&&7\\
1:2& cF12 & 225&&6\\
4:3& mS14 & 12&&6\\
1:1& oP8 & 62&&4\\
2:1& hP6 & 194&&4\\
2:1& mP12 & 14&&4\\
3:1& mP8 & 11&&4\\
2:1& hP12 & 187&&3\\
2:1& hR3 & 160&&3\\
2:1& oP6 & 58&&3\\
3:2& oS20 & 36&&3\\
4:4& mP32 & 14&&3\\
4:5& tI18 & 87&&3\\
5:2& oP28 & 19&&3\\
1:1& hP8 & 187&&2\\
1:1& hP8 & 194&&2\\
1:1& hR4 & 160&&2\\
1:1& mS8 & 12&&2\\
1:2& oP36 & 58&&2\\
2:1& hP12 & 194&&2\\
2:1& oP12 & 62&&2\\
2:1& oP24 & 62&&2\\
2:2& hP12 & 189&&2\\
2:2& mP16 & 14&&2\\
3:1& mP24 & 11&&2\\
3:2& hR5 & 166&&2\\
3:2& mP10 & 11&&2\\
3:2& mS20 & 9&&2\\
4:3& hP14 & 176&&2\\
8:3& hR11 & 148&&2\\
\end{tabular}}
\label{tab:art130:selenide_binary_data}
\etab

\clearpage

\tab
\mycaption{Prevalence of the 40 most common structure types among the ternary oxide compounds.}
\tabvspace
\resizebox{\linewidth}{!}{
\begin{tabular}{L{2.5cm}|L{2.5cm}|L{2.5cm}|L{5cm}|R{3.25cm}}
stoichiometry & Pearson & symmetry  & name & compounds count \\
\hline
7:2:2& cF88 & 227&Pyrochlore&116\\
3:1:1& oP20 & 62&Perovskite-GdFeO$_{3}$ (mostly)&115\\
3:1:1& cP5 & 221&Perovskite-CaTiO$_{3}$&78\\
2:1:1& hR4 & 166&Delafossite-NaCrS$_{2}$&72\\
4:1:1& tI24 & 141&Zircon-ZrSiO$_{4}$&66\\
4:1:2& cF56 & 227&Spinel-Al$_{2}$MgO$_{4}$&66\\
4:1:1& tI24 & 88&Scheelite-CaWO$_{4}$&47\\
4:1:2& oP28 & 62&CaFe$_{2}$O$_{4}$&44\\
4:1:1& mP24 & 14&AgMnO4&43\\
3:1:1& hR10 & 167&Perovskite-NdAlO$_{3}$&36\\
4:1:1& oP24 & 62&Barite-BaSO$_{4}$&34\\
3:1:1& mP20 & 14&&33\\
7:1:3& oS44 & 63&&33\\
2:1:2& hP5 & 164&&32\\
4:1:2& tI14 & 139&&32\\
2:1:2& tI10 & 139&&31\\
4:1:1& oS24 & 63&&31\\
12:3:5& cI160 & 230&&30\\
3:1:1& hR10 & 148&&28\\
2:1:1& hP8 & 194&&27\\
4:1:1& mP12 & 13&&26\\
1:1:1& tP6 & 129&&25\\
5:1:2& mP32 & 14&&25\\
7:2:2& mS22 & 12&&24\\
6:1:2& tP18 & 136&&22\\
11:2:4& mS68 & 15&&20\\
3:1:1& hR10 & 161&&19\\
4:1:2& mP28 & 14&&19\\
7:2:2& mP44 & 14&&19\\
1:1:3& cP5 & 221&&18\\
3:1:1& hR5 & 160&&18\\
1:2:4& tI14 & 139&&17\\
3:1:1& mP40 & 14&&17\\
6:1:2& hP9 & 162&&16\\
7:2:2& aP22 & 2&&16\\
7:2:2& aP44 & 2&&16\\
9:1:3& mP52 & 14&&16\\
1:4:6& hP22 & 186&&15\\
3:1:1& hP30 & 185&&15\\
5:1:2& oP32 & 55&&15\\
\end{tabular}}
\label{tab:art130:oxide_ternary_data}
\etab

\clearpage

\tab
\mycaption{Prevalence of the 40 most common structure types among the ternary sulfide compounds.}
\tabvspace
\resizebox{\linewidth}{!}{
\begin{tabular}{L{2.5cm}|L{2.5cm}|L{2.5cm}|L{5cm}|R{3.25cm}}
stoichiometry & Pearson & symmetry  & name & compounds count \\
\hline
2:1:1& hR4 & 166&Delafossite-NaCrS$_{2}$&88\\
4:1:2& oP28 & 62&CaFe$_{2}$O$_{4}$&77\\
4:1:2& cF56 & 227&Spinel-Al$_{2}$MgO$_{4}$&53\\
5:1:2& oP32 & 62&U$_{3}$S$_{5}$&37\\
3:1:1& oP20 & 62&SrZrS$_{3}$&32\\
8:1:6& hR15 & 148&Mo$_{6}$PbS$_{8}$&32\\
1:1:1& mP12 & 14&CeAsS&27\\
2:1:1& hP8 & 194&SnTaS$_{2}$&26\\
6:1:3& mP20 & 11&Tm$_{2}$S$_{3}$&26\\
7:1:4& hP24 & 173&La$_{3}$CuSiS$_{7}$&25\\
1:1:1& tP6 & 129&&23\\
3:1:1& oP20 & 33&&21\\
22:4:11& mS74 & 12&&20\\
1:2:2& hP5 & 164&&19\\
6:1:3& oP40 & 18&&15\\
1:1:1& cP12 & 198&&14\\
2:2:3& hR7 & 166&&14\\
4:1:3& oP32 & 62&&14\\
1:1:1& oP12 & 62&&12\\
4:1:1& tI96 & 142&&12\\
1:1:1& oP24 & 62&&11\\
2:1:1& mP16 & 14&&11\\
2:1:1& oP16 & 62&&11\\
3:1:1& mP40 & 11&&11\\
6:1:3& hP20 & 182&&11\\
12:3:4& hR38 & 161&&11\\
13:4:5& oP44 & 55&&11\\
1:1:4& oP24 & 62&&10\\
2:1:1& hR4 & 160&&10\\
2:1:1& tI16 & 122&&10\\
2:1:2& tI10 & 139&&10\\
3:1:2& oP24 & 62&&10\\
4:1:2& oS28 & 66&&10\\
8:1:5& mS28 & 12&&10\\
3:1:3& oP28 & 62&&9\\
4:1:2& mS14 & 12&&9\\
4:1:2& oF224 & 70&&9\\
4:1:3& cP16 & 223&&9\\
2:1:1& mS64 & 15&&8\\
2:1:2& mP20 & 14&&8\\
\end{tabular}}
\label{tab:art130:sulfide_ternary_data}
\etab

\tab
\mycaption{Prevalence of the 40 most structure types among the ternary selenide compounds.}
\tabvspace
\resizebox{\linewidth}{!}{
\begin{tabular}{L{2.5cm}|L{2.5cm}|L{2.5cm}|L{5cm}|R{3.25cm}}
stoichiometry & Pearson & symmetry  & name & compounds count \\
\hline
2:1:1& hR4 & 166&Delafossite&51\\
4:1:2& oP28 & 62&CaFe$_{2}$O$_{4}$&48\\
4:1:2& cF56 & 227&Spinel-Al$_{2}$MgO$_{4}$&30\\
8:1:6& hR15 & 148&Mo$_{6}$PbS$_{8}$&25\\
1:1:1& mP12 & 14&CeAsS&22\\
4:1:2& mS14 & 12&CrNb$_{2}$Se$_{4}$-Cr$_{3}$S$_{4}$&18\\
8:1:5& mS28 & 12&Cr$_{5}$CsS$_{8}$&18\\
1:1:1& tP6 & 129&PbClF/Cu$_{2}$Sb&16\\
1:1:1& cP12 & 198&NiSSb&14\\
3:1:1& oP20 & 62&NH$_{4}$CdCl$_{3}$/Sn$_{2}$S$_{3}$&14\\
4:1:2& tI14 & 82&&13\\
5:1:2& oP32 & 62&&13\\
4:1:3& oP32 & 62&&12\\
1:2:2& hP5 & 164&&11\\
2:1:1& mP16 & 14&&10\\
2:1:2& tI10 & 139&&10\\
3:1:1& oS20 & 63&&10\\
3:1:3& cP28 & 198&&10\\
6:1:3& hP20 & 182&&10\\
1:1:1& oP12 & 62&&9\\
1:1:3& oP20 & 62&&9\\
2:1:1& tI16 & 122&&9\\
2:1:2& oI20 & 72&&9\\
3:1:3& hP14 & 176&&9\\
4:1:2& oS28 & 66&&9\\
19:2:15& hR72 & 167&&9\\
2:1:1& oP16 & 19&&8\\
6:1:3& oP40 & 58&&8\\
17:1:8& mS52 & 12&&8\\
4:1:6& hP22 & 186&&7\\
6:2:2& mP20 & 14&&7\\
6:2:6& mP28 & 14&&7\\
2:1:1& mS64 & 15&&6\\
2:1:1& tI16 & 140&&6\\
4:1:2& oF224 & 70&&6\\
1:1:4& oS24 & 63&&5\\
2:1:1& hR4 & 160&&5\\
2:1:6& mP18 & 14&&5\\
2:1:12& oF120 & 43&&5\\
2:2:3& hR7 & 166&&5\\
\end{tabular}}
\label{tab:art130:selenide_ternary_data}
\etab

\clearpage

\subsection{Prevalence of stoichiometries}
\label{subsec:art130:prev_stoich_supp}

Tables~\ref{tab:art130:Prevalence_of_Binaries_Stoichiometries}-\ref{tab:art130:Prevalence_of_Binaries_Stoichiometries_3} list all the binary
stoichiometries among the three compound families examined in this paper.
An interesting finding is that the stoichiometry
$A_1$O$_2$ has 356 unique compounds, a number that is significantly
larger than the number of atoms in the periodic table. This is because
a given chemical composition can have many different structure type
realizations.
The most prominent example is SiO$_{2}$ which has 185 different
reported structures, representing the majority of the 356 compounds
and 244 structure types of this stoichiometry.
In contrast, SiS$_{2}$ has only two reported structures, and SiSe$_{2}$
has only one.
Checking other atoms from the same column of Si in the periodic
table, we find that GeO$_{2}$ has seven structures and CO$_{2}$ has
nine.
The last observation means that, since CO$_{2}$ is
gaseous in atmospheric conditions, the \ICSD\ compounds of
CO$_{2}$ are not in atmospheric conditions (either temperature or
pressure or both).
Examining Tables~\ref{tab:art130:Prevalence_of_Elements_in_Binaries_Stoichiometries}-\ref{tab:art130:Prevalence_of_Elements_in_Binaries_Stoichiometries_3}
we also observe that the $A_xO_y$ set of compounds exhibits
several gaps (missing ratios) along the axis of $y/(x+y)$. There are
no reported binary oxides
from 0.51 to (and not including) 0.55, from 0.34 to 0.4,
from 0.26 to 0.3, and from 0.44 to 0.5.
Those gaps {\it do not exist} in the sulfides and
selenides.
Most of the gaps in the sulfides appear above 0.6, and no
selenide compounds are reported above 0.65.
The maximal ratio for the oxides is 0.84, while the maximal ratio for the sulfides is 0.93.
Tables~\ref{tab:art130:Elements_stoichiometries}-\ref{tab:art130:Elements_stoichiometries_3} show the leading stoichiometry for each element
as well as the number of stoichiometries and unique compounds for this element.
While SiO$_2$ is the {\it only} stoichiometry of silicon oxide (with 185 structure types),
vanadium has 18 different stoichiometries and 42 unique compounds,
VO$_2$ is the stoichiometry with the largest number (10) of structure types.

\tab
\mycaption{Prevalence of binary stoichiometries (1/3).}
\tabvspace
\begin{tabular}{l|r|r|r}
stoichiometry & oxides & sulfides & selenides \\
\hline
1:2& 356 & 123 & 79\\
1:1& 99 & 165 & 108\\
2:3& 146 & 58 & 38\\
1:3& 41 & 12 & 10\\
2:1& 27 & 30 & 21\\
2:5& 26 & 8 & 5\\
3:4& 22 & 25 & 21\\
3:1& 9 & 7 & 1\\
6:11& 9 & 0 & 0\\
3:8& 9 & 2 & 2\\
6:1& 4 & 7 & 0\\
3:5& 7 & 2 & 2\\
5:9& 7 & 0 & 0\\
5:7& 0 & 6 & 0\\
4:7& 6 & 1 & 0\\
4:1& 2 & 1 & 5\\
4:3& 2 & 5 & 4\\
6:13& 5 & 0 & 0\\
4:9& 5 & 2 & 1\\
5:4& 0 & 2 & 4\\
12:29& 4 & 0 & 0\\
2:7& 4 & 2 & 0\\
1:4& 4 & 1 & 2\\
8:1& 3 & 1 & 0\\
3:2& 1 & 3 & 2\\
7:8& 0 & 3 & 2\\
4:5& 3 & 3 & 1\\
7:12& 3 & 0 & 0\\
8:15& 3 & 0 & 1\\
4:11& 3 & 0 & 0\\
\end{tabular}
\label{tab:art130:Prevalence_of_Binaries_Stoichiometries}
\etab

\clearpage

\tab
\mycaption{Prevalence of binary stoichiometries continued (2/3).}
\tabvspace
\begin{tabular}{l|r|r|r}
stoichiometry & oxides & sulfides & selenides \\
\hline
21:8& 0 & 2 & 0\\
7:3& 2 & 0 & 0\\
5:3& 0 & 0 & 2\\
9:8& 0 & 2 & 1\\
6:7& 0 & 1 & 2\\
9:11& 0 & 0 & 2\\
11:20& 2 & 0 & 0\\
7:13& 2 & 0 & 0\\
9:17& 2 & 0 & 0\\
8:21& 2 & 0 & 0\\
17:47& 2 & 0 & 0\\
5:14& 2 & 0 & 0\\
8:23& 2 & 0 & 0\\
9:26& 2 & 0 & 0\\
2:9& 1 & 0 & 2\\
61:2& 1 & 0 & 0\\
12:1& 0 & 1 & 0\\
7:1& 1 & 0 & 0\\
16:3& 1 & 0 & 0\\
9:2& 1 & 1 & 1\\
15:4& 0 & 1 & 0\\
11:3& 1 & 0 & 0\\
7:2& 0 & 0 & 1\\
34:11& 0 & 0 & 1\\
45:16& 0 & 0 & 1\\
14:5& 0 & 1 & 0\\
11:4& 0 & 0 & 1\\
8:3& 0 & 1 & 1\\
5:2& 0 & 0 & 1\\
16:7& 0 & 1 & 0\\
\end{tabular}
\label{tab:art130:Prevalence_of_Binaries_Stoichiometries_2}
\etab

\clearpage

\tab
\mycaption{Prevalence of binary stoichiometries continued (3/3).}
\tabvspace
\begin{tabular}{l|r|r|r}
stoichiometry & oxides & sulfides & selenides \\
\hline
31:16& 0 & 1 & 0\\
29:16& 0 & 1 & 0\\
9:5& 0 & 1 & 0\\
7:4& 0 & 1 & 1\\
6:5& 0 & 1 & 1\\
8:7& 0 & 0 & 1\\
17:15& 0 & 1 & 1\\
17:18& 0 & 1 & 0\\
8:9& 0 & 1 & 1\\
5:6& 0 & 1 & 0\\
13:16& 1 & 0 & 0\\
15:19& 0 & 1 & 1\\
8:11& 0 & 1 & 0\\
15:22& 0 & 1 & 0\\
5:8& 1 & 1 & 1\\
9:16& 1 & 0 & 0\\
16:35& 1 & 0 & 0\\
3:7& 1 & 0 & 0\\
5:12& 1 & 0 & 0\\
13:34& 1 & 0 & 0\\
18:49& 1 & 0 & 0\\
25:73& 1 & 0 & 0\\
4:21& 1 & 0 & 0\\
1:8& 0 & 1 & 0\\
1:14& 0 & 1 & 0\\
\end{tabular}
\label{tab:art130:Prevalence_of_Binaries_Stoichiometries_3}
\etab

\clearpage

These differences carry over into the ternary oxide compounds involving those elements,
where the stoichiometry distribution of silicon ternary oxides is much tilted towards the
silicon poor compounds than those of vanadium and titanium as shown in Figure~\ref{fig:art130:specific_triangle_stoichiometries}.

\fig
\includegraphics[width=1.0\linewidth]{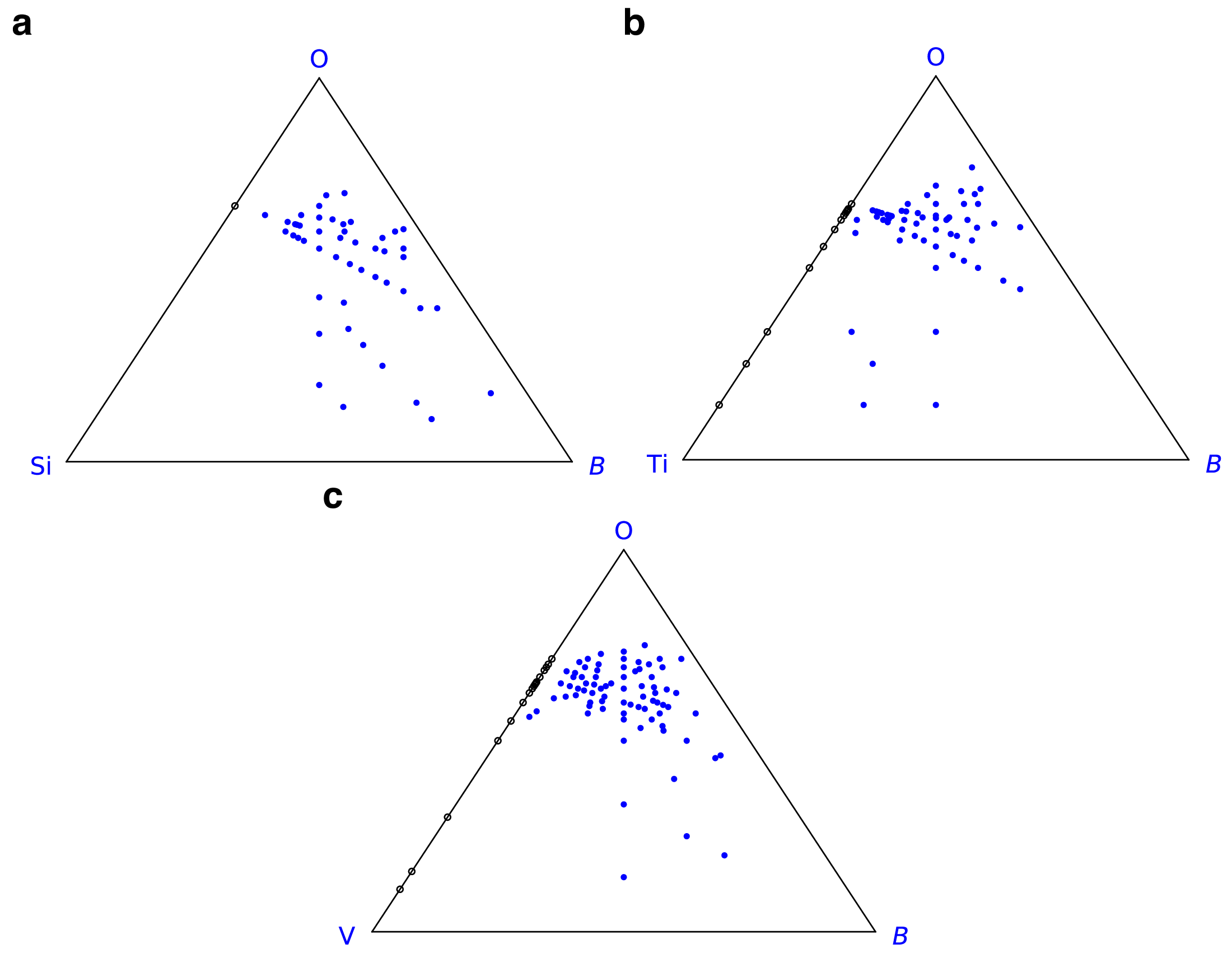}
\mycaption{Comparison of ternary oxide stoichiometries containing (\textbf{a}) silicon, (\textbf{b}) titanium, and (\textbf{c}) vanadium.}
\label{fig:art130:specific_triangle_stoichiometries}
\efig

\clearpage

\tab
\mycaption{Prevalence of ternary stoichiometries ($A_xB_yC_z,~C=$ O, S, Se; top 120) (1/4).}
\tabvspace
\begin{tabular}{l|r|r|r}
stoichiometry & oxides & sulfides & selenides \\
\hline
1:1:3& 718 & 147 & 70\\
1:1:4& 428 & 28 & 8\\
1:2:4& 396 & 242 & 158\\
2:2:7& 304 & 2 & 0\\
1:1:2& 269 & 242 & 145\\
1:2:6& 237 & 8 & 8\\
1:2:5& 149 & 57 & 22\\
1:1:1& 113 & 140 & 90\\
1:2:2& 131 & 64 & 38\\
1:2:3& 100 & 55 & 42\\
2:3:8& 87 & 8 & 4\\
1:3:6& 83 & 63 & 22\\
1:4:4& 78 & 17 & 13\\
1:3:9& 78 & 0 & 0\\
1:3:7& 67 & 1 & 0\\
2:2:5& 64 & 32 & 23\\
2:4:9& 62 & 7 & 3\\
1:3:3& 62 & 50 & 39\\
1:3:4& 58 & 54 & 31\\
1:3:8& 55 & 1 & 0\\
1:2:7& 49 & 3 & 2\\
2:4:11& 46 & 8 & 2\\
3:5:12& 46 & 4 & 0\\
2:3:12& 45 & 0 & 0\\
1:4:1& 15 & 44 & 12\\
1:3:1& 43 & 13 & 30\\
1:2:8& 41 & 2 & 3\\
1:6:8& 4 & 39 & 30\\
1:3:5& 37 & 15 & 6\\
1:1:5& 37 & 2 & 2\\
\end{tabular}
\label{tab:art130:Prevalence_of_Ternary_Stoichiometries}
\etab

\clearpage

\tab
\mycaption{Prevalence of ternary stoichiometries ($A_xB_yC_z,~C=$ O, S, Se; top 120) continued (2/4).}
\tabvspace
\begin{tabular}{l|r|r|r}
stoichiometry & oxides & sulfides & selenides \\
\hline
2:3:9& 34 & 2 & 3\\
1:3:2& 33 & 19 & 11\\
2:3:2& 9 & 32 & 11\\
1:4:7& 32 & 29 & 2\\
2:3:6& 31 & 11 & 11\\
2:2:1& 29 & 30 & 17\\
1:5:8& 29 & 14 & 22\\
2:4:7& 28 & 6 & 0\\
1:2:1& 27 & 12 & 5\\
2:3:4& 17 & 27 & 17\\
2:2:9& 26 & 2 & 2\\
1:5:14& 25 & 0 & 0\\
2:2:3& 25 & 8 & 3\\
2:3:7& 24 & 10 & 3\\
1:4:8& 15 & 24 & 11\\
2:4:1& 24 & 15 & 12\\
2:4:13& 22 & 1 & 1\\
4:6:1& 22 & 5 & 3\\
1:5:4& 21 & 6 & 1\\
4:11:22& 0 & 20 & 3\\
2:6:7& 19 & 5 & 4\\
1:4:6& 18 & 4 & 3\\
3:4:10& 17 & 2 & 0\\
1:4:5& 16 & 0 & 0\\
1:5:2& 3 & 1 & 16\\
1:4:11& 15 & 0 & 0\\
2:4:5& 15 & 4 & 4\\
1:4:3& 14 & 8 & 8\\
3:4:12& 11 & 14 & 0\\
2:4:15& 14 & 0 & 0\\
\end{tabular}
\etab

\clearpage

\tab
\mycaption{Prevalence of ternary stoichiometries ($A_xB_yC_z,~C=$ O, S, Se; top 120) continued (3/4).}
\tabvspace
\begin{tabular}{l|r|r|r}
stoichiometry & oxides & sulfides & selenides \\
\hline
1:6:12& 14 & 0 & 0\\
1:5:5& 13 & 3 & 2\\
2:2:11& 13 & 0 & 0\\
4:5:13& 1 & 12 & 0\\
2:12:3& 0 & 12 & 4\\
1:4:9& 12 & 1 & 0\\
1:6:11& 12 & 0 & 0\\
1:4:12& 12 & 1 & 0\\
4:4:11& 11 & 1 & 1\\
3:4:9& 11 & 5 & 3\\
3:5:14& 11 & 0 & 0\\
2:3:10& 11 & 0 & 0\\
1:6:2& 6 & 2 & 11\\
1:6:4& 9 & 10 & 11\\
2:4:3& 5 & 10 & 7\\
1:12:20& 10 & 0 & 0\\
10:14:1& 10 & 1 & 0\\
2:5:13& 10 & 0 & 0\\
2:15:19& 0 & 4 & 10\\
1:8:6& 10 & 6 & 4\\
1:8:14& 10 & 0 & 0\\
1:7:12& 9 & 0 & 0\\
1:5:7& 8 & 1 & 1\\
1:6:6& 8 & 0 & 0\\
2:9:3& 0 & 0 & 8\\
2:6:13& 8 & 1 & 1\\
1:3:12& 8 & 0 & 3\\
1:8:17& 0 & 8 & 8\\
1:12:19& 8 & 0 & 0\\
3:4:8& 8 & 2 & 1\\
\end{tabular}
\etab

\clearpage

\tab
\mycaption{Prevalence of ternary stoichiometries ($A_xB_yC_z,~C=$ O, S, Se; top 120) continued (4/4).}
\tabvspace
\begin{tabular}{l|r|r|r}
stoichiometry & oxides & sulfides & selenides \\
\hline
1:5:6& 7 & 1 & 0\\
2:7:2& 0 & 6 & 7\\
1:5:1& 2 & 7 & 5\\
3:4:4& 7 & 2 & 0\\
2:3:1& 2 & 5 & 7\\
2:5:10& 7 & 0 & 0\\
2:5:12& 7 & 0 & 0\\
2:3:11& 4 & 7 & 2\\
4:6:19& 7 & 0 & 0\\
1:1:6& 7 & 3 & 2\\
4:6:13& 5 & 6 & 3\\
1:7:1& 0 & 5 & 6\\
1:10:14& 0 & 6 & 4\\
4:5:15& 6 & 0 & 0\\
3:3:1& 6 & 0 & 0\\
1:12:2& 0 & 1 & 6\\
1:3:10& 6 & 0 & 0\\
2:6:1& 2 & 6 & 4\\
5:9:5& 6 & 0 & 0\\
2:6:15& 5 & 0 & 0\\
2:9:6& 1 & 5 & 0\\
4:4:3& 1 & 1 & 5\\
4:5:12& 5 & 1 & 0\\
2:8:7& 5 & 0 & 0\\
3:6:1& 0 & 5 & 0\\
1:8:2& 0 & 5 & 1\\
1:7:6& 3 & 5 & 4\\
1:8:8& 1 & 5 & 2\\
2:9:2& 1 & 5 & 1\\
3:5:2& 5 & 3 & 0\\
\end{tabular}
\etab

\clearpage

\tab
\mycaption{Prevalence of ternary stoichiometries ($A_xB_yC_z,~C=$ O, S, Se, with $M_A>M_B$ when $x\neq y$; top 120) (1/4).}
\tabvspace
\begin{tabular}{l|r|r|r}
stoichiometry & oxides & sulfides & selenides \\
\hline
1:1:3& 718 & 147 & 70\\
1:1:4& 428 & 28 & 8\\
2:2:7& 304 & 2 & 0\\
1:1:2& 269 & 242 & 145\\
2:1:4& 206 & 101 & 70\\
1:2:4& 190 & 141 & 88\\
1:1:1& 113 & 140 & 90\\
2:1:6& 122 & 7 & 3\\
1:2:6& 115 & 1 & 5\\
1:2:5& 90 & 32 & 14\\
2:1:2& 72 & 37 & 24\\
2:2:5& 64 & 32 & 23\\
3:1:9& 62 & 0 & 0\\
2:3:8& 60 & 6 & 2\\
2:1:5& 59 & 25 & 8\\
1:2:2& 59 & 27 & 14\\
2:1:3& 59 & 23 & 22\\
1:3:6& 34 & 54 & 19\\
3:1:6& 49 & 9 & 3\\
4:1:1& 15 & 43 & 12\\
5:3:12& 43 & 2 & 0\\
3:1:3& 42 & 26 & 23\\
1:2:3& 41 & 32 & 20\\
4:1:4& 40 & 8 & 6\\
4:2:9& 40 & 6 & 1\\
1:4:4& 38 & 9 & 7\\
1:1:5& 37 & 2 & 2\\
2:1:7& 36 & 3 & 1\\
3:1:7& 34 & 1 & 0\\
3:1:4& 32 & 33 & 13\\
\end{tabular}
\label{tab:art130:Prevalence_of_Ternary_Stoichiometries_2}
\etab

\clearpage

\tab
\mycaption{Prevalence of ternary stoichiometries ($A_xB_yC_z,~C=$ O, S, Se, with $M_A>M_B$ when $x\neq y$; top 120) continued (2/4).}
\tabvspace
\begin{tabular}{l|r|r|r}
stoichiometry & oxides & sulfides & selenides \\
\hline
1:3:7& 33 & 0 & 0\\
4:2:11& 31 & 5 & 0\\
2:2:1& 29 & 30 & 17\\
3:1:1& 30 & 12 & 29\\
3:1:8& 30 & 1 & 0\\
6:1:8& 2 & 29 & 21\\
4:1:7& 28 & 3 & 2\\
3:2:2& 1 & 27 & 7\\
3:2:8& 27 & 2 & 2\\
3:2:9& 27 & 2 & 3\\
2:2:9& 26 & 2 & 2\\
1:3:4& 26 & 21 & 18\\
2:1:8& 26 & 2 & 1\\
1:4:7& 4 & 26 & 0\\
5:1:14& 25 & 0 & 0\\
2:2:3& 25 & 8 & 3\\
1:3:8& 25 & 0 & 0\\
2:3:6& 24 & 8 & 8\\
3:1:2& 24 & 12 & 4\\
1:3:3& 20 & 24 & 16\\
1:3:5& 23 & 9 & 5\\
3:2:12& 23 & 0 & 0\\
2:3:12& 22 & 0 & 0\\
2:4:9& 22 & 1 & 2\\
6:4:1& 18 & 1 & 2\\
5:1:4& 17 & 6 & 1\\
2:3:7& 16 & 10 & 2\\
4:11:22& 0 & 16 & 3\\
4:1:5& 16 & 0 & 0\\
1:3:9& 16 & 0 & 0\\
\end{tabular}
\etab

\clearpage

\tab
\mycaption{Prevalence of ternary stoichiometries ($A_xB_yC_z,~C=$ O, S, Se, with $M_A>M_B$ when $x\neq y$; top 120) continued (3/4).}
\tabvspace
\begin{tabular}{l|r|r|r}
stoichiometry & oxides & sulfides & selenides \\
\hline
1:2:8& 15 & 0 & 2\\
2:4:11& 15 & 3 & 2\\
1:4:8& 9 & 15 & 9\\
2:4:7& 15 & 4 & 0\\
5:1:8& 15 & 10 & 15\\
3:2:4& 10 & 14 & 9\\
4:2:13& 14 & 1 & 1\\
1:5:8& 14 & 4 & 7\\
3:4:12& 10 & 14 & 0\\
2:1:1& 14 & 1 & 3\\
4:2:1& 14 & 11 & 10\\
3:1:5& 14 & 6 & 1\\
1:2:1& 13 & 11 & 2\\
4:2:7& 13 & 2 & 0\\
4:2:15& 13 & 0 & 0\\
4:1:6& 13 & 0 & 0\\
2:3:4& 7 & 13 & 8\\
1:3:1& 13 & 1 & 1\\
6:1:12& 13 & 0 & 0\\
2:2:11& 13 & 0 & 0\\
1:2:7& 13 & 0 & 1\\
4:1:3& 12 & 5 & 5\\
5:1:5& 12 & 1 & 0\\
5:1:2& 3 & 1 & 12\\
4:1:11& 12 & 0 & 0\\
12:2:3& 0 & 12 & 3\\
2:6:7& 11 & 2 & 0\\
6:1:2& 6 & 2 & 11\\
4:4:11& 11 & 1 & 1\\
5:4:13& 0 & 11 & 0\\
\end{tabular}
\etab

\clearpage

\tab
\mycaption{Prevalence of ternary stoichiometries ($A_xB_yC_z,~C=$ O, S, Se, with $M_A>M_B$ when $x\neq y$; top 120) continued (4/4).}
\tabvspace
\begin{tabular}{l|r|r|r}
stoichiometry & oxides & sulfides & selenides \\
\hline
4:1:12& 11 & 1 & 0\\
6:1:4& 7 & 10 & 11\\
8:1:6& 10 & 2 & 2\\
3:4:10& 10 & 1 & 0\\
1:6:8& 2 & 10 & 9\\
1:12:20& 10 & 0 & 0\\
14:10:1& 10 & 1 & 0\\
2:4:1& 10 & 4 & 2\\
4:1:9& 9 & 1 & 0\\
3:4:9& 9 & 2 & 1\\
1:3:2& 9 & 7 & 7\\
2:4:5& 9 & 2 & 1\\
4:1:8& 6 & 9 & 2\\
3:2:7& 8 & 0 & 1\\
6:2:7& 8 & 3 & 4\\
9:2:3& 0 & 0 & 8\\
2:4:13& 8 & 0 & 0\\
5:2:13& 8 & 0 & 0\\
12:1:19& 8 & 0 & 0\\
2:3:2& 8 & 5 & 4\\
1:8:17& 0 & 8 & 8\\
8:1:14& 8 & 0 & 0\\
3:2:6& 7 & 3 & 3\\
6:1:11& 7 & 0 & 0\\
2:3:9& 7 & 0 & 0\\
3:4:4& 7 & 1 & 0\\
3:2:10& 7 & 0 & 0\\
6:1:6& 7 & 0 & 0\\
5:1:1& 2 & 7 & 4\\
4:3:10& 7 & 1 & 0\\
\end{tabular}
\etab

\clearpage

\tab
\mycaption[Prevalence of binary stoichiometries (1/3).]
{The entries for each element column denote the total number of structure types,
total number of unique compounds and then the leading atom with the total
number of structure types of this stoichiometry in which it appears.
The second column shows the stoichiometry $(x:y)$ for $A_xZ_y$, $Z=$ O, S, Se, respectively.}
\tabvspace
\begin{tabular}{l|l|r|r|r}
ratio $y/(x+y)$ & stoichiometry & oxides & sulfides & selenides \\
\hline
0.032 & (2:61) & 1 1 C(1) &    &   \\
0.077 & (1:12) &    & 1 1 B(1) &   \\
0.11 & (1:8) & 3 3 V(1) & 1 1 Ag(1) &   \\
0.12 & (1:7) & 1 1 Cs(1) &    &   \\
0.14 & (1:6) & 4 4 Ti(2) & 7 7 F(5) &   \\
0.16 & (3:16) & 1 1 V(1) &    &   \\
0.18 & (2:9) & 1 1 Rb(1) & 1 1 Zr(1) & 1 1 Ti(1)\\
0.2 & (1:4) & 2 2 Ta(1) & 1 1 Pd(1) & 5 5 Cl(2)\\
0.21 & (4:15) &    & 1 1 C(1) &   \\
0.21 & (3:11) & 1 1 Cs(1) &    &   \\
0.22 & (2:7) &    &    & 1 1 Pd(1)\\
0.24 & (11:34) &    &    & 1 1 Pd(1)\\
0.25 & (1:3) & 8 9 Zr(3) & 7 7 O(3) & 1 1 O(1)\\
0.26 & (16:45) &    &    & 1 1 Ti(1)\\
0.26 & (5:14) &    & 1 1 Nb(1) &   \\
0.27 & (4:11) &    &    & 1 1 Ti(1)\\
0.27 & (3:8) &    & 1 1 Ti(1) & 1 1 Ti(1)\\
0.28 & (8:21) &    & 1 2 Zr(1) &   \\
0.29 & (2:5) &    &    & 1 1 O(1)\\
0.3 & (3:7) & 2 2 V(2) &    &   \\
0.3 & (7:16) &    & 1 1 Pd(1) &   \\
0.33 & (1:2) & 17 27 H(6) & 18 30 Cu(4) & 15 21 O(4)\\
0.34 & (16:31) &    & 1 1 Cu(1) &   \\
0.36 & (16:29) &    & 1 1 Cu(1) &   \\
0.36 & (5:9) &    & 1 1 Cu(1) &   \\
0.36 & (4:7) &    & 1 1 Cu(1) & 1 1 Pd(1)\\
0.38 & (3:5) &    &    & 2 2 Tl(2)\\
0.4 & (2:3) & 1 1 C(1) & 3 3 Ni(2) & 2 2 Ni(1)\\
0.43 & (3:4) & 2 2 Tl(1) & 4 5 P(2) & 4 4 P(1)\\
0.44 & (4:5) &    & 2 2 V(1) & 2 4 V(1)\\
\end{tabular}
\label{tab:art130:Prevalence_of_Elements_in_Binaries_Stoichiometries}
\etab

\clearpage

\tab
\mycaption[Prevalence of binary stoichiometries continued (2/3).]
{The entries for each element column denote the total number of structure types,
total number of unique compounds and then the leading atom with the total
number of structure types of this stoichiometry in which it appears.
The second column shows the stoichiometry $(x:y)$ for $A_xZ_y$, $Z=$ O, S, Se, respectively.}
\tabvspace
\begin{tabular}{l|l|r|r|r}
ratio $y/(x+y)$ & stoichiometry & oxides & sulfides & selenides \\
\hline
0.45 & (5:6) &    & 1 1 N(1) & 1 1 Ni(1)\\
0.47 & (7:8) &    &    & 1 1 Bi(1)\\
0.47 & (15:17) &    & 1 1 Rh(1) & 1 1 Pd(1)\\
0.47 & (8:9) &    & 2 2 Ni(1) & 1 1 Co(1)\\
0.5 & (1:1) & 51 99 Mg(12) & 88 165 Zn(39) & 38 108 Ga(5)\\
0.51 & (18:17) &    & 1 1 Ni(1) &   \\
0.53 & (9:8) &    & 1 1 As(1) & 1 1 Bi(1)\\
0.53 & (8:7) &    & 3 3 Fe(3) & 2 2 Fe(2)\\
0.54 & (7:6) &    & 1 1 In(1) & 2 2 In(2)\\
0.55 & (6:5) &    & 1 1 Cr(1) &   \\
0.55 & (11:9) &    &    & 2 2 Mo(2)\\
0.55 & (16:13) & 1 1 V(1) &    &   \\
0.56 & (5:4) & 3 3 Ti(1) & 2 3 P(2) & 1 1 P(1)\\
0.56 & (19:15) &    & 1 1 Mo(1) & 1 1 Mo(1)\\
0.57 & (4:3) & 18 22 Fe(8) & 13 25 Fe(3) & 7 21 Ti(2)\\
0.58 & (11:8) &    & 1 1 Tm(1) &   \\
0.58 & (7:5) &    & 1 6 Y(1) &   \\
0.59 & (22:15) &    & 1 1 Tm(1) &   \\
0.6 & (3:2) & 43 146 Bi(16) & 23 58 Yb(6) & 18 38 In(6)\\
0.62 & (8:5) & 1 1 Mn(1) & 1 1 Cr(1) & 1 1 Cr(1)\\
0.62 & (5:3) & 6 7 V(4) & 2 2 U(2) & 2 2 U(2)\\
0.63 & (12:7) & 1 3 Tb(1) &    &   \\
0.64 & (7:4) & 4 6 Ti(3) & 1 1 P(1) &   \\
0.64 & (16:9) & 1 1 Pr(1) &    &   \\
0.64 & (9:5) & 6 7 Ti(3) &    &   \\
0.65 & (20:11) & 1 2 Tb(1) &    &   \\
0.65 & (11:6) & 7 9 Ti(4) &    &   \\
0.65 & (13:7) & 1 2 V(1) &    &   \\
0.65 & (15:8) & 2 3 Ti(2) &    & 1 1 Gd(1)\\
0.65 & (17:9) & 1 2 V(1) &    &   \\
\end{tabular}
\label{tab:art130:Prevalence_of_Elements_in_Binaries_Stoichiometries_2}
\etab

\clearpage

\tab
\mycaption[Prevalence of binary stoichiometries continued (3/3).]
{The entries for each element column denote the total number of structure types,
total number of unique compounds and then the leading atom with the total
number of structure types of this stoichiometry in which it appears.
The second column shows the stoichiometry $(x:y)$ for $A_xZ_y$, $Z=$ O, S, Se, respectively.}
\tabvspace
\begin{tabular}{l|l|r|r|r}
ratio $y/(x+y)$ & stoichiometry & oxides & sulfides & selenides \\
\hline
0.67 & (2:1) & 244 356 Si(185) & 50 123 Ti(9) & 34 79 Ta(10)\\
0.68 & (13:6) & 5 5 V(5) &    &   \\
0.69 & (35:16) & 1 1 U(1) &    &   \\
0.69 & (9:4) & 5 5 V(2) & 2 2 P(2) & 1 1 Ge(1)\\
0.7 & (7:3) & 1 1 V(1) &    &   \\
0.71 & (12:5) & 1 1 Cr(1) &    &   \\
0.71 & (29:12) & 4 4 Nb(4) &    &   \\
0.71 & (5:2) & 20 26 Nb(6) & 4 8 U(1) & 3 5 Th(1)\\
0.72 & (34:13) & 1 1 U(1) &    &   \\
0.72 & (21:8) & 2 2 W(1) &    &   \\
0.73 & (8:3) & 8 9 U(6) & 2 2 Ir(1) & 1 2 Rh(1)\\
0.73 & (49:18) & 1 1 W(1) &    &   \\
0.73 & (11:4) & 3 3 Mo(3) &    &   \\
0.73 & (47:17) & 2 2 W(1) &    &   \\
0.74 & (14:5) & 2 2 W(1) &    &   \\
0.74 & (23:8) & 2 2 Mo(2) &    &   \\
0.74 & (26:9) & 2 2 Mo(2) &    &   \\
0.74 & (73:25) & 1 1 W(1) &    &   \\
0.75 & (3:1) & 33 41 W(13) & 8 12 Ti(2) & 6 10 Ta(3)\\
0.78 & (7:2) & 4 4 Tc(1) & 2 2 P(2) &   \\
0.8 & (4:1) & 3 4 Ru(2) & 1 1 V(1) & 2 2 Nb(1)\\
0.82 & (9:2) & 1 1 P(1) &    & 2 2 V(1)\\
0.84 & (21:4) & 1 1 U(1) &    &   \\
0.89 & (8:1) &    & 1 1 O(1) &   \\
0.93 & (14:1) &    & 1 1 C(1) &   \\
\end{tabular}
\label{tab:art130:Prevalence_of_Elements_in_Binaries_Stoichiometries_3}
\etab

\clearpage

\tab
\mycaption[Prevalence of stoichiometries for the elements in the binary compounds, $A_xB_y,~B=$ O, S, Se (1/3).]
{The data presented is:
$y:x(n_1)$, $n_2$, $n_3$, where $y:x$ is the leading stoichiometry, $n_1$ is number of compounds
for this stoichiometry, $n_2$ is number of stoichiometries and
$n_3$ is number of unique compounds for this element.}
\tabvspace
\begin{tabular}{l|r|r|r}
atom  &  oxides & sulfides & selenides \\
\hline
Ac & 3:2(1),1,1&&\\
Ag & 1:1(5),5,11&1:2(3),2,4&1:2(2),2,3\\
Al & 3:2(7),3,9&3:2(3),1,3&3:2(1),1,1\\
As & 3:2(5),3,8&1:1(4),6,10&1:1(2),3,5\\
Au & 3:2(1),1,1&1:2(1),1,1&1:1(3),1,3\\
B & 3:2(2),3,4&1:12(1),3,3&\\
Ba & 1:1(3),2,5&3:1(2),4,5&1:1(2),3,4\\
Be & 1:1(4),1,4&1:1(1),1,1&1:1(2),1,2\\
Bi & 3:2(16),4,20&3:2(1),1,1&1:1(4),6,11\\
Br & 1:2(1),2,2&1:1(1),1,1&1:1(2),2,3\\
C & 2:1(9),4,13&4:15(1),5,5&2:1(1),1,1\\
Ca & 1:1(2),2,3&1:1(1),1,1&1:1(2),1,2\\
Cd & 1:1(2),2,3&1:1(4),2,5&1:1(3),2,4\\
Ce & 2:1(5),5,12&2:1(4),4,8&2:1(3),3,6\\
Cl & 1:2(1),4,4&1:2(1),2,2&1:4(2),2,3\\
Co & 1:1(5),3,10&8:9(1),4,4&4:3(2),4,6\\
Cr & 2:1(5),8,12&1:1(2),5,7&1:1(2),5,7\\
Cs & 1:7(1),8,8&1:2(1),5,5&1:2(2),4,5\\
Cu & 1:1(3),5,9&1:2(4),7,14&1:2(2),4,7\\
Dy & 3:2(4),1,4&2:1(2),5,6&1:1(2),4,5\\
Er & 3:2(4),1,4&3:2(3),4,8&2:1(2),3,4\\
Eu & 3:2(4),4,7&1:1(3),3,6&1:1(2),2,3\\
F & &1:6(5),1,5&1:4(1),1,1\\
Fe & 4:3(8),4,19&1:1(6),5,18&1:1(4),4,9\\
Ga & 3:2(3),1,3&1:1(2),2,3&1:1(5),3,7\\
Gd & 3:2(4),3,6&2:1(3),4,6&1:1(1),5,5\\
Ge & 2:1(7),1,7&2:1(5),2,7&2:1(5),3,8\\
H & 1:2(6),2,7&1:2(3),1,3&1:2(1),1,1\\
Hf & 2:1(5),1,5&1:2(1),3,3&2:1(1),2,2\\
Hg & 1:1(5),2,7&1:1(4),1,4&1:1(3),1,3\\
\end{tabular}
\label{tab:art130:Elements_stoichiometries}
\etab

\clearpage

\tab
\mycaption[Prevalence of stoichiometries for the elements in the binary compounds, $A_xB_y,~B=$ O, S, Se continued (2/3).]
{The data presented is:
$y:x(n_1)$, $n_2$, $n_3$, where $y:x$ is the leading stoichiometry, $n_1$ is number of compounds
for this stoichiometry, $n_2$ is number of stoichiometries and
$n_3$ is number of unique compounds for this element.}
\tabvspace
\begin{tabular}{l|r|r|r}
atom  &  oxides & sulfides & selenides \\
\hline
Ho & 3:2(4),1,4&1:1(2),4,7&1:1(1),3,3\\
I & 3:1(2),3,4&&\\
In & 3:2(5),1,5&1:1(2),5,7&3:2(6),5,14\\
Ir & 2:1(1),1,1&2:1(2),3,4&2:1(1),2,2\\
K & 2:1(2),4,5&1:2(2),4,6&1:2(1),4,4\\
La & 3:2(4),1,4&1:1(5),4,12&1:1(4),3,7\\
Li & 1:2(2),4,6&1:2(3),2,4&1:2(1),1,1\\
Lu & 3:2(4),1,4&3:2(3),3,6&1:1(1),3,3\\
Mg & 1:1(12),2,13&1:1(2),1,2&1:1(3),2,4\\
Mn & 3:2(4),6,14&1:1(4),2,6&1:1(4),2,5\\
Mo & 2:1(3),7,15&4:3(2),4,6&11:9(2),4,6\\
N & 2:1(4),5,8&1:1(4),3,6&1:1(2),1,2\\
Na & 2:1(2),4,5&1:1(4),4,9&1:2(1),3,3\\
Nb & 5:2(6),6,18&2:1(7),6,14&2:1(7),8,15\\
Nd & 3:2(4),2,5&2:1(3),4,6&2:1(2),3,4\\
Ni & 1:1(4),2,6&2:1(8),6,16&1:1(2),5,6\\
O & &1:3(3),3,5&1:2(4),3,6\\
Os & 4:1(2),2,3&2:1(1),1,1&2:1(1),1,1\\
P & 5:2(3),5,7&3:4(2),8,13&3:4(1),4,4\\
Pa & 2:1(2),2,3&&\\
Pb & 1:1(4),5,14&1:1(13),1,13&1:1(3),2,4\\
Pd & 1:1(4),3,6&1:4(1),5,5&1:1(2),7,8\\
Pm & 3:2(3),1,3&&\\
Pr & 3:2(3),7,12&2:1(3),4,6&2:1(2),3,4\\
Pt & 2:1(5),3,9&1:1(2),2,3&4:5(1),2,2\\
Pu & 3:2(2),3,4&2:1(2),2,3&1:1(1),3,3\\
Rb & 3:2(2),7,8&1:2(3),4,7&1:2(1),4,4\\
Re & 3:1(4),3,8&2:1(2),1,2&2:1(2),1,2\\
Rh & 3:2(4),2,5&15:17(1),4,4&2:1(2),4,5\\
Ru & 2:1(4),2,6&2:1(1),1,1&2:1(1),1,1\\
\end{tabular}
\label{tab:art130:Elements_stoichiometries_2}
\etab

\clearpage

\tab
\mycaption[Prevalence of stoichiometries for the elements in the binary compounds, $A_xB_y,~B=$ O, S, Se continued (3/3).]
{The data presented is:
$y:x(n_1)$, $n_2$, $n_3$, where $y:x$ is the leading stoichiometry, $n_1$ is number of compounds
for this stoichiometry, $n_2$ is number of stoichiometries and
$n_3$ is number of unique compounds for this element.}
\tabvspace
\begin{tabular}{l|r|r|r}
atom  &  oxides & sulfides & selenides \\
\hline
S & 3:1(3),3,5&&\\
Sb & 3:2(5),3,11&3:2(1),1,1&3:2(1),1,1\\
Sc & 3:2(4),1,4&3:2(2),2,3&1:1(1),2,2\\
Se & 2:1(4),3,6&&\\
Si & 2:1(185),1,185&2:1(2),1,2&2:1(1),1,1\\
Sm & 3:2(4),2,5&1:1(2),3,4&1:1(2),4,5\\
Sn & 2:1(8),3,12&1:1(5),3,8&1:1(3),2,4\\
Sr & 1:1(2),2,3&1:1(2),3,4&1:1(1),1,1\\
Ta & 5:2(3),6,9&2:1(7),6,13&2:1(10),4,15\\
Tb & 3:2(4),5,9&2:1(4),4,7&1:1(2),3,4\\
Tc & 2:1(1),2,2&2:1(1),1,1&\\
Te & 2:1(9),4,12&&\\
Th & 2:1(2),1,2&1:1(1),4,4&1:1(2),5,6\\
Ti & 2:1(14),14,42&2:1(9),5,16&1:1(3),9,13\\
Tl & 1:2(2),3,5&1:1(6),4,9&1:1(3),3,6\\
Tm & 3:2(4),1,4&3:2(6),6,12&1:1(1),3,3\\
U & 8:3(6),9,22&5:3(2),7,9&5:3(2),6,8\\
V & 2:1(10),18,42&2:1(3),6,11&4:5(1),5,5\\
W & 3:1(13),9,24&2:1(2),1,2&2:1(1),1,1\\
Xe & 3:1(1),1,1&&\\
Y & 3:2(5),1,5&3:2(2),4,6&1:1(1),2,2\\
Yb & 3:2(2),3,4&3:2(6),4,12&1:1(2),4,5\\
Zn & 1:1(4),2,5&1:1(39),2,40&1:1(2),2,3\\
Zr & 2:1(7),4,12&1:1(2),7,8&1:2(1),3,3\\
\end{tabular}
\label{tab:art130:Elements_stoichiometries_3}
\etab

\clearpage

\subsection{Correlation between ternary and binary stoichiometries for sulfides and selenides}
In this section we analyze the correlation between ternary and binary stoichiometries for sulfides and selenides.
Figure~\ref{fig:art130:tern_bin_stoichiometries} shows that, like in the oxides, in both the sulfides and selenides
we see a quite scattered pattern. However unlike in the oxides many atoms show points below the line $y=4x$.

\fig
\includegraphics[width=1.0\linewidth]{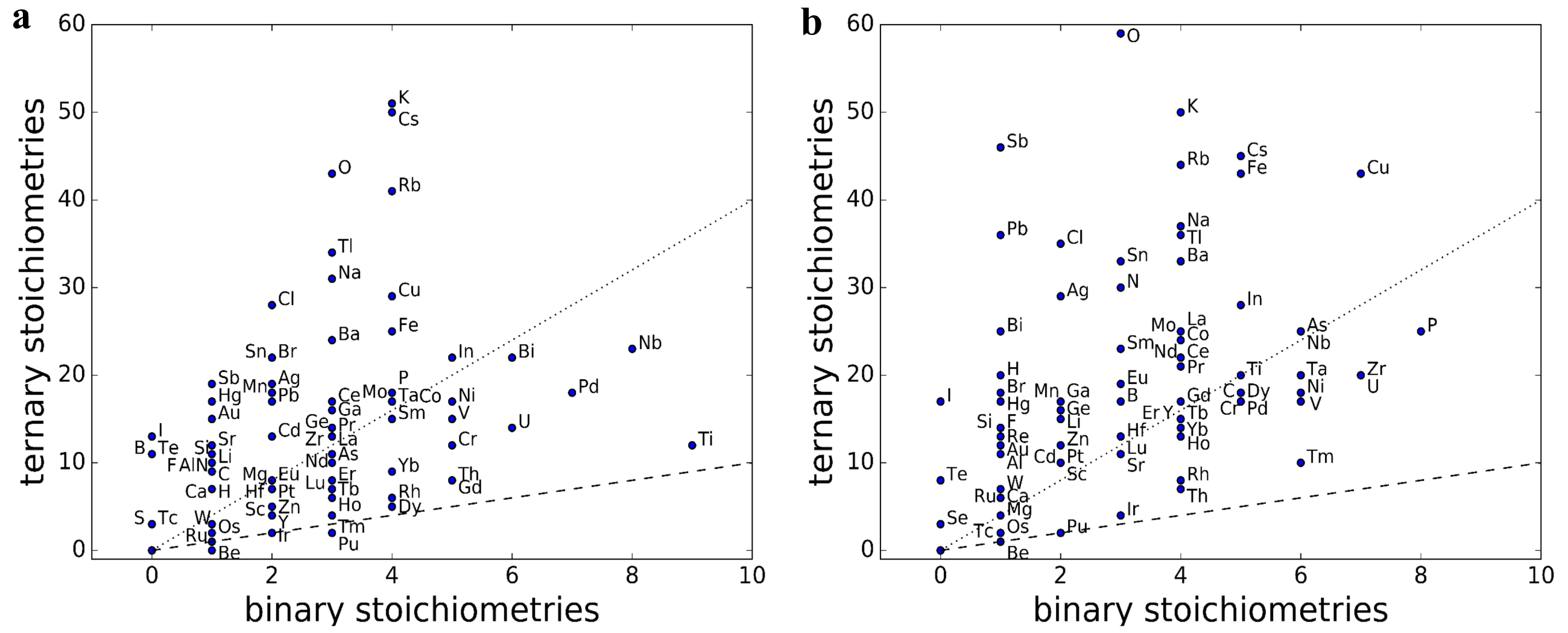}
\mycaption[The number of ternary (\textbf{a}) sulfide and (\textbf{b}) selenide stoichiometries
per element as a function of the count of its respective binary stoichiometries.]
{The dashed line marks perfect similarity $y=x$, and the dotted line marks the ratio $y=4x$.}
\label{fig:art130:tern_bin_stoichiometries}
\efig

We next analyze in Figure~\ref{fig:art130:tern_binproduct_stoichiometries} the number of ternary stoichiometries
as a function of the product of the numbers of the binary stoichiometries of participating atoms.
As for the oxides (Figure~\ref{fig:art130:mendeleev_distribution_all_in_one})
we see a trend of inverse correlation, \ie, as the product of the numbers of binary
stoichiometries increases, the number of ternaries decreases.

\fig
\includegraphics[width=1.0\linewidth]{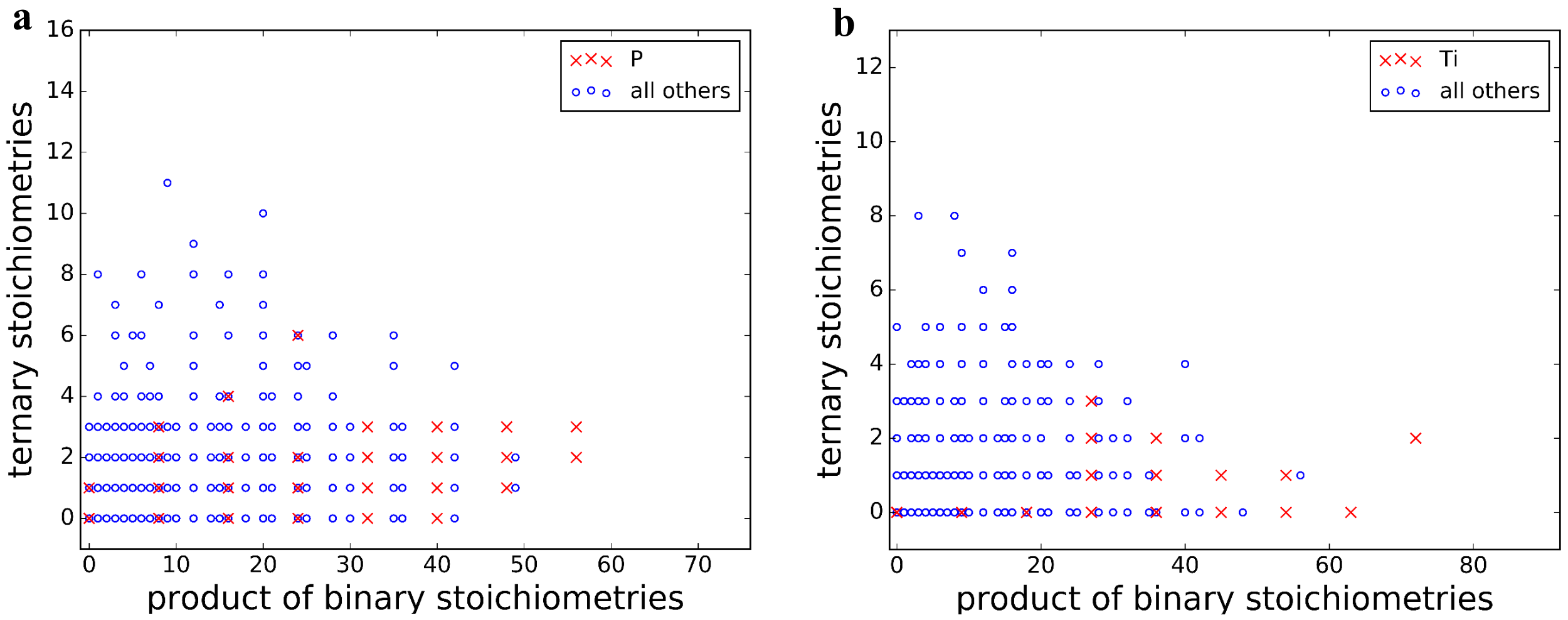}
\mycaption[The number of ternary (\textbf{a}) sulfide and (\textbf{b}) selenide stoichiometries
as a function of the product of the number of the binary stoichiometries of participating elements.]
{The element with the most binary sulfide/selenide stoichiometries (P/Ti) is shown with red ``x'' symbols.
All other compounds are shown with blue circles.}
\label{fig:art130:tern_binproduct_stoichiometries}
\efig

\subsection{Prevalence of unit cell sizes}
In Tables~\ref{tab:art130:Number_of_atoms_in_Binaries_unit_cells} and
\ref{tab:art130:Number_of_atoms_in_ternary_unit_cells}, the number of atoms per unit cell in
binary and ternary compounds is shown for systems of up to 100 atoms in the unit cell.
In the binary oxides, there is higher prevalence for numbers that are multiples of
4, 6, and also 12 ---
for example --- 12(102), 24(58), 80(47) and 72(20). In addition, 5(24) and a few of its multiples are also common.
Prime numbers of atoms per unit cell above 10 are very rare ---
11(2), 19(3), 29(1), 31(2), 67(1) and all the rest do not appear at all.
In the ternary oxides, we see a similar behavior: a high prevalence for numbers that are multiple of
4, 6 and 12 --- for example --- 12 (119), 18 (140), 24 (465), 30(106), 72(102), 80(83), 88(178), 96(51).
Prime numbers, between 10 to 20 do appear --- 11(15), 13(30), 17(6), 19(15), but those
above 20 are very rare.

\tab
\mycaption{Prevalence of unit cell sizes among the binary compounds (1/3).}
\tabvspace
\begin{tabular}{l|r|r|r}
number of atoms & oxides & sulfides & selenides \\
\hline
1 & 0 & 0 & 0\\
2 & 7 & 12 & 8\\
3 & 7 & 15 & 13\\
4 & 24 & 20 & 22\\
5 & 24 & 1 & 5\\
6 & 60 & 48 & 24\\
7 & 0 & 3 & 2\\
8 & 63 & 70 & 60\\
9 & 10 & 1 & 1\\
10 & 15 & 9 & 6\\
11 & 2 & 1 & 2\\
12 & 102 & 63 & 50\\
13 & 0 & 1 & 0\\
14 & 19 & 11 & 9\\
15 & 2 & 1 & 0\\
16 & 23 & 22 & 11\\
17 & 0 & 0 & 1\\
18 & 12 & 6 & 5\\
19 & 3 & 0 & 0\\
20 & 37 & 38 & 20\\
21 & 0 & 2 & 0\\
22 & 8 & 2 & 3\\
23 & 0 & 0 & 0\\
24 & 58 & 39 & 14\\
25 & 1 & 0 & 0\\
26 & 2 & 1 & 2\\
27 & 0 & 1 & 0\\
28 & 27 & 24 & 18\\
29 & 1 & 0 & 0\\
30 & 17 & 10 & 2\\
31 & 2 & 0 & 0\\
32 & 18 & 14 & 10\\
33 & 0 & 0 & 0\\
34 & 2 & 1 & 0\\
35 & 0 & 0 & 0\\
\end{tabular}
\label{tab:art130:Number_of_atoms_in_Binaries_unit_cells}
\etab

\clearpage

\tab
\mycaption{Prevalence of unit cell sizes among the binary compounds continued (2/3).}
\tabvspace
\begin{tabular}{l|r|r|r}
number of atoms & oxides & sulfides & selenides \\
\hline
36 & 25 & 8 & 4\\
37 & 0 & 0 & 0\\
38 & 4 & 0 & 0\\
39 & 0 & 0 & 1\\
40 & 15 & 6 & 2\\
41 & 0 & 0 & 0\\
42 & 0 & 2 & 0\\
43 & 0 & 0 & 0\\
44 & 5 & 5 & 1\\
45 & 0 & 2 & 3\\
46 & 2 & 1 & 0\\
47 & 0 & 0 & 0\\
48 & 27 & 10 & 3\\
49 & 0 & 0 & 0\\
50 & 0 & 0 & 0\\
51 & 0 & 0 & 0\\
52 & 3 & 3 & 2\\
53 & 0 & 0 & 0\\
54 & 2 & 0 & 0\\
55 & 0 & 0 & 0\\
56 & 9 & 9 & 2\\
57 & 0 & 0 & 0\\
58 & 0 & 2 & 0\\
59 & 0 & 0 & 0\\
60 & 5 & 1 & 0\\
61 & 0 & 0 & 0\\
62 & 1 & 0 & 0\\
63 & 0 & 0 & 0\\
64 & 2 & 4 & 1\\
65 & 0 & 0 & 0\\
66 & 0 & 0 & 0\\
67 & 1 & 0 & 0\\
68 & 7 & 3 & 2\\
69 & 0 & 0 & 0\\
70 & 0 & 0 & 0\\
\end{tabular}
\label{tab:art130:Number_of_atoms_in_Binaries_unit_cells_2}
\etab

\clearpage

\tab
\mycaption{Prevalence of unit cell sizes among the binary compounds continued (3/3).}
\tabvspace
\begin{tabular}{l|r|r|r}
number of atoms & oxides & sulfides & selenides \\
\hline
71 & 0 & 0 & 0\\
72 & 20 & 2 & 1\\
73 & 0 & 0 & 0\\
74 & 0 & 1 & 0\\
75 & 0 & 0 & 0\\
76 & 3 & 2 & 0\\
77 & 0 & 0 & 0\\
78 & 0 & 0 & 0\\
79 & 0 & 0 & 0\\
80 & 47 & 6 & 12\\
81 & 0 & 0 & 0\\
82 & 2 & 0 & 0\\
83 & 0 & 0 & 0\\
84 & 2 & 0 & 0\\
85 & 0 & 0 & 0\\
86 & 0 & 0 & 0\\
87 & 0 & 0 & 0\\
88 & 1 & 3 & 2\\
89 & 0 & 0 & 0\\
90 & 0 & 0 & 2\\
91 & 0 & 0 & 0\\
92 & 1 & 0 & 0\\
93 & 0 & 0 & 0\\
94 & 1 & 0 & 0\\
95 & 0 & 0 & 0\\
96 & 22 & 1 & 0\\
97 & 0 & 0 & 0\\
98 & 1 & 0 & 0\\
99 & 0 & 0 & 0\\
100 & 0 & 0 & 0\\
\end{tabular}
\label{tab:art130:Number_of_atoms_in_Binaries_unit_cells_3}
\etab

\clearpage

\tab
\mycaption{Prevalence of unit cell sizes among the ternary compounds (1/3).}
\tabvspace
\begin{tabular}{l|r|r|r}
number of atoms & oxides & sulfides & selenides \\
\hline
1 & 0 & 0 & 0\\
2 & 0 & 0 & 0\\
3 & 1 & 1 & 2\\
4 & 81 & 112 & 64\\
5 & 173 & 36 & 16\\
6 & 62 & 35 & 23\\
7 & 10 & 29 & 16\\
8 & 64 & 48 & 18\\
9 & 38 & 8 & 4\\
10 & 186 & 33 & 35\\
11 & 15 & 2 & 0\\
12 & 119 & 104 & 76\\
13 & 30 & 5 & 1\\
14 & 116 & 44 & 60\\
15 & 12 & 40 & 30\\
16 & 143 & 106 & 69\\
17 & 6 & 3 & 0\\
18 & 140 & 31 & 26\\
19 & 15 & 0 & 0\\
20 & 363 & 179 & 88\\
21 & 10 & 1 & 0\\
22 & 142 & 25 & 20\\
23 & 1 & 0 & 1\\
24 & 465 & 146 & 57\\
25 & 7 & 0 & 0\\
26 & 65 & 26 & 14\\
27 & 16 & 1 & 0\\
28 & 287 & 190 & 130\\
29 & 1 & 0 & 0\\
30 & 106 & 22 & 12\\
31 & 0 & 0 & 0\\
32 & 181 & 96 & 67\\
33 & 4 & 0 & 0\\
34 & 38 & 16 & 8\\
35 & 0 & 1 & 0\\
\end{tabular}
\label{tab:art130:Number_of_atoms_in_ternary_unit_cells}
\etab

\clearpage

\tab
\mycaption{Prevalence of unit cell sizes among the ternary compounds continued (2/3).}
\tabvspace
\begin{tabular}{l|r|r|r}
number of atoms & oxides & sulfides & selenides \\
\hline
36 & 211 & 67 & 46\\
37 & 5 & 0 & 0\\
38 & 26 & 19 & 3\\
39 & 1 & 2 & 0\\
40 & 216 & 65 & 30\\
41 & 2 & 0 & 0\\
42 & 40 & 4 & 5\\
43 & 3 & 0 & 0\\
44 & 193 & 40 & 22\\
45 & 6 & 2 & 2\\
46 & 24 & 2 & 5\\
47 & 1 & 0 & 0\\
48 & 118 & 28 & 17\\
49 & 6 & 0 & 0\\
50 & 12 & 0 & 0\\
51 & 0 & 0 & 0\\
52 & 114 & 27 & 16\\
53 & 0 & 0 & 0\\
54 & 17 & 8 & 2\\
55 & 1 & 0 & 0\\
56 & 171 & 109 & 56\\
57 & 6 & 0 & 0\\
58 & 14 & 8 & 3\\
59 & 1 & 0 & 0\\
60 & 104 & 31 & 10\\
61 & 1 & 0 & 0\\
62 & 7 & 0 & 3\\
63 & 5 & 0 & 0\\
64 & 86 & 31 & 31\\
65 & 0 & 0 & 0\\
66 & 17 & 0 & 1\\
67 & 0 & 0 & 0\\
68 & 99 & 28 & 14\\
69 & 0 & 0 & 0\\
70 & 6 & 0 & 2\\
\end{tabular}
\label{tab:art130:Number_of_atoms_in_ternary_unit_cells_2}
\etab

\clearpage

\tab
\mycaption{Prevalence of unit cell sizes among the ternary compounds continued (3/3).}
\tabvspace
\begin{tabular}{l|r|r|r}
number of atoms & oxides & sulfides & selenides \\
\hline
71 & 0 & 0 & 0\\
72 & 102 & 48 & 39\\
73 & 0 & 0 & 0\\
74 & 2 & 21 & 3\\
75 & 0 & 0 & 0\\
76 & 48 & 6 & 5\\
77 & 0 & 0 & 0\\
78 & 8 & 1 & 0\\
79 & 0 & 0 & 0\\
80 & 83 & 8 & 8\\
81 & 0 & 0 & 0\\
82 & 3 & 1 & 0\\
83 & 0 & 0 & 0\\
84 & 30 & 17 & 7\\
85 & 0 & 0 & 0\\
86 & 7 & 0 & 0\\
87 & 1 & 0 & 0\\
88 & 178 & 12 & 20\\
89 & 0 & 0 & 0\\
90 & 9 & 2 & 1\\
91 & 0 & 0 & 0\\
92 & 20 & 8 & 6\\
93 & 0 & 0 & 0\\
94 & 3 & 0 & 0\\
95 & 0 & 0 & 0\\
96 & 51 & 23 & 6\\
97 & 0 & 0 & 0\\
98 & 1 & 1 & 2\\
99 & 3 & 0 & 0\\
100 & 14 & 3 & 2\\
\end{tabular}
\label{tab:art130:Number_of_atoms_in_ternary_unit_cells_3}
\etab

\clearpage

\subsection{Additional Mendeleev plots}
The Mendeleev map for the 1:1:2 stoichiometry are shown in~\ref{fig:art130:mend_211_stoichiometries}.
The maps of the sulfides and selenides cover nearly identical regions, while that of the oxides
includes an additional row for hydrogen (Mendeleev number 103).

\fig
\includegraphics[width=1.0\linewidth]{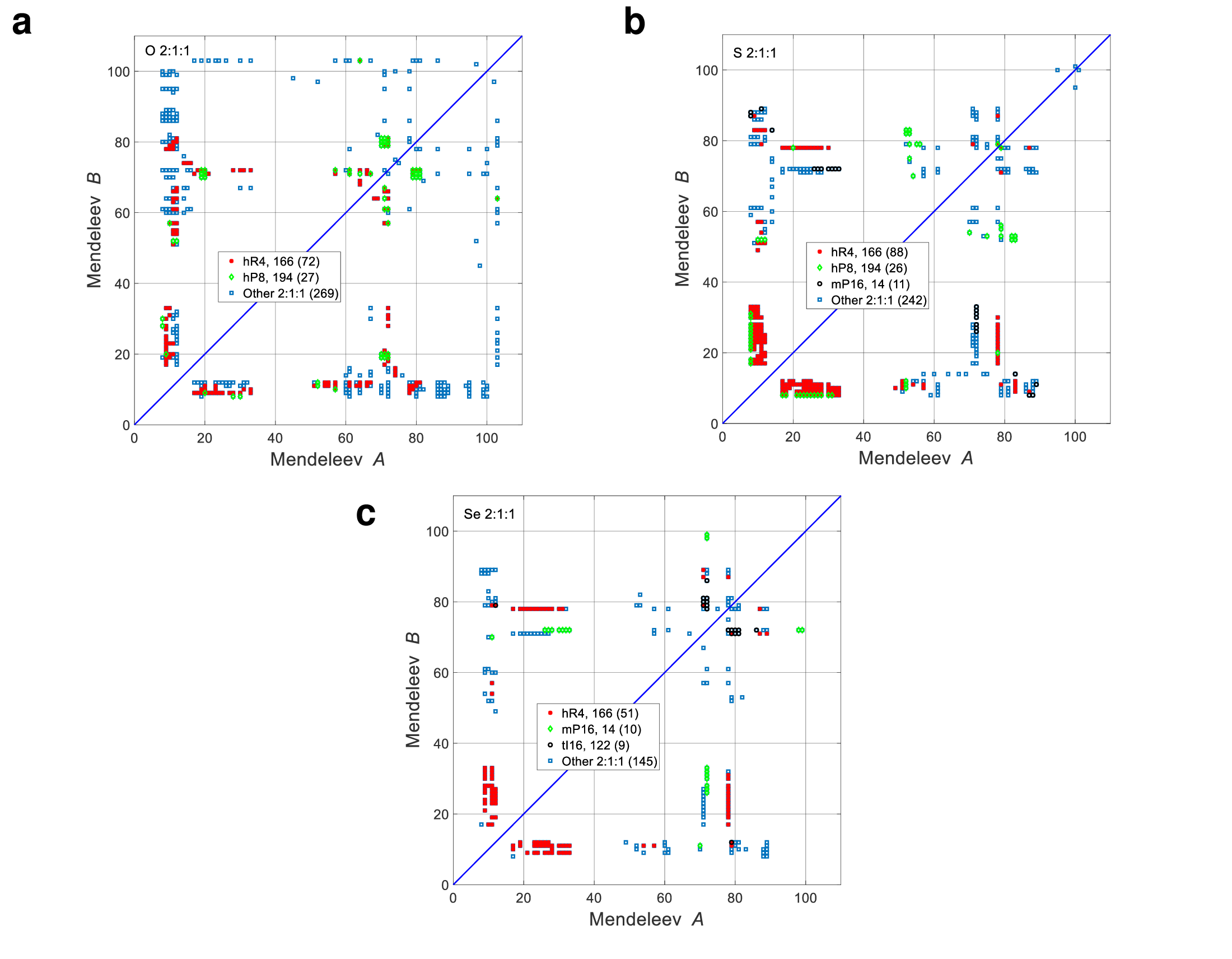}
\mycaption[Comparison of Mendeleev maps for the 211 (\textbf{a}) oxide, (\textbf{b}) sulfide and (\textbf{c}) selenide stoichiometries.]
{The number in parenthesis is the number of compounds for this structure type, for ``Other'',
it refers to the total number of compounds with this stoichiometry.}
\label{fig:art130:mend_211_stoichiometries}
\efig

\subsection{Summary}
We present a comprehensive analysis of the statistics of the binary
and ternary compounds of oxygen, sulfur and selenium. This analysis and the visualization tools presented here are
valuable to finding trends as well as exceptions and peculiar phenomena.

Oxygen has a higher electronegativity (3.44) than sulfur
(2.58) and selenium (2.55), which are similar to each other.
Therefore, one can expect that oxygen will form compounds with a stronger ionic character.
Oxygen is 1000 times more abundant than sulfur, and more than $10^6$ times than selenium~\cite{wedepohl1995composition},
however, it has less than two times the number of binary compounds compared to sulfur and $2.5$ that of selenium.
Hence, the abundance of those elements plays a little role in the relative numbers of their known compounds.
These important differences are reflected in our analysis by
the significantly larger fraction of oxygen rich compounds compared to
those that are sulfur or selenium rich.
Structure type classification also shows that there is little overlap between the oxygen
structure types to sulfur or selenium structure types, while
sulfur and selenium present a much higher overlap. The gaps in these overlaps, especially between
the sulfides and selenides, indicate that favorable candidates for new compounds
may be obtained by simple element substitution in the corresponding structures.
In particular, structures that are significantly more common in one family,
such as KrF$_{2}$ in the oxides, may be good candidates for new compounds in another.
Comparison of these three 6A elements binary and ternary
compounds shows significant differences but also some similarities in the symmetry
distributions among the various Bravais lattices and their
corresponding space groups. In particular, the majority of structure types in all three families have a few or single
compound realizations. This prevalence of unique structure types suggests a ripe field for identification of currently unknown compounds,
by substitution of elements of similar chemical characteristics.
In addition, the analysis of the distribution of known compounds among symmetry space groups and,
in particular, their apparent concentration in specific hot spots of this
symmetry space may be serve as a useful insight for searches of potential new compounds.

An important observation is the existence of different gaps (missing stoichiometries) in the stoichiometry distribution of the oxide
binary compounds compared to the sulfides and selenides (Figures~\ref{fig:triangle_stoichiometries} and \ref{fig:art130:specific_triangle_stoichiometries},
and Tables~\ref{tab:art130:Prevalence_of_Elements_in_Binaries_Stoichiometries}-\ref{tab:art130:Prevalence_of_Elements_in_Binaries_Stoichiometries_3}).
Stoichiometries such as 5:7 appear in the oxides but are missing in the sulfides and selenides.
More rare are non-overlapping gaps between the selenides and sulfides, \eg, 6:1 and 5:7.
These should be prime candidates for new compounds by element substitution between the two families.
Future work would be directed at exploiting these discrepancies to search for new compounds within different subsets of those compound families.

Specific elements tend to present very different stoichiometry distributions, for example, silicon forms only
one oxide stoichiometry (SiO$_2$) while transition metals such as titanium and vanadium present
14 and 18 different stoichiometries respectively.
These differences clearly reflect the different chemistry of those elements, while the large number of reported
SiO$_2$ structures might reflect research bias into silicon compounds.

Another important finding is that there is an inverse correlation between the number of ternary stoichiometries
to the product of binary stoichiometries of participating elements.
This can be caused by the fact that there are too many binary phases and hence it becomes
difficult to create a stable ternary that competes with all of them.

A Mendeleev analysis of the common structure types of these
families shows accumulation of different structures at
well defined regions of their respective maps, similar to the well-known Pettifor maps of binary structure types.
Furthermore, at least for some of the stoichiometries, similarity of the maps for a
given stoichiometry is demonstrated across all three elements.
These maps should therefore prove useful for predictive purposes regarding the existence
of yet unknown compounds of the corresponding structure types.
Future work will be directed at exploiting identified non-overlapping gaps in the
Mendeleev maps for a directed search of new compounds in these families.
Complementary properties (\eg, partial charges, bond analysis, electronic properties)
should be incorporated in the analysis to reveal additional insights of the aforementioned trends among the three elements.
\clearpage
\section{AFLOW Standard for High-Throughput Materials Science Calculations}
\label{sec:art104}

This study follows from a collaborative effort described in Reference~\cite{curtarolo:art104},
which was awarded with Comput. Mater. Sci. Editor's Choice.

\subsection{Introduction} \label{subsec:art104:intro}

The emergence of computational materials science over the last two decades has been inextricably linked to the
development of complex quantum-mechanical codes that enable accurate evaluation of the electronic and
thermodynamic properties of a wide range of materials. The continued advancement of this field entails the
construction of large open databases of materials properties that can be easily reproduced and extended.
One obstacle to the reproducibility of the data is the unavoidable complexity of the codes used to obtain
it. Published data usually includes basic information about the underlying calculations that  allows rough
reproduction. However, exact duplication depends on many details, that are seldom reported, and is therefore
difficult to achieve.

These difficulties might limit the utility of the databases currently being created by high-throughput frameworks,
such as \AFLOW~\cite{aflowPAPER, aflowlibPAPER, aflowAPI} and the Materials Project~\cite{APL_Mater_Jain2013,CMS_Ong2012b}.
For maximal impact, the data stored in these repositories must be generated and represented in a consistent and robust manner,
and shared through standardized calculation and communication protocols. Following these guidelines would promote
optimal use of the results generated by the entire community.

The \AFLOW\ (Automatic FLOW) code is a framework for high-throughput computational materials
discovery~\cite{aflowPAPER, aflowlibPAPER, aflowAPI, aflowlib.org},
using separate \DFT\ packages to calculate electronic
structure and optimize the atomic geometry. The \AFLOW\ framework works with the
\VASP~\citeVASP\ \DFT\ package,
and integration with the \QUANTUMESPRESSO\ software~\cite{quantum_espresso_2009}
is currently in progress.
The \AFLOW\ framework includes
preprocessing functions for generating input files for the \DFT\ package; obtaining the initial geometric structures
by extracting the relevant data from crystallographic information files or by generating them using inbuilt prototype
databases, and then transforming them into standard forms which are easiest to calculate. It then runs and monitors
the \DFT\ calculations automatically, detecting and responding to calculation failures, whether they are due to insufficient
hardware resources or to runtime errors of the \DFT\ calculation itself. Finally, \AFLOW\ contains postprocessing
routines to extract specific properties from the results of one or more of the \DFT\ calculations, such as the band
structure or thermal properties~\cite{curtarolo:art96}.

The \AFLOWorg\ repository~\cite{aflowlibPAPER, aflowAPI, aflowlib.org} was built according to these principles of consistency and reproducibility,
and the data it contains can be easily accessed through a representational state transfer application programming
interface (\RESTAPI)~\cite{aflowAPI}. In this study we present a detailed description of the \AFLOW\ standard
for high-throughput (HT) materials science calculations by which the data in this repository was created.

\subsection{AFLOW calculation types} \label{subsec:art104:AFLOWtypes}
The \AFLOWorg\ repository~\cite{aflowlibPAPER} is divided into databases containing calculated
properties of over 625,000 materials:
the Binary Alloy Project, the Electronic Structure database, the Heusler database, and the Elements database.
These are freely accessible online via the \AFLOWorg~\cite{aflowlib.org}, as well as through the
\API~\cite{aflowAPI}. The Electronic Structure database consists of entries found in the Inorganic Crystal
Structures Database, \ICSD~\cite{ICSD, ICSD3}, and will thus be referred to as ``\ICSD'' throughout this publication.
The Heusler database consists of ternary compounds, primarily based on the Heusler structure but with other
structure types now being added.

The high-throughput construction of these materials databases relies on a pre-defined set of standard {\textit{calculation
types}}. These are designed to accommodate the interest in various properties of a given material (\eg, the ground
state ionic configuration, thermodynamic quantities, electronic and
magnetic properties), the program flow of the HT framework that
envelopes the \DFT\ portions of the calculations, as well as the practical
need for computational robustness. The \AFLOW\ standard thus deals with the parameters involved in the following
calculation types:

\begin{enumerate}
  \item {{\verb!RELAX!}.} Geometry optimizations using algorithms implemented within the \DFT\ package. This calculation
  type is concerned with obtaining the ionic configuration and cell
  shape and volume that correspond to a minimum in the
  total energy. It consists of two sequential relaxation steps. The starting point for the first step, {\verb!RELAX1!},
  can be an entry taken from an external source, such as a library of alloy
  prototypes~\cite{Massalski, curtarolo:calphad_2005_monster}, the \ICSD\ database, or the Pauling
  File~\cite{PaulingFile}. These initial entries are preprocessed by
  \AFLOW, and cast into a unit cell that is most convenient
  for calculation, usually the standard primitive cell, in the format appropriate for the \DFT\ package in use. The second step, {\verb!RELAX2!},
  uses the final ionic positions from the first step as its starting point, and serves as a type of annealing step.
  This is used for jumping out of possible local minima resulting from wavefunction artifacts.
  \item {{\verb!STATIC!}.} A single-point energy calculation. The starting point is the set of final ionic positions,
  as produced by the {\verb!RELAX2!} step. The outcome of this calculation is used in the determination of most
  of the thermodynamic and electronic properties included in the various \AFLOWorg\ database.
 It therefore applies a more demanding set of parameters than those used on the {\verb!RELAX!}
  set of runs.
  \item {{\verb!BANDS!}.} Electronic band structure generation. The converged {\verb!STATIC!} charge
  density and ionic positions are used as the starting points, and the wavefunctions are reoptimized along standardized
  high symmetry lines connecting special {\bf k}-points in the irreducible Brillouin zone (IBZ)~\cite{aflowBZ}.
\end{enumerate}

These calculation types are performed in the order shown above (\ie, {\verb!RELAX1!} $\rightarrow$ {\verb!RELAX2!}
$\rightarrow$ {\verb!STATIC!} $\rightarrow$ {\verb!BANDS!}) on all materials found in the Elements,
\ICSD, and Heusler databases. Those found in the Binary Alloy database contain data produced only by the two
{\verb!RELAX!} calculations.
Sets of these calculation types can be combined to describe more complex
phenomena than can be obtained from a single calculation. For
example, sets of {\verb!RELAX!} and  {\verb!STATIC!} calculations for different cell
volumes and/or atomic configurations are used to calculate
thermal and mechanical properties by the {Automatic Gibbs Library}, \AGL~\cite{curtarolo:art96},
and {Automatic Phonon Library}, \APL~\cite{aflowPAPER}, methods
implemented within the \AFLOW\ framework.
In the following, we describe the parameter sets used to address the
particular challenges of the calculations included in each \AFLOWorg\ repository.

\subsection{The AFLOW Standard parameter set} \label{subsec:art104:AFLOWstandard}
The standard parameters described in this work are classified according to the wide variety of tasks that a typical solid
state \DFT\ calculation involves: Brillouin zone sampling, Fourier transform meshes, basis sets, potentials,
self-interaction error (SIE) corrections, electron spin, algorithms guiding SCF convergence and ionic relaxation, and
output options.

Due to the intrinsic complexity of the \DFT\ codes it is impractical to
specify the full set of \DFT\ calculation parameters within an HT framework. Therefore, the \AFLOW\ standard
adopts many, but not all, of the internal defaults set by the \DFT\ software package. This is most notable in the description of the
Fourier transform meshes, which rely on a discretization scheme that depends on the applied basis and crystal
geometry for its specification. Those internal default settings are cast aside when
error corrections of failed \DFT\ runs, an integral part of \AFLOW{}'s functionality, take place. The settings
described in this work are nevertheless prescribed as fully as is practicable, in the interest of providing as
much information as possible to anyone interested in reproducing or building on our results.

\subsubsection{{\bf k}-point sampling} \label{subsubsec:art104:kpointgrid}
Two approaches are used when sampling the IBZ: the first consists of uniformly distributing a large number
of {\bf k}-points in the IBZ, while the second relies on the construction of paths connecting high symmetry (special)
{\bf k}-points in the IBZ. Within \AFLOW, the second sampling method corresponds to the {\verb!BANDS!}
calculation type, whereas the other calculation types (non-{\verb!BANDS!}) are performed using the first sampling
method.

Sampling in non-{\verb!BANDS!} calculations is obtained by defining and setting $N_{\mathrm{KPPRA}}$, the number of
{\bf k}-points per atom. This quantity determines the total number of {\bf k}-points in the IBZ,
taking into account the {\bf k}-points density along each reciprocal lattice vector as well as the number of atoms
in the simulation cell, via the relation:
\begin{equation} \label{eq:art104:kppra}
  { N_{\mathrm{KPPRA}} \leq \min \left[ \prod\limits_{i=1}^3 N_i \right] \times N_{\mathrm{a} } }
\end{equation}
$N_{\mathrm{a}}$ is the number of atoms in the cell, and the $N_{i}$ factors correspond to  the number
of sampling points along each reciprocal lattice vector,
$\vec{b_{i}}$, respectively. These factors define the grid resolution,
${\it \delta} k{_i} {\| \vec{b{_i}} \|}/{N{_i} }$, which is made as uniform as possible
under the constraint of Equation~\ref{eq:art104:kppra}. The {\bf k}-point meshes are then
generated within the Monkhorst-Pack scheme~\cite{Monkhorst1976}, unless the material belongs to the
{\textit{hP}}, or {\textit{hR}} Bravais lattices, in which case the hexagonal symmetry is preserved by centering the mesh
at the $\Gamma$-point.

Default $N_{\mathrm{KPPRA}}$ values depend on the calculation type and the
database. The $N_{\mathrm{KPPRA}}$ values used for the entries in the Elements
database are material specific and set manually due to convergence of
the total energy calculation. The defaults applied to the
{\verb!RELAX!} and {\verb!STATIC!} calculations are summarized in
Table~\ref{tab:art104:kgridnonbands}.
These defaults ensure proper convergence of the calculations. They
may be too stringent for some cases but enable reliable
application within the HT framework, thus presenting a practicable
balance between accuracy and calculation cost.

\tab
\cprotect\mycaption{Default $N_{\mathrm{KPPRA}}$ values used in non-{\verb!BANDS!} calculations.}
\tabvspace
\begin{tabular}{l | r r}
  database      & {\verb!STATIC!} & {\verb!RELAX!} \\
  \hline
  binary alloy  &  N.A.           &  6000          \\
  Heusler       & 10000           &  6000          \\
  \ICSD\          & 10000           &  8000          \\
\end{tabular}
\label{tab:art104:kgridnonbands}
\etab

For {\verb!BANDS!} calculations \AFLOW\ generates Brillouin zone integration
paths in the manner described in a previous
publication~\cite{aflowBZ}.
The {\bf k}-point sampling density is the {\textit{line
density}} of {\bf k}-points along each of the straight-line
segments of the path in the IBZ. The default setting
of \AFLOW\ is 128 {\bf k}-points along each segment connecting high-symmetry {\bf k}-points in
the IBZ for single element structures, and 20 {\textit
  k}-points for compounds.

The occupancies at the Fermi edge in all non-{\verb!RELAX!} type runs are handled via the tetrahedron method with
Bl{\"o}chl corrections~\cite{Bloechl1994a}. This involves the $N_{\mathrm{KPPRA}}$ parameter, as described above. In
{\verb!RELAX!} type calculations, where the determination of accurate forces is important, some type of
smearing must be performed. In cases where the material is assumed to be a metal, the
Methfessel-Paxton approach~\cite{Methfessel_prb_1989} is adopted, with a smearing width of 0.10~eV.
Gaussian smearing is used in all other types of materials, with a smearing width of 0.05~eV.

\subsubsection{Potentials and basis set} \label{subsubsec:art104:pseudopot}

The interactions involving the valence electron shells are handled with the potentials provided with the \DFT\ software
package. In \VASP, these include ultra-soft pseudopotentials (USPP)~\cite{Vanderbilt, vasp_JPCM_1994} and
projector-augmented wavefunction (\PAW) potentials~\cite{PAW,kresse_vasp_paw}, which are constructed according to the Local
Density Approximation (\LDA)~\cite{Ceperley_prl_1980, Perdew_prb_1981}, and the Generalized Gradient Approximation
(\GGA) PW91~\cite{VASP_PW91_1,VASP_PW91_2} and \PBE~\cite{PBE, PBE2} exchange-correlation (XC) functionals.
The \ICSD, Binary Alloy and Heusler databases built according to the \AFLOW\ standard use the \PBE\ functional combined with
the \PAW\ potential as the default. The \PBE\ functional is among the best studied \GGA\ functionals used in crystalline systems, while the \PAW\ potentials
are preferred due to their advantages over the USPP methodology. Nevertheless, defaults have been defined for a number of potential / XC functional
combinations, and in the case of the Elements database, results are available for \LDA, \GGA-PW91 and \GGA-\PBE\ functionals with both USPP and \PAW\ potentials.
Additionally, there are a small number of entries in the \ICSD\ and Binary Alloy databases (less than 1\% of the total) which have been calculated with the \GGA-PW91
functional using either the USPP or \PAW\ potential. The exact combination of exchange-correlation functional and potential used for a specific entry
in the \AFLOWorg\ database can always be determined by querying the keyword \verb|dft_type| using the \AFLOW\
\RESTAPI~\cite{aflowAPI}.

\DFT\ packages often provide more than one potential of each type per element. The \AFLOW\
standardized lists of \PAW\ and USPP potentials are presented in
Tables~\ref{tab:art104:tab:pot_paw} and \ref{tab:art104:pot_uspp}, respectively.
The ``Label'' column in these tables corresponds to the naming convention adopted
by \VASP. The checksum of each file listed in the tables is included in the accompanying supplement
for verification purposes.

Each potential provided with the \VASP\ package has two recommended plane-wave kinetic energy cut-off ($E_{\mathrm{cut}}$)
values, the smaller of which ensures the reliability of a calculation to within a well-defined error. Additionally,
materials with more than one element type will have two or more sets of recommended $E_{\mathrm{cut}}$ values.
In the \AFLOW\ standard, the applied $E_{\mathrm{cut}}$ value is the largest found among the recommendations for all
species involved in the calculation, increased by a factor of 1.4.

It is possible to evaluate the the non-local parts of the potentials in real space, rather than in the more computationally
intensive reciprocal space. This approach is prone to aliasing errors, and requires the optimization of real-space
projectors if these are to be avoided. The real-space projection scheme is most appropriate for larger systems, \eg, surfaces,
and is therefore not used in the construction of the databases found in the \AFLOWorg\ repository.

\tab
\mycaption[Projector-Augmented Wavefunction (\PAW) potentials, parameterized for the \LDA, PW91, and \PBE\
functionals, included in the \AFLOW\ standard.]
{The \PAW-\PBE\ combination is used as the default for \ICSD\, Binary Alloy and Heusler databases.
$\dagger$: \PBE\ potentials only.
$\ddagger$: \LDA\ and PW91 potentials only.}
\tabvspace
{\small
\begin{tabular}{l r | l r | l r}
  element & label  & element       & label      & element       & label  \\
  \hline
  H       & H      & Se            & Se         & Gd $\ddagger$ & Gd\_3  \\
  He      & He     & Br            & Br         & Tb            & Tb\_3  \\
  Li      & Li\_sv & Kr            & Kr         & Dy            & Dy\_3  \\
  Be      & Be\_sv & Rb            & Rb\_sv     & Ho            & Ho\_3  \\
  B       & B\_h   & Sr            & Sr\_sv     & Er            & Er\_3  \\
  C       & C      & Y             & Y\_sv      & Tm            & Tm     \\
  N       & N      & Zr            & Zr\_sv     & Yb            & Yb     \\
  O       & O      & Nb            & Nb\_sv     & Lu            & Lu     \\
  F       & F      & Mo            & Mo\_pv     & Hf            & Hf     \\
  Ne      & Ne     & Tc            & Tc\_pv     & Ta            & Ta\_pv \\
  Na      & Na\_pv & Ru            & Ru\_pv     & W             & W\_pv  \\
  Mg      & Mg\_pv & Rh            & Rh\_pv     & Re            & Re\_pv \\
  Al      & Al     & Pd            & Pd\_pv     & Os            & Os\_pv \\
  Si      & Si     & Ag            & Ag         & Ir            & Ir     \\
  P       & P      & Cd            & Cd         & Pt            & Pt     \\
  S       & S      & In            & In\_d      & Au            & Au     \\
  Cl      & Cl     & Sn            & Sn         & Hg            & Hg     \\
  Ar      & Ar     & Sb            & Sb         & Tl            & Tl\_d  \\
  K       & K\_sv  & Te            & Te         & Pb            & Pb\_d  \\
  Ca      & Ca\_sv & I             & I          & Bi            & Bi\_d  \\
  Sc      & Sc\_sv & Xe            & Xe         & Po            & Po     \\
  Ti      & Ti\_sv & Cs            & Cs\_sv     & At            & At     \\
  V       & V\_sv  & Ba            & Ba\_sv     & Rn            & Rn     \\
  Cr      & Cr\_pv & La            & La         & Fr            & Fr     \\
  Mn      & Mn\_pv & Ce            & Ce         & Ra            & Ra     \\
  Fe      & Fe\_pv & Pr            & Pr         & Ac            & Ac     \\
  Co      & Co     & Nd            & Nd         & Th            & Th\_s  \\
  Ni      & Ni\_pv & Pm            & Pm         & Pa            & Pa     \\
  Cu      & Cu\_pv & Sm $\dagger$  & Sm         & U             & U      \\
  Zn      & Zn     & Sm $\ddagger$ & Sm\_3      & Np            & Np\_s  \\
  Ga      & Ga\_h  & Eu            & Eu         & Pu            & Pu\_s  \\
  As      & As     & Gd $\dagger$  & Gd         &               &        \\
\end{tabular}}
\label{tab:art104:tab:pot_paw}
\etab

\tab
\mycaption{Ultra-Soft Pseudopotentials (USPP), parameterized for
the \LDA\ and PW91 functionals, included in the \AFLOW\ standard.}
\tabvspace
{\small
\begin{tabular}{l r | l r | l r}
  element & label   & element & label  & element & label \\
  \hline
  H       & H\_soft & As      & As     & Tb      & Tb\_3 \\
  He      & He      & Se      & Se     & Dy      & Dy\_3 \\
  Li      & Li\_pv  & Br      & Br     & Ho      & Ho\_3 \\
  Be      & Be      & Kr      & Kr     & Er      & Er\_3 \\
  B       & B       & Rb      & Rb\_pv & Tm      & Tm    \\
  C       & C       & Sr      & Sr\_pv & Yb      & Yb    \\
  N       & N       & Y       & Y\_pv  & Lu      & Lu    \\
  O       & O       & Zr      & Zr\_pv & Hf      & Hf    \\
  F       & F       & Nb      & Nb\_pv & Ta      & Ta    \\
  Ne      & Ne      & Mo      & Mo\_pv & W       & W     \\
  Na      & Na\_pv  & Tc      & Tc     & Re      & Re    \\
  Mg      & Mg\_pv  & Ru      & Ru     & Os      & Os    \\
  Al      & Al      & Rh      & Rh     & Ir      & Ir    \\
  Si      & Si      & Pd      & Pd     & Pt      & Pt    \\
  P       & P       & Ag      & Ag     & Au      & Au    \\
  S       & S       & Cd      & Cd     & Hg      & Hg    \\
  Cl      & Cl      & In      & In\_d  & Tl      & Tl\_d \\
  Ar      & Ar      & Sn      & Sn     & Pb      & Pb    \\
  K       & K\_pv   & Sb      & Sb     & Bi      & Bi    \\
  Ca      & Ca\_pv  & Te      & Te     & Po      & Po    \\
  Sc      & Sc\_pv  & I       & I      & At      & At    \\
  Ti      & Ti\_pv  & Xe      & Xe     & Rn      & Rn    \\
  V       & V\_pv   & Cs      & Cs\_pv & Fr      & Fr    \\
  Cr      & Cr      & Ba      & Ba\_pv & Ra      & Ra    \\
  Mn      & Mn      & La      & La     & Ac      & Ac    \\
  Fe      & Fe      & Ce      & Ce     & Th      & Th\_s \\
  Co      & Co      & Pr      & Pr     & Pa      & Pa    \\
  Ni      & Ni      & Nd      & Nd     & U       & U     \\
  Cu      & Cu      & Pm      & Pm     & Np      & Np\_s \\
  Zn      & Zn      & Sm      & Sm\_3  & Pu      & Pu\_s \\
  Ga      & Ga\_d   & Eu      & Eu     &         &       \\
  Ge      & Ge      & Gd      & Gd     &         &       \\
\end{tabular}}
\label{tab:art104:pot_uspp}
\etab

\subsubsection{Fourier transform meshes} \label{subsubsec:art104:fftmesh}

As mentioned previously, it is not practical to describe the precise default settings that are applied by the \AFLOW\
standard in the specification of the Fourier transform meshes. We
shall just note that they are defined in terms of the grid
spacing along each of the reciprocal lattice vectors, $\vec{b}_i$. These are obtained from the set of real space lattice
vectors, $\vec{a}_i$, via $ [\vec{b}_1 \vec{b}_2 \vec{b}_3]^T = 2 \pi [\vec{a}_1 \vec{a}_2 \vec{a}_3]^{-1} $. A distance
in reciprocal space is then defined by $d_i={\|\vec{b{_i}}\|} /n_i$, where the set of $n_i$ are the number
of grid points along each reciprocal lattice vector, and where the total number of points in the simulation is
$n_1 \times n_2 \times n_3$.

The \VASP\ package relies primarily on the so-called {\textit{dual grid technique}}, which consists of two overlapping
meshes with different coarseness. The least dense of the two is directly dependent on the applied plane-wave basis, $E_{\mathrm{cut}}$,
while the second is a finer mesh onto which the charge density is mapped. The \AFLOW\ standard relies on placing
sufficient points in the finer mesh such that wrap-around (``aliasing'') errors are avoided. In terms of the quantity $d_i$,
defined above, the finer grid is characterized by $d_i \approx 0.10${\textit{ \r{A}}$^{-1}$}, while the coarse grid results
in $d_i \approx 0.15${\textit{ \r{A}}$^{-1}$}. These two values are approximate, as there is significant dispersion in
these quantities across the various databases.

\subsubsection{DFT$+U$ corrections} \label{subsubsec:art104:Hubbard}

Extended systems containing {\textit d} and {\textit f} block elements are often poorly represented within \DFT\ due to
the well known self interaction error (SIE)~\cite{Perdew_prb_1981}. The influence that the SIE has on the energy gap of
insulators has long been recognized, and several methods that account for it are available. These include the
{\textit{GW}} approximation~\cite{Hedin_GW_1965}, the rotationally invariant approach introduced by
Dudarev~\cite{Dudarev_dftu} and Liechtenstein~\cite{Liechtenstein1995} (denoted here as \DFT$+U$), as well as the recently
developed ACBN0 pseudo-hybrid density functional~\cite{curtarolo:art93}.

The \DFT$+U$ approach is currently the best suited for high-throughput investigations, and is therefore included in
the \AFLOW\ standard for the entire \ICSD\ database, and is also used for certain entries in the Heusler
database containing the elements O, S, Se, and F. It is not used for the Binary Alloy database.
This method has a significant dependence on parameters, as each atom is associated with
two numbers, the screened Coulomb parameter, $U$, and the Stoner exchange parameter, $J$. These are usually reported
as a single factor, combined via $U_{\mathrm{eff}}=U-J$. The set of $U_{\mathrm{eff}}$ values associated with the
{\textit d} block elements~\cite{aflowBZ,curtarolo:art68} are presented in Table~\ref{tab:art104:Ud}, to which the
elements In and Sn have been added.

A subset of the {\textit f}-block elements can be found among the systems included in
the \AFLOWorg\ databases. We are not aware of the existence of a systematic search for the best set
of $U$ and $J$ parameters for this region of the periodic table, so we have relied on an in-house
parameterization~\cite{aflowBZ} in the construction of the databases. The values used are reproduced
in Table~\ref{tab:art104:Uf}. Note that by construction the SIE correction must be applied to a pre-selected value of the
$\ell$-quantum number, and all elements listed in Table~\ref{tab:art104:Ud} correspond to $\ell=2$, while those
found in Table~\ref{tab:art104:Uf} correspond to $\ell=3$.

\tab
\mycaption{$U_{\mathrm{eff}}$ parameters applied to {\textit d} orbitals.}
\tabvspace
  \begin{tabular}{l r | l r}
  element & $U_{\mathrm{eff}}$ & element & $U_{\mathrm{eff}}$ \\
  \hline
  Sc~\cite{ScUJ} & 2.9 & W~\cite{NbUJ}  & 2.2 \\
  Ti~\cite{TiUJ} & 4.4 & Tc~\cite{NbUJ} & 2.7 \\
  V~\cite{VUJ}   & 2.7 & Ru~\cite{NbUJ} & 3.0 \\
  Cr~\cite{CrUJ} & 3.5 & Rh~\cite{NbUJ} & 3.3 \\
  Mn~\cite{CrUJ} & 4.0 & Pd~\cite{NbUJ} & 3.6 \\
  Fe~\cite{FeUJ} & 4.6 & Ag~\cite{AgUJ} & 5.8 \\
  Co~\cite{VUJ}  & 5.0 & Cd~\cite{ZnUJ} & 2.1 \\
  Ni~\cite{VUJ}  & 5.1 & In~\cite{ZnUJ} & 1.9 \\
  Cu~\cite{CrUJ} & 4.0 & Sn~\cite{SnUJ} & 3.5 \\
  Zn~\cite{ZnUJ} & 7.5 & Ta~\cite{NbUJ} & 2.0 \\
  Ga~\cite{GaUJ} & 3.9 & Re~\cite{NbUJ} & 2.4 \\
  Sn~\cite{SnUJ} & 3.5 & Os~\cite{NbUJ} & 2.6 \\
  Nb~\cite{NbUJ} & 2.1 & Ir~\cite{NbUJ} & 2.8 \\
  Mo~\cite{NbUJ} & 2.4 & Pt~\cite{NbUJ} & 3.0 \\
  Ta~\cite{SnUJ} & 2.0 & Au & 4.0 \\
\end{tabular}
\label{tab:art104:Ud}
\etab

\tab
\mycaption{$U$ and $J$ parameters applied to selected {\textit f}-block elements.}
\tabvspace
  \begin{tabular}{l r r | l r r}
   element & {\textit U} & {\textit J} & element & {\textit U} & {\textit J} \\
   \hline
   La~\cite{LaUJ} & 8.1 & 0.6 & Dy~\cite{DyUJ} & 5.6 & 0.0  \\
   Ce~\cite{CeUJ} & 7.0 & 0.7 & Tm~\cite{TmUJ} & 7.0 & 1.0  \\
   Pr~\cite{PrUJ} & 6.5 & 1.0 & Yb~\cite{YbUJ} & 7.0 & 0.67 \\
   Nd~\cite{aflowSCINT}& 7.2 & 1.0 & Lu~\cite{LaUJ} & 4.8 & 0.95 \\
   Sm~\cite{aflowSCINT}& 7.4 & 1.0 & Th~\cite{ThUJ} & 5.0 & 0.0  \\
   Eu~\cite{aflowSCINT}& 6.4 & 1.0 & U~\cite{UUJ}   & 4.0 & 0.0  \\
   Gd~\cite{GdUJ} & 6.7 & 0.1 &    &     &      \\
\end{tabular}
\label{tab:art104:Uf}
\etab

\subsubsection{Spin polarization} \label{subsubsec:art104:SpinPol}

The first of the two {\verb!RELAX!} calculations is always performed in a collinear spin-polarized fashion.
The initial magnetic moments in this step are set to the number of atoms in the system, \eg, 1.0 $\mu B/$atom. If
the magnetization resulting from the {\verb!RELAX1!} step is found to be below 0.025 $\mu B/$atom, \AFLOW\
economizes computational resources by turning spin polarization off in all ensuing calculations. Spin-orbit coupling
is not used in the current \AFLOW\ standard, since it is still
too expensive to include in a HT framework.

\subsubsection{Calculation methods and convergence criteria} \label{subsubsec:art104:convergence}

Two nested loops are involved in the \DFT\ calculations used by \AFLOW\ in the construction of the databases.
The inner loop contains routines that iteratively optimize the electronic degrees of freedom (EDOF), and features
a number of algorithms that are concerned with diagonalizing the Kohn-Sham (KS) Hamiltonian at each iteration.
The outer loop performs adjustments to the system geometry (ionic degrees of freedom, IDOF) until the forces acting
on the system are minimized.

The convergence condition for each loop has been defined in terms of an energy difference, $\delta E$. If successive
energies resulting from the completion of a loop are denoted as $E_{i-1}$ and $E_i$, then
convergence is met when the condition $\delta E \geqslant E_i - E_{i-1}$ is fulfilled. Note that $E_i$ can either be
the electronic energy resulting from the inner loop, or the configurational energy resulting from the outer loop.
The electronic convergence criteria will be denoted  as $\delta E_{\mathrm{elec}}$, and the ionic criteria as $\delta E_{\mathrm{ion}}$.
The \AFLOW\ standard relies on $\delta E_{\mathrm{elec}} = 10^{-5}$~eV and $\delta E_{\mathrm{ion}} = 10^{-4}$~eV for entries in the
Elements database. All other databases include calculations performed with $\delta E_{\mathrm{elec}} = 10^{-3}$~eV and $\delta E_{\mathrm{ion}} = 10^{-2}$~eV.

Optimizations of the EDOF depend on sets of parameters that fall under three general themes: initial guesses, diagonalization
methods, and charge mixing. The outer loop (optimizations of the IDOF) is concerned with the lattice vectors and the ionic
positions, and is not as dependent on user input as the inner
loops. These are described in the following paragraphs.

\boldsection{Electronic degrees of freedom.}
The first step in the process of optimizing the EDOF consists of choosing a trial charge density and a trial
wavefunction. In the case of the non-{\verb!BANDS!}-type calculations, the trial wavefunctions are initialized
using random numbers, while the trial charge density is obtained from the superposition
of atomic charge densities. The {\verb!BANDS!} calculations are not self-consistent, and thus do not feature
a charge density optimization. In these cases the charge density obtained from the previously performed {\verb!STATIC!}
calculation is used in the generation of the starting wavefunctions.

Two iterative methods are used for diagonalizing the KS Hamiltonian: the Davidson blocked scheme
(DBS)~\cite{Liu_rep_1978,Davidson_1983}, and the preconditioned residual minimization method -- direct inversion in
the iterative subspace (RMM--DIIS)~\cite{vasp_prb1996}. Of the two, DBS is known to be the slower and more stable option.
Additionally, the subspace rotation matrix is always optimized. These methods are applied in a manner that is dependent
on the calculation type:

\begin{enumerate}
  \item {\verb!RELAX!} calculations. Geometry optimizations contain at least one determination of the system
  forces. The initial determination consists of 5 initial DBS steps,
  followed by as many RMM-DIIS steps as needed to
  fulfill the $\delta E_{\mathrm{elec}}$ condition. Later determinations of
  system forces are performed by a similar
  sequence, but only a single DBS step is applied at the outset of the process. Across all
  databases the minimum of number of electronic iterations for {\verb!RELAX!} calculations is 2. The maximum number is set
  to 120 for entries in the \ICSD, and 60 for all others.
  \item non-{\verb!RELAX!} calculations. In {\verb!STATIC!} or
  {\verb!BANDS!} calculations, the diagonalizations are always performed using RMM--DIIS. The minimum number of electronic
  iterations performed during non-{\verb!RELAX!} calculations is 2, and the maximum is 120.
\end{enumerate}

If the number of iterations in the inner loop somehow exceed the limits listed above, the calculation breaks
out of this loop, and the system forces and energy are determined. If the $\delta E_{\mathrm{ion}}$ convergence condition is
not met the calculation re-enters the inner loop, and proceeds normally.

Charge mixing is performed via Pulay's method~\cite{Pulay_cpl_1980}. The implementation of this charge mixing
approach in the \VASP\ package depends on a series of parameters, of which all but the maximum $\ell$-quantum number
handled by the mixer have been left in their default state. This parameter is modified
only in systems included in the \ICSD\ database which contain the elements
listed in Tables~\ref{tab:art104:Ud} and \ref{tab:art104:Uf}. In practical terms, the value applied in these cases is the maximum
$\ell$-quantum number found in the \PAW\ potential, multiplied by 2.

\boldsection{Ionic degrees of freedom and lattice vectors.}
The {\verb!RELAX!} calculation type contains determinations of the forces acting on the ions, as well as the full system
stress tensor. The applied algorithm is the conjugate gradients (CG) approach~\cite{press1992numerical}, which depends on
these quantities for the full optimization of the system geometry, \ie, the ionic positions, the lattice vectors, as well
as modifications of the cell volume. The implementation of CG in \VASP\ requires minimal
user input, where the only independent parameter is the initial scaling factor which is always left at its
default value. Convergence of the IDOF, as stated above, depends on the value for the $\delta E_{\mathrm{ion}}$ parameter,
as applied across the various databases. The adopted $E_{\mathrm{cut}}$ (see discussion on ``Potentials and basis set'',
section~\ref{subsubsec:art104:pseudopot}) makes corrections for Pulay stresses unnecessary.

Forces acting on the ions and stress tensor are subjected to Harris-Foulkes~\cite{Harris_prb_1985} corrections.
Molecular dynamics based relaxations are not performed in the construction of the databases found in the
\AFLOWorg\ repository, so any related settings are not applicable to this work.

\subsubsection{Output options} \label{subsubsec:art104:output}

The reproduction of the results presented on \AFLOWorg\ also depends on a select few parameters that
govern the output of the \DFT\ package. The density of states plots are generated from the {\verb!STATIC!}
calculation. States are plotted with a range of -30~eV to 45~eV, and with a resolution of 5000 points. The band
structures are plotted according to the paths of {\bf k}-points generated for a {\verb!BANDS!}
calculation~\cite{aflowBZ}. All bands found between -10~eV and 10~eV are included in the plots.

\subsection{Conclusion} \label{subsec:art104:conclusion}

The \AFLOW\ standard described here has been applied in the automated creation of the \AFLOWorg\ database of
material properties in a consistent and reproducible manner. The use of standardized parameter sets facilitates
the direct comparison of properties between different materials, so that specific trends can be identified to assist
in the formulation of design rules for accelerated materials development. Following this \AFLOW\ standard should
allow materials science researchers to reproduce the results reported by the \AFLOW\ consortium, as well as to
extend on the database and make meaningful comparisons with their own results.
\clearpage
\section{Combining the AFLOW GIBBS and Elastic Libraries for Efficiently and Robustly Screening Thermomechanical Properties of Solids}
\label{sec:art115}

This study follows from a collaborative effort described in Reference~\cite{curtarolo:art115}.

\subsection{Introduction}

Calculating the thermal and elastic properties of materials is
important for predicting the thermodynamic and mechanical stability of structural
phases~\cite{Greaves_Poisson_NMat_2011, Poirier_Earth_Interior_2000,Mouhat_Elastic_PRB_2014, curtarolo:art106}
and assessing their importance for a variety of applications.
Elastic and mechanical properties such as the shear and bulk moduli are important for predicting the
hardness of materials~\cite{Chen_hardness_Intermetallics_2011}, and thus their resistance to
wear and distortion.
Thermal properties, such as specific heat capacity and lattice thermal conductivity, are important for applications including thermal barrier coatings,
thermoelectrics~\cite{zebarjadi_perspectives_2012, aflowKAPPA, Garrity_thermoelectrics_PRB_2016}, and heat sinks~\cite{Watari_MRS_2001, Yeh_2002}.

\boldsection{Elasticity.} There are two main methods for calculating the elastic constants,
based on the response of either the stress tensor or the total energy to a set of
applied strains~\cite{Mehl_TB_Elastic_1996, Mehl_Elastic_1995, Golesorkhtabar_ElaStic_CPC_2013, curtarolo:art100, Silveira_Elastic_CPC_2008, Silveira_Elastic_CPC_2008, Silva_Elastic_PEPI_2007}.
In this study, we obtain the elastic constants from the calculated stress tensors for a set of independent deformations of the crystal lattice.
This method is implemented within the \AFLOW\ framework for
computational materials design
\cite{aflowPAPER,curtarolo:art49,monsterPGM}, where it is referred to as the
\underline{A}utomatic \underline{E}lasticity \underline{L}ibrary (\AEL).
{A similar} implementation within the Materials
Project~\cite{curtarolo:art100} {allows} extensive
screening studies by combining data from these two large
repositories of computational materials data.

\boldsection{Thermal properties.} The determination of the thermal conductivity of materials from first principles requires either calculation of anharmonic
\underline{i}nteratomic \underline{f}orce \underline{c}onstants (IFCs) for use in the
\underline{B}oltzmann \underline{T}ransport \underline{E}quation (BTE)~\cite{Broido2007, Wu_PRB_2012, ward_ab_2009, ward_intrinsic_2010,
Zhang_JACS_2012, Li_PRB_2012, Lindsay_PRL_2013, Lindsay_PRB_2013}, {or molecular dynamics} simulations in combination with
the Green-Kubo formula~\cite{Green_JCP_1954,Kubo_JPSJ_1957}, both of
which are highly demanding computationally even within multiscale approaches~\cite{curtarolo:art12}.
These methods are  unsuitable for rapid generation and screening of large databases of materials properties in order to identify trends
and simple descriptors~\cite{nmatHT}.
Previously, we have implemented the ``\GIBBS'' quasi-harmonic Debye model
\cite{Blanco_CPC_GIBBS_2004, Blanco_jmolstrthch_1996} within both the
\underline{A}utomatic \underline{{\small G}}{\small IBBS} \underline{L}ibrary (\AGL)~\cite{curtarolo:art96} of the
\AFLOW~\cite{aflowPAPER, aflowlibPAPER, aflowAPI, curtarolo:art104,curtarolo:art110} and
Materials Project~\cite{materialsproject.org,APL_Mater_Jain2013,CMS_Ong2012b} frameworks.
This approach does not require large supercell calculations since it
relies merely on first-principles calculations of the energy as a function of unit cell volume. It is thus
much more tractable computationally and eminently suited to investigating the thermal properties of
entire classes of materials in a highly-automated {fashion
to identify} promising candidates for more in-depth experimental and computational analysis.

The data set of computed thermal and elastic properties
produced for this study is available in the \AFLOW\
\cite{aflowlibPAPER} online data repository, either using the \AFLOW\
\underline{RE}presentational \underline{S}tate \underline{T}ransfer \underline{A}pplication \underline{P}rogramming \underline{I}nterface
(\RESTAPI)~\cite{aflowAPI} or via the \AFLOWorg\ web portal~\cite{aflowlibPAPER,aflowBZ}.

\subsection{The AEL-AGL methodology}

The \AEL-\AGL\ methodology combines elastic constants calculations, in
the Automatic Elasticity Library (\AEL), with the calculation of
thermal properties within the Automatic \GIBBS\ Library (\AGL\
\cite{curtarolo:art96}) - ``\GIBBS''~\cite{Blanco_CPC_GIBBS_2004} implementation of the Debye model.
This integrated software library includes automatic {error correction} to facilitate high-throughput
computation of thermal and elastic materials properties within the
\AFLOW\ framework~\cite{aflowPAPER, aflowlibPAPER, aflowAPI, curtarolo:art104,curtarolo:art53,curtarolo:art57,curtarolo:art63,curtarolo:art67,curtarolo:art54}.
The principal ingredients of the calculation are described in the following Sections.

\subsubsection{Elastic properties}
\label{subsubsec:art115:aelmethod}

The elastic constants are evaluated from the stress-strain relations
\begin{equation}
\left( \begin{array}{l} s_{11} \\ s_{22} \\ s_{33} \\ s_{23} \\ s_{13} \\ s_{12} \end{array} \right) =
\left( \begin{array}{l l l l l l} c_{11}\ c_{12}\ c_{13}\ c_{14}\ c_{15}\ c_{16} \\
c_{12}\ c_{22}\ c_{23}\ c_{24}\ c_{25}\ c_{26} \\
c_{13}\ c_{23}\ c_{33}\ c_{34}\ c_{35}\ c_{36} \\
c_{14}\ c_{24}\ c_{34}\ c_{44}\ c_{45}\ c_{46} \\
c_{15}\ c_{25}\ c_{35}\ c_{45}\ c_{55}\ c_{56} \\
c_{16}\ c_{26}\ c_{36}\ c_{46}\ c_{56}\ c_{66} \end{array} \right)
\left( \begin{array}{c} \epsilon_{11} \\ \epsilon_{22} \\ \epsilon_{33} \\ 2\epsilon_{23} \\ 2\epsilon_{13} \\ 2\epsilon_{12} \end{array} \right)
\end{equation}
with stress tensor elements $s_{ij}$ calculated
for a set of independent normal and shear strains $\epsilon_{ij}$. The elements of the
elastic stiffness tensor $c_{ij}$, written in the 6x6 Voigt notation using the mapping~\cite{Poirier_Earth_Interior_2000}:
$11 \mapsto 1$, $22 \mapsto 2$, $33 \mapsto 3$, $23 \mapsto 4$, $13 \mapsto 5$, $12 \mapsto 6$;
are derived from polynomial fits for each independent strain, where the polynomial degree
is automatically set to be less than the number of strains applied in each independent {direction to} avoid overfitting.
The elastic constants are then used to compute the bulk and shear
moduli, using either the Voigt approximation
\begin{equation}
\label{eq:art115:bulkmodvoigt}
B_{\sVoigt} = \frac{1}{9} \left[ (c_{11} + c_{22} + c_{33}) + 2 (c_{12} + c_{23} + c_{13}) \right]
\end{equation}
for the bulk modulus, and
\begin{multline}
\label{eq:art115:shearmodvoigt}
G_{\sVoigt} = \frac{1}{15} \left[ (c_{11} + c_{22} + c_{33}) -  (c_{12} + c_{23} + c_{13}) \right]
+ \frac{1}{5} (c_{44} + c_{55} + c_{66})
\end{multline}
for the shear modulus; or the Reuss approximation, which uses the elements of the compliance tensor $s_{ij}$ (the inverse of the stiffness tensor),
where the bulk modulus is given by
\begin{equation}
\label{eq:art115:bulkmodreuss}
\frac{1}{B_{\sReuss}} =  (s_{11} + s_{22} + s_{33}) + 2 (s_{12} + s_{23} + s_{13})
\end{equation}
and the shear modulus is
\begin{multline}
\label{eq:art115:shearmodreuss}
\frac{15}{G_{\sReuss}} = 4(s_{11} + s_{22} + s_{33}) - 4 (s_{12} + s_{23} + s_{13})
+ 3 (s_{44} + s_{55} + s_{66}).
\end{multline}
For polycrystalline materials, the Voigt approximation {corresponds to assuming that the strain is uniform and that the stress is supported by the individual grains in parallel, giving} the upper bound on the elastic moduli{;} while the Reuss approximation {assumes that the stress is uniform and that the strain is the sum of the strains of the individual grains in series, giving} the lower bound {on the elastic moduli~\cite{Poirier_Earth_Interior_2000}}.
The two approximations can be combined in the \underline{V}oigt-\underline{R}euss-\underline{H}ill (\VRH)~\cite{Hill_elastic_average_1952} averages for the bulk modulus
\begin{equation}
\label{eq:art115:bulkmodvrh}
B_{\sVRH} = \frac{B_{\sVoigt} + B_{\sReuss}}{2};
\end{equation}
and the shear modulus
\begin{equation}
\label{eq:art115:shearmodvrh}
G_{\sVRH} = \frac{G_{\sVoigt} + G_{\sReuss}}{2}.
\end{equation}
The Poisson ratio $\sigma$ is then obtained by:
\begin{equation}
\label{eq:art115:Poissonratio}
\sigma = \frac{3 B_{\sVRH} - 2 G_{\sVRH}}{6 B_{\sVRH} + 2 G_{\sVRH}}
\end{equation}

These elastic moduli can also be used to compute the speed of sound for the transverse and longitudinal waves, as well as the
average speed of sound in the material~\cite{Poirier_Earth_Interior_2000}.
The speed of sound for the longitudinal waves is
\begin{equation}
\label{eq:art115:longitudinalsoundspeed}
v_\sL = \left(\frac{B + \frac{4}{3}G}{\rho}\right)^{\frac{1}{2}}\!\!\!,
\end{equation}
and for the transverse waves
\begin{equation}
\label{eq:art115:transversesoundspeed}
v_\sT = \left(\frac{G}{\rho}\right)^{\frac{1}{2}}\!\!\!,
\end{equation}
where $\rho$ is the mass density of the material. The average speed of
sound is then evaluated by
\begin{equation}
\label{eq:art115:speedsound}
{\overline v} = \left[\frac{1}{3} \left( \frac{2}{v_\sT^3} + \frac{1}{v_\sL^3} \right) \right]^{-\frac{1}{3}}\!\!\!.
\end{equation}

\subsubsection{The \AGL\ quasi-harmonic Debye-Gr{\"u}neisen model}

The Debye temperature of a solid can be written as~\cite{Poirier_Earth_Interior_2000}
\begin{equation}
\label{eq:art115:debyetempv}
\theta_\sDebye = \frac{\hbar}{k_\sB}\left[\frac{6 \pi^2 n}{V}\right]^{1/3} \!\! {\overline v},
\end{equation}
where $n$ is the number of atoms in the cell, $V$ is its volume, and
${\overline v}$ is the average speed of sound of Equation~\ref{eq:art115:speedsound}.
It can be shown by combining Equations~\ref{eq:art115:Poissonratio}, \ref{eq:art115:longitudinalsoundspeed}, \ref{eq:art115:transversesoundspeed} and \ref{eq:art115:speedsound}
that ${\overline v}$ is equivalent to~\cite{Poirier_Earth_Interior_2000}
\begin{equation}
\label{eq:art115:speedsoundB}
{\overline v}  =  \sqrt{\frac{B_\sS}{\rho}} f(\sigma).
\end{equation}
where $B_\sS$ is the adiabatic bulk modulus, $\rho$ is the density, and $f(\sigma)$ is a function of the Poisson ratio $\sigma$:
\begin{equation}
\label{eq:art115:fpoisson}
f(\sigma) = \left\{ 3 \left[ 2 \left( \frac{2}{3} \!\cdot\! \frac{1 + \sigma}{1 - 2 \sigma} \right)^{3/2} \!\!\!\!\!\!\!+ \left( \frac{1}{3} \!\cdot\! \frac{1 + \sigma}{1 - \sigma} \right)^{3/2} \right]^{-1} \right\}^{\frac{1}{3}}\!\!\!\!,
\end{equation}
In an earlier version of \AGL~\cite{curtarolo:art96}, the Poisson ratio in Equation~\ref{eq:art115:fpoisson} was assumed to have the {constant
value $\sigma = 0.25$ which} is the ratio for a Cauchy solid. This was found to be a reasonable approximation, producing
good correlations with experiment.
The \AEL\ approach, Equation~\ref{eq:art115:Poissonratio}, directly evaluates $\sigma$ assuming only that it is independent of temperature and pressure.
Substituting Equation~\ref{eq:art115:speedsoundB} into Equation~\ref{eq:art115:debyetempv}, the
Debye temperature is obtained as
\begin{equation}
\label{eq:art115:debyetemp}
\theta_\sDebye = \frac{\hbar}{k_\sB}[6 \pi^2 V^{1/2} n]^{1/3} f(\sigma) \sqrt{\frac{B_\sS}{M}},
\end{equation}
where $M$ is the mass of the unit cell.
The bulk modulus $B_\sS$ is obtained from a set of DFT calculations for different volume cells, either by fitting the resulting $E_\sDFT(V)$
data to a phenomenological equation of state or by taking the numerical second derivative of
a polynomial fit
\begin{eqnarray}
\label{eq:art115:bulkmod}
B_\sS (V) &\approx& B_{\mathrm{static}} (\vec{x}) \approx B_{\mathrm{static}}(\vec{x}_\sopt(V)) \\ \nonumber
          &=&V \left( \frac{\partial^2 E(\vec{x}_\sopt (V))}{\partial V^2} \right) = V \left( \frac{\partial^2 E(V)}{\partial V^2} \right).
\end{eqnarray}
Inserting Equation~\ref{eq:art115:bulkmod} into Equation~\ref{eq:art115:debyetemp} gives the Debye temperature as a function of volume $\theta_\sDebye(V)$, for each value of
pressure, $p$, and temperature, $T$.

The equilibrium volume at any particular $(p, T)$ point is obtained by minimizing the Gibbs free energy with
respect to volume. First, the vibrational Helmholtz free energy, $F_\svib(\vec{x}; T)$, is calculated in the quasi-harmonic approximation
\begin{equation}
F_\svib(\vec{x}; T) \!=\!\! \int_0^{\infty} \!\!\left[\frac{\hbar \omega}{2} \!+\! k_\sB T\ \mathrm{log}\!\left(1\!-\!{\mathrm e}^{- \hbar \omega / k_\sB T}\right)\!\right]\!g(\vec{x}; \omega) d\omega,
\end{equation}
where $g(\vec{x}; \omega)$ is the phonon density of states and $\vec{x}$ describes the geometrical configuration of the system. In the Debye-Gr{\"u}neisen model, $F_\svib$ can be expressed
in terms of the Debye temperature $\theta_\sDebye$
\begin{equation}
\label{eq:art115:helmholtzdebye}
F_\svib(\theta_\sDebye; T) \!=\! n k_\sB T \!\left[ \frac{9}{8} \frac{\theta_\sDebye}{T} \!+\! 3\ \mathrm{log}\!\left(1 \!-\! {\mathrm e}^{- \theta_\sDebye / T}\!\right) \!\!-\!\! D\left(\frac{\theta_\sDebye}{T}\right)\!\!\right],
\end{equation}
where $D(\theta_\sDebye / T)$ is the Debye integral
\begin{equation}
D \left(\theta_\sDebye/T \right) = 3 \left( \frac{T}{\theta_\sDebye} \right)^3 \int_0^{\theta_\sDebye/T} \frac{x^3}{e^x - 1} dx.
\end{equation}
The Gibbs free energy is calculated as
\begin{equation}
\label{eq:art115:gibbsdebye}
{\sf G}(V; p, T) = E_\sDFT(V) + F_\svib (\theta_\sDebye(V); T)  + pV,
\end{equation}
and fitted by a polynomial of $V$. The equilibrium volume, $V_{\mathrm{eq}}$, is that which minimizes ${\sf G}(V; p, T)$.

Once $V_{\mathrm{eq}}$ has been determined, $\theta_\sDebye$ can be determined, and then other thermal properties including the Gr{\"u}neisen parameter and thermal
conductivity can be calculated as described in the following Sections.

\subsubsection{Equations of state}
\label{subsubsec:art115:eqnsofstate}

Within \AGL\, the bulk modulus can be determined either numerically from the second derivative of the polynomial fit of $E_\sDFT(V)$,
Equation~\ref{eq:art115:bulkmod}, or by fitting the $(p,V)$ data to a
phenomenological equation of state (\EOS). Three different analytic \EOS\ have been implemented within
\AGL: the Birch-Murnaghan \EOS~\cite{Birch_Elastic_JAP_1938, Poirier_Earth_Interior_2000, Blanco_CPC_GIBBS_2004}; the Vinet \EOS~\cite{Vinet_EoS_JPCM_1989, Blanco_CPC_GIBBS_2004};
and the Baonza-C{\'a}ceres-N{\'u}{\~n}ez spinodal \EOS~\cite{Baonza_EoS_PRB_1995, Blanco_CPC_GIBBS_2004}.

The Birch-Murnaghan \EOS\ is
\begin{equation}
\label{eq:art115:birch}
\frac{p}{3 f (1 + 2 f)^\frac{5}{2}} = \sum_{i=0}^2 a_i f^i ,
\end{equation}
where $p$ is the pressure, $a_i$ are polynomial coefficients, and $f$ is the ``compression''  given by
\begin{equation}
\label{eq:art115:birchf}
f = \frac{1}{2} \left[\left(\frac{V}{V_0} \right)^{-\frac{2}{3}}- 1 \right].
\end{equation}
The zero pressure bulk modulus is equal to the coefficient $a_0$.

The Vinet \EOS\ is~\cite{Vinet_EoS_JPCM_1989, Blanco_CPC_GIBBS_2004}
\begin{equation}
\label{eq:art115:vinet}
\log \left[ \frac{p x^2}{3 (1 - x)} \right] = \log B_0 + a (1 - x),
\end{equation}
where $a$ and $\log B_0$ are fitting parameters and
\begin{equation}
\label{eq:art115:vinetx}
x = \left(\frac{V}{V_0} \right)^{\frac{1}{3}}\!\!\!, \
a = 3 (B_0' - 1) / 2.
\end{equation}
The isothermal bulk modulus $B_\sT$ is given by~\cite{Vinet_EoS_JPCM_1989, Blanco_CPC_GIBBS_2004}
\begin{equation}
\label{eq:art115:vinetBT}
B_\sT = - x^{-2} B_0 e^{a(1-x)} f(x),
\end{equation}
where
\begin{equation*}
\label{eq:art115:vinetfx}
f(x)  = x - 2 - ax (1 - x).
\end{equation*}

The Baonza-C{\'a}ceres-N{\'u}{\~n}ez spinodal equation of state has the form~\cite{Baonza_EoS_PRB_1995, Blanco_CPC_GIBBS_2004}
\begin{equation}
\label{eq:art115:bcn}
V = V_{\mathrm{sp}} \exp \left[ - \left(\frac{K^*}{1 - \beta} \right) (p - p_{\mathrm{sp}})^{1 - \beta} \right],
\end{equation}
where $K^*$, $p_{\mathrm{sp}}$ and $\beta$ are the fitting parameters, and $V_{\mathrm{sp}} $ is given by
\begin{equation*}
V_{\mathrm{sp}}  = V_0 \exp \left[ \frac{\beta}{\left(1 - \beta \right) B_0'} \right],
\end{equation*}
where $B_0 = [K^*]^{-1} (-p_{\mathrm{sp}})^{\beta}$ and $B_0' = (-p_{\mathrm{sp}})^{-1}\beta B_0$.
The isothermal bulk modulus $B_\sT$ is then given by~\cite{Baonza_EoS_PRB_1995, Blanco_CPC_GIBBS_2004}
\begin{equation}
\label{eq:art115:bcnBT}
B_\sT = \frac{(p - p_{\mathrm{sp}})^{\beta}}{K^*}.
\end{equation}

{Note that \AGL\ uses $B_\sT$ instead of $B_\sS$ in Equation~\ref{eq:art115:debyetemp} when one of these phenomenological \EOS\
is selected. $B_\sS$ can then be calculated as
\begin{equation}
\label{eq:art115:BsBT}
B_\sS = B_\sT(1 + \alpha \gamma T),
\end{equation}
where $\gamma$ is the Gr{\"u}neisen parameter (described in Section~\ref{subsubsec:gruneisen} below), and $\alpha$ is the thermal expansion
\begin{equation}
\label{eq:art115:thermal_expansion}
\alpha = \frac{\gamma C_\sV}{B_\sT V},
\end{equation}
where $C_\sV$ is the heat capacity at constant volume, given by
\begin{equation}
 \label{eq:art115:heat_capacity}
C_\sV = 3 n k_\sB \left[4 D\left(\frac{\theta_\sD}{T}\right) - \frac{3 \theta_\sD / T}{\exp(\theta_\sD / T) - 1} \right].
\end{equation}
}

\subsubsection{The Gr{\"u}neisen parameter}
\label{subsubsec:gruneisen}

The Gr{\"u}neisen parameter describes the variation of the thermal properties of a material with the unit cell size, and contains
information about higher order phonon scattering which is important
for calculating the lattice thermal conductivity
\cite{Leibfried_formula_1954, slack, Morelli_Slack_2006, Madsen_PRB_2014, curtarolo:art96},
and thermal expansion~\cite{Poirier_Earth_Interior_2000, Blanco_CPC_GIBBS_2004, curtarolo:art114}.
It is defined as the phonon frequencies dependence on the unit cell volume
\begin{equation}
\label{eq:art115:gamma_micro}
\gamma_i = - \frac{V}{\omega_i} \frac{\partial \omega_i}{\partial V}.
\end{equation}
Debye's theory assumes that the volume dependence of all mode
frequencies is the same as that of the cut-off Debye frequency, so the Gr{\"u}neisen parameter can be expressed in terms of $\theta_\sDebye$
\begin{equation}
\label{eq:art115:gruneisen_theta}
\gamma = - \frac{\partial \ \mathrm{log} (\theta_\sDebye(V))}{\partial \ \mathrm{log} V}.
\end{equation}

This macroscopic definition of the Debye temperature is a weighted
average of Equation~\ref{eq:art115:gamma_micro} with the heat capacities for each branch of the phonon spectrum
\begin{equation}
\gamma = \frac{\sum_i \gamma_i C_{V, i}} {\sum_i C_{V,i}}.
\end{equation}

{
Within \AGL~\cite{curtarolo:art96}, the Gr{\"u}neisen parameter can
be calculated in several different ways, including direct evaluation of Equation~\ref{eq:art115:gruneisen_theta},
by using the more stable Mie-Gr{\"u}neisen equation~\cite{Poirier_Earth_Interior_2000},
\begin{equation}
\label{eq:art115:miegruneisen}
p - p_{T=0} = \gamma \frac{U_\svib}{V},
\end{equation}
where $U_\svib$ is the vibrational internal energy~\cite{Blanco_CPC_GIBBS_2004}
\begin{equation}
\label{eq:art115:Uvib}
U_\svib = n k_\sB T\left[ \frac{9}{8} \frac{\theta_\sDebye}{T} + 3D \left( \frac{\theta_\sDebye}{T} \right)\right].
\end{equation}
The ``Slater gamma'' expression~\cite{Poirier_Earth_Interior_2000}
\begin{equation}
\label{eq:art115:slatergamma}
\gamma = - \frac{1}{6} + \frac{1}{2} \frac{\partial B_\sS}{\partial p}
\end{equation}
is the default method in the automated workflow used
for the \AFLOW\ database.
}

\subsubsection{Thermal conductivity}

In the \AGL\ framework, the thermal conductivity is calculated using the
Leibfried-Schl{\"o}mann equation~\cite{Leibfried_formula_1954, slack, Morelli_Slack_2006}
\begin{eqnarray}
\label{eq:art115:thermal_conductivity}
\kappa_{\mathrm l} (\theta_\acoustic) &=& \frac{0.849 \times 3 \sqrt[3]{4}}{20 \pi^3(1 - 0.514\gamma_\acoustic^{-1} + 0.228\gamma_\acoustic^{-2})}
                                      \left( \frac{k_\sB \theta_\acoustic}{\hbar} \right)^2 \frac{k_\sB m V^{\frac{1}{3}}}{\hbar \gamma_\acoustic^2}.
\end{eqnarray}
where $V$ is the volume of the unit cell and $m$ is the average atomic mass.
It should be noted that the Debye temperature and Gr{\"u}neisen parameter in this formula, $\theta_\acoustic$ and $\gamma_\acoustic$, are slightly
different {from} the traditional Debye temperature, $\theta_\sDebye$, calculated in Equation~\ref{eq:art115:debyetemp} and Gr{\"u}neisen parameter, $\gamma$, obtained from
Equation~\ref{eq:art115:slatergamma}. Instead, $\theta_\acoustic$ and $\gamma_\acoustic$  are obtained by only considering the acoustic modes, based on the assumption that the optical
phonon modes in crystals do not contribute to heat transport~\cite{slack}. This $\theta_\acoustic$ is referred to as the ``acoustic'' Debye temperature
\cite{slack, Morelli_Slack_2006}. It can be derived directly from the phonon DOS by integrating only over the acoustic modes~\cite{slack,
Wee_Fornari_TiNiSn_JEM_2012}. Alternatively, it can be calculated from the traditional Debye temperature $\theta_\sDebye$~\cite{slack, Morelli_Slack_2006}
\begin{equation}
\label{eq:art115:acousticdebyetemp}
\theta_\acoustic = \theta_\sDebye n^{-\frac{1}{3}}.
\end{equation}

{There is no simple way to extract the ``acoustic''  Gr{\"u}neisen parameter from the traditional Gr{\"u}neisen parameter.}
Instead, it must be calculated from Equation~\ref{eq:art115:gamma_micro} for each phonon branch separately and summed over the acoustic branches~\cite{curtarolo:art114, curtarolo:art119}.
This requires using the quasi-harmonic phonon approximation which involves calculating the full phonon spectrum for different
volumes~\cite{Wee_Fornari_TiNiSn_JEM_2012, curtarolo:art114, curtarolo:art119}, and is therefore too computationally demanding to be used for
high-throughput screening, particularly for large, low symmetry systems. Therefore, we use the approximation
$\gamma_\acoustic = \gamma$ in the \AEL-\AGL\ approach to {calculate} the thermal conductivity. The dependence of the expression in
Equation~\ref{eq:art115:thermal_conductivity} on $\gamma$ is weak~\cite{curtarolo:art96, Morelli_Slack_2006}, thus
the evaluation of $\kappa_l$ using the traditional  Gr{\"u}neisen parameter introduces just a small systematic error which is insignificant for
screening purposes~\cite{curtarolo:art119}.

The thermal conductivity at temperatures other than $\theta_\acoustic$ is estimated by~\cite{slack, Morelli_Slack_2006, Madsen_PRB_2014}:
\begin{equation}
\label{eq:art115:kappa_temperature}
\kappa_{\mathrm l} (T) = \kappa_{\mathrm l}(\theta_\acoustic) \frac{\theta_\acoustic}{T}.
\end{equation}

\subsubsection{\DFT\ calculations and workflow details}

The \DFT\ calculations to obtain $E(V)$ and the strain tensors were performed using
the \VASP\ software~\cite{kresse_vasp} with projector-augmented-wave
pseudopotentials~\cite{PAW} and the \PBE\ parameterization of the
generalized gradient approximation to the exchange-correlation
functional~\cite{PBE}, using the {parameters described} in the \AFLOW\
Standard~\cite{curtarolo:art104}. The energies were calculated at zero
temperature and pressure, with spin polarization and without zero-point motion or lattice
vibrations. The initial crystal structures were fully relaxed (cell
volume and shape and the basis atom coordinates inside the cell).

For the \AEL\ calculations, 4 strains were applied in each independent lattice direction
(two compressive and two expansive) with a maximum strain of 1\% in each direction,
for a total of 24 configurations~\cite{curtarolo:art100}. For cubic systems,
the crystal symmetry was used to reduce the number of required strain configurations
to 8. For each configuration, two ionic positions \AFLOW\ Standard {\verb!RELAX!}~\cite{curtarolo:art104}
calculations at fixed cell volume and shape were followed by a single \AFLOW\ Standard  {\verb!STATIC!}~\cite{curtarolo:art104}
calculation.
The elastic constants are then calculated by fitting the elements of stress tensor obtained for each independent strain.
The stress tensor from the zero-strain configuration
(\ie, the initial unstrained relaxed structure) can also be {included in the set of fitted strains}, although this was found to have negligible effect on the results.
Once these calculations are complete, it is verified that the eigenvalues of the stiffness tensor are all positive,
that the stiffness tensor obeys the appropriate symmetry rules for the lattice type~\cite{Mouhat_Elastic_PRB_2014}, and
that the applied strain is still within the linear regime, using the method described by de Jong~\etal~\cite{curtarolo:art100}.
If any of these conditions fail, the calculation is repeated with
adjusted applied strain.

The \AGL\ calculation of $E(V)$ is fitted to the energy at 28 different
volumes of the unit cell obtained by increasing or decreasing the relaxed lattice parameters in fractional
increments of 0.01, with a single \AFLOW\ Standard
{\verb!STATIC!}~\cite{curtarolo:art104} calculation at each volume.
The resulting $E(V)$ data is checked for convexity and to verify that the minimum energy is at the
initial volume (\ie, at the properly relaxed cell size). If any of these
conditions fail, the calculation is repeated with adjusted parameters,
\eg, increased k-point grid density.

\subsubsection{Correlation analysis}

Pearson and Spearman correlations {are used to}
analyze the results for entire sets of materials. The {Pearson coefficient} $r$ is a measure of the linear
correlation between two variables, $X$ and $Y$. It is calculated by
\begin{equation}
\label{eq:art115:Pearson}
r = \frac{\sum_{i=1}^{n} \left(X_i - \overline{X} \right) \left(Y_i - \overline{Y} \right) }{ \sqrt{\sum_{i=1}^{n} \left(X_i - \overline{X} \right)^2} \sqrt{\sum_{i=1}^{n} \left(Y_i - \overline{Y} \right)^2}},
\end{equation}
where $\overline{X}$ and $\overline{Y}$ are the mean values of $X$ and $Y$.

The {Spearman coefficient} $\rho$ is a measure of the monotonicity of the relation between two variables.
The raw values of the two variables $X_i$ and $Y_i$ are sorted in ascending order, and are assigned rank values $x_i$ and $y_i$ which
are equal to their position in the sorted list. If there is more than one variable with the same value, the average of the position values
are assigned to {all duplicate entries}. The correlation coefficient is then given by
\begin{equation}
\label{eq:art115:Spearman}
\rho = \frac{\sum_{i=1}^{n} \left(x_i - \overline{x} \right) \left(y_i - \overline{y} \right) }{ \sqrt{\sum_{i=1}^{n} \left(x_i - \overline{x} \right)^2} \sqrt{\sum_{i=1}^{n} \left(y_i - \overline{y} \right)^2}}.
\end{equation}
It is useful for determining how well the ranking order of the values of one variable predict the ranking order of the values of the other variable.

The discrepancy between the \AEL-\AGL\ predictions and experiment is
evaluated in terms  normalized root-mean-square relative deviation
\begin{equation}
\label{eq:art115:RMSD}
{\mathrm{RMSrD}}  = \sqrt{\frac{ \sum_{i=1}^{n} \left( \frac{X_i - Y_i}{X_i} \right)^2 }{N - 1}} ,
\end{equation}
{In contrast} to the correlations described above, lower values of the \RMSrD\ indicate better agreement with experiment. This measure is particularly useful for
comparing predictions of the same property using different
methodologies that may have very similar correlations with, but different
deviations from, the experimental results.

\subsection{Results}

We used the \AEL-\AGL\ methodology to calculate the mechanical and thermal properties, including the bulk modulus,
shear modulus, Poisson ratio,  Debye temperature, Gr{\"u}neisen parameter and thermal conductivity for a set of 74 materials
with structures including diamond, zincblende, rocksalt, wurtzite, rhombohedral and body-centered tetragonal.
The results have been compared to experimental values (where available), and the correlations between the calculated and
experimental values were deduced.
In cases where multiple experimental values are present in the literature, we used the most recently reported
value, unless otherwise specified.

In Section~\ref{subsubsec:art115:aelmethod}, three different approximations for the bulk and shear moduli are described: Voigt (Equations~\ref{eq:art115:bulkmodvoigt}, \ref{eq:art115:shearmodvoigt}),
Reuss (Equations~\ref{eq:art115:bulkmodreuss}, \ref{eq:art115:shearmodreuss}), and the Voigt-Reuss-Hill (\VRH) average (Equations~\ref{eq:art115:bulkmodvrh}, \ref{eq:art115:shearmodvrh}).
These approximation{s give very similar values for the
bulk modulus} for the set of materials included in this work, particularly those with cubic symmetry.
Therefore only {$B_\sVRH^\sAEL$}
is explicitly cited in the following listed results
(the values obtained for all three  approximations are available in the \AFLOW\ database entries for
these materials). The values for the shear modulus in these three
approximations exhibit larger variations, and are therefore all listed and compared to experiment.
In several cases, the experimental values of the bulk and shear moduli have been calculated
from the measured  elastic constants using Equations~\ref{eq:art115:bulkmodvoigt} through \ref{eq:art115:shearmodvrh}, and an experimental Poisson ratio $\sigma^\EXP$
was calculated from these values using Equation~\ref{eq:art115:Poissonratio}.

As described in Section~\ref{subsubsec:art115:eqnsofstate}, the bulk modulus in \AGL\ can be calculated from a polynomial fit of the $E(V)$ data as shown in Equation~\ref{eq:art115:bulkmod},
or by fitting the $E(V)$ data to one of three empirical equations
of state: Birch-Murnaghan (Equation~\ref{eq:art115:birch}), Vinet (Equation~\ref{eq:art115:vinet}), and the Baonza-C{\'a}ceres-N{\'u}{\~n}ez
(Equation~\ref{eq:art115:bcn}). We compare the results of these four methods, labeled $B_\sStatic^\sAGL$,  $B_\sStatic^\sBM$, $B_\sStatic^\sVIN$, and
$B_\sStatic^\sBCN$, respectively,  with the experimental values $B^\EXP$ and those obtained from the
elastic calculations $B_\sVRH^\sAEL$.
The Debye temperatures, Gr{\"u}neisen parameters and thermal conductivities depend on the calculated bulk modulus and are
therefore also cited below for each of the equations of state.
Also included are the Debye temperatures derived from the calculated
elastic constants and speed of sound as given by Equation~\ref{eq:art115:speedsound}.
The Debye temperatures, $\theta_\sDebye^\sBM$
(Equation~\ref{eq:art115:birch}), $\theta_\sDebye^\sVIN$ (Equation~\ref{eq:art115:vinet}),
$\theta_\sDebye^\sBCN$ (Equation~\ref{eq:art115:bcn}), calculated using the Poisson ratio $\sigma^\sAEL$ obtained from
Equation~\ref{eq:art115:Poissonratio}, are compared to $\theta_\sDebye^\sAGL$, obtained from the numerical fit
of $E(V)$ (Equation~\ref{eq:art115:bulkmod}) using both $\sigma^\sAEL$ and the approximation $\sigma =
0.25$ used in Reference~\onlinecite{curtarolo:art96}, to
$\theta_\sDebye^\sAEL$, calculated with the speed of sound obtained
using Equation~\ref{eq:art115:speedsound},
and to the experimental values $\theta^\EXP$.
The values of the acoustic Debye temperature ($\theta_\acoustic$, Equation~\ref{eq:art115:acousticdebyetemp})
are shown, where available, in parentheses below the traditional Debye temperature value.

The experimental Gr{\"u}neisen parameter, $\gamma^\EXP$, is compared to $\gamma^\sAGL$ (Equation~\ref{eq:art115:bulkmod}),  obtained using the numerical
polynomial fit of $E(V)$ and both values of the Poisson ratio
($\sigma^\sAEL$ and the approximation $\sigma = 0.25$ from
Reference~\onlinecite{curtarolo:art96}), and to $\gamma^\sBM$
(Equation~\ref{eq:art115:birch}), $\gamma^\sVIN$ (Equation~\ref{eq:art115:vinet}), and $\gamma^\sBCN$ (Equation~\ref{eq:art115:bcn}), calculated
using $\sigma^\sAEL$ only. Similarly, the experimental lattice thermal
conductivity $\kappa^\EXP$ is compared to $\kappa^\sAGL$ (Equation~\ref{eq:art115:bulkmod}),
obtained using the numerical polynomial fit and both the calculated
and approximated values of $\sigma$, and to $\kappa^\sBM$
(Equation~\ref{eq:art115:birch}), $\kappa^\sVIN$ (Equation~\ref{eq:art115:vinet}), and $\kappa^\sBCN$
(Equation~\ref{eq:art115:bcn}), calculated using only $\sigma^\sAEL$.

The \AEL\ method has been been previously implemented in the Materials Project framework for calculating
elastic constants~\cite{curtarolo:art100}. {Data from} the Materials Project database are included
in the tables below for comparison {for the bulk modulus $B_\sVRH^\sMP$, shear modulus $G_\sVRH^\sMP$, and Poisson ratio $\sigma^\sMP$.}

\subsubsection{Zincblende and diamond structure materials}

The mechanical and thermal properties were calculated for a set of materials with the
zincblende(spacegroup: $F\overline{4}3m$,\ $\#$216; Pearson symbol: cF8; \AFLOW\ prototype: {\sf AB\_cF8\_216\_c\_a}~\cite{aflowANRL}\footnote{\url{http://aflow.org/CrystalDatabase/AB_cF8_216_c_a.html}})
and diamond ($Fd\overline{3}m$,\ $\#$227; cF8; {\sf A\_cF8\_227\_a}~\cite{aflowANRL}\footnote{\url{http://aflow.org/CrystalDatabase/A_cF8_227_a.html}}) structures.
This {is the same set of materials
as} in Table I of Reference~\onlinecite{curtarolo:art96}, which in {turn are from} Table II of
Reference~\onlinecite{slack} and Table 2.2 of Reference~\onlinecite{Morelli_Slack_2006}.

The elastic {properties bulk modulus}, shear modulus and Poisson {ratio calculated} using \AEL\ and \AGL\ are shown
in Table~\ref{tab:art115:zincblende_elastic} and Figure~\ref{fig:art115:zincblende_thermal_elastic}, together
with experimental values from the literature where available. As can be seen
from the results in Table~\ref{tab:art115:zincblende_elastic} and Figure~\ref{fig:art115:zincblende_thermal_elastic}(a), the $B_\sVRH^\sAEL$ values are
generally closest to experiment as shown by the \RMSrD\ value of $0.13$, producing an underestimate of the order of 10\%. The \AGL\ values from both the numerical
fit and the empirical equations of state are generally very similar to each other, while being slightly less than the $B_\sVRH^\sAEL$
values.

\tab
\mycaption[Bulk modulus, shear modulus and Poisson ratio of
zincblende and diamond structure semiconductors.]
{The zincblende structure is designated \AFLOW\ prototype {\sf AB\_cF8\_216\_c\_a}~\cite{aflowANRL}
and the diamond structure {\sf A\_cF8\_227\_a}~\cite{aflowANRL}.
``N/A'' = Not available for that source.
Units: $B$ and $G$ in \GPa.}
\tabvspace
\resizebox{\linewidth}{!}{
\begin{tabular}{l|r|r|r|r|r|r|r|r|r|r|r|r|r|r|r}
comp. & $B^\EXP$  & $B_\sVRH^\sAEL$ & $B_\sVRH^\sMP$ & $B_\sStatic^\sAGL$ & $B_\sStatic^\sBM$ & $B_\sStatic^\sVIN$ &  $B_\sStatic^\sBCN$ & $G^\EXP$ & $G_\sVoigt^\sAEL$ & $G_\sReuss^\sAEL$ &  $G_\sVRH^\sAEL$ & $G_\sVRH^\sMP$ & $\sigma^\EXP$ & $\sigma^\sAEL$ & $\sigma^\sMP$      \\
\hline
C & 442~\cite{Semiconductors_BasicData_Springer,
    Lam_BulkMod_PRB_1987, Grimsditch_ElasticDiamond_PRB_1975} & 434 & N/A & 408 & 409 & 403 & 417 & 534~\cite{Semiconductors_BasicData_Springer, Grimsditch_ElasticDiamond_PRB_1975}  & 520 & 516 & 518 & N/A & 0.069~\cite{Semiconductors_BasicData_Springer, Grimsditch_ElasticDiamond_PRB_1975} & 0.073 & N/A \\
SiC & 248~\cite{Strossner_ElasticSiC_SSC_1987} & 212 & 211 & 203 & 207 & 206 & 206 & 196~\cite{Fate_ShearSiC_JACeramS_1974} & 195 & 178 & 187 & 187 & 0.145~\cite{Lam_BulkMod_PRB_1987, Fate_ShearSiC_JACeramS_1974} & 0.160 & 0.16 \\
      & 211~\cite{Semiconductors_BasicData_Springer, Lam_BulkMod_PRB_1987} & & & & & & & 170~\cite{Semiconductors_BasicData_Springer} & & & & & 0.183~\cite{Semiconductors_BasicData_Springer} & & \\
Si & 97.8~\cite{Semiconductors_BasicData_Springer, Hall_ElasticSi_PR_1967} & 89.1 & 83.0 & 84.2 & 85.9 & 85.0 & 86.1 & 66.5~\cite{Semiconductors_BasicData_Springer, Hall_ElasticSi_PR_1967} & 64 & 61 & 62.5 & 61.2 & 0.223~\cite{Semiconductors_BasicData_Springer, Hall_ElasticSi_PR_1967} & 0.216 & 0.2  \\
      & 98~\cite{Lam_BulkMod_PRB_1987}  & & & & & & & & & & & & & \\
Ge & 75.8~\cite{Semiconductors_BasicData_Springer, Bruner_ElasticGe_PRL_1961} & 61.5 & 59.0 & 54.9 & 55.7 & 54.5 & 56.1 & 55.3~\cite{Semiconductors_BasicData_Springer, Bruner_ElasticGe_PRL_1961} & 47.7 & 44.8 & 46.2 & 45.4 & 0.207~\cite{Semiconductors_BasicData_Springer, Bruner_ElasticGe_PRL_1961} & 0.199 & 0.19 \\
      & 77.2~\cite{Lam_BulkMod_PRB_1987}  & & & & & & & & & & & & & \\
BN & 367.0~\cite{Lam_BulkMod_PRB_1987} & 372 & N/A & 353 & 356 & 348 & 359 & N/A & 387 & 374 & 380 & N/A & N/A & 0.119 & N/A \\
BP & 165.0~\cite{Semiconductors_BasicData_Springer, Lam_BulkMod_PRB_1987} & 162 & 161 & 155 & 157 & 156 & 157 & 136~\cite{Semiconductors_BasicData_Springer, Wettling_ElasticBP_SSC_1984} & 164 & 160 & 162 & 162 & 0.186~\cite{Semiconductors_BasicData_Springer, Wettling_ElasticBP_SSC_1984} & 0.125 & 0.12 \\
      & 267~\cite{Semiconductors_BasicData_Springer, Suzuki_ElasticBP_JAP_1983}  & & & & & & & & & & & & & \\
      & 172~\cite{Semiconductors_BasicData_Springer, Wettling_ElasticBP_SSC_1984} & & & & & & & & & & & & & \\
AlP & 86.0~\cite{Lam_BulkMod_PRB_1987} & 82.9 & 85.2 & 78.9 & 80.4 & 79.5 & 80.4 & N/A & 48.6 & 44.2 & 46.4 & 47.2 & N/A & 0.264 & 0.27 \\
AlAs & 77.0~\cite{Lam_BulkMod_PRB_1987} & 67.4 & 69.8 & 63.8 & 65.1 & 64.0 & 65.3 & N/A & 41.1 & 37.5 & 39.3 & 39.1 & N/A & 0.256 & 0.26 \\
      & 74~\cite{Greene_ElasticAlAs_PRL_1994}  & & & & & & & & & & & & & \\
AlSb & 58.2~\cite{Lam_BulkMod_PRB_1987, Semiconductors_BasicData_Springer, Bolef_ElasticAlSb_JAP_1960, Weil_ElasticAlSb_JAP_1972} &  49.4 & 49.2 & 46.5 & 47.8 & 46.9 & 47.8 &  31.9~\cite{Semiconductors_BasicData_Springer, Bolef_ElasticAlSb_JAP_1960, Weil_ElasticAlSb_JAP_1972} & 29.7 & 27.4 & 28.5 & 29.6 & 0.268~\cite{Semiconductors_BasicData_Springer, Bolef_ElasticAlSb_JAP_1960, Weil_ElasticAlSb_JAP_1972}  & 0.258 & 0.25 \\
GaP & 88.7~\cite{Lam_BulkMod_PRB_1987} & 78.8 & 76.2 & 71.9 & 73.4 & 72.2 & 73.8 & 55.3~\cite{Boyle_ElasticGaPSb_PRB_1975} & 53.5 & 49.1 & 51.3 & 51.8 & 0.244~\cite{Boyle_ElasticGaPSb_PRB_1975} & 0.232 & 0.22 \\
      & 89.8~\cite{Boyle_ElasticGaPSb_PRB_1975} & & & & & & & & & & & & & \\
GaAs & 74.8~\cite{Lam_BulkMod_PRB_1987} & 62.7 & 60.7 & 56.8 & 57.7 & 56.6 & 58.1 & 46.6~\cite{Bateman_ElasticGaAs_JAP_1975} & 42.6 & 39.1 & 40.8 & 40.9 & 0.244~\cite{Bateman_ElasticGaAs_JAP_1975} & 0.233 & 0.23 \\
      & 75.5~\cite{Bateman_ElasticGaAs_JAP_1975} & & & & & & & & & & & & & \\
GaSb & 57.0~\cite{Lam_BulkMod_PRB_1987} & 47.0 & 44.7 & 41.6 & 42.3 & 41.2 & 42.6 & 34.2~\cite{Boyle_ElasticGaPSb_PRB_1975} & 30.8 & 28.3 & 29.6 & 30.0 &  0.248~\cite{Boyle_ElasticGaPSb_PRB_1975} & 0.240 & 0.23 \\
      & 56.3~\cite{Boyle_ElasticGaPSb_PRB_1975} & & & & & & & & & & & & & \\
InP  & 71.1~\cite{Lam_BulkMod_PRB_1987, Nichols_ElasticInP_SSC_1980} & 60.4 & N/A & 56.4 & 57.6 & 56.3 & 57.8 & 34.3~\cite{Nichols_ElasticInP_SSC_1980}  & 33.6 & 29.7 & 31.6 & N/A & 0.292~\cite{Nichols_ElasticInP_SSC_1980} & 0.277 & N/A \\
InAs & 60.0~\cite{Lam_BulkMod_PRB_1987} & 50.1 & 49.2 & 45.7 & 46.6 & 45.4 & 46.9 & 29.5~\cite{Semiconductors_BasicData_Springer, Gerlich_ElasticAlSb_JAP_1963} & 27.3 & 24.2 & 25.7 & 25.1 & 0.282~\cite{Semiconductors_BasicData_Springer, Gerlich_ElasticAlSb_JAP_1963} & 0.281 & 0.28 \\
      & 57.9~\cite{Semiconductors_BasicData_Springer, Gerlich_ElasticAlSb_JAP_1963} & & & & & & & & & & & & & \\
InSb & 47.3~\cite{Lam_BulkMod_PRB_1987, DeVaux_ElasticInSb_PR_1956} & 38.1 & N/A & 34.3 & 35.0 & 34.1 & 35.2 & 22.1~\cite{DeVaux_ElasticInSb_PR_1956} & 21.3 & 19.0 & 20.1 & N/A & 0.298~\cite{DeVaux_ElasticInSb_PR_1956} & 0.275 & N/A \\
      & 48.3~\cite{Semiconductors_BasicData_Springer, Slutsky_ElasticInSb_PR_1959} & & & & & & & 23.7~\cite{Semiconductors_BasicData_Springer, Slutsky_ElasticInSb_PR_1959} & & & & & 0.289~\cite{Semiconductors_BasicData_Springer, Slutsky_ElasticInSb_PR_1959} & \\
      & 46.5~\cite{Vanderborgh_ElasticInSb_PRB_1990} & & & & & & & & & & & & & \\
ZnS & 77.1~\cite{Lam_BulkMod_PRB_1987} & 71.2 & 68.3 & 65.8 & 66.1 & 65.2 & 66.6 & 30.9~\cite{Semiconductors_BasicData_Springer} & 36.5 & 31.4 & 33.9 & 33.2 & 0.318~\cite{Semiconductors_BasicData_Springer} & 0.294 & 0.29 \\
      & 74.5~\cite{Semiconductors_BasicData_Springer} & & & & & & & & & & & & & \\
ZnSe & 62.4~\cite{Lam_BulkMod_PRB_1987, Lee_ElasticZnSeTe_JAP_1970} & 58.2 & 58.3 & 53.3 & 53.8 & 52.8 & 54.1 & 29.1~\cite{Lee_ElasticZnSeTe_JAP_1970} & 29.5 & 25.6 & 27.5 & 27.5 & 0.298~\cite{Lee_ElasticZnSeTe_JAP_1970} & 0.296 & 0.3\\
ZnTe & 51.0~\cite{Lam_BulkMod_PRB_1987, Lee_ElasticZnSeTe_JAP_1970} & 43.8 & 46.0 & 39.9 & 40.5 & 39.4 & 40.7 & 23.4~\cite{Lee_ElasticZnSeTe_JAP_1970} & 23.3 & 20.8 & 22.1 & 22.4 & 0.30~\cite{Lee_ElasticZnSeTe_JAP_1970} & 0.284 & 0.29 \\
CdSe & 53.0~\cite{Lam_BulkMod_PRB_1987} & 46.7 & 44.8 & 41.5 & 42.1 & 41.1 & 42.3 & N/A & 16.2 & 13.1 & 14.7 & 15.3 & N/A & 0.358 & 0.35 \\
CdTe & 42.4~\cite{Lam_BulkMod_PRB_1987} & 36.4 & 35.3 & 32.2 & 32.7 & 31.9 & 32.8 & N/A & 14.2 & 11.9 & 13.0 & 13.6 & N/A & 0.340 & 0.33 \\
HgSe & 50.0~\cite{Lam_BulkMod_PRB_1987} & 43.8 & 41.2 & 39.0 & 39.7 & 38.5 & 39.9 & 14.8~\cite{Lehoczky_ElasticHgSe_PR_1969} & 15.6 & 11.9 & 13.7 & 13.3 & 0.361~\cite{Lehoczky_ElasticHgSe_PR_1969} & 0.358 & 0.35 \\
      & 48.5~\cite{Lehoczky_ElasticHgSe_PR_1969} & & & & & & & & & & & & & \\
HgTe & 42.3~\cite{Lam_BulkMod_PRB_1987, Semiconductors_BasicData_Springer, Cottam_ElasticHgTe_JPCS_1975} & 35.3 & N/A & 31.0 & 31.6 & 30.8 & 31.9  & 14.7~\cite{Semiconductors_BasicData_Springer, Cottam_ElasticHgTe_JPCS_1975} & 14.4 & 11.6 & 13.0 & N/A & 0.344~\cite{Semiconductors_BasicData_Springer, Cottam_ElasticHgTe_JPCS_1975} & 0.335 & N/A \\
\end{tabular}}
\label{tab:art115:zincblende_elastic}
\etab

For the shear modulus, the experimental values $G^\EXP$  are compared to the \AEL\ values $G_\sVoigt^\sAEL$,
$G_\sReuss^\sAEL$ and $G_\sVRH^\sAEL$. As can be seen from the values in
Table~\ref{tab:art115:zincblende_elastic} and Figure~\ref{fig:art115:zincblende_thermal_elastic}(b), the agreement with the experimental values is generally
good with a very low \RMSrD\ of 0.111 for $G_\sVRH^\sAEL$, with the Voigt approximation tending to overestimate and the Reuss approximation tending to underestimate, as would be
expected. The experimental values of the Poisson ratio $\sigma^\EXP$ and the \AEL\ values $\sigma^\sAEL$ (Equation~\ref{eq:art115:Poissonratio}) are
also shown in Table~\ref{tab:art115:zincblende_elastic} and Figure~\ref{fig:art115:zincblende_thermal_elastic}(c), and the values are generally in good
agreement. The Pearson (\ie, linear, Equation~\ref{eq:art115:Pearson}) and Spearman (\ie, rank order, Equation~\ref{eq:art115:Spearman}) correlations between all of
the \AEL-\AGL\ elastic property values and experiment are shown in Table~\ref{tab:art115:zincblende_correlation}, and are generally
very high for all of these properties, ranging from 0.977 and 0.982 respectively for $\sigma^\EXP$ \vs\ $\sigma^\sAEL$, up to 0.999
and 0.992 for $B^\EXP$ \vs\ $B_\sVRH^\sAEL$. These very high correlation values demonstrate the validity of using the \AEL-\AGL\
methodology to predict the elastic and mechanical properties of
materials.

The Materials Project values of $B_\sVRH^\sMP$, $G_\sVRH^\sMP$ and $\sigma^\sMP$ for diamond and zincblende structure materials are also shown in
Table~\ref{tab:art115:zincblende_elastic}, where available. The Pearson correlations values for the experimental results with the available values of
$B_\sVRH^\sMP$, $G_\sVRH^\sMP$ and $\sigma^\sMP$ were calculated to be 0.995, 0.987 and 0.952, respectively, while the respective Spearman correlations
were 0.963, 0.977 and 0.977, and the \RMSrD\ values were 0.149, 0.116 and 0.126. For comparison, the corresponding Pearson correlations for the same
subset of materials for $B_\sVRH^\sAEL$, $G_\sVRH^\sAEL$ and $\sigma^\sAEL$ are 0.997, 0.987, and 0.957 respectively,  while the respective Spearman correlations
were 0.982, 0.977 and 0.977, and the \RMSrD\ values were 0.129, 0.114 and 0.108. These correlation values are very similar, and the general close agreement
{for $B_\sVRH^\sAEL$, $G_\sVRH^\sAEL$ and $\sigma^\sAEL$ with $B_\sVRH^\sMP$, $G_\sVRH^\sMP$ and $\sigma^\sMP$}
demonstrate that the small differences in the parameters used for the \DFT\ calculations make little difference to the results,
indicating that the parameter set used here is robust for high-throughput calculations.

\fig
\includegraphics[width=0.98\linewidth]{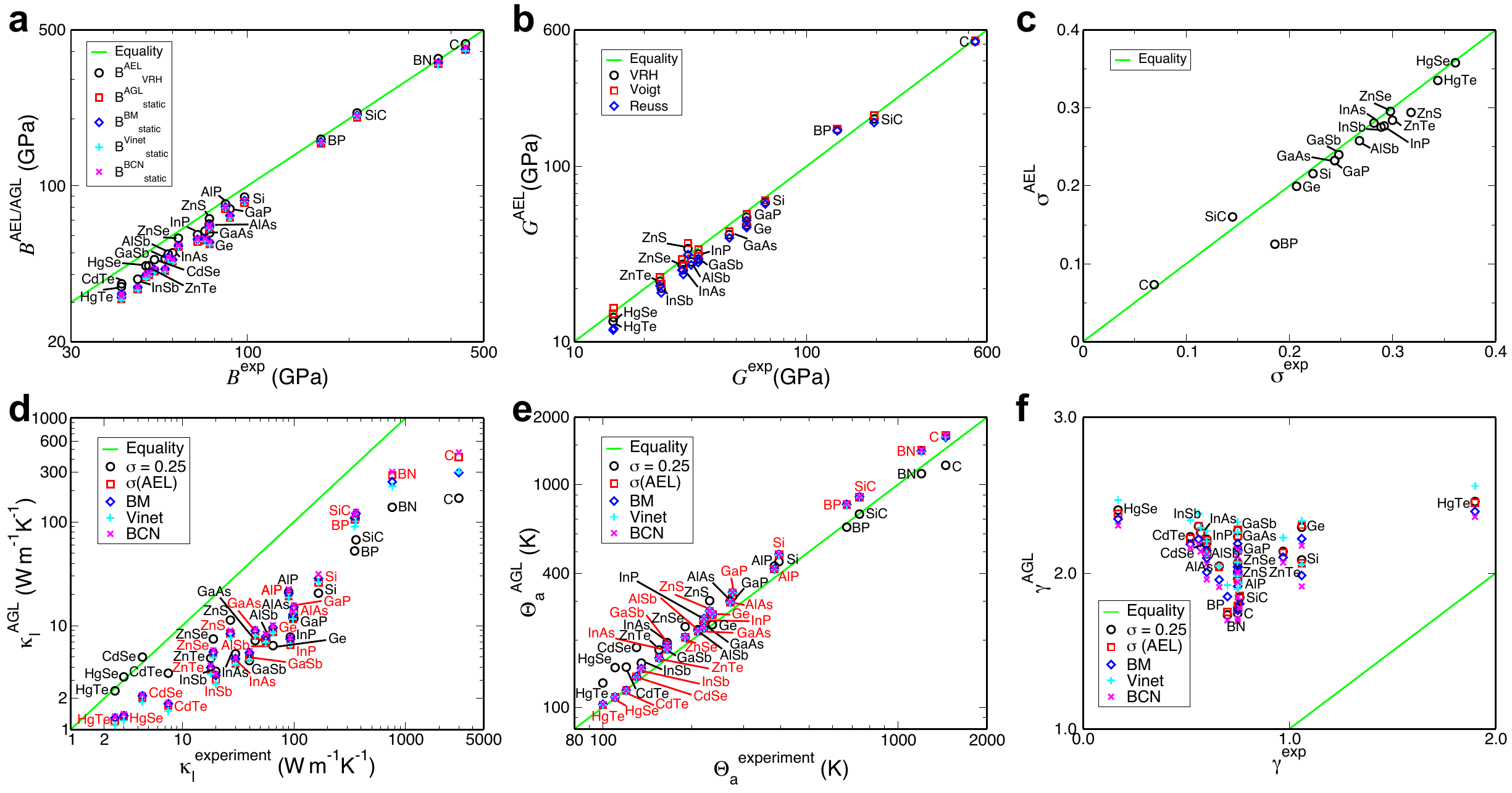}
\mycaption[
({\bf a}) Bulk modulus,
({\bf b}) shear modulus,
({\bf c}) Poisson ratio,
({\bf d}) lattice thermal conductivity at 300~K,
({\bf e}) acoustic Debye temperature and
({\bf f}) Gr{\"u}neisen parameter of zincblende and
diamond structure semiconductors.]
{The zincblende structure is designated \AFLOW\ prototype {\sf AB\_cF8\_216\_c\_a}~\cite{aflowANRL}
and the diamond structure {\sf A\_cF8\_227\_a}~\cite{aflowANRL}.}
\label{fig:art115:zincblende_thermal_elastic}
\efig

The thermal {properties Debye} temperature, Gr{\"u}neisen parameter and thermal conductivity calculated using \AGL\ for this set of materials are
compared to the experimental values  taken from the literature in Table~\ref{tab:art115:zincblende_thermal} and are also plotted in Figure~\ref{fig:art115:zincblende_thermal_elastic}.
For the Debye temperature, the experimental values $\theta^\EXP$ are compared {to
$\theta_\sDebye^\sAGL$, $\theta_\sDebye^\sBM$, $\theta_\sDebye^\sVIN$ and $\theta_\sDebye^\sBCN$} in Figure~\ref{fig:art115:zincblende_thermal_elastic}(e), while {the values} for
the empirical equations of state are provided in Table~\ref{tab:art115:zincblende_thermal_eos}.
Note that the $\theta^\EXP$ values taken from Reference~\onlinecite{slack} and
Reference~\onlinecite{Morelli_Slack_2006} are for $\theta_\acoustic$, and generally are in good agreement with the $\theta_\acoustic^\sAGL$ values. The
values obtained using the numerical $E(V)$ fit and the three different equations of state are also in good agreement with each other, whereas
the values of $\theta_\sDebye^\sAGL$ calculated using different $\sigma$ values differ significantly, indicating that for this property the value
of $\sigma$ used is far more important than the equation of state used.  The correlation
between $\theta^\EXP$ and the various \AGL\ values is also very high,
of the order of 0.999, and the \RMSrD\ is low, of the order of 0.13.

\tab
\mycaption[Thermal properties lattice thermal conductivity at
300~K, Debye temperature and Gr{\"u}neisen parameter of
zincblende and diamond structure semiconductors, comparing the effect of using the
calculated value of the Poisson ratio to the previous approximation of $\sigma = 0.25$.]
{The zincblende structure is designated \AFLOW\ prototype {\sf AB\_cF8\_216\_c\_a}~\cite{aflowANRL}
and the diamond structure {\sf A\_cF8\_227\_a}~\cite{aflowANRL}.
The values listed for $\theta^{\mathrm{exp}}$ are
$\theta_\acoustic$, except 141K for HgTe which is $\theta_{\mathrm D}$~\cite{Snyder_jmatchem_2011}.
Units: $\kappa$ in \WmK, $\theta$ in \K.}
\tabvspace
\resizebox{\linewidth}{!}{
\begin{tabular}{l|r|r|r|r|r|r|r|r|r|r}
comp. & $\kappa^\EXP$  & $\kappa^\sAGL $ & $\kappa^\sAGL$ & $\theta^\EXP$  & $\theta_\sDebye^\sAGL$ & $\theta_\sDebye^\sAGL$ & $\theta_\sDebye^\sAEL$ & $\gamma^\EXP$ & $\gamma^\sAGL$ & $\gamma^\sAGL$  \\
      &  & & & &  ($\theta_\acoustic^\sAGL$) & ($\theta_\acoustic^\sAGL$) & & & \\
      & & ($\sigma = 0.25$)\cite{curtarolo:art96} & & & ($\sigma = 0.25$)~\cite{curtarolo:art96} & & & & ($\sigma = 0.25$)\cite{curtarolo:art96} &   \\
\hline
C &  3000~\cite{Morelli_Slack_2006} & 169.1 & 419.9  & 1450~\cite{slack, Morelli_Slack_2006} & 1536 & 2094 & 2222 & 0.75~\cite{Morelli_Slack_2006} & 1.74 & 1.77 \\
      & & & & & (1219) & (1662) & & 0.9~\cite{slack} & & \\
SiC & 360~\cite{Ioffe_Inst_DB} & 67.19 & 113.0 & 740~\cite{slack} & 928 & 1106 & 1143 & 0.76~\cite{slack} & 1.84 & 1.85	\\
      & & & & & (737) & (878) & & & & \\
Si & 166~\cite{Morelli_Slack_2006} & 20.58 & 26.19 & 395~\cite{slack, Morelli_Slack_2006} & 568 & 610 & 624 & 1.06~\cite{Morelli_Slack_2006} & 2.09 & 2.06	 \\
      & & & & & (451) & (484) &  & 0.56~\cite{slack} &  & \\
Ge &  65~\cite{Morelli_Slack_2006} &  6.44 & 8.74 & 235~\cite{slack, Morelli_Slack_2006} & 296 & 329 & 342 & 1.06~\cite{Morelli_Slack_2006} & 2.3 & 2.31 	 \\
      & & & & & (235) & (261) &  & 0.76~\cite{slack} & &   \\
BN & 760~\cite{Morelli_Slack_2006} & 138.4 & 281.6 & 1200~\cite{Morelli_Slack_2006} & 1409 & 1793 & 1887 & 0.7~\cite{Morelli_Slack_2006} & 1.73 & 1.75	\\
      & & & & & (1118) & (1423) & & & & \\
BP & 350~\cite{Morelli_Slack_2006} & 52.56 & 105.0 & 670~\cite{slack, Morelli_Slack_2006} & 811 & 1025 & 1062 & 0.75~\cite{Morelli_Slack_2006} & 1.78 & 1.79	\\
      & & & & & (644) & (814) & & & & \\
AlP & 90~\cite{Landolt-Bornstein, Spitzer_JPCS_1970} & 21.16 & 19.34 & 381~\cite{Morelli_Slack_2006} & 542 & 525 & 531 & 0.75~\cite{Morelli_Slack_2006} & 1.96 & 1.96	 \\
      & & & & & (430) & (417) & & & & \\
AlAs & 98~\cite{Morelli_Slack_2006} &  12.03 & 11.64 & 270~\cite{slack, Morelli_Slack_2006} & 378 & 373 & 377 & 0.66~\cite{slack, Morelli_Slack_2006} & 2.04 & 2.04	 \\
      & & & & &  (300) & (296) & & & & \\
AlSb & 56~\cite{Morelli_Slack_2006} & 7.22 & 6.83 & 210~\cite{slack, Morelli_Slack_2006} & 281 & 276 & 277 & 0.6~\cite{slack, Morelli_Slack_2006} & 2.12 & 2.13 	 \\
      & & & & & (223) & (219) & & & & \\
GaP & 100~\cite{Morelli_Slack_2006} & 11.76 & 13.34 & 275~\cite{slack, Morelli_Slack_2006} & 396 & 412 & 423 & 0.75~\cite{Morelli_Slack_2006} & 2.15 & 2.15 	\\
      & & & &  & (314) & (327) & &  0.76~\cite{slack} &  &  \\
GaAs & 45~\cite{Morelli_Slack_2006} & 7.2 & 8.0 & 220~\cite{slack, Morelli_Slack_2006} & 302 & 313	& 322 & 0.75~\cite{slack, Morelli_Slack_2006} & 2.23 & 2.24 \\
      & & & & & (240) & (248) & & & & \\
GaSb & 40~\cite{Morelli_Slack_2006} & 4.62 & 4.96 & 165~\cite{slack, Morelli_Slack_2006} & 234 & 240 & 248 & 0.75~\cite{slack, Morelli_Slack_2006} & 2.27 & 2.28 	 \\
      & & & & & (186) & (190) & & & & \\
InP & 93~\cite{Morelli_Slack_2006} & 7.78 & 6.53 & 220~\cite{slack, Morelli_Slack_2006} & 304 & 286 & 287 & 0.6~\cite{slack, Morelli_Slack_2006} & 2.22 & 2.21 	 \\
      & & & & & (241) & (227) & & & & \\
InAs & 30~\cite{Morelli_Slack_2006} & 5.36 & 4.33 & 165~\cite{slack, Morelli_Slack_2006} &  246 & 229 & 231 & 0.57~\cite{slack, Morelli_Slack_2006} & 2.26 & 2.26	 \\
      & & & & & (195) & (182) & & & & \\
InSb & 20~\cite{Morelli_Slack_2006} & 3.64 & 3.02 & 135~\cite{slack, Morelli_Slack_2006} & 199 & 187 & 190 & 0.56~\cite{slack, Morelli_Slack_2006} & 2.3 & 2.3 	 \\
      & 16.5~\cite{Snyder_jmatchem_2011} & & & & (158) & (148) &  & & &  \\
ZnS & 27~\cite{Morelli_Slack_2006} & 11.33 & 8.38 & 230~\cite{slack, Morelli_Slack_2006} & 379 & 341 & 346 & 0.75~\cite{slack, Morelli_Slack_2006} & 2.01 & 2.00 	 \\
      & & & & & (301) & (271) & & & & \\
ZnSe & 19~\cite{Morelli_Slack_2006} & 7.46 & 5.44 & 190~\cite{slack, Morelli_Slack_2006} & 290 & 260	& 263 & 0.75~\cite{slack, Morelli_Slack_2006} & 2.07 & 2.06 	\\
      & 33~\cite{Snyder_jmatchem_2011} & & &  & (230) & (206) & & & &  \\
ZnTe & 18~\cite{Morelli_Slack_2006} &  4.87 & 3.83 & 155~\cite{slack, Morelli_Slack_2006} & 228 & 210 & 212 & 0.97~\cite{slack, Morelli_Slack_2006} & 2.14 & 2.13  \\
      & & & & & (181) & (167) & & & & \\
CdSe & 4.4~\cite{Snyder_jmatchem_2011} & 4.99 & 2.04 & 130~\cite{Morelli_Slack_2006} & 234 & 173 & 174 & 0.6~\cite{Morelli_Slack_2006} & 2.19 & 2.18 \\
      & & & & & (186) & (137) & & & & \\
CdTe & 7.5~\cite{Morelli_Slack_2006} & 3.49 & 1.71 & 120~\cite{slack, Morelli_Slack_2006} & 191 & 150 & 152 & 0.52~\cite{slack, Morelli_Slack_2006} & 2.23 & 2.22	 \\
      & & & & & (152) & (119) & & & & \\
HgSe & 3~\cite{Whitsett_PRB_1973} & 3.22 & 1.32 & 110~\cite{slack} & 190 & 140	& 140 & 0.17~\cite{slack} & 2.4 & 2.38	 \\
      & & & & &  (151) & (111) & & & & \\
HgTe & 2.5~\cite{Snyder_jmatchem_2011}  & 2.36 & 1.21 & 141~\cite{Snyder_jmatchem_2011}  & 162 & 129 & 130 & 1.9~\cite{Snyder_jmatchem_2011}  & 2.46 & 2.45 \\
      & & & & (100)~\cite{slack} & (129) & (102) & & & & \\
\end{tabular}}
\label{tab:art115:zincblende_thermal}
\etab

The experimental values $\gamma^\EXP$ of the Gr{\"u}neisen parameter are plotted {against
$\gamma^\sAGL$, $\gamma^\sBM$, $\gamma^\sVIN$ and $\gamma^\sBCN$} in Figure~\ref{fig:art115:zincblende_thermal_elastic}(f), and the values
are listed in Table~\ref{tab:art115:zincblende_thermal} and in Table~\ref{tab:art115:zincblende_thermal_eos}.
The very high \RMSrD\ values (see Table~\ref{tab:art115:zincblende_correlation}) show that \AGL\ has problems accurately predicting
the Gr{\"u}neisen parameter for this set of materials, as the calculated value is often 2 to 3 times larger than the experimental one.
Note also that there are quite large differences between the values obtained for different equations of state, with $\gamma^\sBCN$ generally
having the lowest values while $\gamma^\sVIN$ has the highest values.
On the other hand, in contrast to the case of $\theta_\sDebye^\sAGL$, the value of $\sigma$ used makes little difference to the value
of $\gamma^\sAGL$. The {correlations} between $\gamma^\EXP$ and the \AGL\ values, as shown in Table~\ref{tab:art115:zincblende_correlation},
are also quite poor, with no value higher than 0.2 for the Pearson correlations, and {negative Spearman} correlations.

The experimental thermal conductivity $\kappa^\EXP$ is compared in Figure~\ref{fig:art115:zincblende_thermal_elastic}(d) to the
thermal conductivities calculated with \AGL\ using the Leibfried-Schl{\"o}mann equation (Equation~\ref{eq:art115:thermal_conductivity}): $\kappa^\sAGL$, $\kappa^\sBM$, $\kappa^\sVIN$ and $\kappa^\sBCN$,
while the values are listed in  Table~\ref{tab:art115:zincblende_thermal} and in Table~\ref{tab:art115:zincblende_thermal_eos}.
The absolute agreement between the \AGL\ values and $\kappa^\EXP$ is quite poor, with \RMSrD\ values of the order of 0.8 and discrepancies of tens, or even hundreds, of percent
quite common. Considerable disagreements also exist between different experimental reports of these properties, in
almost all cases where they exist. Unfortunately, the scarcity of experimental data from different sources on the thermal properties of these materials
prevents reaching definite conclusions regarding the true values of these properties. The available data can thus only be considered as a rough indication
of their order of magnitude.

{The Pearson} correlations between the \AGL\ calculated thermal conductivity values and the experimental
values are high, ranging from $0.871$ to $0.932$, while the Spearman correlations are even higher, ranging from $0.905$
to $0.954$, as shown in Table~\ref{tab:art115:zincblende_correlation}. In particular, note that using the $\sigma^\sAEL$ in the \AGL\ calculations
improves the correlations by about 5\%, from $0.878$ to $0.927$ and from $0.905$ to $0.954$. For the different equations of state,
$\kappa^\sAGL$ and $\kappa^\sBCN$ appear to correlate better with $\kappa^\EXP$ than $\kappa^\sBM$ and $\kappa^\sVIN$ for this set of
materials.

\tab
\mycaption{Correlations and deviations between experimental values and \AEL\ and \AGL\ results for
elastic and thermal properties for zincblende and diamond structure semiconductors.}
\tabvspace
\begin{tabular}{l|r|r|r}
property  & Pearson & Spearman & \RMSrD\ \\
          & (linear) & (rank order) \\
\hline
$\kappa^\EXP$ \vs\ $\kappa^\sAGL$  ($\sigma = 0.25$)~\cite{curtarolo:art96} & 0.878 & 0.905 & 0.776 \\
$\kappa^\EXP$ \vs\ $\kappa^\sAGL$ & 0.927 & 0.95 & 0.796 \\
$\kappa^\EXP$ \vs\ $\kappa^\sBM$ & 0.871 & 0.954 & 0.787  \\
$\kappa^\EXP$ \vs\ $\kappa^\sVIN$ & 0.908 &  0.954 & 0.815 \\
$\kappa^\EXP$ \vs\ $\kappa^\sBCN$ & 0.932 & 0.954 & 0.771 \\
$\theta_\acoustic^\EXP$ \vs\ $\theta_\acoustic^\sAGL$  ($\sigma = 0.25$)~\cite{curtarolo:art96} & 0.995 & 0.984 & 0.200 \\
$\theta_\acoustic^\EXP$ \vs\ $\theta_\acoustic^\sAGL$ & 0.999 & 0.998 & 0.132 \\
$\theta_\acoustic^\EXP$ \vs\ $\theta_\acoustic^\sBM$ & 0.999 & 0.998  & 0.132 \\
$\theta_\acoustic^\EXP$ \vs\ $\theta_\acoustic^\sVIN$ & 0.999 &  0.998 & 0.127 \\
$\theta_\acoustic^\EXP$ \vs\ $\theta_\acoustic^\sBCN$ & 0.999 & 0.998 & 0.136 \\
$\gamma^\EXP$ \vs\ $\gamma^\sAGL$  ($\sigma = 0.25$)~\cite{curtarolo:art96} & 0.137 & -0.187 & 3.51 \\
$\gamma^\EXP$ \vs\ $\gamma^\sAGL$ & 0.145 & -0.165 & 3.49 \\
$\gamma^\EXP$ \vs\ $\gamma^\sBM$ & 0.169 & -0.178 & 3.41  \\
$\gamma^\EXP$ \vs\ $\gamma^\sVIN$ & 0.171 &  -0.234  & 3.63 \\
$\gamma^\EXP$ \vs\ $\gamma^\sBCN$ & 0.144 & -0.207 & 3.32 \\
$B^\EXP$ \vs\ $B_\sVRH^\sAEL$ & 0.999 & 0.992  & 0.130 \\
$B^\EXP$ \vs\ $B_\sStatic^\sAGL$ & 0.999 & 0.986 & 0.201 \\
$B^\EXP$ \vs\ $B_\sStatic^\sBM$ & 0.999 & 0.986 & 0.189 \\
$B^\EXP$ \vs\ $B_\sStatic^\sVIN$ & 0.999 & 0.986 & 0.205 \\
$B^\EXP$ \vs\ $B_\sStatic^\sBCN$ & 0.999 & 0.986 & 0.185 \\
$G^\EXP$ \vs\ $G_\sVRH^\sAEL$ & 0.998 & 0.980 & 0.111  \\
$G^\EXP$ \vs\ $G_\sVoigt^\sAEL$ & 0.998 & 0.980 & 0.093 \\
$G^\EXP$ \vs\ $G_\sReuss^\sAEL$ & 0.998 & 0.980 & 0.152 \\
$\sigma^\EXP$ \vs\ $\sigma^\sAEL$ & 0.977 & 0.982 & 0.095 \\
\end{tabular}
\label{tab:art115:zincblende_correlation}
\etab

As we noted in our previous work on \AGL~\cite{curtarolo:art96}, some of the inaccuracy in the thermal conductivity results may be due to the inability of the Leibfried-Schl{\"o}mann equation to fully
describe effects such as the suppression of phonon-phonon scattering due to large gaps between the branches of
the phonon dispersion~\cite{Lindsay_PRL_2013}. This can be seen from the thermal conductivity values shown in Table 2.2 of Reference~\onlinecite{Morelli_Slack_2006}
calculated using the experimental values of $\theta_\acoustic$ and $\gamma$ in the Leibfried-Schl{\"o}mann equation. There are large discrepancies in certain cases such as diamond,
while the Pearson and Spearman correlations of $0.932$ and $0.941$ respectively are very similar to the correlations we calculated using the \AGL\ evaluations of
$\theta_\acoustic$ and $\gamma$.

Thus, the unsatisfactory quantitative reproduction of these quantities by the Debye quasi-harmonic model
has little impact on its effectiveness as a screening tool for identifying high or
low thermal conductivity materials. The model can be used when these
experimental values are unavailable to help determine the relative values of these quantities and for
ranking {materials conductivity}.

\subsubsection{Rocksalt structure materials}

The mechanical and thermal properties were calculated for a set of materials with the rocksalt structure
(spacegroup: $Fm\overline{3}m$,\ $\#$225; Pearson symbol: cF8;
\AFLOW\ prototype: {\sf AB\_cF8\_225\_a\_b}~\cite{aflowANRL}\footnote{\url{http://aflow.org/CrystalDatabase/AB_cF8_225_a_b.html}}).
 This {is the same set of materials
as} in Table II of Reference~\onlinecite{curtarolo:art96}, which in turn {are from} the
sets in Table III of Reference~\onlinecite{slack} and Table 2.1 of Reference~\onlinecite{Morelli_Slack_2006}.

The elastic properties of bulk modulus, shear modulus and Poisson ratio, as calculated using \AEL\ and \AGL\ are shown
in Table~\ref{tab:art115:rocksalt_elastic} and Figure~\ref{fig:art115:rocksalt_thermal_elastic}, together
with experimental values from the literature where available. As can be seen
from the results in Table~\ref{tab:art115:rocksalt_elastic} and Figure~\ref{fig:art115:rocksalt_thermal_elastic}(a), for this set of materials the
$B_\sVRH^\sAEL$ values are closest to experiment, with an \RMSrD\ of 0.078. The \AGL\ values from both the numerical
fit and the empirical equations of state are generally very similar to each other, while being slightly less than the $B_\sVRH^\sAEL$
values.

\tab
\mycaption[Mechanical properties bulk modulus, shear modulus
and Poisson ratio of rocksalt structure semiconductors.]
{The rocksalt structure is designated \AFLOW\ Prototype {\sf AB\_cF8\_225\_a\_b}~\cite{aflowANRL}.
``N/A'' = Not available for that source.
Units: $B$ and $G$ in \GPa.}
\tabvspace
\resizebox{\linewidth}{!}{
\begin{tabular}{l|r|r|r|r|r|r|r|r|r|r|r|r|r|r|r}
comp. & $B^\EXP$  & $B_\sVRH^\sAEL$ & $B_\sVRH^\sMP$ & $B_\sStatic^\sAGL$ & $B_\sStatic^\sBM$ & $B_\sStatic^\sVIN$ &  $B_\sStatic^\sBCN$  & $G^\EXP$ & $G_\sVoigt^\sAEL$ & $G_\sReuss^\sAEL$ &  $G_\sVRH^\sAEL$ & $G_\sVRH^\sMP$ & $\sigma^\EXP$ & $\sigma^\sAEL$ & $\sigma^\sMP$    \\
\hline
LiH & 33.7~\cite{Laplaze_ElasticLiH_SSC_1976} & 37.7 & 36.1 & 29.5 & 29.0 & 27.7 & 31.4 & 36.0~\cite{Laplaze_ElasticLiH_SSC_1976} & 43.4 & 42.3 & 42.8 & 42.9 & 0.106~\cite{Laplaze_ElasticLiH_SSC_1976} & 0.088 & 0.07 \\
LiF & 69.6~\cite{Haussuhl_ElasticRocksalt_ZP_1960} & 70.4 & 69.9 & 58.6 & 59.9 & 57.5 & 61.2 & 48.8~\cite{Haussuhl_ElasticRocksalt_ZP_1960} & 46.4 & 45.8 & 46.1 & 50.9 & 0.216~\cite{Haussuhl_ElasticRocksalt_ZP_1960} & 0.231 & 0.21 \\
NaF & 48.5~\cite{Haussuhl_ElasticRocksalt_ZP_1960} & 46.9 & 47.6 & 38.7 & 38.6 & 36.8 & 39.3 & 31.2~\cite{Haussuhl_ElasticRocksalt_ZP_1960} & 29.5 & 28.4 & 28.9 & 30.0 & 0.236~\cite{Haussuhl_ElasticRocksalt_ZP_1960} & 0.244 & 0.24 \\
NaCl & 25.1~\cite{Haussuhl_ElasticRocksalt_ZP_1960} & 24.9 & 22.6 & 20.0 & 20.5 & 19.2 & 20.7 & 14.6~\cite{Haussuhl_ElasticRocksalt_ZP_1960} & 14.0 & 12.9 & 13.5 & 14.3 & 0.255~\cite{Haussuhl_ElasticRocksalt_ZP_1960} & 0.271 & 0.24 \\
NaBr & 20.6~\cite{Haussuhl_ElasticRocksalt_ZP_1960} & 20.5 & 27.1 & 16.3 & 16.9 & 15.7 & 16.9 & 11.6~\cite{Haussuhl_ElasticRocksalt_ZP_1960} & 11.0 & 9.9 & 10.4 & 11.6 & 0.264~\cite{Haussuhl_ElasticRocksalt_ZP_1960} & 0.283 & 0.31 \\
NaI & 15.95~\cite{Haussuhl_ElasticRocksalt_ZP_1960} & 16.4 & 15.8 & 12.6 & 13.2 & 12.2 & 13.1 & 8.59~\cite{Haussuhl_ElasticRocksalt_ZP_1960} & 8.35 & 7.31 & 7.83 & 8.47 & 0.272~\cite{Haussuhl_ElasticRocksalt_ZP_1960} & 0.295 & 0.27 \\
KF & 31.6~\cite{Haussuhl_ElasticRocksalt_ZP_1960} & 29.9 & 28.9 & 25.1 & 24.2 & 22.9  & 24.7 & 16.7~\cite{Haussuhl_ElasticRocksalt_ZP_1960} & 16.5 & 15.4 & 15.9 & 16.5 & 0.275~\cite{Haussuhl_ElasticRocksalt_ZP_1960} & 0.274 & 0.26 \\
KCl & 18.2~\cite{Haussuhl_ElasticRocksalt_ZP_1960} & 16.7 & 15.8 & 13.8 & 13.7 & 12.7  & 13.6 & 9.51~\cite{Haussuhl_ElasticRocksalt_ZP_1960} & 10.1 & 8.51 & 9.30 & 9.24 & 0.277~\cite{Haussuhl_ElasticRocksalt_ZP_1960} & 0.265 & 0.26 \\
KBr & 15.4~\cite{Haussuhl_ElasticRocksalt_ZP_1960} & 13.8 & 21.6 & 11.1 & 11.4 & 10.5 & 11.2 & 7.85~\cite{Haussuhl_ElasticRocksalt_ZP_1960} & 8.14 & 6.46 & 7.30 & 7.33 & 0.282~\cite{Haussuhl_ElasticRocksalt_ZP_1960} & 0.276 & 0.35 \\
KI & 12.2~\cite{Haussuhl_ElasticRocksalt_ZP_1960} & 10.9 & 9.52 & 8.54 & 9.03 &  8.28 & 8.84 & 5.96~\cite{Haussuhl_ElasticRocksalt_ZP_1960} & 6.05 & 4.39 & 5.22 & 5.55 & 0.290~\cite{Haussuhl_ElasticRocksalt_ZP_1960} & 0.294 & 0.26 \\
RbCl & 16.2~\cite{Haussuhl_ElasticRocksalt_ZP_1960} & 14.3 & 14.6 & 12.1 & 11.8 & 11.0 & 11.8 & 7.63~\cite{Haussuhl_ElasticRocksalt_ZP_1960} & 8.06 & 6.41 & 7.24 & 7.67 & 0.297~\cite{Haussuhl_ElasticRocksalt_ZP_1960} & 0.284 & 0.28 \\
RbBr & 13.8~\cite{Haussuhl_ElasticRocksalt_ZP_1960} & 12.6 & 13.8 & 10.3 & 9.72 & 9.06 & 9.67 & 6.46~\cite{Haussuhl_ElasticRocksalt_ZP_1960} & 7.12 & 5.24 & 6.18 & 6.46 & 0.298~\cite{Haussuhl_ElasticRocksalt_ZP_1960} & 0.289 & 0.3 \\
RbI & 11.1~\cite{Haussuhl_ElasticRocksalt_ZP_1960} & 9.90 & 9.66 & 8.01 & 7.74 & 7.12 & 7.54 & 5.03~\cite{Haussuhl_ElasticRocksalt_ZP_1960} & 5.50 & 3.65 & 4.57 & 4.63 & 0.303~\cite{Haussuhl_ElasticRocksalt_ZP_1960} & 0.300 & 0.29 \\
AgCl & 44.0~\cite{Hughes_ElasticAgCl_PRB_1996} & 40.6 & N/A & 33.7 & 34.1 & 33.0 & 34.7 & 8.03~\cite{Hughes_ElasticAgCl_PRB_1996} & 8.68 & 8.66 & 8.67 & N/A & 0.414~\cite{Hughes_ElasticAgCl_PRB_1996} & 0.400 & N/A \\
MgO  & 164~\cite{Sumino_ElasticMgO_JPE_1976} & 152 & 152 & 142 & 142 & 140 & 144 & 131~\cite{Sumino_ElasticMgO_JPE_1976}  & 119 & 115 & 117 & 119 & 0.185~\cite{Sumino_ElasticMgO_JPE_1976} & 0.194 & 0.19 \\
CaO & 113~\cite{Chang_ElasticCaSrBaO_JPCS_1977} & 105 & 105 & 99.6 & 100 & 98.7 & 101 & 81.0~\cite{Chang_ElasticCaSrBaO_JPCS_1977} & 73.7 & 73.7 & 73.7 & 74.2 & 0.210~\cite{Chang_ElasticCaSrBaO_JPCS_1977} & 0.216 & 0.21 \\
SrO & 91.2~\cite{Chang_ElasticCaSrBaO_JPCS_1977} & 84.7 & 87.4 &80.0 & 80.2 & 79.1 & 80.8 & 58.7~\cite{Chang_ElasticCaSrBaO_JPCS_1977} & 55.1 & 55.0 & 55.1 & 56.0 &  0.235~\cite{Chang_ElasticCaSrBaO_JPCS_1977} & 0.233 & 0.24 \\
BaO & 75.4~\cite{Chang_ElasticCaSrBaO_JPCS_1977} & 69.1 & 68.4 & 64.6 & 64.3 & 63.0 & 64.6 & 35.4~\cite{Chang_ElasticCaSrBaO_JPCS_1977} & 36.4 & 36.4 & 36.4 & 37.8 & 0.297~\cite{Chang_ElasticCaSrBaO_JPCS_1977} & 0.276 & 0.27 \\
PbS & 52.9~\cite{Semiconductors_BasicData_Springer, Peresada_ElasticPbS_PSSa_1976} & 53.5 & N/A & 49.9 & 50.8 & 50.0 & 51.0 & 27.9~\cite{Semiconductors_BasicData_Springer, Peresada_ElasticPbS_PSSa_1976} & 34.0 & 26.8 & 30.4 & N/A & 0.276~\cite{Semiconductors_BasicData_Springer, Peresada_ElasticPbS_PSSa_1976} & 0.261 & N/A \\
PbSe & 54.1~\cite{Semiconductors_BasicData_Springer, Lippmann_ElasticPbSe_PSSa_1971} & 47.7 & N/A & 43.9 & 44.8 & 43.9 & 44.9 & 26.2~\cite{Semiconductors_BasicData_Springer, Lippmann_ElasticPbSe_PSSa_1971} & 31.7 & 23.6 & 27.6 &  N/A & 0.291~\cite{Semiconductors_BasicData_Springer, Lippmann_ElasticPbSe_PSSa_1971} & 0.257 & N/A \\
PbTe & 39.8~\cite{Semiconductors_BasicData_Springer, Miller_ElasticPbTe_JPCSS_1981} & 39.5 & N/A & 36.4 & 36.6 & 35.8 & 36.8 & 23.1~\cite{Semiconductors_BasicData_Springer, Miller_ElasticPbTe_JPCSS_1981} & 28.7 & 19.8 & 24.3 & N/A & 0.256~\cite{Semiconductors_BasicData_Springer, Miller_ElasticPbTe_JPCSS_1981} & 0.245 & N/A \\
SnTe & 37.8~\cite{Semiconductors_BasicData_Springer, Seddon_ElasticSnTe_SSC_1976} & 40.4 & 39.6 & 38.1 & 38.4 & 37.6 & 38.6 & 20.8~\cite{Semiconductors_BasicData_Springer, Seddon_ElasticSnTe_SSC_1976}  & 31.4 & 22.0 & 26.7 & 27.6 & 0.267~\cite{Semiconductors_BasicData_Springer, Seddon_ElasticSnTe_SSC_1976}  & 0.229 & 0.22 \\
\end{tabular}}
\label{tab:art115:rocksalt_elastic}
\etab

\fig
\includegraphics[width=0.98\linewidth]{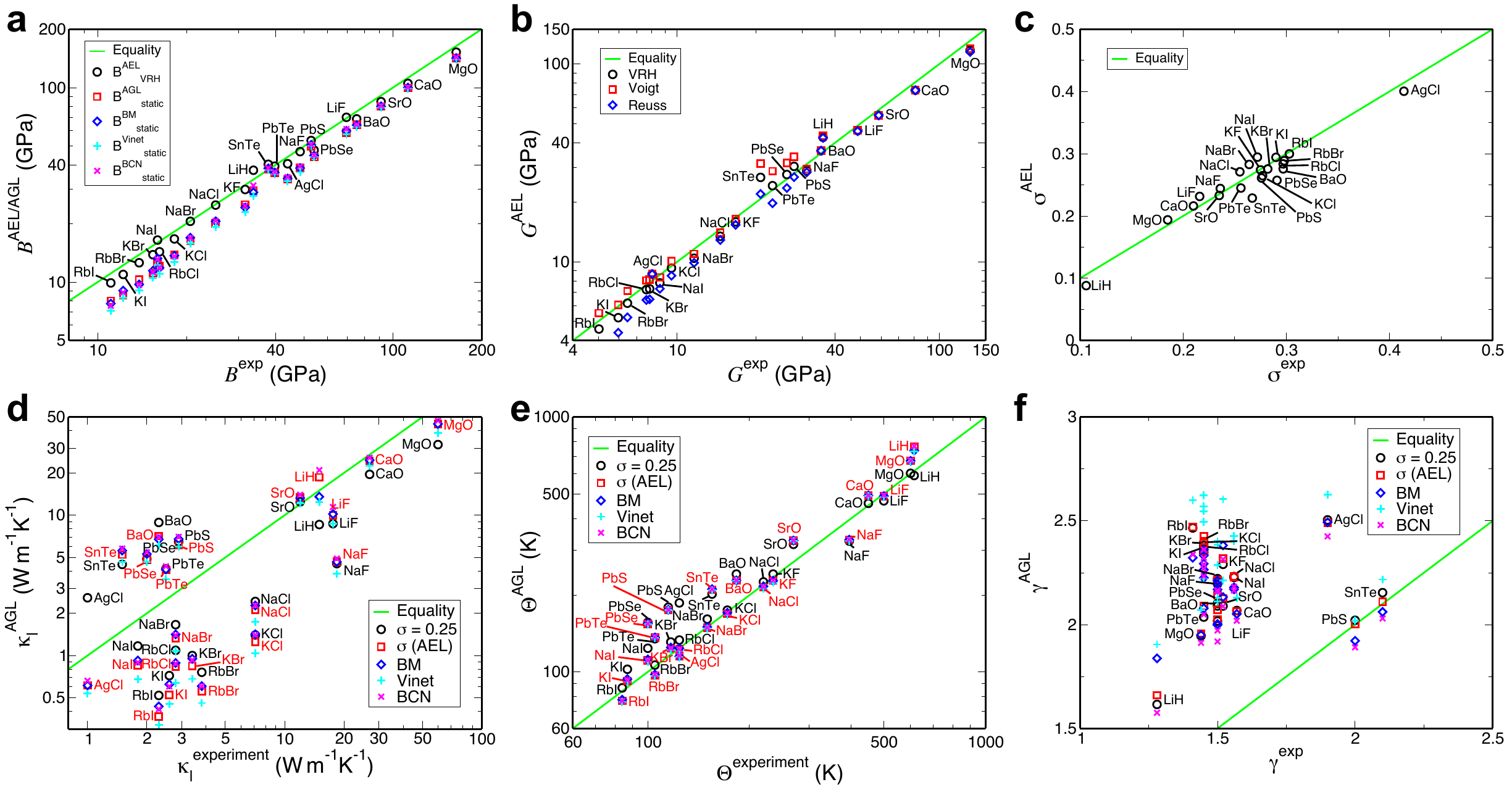}
\mycaption[
({\bf a}) Bulk modulus,
({\bf b}) shear modulus,
({\bf c}) Poisson ratio,
({\bf d}) lattice thermal conductivity at 300~K,
({\bf e}) Debye temperature and
({\bf f}) Gr{\"u}neisen parameter of rocksalt structure
semiconductors.]
{The rocksalt structure is designated \AFLOW\ Prototype {\sf AB\_cF8\_225\_a\_b}~\cite{aflowANRL}.
The Debye temperatures plotted in ({\bf b}) are
$\theta_\acoustic$, except for SnTe where $\theta_{\mathrm D}$ is
quoted in Reference~\onlinecite{Snyder_jmatchem_2011}.}
\label{fig:art115:rocksalt_thermal_elastic}
\efig

For the shear modulus, the experimental values $G^\EXP$  are compared to the \AEL\ values $G_\sVoigt^\sAEL$,
$G_\sReuss^\sAEL$ and $G_\sVRH^\sAEL$. As can be seen from the values in
Table~\ref{tab:art115:rocksalt_elastic} and Figure~\ref{fig:art115:rocksalt_thermal_elastic}(b), the agreement with the experimental values is generally
good with an \RMSrD\ of 0.105 for $G_\sVRH^\sAEL$, with the Voigt approximation tending to overestimate and the Reuss approximation tending to underestimate, as would be
expected. The experimental values of the Poisson ratio $\sigma^\EXP$ and the \AEL\ values $\sigma^\sAEL$ (Equation~\ref{eq:art115:Poissonratio}) are
also shown in Table~\ref{tab:art115:rocksalt_elastic} and Figure~\ref{fig:art115:rocksalt_thermal_elastic}(c), and the values are generally in good
agreement. The Pearson (\ie, linear, Equation~\ref{eq:art115:Pearson}) and Spearman (\ie, rank order, Equation~\ref{eq:art115:Spearman}) correlations between all of
the the \AEL-\AGL\ elastic property values and experiment are shown in Table~\ref{tab:art115:rocksalt_correlation}, and are generally
very high for all of these properties, ranging from 0.959 and 0.827 respectively for $\sigma^\EXP$ \vs\ $\sigma^\sAEL$, up to 0.998
and 0.995 for $B^\EXP$ \vs\ $B_\sVRH^\sAEL$. These very high correlation values demonstrate the validity of using the \AEL-\AGL\
methodology to predict the elastic and mechanical properties of materials.

{The values} of $B_\sVRH^\sMP$, $G_\sVRH^\sMP$ and $\sigma^\sMP$ for rocksalt structure materials are also shown in
Table~\ref{tab:art115:rocksalt_elastic}, where available. The Pearson {correlations for} the experimental results with the available values of
$B_\sVRH^\sMP$, $G_\sVRH^\sMP$ and $\sigma^\sMP$ were calculated to be 0.997, 0.994 and 0.890, respectively, while the respective Spearman correlations
were 0.979, 0.998 and 0.817, and the \RMSrD\ values were 0.153, 0.105 and 0.126. For comparison, the corresponding Pearson correlations for the same
subset of materials for $B_\sVRH^\sAEL$, $G_\sVRH^\sAEL$ and $\sigma^\sAEL$ are 0.998, 0.995, and 0.951 respectively,  while the respective Spearman correlations
were 0.996, 1.0 and 0.843, and the \RMSrD\ values were 0.079, 0.111 and 0.071. These correlation values are very similar, and the general close agreement
for the results for the values of $B_\sVRH^\sAEL$, $G_\sVRH^\sAEL$ and $\sigma^\sAEL$ with those of $B_\sVRH^\sMP$, $G_\sVRH^\sMP$ and $\sigma^\sMP$
demonstrate that the small differences in the parameters used for the \DFT\ calculations make little difference to the results,
indicating that the parameter set used here is robust for high-throughput calculations.

The thermal properties of Debye temperature, Gr{\"u}neisen parameter and thermal conductivity calculated using \AGL\ are
compared to the experimental values  taken from the literature in Table~\ref{tab:art115:rocksalt_thermal} and are also plotted in Figure~\ref{fig:art115:rocksalt_thermal_elastic}.
For the Debye temperature, the experimental values $\theta^\EXP$ are compared {to
$\theta_\sDebye^\sAGL$}, $\theta_\sDebye^\sBM$, $\theta_\sDebye^\sVIN$ and $\theta_\sDebye^\sBCN$ in Figure~\ref{fig:art115:rocksalt_thermal_elastic}(e), while the actual values for
the empirical equations of state are provided in Table~\ref{tab:art115:rocksalt_thermal_eos}.
Note that the $\theta^\EXP$ values taken from Reference~\onlinecite{slack} and
Reference~\onlinecite{Morelli_Slack_2006} are for $\theta_\acoustic$, and generally are in good agreement with the $\theta_\acoustic^\sAGL$ values. The
values obtained using the numerical $E(V)$ fit and the three different equations of state are also in good agreement with each other, whereas
the values of $\theta_\sDebye^\sAGL$ calculated using different $\sigma$ values differ significantly, indicating that, as in the case of the zincblende
and diamond structures, the value of $\sigma$ used is far more important  for this property than the equation of state used.  The correlation
between $\theta^\EXP$ and the various \AGL\ values is also quite high, of the order of 0.98 for the Pearson correlation and 0.92 for the Spearman
correlation.

\tab
\mycaption[Thermal properties lattice thermal conductivity at 300~K, Debye temperature and Gr{\"u}neisen parameter of rocksalt
structure semiconductors, comparing the effect of using the
calculated value of the Poisson ratio to the previous approximation of $\sigma = 0.25$.]
{The rocksalt structure is designated \AFLOW\ Prototype {\sf AB\_cF8\_225\_a\_b}~\cite{aflowANRL}.
The values listed for $\theta^{\mathrm{exp}}$ are
$\theta_\acoustic$, except 155K for SnTe which is $\theta_{\mathrm D}$~\cite{Snyder_jmatchem_2011}.
``N/A'' = Not available for that source.
Units: $\kappa$ in \WmK, $\theta$ in \K.}
\tabvspace
\resizebox{\linewidth}{!}{
\begin{tabular}{l|r|r|r|r|r|r|r|r|r|r}
comp. & $\kappa^\EXP$  & $\kappa^\sAGL $ & $\kappa^\sAGL$  & $\theta^\EXP$  & $\theta_\sDebye^\sAGL$ & $\theta_\sDebye^\sAGL$ & $\theta_\sDebye^\sAEL$ & $\gamma^\EXP$ & $\gamma^\sAGL$ & $\gamma^\sAGL$ \\
      & & & & & ($\theta_\acoustic^\sAGL$) & ($\theta_\acoustic^\sAGL$) & & & &    \\
      & & ($\sigma = 0.25$)~\cite{curtarolo:art96}  & & & ($\sigma = 0.25$)~\cite{curtarolo:art96} & & & & ($\sigma = 0.25$)~\cite{curtarolo:art96} & \\
\hline
LiH & 15~\cite{Morelli_Slack_2006} & 8.58 & 18.6 & 615~\cite{slack, Morelli_Slack_2006} & 743 & 962 & 1175 & 1.28~\cite{slack, Morelli_Slack_2006} & 1.62 & 1.66 \\
      & & & & & (590) & (764) & & & &  \\
LiF & 17.6~\cite{Morelli_Slack_2006} & 8.71 & 9.96 & 500~\cite{slack, Morelli_Slack_2006} & 591 & 617 & 681 & 1.5~\cite{slack, Morelli_Slack_2006} & 2.02 & 2.03 	 \\
      & & & & & (469) &  (490) & & &  & \\
NaF &  18.4~\cite{Morelli_Slack_2006} &  4.52 & 4.67 & 395~\cite{slack, Morelli_Slack_2006} & 411 & 416 & 455 & 1.5~\cite{slack, Morelli_Slack_2006} & 2.2 &  2.21 	  \\
      & & & & & (326) & (330) & & &  & \\
NaCl & 7.1~\cite{Morelli_Slack_2006} & 2.43 & 2.12 & 220~\cite{slack, Morelli_Slack_2006} & 284 & 271 & 289 & 1.56~\cite{slack, Morelli_Slack_2006} & 2.23 & 2.23 	 \\
      & & & & & (225) & (215)  & & & & \\
NaBr & 2.8~\cite{Morelli_Slack_2006} & 1.66 & 1.33 & 150~\cite{slack, Morelli_Slack_2006} & 203 & 188 & 198 & 1.5~\cite{slack, Morelli_Slack_2006} & 2.22 & 2.22 	 \\
      & & & & & (161) & (149) & & & &\\
NaI & 1.8~\cite{Morelli_Slack_2006} & 1.17 & 0.851 & 100~\cite{slack, Morelli_Slack_2006} & 156 & 140 & 147 & 1.56~\cite{slack, Morelli_Slack_2006} & 2.23 & 2.23 	 \\
      & & & & & (124) & (111) & & & &\\
KF & N/A & 2.68 & 2.21 & 235~\cite{slack, Morelli_Slack_2006} & 305 & 288 & 309	& 1.52~\cite{slack, Morelli_Slack_2006} & 2.29 & 2.32 	\\
      & & & & & (242) & (229) & & & &\\
KCl & 7.1~\cite{Morelli_Slack_2006} & 1.4 & 1.25 & 172~\cite{slack, Morelli_Slack_2006} & 220 & 213 & 226 & 1.45~\cite{slack, Morelli_Slack_2006} & 2.38 & 2.40 	 \\
      & & & & & (175) & (169) & & & &\\
KBr & 3.4~\cite{Morelli_Slack_2006} &  1.0 & 0.842 & 117~\cite{slack, Morelli_Slack_2006} & 165 & 156 & 162 & 1.45~\cite{slack, Morelli_Slack_2006} & 2.37 & 2.37 \\
      & & & & & (131) & (124) & & & &\\
KI & 2.6~\cite{Morelli_Slack_2006} & 0.72 & 0.525 & 87~\cite{slack, Morelli_Slack_2006} & 129 & 116 & 120 & 1.45~\cite{slack, Morelli_Slack_2006} & 2.35 & 2.35 	 \\
      & & & & & (102) & (92) & & & &\\
RbCl & 2.8~\cite{Morelli_Slack_2006} & 1.09 & 0.837 & 124~\cite{slack, Morelli_Slack_2006} & 168 & 155 & 160 & 1.45~\cite{slack, Morelli_Slack_2006} & 2.34 & 2.37 	 \\
      & & & & & (133) & (123) & & & &\\
RbBr & 3.8~\cite{Morelli_Slack_2006} & 0.76 & 0.558 & 105~\cite{slack, Morelli_Slack_2006} & 134 & 122 & 129 & 1.45~\cite{slack, Morelli_Slack_2006} & 2.40 & 2.43 \\
      & & & & & (106) & (97) & & & &\\
RbI & 2.3~\cite{Morelli_Slack_2006} & 0.52 & 0.368 & 84~\cite{slack, Morelli_Slack_2006} & 109 & 97 & 102 & 1.41~\cite{slack, Morelli_Slack_2006} & 2.47 & 2.47 	 \\
      & & & & & (87) & (77) & & & &\\
AgCl & 1.0~\cite{Landolt-Bornstein, Maqsood_IJT_2003}  & 2.58 & 0.613 & 124~\cite{slack} & 235 & 145 & 148 & 1.9~\cite{slack} & 2.5 & 2.49 	 \\
      & & & & & (187) & (115) & & & &\\
MgO  & 60~\cite{Morelli_Slack_2006} & 31.9 & 44.5 & 600~\cite{slack, Morelli_Slack_2006} & 758 & 849 & 890	& 1.44~\cite{slack, Morelli_Slack_2006} & 1.95 & 1.96 \\
      & & & & & (602) & (674) & & & &\\
CaO & 27~\cite{Morelli_Slack_2006} & 19.5 & 24.3 & 450~\cite{slack, Morelli_Slack_2006} & 578 & 620 & 638 & 1.57~\cite{slack, Morelli_Slack_2006} & 2.07 & 2.06 	 \\
      & & & & & (459) & (492) & & & &\\
SrO & 12~\cite{Morelli_Slack_2006} & 12.5 & 13.4 & 270~\cite{slack, Morelli_Slack_2006} & 399 & 413 & 421 & 1.52~\cite{slack, Morelli_Slack_2006} & 2.09 & 2.13 	 \\
      & & & & & (317) & (328) &  & & &\\
BaO & 2.3~\cite{Morelli_Slack_2006} & 8.88 & 7.10 & 183~\cite{slack, Morelli_Slack_2006} & 305 & 288 & 292 & 1.5~\cite{slack, Morelli_Slack_2006} & 2.09 & 2.14 \\
      & & & & & (242) & (229) & & & &\\
PbS & 2.9~\cite{Morelli_Slack_2006} & 6.48 & 6.11 & 115~\cite{slack, Morelli_Slack_2006} & 226 & 220 & 221 & 2.0~\cite{slack, Morelli_Slack_2006} & 2.02 & 2.00 	\\
      & & & & & (179) & (175) & & & &\\
PbSe & 2.0~\cite{Morelli_Slack_2006} & 4.88 & 4.81 & 100~\cite{Morelli_Slack_2006} & 197 & 194 & 196 & 1.5~\cite{Morelli_Slack_2006} & 2.1 & 2.07 	 \\
      & & & & & (156) & (154) & & & &\\
PbTe & 2.5~\cite{Morelli_Slack_2006} & 4.15 & 4.07 & 105~\cite{slack, Morelli_Slack_2006} & 170 & 172 & 175 & 1.45~\cite{slack, Morelli_Slack_2006} & 2.04 & 2.09 	 \\
      & & & & & (135) & (137) & & & &\\
SnTe & 1.5~\cite{Snyder_jmatchem_2011} & 4.46 & 5.24 & 155~\cite{Snyder_jmatchem_2011} & 202 & 210 & 212 & 2.1~\cite{Snyder_jmatchem_2011} & 2.15 & 2.11 	 \\
      & & & & & (160) & (167) & & & &\\
\end{tabular}}
\label{tab:art115:rocksalt_thermal}
\etab

The experimental values $\gamma^\EXP$ of the Gr{\"u}neisen parameter are plotted {against
$\gamma^\sAGL$}, $\gamma^\sBM$, $\gamma^\sVIN$ and $\gamma^\sBCN$ in Figure~\ref{fig:art115:rocksalt_thermal_elastic}(f), and the values
are listed in Table~\ref{tab:art115:rocksalt_thermal} and in Table~\ref{tab:art115:rocksalt_thermal_eos}.
These results show that \AGL\ has problems accurately predicting the Gr{\"u}neisen parameter for this set of materials as well, as the calculated values
are often 30\% to 50\% larger than the experimental ones and the \RMSrD\ values are of the order of 0.5. Note also that there are quite large differences between the values
obtained for different equations of state, with $\gamma^\sBCN$ generally having the lowest values while
$\gamma^\sVIN$ has the highest values, a similar pattern to that seen above for the zincblende and diamond structure materials. On the other hand, in contrast to the case of $\theta_\sDebye^\sAGL$,
the value of $\sigma$ used makes little difference to the value of $\gamma^\sAGL$. The correlation values between $\gamma^\EXP$
and the \AGL\ values, as shown in Table~\ref{tab:art115:rocksalt_correlation}, are also quite poor, with values ranging from -0.098 to
0.118 for the Pearson correlations, and negative values for the Spearman correlations.

The experimental thermal conductivity $\kappa^\EXP$ is compared in Figure~\ref{fig:art115:rocksalt_thermal_elastic}(d) to the
thermal conductivities calculated with \AGL\ using the Leibfried-Schl{\"o}mann equation (Equation~\ref{eq:art115:thermal_conductivity}): $\kappa^\sAGL$, $\kappa^\sBM$, $\kappa^\sVIN$ and $\kappa^\sBCN$,
while the values are listed in  Table~\ref{tab:art115:rocksalt_thermal} and in Table~\ref{tab:art115:rocksalt_thermal_eos}.
The linear correlation between the \AGL\ values and $\kappa^\EXP$ is somewhat better than for the zincblende materials set, with a Pearson
correlation as high as $0.94$, although the Spearman correlations are somewhat lower, ranging from $0.445$
to $0.556$. In particular, note that using the $\sigma^\sAEL$ in the \AGL\ calculations improves the correlations by about
2\% to 8\%, from $0.910$ to $0.932$ and from $0.445$ to $0.528$. For the different equations of state, the results for $\kappa^\sBM$
appear to correlate best with $\kappa^\EXP$ for this set of materials.

\tab
\mycaption{Correlations between experimental values and \AEL\ and \AGL\ results for
elastic and thermal properties for rocksalt structure semiconductors.}
\tabvspace
\begin{tabular}{l|r|r|r}
property  & Pearson & Spearman & \RMSrD\ \\
          & (linear) & (rank order) \\
\hline
$\kappa^\EXP$ \vs\ $\kappa^\sAGL$  ($\sigma = 0.25$)~\cite{curtarolo:art96} & 0.910 & 0.445 & 1.093  \\
$\kappa^\EXP$ \vs\ $\kappa^\sAGL$ & 0.932 & 0.528 & 1.002  \\
$\kappa^\EXP$ \vs\ $\kappa^\sBM$ & 0.940 & 0.556 & 1.038   \\
$\kappa^\EXP$ \vs\ $\kappa^\sVIN$ & 0.933 &  0.540 & 0.920  \\
$\kappa^\EXP$ \vs\ $\kappa^\sBCN$ & 0.930 & 0.554 & 1.082  \\
$\theta_\acoustic^\EXP$ \vs\ $\theta_\acoustic^\sAGL$  ($\sigma = 0.25$)~\cite{curtarolo:art96} & 0.985 & 0.948 & 0.253  \\
$\theta_\acoustic^\EXP$ \vs\ $\theta_\acoustic^\sAGL$ & 0.978 & 0.928 & 0.222 \\
$\theta_\acoustic^\EXP$ \vs\ $\theta_\acoustic^\sBM$ & 0.980 & 0.926 & 0.222  \\
$\theta_\acoustic^\EXP$ \vs\ $\theta_\acoustic^\sVIN$ & 0.979 &  0.925 & 0.218 \\
$\theta_\acoustic^\EXP$ \vs\ $\theta_\acoustic^\sBCN$ & 0.978 & 0.929 & 0.225 \\
$\gamma^\EXP$ \vs\ $\gamma^\sAGL$  ($\sigma = 0.25$)~\cite{curtarolo:art96} & 0.118 & -0.064 & 0.477 \\
$\gamma^\EXP$ \vs\ $\gamma^\sAGL$ & 0.036 & -0.110 & 0.486 \\
$\gamma^\EXP$ \vs\ $\gamma^\sBM$ & -0.019 & -0.088 &  0.462 \\
$\gamma^\EXP$ \vs\ $\gamma^\sVIN$ & -0.098 &  -0.086 & 0.591 \\
$\gamma^\EXP$ \vs\ $\gamma^\sBCN$ & 0.023 & -0.110 & 0.443 \\
$B^\EXP$ \vs\ $B_\sVRH^\sAEL$ & 0.998 & 0.995 & 0.078 \\
$B^\EXP$ \vs\ $B_\sStatic^\sAGL$ & 0.998 & 0.993 & 0.201 \\
$B^\EXP$ \vs\ $B_\sStatic^\sBM$ & 0.997 & 0.993 & 0.199  \\
$B^\EXP$ \vs\ $B_\sStatic^\sVIN$ & 0.997 & 0.990 & 0.239  \\
$B^\EXP$ \vs\ $B_\sStatic^\sBCN$ & 0.998 & 0.993 & 0.197 \\
$G^\EXP$ \vs\ $G_\sVRH^\sAEL$ & 0.994 & 0.997 & 0.105 \\
$G^\EXP$ \vs\ $G_\sVoigt^\sAEL$ & 0.991 & 0.990 & 0.157 \\
$G^\EXP$ \vs\ $G_\sReuss^\sAEL$ & 0.995 & 0.995 & 0.142 \\
$\sigma^\EXP$ \vs\ $\sigma^\sAEL$ & 0.959 & 0.827 & 0.070 \\
\end{tabular}
\label{tab:art115:rocksalt_correlation}
\etab

As in the case of the diamond and zincblende structure materials discussed in the previous Section,
Reference~\onlinecite{Morelli_Slack_2006} includes values of the thermal conductivity at 300~K for rocksalt structure materials,
calculated using the experimental values of $\theta_\acoustic$ and $\gamma$ in the Leibfried-Schl{\"o}mann equation, in Table 2.1.
The correlation values of $0.986$ and $0.761$ with experiment are
better than those obtained for the \AGL\ results by a larger margin than for the zincblende materials.
Nevertheless, the Pearson correlation between the calculated and
experimental conductivities is high in both calculations, indicating that the \AGL\
approach may be used as a screening tool for high or low conductivity
compounds in cases where gaps exist in the experimental data for these
materials.

\subsubsection{Hexagonal structure materials}

The experimental data for this set of materials appears in Table III of Reference~\onlinecite{curtarolo:art96}, taken from Table 2.3 of
Reference~\onlinecite{Morelli_Slack_2006}. Most of these materials have the wurtzite structure ($P6_3mc$,\ $\#$186;
Pearson symbol: hP4; \AFLOW\ prototype: {\sf AB\_hP4\_186\_b\_b}~\cite{aflowANRL}\footnote{\url{http://aflow.org/CrystalDatabase/AB_hP4_186_b_b.html}}) except InSe which is $P6_3mmc$,\ $\#$194,
Pearson symbol: hP8.

The calculated elastic properties are shown in Table~\ref{tab:art115:wurzite_elastic} and Figure~\ref{fig:art115:wurzite_thermal_elastic}. The bulk moduli
values obtained from a direct calculation of the elastic tensor, $B_\sVRH^\sAEL$, are usually slightly higher than those obtained from the
$E(V)$ curve and are also closer to experiment (Table~\ref{tab:art115:wurzite_elastic} and Figure~\ref{fig:art115:wurzite_thermal_elastic}(a)), with the exception of
InSe where it is noticeably lower.

\tab
\mycaption[Bulk modulus, shear modulus and Poisson ratio of hexagonal structure semiconductors.]
{``N/A'' = Not available for that source.
Units: $B$ and $G$ in \GPa.}
\tabvspace
\resizebox{\linewidth}{!}{
\begin{tabular}{l|r|r|r|r|r|r|r|r|r|r|r|r|r|r|r}
comp. & $B^\EXP$  & $B_\sVRH^\sAEL$ & $B_\sVRH^\sMP$ & $B_\sStatic^\sAGL$ & $B_\sStatic^\sBM$ & $B_\sStatic^\sVIN$ &  $B_\sStatic^\sBCN$ & $G^\EXP$ & $G_\sVoigt^\sAEL$ & $G_\sReuss^\sAEL$ &  $G_\sVRH^\sAEL$ & $G_\sVRH^\sMP$ & $\sigma^\EXP$ & $\sigma^\sAEL$  & $\sigma^\sMP$     \\
\hline
SiC & 219~\cite{Arlt_ELasticSiC_JAAcS_1965} & 213 & 213 & 204 & 208 & 207 & 207 & 198~\cite{Arlt_ELasticSiC_JAAcS_1965} & 188 & 182 & 185 & 187 & 0.153~\cite{Arlt_ELasticSiC_JAAcS_1965} & 0.163 & 0.16 \\
AlN & 211~\cite{Landolt-Bornstein, McNeil_ElasticAlN_JACerS_1993} & 195 & 194 & 187 & 190 & 189 & 189 & 135~\cite{Landolt-Bornstein, McNeil_ElasticAlN_JACerS_1993} & 123 & 122 & 122 & 122 & 0.237~\cite{Landolt-Bornstein, McNeil_ElasticAlN_JACerS_1993} & 0.241 & 0.24 \\
      & 200~\cite{Dodd_BulkmodAlN_JMS_2001} & & & & & & & 130~\cite{Dodd_BulkmodAlN_JMS_2001} & & & & & 0.234~\cite{Dodd_BulkmodAlN_JMS_2001} &\\
GaN & 195~\cite{Semiconductors_BasicData_Springer, Savastenko_ElasticGaN_PSSa_1978} & 175 & 172 & 166 & 167 & 166 & 168 & 51.6~\cite{Semiconductors_BasicData_Springer, Savastenko_ElasticGaN_PSSa_1978}  & 107 & 105 & 106 & 105 & 0.378~\cite{Semiconductors_BasicData_Springer, Savastenko_ElasticGaN_PSSa_1978}  & 0.248 & 0.25 \\
      & 210~\cite{Polian_ElasticGaN_JAP_1996} & & & & & & & 123~\cite{Polian_ElasticGaN_JAP_1996} & & & & & 0.255~\cite{Polian_ElasticGaN_JAP_1996} & \\
ZnO & 143~\cite{Semiconductors_BasicData_Springer, Kobiakov_ElasticZnOCdS_SSC_1980} & 137 & 130 & 128 & 129 & 127 & 129 & 49.4~\cite{Semiconductors_BasicData_Springer, Kobiakov_ElasticZnOCdS_SSC_1980} & 51.7 & 51.0 & 51.4 & 41.2 & 0.345~\cite{Semiconductors_BasicData_Springer, Kobiakov_ElasticZnOCdS_SSC_1980} & 0.334 & 0.36 \\
BeO & 224.4~\cite{Cline_JAP_1967} & 206 & 208 & 195 & 195 & 192 & 198 & 168~\cite{Cline_JAP_1967} & 157 & 154 & 156 & 156 & 0.201~\cite{Cline_JAP_1967} & 0.198 & 0.2 \\
CdS & 60.7~\cite{Semiconductors_BasicData_Springer, Kobiakov_ElasticZnOCdS_SSC_1980} & 55.4 & 53.3 & 49.7 & 50.3 & 49.4 & 50.6 & 18.2~\cite{Semiconductors_BasicData_Springer, Kobiakov_ElasticZnOCdS_SSC_1980}  & 17.6 & 17.0 & 17.3 & 17.6 & 0.364~\cite{Semiconductors_BasicData_Springer, Kobiakov_ElasticZnOCdS_SSC_1980} & 0.358 & 0.35 \\
InSe & 37.1~\cite{Gatulle_ElasticInSe_PSSb_1983} & 19.2 & N/A & 39.8 & 40.8 & 39.7 & 41.0 & 14.8~\cite{Gatulle_ElasticInSe_PSSb_1983} & 14.9 & 12.3 & 13.6 & N/A & 0.324~\cite{Gatulle_ElasticInSe_PSSb_1983} & 0.214 & N/A \\
InN & 126~\cite{Ueno_BulkmodInN_PRB_1994} & 124 & N/A & 118 & 120 & 119 & 119 & N/A & 55.4 & 54.4 & 54.9 & N/A & N/A & 0.308 & N/A \\
\end{tabular}}
\label{tab:art115:wurzite_elastic}
\etab

\fig
\includegraphics[width=0.98\linewidth]{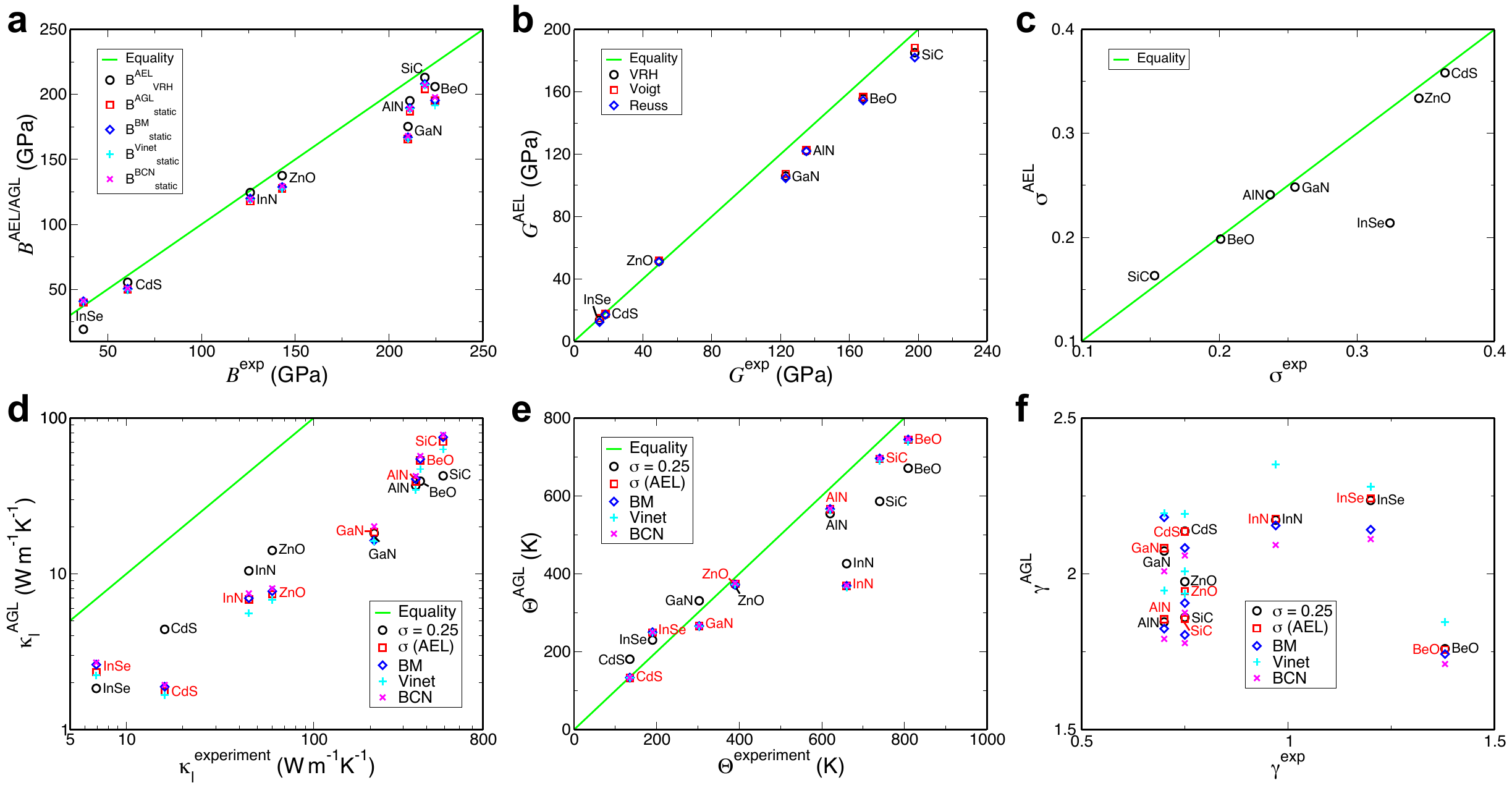}
\mycaption[
({\bf a}) Bulk modulus,
({\bf b}) shear modulus,
({\bf c}) Poisson ratio,
({\bf d}) lattice thermal conductivity,
({\bf e}) Debye temperature and
({\bf f}) Gr{\"u}neisen parameter of hexagonal structure
semiconductors.]
{The Debye temperatures plotted in ({\bf e}) are
$\theta_\acoustic$, except for InSe and InN where $\theta_{\mathrm D}$
values are quoted in References~\onlinecite{Snyder_jmatchem_2011, Ioffe_Inst_DB, Krukowski_jphyschemsolids_1998}.}
\label{fig:art115:wurzite_thermal_elastic}
\efig

For the shear modulus, the experimental values $G^\EXP$  are compared to the \AEL\ values $G_\sVoigt^\sAEL$,
$G_\sReuss^\sAEL$ and $G_\sVRH^\sAEL$. As can be seen in
Table~\ref{tab:art115:wurzite_elastic} and Figure~\ref{fig:art115:wurzite_thermal_elastic}(b), the agreement with the experimental values is very
good. Similarly good agreement is obtained for the Poisson ratio of most materials (Table~\ref{tab:art115:wurzite_elastic}
and Figure~\ref{fig:art115:wurzite_thermal_elastic}(c)), with
a single exception for InSe where the calculation deviates significantly from the experiment.
The Pearson (\ie, linear, Equation~\ref{eq:art115:Pearson}) and Spearman (\ie, rank order, Equation~\ref{eq:art115:Spearman}) correlations between the calculated
elastic properties and their experimental values are generally
quite high (Table~\ref{tab:art115:wurzite_correlation}), ranging from 0.851 and 0.893 respectively for $\sigma^\EXP$ \vs\ $\sigma^\sAEL$, up to 0.998
and 1.0 for $G^\EXP$ \vs\ $G_\sVRH^\sAEL$.

The Materials Project values of $B_\sVRH^\sMP$, $G_\sVRH^\sMP$ and $\sigma^\sMP$ for hexagonal structure materials are also shown in
Table~\ref{tab:art115:wurzite_elastic}, where available. The Pearson correlations values for the experimental results with the available values of
$B_\sVRH^\sMP$, $G_\sVRH^\sMP$ and $\sigma^\sMP$ were calculated to be 0.984, 0.998 and 0.993, respectively, while the respective Spearman correlations
were 0.943, 1.0 and 0.943, and the \RMSrD\ values were 0.117, 0.116 and 0.034. For comparison, the corresponding Pearson correlations for the same
subset of materials for $B_\sVRH^\sAEL$, $G_\sVRH^\sAEL$ and $\sigma^\sAEL$ are 0.986, 0.998, and 0.998 respectively,  while the respective Spearman correlations
were 0.943, 1.0 and 1.0, and the \RMSrD\ values were 0.100, 0.091 and 0.036. These correlation values are very similar, and the general close agreement
for the results for the values of $B_\sVRH^\sAEL$, $G_\sVRH^\sAEL$ and $\sigma^\sAEL$ with those of $B_\sVRH^\sMP$, $G_\sVRH^\sMP$ and $\sigma^\sMP$
demonstrate that the small differences in the parameters used for the \DFT\ calculations make little difference to the results,
indicating that the parameter set used here is robust for high-throughput calculations.

The thermal properties calculated using \AGL\ are
compared to the experimental values in Table~\ref{tab:art115:wurzite_thermal} and are also plotted in Figure~\ref{fig:art115:wurzite_thermal_elastic}.
For the Debye temperature, the $\theta^\EXP$ values taken from Reference~\onlinecite{Morelli_Slack_2006} are for $\theta_\acoustic$,
and are mostly in good agreement with the calculated $\theta_\acoustic^\sAGL$ values.  As in the case of the other materials sets,
the values obtained using the numerical $E(V)$ fit and the three different
equations of state are very similar to each other, whereas $\theta_\sDebye^\sAGL$ calculated using $\sigma=0.25$ differs significantly.
In fact, the values of $\theta_\sDebye^\sAGL$ calculated with $\sigma^\sAEL$ have a lower the correlation with $\theta^\EXP$ than the values calculated with
$\sigma = 0.25$ do, although the \RMSrD\ values are lower when $\sigma^\sAEL$ is used. However, most of this discrepancy appears to be due to the clear
outlier value for the material InN. When the values for this material are removed from the data set, the Pearson correlation values become very similar
when both the $\sigma = 0.25$ and $\sigma = \sigma^\sAEL$ values are used,  increasing to 0.995 and 0.994 respectively.

\tab
\mycaption[Lattice thermal conductivity, Debye temperature and Gr{\"u}neisen parameter
of hexagonal structure semiconductors, comparing the effect of using the
calculated value of the Poisson ratio to the previous approximation of $\sigma = 0.25$.]
{The values listed for $\theta^{\mathrm{exp}}$ are $\theta_\acoustic$,
except 190K for InSe~\cite{Snyder_jmatchem_2011} and 660K for InN~\cite{Ioffe_Inst_DB,Krukowski_jphyschemsolids_1998}
which are $\theta_{\mathrm D}$.
``N/A'' = Not available for that source.
Units: $\kappa$ in \WmK, $\theta$ in \K.}
\tabvspace
\resizebox{\linewidth}{!}{
\begin{tabular}{l|r|r|r|r|r|r|r|r|r|r}
comp. & $\kappa^\EXP$  & $\kappa^\sAGL $ & $\kappa^\sAGL$  & $\theta^\EXP$  & $\theta_\sDebye^\sAGL$ & $\theta_\sDebye^\sAGL$ & $\theta_\sDebye^\sAEL$ & $\gamma^\EXP$ & $\gamma^\sAGL$ & $\gamma^\sAGL$ \\
      & & & & & ($\theta_\acoustic^\sAGL$) & ($\theta_\acoustic^\sAGL$) & & & &    \\
      & & ($\sigma = 0.25$)~\cite{curtarolo:art96}  & & & ($\sigma = 0.25$)~\cite{curtarolo:art96} & & & & ($\sigma = 0.25$)~\cite{curtarolo:art96} & \\
\hline
SiC & 490~\cite{Morelli_Slack_2006} & 42.49 & 70.36 & 740~\cite{Morelli_Slack_2006} & 930 & 1103 & 1138 & 0.75~\cite{Morelli_Slack_2006} & 1.86 & 1.86 \\
      & & & & & (586) & (695) & & & &  \\
AlN & 350~\cite{Morelli_Slack_2006} & 36.73 & 39.0 & 620~\cite{Morelli_Slack_2006} & 880 & 898 & 917 & 0.7~\cite{Morelli_Slack_2006} & 1.85 & 1.85 	 \\
      & & & & & (554) & (566) & & & & \\
GaN &  210~\cite{Morelli_Slack_2006} &  18.17 & 18.54 & 390~\cite{Morelli_Slack_2006} & 592 & 595 & 606 & 0.7~\cite{Morelli_Slack_2006} & 2.07 & 2.08 	 \\
      & & & & & (373) & (375) &  & & & \\
ZnO & 60~\cite{Morelli_Slack_2006} & 14.10 & 7.39 & 303~\cite{Morelli_Slack_2006} & 525 & 422 & 427 & 0.75~\cite{Morelli_Slack_2006} & 1.97 & 1.94 	 \\
      & & & & & (331) & (266) &  & & & \\
BeO & 370~\cite{Morelli_Slack_2006} & 39.26 & 53.36 & 809~\cite{Morelli_Slack_2006} & 1065 & 1181 & 1235 & 1.38~\cite{Slack_JAP_1975, Cline_JAP_1967, Morelli_Slack_2006} & 1.76 & 1.76 	 \\
      & & & & & (671) & (744) & & & & \\
CdS & 16~\cite{Morelli_Slack_2006} & 4.40 & 1.76 & 135~\cite{Morelli_Slack_2006} & 287 & 211 & 213 & 0.75~\cite{Morelli_Slack_2006} & 2.14 & 2.14 	 \\
      & & & & & (181) & (133) & & & & \\
InSe & 6.9~\cite{Snyder_jmatchem_2011} &  1.84 &  2.34 & 190~\cite{Snyder_jmatchem_2011} & 230 & 249 & 168 & 1.2~\cite{Snyder_jmatchem_2011} & 2.24 & 2.24 	 \\
      & & & & & (115) & (125) &  & & & \\
InN & 45~\cite{Ioffe_Inst_DB, Krukowski_jphyschemsolids_1998} & 10.44 & 6.82 & 660~\cite{Ioffe_Inst_DB, Krukowski_jphyschemsolids_1998} & 426 & 369 & 370 & 0.97~\cite{Krukowski_jphyschemsolids_1998} & 2.17 & 2.18 	 \\
      & & & & & (268) & (232)   & & & & \\
\end{tabular}}
\label{tab:art115:wurzite_thermal}
\etab

The experimental and calculated values of the Gr{\"u}neisen parameter are listed in Table~\ref{tab:art115:wurzite_thermal}
and in Table~\ref{tab:art115:wurzite_thermal_eos}, and are plotted in Figure~\ref{fig:art115:wurzite_thermal_elastic}(f).
Again, the Debye model does not reproduce the experimental data, as the calculated values
are often 2 to 3 times too large and the \RMSrD\ is larger than 1.5.
The corresponding correlation, shown in Table~\ref{tab:art115:wurzite_correlation}, are also quite poor, with no value higher than 0.160 for
the Spearman correlations, and negative values for the Pearson correlations.

The comparison between the experimental thermal conductivity $\kappa^\EXP$ and the calculated values is also quite poor
(Figure~\ref{fig:art115:wurzite_thermal_elastic}(d) and Table~\ref{tab:art115:wurzite_thermal}), with \RMSrD\ values of the order of 0.9.
Considerable disagreements also exist between different experimental reports for most materials.
Nevertheless, the Pearson correlations between the \AGL\ calculated thermal conductivity values and the experimental
values are high, ranging from $0.974$ to $0.980$, while the Spearman correlations are even higher, ranging from $0.976$
to $1.0$.

\tab
\mycaption{Correlations between experimental values and \AEL\ and \AGL\ results for
elastic and thermal properties for hexagonal structure semiconductors.}
\begin{tabular}{l|r|r|r}
property  & Pearson & Spearman & \RMSrD\ \\
          & (linear) & (rank order) \\
\hline
$\kappa^\EXP$ \vs\ $\kappa^\sAGL$  ($\sigma = 0.25$)~\cite{curtarolo:art96} & 0.977 & 1.0 & 0.887  \\
$\kappa^\EXP$ \vs\ $\kappa^\sAGL$ & 0.980 & 0.976 & 0.911 \\
$\kappa^\EXP$ \vs\ $\kappa^\sBM$ & 0.974 & 0.976 & 0.904  \\
$\kappa^\EXP$ \vs\ $\kappa^\sVIN$ & 0.980 &  0.976 & 0.926  \\
$\kappa^\EXP$ \vs\ $\kappa^\sBCN$ & 0.980 & 0.976 & 0.895  \\
$\theta_\acoustic^\EXP$ \vs\ $\theta_\acoustic^\sAGL$  ($\sigma = 0.25$)~\cite{curtarolo:art96} & 0.960 & 0.976  & 0.233 \\
$\theta_\acoustic^\EXP$ \vs\ $\theta_\acoustic^\sAGL$ & 0.921 & 0.929 & 0.216 \\
$\theta_\acoustic^\EXP$ \vs\ $\theta_\acoustic^\sBM$ & 0.921 & 0.929 & 0.217  \\
$\theta_\acoustic\EXP$ \vs\ $\theta_\acoustic^\sVIN$ & 0.920 &  0.929 & 0.218 \\
$\theta_\acoustic^\EXP$ \vs\ $\theta_\acoustic^\sBCN$ & 0.921 & 0.929 & 0.216 \\
$\gamma^\EXP$ \vs\ $\gamma^\sAGL$  ($\sigma = 0.25$)~\cite{curtarolo:art96} & -0.039 & 0.160 & 1.566 \\
$\gamma^\EXP$ \vs\ $\gamma^\sAGL$ & -0.029 & 0.160 & 1.563 \\
$\gamma^\EXP$ \vs\ $\gamma^\sBM$ & -0.124 & -0.233 & 1.547  \\
$\gamma^\EXP$ \vs\ $\gamma^\sVIN$ & -0.043 &  0.012 & 1.677 \\
$\gamma^\EXP$ \vs\ $\gamma^\sBCN$ & -0.054 & 0.098 & 1.467 \\
$B^\EXP$ \vs\ $B_\sVRH^\sAEL$ & 0.990 & 0.976 & 0.201 \\
$B^\EXP$ \vs\ $B_\sStatic^\sAGL$ & 0.990 & 0.976 & 0.138 \\
$B^\EXP$ \vs\ $B_\sStatic^\sBM$ & 0.988 & 0.976 & 0.133  \\
$B^\EXP$ \vs\ $B_\sStatic^\sVIN$ & 0.988 & 0.976 & 0.139 \\
$B^\EXP$ \vs\ $B_\sStatic^\sBCN$ & 0.990 & 0.976 & 0.130 \\
$G^\EXP$ \vs\ $G_\sVRH^\sAEL$ & 0.998 & 1.0 & 0.090 \\
$G^\EXP$ \vs\ $G_\sVoigt^\sAEL$ & 0.998 & 1.0 & 0.076 \\
$G^\EXP$ \vs\ $G_\sReuss^\sAEL$ & 0.998 & 1.0 & 0.115 \\
$\sigma^\EXP$ \vs\ $\sigma^\sAEL$ & 0.851 & 0.893 & 0.143 \\
\end{tabular}
\label{tab:art115:wurzite_correlation}
\etab

As for the rocksalt and zincblende material sets, Reference~\onlinecite{Morelli_Slack_2006} (Table 2.3) includes
values of the thermal conductivity at 300~K for wurtzite structure materials, calculated using the
experimental values of the Debye temperature and Gr{\"u}neisen parameter in the Leibfried-Schl{\"o}mann equation.
The Pearson and Spearman correlations are $0.996$ and $1.0$ respectively, which are slightly higher than the correlations obtained using
the \AGL\ calculated quantities. The difference is insignificant since all of these
correlations are very high and
could reliably serve as a screening tool of the thermal conductivity.
However, as we noted in our previous work on \AGL~\cite{curtarolo:art96}, the high correlations calculated with the
experimental $\theta_\acoustic$ and $\gamma$ were obtained using
$\gamma=0.75$ for BeO. Table 2.3 of
Reference~\onlinecite{Morelli_Slack_2006} also cites an alternative value
of $\gamma=1.38$ for BeO (Table~\ref{tab:art115:wurzite_thermal}). Using this outlier
value would severely degrade the results down to $0.7$, for the
Pearson correlation, and $0.829$, for the Spearman correlation.
These values are too low for a reliable screening tool. This
demonstrates the ability of the
\AEL-\AGL\ calculations to compensate for anomalies in the
experimental data when
they exist and still provide a reliable screening method for the
thermal conductivity.

\subsubsection{Rhombohedral materials}

The elastic properties of a few materials with rhombohedral structures
(spacegroups: $R\overline{3}mR$,\ $\#$166, $R\overline{3}mH$,\ $\#$166; Pearson symbol: hR5; \AFLOW\ prototype: {\sf A2B3\_hR5\_166\_c\_ac}~\cite{aflowANRL}\footnote{\url{http://aflow.org/CrystalDatabase/A2B3_hR5_166_c_ac.html}};
and spacegroup: $R\overline{3}cH$,\ $\#$167; Pearson symbol: hR10; \AFLOW\ prototype: {\sf A2B3\_hR10\_167\_c\_e}~\cite{aflowANRL}\footnote{\url{http://aflow.org/CrystalDatabase/A2B3_hR10_167_c_e.html}})
are shown in Table~\ref{tab:art115:rhombo_elastic} (we have left out the material Fe$_2$O$_3$ which was included in
the data set in Table IV of Reference~\onlinecite{curtarolo:art96}, due to convergence issues with some of the
strained structures required for the calculation of the elastic tensor).
The comparison between experiment and calculation is qualitatively reasonable, but the scarcity of experimental results
does not allow for a proper correlation analysis.

\tab
\mycaption[Bulk modulus, shear modulus and Poisson ratio  of rhombohedral semiconductors.]
{``N/A'' = Not available for that source.
Units: $B$ and $G$ in \GPa.}
\tabvspace
\resizebox{\linewidth}{!}{
\begin{tabular}{l|r|r|r|r|r|r|r|r|r|r|r|r|r|r|r}
comp. & $B^\EXP$  & $B_\sVRH^\sAEL$ & $B_\sVRH^\sMP$ & $B_\sStatic^\sAGL$ & $B_\sStatic^\sBM$ & $B_\sStatic^\sVIN$ &  $B_\sStatic^\sBCN$ & $G^\EXP$ & $G_\sVoigt^\sAEL$ & $G_\sReuss^\sAEL$ &  $G_\sVRH^\sAEL$ & $G_\sVRH^\sMP$ & $\sigma^\EXP$ & $\sigma^\sAEL$ & $\sigma^\sMP$      \\
\hline
Bi$_2$Te$_3$ & 37.0~\cite{Semiconductors_BasicData_Springer, Jenkins_ElasticBi2Te3_PRB_1972} & 28.8 & 15.0 & 43.7 & 44.4 & 43.3 & 44.5 & 22.4~\cite{Semiconductors_BasicData_Springer, Jenkins_ElasticBi2Te3_PRB_1972} & 23.5 & 16.3 & 19.9 & 10.9 & 0.248~\cite{Semiconductors_BasicData_Springer, Jenkins_ElasticBi2Te3_PRB_1972} & 0.219 & 0.21 \\
Sb$_2$Te$_3$ & N/A &  22.9 & N/A & 45.3 & 46.0 & 45.2 & 46.0 & N/A & 20.6 & 14.5 & 17.6 & N/A & N/A & 0.195 & N/A \\
Al$_2$O$_3$ & 254~\cite{Goto_ElasticAl2O3_JGPR_1989} & 231 & 232 & 222 & 225 & 224 & 224 & 163.1~\cite{Goto_ElasticAl2O3_JGPR_1989} & 149 & 144 & 147 & 147 & 0.235~\cite{Goto_ElasticAl2O3_JGPR_1989} & 0.238 & 0.24 \\
Cr$_2$O$_3$ & 234~\cite{Alberts_ElasticCr2O3_JMMM_1976} & 203 & 203 & 198 & 202 & 201 & 201 & 129~\cite{Alberts_ElasticCr2O3_JMMM_1976} & 115 & 112 & 113 & 113 & 0.266~\cite{Alberts_ElasticCr2O3_JMMM_1976} & 0.265 & 0.27 \\
Bi$_2$Se$_3$ & N/A & 93.9 & N/A & 57.0 & 57.5 & 56.4 & 57.9 & N/A & 53.7 & 28.0 & 40.9 & N/A & N/A & 0.310 & N/A \\
\end{tabular}}
\label{tab:art115:rhombo_elastic}
\etab

The thermal properties calculated using \AGL\ are
compared to the experimental values in Table~\ref{tab:art115:rhombo_thermal} and the thermal conductivity is also plotted in
Figure~\ref{fig:art115:mixed_thermal}(a).
The experimental Debye temperatures are $\theta_\sDebye$ for Bi$_2$Te$_3$ and Sb$_2$Te$_3$, and
$\theta_\acoustic$ for Al$_2$O$_3$. The values obtained using the numerical $E(V)$ fit and the three different equations of state
(see Table~\ref{tab:art115:rhombo_thermal_eos})
are very similar, but just roughly reproduce the experiments.

\tab
\mycaption[Lattice thermal conductivity, Debye temperatures and Gr{\"u}neisen parameter of rhombohedral
semiconductors, comparing the effect of using the
calculated value of the Poisson ratio to the previous approximation of $\sigma = 0.25$.]
{The experimental Debye temperatures are $\theta_{\mathrm D}$ for
Bi$_2$Te$_3$ and Sb$_2$Te$_3$, and  $\theta_\acoustic$ for Al$_2$O$_3$.
``N/A'' = Not available for that source.
Units: $\kappa$ in \WmK, $\theta$ in \K.}
\tabvspace
\resizebox{\linewidth}{!}{
\begin{tabular}{l|r|r|r|r|r|r|r|r|r|r}
comp. & $\kappa^\EXP$  & $\kappa^\sAGL $ & $\kappa^\sAGL$  & $\theta^\EXP$  & $\theta_\sDebye^\sAGL$ & $\theta_\sDebye^\sAGL$ & $\theta_\sDebye^\sAEL$ & $\gamma^\EXP$ & $\gamma^\sAGL$ & $\gamma^\sAGL$ \\
      & & & & & ($\theta_\acoustic^\sAGL$) & ($\theta_\acoustic^\sAGL$) & & & &    \\
      & & ($\sigma = 0.25$)~\cite{curtarolo:art96}  & & & ($\sigma = 0.25$)~\cite{curtarolo:art96} & & & & ($\sigma = 0.25$)~\cite{curtarolo:art96} & \\
\hline
Bi$_2$Te$_3$ & 1.6~\cite{Snyder_jmatchem_2011} & 2.79 & 3.35 & 155~\cite{Snyder_jmatchem_2011} & 191 & 204 & 161 & 1.49~\cite{Snyder_jmatchem_2011} & 2.13 & 2.14 	 \\
      & & & & & (112) & (119) & & & & \\
Sb$_2$Te$_3$ & 2.4~\cite{Snyder_jmatchem_2011} & 2.90 & 4.46 & 160~\cite{Snyder_jmatchem_2011} & 217 & 243 & 170 & 1.49~\cite{Snyder_jmatchem_2011} & 2.2 & 2.11 	\\
      & & & & & (127) & (142) & & & & \\
Al$_2$O$_3$ & 30~\cite{Slack_PR_1962} &  20.21 & 21.92 & 390~\cite{slack} & 927 & 952 & 975 & 1.32~\cite{slack} & 1.91 & 1.91 	 \\
      & & & & & (430) & (442) & & & & \\
Cr$_2$O$_3$  & 16~\cite{Landolt-Bornstein, Bruce_PRB_1977} & 10.87 & 12.03 & N/A & 733 &  717 & 720 & N/A & 2.26 & 2.10 	\\
      & & & & &  (340) & (333) & & & & \\
Bi$_2$Se$_3$ & 1.34~\cite{Landolt-Bornstein} & 3.60 & 2.41 & N/A & 223 & 199 & 241 & N/A & 2.08 & 2.12 	\\
      & & & & & (130) & (116) & & & & \\
\end{tabular}}
\label{tab:art115:rhombo_thermal}
\etab

The calculated Gr{\"u}neisen parameters are about 50\% larger than the experimental ones, and
the value of $\sigma$ used makes a little difference in the calculation.
The absolute agreement between the \AGL\ values and $\kappa^\EXP$ is also quite poor (Figure~\ref{fig:art115:mixed_thermal}(a)).
However, despite all these discrepancies,
the Pearson correlations between the calculated thermal conductivities and the experimental
values are all high, of the order of $0.998$, while the Spearman correlations range from $0.7$ to $1.0$,
with all of the different equations of state having very similar correlations with experiment.
Using the calculated $\sigma^\sAEL$, \vs\ the rough Cauchy approximation, improves the Spearman correlation from $0.7$ to $1.0$.

\fig
\includegraphics[width=\linewidth]{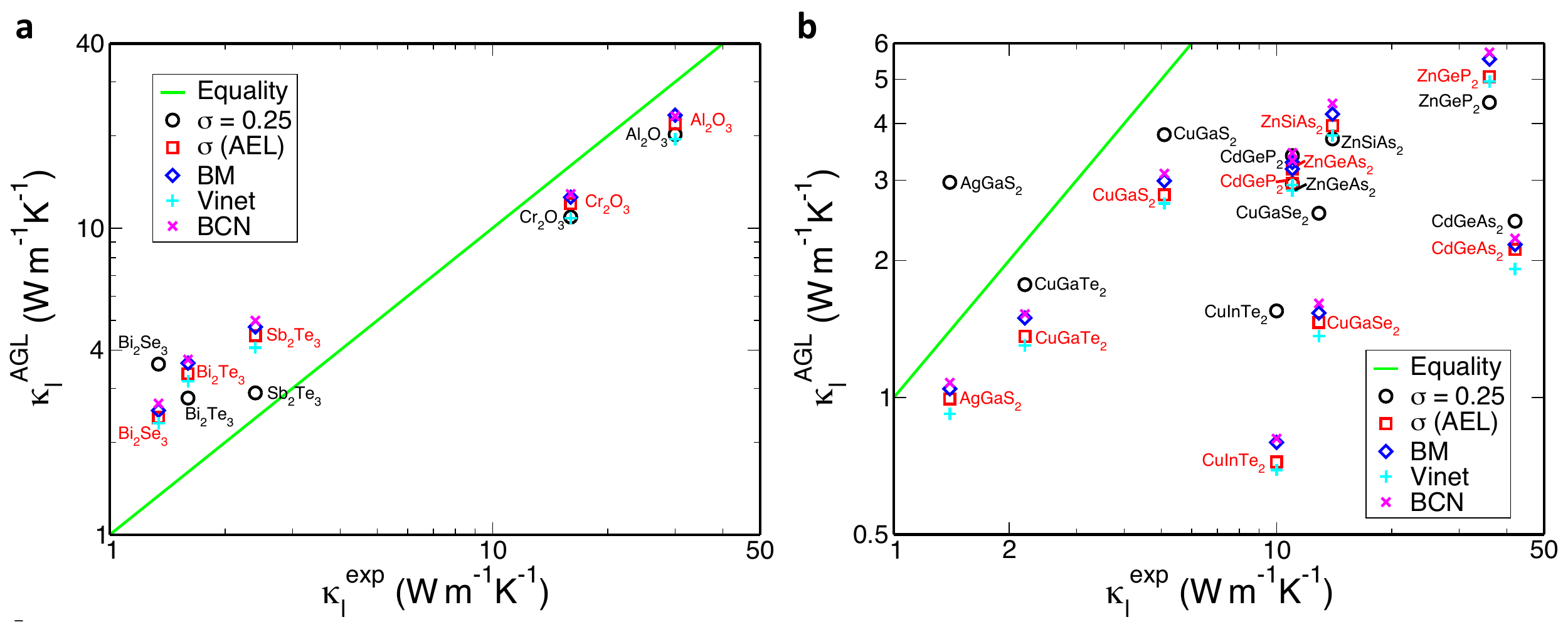}
\mycaption{
({\bf a})  Lattice thermal conductivity of rhombohedral semiconductors at 300~K.
({\bf b})  Lattice thermal conductivity of body-centered tetragonal semiconductors at 300~K.
}
\label{fig:art115:mixed_thermal}
\efig

\tab
\mycaption{Correlations between experimental values and \AEL\ and \AGL\ results for
elastic and thermal properties for rhombohedral structure semiconductors.}
\tabvspace
\begin{tabular}{l|r|r|r}
property  & Pearson & Spearman & \RMSrD\ \\
          & (linear) & (rank order) \\
\hline
$\kappa^\EXP$ \vs\ $\kappa^\sAGL$  ($\sigma = 0.25$)~\cite{curtarolo:art96} & 0.997 & 0.7 & 0.955  \\
$\kappa^\EXP$ \vs\ $\kappa^\sAGL$ & 0.998 & 1.0 & 0.821 \\
$\kappa^\EXP$ \vs\ $\kappa^\sBM$ & 0.997 & 1.0 & 0.931  \\
$\kappa^\EXP$ \vs\ $\kappa^\sVIN$ & 0.998 &  1.0 & 0.741 \\
$\kappa^\EXP$ \vs\ $\kappa^\sBCN$ & 0.997 & 1.0 & 1.002 \\
\end{tabular}
\label{tab:art115:rhombo_correlation}
\etab

\subsubsection{Body-centered tetragonal materials}

\tab
\mycaption[Bulk modulus, shear modulus and Poisson ratio of body-centered tetragonal semiconductors.]
{Note that there appears to be an error in Table 1 of Reference~\onlinecite{Fernandez_ElasticCuInTe_PSSa_1990}
where the bulk modulus values are stated to be in units of $10^{12}$ Pa.
This seems unlikely, as that would give a bulk modulus for CuInTe$_2$ an order of magnitude larger than
that for diamond.
Also, units of $10^{12}$ Pa would be inconsistent with the experimental results listed in
Reference~\onlinecite{Neumann_ElasticCuInTe_PSSa_1986},
so therefore it seems that these values are in units of
$10^{10}$ Pa, which are the values shown here.
``N/A'' = Not available for that source.
Units: $B$ and $G$ {in} \GPa.}
\tabvspace
\resizebox{\linewidth}{!}{
\begin{tabular}{l|r|r|r|r|r|r|r|r|r|r|r|r|r|r|r}
comp. & $B^\EXP$  & $B_\sVRH^\sAEL$ & $B_\sVRH^\sMP$ & $B_\sStatic^\sAGL$ & $B_\sStatic^\sBM$ & $B_\sStatic^\sVIN$ &  $B_\sStatic^\sBCN$ & $G^\EXP$ & $G_\sVoigt^\sAEL$ & $G_\sReuss^\sAEL$ & $G_\sVRH^\sAEL$ & $G_\sVRH^\sMP$ & $\sigma^\EXP$ & $\sigma^\sAEL$  & $\sigma^\sMP$     \\
\hline
CuGaTe$_2$ & N/A & 47.0 & N/A & 42.5 & 43.2 & 42.0 & 43.5 & N/A & 25.1 & 22.1 & 23.6 & N/A & N/A & 0.285 & N/A \\
ZnGeP$_2$ & N/A & 73.1 & 74.9 & 70.1 & 71.1 & 70.0 & 71.4 & N/A & 50.5 & 46.2 & 48.4 & 48.9 & N/A & 0.229 & 0.23 \\
ZnSiAs$_2$ & N/A & 67.4 & 65.9 & 63.4 & 64.3 & 63.1 & 64.6 & N/A & 44.4 & 40.4 & 42.4 & 42.2 & N/A & 0.240 & 0.24 \\
CuInTe$_2$ & 36.0~\cite{Neumann_ElasticCuInTe_PSSa_1986} & 53.9 & N/A & 38.6 & 39.2 & 38.2 & 39.4 & N/A & 20.4 & 17.2 & 18.8 & N/A & 0.313~\cite{Fernandez_ElasticCuInTe_PSSa_1990} & 0.344 & N/A \\
      & 45.4~\cite{Fernandez_ElasticCuInTe_PSSa_1990} & & & & & & & & & & & & & \\
AgGaS$_2$ & 67.0~\cite{Grimsditch_ElasticAgGaS2_PRB_1975} & 70.3 & N/A & 56.2 & 57.1 & 56.0 & 57.4 & 20.8~\cite{Grimsditch_ElasticAgGaS2_PRB_1975} & 20.7 & 17.4 & 19.1 & N/A & 0.359~\cite{Grimsditch_ElasticAgGaS2_PRB_1975} & 0.375 & N/A \\
CdGeP$_2$ & N/A & 65.3 & 65.2 & 60.7 & 61.6 & 60.4 & 61.9  & N/A & 37.7 & 33.3 & 35.5 & 35.0 & N/A & 0.270 & 0.27 \\
CdGeAs$_2$ & 69.9~\cite{Hailing_ElasticCdGeAs_JPCSS_1982} & 52.6 & N/A & 49.2 & 49.6 & 48.3 & 49.9 & 29.5~\cite{Hailing_ElasticCdGeAs_JPCSS_1982} & 30.9 & 26.2 & 28.6 & N/A & 0.315~\cite{Hailing_ElasticCdGeAs_JPCSS_1982} & 0.270 & N/A \\
CuGaS$_2$ & 94.0~\cite{Bettini_ElasticCuGaS_SSC_1975} & 73.3 & N/A & 69.0 & 69.9 & 68.7 & 70.6 & N/A & 37.8 & 32.4 & 35.1 & N/A & N/A  & 0.293 & N/A  \\
CuGaSe$_2$ & N/A & 69.9 & N/A & 54.9 & 55.6 & 54.4 & 56.0 & N/A & 30.3 & 26.0 & 28.1 & N/A & N/A & 0.322 & N/A \\
ZnGeAs$_2$ & N/A & 59.0 & N/A & 56.2 & 56.7 & 55.5 & 57.1 & N/A & 39.0 & 35.6 & 37.3 & N/A & N/A & 0.239 & N/A \\
\end{tabular}}
\label{tab:art115:bct_elastic}
\etab

The mechanical properties of the body-centered tetragonal materials (spacegroup:
$I\overline{4}2d$,\ $\#$122; Pearson symbol: tI16; \AFLOW\ prototype: {\sf ABC2\_tI16\_122\_a\_b\_d}~\cite{aflowANRL}\footnote{\url{http://aflow.org/CrystalDatabase/ABC2_tI16_122_a_b_d.html}})
of Table V of Reference~\onlinecite{curtarolo:art96} are reported in Table~\ref{tab:art115:bct_elastic}.
The calculated bulk moduli miss considerably the few available experimental results, while the shear moduli
are well reproduced. Reasonable estimates are also obtained for the Poisson ratio.

The thermal properties are reported in Table~\ref{tab:art115:bct_thermal} and Figure~\ref{fig:art115:mixed_thermal}(b).
The $\theta^\EXP$ values are all for $\theta_\sDebye$, and in most cases are in good agreement with the values obtained
with the \AEL\ calculated $\sigma$. The
values from the numerical $E(V)$ fit and the three different equations of state are again very similar, but differ significantly
from {$\theta_\sDebye^\sAGL$} calculated with $\sigma=0.25$.

\tab
\mycaption[Lattice thermal conductivity at 300~K, Debye temperatures and Gr{\"u}neisen parameter of  body-centered tetragonal
semiconductors, comparing the effect of using the
calculated value of the Poisson ratio to the previous approximation of $\sigma = 0.25$.]
{``N/A'' = Not available for that source.
Units: $\kappa$ in \WmK, $\theta$ in \K.}
\tabvspace
\resizebox{\linewidth}{!}{
\begin{tabular}{l|r|r|r|r|r|r|r|r|r|r}
comp. & $\kappa^\EXP$  & $\kappa^\sAGL $ & $\kappa^\sAGL$  & $\theta^\EXP$  & $\theta_\sDebye^\sAGL$ & $\theta_\sDebye^\sAGL$ & $\theta_\sDebye^\sAEL$ & $\gamma^\EXP$ & $\gamma^\sAGL$ & $\gamma^\sAGL$ \\
      & & & & & ($\theta_\acoustic^\sAGL$) & ($\theta_\acoustic^\sAGL$) & & & &    \\
      & & ($\sigma = 0.25$)~\cite{curtarolo:art96}  & & & ($\sigma = 0.25$)~\cite{curtarolo:art96} & & & & ($\sigma = 0.25$)~\cite{curtarolo:art96} & \\
\hline
CuGaTe$_2$ & 2.2~\cite{Snyder_jmatchem_2011} & 1.77 & 1.36 & 226~\cite{Snyder_jmatchem_2011} & 234 & 215 & 218 & 1.46~\cite{Snyder_jmatchem_2011} & 2.32 & 2.32 	 \\
      & & & & & (117) & (108) & & & & \\
ZnGeP$_2$ & 35~\cite{Landolt-Bornstein, Beasley_AO_1994} & 4.45 & 5.07 & 500~\cite{Landolt-Bornstein} & 390 & 408 & 411 & N/A & 2.13 & 2.14 	   \\
      & 36~\cite{Landolt-Bornstein, Beasley_AO_1994} & & & 428~\cite{Abrahams_JCP_1975} & (195) & (204) & & & & \\
      & 18~\cite{Landolt-Bornstein, Shay_1975, Masumoto_JPCS_1966} & & & & & & & & & \\
ZnSiAs$_2$ & 14\cite{Landolt-Bornstein, Shay_1975, Masumoto_JPCS_1966} & 3.70 & 3.96  & 347~\cite{Landolt-Bornstein, Bohnhammel_PSSa_1981} & 342 & 350 & 354 & N/A & 2.15 & 2.15  	 \\
      & & & & & (171) & (175) & & & & \\
CuInTe$_2$ & 10\cite{Landolt-Bornstein, Rincon_PSSa_1995} & 1.55 & 0.722 & 185~\cite{Landolt-Bornstein, Rincon_PSSa_1995} & 215 & 166 & 185 & 0.93~\cite{Rincon_PSSa_1995} & 2.33 & 2.32 	 \\
      & & & & 195~\cite{Landolt-Bornstein, Bohnhammel_PSSa_1982}  & (108) & (83) & & & &\\
AgGaS$_2$ & 1.4\cite{Landolt-Bornstein, Beasley_AO_1994} & 2.97 & 0.993 & 255~\cite{Landolt-Bornstein, Abrahams_JCP_1975} & 324 & 224 & 237 & N/A & 2.20 & 2.20 	\\
      & & & & & (162) & (112) & & & & \\
CdGeP$_2$ & 11~\cite{Landolt-Bornstein, Shay_1975, Masumoto_JPCS_1966} & 3.40 & 2.96 & 340~\cite{Landolt-Bornstein, Abrahams_JCP_1975} & 335 & 320 & 324 & N/A & 2.20 & 2.21 \\
      & & & & & (168) & (160) & & & & \\
CdGeAs$_2$ & 42~\cite{Landolt-Bornstein, Shay_1975} & 2.44 & 2.11 & 241~\cite{Bohnhammel_PSSa_1981} & 266 & 254 & 255 & N/A & 2.20 & 2.20 	\\
      & & & & & (133) & (127) & & & &\\
CuGaS$_2$ & 5.09~\cite{Landolt-Bornstein} & 3.78 & 2.79 & 356~\cite{Landolt-Bornstein, Abrahams_JCP_1975} & 387 & 349 & 349 & N/A & 2.24 & 2.24 	 \\
      & & & & & (194) & (175) & & & &\\
CuGaSe$_2$ & 12.9~\cite{Landolt-Bornstein, Rincon_PSSa_1995} & 2.54 & 1.46 & 262~\cite{Landolt-Bornstein, Bohnhammel_PSSa_1982} & 294 & 244 & 265 & N/A & 2.27 & 2.26 	 \\
      & & & & & (147) & (122) & & & &\\
ZnGeAs$_2$ & 11\cite{Landolt-Bornstein, Shay_1975} & 2.95 & 3.18 & N/A & 299 & 307 & 308 & N/A & 2.16 & 2.17 	 \\
      & & & & & (150) & (154) & & & &\\
\end{tabular}}
\label{tab:art115:bct_thermal}
\etab

The comparison of the experimental thermal conductivity $\kappa^\EXP$ to the calculated values, in Figure~\ref{fig:art115:mixed_thermal}(b),
shows poor reproducibility. The available data can thus only be considered a rough indication of their order of magnitude.
The Pearson and Spearman correlations are also quite low for all types of calculation,
but somewhat better when the calculated $\sigma^\sAEL$ is used instead of the Cauchy approximation.

\tab
\mycaption{Correlations between experimental values and \AEL\ and \AGL\ results for
elastic and thermal properties for body-centered tetragonal structure semiconductors.}
\tabvspace
\begin{tabular}{l|r|r|r}
property  & Pearson & Spearman & \RMSrD\ \\
          & (linear) & (rank order) \\
\hline
$\kappa^\EXP$ \vs\ $\kappa^\sAGL$  ($\sigma = 0.25$)~\cite{curtarolo:art96} & 0.265 & 0.201 & 0.812 \\
$\kappa^\EXP$ \vs\ $\kappa^\sAGL$ & 0.472 & 0.608 & 0.766 \\
$\kappa^\EXP$ \vs\ $\kappa^\sBM$ & 0.467 & 0.608 & 0.750 \\
$\kappa^\EXP$ \vs\ $\kappa^\sVIN$ & 0.464 &  0.608 & 0.778 \\
$\kappa^\EXP$ \vs\ $\kappa^\sBCN$ & 0.460 & 0.608 & 0.741 \\
\end{tabular}
\label{tab:art115:bct_correlation}
\etab

\subsubsection{Miscellaneous materials}

In this Section we consider materials with various other structures, as in Table VI of Reference~\onlinecite{curtarolo:art96}:
CoSb$_3$ and IrSb$_3$
(spacegroup: $Im\overline{3}$,\ $\#$204; Pearson symbol: cI32; \AFLOW\ prototype: {\sf A3B\_cI32\_204\_g\_c}~\cite{aflowANRL}\footnote{\url{http://aflow.org/CrystalDatabase/A3B_cI32_204_g_c.html}}),
ZnSb ($Pbca$,\ $\#$61; oP16; \AFLOW\ prototype: {\sf AB\_oP16\_61\_c\_c}~\cite{aflowANRL}\footnote{\url{http://aflow.org/CrystalDatabase/AB_oP16_61_c_c.html}}),
Sb$_2$O$_3$ ($Pccn$,\ $\#$56; oP20), InTe ($Pm\overline{3}m$,\ $\#$221; cP2; \AFLOW\ prototype: {\sf AB\_cP2\_221\_b\_a}~\cite{aflowANRL}\footnote{\url{http://aflow.org/CrystalDatabase/AB_cP2_221_b_a.html}},
and $I4/mcm$,\ $\#$140; tI16), Bi$_2$O$_3$  ($P121/c1,\ \#14$; mP20); and SnO$_2$ ($P42/mnm,\ \#136$; tP6; {\sf A2B\_tP6\_136\_f\_a}~\cite{aflowANRL}\footnote{\url{http://aflow.org/CrystalDatabase/A2B_tP6_136_f_a.html}}).
Two different structures are listed for InTe. In Reference~\onlinecite{curtarolo:art96}, we
considered its simple cubic structure, but this is a high-pressure phase~\cite{Chattopadhyay_BulkModInTe_JPCS_1985}, while the ambient
pressure phase is body-centered tetragonal. It appears that the thermal conductivity results should be for the body-centered tetragonal
phase~\cite{Spitzer_JPCS_1970}, therefore both sets of results are reported here. The correlation values shown in the tables below
were calculated for the body-centered tetragonal structure.

The elastic properties are shown
in Table~\ref{tab:art115:misc_elastic}. Large discrepancies appear between the results of all calculations
and the few available experimental results.

\tab
\mycaption[Bulk modulus, shear modulus and Poisson ratio of materials with various
structures.]
{``N/A'' = Not available for that source.
Units: $B$ and $G$ {in} \GPa.}
\tabvspace
\resizebox{\linewidth}{!}{
\begin{tabular}{l|r|r|r|r|r|r|r|r|r|r|r|r|r|r|r|r}
comp. & Pearson  & $B^\EXP$  & $B_\sVRH^\sAEL$ & $B_\sVRH^\sMP$ & $B_\sStatic^\sAGL$ & $B_\sStatic^\sBM$ & $B_\sStatic^\sVIN$ &  $B_\sStatic^\sBCN$ & $G^\EXP$ & $G_\sVoigt^\sAEL$ & $G_\sReuss^\sAEL$ &  $G_\sVRH^\sAEL$ &  $G_\sVRH^\sMP$ & $\sigma^\EXP$ & $\sigma^\sAEL$ & $\sigma^\sMP$      \\
\hline
CoSb$_3$ & $cI32$ & N/A & 78.6 & 82.9 & 75.6 & 76.1 & 75.1 & 76.3 & N/A & 57.2 & 55.1 & 56.2 & 57.0 & N/A & 0.211 & 0.22 \\
IrSb$_3$ & $cI32$ & N/A & 97.5 & 98.7 & 94.3 & 94.8 & 93.8 & 95.5 & N/A & 60.9 & 59.4 & 60.1 & 59.7& N/A & 0.244 & 0.25 \\
ZnSb & $oP16$ & N/A & 47.7 & 47.8 & 46.7 & 47.0 & 46.0 & 47.7 & N/A & 29.2 & 27.0 & 28.1 & 28.2 & N/A & 0.253 & 0.25 \\
Sb$_2$O$_3$ & $oP20$ & N/A & 16.5 & 19.1 & 97.8 & 98.7 & 97.8 & 98.7 & N/A & 22.8 & 16.4 & 19.6 & 20.4 & N/A & 0.0749 & 0.11 \\
InTe & $cP2$ & 90.2~\cite{Chattopadhyay_BulkModInTe_JPCS_1985} & 41.7 & N/A & 34.9 & 34.4 & 33.6 & 34.7 & N/A & 8.41 & 8.31 & 8.36 & N/A& N/A & 0.406 & N/A \\
InTe & $tI16$ & 46.5~\cite{Chattopadhyay_BulkModInTe_JPCS_1985} & 20.9 & N/A & 32.3 & 33.1 & 32.2 & 33.2 & N/A & 13.4 & 13.0 & 13.2 & N/A & N/A & 0.239 & N/A \\
Bi$_2$O$_3$ & $mP20$ & N/A & 48.0 & 54.5 & 108 & 110 & 109 & 109 & N/A & 30.3 & 25.9 & 28.1 & 29.9 & N/A & 0.255 & 0.27 \\
SnO$_2$ & $tP6$ & 212~\cite{Chang_ElasticSnO2_JGPR_1975} & 159 & N/A & 158 & 162 & 161 & 161 & 106~\cite{Chang_ElasticSnO2_JGPR_1975} & 86.7 & 65.7 & 76.2 & N/A & 0.285~\cite{Chang_ElasticSnO2_JGPR_1975} & 0.293 & N/A \\
\end{tabular}}
\label{tab:art115:misc_elastic}
\etab

The thermal properties are
compared to the experimental values in Table~\ref{tab:art115:misc_thermal}.
The experimental Debye temperatures are for $\theta_\sDebye$, except ZnSb for which it is $\theta_\acoustic$. Good agreement
is found between calculation and the few available experimental values. Again, the numerical $E(V)$ fit and the three different
equations of state give similar results.
For the Gr{\"u}neisen parameter, experiment and calculations again differ considerably, while the changes due to the different
values of $\sigma$ used in the
calculations are negligible.

\tab
\mycaption[Lattice thermal conductivity at 300~K, Debye temperatures
and Gr{\"u}neisen parameter of materials with various structures, comparing the effect of using the
calculated value of the Poisson ratio to the previous approximation of $\sigma = 0.25$.]
{The experimental Debye temperatures are $\theta_{\mathrm D}$,
except ZnSb for which it is $\theta_\acoustic$.
``N/A'' = Not available for that source.
Units: $\kappa$ in \WmK, $\theta$ in \K.}
\tabvspace
\resizebox{\linewidth}{!}{
\begin{tabular}{l|r|r|r|r|r|r|r|r|r|r|r}
comp. & Pearson  & $\kappa^\EXP$  & $\kappa^\sAGL $ & $\kappa^\sAGL$  & $\theta^\EXP$  & $\theta_\sDebye^\sAGL$ & $\theta_\sDebye^\sAGL$ & $\theta_\sDebye^\sAEL$ & $\gamma^\EXP$ & $\gamma^\sAGL$ & $\gamma^\sAGL$ \\
      & & & & & & ($\theta_\acoustic^\sAGL$) & ($\theta_\acoustic^\sAGL$) & & & &    \\
      & & & ($\sigma = 0.25$)~\cite{curtarolo:art96}  & & & ($\sigma = 0.25$)~\cite{curtarolo:art96} & & & & ($\sigma = 0.25$)~\cite{curtarolo:art96} & \\
\hline
CoSb$_3$ & $cI32$ & 10~\cite{Snyder_jmatchem_2011} & 1.60 & 2.60 & 307~\cite{Snyder_jmatchem_2011} & 284 & 310 & 312 & 0.95~\cite{Snyder_jmatchem_2011} & 2.63 & 2.33 \\
      & & & & & & (113) & (123) & & & & \\
IrSb$_3$ & $cI32$ & 16~\cite{Snyder_jmatchem_2011} & 2.64 & 2.73 & 308~\cite{Snyder_jmatchem_2011} & 283 & 286 & 286 & 1.42~\cite{Snyder_jmatchem_2011} & 2.34 & 2.34  \\
      & & & & & & (112) & (113) & & & & \\
ZnSb & $oP16$ &  3.5~\cite{Madsen_PRB_2014, Bottger_JEM_2010} &  1.24 & 1.23 & 92~\cite{Madsen_PRB_2014} & 244 & 242 & 237 &  0.76~\cite{Madsen_PRB_2014, Bottger_JEM_2010} &  2.24 & 2.23 	 \\
      & & & & & & (97) & (96) & & & & \\
Sb$_2$O$_3$ & $oP20$ & 0.4~\cite{Landolt-Bornstein} & 3.45 & 8.74 & N/A & 418 & 572 & 238 & N/A & 2.13 & 2.12 	\\
      & & & & & & (154) & (211) & & & & \\
InTe & $cP2$ & N/A & 3.12 & 0.709 & N/A & 191 & 113 & 116 & N/A & 2.28 & 2.19 	\\
      & & & & & & (152) & (90) & & & & \\
InTe & $tP16$ & 1.7~\cite{Snyder_jmatchem_2011, Spitzer_JPCS_1970} & 1.32 & 1.40 & 186~\cite{Snyder_jmatchem_2011} & 189 & 193 & 150 & 1.0~\cite{Snyder_jmatchem_2011} & 2.23 & 2.24 	\\
      & & & & & & (95) & (97) & & & & \\
Bi$_2$O$_3$ & $mP20$ & 0.8~\cite{Landolt-Bornstein} & 3.04 & 2.98 & N/A & 345 & 342 & 223 & N/A & 2.10 & 2.10 	 \\
      & & & & & & (127) & (126) & & & & \\
SnO$_2$ & $tP6$ & 98~\cite{Turkes_jpcss_1980} & 9.56 & 6.98 & N/A & 541 & 487 & 480 & N/A & 2.48 & 2.42 	 \\
      &  & 55~\cite{Turkes_jpcss_1980} & & & & (298) & (268) & & & & \\
\end{tabular}}
\label{tab:art115:misc_thermal}
\etab

The experimental thermal conductivity $\kappa^\EXP$ is compared in Table~\ref{tab:art115:misc_thermal} to the thermal conductivity
calculated with \AGL\ using the
Leibfried-Schl{\"o}mann equation (Equation~\ref{eq:art115:thermal_conductivity}) for $\kappa^\sAGL$, while the values obtained for $\kappa^\sBM$, $\kappa^\sVIN$
and $\kappa^\sBCN$ are listed in Table~\ref{tab:art115:misc_thermal_eos}.
The absolute agreement between the \AGL\ values and $\kappa^\EXP$ is quite poor.
The scarcity of
experimental data from different sources
on the thermal properties of these materials prevents reaching definite conclusions regarding the true values of these
properties. The available data can thus
only be considered as a rough indication of their order of magnitude.

\tab
\mycaption{Correlations between experimental values and \AEL\ and \AGL\ results for
elastic and thermal properties for materials with miscellaneous structures.}
\tabvspace
\begin{tabular}{l|r|r|r}
property  & Pearson & Spearman & \RMSrD\ \\
          & (linear) & (rank order) \\
\hline
$\kappa^\EXP$ \vs\ $\kappa^\sAGL$  ($\sigma = 0.25$)~\cite{curtarolo:art96} & 0.937 & 0.071 & 3.38 \\
$\kappa^\EXP$ \vs\ $\kappa^\sAGL$ & 0.438 & -0.143 & 8.61 \\
$\kappa^\EXP$ \vs\ $\kappa^\sBM$ & 0.498 & -0.143 &  8.81 \\
$\kappa^\EXP$ \vs\ $\kappa^\sVIN$ & 0.445 &  0.0 & 8.01 \\
$\kappa^\EXP$ \vs\ $\kappa^\sBCN$ & 0.525 & -0.143 & 9.08 \\
\end{tabular}
\label{tab:art115:misc_correlation}
\etab

For these materials, the Pearson  correlation between the calculated
and experimental values of the thermal conductivity ranges from $0.438$ to $0.937$, while the corresponding
Spearman correlations range from $-0.143$ to $0.071$. In this case, using $\sigma^\sAEL$ in the \AGL\
calculations does not improve the correlations, instead actually lowering the values somewhat.
However, it should be noted that the Pearson correlation is heavily influenced by the values for SnO$_2$.
When this entry is removed from the list, the Pearson correlation values fall to $-0.471$ and $-0.466$
when the $\sigma = 0.25$ and $\sigma = \sigma^\sAEL$ values are used, respectively.
The low correlation values, particularly for the Spearman correlation, for this set of materials demonstrates the
importance of the information about the material structure when interpreting results obtained using the \AGL\ method
in order to identify candidate materials for specific thermal applications. This is partly due to the fact that the Gr{\"u}neisen
parameter values tend to be similar for materials with the same
structure. Therefore, the effect of the Gr{\"u}neisen parameter on the ordinal ranking of
the lattice thermal conductivity of materials with the same structure
is small.

\subsubsection{Thermomechanical properties from LDA}

{
The thermomechanical properties of a randomly-selected subset of the materials investigated in this work were calculated using \LDA\
in order to check the impact of the choice of exchange-correlation functional on the results. For the \LDA\ calculations, all structures were
first re-relaxed using the \LDA\ exchange-correlation functional with \VASP\ using the appropriate parameters and potentials as
described in the \AFLOW\ standard~\cite{curtarolo:art104}, and then the appropriate strained structures were calculated using \LDA.
These calculations were restricted to a subset of materials to limit the total number of additional first-principles calculations required, and the materials were
selected randomly from each of the sets in the previous sections so as to cover as wide a range of different structure types as possible, given the available experimental data.
Results for elastic properties obtained using \LDA, \GGA\ and experimental measurements are shown in Table~\ref{tab:art115:LDA_elastic}, while the thermal properties are shown in
Table~\ref{tab:art115:LDA_thermal}. All thermal properties listed in Table~\ref{tab:art115:LDA_thermal} were calculated using $\sigma^\sAEL$ in the expression
for the Debye temperature.}

{
In general, the \LDA\ values for elastic and thermal properties are slightly higher than the \GGA\ values, as would be generally expected
due to their relative tendencies to overbind and underbind, respectively~\cite{He_GGA_LDA_PRB_2014, Saadaoui_GGA_LDA_EPJB_2015}.
The correlations and \RMSrD\ of both the \LDA\ and \GGA\ results with experiment for this set of materials are listed in Table~\ref{tab:art115:LDA_correlation}.
The Pearson and Spearman correlation values for \LDA\ and \GGA\ are very close to each other for most of the listed properties. The \RMSrD\ values show
greater differences, although it isn't clear that one of the exchange-correlation functionals consistently gives better predictions than the other.
Therefore, the choice of exchange-correlation functional will make little difference to the predictive capability of the workflow, so we choose to
use \GGA-\PBE\ as it is the functional used for performing the structural relaxation for the entries in the \AFLOW\ data repository.
}

\tab
\mycaption[Bulk modulus, shear modulus and Poisson ratio of a subset of the materials investigated in this work,
comparing the effect of using different exchange-correlation functionals.]
{``N/A'' = Not available for that source.
Units: $B$ and $G$ in \GPa.}
\tabvspace
\resizebox{\linewidth}{!}{
\begin{tabular}{l|r|r|r|r|r|r|r|r|r|r|r|r|r|r|r}
comp. & $B^\EXP$  & $B_\sVRH^\sGGA$ & $B_\sVRH^\sLDA$ & $B_\sStatic^\sGGA$ & $B_\sStatic^\sLDA$  & $G^\EXP$ & $G_\sVoigt^\sGGA$ & $G_\sVoigt^\sLDA$ & $G_\sReuss^\sGGA$ & $G_\sReuss^\sLDA$ &  $G_\sVRH^\sGGA$ & $G_\sVRH^\sLDA$ & $\sigma^\EXP$ & $\sigma^\sGGA$ & $\sigma^\sLDA$      \\
\hline
Si & 97.8~\cite{Semiconductors_BasicData_Springer, Hall_ElasticSi_PR_1967} & 89.1& 96.9 & 84.2 & 92.1 & 66.5~\cite{Semiconductors_BasicData_Springer, Hall_ElasticSi_PR_1967} & 64 & 65 & 61 & 61.9 & 62.5 & 63.4 & 0.223~\cite{Semiconductors_BasicData_Springer, Hall_ElasticSi_PR_1967} & 0.216 & 0.231  \\
BN & 367.0~\cite{Lam_BulkMod_PRB_1987} & 372 & 402 & 353 & 382 & N/A & 387 & 411 & 374 & 395 & 380 & 403 & N/A & 0.119 & 0.124 \\
GaSb & 57.0~\cite{Lam_BulkMod_PRB_1987} & 47.0 & 58.3 & 41.6 & 52.3 & 34.2~\cite{Boyle_ElasticGaPSb_PRB_1975} & 30.8 & 35.3 & 28.3 & 32.2 & 29.6 & 33.7  &  0.248~\cite{Boyle_ElasticGaPSb_PRB_1975} & 0.240 & 0.258 \\
InAs & 60.0~\cite{Lam_BulkMod_PRB_1987} & 50.1 & 62.3 & 45.7 & 57.4 & 29.5~\cite{Semiconductors_BasicData_Springer, Gerlich_ElasticAlSb_JAP_1963} & 27.3 & 30.1 & 24.2 & 26.4 & 25.7 & 28.2 & 0.282~\cite{Semiconductors_BasicData_Springer, Gerlich_ElasticAlSb_JAP_1963} & 0.281 & 0.303 \\
ZnS & 77.1~\cite{Lam_BulkMod_PRB_1987} & 71.2 & 88.4 & 65.8 & 83.3 & 30.9~\cite{Semiconductors_BasicData_Springer} & 36.5 & 42.1 & 31.4 & 35.7 & 33.9 & 38.9 & 0.318~\cite{Semiconductors_BasicData_Springer} & 0.294 & 0.308 \\
NaCl & 25.1~\cite{Haussuhl_ElasticRocksalt_ZP_1960} & 24.9 & 33.3 & 20.0 & 27.6 & 14.6~\cite{Haussuhl_ElasticRocksalt_ZP_1960} & 14.0 & 19.8 & 12.9 & 16.6 & 13.5 & 18.2 & 0.255~\cite{Haussuhl_ElasticRocksalt_ZP_1960} & 0.271 & 0.269 \\
KI & 12.2~\cite{Haussuhl_ElasticRocksalt_ZP_1960} & 10.9 & 16.3 & 8.54 & 13.3 & 5.96~\cite{Haussuhl_ElasticRocksalt_ZP_1960} & 6.05 & 9.39 & 4.39 & 5.3 & 5.22 & 7.35 & 0.290~\cite{Haussuhl_ElasticRocksalt_ZP_1960} & 0.294 & 0.305 \\
RbI & 11.1~\cite{Haussuhl_ElasticRocksalt_ZP_1960} & 9.90 & 14.8 & 8.01 & 12.1 & 5.03~\cite{Haussuhl_ElasticRocksalt_ZP_1960} & 5.50 & 8.54 & 3.65 & 3.94 & 4.57 & 6.24 & 0.303~\cite{Haussuhl_ElasticRocksalt_ZP_1960} & 0.300 & 0.315 \\
MgO  & 164~\cite{Sumino_ElasticMgO_JPE_1976} & 152 & 164 & 142 & 163 & 131~\cite{Sumino_ElasticMgO_JPE_1976}  & 119 & 138 & 115 & 136 & 117 & 137 & 0.185~\cite{Sumino_ElasticMgO_JPE_1976} & 0.194 & 0.173 \\
CaO & 113~\cite{Chang_ElasticCaSrBaO_JPCS_1977} & 105 & 129 & 99.6 & 122 & 81.0~\cite{Chang_ElasticCaSrBaO_JPCS_1977} & 73.7 & 87.4 & 73.7 & 86.3 & 73.7 & 86.9 & 0.210~\cite{Chang_ElasticCaSrBaO_JPCS_1977} & 0.216 & 0.225 \\
GaN & 195~\cite{Semiconductors_BasicData_Springer, Savastenko_ElasticGaN_PSSa_1978} & 175 & 202 & 166 & 196 & 51.6~\cite{Semiconductors_BasicData_Springer, Savastenko_ElasticGaN_PSSa_1978}  & 107 & 116 & 105 & 113 & 106 & 114 & 0.378~\cite{Semiconductors_BasicData_Springer, Savastenko_ElasticGaN_PSSa_1978}  & 0.248 & 0.262 \\
     & 210~\cite{Polian_ElasticGaN_JAP_1996} & & & & & 123~\cite{Polian_ElasticGaN_JAP_1996} & & & & & & & 0.255~\cite{Polian_ElasticGaN_JAP_1996} & & \\
CdS & 60.7~\cite{Semiconductors_BasicData_Springer, Kobiakov_ElasticZnOCdS_SSC_1980} & 55.4 & 68.2 & 49.7 & 64.1 & 18.2~\cite{Semiconductors_BasicData_Springer, Kobiakov_ElasticZnOCdS_SSC_1980}  & 17.6 & 18.4 & 17.0 & 17.8 & 17.3 & 18.1 & 0.364~\cite{Semiconductors_BasicData_Springer, Kobiakov_ElasticZnOCdS_SSC_1980} & 0.358 & 0.378 \\
Al$_2$O$_3$ & 254~\cite{Goto_ElasticAl2O3_JGPR_1989} & 231 & 259 & 222 & 250 & 163.1~\cite{Goto_ElasticAl2O3_JGPR_1989} & 149 & 166 & 144 & 163 & 147 & 165 & 0.235~\cite{Goto_ElasticAl2O3_JGPR_1989} & 0.238 & 0.238 \\
CdGeP$_2$ & N/A & 65.3 & 78.4 & 60.7 & 74.5 & N/A & 37.7 & 42.1 & 33.3 & 36.8 & 35.5 & 39.4 & N/A & 0.270 & 0.285 \\
CuGaSe$_2$ & N/A & 69.9 & 76.4 & 54.9 & 72.1 & N/A & 30.3 & 34.7 & 26.0 & 30.0 & 28.1 & 32.3  & N/A & 0.322 & 0.315 \\
CoSb$_3$ & N/A & 78.6 & 99.6 & 75.6 & 96.1 & N/A & 57.2 & 67.1 & 55.1 & 64.2 & 56.2 & 65.7 & N/A & 0.211 & 0.23 \\
\end{tabular}}
\label{tab:art115:LDA_elastic}
\etab

\tab
\mycaption[Thermal properties lattice thermal conductivity at
300~K, Debye temperature and Gr{\"u}neisen parameter of
a subset of materials, comparing the effect of using different exchange-correlation functionals.]
{The values listed for $\theta^{\mathrm{exp}}$ are $\theta_\acoustic$, except 340~K for CdGeP$_2$~\cite{Landolt-Bornstein, Abrahams_JCP_1975}, 262K for CuGaSe$_2$
\cite{Landolt-Bornstein, Bohnhammel_PSSa_1982} and 307~K for CoSb$_3$~\cite{Snyder_jmatchem_2011} which are $\theta_{\mathrm D}$.
Units: $\kappa$ in \WmK, $\theta$ in \K.}
\tabvspace
\resizebox{\linewidth}{!}{
\begin{tabular}{l|r|r|r|r|r|r|r|r|r}
comp. & $\kappa^\EXP$  & $\kappa^\sGGA $ & $\kappa^\sLDA$ & $\theta^\EXP$  & $\theta_\sDebye^\sGGA$ & $\theta_\sDebye^\sLDA$ & $\gamma^\EXP$ & $\gamma^\sGGA$ & $\gamma^\sLDA$  \\
      & & & & &  ($\theta_\acoustic^\sGGA$) & ($\theta_\acoustic^\sLDA$) & & \\
\hline
Si & 166~\cite{Morelli_Slack_2006} & 26.19 & 27.23 & 395~\cite{slack, Morelli_Slack_2006} & 610 & 614 & 1.06~\cite{Morelli_Slack_2006} & 2.06 & 2.03	 \\
      & & & & & (484) & (487) & 0.56~\cite{slack} &  \\
BN & 760~\cite{Morelli_Slack_2006} & 281.6 & 312.9 & 1200~\cite{Morelli_Slack_2006} & 1793 & 1840 & 0.7~\cite{Morelli_Slack_2006} & 1.75 & 1.72	\\
      & & & & & (1423) & (1460) & & & \\
GaSb & 40~\cite{Morelli_Slack_2006} & 4.96 & 5.89 & 165~\cite{slack, Morelli_Slack_2006} & 240 & 254 & 0.75~\cite{slack, Morelli_Slack_2006} & 2.28 & 2.25 	 \\
      & & & & & (190) & (202) & & & \\
InAs & 30~\cite{Morelli_Slack_2006} & 4.33 & 4.92 & 165~\cite{slack, Morelli_Slack_2006} & 229 & 238 & 0.57~\cite{slack, Morelli_Slack_2006} & 2.26 & 2.22	 \\
      & & & & & (182) & (189) & & & \\
ZnS & 27~\cite{Morelli_Slack_2006} & 8.38 & 9.58 & 230~\cite{slack, Morelli_Slack_2006} & 341 & 363 & 0.75~\cite{slack, Morelli_Slack_2006} & 2.00 & 2.02 	 \\
        & & & & & (271) & (288) & & & \\
NaCl & 7.1~\cite{Morelli_Slack_2006} & 2.12 & 2.92 & 220~\cite{slack, Morelli_Slack_2006} & 271 & 312 & 1.56~\cite{slack, Morelli_Slack_2006} & 2.23 & 2.29 	 \\
      & & & & & (215) & (248) & & & \\
KI & 2.6~\cite{Morelli_Slack_2006} & 0.525 & 0.811 & 87~\cite{slack, Morelli_Slack_2006} & 116 & 137 & 1.45~\cite{slack, Morelli_Slack_2006} & 2.35 & 2.37 	 \\
      & & & & & (92) & (109) & & & \\
RbI & 2.3~\cite{Morelli_Slack_2006} & 0.368 & 0.593 & 84~\cite{slack, Morelli_Slack_2006} & 97 & 115 & 1.41~\cite{slack, Morelli_Slack_2006} & 2.47 & 2.45 	 \\
      & & & & & (77) & (91) & & & \\
MgO  & 60~\cite{Morelli_Slack_2006} & 44.5 & 58.4 & 600~\cite{slack, Morelli_Slack_2006} & 849 & 935 & 1.44~\cite{slack, Morelli_Slack_2006} & 1.96 & 1.95 \\
      & & & & & (674) & (742) & & & \\
CaO & 27~\cite{Morelli_Slack_2006} & 24.3 & 28.5 & 450~\cite{slack, Morelli_Slack_2006} & 620 & 665 & 1.57~\cite{slack, Morelli_Slack_2006} & 2.06 & 2.09 	 \\
      & & & & & (492) & (528) & & & \\
GaN &  210~\cite{Morelli_Slack_2006} & 18.54 & 21.34 & 390~\cite{Morelli_Slack_2006} & 595 & 619 & 0.7~\cite{Morelli_Slack_2006} & 2.08 & 2.04 	 \\
      & & & & & (375) & (390) & & &  \\
CdS & 16~\cite{Morelli_Slack_2006} & 1.76 & 1.84 & 135~\cite{Morelli_Slack_2006} & 211 & 217 & 0.75~\cite{Morelli_Slack_2006} & 2.14 & 2.14 	 \\
      & & & & & (133) & (137) & & &  \\
Al$_2$O$_3$ & 30~\cite{Slack_PR_1962} & 21.92 & 25.36 & 390~\cite{slack} & 952 & 1002 & 1.32~\cite{slack} & 1.91 & 1.91 	 \\
      & & & & & (442) & (465) & & & \\
CdGeP$_2$ & 11~\cite{Landolt-Bornstein, Shay_1975, Masumoto_JPCS_1966} & 2.96 & 3.47 & 340~\cite{Landolt-Bornstein, Abrahams_JCP_1975} & 320 & 337 & N/A & 2.21 & 2.18 \\
      & & & & & (160) & (169) & & & \\
CuGaSe$_2$ & 12.9~\cite{Landolt-Bornstein, Rincon_PSSa_1995} & 1.46 & 2.23 & 262~\cite{Landolt-Bornstein, Bohnhammel_PSSa_1982} & 244 & 281 & N/A & 2.26 & 2.23 	 \\
      & & & & & (122) & (141) & & & \\
CoSb$_3$ & 10~\cite{Snyder_jmatchem_2011} & 2.60 & 3.25 & 307~\cite{Snyder_jmatchem_2011} & 310 & 332 & 0.95~\cite{Snyder_jmatchem_2011} & 2.33 & 2.28 \\
      & & & & & (123) & (132) & & & \\
\end{tabular}}
\label{tab:art115:LDA_thermal}
\etab

\tab
\mycaption{Correlations between experimental values and \AEL\ and \AGL\ results for
elastic and thermal properties comparing the \LDA\ and \GGA\ exchange-correlation functionals
for this subset of materials.}
\tabvspace
\begin{tabular}{l|r|r|r}
property  & Pearson & Spearman & \RMSrD\ \\
          & (linear) & (rank order) \\
\hline
$\kappa^\EXP$ \vs\ $\kappa^\sGGA$ & 0.963 & 0.867 & 0.755  \\
$\kappa^\EXP$ \vs\ $\kappa^\sLDA$ & 0.959 & 0.848 & 0.706 \\
$\theta^\EXP$ \vs\ $\theta^\sGGA$ & 0.996 & 0.996  & 0.119 \\
$\theta^\EXP$ \vs\ $\theta^\sLDA$ & 0.996 & 0.996 & 0.174 \\
$\gamma^\EXP$ \vs\ $\gamma^\sGGA$ & 0.172 & 0.130 & 1.514 \\
$\gamma^\EXP$ \vs\ $\gamma^\sLDA$ & 0.265 & 0.296 & 1.490 \\
$B^\EXP$ \vs\ $B_\sVRH^\sGGA$ & 0.995 & 1.0 & 0.111 \\
$B^\EXP$ \vs\ $B_\sVRH^\sLDA$ & 0.996 & 1.0 & 0.185 \\
$B^\EXP$ \vs\ $B_\sStatic^\sGGA$ & 0.996 & 1.0 & 0.205 \\
$B^\EXP$ \vs\ $B_\sStatic^\sLDA$ & 0.998 & 1.0 & 0.072 \\
$G^\EXP$ \vs\ $G_\sVRH^\sGGA$ & 0.999 & 0.993 & 0.108 \\
$G^\EXP$ \vs\ $G_\sVRH^\sLDA$ & 0.997 & 0.986 & 0.153 \\
$G^\EXP$ \vs\ $G_\sVoigt^\sGGA$ & 0.998 & 0.993 & 0.096 \\
$G^\EXP$ \vs\ $G_\sVoigt^\sLDA$ & 0.996 & 0.986 & 0.315 \\
$G^\EXP$ \vs\ $G_\sReuss^\sGGA$ & 0.999 & 0.993 & 0.163 \\
$G^\EXP$ \vs\ $G_\sReuss^\sLDA$ & 0.997 & 0.993 & 0.111 \\
$\sigma^\EXP$ \vs\ $\sigma^\sGGA$ & 0.982 & 0.986 & 0.037 \\
$\sigma^\EXP$ \vs\ $\sigma^\sLDA$ & 0.983 & 0.993 & 0.052 \\
\end{tabular}
\label{tab:art115:LDA_correlation}
\etab

\subsubsection{AGL predictions for thermal conductivity}

The \AEL-\AGL\ methodology has been applied for
high-throughput screening of the elastic and thermal properties of
over 3000 materials included in the  \AFLOW\ database~\cite{aflowAPI}.
Tables~\ref{tab:art115:highkappa} and \ref{tab:art115:lowkappa} {list those} found
to have the highest and lowest thermal conductivities, respectively.
The high conductivity list is unsurprisingly dominated by various phases of elemental
carbon{, boron nitride, boron carbide and boron carbon nitride,} while {all other}
high-conductivity materials also contain at least one of the elements C, B or N.

\tab
\mycaption[Materials from \AFLOW\ database with highest thermal conductivities as predicted using
the \AEL-\AGL\ methodology.]
{The \AFLOW\ \underline{u}nique \underline{id}entifier (\AUID) is a permanent, server-independent identifier for each entry in the  \AFLOW\ database~\cite{aflowAPI}.
This identifier allows any of these entries to be retrieved from the repository, and ensures the retrievability and reproducibility of the data
irrespective of changes in the underlying database structure or hosting location.
Units: $\kappa$ in \WmK.}
\tabvspace
\begin{tabular}{l|r|r|r|r}
comp. & Pearson & space group \# & $\kappa^\sAGL $  & \AUID    \\
\hline
C & cF8 & 227 & 420 & 3ab7e139e1c29c9f \\
BN & cF8 & 216 & 282 & fd5539a4f79db51c \\
C & hP4 & 194 & 272 & 440c4eee274b61b6  \\
C & tI8 & 139 & 206 & b2688e84030188b8 \\
BC$_2$N & oP4 & 25 & 188 & c0e7523ff8d34297 \\
BN & hP4 & 186 & 178 & 56d00a95d21b5c3a \\
C & hP8 & 194 & 167 & c42dc8ec018245e5 \\
C & cI16 & 206 & 162 & c969067f8a3bbde9 \\
C & oS16 & 65 & 147 & bdc82cca41c811c6 \\
C & mS16 & 12 & 145 & a59baaad49eb5ab9 \\
BC$_7$ & tP8 & 115 & 145 & 0401731cb29df494 \\
BC$_5$ & oI12 & 44 & 137 & f759c5600121a9e9 \\
Be$_2$C & cF12 & 225 & 129 & 378e092c24555651 \\
CN$_2$ & tI6 & 119 & 127 & 6852d98ddee59417 \\
C & hP12 & 194 & 127 &	bd79f9fa8154aa95 \\
BC$_7$ & oP8 & 25 & 125 & 4d13f06b9fe563ef \\
B$_2$C$_4$N$_2$ & oP8 & 17 & 120 & 9e325d34d65bd890\\
MnB$_2$ & hP3 & 191 & 117 & 0e5997687be5d3dc \\
C & hP4 & 194 & 117 & 2be120d88682ee01 \\
SiC & cF8 & 216 & 113 & 2cab0c35952c733f \\
TiB$_2$ & hP3 & 191 & 110 & 32d72b1701a0a640 \\
AlN & cF8 & 225 & 107 & 06c4f5b0f1a49096 \\
BP & cF8 & 216 & 105 & 598a7a7328a47d85 \\
C & hP16 & 194 & 105 &	c9d6a8b917d502f0 \\
VN & hP2 & 187 & 101 &	aa89372868af03a8 \\
\end{tabular}
\label{tab:art115:highkappa}
\etab

The low thermal conductivity list tends to contain materials
with large unit cells and heavier elements such as Hg, Tl, Pb and Au.

\tab
\mycaption[Materials from \AFLOW\ database with lowest thermal conductivities as predicted using
the \AEL-\AGL\ methodology.]
{The \AFLOW\ \underline{u}nique \underline{id}entifier (\AUID) is a permanent, server-independent identifier for each entry in the  \AFLOW\ database~\cite{aflowAPI}.
This identifier allows any of these entries to be retrieved from the repository, and ensures the retrievability and reproducibility of the data
irrespective of changes in the underlying database structure or hosting location.
Units: $\kappa$ in \WmK.}
\tabvspace
\begin{tabular}{l|r|r|r|r}
comp. & Pearson & space group \# & $\kappa^\sAGL $  & \AUID    \\
\hline
Hg$_{33}$Rb$_3$ & cP36 & 221 & 0.0113 & 3a84e674e05ac4e6 \\
Hg$_{33}$K$_3$ & cP36 & 221 & 0.0116 & ac7610d35123f5c5 \\
Cs$_6$Hg$_{40}$ & cP46 & 223 & 0.0136 & 978182b72d30a019 \\
Ca$_{16}$Hg$_{36}$ & cP52 & 215 & 0.0751 & fe8eeb1e2af8df90 \\
CrTe & cF8 & 216 & 0.081 & 53c8683bd5648144 \\
Hg$_4$K$_2$ & oI12 & 74 & 0.086 & 50b2883feb14cd6e \\
Sb$_6$Tl$_{21}$ & cI54 & 229 & 0.089 & f7933008a130dc06 \\
Se & cF24 & 227 & 0.093 & 7d6a2e6c742211e5 \\
Cs$_8$I$_{24}$Sn$_4$ & cF36 & 225 & 0.104 &  460691dc51cf5b5d \\
Ag$_2$Cr$_4$Te$_8$ & cF56 & 227 & 0.107 & a30bbe2831fa8a18 \\
AsCdLi & cF12 & 216 &	0.116 & f818510c8952b114 \\
Au$_{36}$In$_{16}$ & cP52 & 215 & 0.117 & bda82cdcf87fa384 \\
Cd$_3$In & cP4 & 221 & 0.128 & 3bc3fc68c58fdd1f \\
AuLiSb & cF12 & 216 & 0.130 & bdab7ec2c162ee22 \\
K$_5$Pb$_{24}$ & cI58 & 217 & 0.135 & 58f4471901eff079 \\
K$_8$Sn$_{46}$ & cP54 & 223 & 0.142 & 6b4795df74caacfc \\
Au$_7$Cd$_{16}$Na$_6$ & cF116 & 225 & 0.145 & ec21f32abca24cbd \\
Cs & cI2 & 229 & 0.148 & 5acbf212d1783298 \\
Cs$_8$Pb$_4$Cl$_{24}$ & cF36 & 225 & 0.157 & 84738cad161f83b3 \\
Au$_{4}$In$_8$Na$_{12}$ & cF96 & 227 & 0.158 & 0393c62d375f5ec6\\
SeTl & cP2 & 221 & 0.164 & 5ebc0f014499d22b \\
Cd$_{33}$Na$_6$ & cP39 & 200 & 0.166 & 0e4a5c866567f309 \\
Au$_{18}$In$_{15}$Na$_6$ & cP39 & 200 & 0.168 & f7355e2e7474fb1c \\
Cd$_{26}$Cs$_2$ & cF112 & 226 & 0.173 & cfe1448550ccd1d1 \\
Ag$_2$I$_2$ & hP4 & 186 & 0.192 & d611e813a85efcb0 \\
\end{tabular}
\label{tab:art115:lowkappa}
\etab

By combining the \AFLOW\ search for thermal conductivity values with other properties such as chemical, electronic or structural factors,
candidate materials for specific engineering applications can be rapidly identified for further in-depth analysis using more accurate
computational methods and for experimental examination. {The full set of thermomechanical properties calculated using
\AEL-\AGL\ for over 3500 entries can be accessed online at \AFLOWorg~\cite{aflowlib.org}, which incorporates search and sort functionality to
generate customized lists of materials.}

\subsubsection{Results for different equations of state}

This section includes the results for the thermal conductivity, Debye temperature and the Gr{\"u}neisen parameter for the set of 74 materials listed in this work as
calculated using the Birch-Murnaghan~\cite{Birch_Elastic_JAP_1938, Poirier_Earth_Interior_2000, Blanco_CPC_GIBBS_2004}, Vinet~\cite{Vinet_EoS_JPCM_1989, Blanco_CPC_GIBBS_2004},
and Baonza-C{\'a}ceres-N{\'u}{\~n}ez~\cite{Baonza_EoS_PRB_1995, Blanco_CPC_GIBBS_2004} equations of state. The experimental values for the lattice thermal conductivity
$\kappa^\EXP$ are compared to the \AGL\ values obtained using the numerical polynomial fit $\kappa^\sAGL$, and the three empirical equations of state:
Birch-Murnaghan, $\kappa^\sBM$; Vinet, $\kappa^\sVIN$; and Baonza-C{\'a}ceres-N{\'u}{\~n}ez, $\kappa^\sBCN$. The experimental values for the Debye temperature
$\theta^\EXP$ are compared to the \AGL\ values obtained using the numerical polynomial fit $\theta_\sDebye^\sAGL$, and the three empirical equations of state:
Birch-Murnaghan, $\theta_\sDebye^\sBM$; Vinet, $\theta_\sDebye^\sVIN$; and Baonza-C{\'a}ceres-N{\'u}{\~n}ez, $\theta_\sDebye^\sBCN$. The \AGL\ values listed are for
$\theta_\sDebye$, while the values for $\theta_\acoustic$ are listed underneath in parentheses. The experimental values for the Gr{\"u}neisen parameter
$\gamma^\EXP$ are compared to the \AGL\ values obtained using the numerical polynomial fit $\gamma^\sAGL$, and the three empirical equations of state:
Birch-Murnaghan, $\gamma_\sDebye^\sBM$; Vinet, $\gamma_\sDebye^\sVIN$; and Baonza-C{\'a}ceres-N{\'u}{\~n}ez, $\gamma_\sDebye^\sBCN$. The results for the
diamond and zincblende structure set of materials are listed in Table~\ref{tab:art115:zincblende_thermal_eos}, the results for the rocksalt structure set of materials
are listed in Table~\ref{tab:art115:rocksalt_thermal_eos}, the results for the hexagonal structure set of materials are listed in Table~\ref{tab:art115:wurzite_thermal_eos}, the results
for the rhombohedral structure set of materials are listed in Table~\ref{tab:art115:rhombo_thermal_eos}, the results for the body-centered tetragonal structure set of materials
are listed in Table~\ref{tab:art115:bct_thermal_eos}, and the results for the miscellaneous structure materials are listed in Table~\ref{tab:art115:misc_thermal_eos}.

\tab
\mycaption[Thermal conductivities, Debye temperatures and Gr{\"u}neisen parameters of
zincblende and diamond structure semiconductors, calculated using the different equations of state.]
{The values listed for $\theta^\EXP$ are
$\theta_\acoustic$, except 141K for HgTe which is $\theta_{\mathrm D}$~\cite{Snyder_jmatchem_2011}.
Units: $\kappa$ in \WmK, $\theta$ in \K.}
\tabvspace
\resizebox{\linewidth}{!}{
\begin{tabular}{l|r|r|r|r|r|r|r|r|r|r|r|r|r|r|r}
comp. & $\kappa^\EXP$  & $\kappa^\sAGL$ & $\kappa^\sBM$ & $\kappa^\sVIN$ & $\kappa^\sBCN$  & $\theta^\EXP$  & $\theta_\sDebye^\sAGL$ & $\theta_\sDebye^\sBM$ & $\theta_\sDebye^\sVIN$ & $\theta_\sDebye^\sBCN$ & $\gamma^\EXP$ & $\gamma^\sAGL$ &  $\gamma^\sBM$ &  $\gamma^\sVIN$  &  $\gamma^\sBCN$  \\
& & & & & & & ($\theta_\acoustic^\sAGL$) & ($\theta_\acoustic^\sBM$) & ($\theta_\acoustic^\sVIN$) & ($\theta_\acoustic^\sBCN$) & & & & & \\
\hline
C &  3000~\cite{Morelli_Slack_2006} & 419.9 & 298.5 & 307.0 & 466.8 & 1450~\cite{slack, Morelli_Slack_2006} & 2094 & 2051 & 2056 & 2103 & 0.75~\cite{Morelli_Slack_2006}  & 1.77 & 2.01 &	1.99 & 1.69 \\
 & & & & & & & (1662) & (1628) & (1632) & (1669) & 0.9~\cite{slack} & & &  & \\
SiC & 360~\cite{Ioffe_Inst_DB} & 113.0 & 120.7 & 101.3 & 125.4 & 740~\cite{slack} & 1106 & 1108 & 1100 & 1110 & 0.76~\cite{slack} & 1.85 & 1.80 & 1.93 & 1.78 	\\
& & & & & & & (878) & (879) & (873) & (881) & & & & & \\
Si & 166~\cite{Morelli_Slack_2006} & 26.19 & 28.61 & 26.23 & 31.39 & 395~\cite{slack, Morelli_Slack_2006} & 610 & 611 & 609 & 614 & 1.06~\cite{Morelli_Slack_2006} & 2.06 & 1.99 & 2.06 & 1.92	 \\
 & & & & & & & (484) & (485) & (483) & (487) & 0.56~\cite{slack} &  & & &  \\
Ge &  65~\cite{Morelli_Slack_2006}  & 8.74 & 9.61 & 8.54 & 10.12 & 235~\cite{slack, Morelli_Slack_2006} & 329 & 330 & 329 & 331 & 1.06~\cite{Morelli_Slack_2006}  & 2.31 & 2.22 & 2.34 & 2.18	 \\
& & & & & & & (261) & (262) & (261) & (263)  & 0.76~\cite{slack} & &  & &  \\
BN & 760~\cite{Morelli_Slack_2006} & 281.6 & 243.1 & 220.5 & 303.4 & 1200~\cite{Morelli_Slack_2006} & 1793 & 1777 & 1769 & 1798 & 0.7~\cite{Morelli_Slack_2006} & 1.75 & 1.85 & 1.92 & 1.70	\\
& & & & & & & (1423) & (1410) & (1404) & (1427)  & & & & &  \\
BP & 350~\cite{Morelli_Slack_2006} & 105.0 & 108.8 & 89.95 & 117.8 & 670~\cite{slack, Morelli_Slack_2006} & 1025 & 1025 & 1016 & 1029 & 0.75~\cite{Morelli_Slack_2006} & 1.79 & 1.76 & 1.90 & 1.71 	\\
& & & & & & & (814) & (814) & (806) & (817)  & & & & & \\
AlP & 90~\cite{Landolt-Bornstein, Spitzer_JPCS_1970} & 19.34 & 20.48 & 18.79 & 22.49 & 381~\cite{Morelli_Slack_2006} & 525 & 526 & 524 & 528 & 0.75~\cite{Morelli_Slack_2006} & 1.96 & 1.92 & 1.98 & 1.84 	 \\
& & & & & & & (417) & (417) & (416) & (419)  & & & & & \\
AlAs & 98~\cite{Morelli_Slack_2006} & 11.64 & 12.84 & 11.59 & 13.64 & 270~\cite{slack, Morelli_Slack_2006} & 373 & 374 & 373 & 375 & 0.66~\cite{slack, Morelli_Slack_2006} & 2.04 & 1.96 & 2.04 & 1.91 	 \\
& & & & & & & (296) & (297) & (296) & (298)  & & & & & \\
AlSb & 56~\cite{Morelli_Slack_2006} & 6.83 & 7.84 & 6.85 & 8.34 & 210~\cite{slack, Morelli_Slack_2006}  & 276 & 277 & 276 & 278 & 0.6~\cite{slack, Morelli_Slack_2006} & 2.13 & 2.01 & 2.12 & 1.96	 \\
& & & & & & & (219) & (220) & (219) & (221)  & & & & & \\
GaP & 100~\cite{Morelli_Slack_2006} & 13.34 & 15.09 & 13.49 & 15.74 & 275~\cite{slack, Morelli_Slack_2006} & 412 & 414 & 412 & 414 & 0.75~\cite{Morelli_Slack_2006} & 2.15 & 2.04 & 2.14 & 2.0	\\
 & & & & & & & (327) & (329) & (327) & (329) &  0.76~\cite{slack} &  &  & &  \\
GaAs & 45~\cite{Morelli_Slack_2006} & 8.0 & 8.95 & 7.85 & 9.30 & 220~\cite{slack, Morelli_Slack_2006} & 313 & 315 & 313 & 315 & 0.75~\cite{slack, Morelli_Slack_2006} & 2.24 & 2.15 & 2.26 & 2.11	 \\
& & & & & & & (248) & (250) & (248) & (250)  & & & & &  \\
GaSb & 40~\cite{Morelli_Slack_2006} & 4.96 & 5.49 & 4.68 & 5.69 & 165~\cite{slack, Morelli_Slack_2006} & 240 & 241 & 239 & 241 & 0.75~\cite{slack, Morelli_Slack_2006} & 2.28 & 2.19 & 2.33 & 2.15  \\
& & & & & & & (190) & (191) & (190) & (191)  & & & & &  \\
InP  & 93~\cite{Morelli_Slack_2006} & 6.53 & 7.40 & 6.57 & 7.71 & 220~\cite{slack, Morelli_Slack_2006} & 286 & 287 & 286 & 287 & 0.6~\cite{slack, Morelli_Slack_2006} & 2.21 & 2.1 & 2.2 & 2.06 	 \\
& & & & & & & (227) & (228) & (227) & (228)  & & & & & \\
InAs & 30~\cite{Morelli_Slack_2006} & 4.33 & 4.80 & 4.20 & 4.93 & 165~\cite{slack, Morelli_Slack_2006} & 229 & 230 & 229 & 230 & 0.57~\cite{slack, Morelli_Slack_2006} & 2.26 & 2.17 & 2.29 & 2.14	 \\
& & & & & & & (182) & (183) & (182) & (183)  & & & & & \\
InSb & 20~\cite{Morelli_Slack_2006} & 3.02 & 3.33 & 2.76 & 3.44 & 135~\cite{slack, Morelli_Slack_2006} & 187 & 188 & 186 & 188 & 0.56~\cite{slack, Morelli_Slack_2006} & 2.3 & 2.22 & 2.38 & 2.18	 \\
 & & & & & & & (148) & (149) & (148) & (149) & 16.5~\cite{Snyder_jmatchem_2011} & & & & \\
ZnS  & 27~\cite{Morelli_Slack_2006} & 8.38 & 8.40 & 7.67 & 8.96 & 230~\cite{slack, Morelli_Slack_2006} & 341 & 341 & 340 & 342 & 0.75~\cite{slack, Morelli_Slack_2006} & 2.0 & 1.99 & 2.07 & 1.94	 \\
& & & & & & & (271) & (271) & (270) & (271)  & & & & & \\
ZnSe & 19~\cite{Morelli_Slack_2006} & 5.44 & 5.55 & 4.93 & 5.80 & 190~\cite{slack, Morelli_Slack_2006} & 260	& 260 & 259 & 261 & 0.75~\cite{slack, Morelli_Slack_2006} & 2.06 & 2.04 & 2.14 & 2.01	\\
 & 33~\cite{Snyder_jmatchem_2011} & & & & & & (206) & (206) & (206) & (207)  & & & & &  \\
ZnTe & 18~\cite{Morelli_Slack_2006} & 3.83 & 3.95 & 3.44 & 4.10 & 155~\cite{slack, Morelli_Slack_2006} & 210 & 210 & 209 & 210  & 0.97~\cite{slack, Morelli_Slack_2006} & 2.13 & 2.1 & 2.23 & 2.07 \\
& & & & & & & (167) & (167) & (166) & (167)  & & & & & \\
CdSe & 4.4~\cite{Snyder_jmatchem_2011} & 2.04 & 2.11 & 1.84 & 2.16 & 130~\cite{Morelli_Slack_2006} & 173 & 173 & 172 & 173 & 0.6~\cite{Morelli_Slack_2006} & 2.18 & 2.15 & 2.27 & 2.12 \\
& & & & & & & (137) & (137) & (137) & (137)  & & & & & \\
CdTe & 7.5~\cite{Morelli_Slack_2006} & 1.71 & 1.77 & 1.50 & 1.81 & 120~\cite{slack, Morelli_Slack_2006} & 150 & 150 & 149 & 150 & 0.52~\cite{slack, Morelli_Slack_2006} & 2.22 & 2.19 & 2.34 & 2.16	 \\
& & & & & & & (119) & (119) & (118) & (119)  & & & & & \\
HgSe & 3~\cite{Whitsett_PRB_1973} & 1.32 & 1.36 & 1.22 & 1.41 & 110~\cite{slack} & 140	& 140 & 140 & 140	& 0.17~\cite{slack} & 2.38 & 2.35 & 2.47  & 2.31 \\
& & & & & & & (111) & (111) & (111) & (111)  & & & & & \\
HgTe & 2.5~\cite{Snyder_jmatchem_2011}  & 1.21 & 1.30 & 1.10 & 1.34 & 141~\cite{Snyder_jmatchem_2011}  &  129	& 130 & 129 & 130	& 1.9~\cite{Snyder_jmatchem_2011}  & 2.45 & 2.40 & 2.56 & 2.36 \\
  & & & & & & (100)~\cite{slack} & (102) & (103) & (102) & (103)  & 0.46\cite{slack}  & & & &  \\
\end{tabular}}
\label{tab:art115:zincblende_thermal_eos}
\etab

\tab
\mycaption[Thermal properties lattice thermal conductivity at 300~K, Debye temperature and Gr{\"u}neisen parameter of rocksalt
structure semiconductors, calculated using the different equations of state.]
{The values listed for $\theta^{\mathrm{exp}}$ are
$\theta_\acoustic$, except 155K for SnTe which is $\theta_{\mathrm D}$~\cite{Snyder_jmatchem_2011}.
``N/A'' = Not available for that source.
Units: $\kappa$ in \WmK, $\theta$ in \K.}
\tabvspace
\resizebox{\linewidth}{!}{
\begin{tabular}{l|r|r|r|r|r|r|r|r|r|r|r|r|r|r|r|r|r|r}
comp. & $\kappa^\EXP$  & $\kappa^\sAGL$ & $\kappa^\sBM$ & $\kappa^\sVIN$ & $\kappa^\sBCN$  & $\theta^\EXP$  & $\theta_\sDebye^\sAGL$ & $\theta_\sDebye^\sBM$ & $\theta_\sDebye^\sVIN$ & $\theta_\sDebye^\sBCN$ & $\gamma^\EXP$ & $\gamma^\sAGL$ &  $\gamma^\sBM$ &  $\gamma^\sVIN$  &  $\gamma^\sBCN$  \\
& & & & & & & ($\theta_\acoustic^\sAGL$) & ($\theta_\acoustic^\sBM$) & ($\theta_\acoustic^\sVIN$) & ($\theta_\acoustic^\sBCN$) & & & & & \\
\hline
LiH & 15~\cite{Morelli_Slack_2006} & 18.6 & 13.54 & 12.37 & 20.98 & 615~\cite{slack, Morelli_Slack_2006} & 962 & 931 & 927 & 968 & 1.28~\cite{slack, Morelli_Slack_2006} & 1.66 & 1.84 & 1.90 & 1.58 \\
& & & & & & &  (764) & (739) & (734) & (768) & & & & & \\
LiF & 17.6~\cite{Morelli_Slack_2006} & 9.96 & 10.19 & 8.68 & 11.45 & 500~\cite{slack, Morelli_Slack_2006} & 617 &	617 & 610 & 623 & 1.5~\cite{slack, Morelli_Slack_2006} & 2.03 & 2.00 & 2.13 & 1.92	 \\
& & & & & & &  (490) & (490) & (485) & (494) & & & & & \\
NaF &  18.4~\cite{Morelli_Slack_2006} &  4.67 & 4.65 & 3.82 & 4.91 & 395~\cite{slack, Morelli_Slack_2006} & 416 & 416 & 411 & 417  & 1.5~\cite{slack, Morelli_Slack_2006} & 2.21 & 2.21 & 2.39 & 2.16	  \\
& & & & & & &  (330) & (330) & (326) & (331) & & & & & \\
NaCl & 7.1~\cite{Morelli_Slack_2006} & 2.12 & 2.27 & 1.74 & 2.28 & 220~\cite{slack, Morelli_Slack_2006} & 271 & 273 & 268 & 272 & 1.56~\cite{slack, Morelli_Slack_2006} & 2.23 & 2.18 & 2.40 & 2.16	 \\
& & & & & & &  (215) & (217) & (213) & (216) & & & & & \\
NaBr & 2.8~\cite{Morelli_Slack_2006} & 1.33 & 1.42 & 1.08 & 1.40 & 150~\cite{slack, Morelli_Slack_2006} & 188 & 189 & 186 & 188 & 1.5~\cite{slack, Morelli_Slack_2006} & 2.22 & 2.17 & 2.40 & 2.16 	 \\
& & & & & & & (149) & (150) & (148) & (149) & & & & & \\
NaI & 1.8~\cite{Morelli_Slack_2006} & 0.851 & 0.922 & 0.679 & 0.892 & 100~\cite{slack, Morelli_Slack_2006} & 140 & 141	& 138 & 140 & 1.56~\cite{slack, Morelli_Slack_2006} & 2.23 & 2.17 & 2.43 & 2.18	 \\
& & & & & & & (111) & (112) & (110) & (111) & & & & & \\
KF & N/A & 2.21 & 2.07 & 1.62 & 2.22 & 235~\cite{slack, Morelli_Slack_2006} & 288 &	287	& 281 & 288 & 1.52~\cite{slack, Morelli_Slack_2006} & 2.32 & 2.38 & 2.60 & 2.32 	\\
& & & & & & & (229) & (228) & (224) & (229) & & & & & \\
KCl & 7.1~\cite{Morelli_Slack_2006} & 1.25 & 1.42 & 1.04 & 1.40 & 172~\cite{slack, Morelli_Slack_2006} & 213 & 215 & 210 & 214 & 1.45~\cite{slack, Morelli_Slack_2006} & 2.40 & 2.29 & 2.57 & 2.29 	 \\
& & & & & & & (169) & (171) & (167) & (170) & & & & & \\
KBr & 3.4~\cite{Morelli_Slack_2006} & 0.842 & 0.949 & 0.682 & 0.928 & 117~\cite{slack, Morelli_Slack_2006} & 156 & 157 & 153 & 156  & 1.45~\cite{slack, Morelli_Slack_2006} & 2.37 & 2.26 & 2.55 & 2.27 \\
& & & & & & & (124) & (125) & (121) & (124) & & & & & \\
KI & 2.6~\cite{Morelli_Slack_2006} & 0.525 & 0.624 & 0.451 & 0.603 & 87~\cite{slack, Morelli_Slack_2006} & 116 & 118 & 115 & 117 & 1.45~\cite{slack, Morelli_Slack_2006} & 2.35 & 2.23 & 2.50 & 2.23	 \\
& & & & & & & (92) & (94) & (91) & (93) & & & & & \\
RbCl & 2.8~\cite{Morelli_Slack_2006} & 0.837 & 0.886 & 0.638 & 0.878 & 124~\cite{slack, Morelli_Slack_2006} & 155 & 156 & 152 & 155 & 1.45~\cite{slack, Morelli_Slack_2006} & 2.37 & 2.33 & 2.62 & 2.32	 \\
& & & & & & & (123) & (124) & (121) & (123) & & & & & \\
RbBr & 3.8~\cite{Morelli_Slack_2006} & 0.558 & 0.606 & 0.459 & 0.606 & 105~\cite{slack, Morelli_Slack_2006} & 122 & 123 & 121 & 123 & 1.45~\cite{slack, Morelli_Slack_2006} & 2.43 & 2.36 & 2.62 & 2.36 \\
& & & & & & & (97) & (98) & (96) & (98) & & & & & \\
RbI & 2.3~\cite{Morelli_Slack_2006} & 0.368 & 0.434 & 0.320 & 0.415 & 84~\cite{slack, Morelli_Slack_2006} & 97 & 98 & 96 & 97 & 1.41~\cite{slack, Morelli_Slack_2006} & 2.47 & 2.32 & 2.60 & 2.34 	 \\
& & & & & & & (77) & (78) & (76) & (77) & & & & & \\
AgCl & 1.0~\cite{Landolt-Bornstein, Maqsood_IJT_2003}  & 0.613 & 0.612 & 0.535 & 0.663 & 124~\cite{slack} & 145 & 145 & 144 & 146 & 1.9~\cite{slack} & 2.49 & 2.49 & 2.63 & 2.43 	 \\
& & & & & & & (115) & (115) & (114) & (116) & & & & &  \\
MgO  & 60~\cite{Morelli_Slack_2006} & 44.5 & 44.7 & 38.5 & 47.1 & 600~\cite{slack, Morelli_Slack_2006} & 849 & 848 & 842 & 851	& 1.44~\cite{slack, Morelli_Slack_2006} & 1.96 & 1.95 & 2.07 & 1.91 \\
& & & & & & & (674) & (673) & (668) & (675) & & & & & \\
CaO & 27~\cite{Morelli_Slack_2006} & 24.3 & 24.7 & 22.5 & 25.7 & 450~\cite{slack, Morelli_Slack_2006} & 620 & 620 & 618 & 621 & 1.57~\cite{slack, Morelli_Slack_2006} & 2.06  & 2.05 & 2.13 & 2.02	 \\
& & & & & & & (492) & (492) & (491) & (493) & & & & & \\
SrO & 12~\cite{Morelli_Slack_2006} & 13.4 & 13.3 & 12.2 & 14.0 & 270~\cite{slack, Morelli_Slack_2006} & 413 & 413 & 412 & 414 & 1.52~\cite{slack, Morelli_Slack_2006} & 2.13 & 2.13 & 2.21	& 2.09 	 \\
& & & & & & & (328) &  (328) & (327) & (329) & & & & & \\
BaO & 2.3~\cite{Morelli_Slack_2006} & 7.10 & 6.73 & 6.10 & 6.98 & 183~\cite{slack, Morelli_Slack_2006} & 288 & 288 & 287 & 288 & 1.5~\cite{slack, Morelli_Slack_2006} & 2.14 & 2.20 & 2.29 & 2.16 \\
& & & & & & & (229) & (229) & (228) & (229) & & & & & \\
PbS & 2.9~\cite{Morelli_Slack_2006} & 6.11 & 6.77 & 5.99 & 7.02 & 115~\cite{slack, Morelli_Slack_2006} & 220 & 221 & 220 & 221 & 2.0~\cite{slack, Morelli_Slack_2006} & 2.00 & 1.92 & 2.02 & 1.89	\\
& & & & & & & (175) & (175) & (175) & (175) & & & & & \\
PbSe & 2.0~\cite{Morelli_Slack_2006} & 4.81 & 5.29 & 4.63 & 5.44 & 100~\cite{Morelli_Slack_2006} & 194 & 195 & 194 & 195 & 1.5~\cite{Morelli_Slack_2006} & 2.07 & 2.00 & 2.11 & 1.97	 \\
& & & & & & & (154) & (155) & (154) & (155) & & & & & \\
PbTe & 2.5~\cite{Morelli_Slack_2006} & 4.07 & 4.11 & 3.50 & 4.32 & 105~\cite{slack, Morelli_Slack_2006} & 172 & 172 & 171 & 173 & 1.45~\cite{slack, Morelli_Slack_2006} & 2.09 & 2.08 & 2.22 & 2.05 	 \\
& & & & & & & (137) & (137) & (136) & (137) & & & & & \\
SnTe & 1.5~\cite{Snyder_jmatchem_2011} & 5.24 & 5.59 & 4.64 & 5.78 & 155~\cite{Snyder_jmatchem_2011} & 210 & 211 & 209 & 211 & 2.1~\cite{Snyder_jmatchem_2011} & 2.11 & 2.06 & 2.22 & 2.03 	 \\
& & & & & & & (167) & (167) & (166) & (167) & & & & & \\
\end{tabular}}
\label{tab:art115:rocksalt_thermal_eos}
\etab

\tab
\mycaption[Lattice thermal conductivity, Debye temperature and Gr{\"u}neisen parameter of hexagonal
structure semiconductors, calculated using the different equations of state.]
{The values listed for $\theta^{\mathrm{exp}}$ are
$\theta_\acoustic$, except 190K for InSe~\cite{Snyder_jmatchem_2011} and 660K for InN~\cite{Ioffe_Inst_DB, Krukowski_jphyschemsolids_1998}
which are $\theta_{\mathrm D}$.
``N/A'' = Not available for that source.
Units: $\kappa$ in \WmK, $\theta$ in \K.}
\tabvspace
\resizebox{\linewidth}{!}{
\begin{tabular}{l|r|r|r|r|r|r|r|r|r|r|r|r|r|r|r}
comp. & $\kappa^\EXP$ & $\kappa^\sAGL$ & $\kappa^\sBM$ & $\kappa^\sVIN$ & $\kappa^\sBCN$  & $\theta^\EXP$  & $\theta_\sDebye^\sAGL$ & $\theta_\sDebye^\sBM$ & $\theta_\sDebye^\sVIN$ & $\theta_\sDebye^\sBCN$ & $\gamma^\EXP$ & $\gamma^\sAGL$ &  $\gamma^\sBM$ &  $\gamma^\sVIN$  &  $\gamma^\sBCN$    \\
& & & & & & & ($\theta_\acoustic^\sAGL$) & ($\theta_\acoustic^\sBM$) & ($\theta_\acoustic^\sVIN$) & ($\theta_\acoustic^\sBCN$) & & & & &  \\
\hline
SiC & 490~\cite{Morelli_Slack_2006} & 70.36 & 75.17 & 62.86 & 77.82 & 740~\cite{Morelli_Slack_2006} & 1103 & 1105 & 1096 & 1106 & 0.75~\cite{Morelli_Slack_2006} & 1.86 & 1.80 & 1.94 & 1.78 \\
& & & & & & & (695) & (696) & (690) & (697) & & & & &   \\
AlN & 350~\cite{Morelli_Slack_2006} & 39.0 & 40.53 & 34.49 & 42.3 & 620~\cite{Morelli_Slack_2006} & 898 & 899 & 893 & 900 & 0.7~\cite{Morelli_Slack_2006} & 1.85 & 1.82 & 1.95 & 1.79 	 \\
& & & & & & & (566) & (566) & (563) & (567) & & & & &  \\
GaN &  210~\cite{Morelli_Slack_2006} & 18.53 & 16.33 & 16.21 & 20.15 & 390~\cite{Morelli_Slack_2006} & 595 & 590 & 591 & 596 & 0.7~\cite{Morelli_Slack_2006} & 2.08 & 2.18 & 2.19 & 2.01	 \\
& & & & & & & (375) & (372) & (372) & (375)  & & & & & \\
ZnO & 60~\cite{Morelli_Slack_2006} & 7.39 & 7.72 & 6.80 & 8.06 & 303~\cite{Morelli_Slack_2006} & 422 & 422 & 420 & 423 & 0.75~\cite{Morelli_Slack_2006}  & 1.94 & 1.91 & 2.01 & 1.87 	 \\
& & & & & & & (266) & (266) & (265) & (266)  & & & & &  \\
BeO & 370~\cite{Morelli_Slack_2006}  & 53.36 & 54.41 & 46.97 & 56.95 & 809~\cite{Morelli_Slack_2006} & 1181 & 1182 & 1173 & 1184 & 1.38~\cite{Slack_JAP_1975, Cline_JAP_1967, Morelli_Slack_2006} & 1.76 & 1.74 & 1.85 & 1.71	 \\
& & & & & & & (744) & (745) & (739) & (746)  & & & & & \\
CdS & 16~\cite{Morelli_Slack_2006} & 1.76 & 1.89 & 1.66 & 1.93 & 135~\cite{Morelli_Slack_2006} & 211 & 212 & 211 & 212 & 0.75~\cite{Morelli_Slack_2006} & 2.14 & 2.08 & 2.19 & 2.06	 \\
& & & & & & & (133) & (134) & (133) & (134)  & & & & & \\
InSe & 6.9~\cite{Snyder_jmatchem_2011} &  2.34 & 2.61 & 2.23 & 2.69 & 190~\cite{Snyder_jmatchem_2011} & 249 & 250 & 248	& 250 & 1.2~\cite{Snyder_jmatchem_2011} & 2.24 & 2.14 & 2.28 & 2.11 	 \\
& & & & & & & (125) & (125) & (124) & (125)  & & & & & \\
InN & 45~\cite{Ioffe_Inst_DB, Krukowski_jphyschemsolids_1998} & 6.82 & 6.97 & 5.59 & 7.49 & 660~\cite{Ioffe_Inst_DB, Krukowski_jphyschemsolids_1998} & 369 & 369 & 365 & 370 & 0.97~\cite{Krukowski_jphyschemsolids_1998} & 2.18 & 2.15 & 2.35 & 2.09 	 \\
& & & & & & & (232) & (232) & (230) & (233)  & & & & & \\
\end{tabular}}
\label{tab:art115:wurzite_thermal_eos}
\etab

\tab
\mycaption[Lattice thermal conductivity, Debye temperatures and Gr{\"u}neisen parameter of rhombohedral
semiconductors, calculated using the different equations of state.]
{The experimental Debye temperatures are $\theta_{\mathrm D}$ for
Bi$_2$Te$_3$ and Sb$_2$Te$_3$, and  $\theta_\acoustic$ for Al$_2$O$_3$.
``N/A'' = Not available for that source.
Units: $\kappa$ in \WmK, $\theta$ in \K.}
\tabvspace
\resizebox{\linewidth}{!}{
\begin{tabular}{l|r|r|r|r|r|r|r|r|r|r|r|r|r|r|r}
comp. & $\kappa^\EXP$ & $\kappa^\sAGL$ & $\kappa^\sBM$ & $\kappa^\sVIN$ & $\kappa^\sBCN$  & $\theta^\EXP$  & $\theta_\sDebye^\sAGL$ & $\theta_\sDebye^\sBM$ & $\theta_\sDebye^\sVIN$ & $\theta_\sDebye^\sBCN$ & $\gamma^\EXP$ & $\gamma^\sAGL$ &  $\gamma^\sBM$ &  $\gamma^\sVIN$  &  $\gamma^\sBCN$    \\
& & & & & & & ($\theta_\acoustic^\sAGL$) & ($\theta_\acoustic^\sBM$) & ($\theta_\acoustic^\sVIN$) & ($\theta_\acoustic^\sBCN$) & & & & &  \\
\hline
Bi$_2$Te$_3$ & 1.6~\cite{Snyder_jmatchem_2011} & 3.35 & 3.63 & 3.17 & 3.73 & 155~\cite{Snyder_jmatchem_2011} & 204 & 205 & 204 & 205 & 1.49~\cite{Snyder_jmatchem_2011} & 2.14 & 2.08 & 2.20 & 2.05	 \\
& & & & & & & (119) & (120) & (119) & (120) & & & & & \\
Sb$_2$Te$_3$ & 2.4~\cite{Snyder_jmatchem_2011} & 4.46 & 4.76 & 4.07 & 4.99 & 160~\cite{Snyder_jmatchem_2011} & 243 & 244 & 242 & 244 & 1.49~\cite{Snyder_jmatchem_2011} & 2.11 & 2.06 & 2.19 & 2.02	\\
& & & & & & & (142) & (143) & (142) & (143) & & & & & \\
Al$_2$O$_3$ & 30~\cite{Slack_PR_1962} & 21.92 & 23.36 & 19.51 & 23.19 & 390~\cite{slack} & 952 & 954 & 947 & 954 & 1.32~\cite{slack} & 1.91 & 1.86 & 2.00 & 1.87	 \\
& & & & & & & (442) & (443) & (440) & (443) & & & & & \\
Cr$_2$O$_3$  & 16~\cite{Landolt-Bornstein, Bruce_PRB_1977} & 12.03 & 12.61 & 10.78 & 12.92 & N/A & 718 & 717 & 713 & 718 & N/A & 2.10 & 2.05 & 2.19 & 2.04 	\\
& & & & & & &  (333) & (333) & (331) & (333) & & & & & \\
Bi$_2$Se$_3$ & 1.34~\cite{Landolt-Bornstein} & 2.41 & 2.54 & 2.31 & 2.68 & N/A & 199 & 199 & 199 & 200 & N/A & 2.12 & 2.07 & 2.16 & 2.03 	\\
& & & & & & & (116) & (116) & (116) & (117) & & & & & \\
\end{tabular}}
\label{tab:art115:rhombo_thermal_eos}
\etab

\tab
\mycaption[Lattice thermal conductivity at 300~K, Debye temperatures and Gr{\"u}neisen parameter of  body-centered tetragonal
semiconductors, calculated using the different equations of state.]
{``N/A'' = Not available for that source.
Units: $\kappa$ in \WmK, $\theta$ in \K.}
\tabvspace
\resizebox{\linewidth}{!}{
\begin{tabular}{l|r|r|r|r|r|r|r|r|r|r|r|r|r|r|r}
comp. & $\kappa^\EXP$ & $\kappa^\sAGL$ & $\kappa^\sBM$ & $\kappa^\sVIN$ & $\kappa^\sBCN$  & $\theta^\EXP$  & $\theta_\sDebye^\sAGL$ & $\theta_\sDebye^\sBM$ & $\theta_\sDebye^\sVIN$ & $\theta_\sDebye^\sBCN$ & $\gamma^\EXP$ & $\gamma^\sAGL$ &  $\gamma^\sBM$ &  $\gamma^\sVIN$  &  $\gamma^\sBCN$    \\
& & & & & & & ($\theta_\acoustic^\sAGL$) & ($\theta_\acoustic^\sBM$) & ($\theta_\acoustic^\sVIN$) & ($\theta_\acoustic^\sBCN$) & & & & &  \\
\hline
CuGaTe$_2$ & 2.2~\cite{Snyder_jmatchem_2011} & 1.36 & 1.49 & 1.30 & 1.53 & 226~\cite{Snyder_jmatchem_2011} & 215 & 216 & 215 & 216 & 1.46~\cite{Snyder_jmatchem_2011} & 2.32 & 2.23 & 2.36 & 2.21 	 \\
& & & & & & & (108) & (108) & (108) & (108) & & & & &\\
ZnGeP$_2$ & 35~\cite{Landolt-Bornstein, Beasley_AO_1994}  & 5.07 & 5.54 & 4.95 & 5.73 & 500~\cite{Landolt-Bornstein} & 408 & 410 & 408 & 410 & N/A & 2.14 & 2.07 & 2.17 & 2.04 	   \\
& 36~\cite{Landolt-Bornstein, Beasley_AO_1994} & & & & & & (204) & (205) & (204) & (205) & & & & & \\
& 18~\cite{Landolt-Bornstein, Shay_1975, Masumoto_JPCS_1966} & & & & & & & & & & & & & & \\
ZnSiAs$_2$ & 14\cite{Landolt-Bornstein, Shay_1975, Masumoto_JPCS_1966} & 3.96 & 4.19 & 3.76 & 4.43 & 347~\cite{Landolt-Bornstein, Bohnhammel_PSSa_1981} & 350 & 350 & 349 & 351 & N/A & 2.15 & 2.10 & 2.20 & 2.05	 \\
& & & & & & & (175) & (175) & (175) & (176) & & & & &\\
CuInTe$_2$ & 10\cite{Landolt-Bornstein, Rincon_PSSa_1995} & 0.722 & 0.797 & 0.693 & 0.812 & 185~\cite{Landolt-Bornstein, Rincon_PSSa_1995} & 166 & 167	& 166 & 167 & 0.93~\cite{Rincon_PSSa_1995} & 2.32 & 2.23 & 2.36 & 2.21 	 \\
& & & & & & 195~\cite{Landolt-Bornstein, Bohnhammel_PSSa_1982}  & (83) & (84) & (83) & (84) & & & & &\\
AgGaS$_2$ & 1.4\cite{Landolt-Bornstein, Beasley_AO_1994} & 0.993 & 1.04 & 0.92 & 1.08 & 255~\cite{Landolt-Bornstein, Abrahams_JCP_1975} & 224 & 224 & 223 & 224 & N/A & 2.20 & 2.14 & 2.26 & 2.11	\\
& & & & & & & (112) & (112) & (112) & (112) & & & & &\\
CdGeP$_2$ & 11~\cite{Landolt-Bornstein, Shay_1975, Masumoto_JPCS_1966} & 2.96 & 3.18 & 2.85 & 3.31 & 340~\cite{Landolt-Bornstein, Abrahams_JCP_1975} & 320 & 321 & 320 & 321 & N/A & 2.21 & 2.14 & 2.25 & 2.10 	 \\
& & & & & & & (160) & (161) & (160) & (161) & & & & &\\
CdGeAs$_2$ & 42~\cite{Landolt-Bornstein, Shay_1975} & 2.11 & 2.17 & 1.92 & 2.24 & N/A & 254 & 254 & 253 & 254 & N/A & 2.20 & 2.17 & 2.29 & 2.14 	\\
& & & & & & & (127) & (127) & (127) & (127) & & & & &\\
CuGaS$_2$ & 5.09~\cite{Landolt-Bornstein} & 2.79 & 2.99 & 2.67 & 3.11 & 356~\cite{Landolt-Bornstein, Abrahams_JCP_1975} & 349 & 350 & 348 & 350 & N/A & 2.24 & 2.18  & 2.28 & 2.14 	 \\
& & & & & & & (175) & (175) & (174) & (175) & & & & &\\
CuGaSe$_2$ & 12.9~\cite{Landolt-Bornstein, Rincon_PSSa_1995} & 1.46 & 1.53 & 1.37 & 1.61 & 262~\cite{Landolt-Bornstein, Bohnhammel_PSSa_1982} & 244 & 244 & 243 & 245 & N/A & 2.26 & 2.21 & 2.32 & 2.17	 \\
& & & & & & & (122) & (122) & (122) & (123) & & & & &\\
ZnGeAs$_2$ & 11\cite{Landolt-Bornstein, Shay_1975} & 3.18 & 3.29 & 2.93 & 3.45 & N/A & 307 & 307 & 306 & 308 & N/A & 2.17 & 2.13 & 2.24 & 2.10	 \\
& & & & & & & (154) & (154) & (153) & (154) & & & & &\\
\end{tabular}}
\label{tab:art115:bct_thermal_eos}
\etab

\tab
\mycaption[Lattice thermal conductivity at 300~K, Debye temperatures and Gr{\"u}neisen parameter of  materials with various
structures, calculated using the different equations of state.]
{The experimental Debye temperatures are $\theta_{\mathrm D}$,
except ZnSb for which it is $\theta_\acoustic$.
``N/A'' = Not available for that source.
Units: $\kappa$ in \WmK, $\theta$ in \K.}
\tabvspace
\resizebox{\linewidth}{!}{
\begin{tabular}{l|r|r|r|r|r|r|r|r|r|r|r|r|r|r|r|r}
comp. & Pearson  & $\kappa^\EXP$ & $\kappa^\sAGL$ & $\kappa^\sBM$ & $\kappa^\sVIN$ & $\kappa^\sBCN$  & $\theta^\EXP$  & $\theta_\sDebye^\sAGL$ & $\theta_\sDebye^\sBM$ & $\theta_\sDebye^\sVIN$ & $\theta_\sDebye^\sBCN$ & $\gamma^\EXP$ & $\gamma^\sAGL$ &  $\gamma^\sBM$ &  $\gamma^\sVIN$  &  $\gamma^\sBCN$    \\
& & & & & & & & ($\theta_\acoustic^\sAGL$) & ($\theta_\acoustic^\sBM$) & ($\theta_\acoustic^\sVIN$) & ($\theta_\acoustic^\sBCN$) & & & & &  \\
\hline
CoSb$_3$ & cI32 & 10~\cite{Snyder_jmatchem_2011} & 2.60 & 2.58 & 2.38 & 2.78 & 307~\cite{Snyder_jmatchem_2011} & 310 & 310 & 309 & 311 & 0.95~\cite{Snyder_jmatchem_2011} & 2.33 & 2.33 & 2.42 & 2.27  \\
& & & & & & & & (123) & (123) & (123) & (123) & & & &  \\
IrSb$_3$ & cI32 & 16~\cite{Snyder_jmatchem_2011} & 2.73 & 2.89 & 2.67 & 3.01 & 308~\cite{Snyder_jmatchem_2011} & 286 & 287 & 286 & 287 & 1.42~\cite{Snyder_jmatchem_2011} & 2.34 & 2.29 & 2.37 & 2.25  \\
& & & & & & & & (113) & (114) & (113) & (114) & & & &  \\
ZnSb & oP16 &  3.5~\cite{Madsen_PRB_2014, Bottger_JEM_2010} & 1.23 & 1.29 & 1.13 & 1.36 & 92~\cite{Madsen_PRB_2014} & 242 & 242 & 241 & 243 &  0.76~\cite{Madsen_PRB_2014, Bottger_JEM_2010} & 2.23 & 2.18 & 2.30 & 2.14 	 \\
& & & & & & & & (96) & (96) & (96) & (96) & & & &  \\
Sb$_2$O$_3$ & oP20 & 0.4~\cite{Landolt-Bornstein} & 8.74 & 8.93 & 8.18 & 9.20 & N/A & 572 & 573 & 571 & 573 & N/A & 2.12 & 2.10 & 2.18 & 2.07	\\
& & & & & & & & (211) & (211) & (210) & (211) & & & &  \\
InTe & cP2 & N/A & 0.709 & 0.602 & 0.524 & 0.626 & N/A & 113 & 112 & 111 & 112 & N/A & 2.19 & 2.33 & 2.45 & 2.29 \\
& & & & & & & & (90) & (89) & (88) & (89) & & & &  \\
InTe & tP16 & 1.7~\cite{Snyder_jmatchem_2011} & 1.40 & 1.53 & 1.27 & 1.55 & 186~\cite{Snyder_jmatchem_2011} & 193 & 194 & 192 & 194 & 1.0~\cite{Snyder_jmatchem_2011} & 2.24 & 2.16 & 2.32 & 2.14	\\
& & & & & & & & (97) & (97) & (96) & (97) & & & &  \\
Bi$_2$O$_3$ & mP20 & 0.8~\cite{Landolt-Bornstein} & 2.98 & 3.05 & 2.49 & 3.14 & N/A & 342 & 342 & 339 & 342 & N/A & 2.10 & 2.08 & 2.26 & 2.05	 \\
& & & & & & & & (126) & (126) & (125) & (126) & & & &  \\
SnO$_2$ & tP6 & 98\cite{Turkes_jpcss_1980} & 6.98 & 7.76 & 6.52 & 8.31 & N/A & 487 & 489 & 485 & 490 & N/A & 2.42 & 2.32 & 2.48 & 2.25	 \\
&  &  55~\cite{Turkes_jpcss_1980} & & & & & & (268) & (269) & (267) & (270) & & & &  \\
\end{tabular}}
\label{tab:art115:misc_thermal_eos}
\etab

\subsubsection{Elastic constant values}

The elastic constant values in the 6x6 Voigt notation are shown for zincblende and diamond structure materials in Table~\ref{tab:art115:zincblende_elastic_supp}, for rocksalt structure materials in Table~\ref{tab:art115:rocksalt_elastic_supp}, for hexagonal structure materials in Table~\ref{tab:art115:wurzite_elastic_supp}, for rhombohedral structure materials in Table~\ref{tab:art115:rhombo_elastic_supp}, for body-centered tetragonal
ternary materials in Table~\ref{tab:art115:bct_elastic_supp}, for body-centered cubic and simple cubic materials in Table~\ref{tab:art115:bcc_elastic}, for orthorhombic structures in Table~\ref{tab:art115:orc_elastic}, and for tetragonal
structure materials in Table~\ref{tab:art115:tet_elastic}.

\tab
\mycaption[Elastic constants $c_{11}$, $c_{12}$ and $c_{44}$ of
zincblende and diamond structure semiconductors.]
{The zincblende structure is designated \AFLOW\ prototype {\sf AB\_cF8\_216\_c\_a}~\cite{aflowANRL}
and the diamond structure {\sf A\_cF8\_227\_a}~\cite{aflowANRL}.
``N/A'' = Not available for that source.
Units: \GPa.}
\tabvspace
\resizebox{\linewidth}{!}{
\begin{tabular}{l|r|r|r|r|r|r}
comp. & $c_{11}^\EXP$  & $c_{11}^\sAEL$ & $c_{12}^\EXP$  & $c_{12}^\sAEL$ & $c_{44}^\EXP$  & $c_{44}^\sAEL$      \\
\hline
C & 1076.4~\cite{Semiconductors_BasicData_Springer} & 1048  & 125.2~\cite{Semiconductors_BasicData_Springer}  & 127 & 577.4~\cite{Semiconductors_BasicData_Springer} & 560 \\
SiC & 352.3~\cite{Semiconductors_BasicData_Springer} & 384 & 140.4~\cite{Semiconductors_BasicData_Springer} & 127 & 232.9~\cite{Semiconductors_BasicData_Springer} & 240 \\
Si & 165.64~\cite{Semiconductors_BasicData_Springer, Hall_ElasticSi_PR_1967} & 153 & 63.94~\cite{Semiconductors_BasicData_Springer, Hall_ElasticSi_PR_1967} & 57.1 & 79.51~\cite{Semiconductors_BasicData_Springer, Hall_ElasticSi_PR_1967} & 74.6 \\
Ge & 129.9~\cite{Semiconductors_BasicData_Springer, Bruner_ElasticGe_PRL_1961} & 107 & 48.73~\cite{Semiconductors_BasicData_Springer, Bruner_ElasticGe_PRL_1961} & 38.8 & 68.0~\cite{Semiconductors_BasicData_Springer, Bruner_ElasticGe_PRL_1961} & 56.7 \\
BN & N/A & 777 & N/A & 170 & N/A & 442 \\
BP & 315.0~\cite{Semiconductors_BasicData_Springer, Wettling_ElasticBP_SSC_1984} & 339 & 100~\cite{Semiconductors_BasicData_Springer, Wettling_ElasticBP_SSC_1984} & 73.3 & 160~\cite{Semiconductors_BasicData_Springer, Wettling_ElasticBP_SSC_1984} & 185 \\
AlP & N/A & 125 & N/A & 61.6 & N/A & 59.7 \\
AlAs & N/A & 104 & N/A & 49.3 & N/A & 50.4 \\
AlSb & 87.69~\cite{Semiconductors_BasicData_Springer, Bolef_ElasticAlSb_JAP_1960, Weil_ElasticAlSb_JAP_1972} & 76.3 & 43.41~\cite{Semiconductors_BasicData_Springer,  Bolef_ElasticAlSb_JAP_1960, Weil_ElasticAlSb_JAP_1972} & 36.0 & 40.76~\cite{Semiconductors_BasicData_Springer, Bolef_ElasticAlSb_JAP_1960, Weil_ElasticAlSb_JAP_1972} & 36.0 \\
GaP & 141.4~\cite{Boyle_ElasticGaPSb_PRB_1975} & 127 & 63.98~\cite{Boyle_ElasticGaPSb_PRB_1975} & 54.9 & 70.28~\cite{Boyle_ElasticGaPSb_PRB_1975} & 65.2 \\
GaAs & 188.8~\cite{Bateman_ElasticGaAs_JAP_1975} & 101 & 53.8~\cite{Bateman_ElasticGaAs_JAP_1975} & 43.7 & 59.4~\cite{Bateman_ElasticGaAs_JAP_1975} & 51.9 \\
GaSb & 88.34~\cite{Boyle_ElasticGaPSb_PRB_1975} & 74.6 & 40.23~\cite{Boyle_ElasticGaPSb_PRB_1975} & 33.2 &  43.22~\cite{Boyle_ElasticGaPSb_PRB_1975} & 37.6 \\
InP  & 101.1~\cite{Nichols_ElasticInP_SSC_1980} & 87.7 & 56.1~\cite{Nichols_ElasticInP_SSC_1980}  & 46.7 & 45.6~\cite{Nichols_ElasticInP_SSC_1980} & 42.3 \\
InAs & 83.29~\cite{Semiconductors_BasicData_Springer, Gerlich_ElasticAlSb_JAP_1963} & 72.4 & 45.26~\cite{Semiconductors_BasicData_Springer, Gerlich_ElasticAlSb_JAP_1963} & 38.9 & 39.59~\cite{Semiconductors_BasicData_Springer, Gerlich_ElasticAlSb_JAP_1963} & 34.3 \\
InSb & 66.0~\cite{DeVaux_ElasticInSb_PR_1956} & 55.8 & 38.0~\cite{DeVaux_ElasticInSb_PR_1956} & 29.3 & 30.0~\cite{DeVaux_ElasticInSb_PR_1956} & 26.7 \\
ZnS & 98.1~\cite{Semiconductors_BasicData_Springer} & 99.2 & 62.7~\cite{Semiconductors_BasicData_Springer} & 57.2 & 44.83~\cite{Semiconductors_BasicData_Springer} & 46.9 \\
ZnSe & 85.9~\cite{Lee_ElasticZnSeTe_JAP_1970} & 81.4 & 50.6~\cite{Lee_ElasticZnSeTe_JAP_1970} & 46.6 & 40.6~\cite{Lee_ElasticZnSeTe_JAP_1970} & 37.5 \\
ZnTe & 71.1~\cite{Lee_ElasticZnSeTe_JAP_1970} & 63.2 & 40.7~\cite{Lee_ElasticZnSeTe_JAP_1970} & 34.1 & 31.3~\cite{Lee_ElasticZnSeTe_JAP_1970} & 29.2 \\
CdSe & N/A & 57.7 & N/A & 41.1 & N/A & 21.5 \\
CdTe & N/A & 46.7 & N/A & 31.2 & N/A & 18.5  \\
HgSe & 59.5~\cite{Lehoczky_ElasticHgSe_PR_1969}  & 53.3 & 43.07~\cite{Lehoczky_ElasticHgSe_PR_1969} & 39.0 & 22.015~\cite{Lehoczky_ElasticHgSe_PR_1969} & 21.2 \\
HgTe & 53.61~\cite{Semiconductors_BasicData_Springer, Cottam_ElasticHgTe_JPCS_1975} & 45.0 & 36.6~\cite{Semiconductors_BasicData_Springer, Cottam_ElasticHgTe_JPCS_1975} & 30.4 & 21.23~\cite{Semiconductors_BasicData_Springer, Cottam_ElasticHgTe_JPCS_1975} & 19.2 \\
\end{tabular}}
\label{tab:art115:zincblende_elastic_supp}
\etab

\tab
\mycaption[Elastic constants $c_{11}$, $c_{12}$ and $c_{44}$  of
rocksalt structure semiconductors.]
{The rocksalt structure is designated \AFLOW\ Prototype {\sf AB\_cF8\_225\_a\_b}~\cite{aflowANRL}.
``N/A'' = Not available for that source.
Units: \GPa.}
\tabvspace
\begin{tabular}{l|r|r|r|r|r|r}
comp. & $c_{11}^\EXP$  & $c_{11}^\sAEL$ & $c_{12}^\EXP$  & $c_{12}^\sAEL$ & $c_{44}^\EXP$  & $c_{44}^\sAEL$  \\
\hline
LiH & 67.1~\cite{Laplaze_ElasticLiH_SSC_1976} & 84.8 & 17.0~\cite{Laplaze_ElasticLiH_SSC_1976} & 14.2 & 46.0~\cite{Laplaze_ElasticLiH_SSC_1976} & 48.8 \\
LiF & 113.55~\cite{Haussuhl_ElasticRocksalt_ZP_1960} & 124 & 47.6~\cite{Haussuhl_ElasticRocksalt_ZP_1960} & 43.7 & 63.5~\cite{Haussuhl_ElasticRocksalt_ZP_1960} & 50.6 \\
NaF & 97.0~\cite{Haussuhl_ElasticRocksalt_ZP_1960} & 96.1 & 24.3~\cite{Haussuhl_ElasticRocksalt_ZP_1960} & 22.3 & 28.1~\cite{Haussuhl_ElasticRocksalt_ZP_1960} & 24.6 \\
NaCl & 49.36~\cite{Haussuhl_ElasticRocksalt_ZP_1960} & 50.5 & 12.9~\cite{Haussuhl_ElasticRocksalt_ZP_1960} & 12.1 & 12.65~\cite{Haussuhl_ElasticRocksalt_ZP_1960} & 10.6 \\
NaBr & 40.12~\cite{Haussuhl_ElasticRocksalt_ZP_1960} & 41.2 & 10.9~\cite{Haussuhl_ElasticRocksalt_ZP_1960} & 10.2 & 9.9~\cite{Haussuhl_ElasticRocksalt_ZP_1960} & 7.97 \\
NaI & 30.25~\cite{Haussuhl_ElasticRocksalt_ZP_1960} & 32.7 & 8.8~\cite{Haussuhl_ElasticRocksalt_ZP_1960} & 8.3 & 7.4~\cite{Haussuhl_ElasticRocksalt_ZP_1960} & 5.77 \\
KF & 65.6~\cite{Haussuhl_ElasticRocksalt_ZP_1960} & 59.3 & 14.6~\cite{Haussuhl_ElasticRocksalt_ZP_1960} & 15.3 & 12.5~\cite{Haussuhl_ElasticRocksalt_ZP_1960} & 12.8 \\
KCl & 40.78~\cite{Haussuhl_ElasticRocksalt_ZP_1960} & 37.2 & 6.9~\cite{Haussuhl_ElasticRocksalt_ZP_1960} & 6.39 & 6.33~\cite{Haussuhl_ElasticRocksalt_ZP_1960} & 6.55 \\
KBr & 34.76~\cite{Haussuhl_ElasticRocksalt_ZP_1960} & 31.3 & 5.7~\cite{Haussuhl_ElasticRocksalt_ZP_1960} & 5.1  & 5.07~\cite{Haussuhl_ElasticRocksalt_ZP_1960} & 4.83 \\
KI & 27.6~\cite{Haussuhl_ElasticRocksalt_ZP_1960} & 24.8 & 4.5~\cite{Haussuhl_ElasticRocksalt_ZP_1960} & 4.02 & 3.7~\cite{Haussuhl_ElasticRocksalt_ZP_1960} & 3.17 \\
RbCl & 36.34~\cite{Haussuhl_ElasticRocksalt_ZP_1960} & 31.6 & 6.15~\cite{Haussuhl_ElasticRocksalt_ZP_1960} & 5.68 & 4.65~\cite{Haussuhl_ElasticRocksalt_ZP_1960} & 4.8 \\
RbBr & 31.57~\cite{Haussuhl_ElasticRocksalt_ZP_1960} & 28.7 & 4.95~\cite{Haussuhl_ElasticRocksalt_ZP_1960} & 4.5 & 3.8~\cite{Haussuhl_ElasticRocksalt_ZP_1960} & 3.8 \\
RbI & 25.83~\cite{Haussuhl_ElasticRocksalt_ZP_1960} & 23.1 & 3.7~\cite{Haussuhl_ElasticRocksalt_ZP_1960} & 3.3 & 2.78~\cite{Haussuhl_ElasticRocksalt_ZP_1960} & 2.57 \\
AgCl & 59.6~\cite{Hughes_ElasticAgCl_PRB_1996} & 52.7 & 36.2~\cite{Hughes_ElasticAgCl_PRB_1996} & 34.6 & 6.21~\cite{Hughes_ElasticAgCl_PRB_1996} & 8.4 \\
MgO  & 297.8~\cite{Sumino_ElasticMgO_JPE_1976} & 276 & 97.0~\cite{Sumino_ElasticMgO_JPE_1976}  & 90.7 & 156.3~\cite{Sumino_ElasticMgO_JPE_1976} & 137 \\
CaO & 221.89~\cite{Chang_ElasticCaSrBaO_JPCS_1977} & 202 & 57.81~\cite{Chang_ElasticCaSrBaO_JPCS_1977} & 57.0 & 80.32~\cite{Chang_ElasticCaSrBaO_JPCS_1977} & 74.6 \\
SrO & 175.47~\cite{Chang_ElasticCaSrBaO_JPCS_1977} & 161 & 49.08~\cite{Chang_ElasticCaSrBaO_JPCS_1977} & 46.7 & 55.87~\cite{Chang_ElasticCaSrBaO_JPCS_1977} & 53.8 \\
BaO & 126.14~\cite{Chang_ElasticCaSrBaO_JPCS_1977} & 118 & 50.03~\cite{Chang_ElasticCaSrBaO_JPCS_1977} & 44.8 & 33.68~\cite{Chang_ElasticCaSrBaO_JPCS_1977} & 36.4 \\
PbS & 126.15~\cite{Semiconductors_BasicData_Springer, Peresada_ElasticPbS_PSSa_1976} & 127 & 16.24~\cite{Semiconductors_BasicData_Springer, Peresada_ElasticPbS_PSSa_1976} & 16.9 & 17.09~\cite{Semiconductors_BasicData_Springer, Peresada_ElasticPbS_PSSa_1976} & 20.0 \\
PbSe & 123.7~\cite{Semiconductors_BasicData_Springer, Lippmann_ElasticPbSe_PSSa_1971} & 119 & 19.3~\cite{Semiconductors_BasicData_Springer, Lippmann_ElasticPbSe_PSSa_1971} & 12.2 & 15.91~\cite{Semiconductors_BasicData_Springer, Lippmann_ElasticPbSe_PSSa_1971} & 17.2 \\
PbTe & 105.3~\cite{Semiconductors_BasicData_Springer, Miller_ElasticPbTe_JPCSS_1981} & 107 & 7.0~\cite{Semiconductors_BasicData_Springer, Miller_ElasticPbTe_JPCSS_1981} & 5.63 & 13.22~\cite{Semiconductors_BasicData_Springer, Miller_ElasticPbTe_JPCSS_1981} & 14.1 \\
SnTe & 109.3~\cite{Semiconductors_BasicData_Springer, Seddon_ElasticSnTe_SSC_1976} & 114 & 2.1~\cite{Semiconductors_BasicData_Springer, Seddon_ElasticSnTe_SSC_1976}  & 3.72 & 9.69~\cite{Semiconductors_BasicData_Springer, Seddon_ElasticSnTe_SSC_1976}  & 15.7 \\
\end{tabular}
\label{tab:art115:rocksalt_elastic_supp}
\etab

\tab
\mycaption[Elastic constants $c_{11}$, $c_{12}$, $c_{13}$, $c_{33}$, $c_{44}$ and $c_{66}$ of hexagonal structure semiconductors.]
{Experimental values, where available, are shown in parentheses underneath the calculated values.
``N/A'' = Not available for that source.
Units: $B$ and $G$ in \GPa.}
\tabvspace
\resizebox{\linewidth}{!}{
\begin{tabular}{l|r|r|r|r|r|r}
comp. & $c_{11}^\sAEL$ & $c_{12}^\sAEL$ & $c_{13}^\sAEL$ & $c_{33}^\sAEL$ & $c_{44}^\sAEL$ & $c_{66}^\sAEL$     \\
& ($c_{11}^\EXP$)  & ($c_{12}^\EXP$)  & ($c_{13}^\EXP$)  & ($c_{33}^\EXP$)  & ($c_{44}^\EXP$)  & ($c_{66}^\EXP$)   \\
\hline
SiC & 494 & 102 & 48.7 & 534 & 151 & 196 \\
& (500~\cite{Arlt_ELasticSiC_JAAcS_1965}) & (92~\cite{Arlt_ELasticSiC_JAAcS_1965}) & (55.8~\cite{Arlt_ELasticSiC_JAAcS_1965}) & (564~\cite{Arlt_ELasticSiC_JAAcS_1965}) & (168~\cite{Arlt_ELasticSiC_JAAcS_1965}) & (204~\cite{Arlt_ELasticSiC_JAAcS_1965}) \\
AlN & 377 & 123 & 97.7 & 356 & 113 & 124 \\
& (410.5~\cite{Landolt-Bornstein, McNeil_ElasticAlN_JACerS_1993}) & (148.5~\cite{Landolt-Bornstein, McNeil_ElasticAlN_JACerS_1993}) & (98.9~\cite{Landolt-Bornstein, McNeil_ElasticAlN_JACerS_1993}) & (388.5~\cite{Landolt-Bornstein, McNeil_ElasticAlN_JACerS_1993}) & (124.6~\cite{Landolt-Bornstein, McNeil_ElasticAlN_JACerS_1993}) & (131.0~\cite{Landolt-Bornstein, McNeil_ElasticAlN_JACerS_1993}) \\
GaN & 329 & 115 & 80.5 & 362 & 90.3 & 109 \\
& (296~\cite{Semiconductors_BasicData_Springer, Savastenko_ElasticGaN_PSSa_1978}) & (130.0~\cite{Semiconductors_BasicData_Springer, Savastenko_ElasticGaN_PSSa_1978})  & (158.0~\cite{Semiconductors_BasicData_Springer, Savastenko_ElasticGaN_PSSa_1978})  & (267~\cite{Semiconductors_BasicData_Springer, Savastenko_ElasticGaN_PSSa_1978}) & (24.0~\cite{Semiconductors_BasicData_Springer, Savastenko_ElasticGaN_PSSa_1978})  & (83.0~\cite{Semiconductors_BasicData_Springer, Savastenko_ElasticGaN_PSSa_1978})  \\
& (390~\cite{Polian_ElasticGaN_JAP_1996}) & (145.0~\cite{Polian_ElasticGaN_JAP_1996})  & (106.0~\cite{Polian_ElasticGaN_JAP_1996})  & (398~\cite{Polian_ElasticGaN_JAP_1996}) & (105.0~\cite{Polian_ElasticGaN_JAP_1996})  & (123.0~\cite{Polian_ElasticGaN_JAP_1996})  \\
ZnO & 210 & 109 & 93.2 & 220 & 46.4 & 51.4 \\
 & (207~\cite{Semiconductors_BasicData_Springer, Kobiakov_ElasticZnOCdS_SSC_1980}) & (117.7~\cite{Semiconductors_BasicData_Springer, Kobiakov_ElasticZnOCdS_SSC_1980}) & (106.1~\cite{Semiconductors_BasicData_Springer, Kobiakov_ElasticZnOCdS_SSC_1980}) & (209.5~\cite{Semiconductors_BasicData_Springer, Kobiakov_ElasticZnOCdS_SSC_1980}) & (44.8~\cite{Semiconductors_BasicData_Springer, Kobiakov_ElasticZnOCdS_SSC_1980}) & (44.6~\cite{Semiconductors_BasicData_Springer, Kobiakov_ElasticZnOCdS_SSC_1980}) \\
BeO & 427 & 110 & 79.4 & 464 & 138 & 158 \\
 & (460.6~\cite{Cline_JAP_1967}) & (126.5~\cite{Cline_JAP_1967}) & (88.48~\cite{Cline_JAP_1967}) & (491.6~\cite{Cline_JAP_1967}) & (147.7~\cite{Cline_JAP_1967}) & (167.0~\cite{Cline_JAP_1967}) \\
CdS & 80.9 & 47.2 & 39.4 & 87.2 & 14.6 & 17.6 \\
 & (83.1~\cite{Semiconductors_BasicData_Springer, Kobiakov_ElasticZnOCdS_SSC_1980}) & (50.4~\cite{Semiconductors_BasicData_Springer, Kobiakov_ElasticZnOCdS_SSC_1980}) & (46.2~\cite{Semiconductors_BasicData_Springer, Kobiakov_ElasticZnOCdS_SSC_1980}) & (94.8~\cite{Semiconductors_BasicData_Springer, Kobiakov_ElasticZnOCdS_SSC_1980}) & (15.33~\cite{Semiconductors_BasicData_Springer, Kobiakov_ElasticZnOCdS_SSC_1980}) & (16.3~\cite{Semiconductors_BasicData_Springer, Kobiakov_ElasticZnOCdS_SSC_1980}) \\
InSe & 58.95 & 18.0 & 7.5 & 19.6 & 9.95 & 20.5 \\
 & (73.0~\cite{Gatulle_ElasticInSe_PSSb_1983}) & (27.0~\cite{Gatulle_ElasticInSe_PSSb_1983}) & (30.0~\cite{Gatulle_ElasticInSe_PSSb_1983}) & (36.0~\cite{Gatulle_ElasticInSe_PSSb_1983}) & (11.7~\cite{Gatulle_ElasticInSe_PSSb_1983}) & (23.0~\cite{Gatulle_ElasticInSe_PSSb_1983}) \\
InN & 205 & 94.7 & 77.2 & 213 & 48.1 & 55.4 \\
& N/A & N/A & N/A & N/A & N/A & N/A  \\
\end{tabular}}
\label{tab:art115:wurzite_elastic_supp}
\etab

\tab
\mycaption[Elastic constants $c_{11}$, $c_{12}$, $c_{13}$, $c_{14}$, $c_{33}$, $c_{44}$ and $c_{66}$ of  rhombohedral
semiconductors.]
{Experimental values, where available, are shown in parentheses underneath the calculated values.
``N/A'' = Not available for that source.
Units: \GPa.}
\tabvspace
\resizebox{\linewidth}{!}{
\begin{tabular}{l|r|r|r|r|r|r|r}
comp. & $c_{11}^\sAEL$ & $c_{12}^\sAEL$ & $c_{13}^\sAEL$ & $c_{14}^\sAEL$ & $c_{33}^\sAEL$ & $c_{44}^\sAEL$ & $c_{66}^\sAEL$     \\
& ($c_{11}^\EXP$)  & ($c_{12}^\EXP$)  & ($c_{13}^\EXP$)  & ($c_{14}^\EXP$) & ($c_{33}^\EXP$) & ($c_{44}^\EXP$) & ($c_{66}^\EXP$)     \\
\hline
Bi$_2$Te$_3$ & 67.6 & 16.6 & 22.05 & 13.9 & 32.7 & 29.25 & 24.6 \\
& (68.47~\cite{Semiconductors_BasicData_Springer, Jenkins_ElasticBi2Te3_PRB_1972}) & (21.77~\cite{Semiconductors_BasicData_Springer, Jenkins_ElasticBi2Te3_PRB_1972}) & (27.04~\cite{Semiconductors_BasicData_Springer, Jenkins_ElasticBi2Te3_PRB_1972}) & (13.25~\cite{Semiconductors_BasicData_Springer, Jenkins_ElasticBi2Te3_PRB_1972}) & (47.68~\cite{Semiconductors_BasicData_Springer, Jenkins_ElasticBi2Te3_PRB_1972}) & (27.38~\cite{Semiconductors_BasicData_Springer, Jenkins_ElasticBi2Te3_PRB_1972}) & (23.35~\cite{Semiconductors_BasicData_Springer, Jenkins_ElasticBi2Te3_PRB_1972}) \\
Sb$_2$Te$_3$ & 67.8 & 11.2 & 19.1 & 9.92 & 23.2 & 21.35 & 28.8  \\
 & (N/A) & (N/A) & (N/A) & (N/A) & (N/A) & (N/A) & (N/A)   \\
Al$_2$O$_3$ & 458 & 133  & 123 & -22.2 & 437 & 138 & 145 \\
& (197.3~\cite{Goto_ElasticAl2O3_JGPR_1989}) & (162.8~\cite{Goto_ElasticAl2O3_JGPR_1989}) & (116.0~\cite{Goto_ElasticAl2O3_JGPR_1989}) & (-21.9~\cite{Goto_ElasticAl2O3_JGPR_1989}) & (500.9~\cite{Goto_ElasticAl2O3_JGPR_1989}) & (146.8~\cite{Goto_ElasticAl2O3_JGPR_1989}) & (17.25~\cite{Goto_ElasticAl2O3_JGPR_1989}) \\
Cr$_2$O$_3$ & 350 & 145 & 131 & 17.1 & 325 & 128 & 111.5 \\
& (374~\cite{Alberts_ElasticCr2O3_JMMM_1976}) & (148~\cite{Alberts_ElasticCr2O3_JMMM_1976}) & (175~\cite{Alberts_ElasticCr2O3_JMMM_1976}) & (-19~\cite{Alberts_ElasticCr2O3_JMMM_1976}) & (362~\cite{Alberts_ElasticCr2O3_JMMM_1976}) & (159~\cite{Alberts_ElasticCr2O3_JMMM_1976}) & (113~\cite{Alberts_ElasticCr2O3_JMMM_1976})  \\
Bi$_2$Se$_3$ & 135 & 85.2 & 69.4  & 43.7 & 145 & 64.7 & 82.9 \\
& (N/A) & (N/A) & (N/A) & (N/A) & (N/A) & (N/A) & (N/A) \\
\end{tabular}}
\label{tab:art115:rhombo_elastic_supp}
\etab

\tab
\mycaption[Elastic constants $c_{11}$, $c_{12}$, $c_{13}$, $c_{33}$, $c_{44}$ and $c_{66}$ of body-centered tetragonal semiconductors.]
{Experimental values, where available, are shown in parentheses underneath the calculated values.
``N/A'' = Not available for that source.
Units: \GPa.}
\tabvspace
\resizebox{\linewidth}{!}{
\begin{tabular}{l|r|r|r|r|r|r}
comp. & $c_{11}^\sAEL$ & $c_{12}^\sAEL$ & $c_{13}^\sAEL$ & $c_{33}^\sAEL$ & $c_{44}^\sAEL$ & $c_{66}^\sAEL$     \\
& ($c_{11}^\EXP$) & ($c_{12}^\EXP$) & ($c_{13}^\EXP$) & ($c_{33}^\EXP$) & ($c_{44}^\EXP$) & ($c_{66}^\EXP$)   \\
\hline
CuGaTe$_2$ & 67.6 & 36.3 & 37.0 & 66.8 & 32.1 & 31.1 \\
& (N/A) & (N/A) & (N/A) & (N/A) & (N/A) & (N/A) \\
ZnGeP$_2$ & 118 & 48.95 & 51.7 & 117 & 62.1 & 61.05 \\
& (N/A) & (N/A) & (N/A) & (N/A) & (N/A) & (N/A) \\
ZnSiAs$_2$ & 108 & 44.7 & 49.3 & 103 & 55.3 & 53.0 \\
& (N/A) & (N/A) & (N/A) & (N/A) & (N/A) & (N/A) \\
CuInTe$_2$ & 71.4 & 42.1 & 49.55 & 64.5 & 26.6 & 26.7 \\
& (N/A) & (N/A) & (N/A) & (N/A) & (N/A) & (N/A)  \\
AgGaS$_2$ & 93.95 & 61.9 & 62.8 & 75.7 & 25.1 & 28.0  \\
& (87.9~\cite{Grimsditch_ElasticAgGaS2_PRB_1975}) & (58.4~\cite{Grimsditch_ElasticAgGaS2_PRB_1975}) & (59.2~\cite{Grimsditch_ElasticAgGaS2_PRB_1975}) & (75.8~\cite{Grimsditch_ElasticAgGaS2_PRB_1975}) & (24.1~\cite{Grimsditch_ElasticAgGaS2_PRB_1975}) & (30.8~\cite{Grimsditch_ElasticAgGaS2_PRB_1975})  \\
CdGeP$_2$ & 102 & 46.25 & 50.6 & 88.2 & 48.0 & 44.9 \\
& (N/A) & (N/A) & (N/A) & (N/A) & (N/A) & (N/A) \\
CdGeAs$_2$ & 80.15 & 38.7 & 41.6 & 69.8 & 36.1 & 46.4 \\
& (94.5~\cite{Hailing_ElasticCdGeAs_JPCSS_1982}) & (59.6~\cite{Hailing_ElasticCdGeAs_JPCSS_1982}) & (59.7~\cite{Hailing_ElasticCdGeAs_JPCSS_1982}) & (83.4~\cite{Hailing_ElasticCdGeAs_JPCSS_1982}) & (42.1~\cite{Hailing_ElasticCdGeAs_JPCSS_1982}) & (40.8~\cite{Hailing_ElasticCdGeAs_JPCSS_1982}) \\
CuGaS$_2$ & 102 & 57.0 & 60.5 & 104 & 48.9 & 47.9 \\
& (N/A) & (N/A) & (N/A) & (N/A) & (N/A) & (N/A) \\
CuGaSe$_2$ & 93.65 & 57.5 & 58.8 & 92.75 & 39.3 & 37.95 \\
& (N/A) & (N/A) & (N/A) & (N/A) & (N/A) & (N/A) \\
ZnGeAs$_2$ & 93.7 & 40.65 & 42.6 & 92.6 & 48.2 & 47.1 \\
& (N/A) & (N/A) & (N/A) & (N/A) & (N/A) & (N/A) \\
\end{tabular}}
\label{tab:art115:bct_elastic_supp}
\etab

\tab
\mycaption[Elastic constants $c_{11}$, $c_{12}$ and $c_{44}$  of materials with BCC and simple
cubic structures.]
{``N/A'' = Not available for that source.
Units: \GPa.}
\tabvspace
\begin{tabular}{l|r|r|r|r|r|r|r}
comp. & Pearson & $c_{11}^\sAEL$ & $c_{11}^\EXP$ & $c_{12}^\sAEL$ & $c_{12}^\EXP$ & $c_{44}^\sAEL$ & $c_{44}^\EXP$     \\
\hline
CoSb$_3$ & cI32 & 173 & N/A & 31.2 & N/A & 48.0 & N/A \\
IrSb$_3$ & cI32 & 195 & N/A & 48.9 & N/A & 52.85 & N/A \\
InTe & cP2 & 54.4 & N/A & 35.3 & N/A & 7.65 & N/A \\
\end{tabular}
\label{tab:art115:bcc_elastic}
\etab

\tab
\mycaption[Elastic constants $c_{11}$, $c_{12}$, $c_{13}$, $c_{23}$, $c_{33}$, $c_{44}$, $c_{55}$ and $c_{66}$  of materials with orthorhombic structures.]
{Experimental values, where available, are shown in parentheses underneath the calculated values.
``N/A'' = Not available for that source.
Units: \GPa.}
\tabvspace
\resizebox{\linewidth}{!}{
\begin{tabular}{l|r|r|r|r|r|r|r|r|r|r}
comp. & Pearson & $c_{11}^\sAEL$ & $c_{12}^\sAEL$ & $c_{13}^\sAEL$ & $c_{22}^\sAEL$ & $c_{23}^\sAEL$ & $c_{33}^\sAEL$ & $c_{44}^\sAEL$ & $c_{55}^\sAEL$ & $c_{66}^\sAEL$     \\
& & ($c_{11}^\EXP$) & ($c_{12}^\EXP$) & ($c_{13}^\EXP$) & ($c_{22}^\EXP$) & ($c_{23}^\EXP$) & ($c_{33}^\EXP$) & ($c_{44}^\EXP$) & ($c_{55}^\EXP$) & ($c_{66}^\EXP$)   \\
\hline
ZnSb & oP16 & 84.1 & 30.5 & 28.4 & 93.1 & 25.3 & 83.2 & 16.9  & 39.3 & 31.4 \\
& & (N/A) & (N/A) & (N/A) & (N/A) & (N/A) & (N/A)  & (N/A) & (N/A) & (N/A)    \\
Sb$_2$O$_3$ & oP20 & 17.4 & 7.17 & 0.0 & 82.7 & -7.08 & 79.35 & 24.9 & 18.4 & 11.1 \\
& & (N/A) & (N/A) & (N/A) & (N/A) & (N/A) & (N/A)  & (N/A) & (N/A) & (N/A)  \\
\end{tabular}}
\label{tab:art115:orc_elastic}
\etab

\tab
\mycaption[Elastic constants $c_{11}$, $c_{12}$, $c_{13}$, $c_{33}$, $c_{44}$ and $c_{66}$ of materials with tetragonal structures.]
{Experimental values, where available, are shown in parentheses underneath the calculated values.
``N/A'' = Not available for that source.
Units: \GPa.}
\tabvspace
\resizebox{\linewidth}{!}{
\begin{tabular}{l|r|r|r|r|r|r|r|r|r|r}
comp. & Pearson & $c_{11}^\sAEL$ & $c_{12}^\sAEL$ & $c_{13}^\sAEL$ & $c_{33}^\sAEL$ & $c_{44}^\sAEL$ & $c_{66}^\sAEL$     \\
& & ($c_{11}^\EXP$) & ($c_{12}^\EXP$) & ($c_{13}^\EXP$) & ($c_{33}^\EXP$) & ($c_{44}^\EXP$) & ($c_{66}^\EXP$)   \\
\hline
InTe & tI16 & 32.4 & 11.55 & 13.4 & 52.8 & 13.4 & 13.45 \\
& & (N/A) & (N/A) & (N/A) & (N/A) & (N/A) & (N/A)   \\
SnO$_2$ & tP6 & 191 & 128 & 123 & 346 & 73.8 & 168 \\
& & (261.7~\cite{Chang_ElasticSnO2_JGPR_1975}) & (177.2~\cite{Chang_ElasticSnO2_JGPR_1975}) & (155.5~\cite{Chang_ElasticSnO2_JGPR_1975}) & (449.6~\cite{Chang_ElasticSnO2_JGPR_1975}) & (103.07~\cite{Chang_ElasticSnO2_JGPR_1975}) & (207.4~\cite{Chang_ElasticSnO2_JGPR_1975}) \\
\end{tabular}}
\label{tab:art115:tet_elastic}
\etab

\subsection{Conclusions}

We have implemented the ``Automatic Elasticity Library'' framework for \abinitio\
elastic constant calculations, and integrated it with the ``Automatic \GIBBS\ Library'' implementation of the \GIBBS\ quasi-harmonic Debye model within
the  \AFLOW\ and Materials Project ecosystems.
We used it
to  automatically calculate the bulk modulus, shear modulus, Poisson ratio, thermal conductivity, Debye temperature and Gr{\"u}neisen parameter of materials with
various structures and compared them with available experimental results.

A major aim of high-throughput calculations is to identify useful
property descriptors for screening large datasets of structures~\cite{nmatHT}.
Here, we have examined whether the {\it inexpensive} Debye model, despite its well known deficiencies, can be usefully leveraged for estimating thermal properties of materials by analyzing
correlations between calculated and corresponding experimental quantities.

It is found that the \AEL\ calculation of the elastic moduli
reproduces the experimental results quite well, within 5\% to 20\%,
particularly for materials with cubic and
hexagonal structures. The \AGL\ method, using an isotropic approximation
for the bulk modulus, tends to provide a slightly worse quantitative
agreement but still reproduces trends equally well.
The correlations are very high, often above $~0.99$.
Using different values of the Poisson ratio mainly affects Debye temperatures,
while having very little effect on Gr{\"u}neisen parameters.
Several different numerical and empirical equations of state have also been investigated. The differences
between the results obtained from them are
small, but in some cases they are found to introduce an additional
source of error compared to a direct evaluation of the bulk modulus
from the elastic tensor or from the $E(V)$ curve.
Using the different equations of state has very little effect on Debye temperatures,
but has more of an effect on Gr{\"u}neisen parameters.
Currently, the values for \AGL\ properties available in the \AFLOW\ repository are those calculated by numerically fitting the $E_\sDFT(V)$
data and calculating the bulk modulus using Equation~\ref{eq:art115:bulkmod}.
{The effect of using different exchange-correlation functionals was investigated for a subset of 16 materials. The results showed that
\LDA\ tended to overestimate thermomechanical properties such as bulk modulus or Debye temperature, compared to \GGA{}'s tendency
to underestimate. However, neither functional was consistently better than the other at predicting trends. We therefore use \GGA-\PBE\ for
the automated \AEL-\AGL\ calculations in order to maintain consistency with the rest of the \AFLOW\ data.}

The \AEL-\AGL\ evaluation of the Debye temperature provides good
agreement with experiment for this set of materials, whereas the predictions of the Gr{\"u}neisen parameter
are quite poor. However, since the Gr{\"u}neisen parameter is slowly varying for materials sharing crystal structures, the \AEL-\AGL\
methodology provides a reliable screening tool for identifying materials with very high or very low thermal conductivity.
The correlations between the experimental values of the thermal conductivity and those calculated with  \AGL\ are summarized in
Table~\ref{tab:art115:kappa_correlation}. For the entire set of materials examined we find high values of the Pearson correlation
between $\kappa^\EXP$  and $\kappa^\sAGL$, ranging from $0.880$ to $0.933$. It is particularly high, above $0.9$, for materials
with high symmetry (cubic, hexagonal or rhombohedral) structures, but significantly lower for anisotropic materials.
In our previous work on \AGL~\cite{curtarolo:art96}, we used an approximated the value of $\sigma = 0.25$ in Equation~\ref{eq:art115:fpoisson}.
Using instead the Poisson ratio calculated in \AEL, $\sigma^\sAEL$, the overall correlations are improved
by about 5\%, from $0.880$ to $0.928$, in the agreement with previous
work on metals~\cite{Liu_Debye_CMS_2015}. The correlations for
anisotropic materials, such as the body-centered tetragonal set
examined here, improved even more, demonstrating the significance of a
direct evaluation of the Poisson ratio.
This combined algorithm demonstrates the advantage of an integrated high-throughput materials design framework such as \AFLOW,
which enables the calculation of interdependent properties within a single automated workflow.

A direct \AEL\ evaluation of the Poisson ratio, instead of assuming a
simple approximation, e.g.\ a Cauchy solid with $\sigma = 0.25$,
consistently improves the correlations of the \AGL-Debye temperatures
with experiments.
However, it has very little effect on the values obtained for the Gr{\"u}neisen parameter.
Simple approximations lead to more numerically-robust and better system-size scaling calculations,
as they avoid the complications inherent in obtaining the elastic tensor.
{Therefore, \AGL\ could also be used on its own for initial rapid screening,
with \AEL\ being performed later for potentially interesting materials to increase the accuracy of the results.}

\tab
\mycaption{Correlations between experimental values and \AEL\ and \AGL\ results for
elastic and thermal properties for the entire set of materials.}
\tabvspace
\begin{tabular}{l|r|r|r}
property  & Pearson & Spearman & \RMSrD\ \\
          & (linear) & (rank order) \\
\hline
$\kappa^\EXP$ \vs\ $\kappa^\sAGL$  ($\sigma = 0.25$)~\cite{curtarolo:art96} & 0.880 & 0.752 & 1.293 \\
$\kappa^\EXP$ \vs\ $\kappa^\sAGL$ & 0.928 & 0.720 & 2.614 \\
$\kappa^\EXP$ \vs\ $\kappa^\sBM$ & 0.879 & 0.735 & 2.673  \\
$\kappa^\EXP$ \vs\ $\kappa^\sVIN$ & 0.912 &  0.737 & 2.443 \\
$\kappa^\EXP$ \vs\ $\kappa^\sBCN$ & 0.933 & 0.733 & 2.751 \\
\end{tabular}
\label{tab:art115:kappa_correlation}
\etab

With respect to rapid estimation of thermal conductivities,
the approximations in the Leibfried-Schl{\"o}mann formalism
miss some of the details affecting the lattice thermal conductivity, such as the suppression of phonon-phonon scattering due to
large gaps between the branches of the phonon dispersion~\cite{Lindsay_PRL_2013}.
Nevertheless, the high correlations between  $\kappa^\EXP$ and
$\kappa^\sAGL$ found for most of the structure families in this study demonstrate the utility of the \AEL-\AGL\ approach
as a screening method for large databases of materials where
experimental data is lacking or ambiguous.
Despite its intrinsic limitations, the synergy presented by the \AEL-\AGL\ approach
provides the right balance between accuracy and complexity in identifying materials with
promising properties for further investigation.

\subsection{AFLOW AEL-AGL REST-API}
\label{subsec:art115:restapi_keywords}

The \AEL-\AGL\ methodology described in this work is being used to calculate the elastic and thermal properties of materials in a high-throughput
fashion by the \AFLOW\ consortium. The results are now available on the \AFLOW\ database~\cite{aflowlib.org, aflowlibPAPER}
via the \AFLOW\ \RESTAPI~\cite{aflowAPI}. The following optional materials keywords have now been added to the  \AFLOW\ \RESTAPI\
to facilitate accessing this data.

\def\description{\item {{\it Description.}\ }}
\def\type{\item {{\it Type.}\ }}
\def\example{\item {{\it Example.}\ }}
\def\units{\item {{\it Units.}\ }}
\def\syntax{\item {{\it Request syntax.}\ }}

\begin{itemize}

\item
\verb|ael_bulk_modulus_reuss|
\begin{itemize}
\description Returns \AEL\ bulk modulus as calculated using the Reuss average.
\type \verb|number|.
\units GPa.
\example \verb|ael_bulk_modulus_reuss=105.315|.
\syntax \verb|$aurl/?ael_bulk_modulus_reuss|.
\end{itemize}

\item
\verb|ael_bulk_modulus_voigt|
\begin{itemize}
\description Returns \AEL\ bulk modulus as calculated using the Voigt average.
\type \verb|number|.
\units GPa.
\example \verb|ael_bulk_modulus_voigt=105.315|.
\syntax \verb|$aurl/?ael_bulk_modulus_voigt|.
\end{itemize}

\item
\verb|ael_bulk_modulus_vrh|
\begin{itemize}
\description Returns \AEL\ bulk modulus as calculated using the
Voigt-Reuss-Hill (\VRH) average.
\type \verb|number|.
\units GPa.
\example \verb|ael_bulk_modulus_vrh=105.315|.
\syntax \verb|$aurl/?ael_bulk_modulus_vrh|.
\end{itemize}

\item
\verb|ael_elastic_anisotropy|
\begin{itemize}
\description Returns \AEL\ elastic anisotropy.
\type \verb|number|.
\units dimensionless.
\example \verb|ael_elastic_anistropy=0.000816153|.
\syntax \verb|$aurl/?ael_elastic_anisotropy|.
\end{itemize}

\item
\verb|ael_poisson_ratio|
\begin{itemize}
\description Returns \AEL\ Poisson ratio.
\type \verb|number|.
\units dimensionless.
\example \verb|ael_poisson_ratio=0.21599|.
\syntax \verb|$aurl/?ael_poisson_ratio|.
\end{itemize}

\item
\verb|ael_shear_modulus_reuss|
\begin{itemize}
\description Returns \AEL\ shear modulus as calculated using the Reuss average.
\type \verb|number|.
\units GPa.
\example \verb|ael_shear_modulus_reuss=73.7868|.
\syntax \verb|$aurl/?ael_shear_modulus_reuss|.
\end{itemize}

\item
\verb|ael_shear_modulus_voigt|
\begin{itemize}
\description Returns \AEL\ shear modulus as calculated using the Voigt average.
\type \verb|number|.
\units GPa.
\example \verb|ael_shear_modulus_voigt=73.7989|.
\syntax \verb|$aurl/?ael_shear_modulus_voigt|.
\end{itemize}

\item
\verb|ael_shear_modulus_vrh|
\begin{itemize}
\description Returns \AEL\ shear modulus as calculated using the
Voigt-Reuss-Hill (\VRH) average.
\type \verb|number|.
\units GPa.
\example \verb|ael_shear_modulus_vrh=73.7929|.
\syntax \verb|$aurl/?ael_shear_modulus_vrh|.
\end{itemize}

\item
\verb|ael_speed_of_sound_average|
\begin{itemize}
\description Returns \AEL\ average speed of sound calculated from the transverse and longitudinal speeds of sound.
\type \verb|number|.
\units m/s.
\example \verb|ael_speed_of_sound_average=500.0|.
\syntax \verb|$aurl/?ael_speed_of_sound_average|.
\end{itemize}

\item
\verb|ael_speed_of_sound_longitudinal|
\begin{itemize}
\description Returns \AEL\ speed of sound in the longitudinal direction.
\type \verb|number|.
\units m/s.
\example \verb|ael_speed_of_sound_longitudinal=500.0|.
\syntax \verb|$aurl/?ael_speed_of_sound_longitudinal|.
\end{itemize}

\item
\verb|ael_speed_of_sound_transverse|
\begin{itemize}
\description Returns \AEL\ speed of sound in the transverse direction.
\type \verb|number|.
\units m/s.
\example \verb|ael_speed_of_sound_transverse=500.0|.
\syntax \verb|$aurl/?ael_speed_of_sound_transverse|.
\end{itemize}

\end{itemize}

\begin{itemize}

\item
\verb|agl_acoustic_debye|
\begin{itemize}
\description Returns \AGL\ acoustic Debye temperature.
\type \verb|number|.
\units K.
\example \verb|agl_acoustic_debye=492|.
\syntax \verb|$aurl/?agl_acoustic_debye|.
\end{itemize}

\item
\verb|agl_bulk_modulus_isothermal_300K|
\begin{itemize}
\description Returns \AGL\ isothermal bulk modulus at 300~K and zero pressure.
\type \verb|number|.
\units GPa.
\example \verb|agl_bulk_modulus_isothermal_300K=96.6|.
\syntax \verb|$aurl/?agl_bulk_modulus_isothermal_300K|.
\end{itemize}

\item
\verb|agl_bulk_modulus_static_300K|
\begin{itemize}
\description Returns \AGL\ static bulk modulus at 300~K and zero pressure.
\type \verb|number|.
\units GPa.
\example \verb|agl_bulk_modulus_static_300K=99.59|.
\syntax \verb|$aurl/?agl_bulk_modulus_static_300K|.
\end{itemize}

\item
\verb|agl_debye|
\begin{itemize}
\description Returns \AGL\ Debye temperature.
\type \verb|number|.
\units K.
\example \verb|agl_debye=620|.
\syntax \verb|$aurl/?agl_debye|.
\end{itemize}

\item
\verb|agl_gruneisen|
\begin{itemize}
\description Returns \AGL\ Gr{\"u}neisen parameter.
\type \verb|number|.
\units dimensionless.
\example \verb|agl_gruneisen=2.06|.
\syntax \verb|$aurl/?agl_gruneisen|.
\end{itemize}

\item
\verb|agl_heat_capacity_Cv_300K|
\begin{itemize}
\description Returns \AGL\ heat capacity at constant volume (C$_V$) at 300~K and zero pressure.
\type \verb|number|.
\units k$_\mathrm{B}$/cell.
\example \verb|agl_heat_capacity_Cv_300K=4.901|.
\syntax \verb|$aurl/?agl_heat_capacity_Cv_300K|.
\end{itemize}

\item
\verb|agl_heat_capacity_Cp_300K|
\begin{itemize}
\description Returns \AGL\ heat capacity at constant pressure (C$_p$) at 300~K and zero pressure.
\type \verb|number|.
\units k$_\mathrm{B}$/cell.
\example \verb|agl_heat_capacity_Cp_300K=5.502|.
\syntax \verb|$aurl/?agl_heat_capacity_Cp_300K|.
\end{itemize}

\item
\verb|agl_poisson_ratio_source|
\begin{itemize}
\description Returns source of Poisson ratio used to calculate Debye temperature in \AGL. Possible sources include \verb|ael_poisson_ratio_<value>|, in
which case the Poisson ratio was calculated from first principles using \AEL; \verb|empirical_ratio_<value>|, in which case the value was taken
from the literature; and \verb|Cauchy_ratio_0.25|, in which case the default value of 0.25 of the Poisson ratio of a Cauchy solid
was used.
\type \verb|string|.
\example \verb|agl_poisson_ratio_source=ael_poisson_ratio_0.193802|.
\syntax \verb|$aurl/?agl_poisson_ratio_source|.
\end{itemize}

\item
\verb|agl_thermal_conductivity_300K|
\begin{itemize}
\description Returns \AGL\ thermal conductivity at 300~K.
\type \verb|number|.
\units W/m*K.
\example \verb|agl_thermal_conductivity_300K=24.41|.
\syntax \verb|$aurl/?agl_thermal_conductivity_300K|.
\end{itemize}

\item
\verb|agl_thermal_expansion_300K|
\begin{itemize}
\description Returns \AGL\ thermal expansion at 300~K and zero pressure.
\type \verb|number|.
\units 1/K.
\example \verb|agl_thermal_expansion_300K=4.997e-05|.
\syntax \verb|$aurl/?agl_thermal_expansion_300K|.
\end{itemize}

\end{itemize}
\clearpage
\chapter{Data-driven Approaches}
\section{AFLOW-CHULL: Cloud-Oriented Platform for Autonomous Phase Stability Analysis}
\label{sec:art146}

This study follows from a collaborative effort described in Reference~\cite{curtarolo:art146}.

\subsection{Introduction}
Accelerating the discovery of new functional materials demands an efficient determination of synthesizability.
In general, materials synthesis is a multifaceted problem, spanning
\textbf{i.} technical challenges, such as experimental apparatus design and growth conditions~\cite{Jansen_AngChemInt_2002,Potyrailo_ACSCombSci_2011},
as well as
\textbf{ii.} economic and environmental obstacles, including accessibility and handling of necessary components~\cite{Kuzmin_JPCM_2014,curtarolo:art109}.
Phase stability is a limiting factor.
Often, it accounts for the gap between
materials prediction and experimental realization.
Addressing stability requires an understanding of how phases compete thermodynamically.
Despite the wealth of available experimental phase diagrams~\cite{ASMAlloyInternational},
the number of systems explored represents a negligible fraction of
all hypothetical structures~\cite{Walsh_NChem_2015,curtarolo:art124}.
Large materials databases~\cite{aflowlibPAPER,aflowAPI,curtarolo:art104,aflux,nomad,APL_Mater_Jain2013,Saal_JOM_2013,cmr_repository,Pizzi_AiiDA_2016}
enable the construction of calculated phase diagrams,
where aggregate structural and energetic materials data is employed.
The analysis delivers many fundamental thermodynamic descriptors,
including stable/unstable classification,
phase coexistence, measures of robust stability, and determination of
decomposition reactions~\cite{curtarolo:art109,curtarolo:art113,Bechtel_PRM_2018,Li_CMS_2018,Balachandran_PRM_2018}.

As with all informatics-based approaches, \abinitio\ phase diagrams require an abundance of data:
well-converged enthalpies from a variety of different phases.
Many thermodynamic descriptors computed
from the \AFLOWorg\ repository
have already demonstrated predictive power in characterizing phase
stability~\cite{curtarolo:art49,curtarolo:art51,curtarolo:art53,curtarolo:art57,curtarolo:art63,curtarolo:art67,curtarolo:art70,curtarolo:art74,monsterPGM,curtarolo:art106,curtarolo:art109,curtarolo:art112,curtarolo:art113,curtarolo:art117,curtarolo:art126,curtarolo:art130},
including one investigation that resulted in the synthesis of
two new magnets --- the first ever discovered by computational approaches~\cite{curtarolo:art109}.
As exploration embraces more complex systems, such analyses are expected to
become increasingly critical in confining the search space.
In fact, prospects for stable ordered phases diminish with every new component (dimension), despite the growing number of combinations.
This is due to increased competition with
\textbf{i.} phases of lower dimensionality, \eg, ternary phases competing with stable binary phases~\cite{curtarolo:art130}, and
\textbf{ii.} disordered (higher entropy) phases~\cite{curtarolo:art99,curtarolo:art122,curtarolo:art139}.

To address the challenge, a new module has been implemented in the autonomous, open-source~\cite{gnu_license}
\AFLOW\ (\underline{A}utomatic \underline{Flow}) framework for \abinitio\ calculations~\citeAFLOW.
\AFLOWHULL\ (\AFLOW\ \underline{c}onvex \underline{hull}) offers a thermodynamic characterization that can be employed
locally from any \UNIX-like machine, including those running Linux and macOS.
Built-in data curation and validation schemes ensure results are well-converged:
adhering to proper hull statistics, performing outlier detection, and determining structural equivalence.
\AFLOWHULL\ is powered by the \AFLUX\ Search-\API\ (\underline{a}pplication \underline{p}rogramming \underline{i}nterface)~\cite{aflux},
which enables access to more than 2 million compounds from the \AFLOWorg\ repository.
With \AFLUX\ integration, data-bindings are flexible enough to serve any materials database,
including large heterogeneous repositories such as \NOMAD~\cite{nomad}.

Several analysis output types have been created for integration
into a variety of design workflows, including plain text and
\JSON\ (\underline{J}ava\underline{S}cript \underline{O}bject \underline{N}otation) file types.
A small set of example scripts have been included demonstrating
how to employ \AFLOWHULL\ from within a Python environment, much in the spirit of \AFLOWSYM~\cite{curtarolo:art135}.
The \JSON\ output also powers an interactive, online web application offering enhanced presentation of thermodynamic descriptors and
visualization of 2-/3-dimensional hulls.
The application can be accessed through the \AFLOWorg\ portal under ``Apps and Docs'' or directly at {\sf aflow.org/aflow-chull}.

As a test-bed, the module is applied to all 2 million compounds available in the \AFLOWorg\ repository.
After enforcing stringent hull convergence criteria, the module resolves a thermodynamic characterization
for more than 1,300 binary and ternary systems.
Stable phases are screened for previously explored systems and ranked by their
relative stability criterion, a dimensionless quantity capturing the
effect of the phase on the minimum energy surface~\cite{curtarolo:art109}.
Several promising candidates are identified, including
\CHULLCountPromisingTernaries\ $C15_{b}$-type structures $\left(F\overline{4}3m~\#216\right)$ and two half-Heuslers.
Hence, screening criteria based on these thermodynamic descriptors can accelerate the
discovery of new stable phases.
More broadly, the design of more challenging materials, including ceramics~\cite{curtarolo:art80} and metallic glasses~\cite{curtarolo:art112},
benefit from autonomous, integrated platforms such as \AFLOWHULL.

\fig
\includegraphics[width=1.00\linewidth]{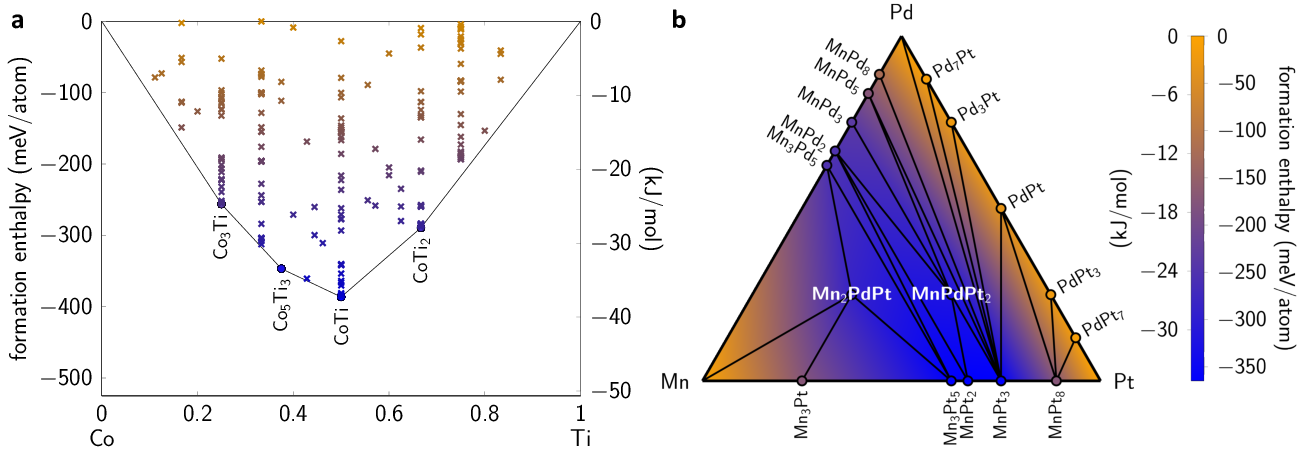}
\mycaption[Example hull illustrations in 2-/3-dimensions as generated by \AFLOWHULL.]
{(\textbf{a}) Co-Ti and (\textbf{b}) Mn-Pd-Pt.}
\label{fig:art146:hull_examples}
\efig

\subsection{Methods}

\fig
\includegraphics[width=1.00\linewidth]{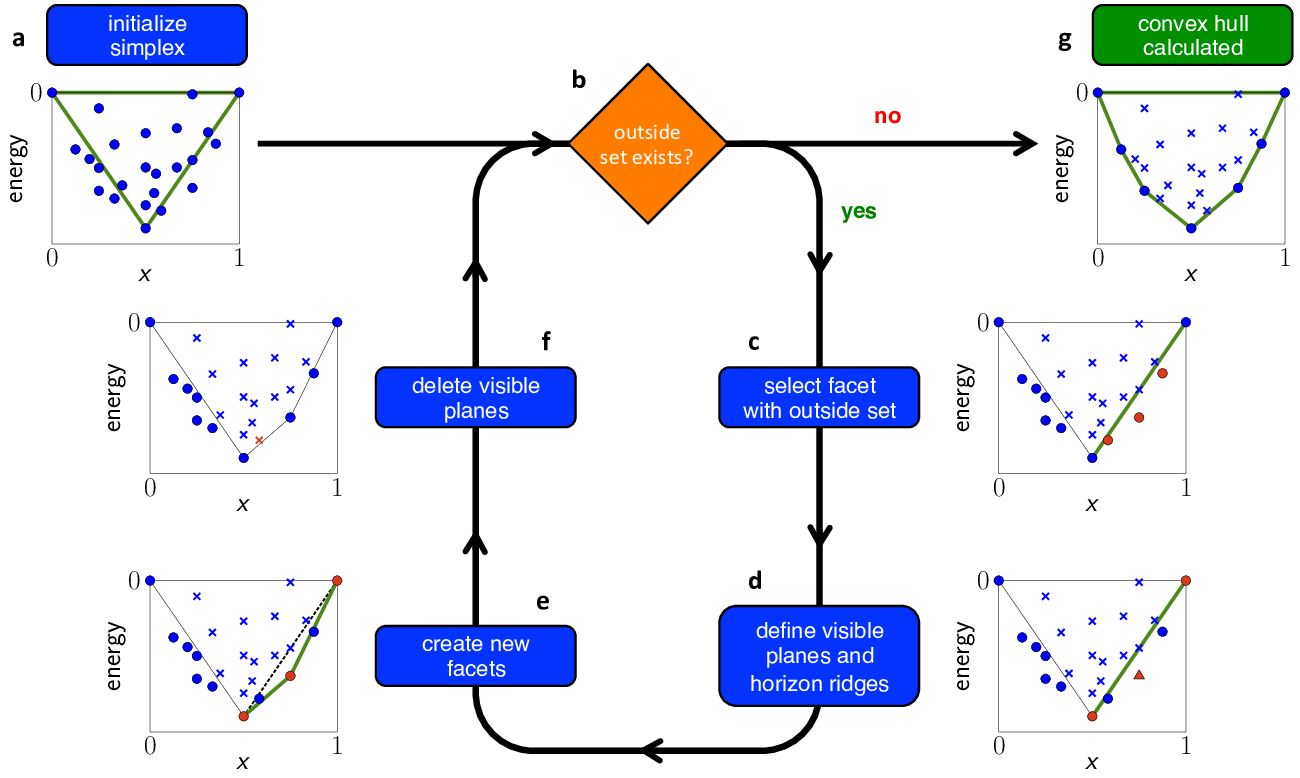}
\mycaption[Illustration of the convex hull construction for a binary system with \AFLOWHULL.]
{The approach is inspired by the \QHULL\ algorithm~\cite{qhull}.
The points on the plot represent structures from the \AFLOWorg\ database~\citeAFLOWLIB.
(\textbf{a}) and (\textbf{g}) denote the beginning and end of the algorithm, respectively.
(\textbf{c}-\textbf{f}) denote the iterative loop that continues until the
condition denoted by (\textbf{b}) is no longer satisfied.
Points are marked with crosses if, by that step in the algorithm, they have been determined to be inside the hull,
and otherwise are marked with circles.
The furthest point from the facet in (\textbf{d}) is marked with a triangle.
Points and facets of interest are highlighted in \textcolor{pranab_red}{{\bf red}} and \textcolor{pranab_green}{{\bf green}}, respectively.}
\label{fig:art146:hull_workflow}
\efig

\boldsection{Defining thermodynamic stability.}
For a multicomponent system at a fixed temperature ($T$) and pressure ($p$),
the minimum Gibbs free energy $G$ (per atom) defines the thermodynamic equilibrium:
\begin{equation}
G(T,p,\{x_{i}\})=H-TS
\label{eq:art146:gibbs_free_energy}
\end{equation}
where $x_{i}$ is the atomic concentration of the $i$-species,
$H$ is the enthalpy, and $S$ is the entropy.
A binary phase $A_{x_{A}}B_{x_{B}}$ is stable at equilibrium with respect to its components
$A$ and $B$ if the corresponding formation reaction releases energy:
\begin{equation}
x_{A} A + x_{B} B \xrightarrow[]{\Delta G<0} A_{x_{A}}B_{x_{B}},
\label{eq:art146:formation_reaction}
\end{equation}
where $\Delta G$ is the energy difference between the mixed phase
and the sum of its components.
Conversely, a positive $\Delta G$ suggests the decomposition of $A_{x_{A}}B_{x_{B}}$ is preferred, and
is thus unstable.
In general, the magnitude of $\Delta G$ quantifies the propensity for the reaction,
and the sign determines the direction.

Relative stability can be visualized on a free-energy-concentration diagram
--- $G$ \vs\ $\left\{ x_i \right\}$ ---
where $\Delta G$ is depicted as the energetic vertical-distance between $A_{x_{A}}B_{x_{B}}$ and the
tie-line connecting $A$ and $B$ end-members (elemental phases).
End-members constitute only a single pathway to formation/decomposition, and
all feasible reactions should be considered for system-wide stability.
{Identification of equilibrium phases} is mathematically equivalent to the construction
of the convex hull --- the set of the most extreme or ``outside'' points (Figure~\ref{fig:art146:hull_examples}(a)).
{The convex hull characterizes the phase stability of the system at equilibrium
and does not include kinetic considerations for synthesis.
Growth conditions affect the final outcome leading to formation of polymorphs and/or metastable phases,
which could differ from the equilibrium phases.
This is a formidable task for high-throughput characterization.
To help identify kinetic pathways for synthesis, \AFLOWHULL\
includes (more in future releases) potential kinetic descriptors,
\eg, chemical decompositions, distance from stability, entropic temperature~\cite{curtarolo:art98},
glass formation ability~\cite{curtarolo:art112}, and spectral entropy analysis for high-entropy systems.}

In the zero temperature limit (as is the case for ground-state density functional theory),
the entropic term of Equation~\ref{eq:art146:gibbs_free_energy} vanishes,
leaving the formation enthalpy term (per atom) as the driving force:
\begin{equation}
  H_\mathrm{f}=H_{A_{x_{A}}B_{x_{B}}}-\left(x_{A} H_{A} + x_{B} H_{B} \right).
\end{equation}
By construction, formation enthalpies of stable elemental phases are zero, restricting
the convex hull to the lower hemisphere.
{Zero-point energies are not yet included in the \AFLOWorg\ repository and thus are neglected from the enthalpy calculations.
Efforts to incorporate vibrational characterizations are underway~\cite{curtarolo:art96,Nath_QHA_2016}.
This contribution could have a large impact on compounds containing light-elements, such as
hydrogen~\cite{Majzoub_PRB_2005}, which comprise a small minority (less than 1\%) of the overall repository.}

By offsetting the enthalpy with that of the elemental phases,
$H_\mathrm{f}$ quantifies the energy gain from forming new bonds between
unlike components,\footnote{The formation enthalpy is not to be confused with the cohesive energy, which quantifies
the energy difference between the phase and its fully gaseous (single atoms) counterpart, \ie,
the energy in all bonds.} \eg, $A-B$.
{Currently, the \AFLOWHULL\ framework does not allow the renormalization of chemical potentials to
improve the calculation of formation enthalpies when gas phases are involved.
A new first-principles approach is being developed and tested in \AFLOW,
and will be implemented in future versions of the \AFLOWHULL\
software together with the available approaches}~\cite{CrUJ,Lany_Zunger_FERE_2012}.

The tie-lines connecting stable phases in Figure~\ref{fig:art146:hull_examples}(a)
define regions of phase separation where the two phases coexist at equilibrium.
The chemical potentials are equal for each component among coexisting phases,
implying the common tangent tie-line construction~\cite{Ganguly_thermo_2008,Darken_pchemmetals_1953}.
{Under thermodynamic equilibrium,} phases above a tie-line will decompose into a linear combination of the stable phases that
define the tie-line (Figure~\ref{fig:art146:hull_analyses}(d)).
The Gibbs phase rule~\cite{McQuarrie} dictates the shape of tie-lines for $N$-ary systems,
which generalizes to $\left(N-1\right)$-dimensional triangles (simplexes) and correspond to facets of the convex hull,
\eg, lines in two dimensions (Figure~\ref{fig:art146:hull_examples}(a)),
triangles in three dimensions (Figure~\ref{fig:art146:hull_examples}(b)),
and tetrahedra in four.
The set of equilibrium facets define the $N$-dimensional minimum energy surface.

\fig
\includegraphics[width=1.00\linewidth]{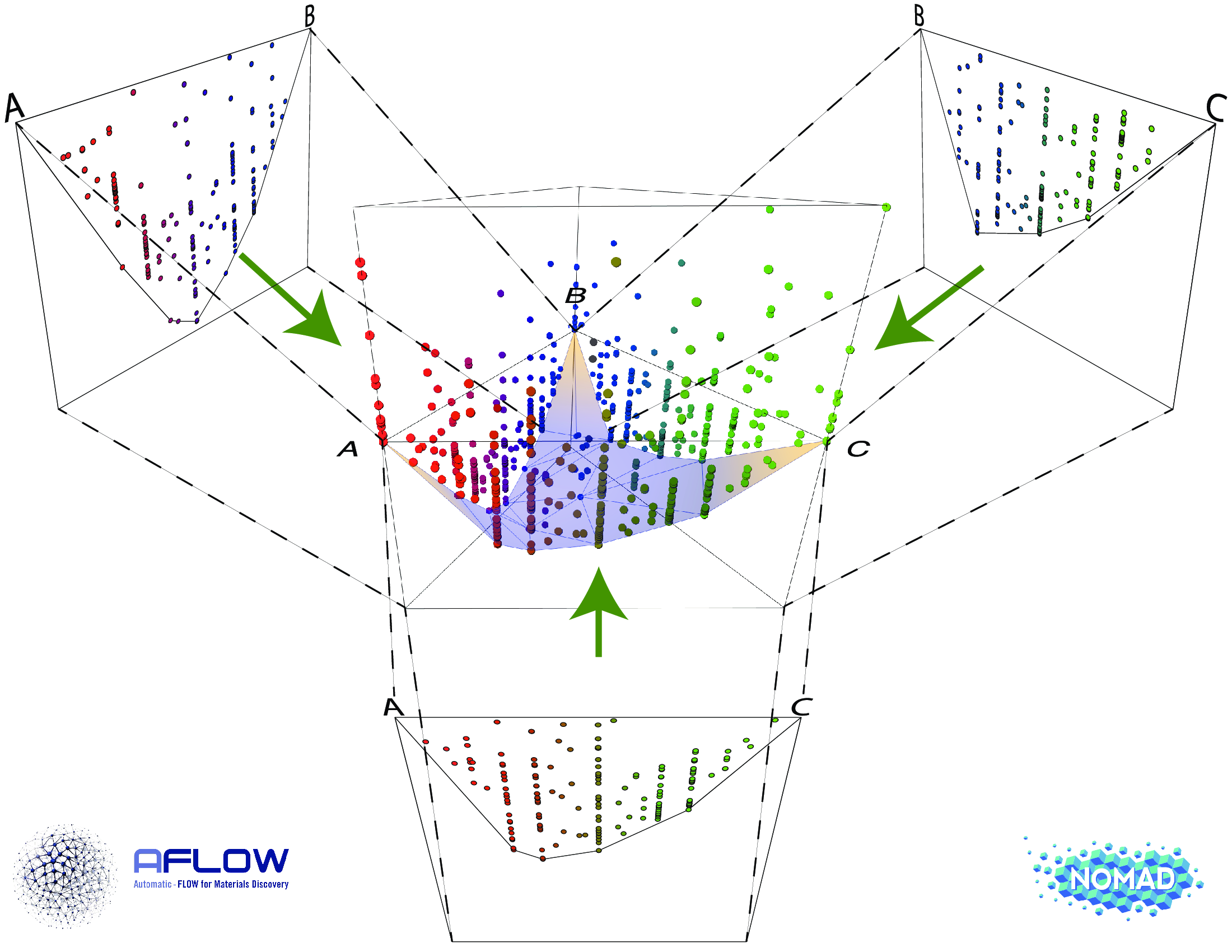}
\mycaption[Illustration of the \AFLOWHULL\ iterative hull scheme.]
{The convex hull and associated properties are first calculated for the binary
hulls, and then propagated to the ternary hull.
This is generalized for $N$-dimensions.}
\label{fig:art146:dimensions}
\efig

\boldsection{Hull construction.}
\AFLOWHULL\ calculates the $N$-dimensional convex hull corresponding to an $N$-ary system
with an algorithm partially inspired by \QHULL~\cite{qhull}.
The algorithm
is efficient in identifying the most important points for construction of facets,
which are treated as hyperplanes instead of boundary-defining inequalities.
\AFLOWHULL\ uniquely accommodates thermodynamic hulls,
\ie, data occupying the lower half hemisphere and
defined by stoichiometric coordinates  $\left(0 \leq x_{i} \leq 1 \right)$.
Points corresponding to individual phases are characterized by their stoichiometric and energetic coordinates:
\begin{equation}
\mathbf{p}=\left[x_{1}, x_{2}, \ldots, x_{N-1}, H_\mathrm{f}\right] = \left[\mathbf{x}, H_\mathrm{f}\right],
\label{eq:art146:point}
\end{equation}
where $x_{N}$ is implicit $\left(\sum_{i}x_i=1\right)$.
Data preparation includes the
\textbf{i.} elimination of phases unstable with respect to end-members (points above the zero $H_{\mathrm{f}}$ tie-line)
and \textbf{ii.} organization of phases by stoichiometry and sorted by energy.
Through this stoichiometry group structure, all but the minimum energy phases are eliminated from
the convex hull calculation.

The workflow is illustrated in Figure~\ref{fig:art146:hull_workflow}.
\AFLOWHULL\ operates by partitioning space, iteratively defining
``inside'' \vs\ ``outside'' half-spaces until all points are either on the hull or inside of it.
First, a simplex is initialized (Figure~\ref{fig:art146:hull_workflow}(a)) with the most extreme points:
stable end-members and the globally stable mixed phase (lowest energy).
A facet is described as:
\begin{equation}
\mathbf{n} \cdot \mathbf{r} + D = 0,
\label{eq:art146:plane_eq}
\end{equation}
where $\mathbf{n}$ is the characteristic normal vector, $\mathbf{r}$ is the position vector,
and $D$ is the offset.
A general hyperplane is defined by $N$ points and $k=\left(N-1\right)$ corresponding edges
$\mathbf{v}_{k}=\mathbf{p}_{k}-\mathbf{p}_{\mathrm{origin}}$.
To construct $\mathbf{n}$, \AFLOWHULL\ employs a generalized cross product approach~\cite{Massey_AMM_1983},
where $n_{i \in \{1,\ldots,N\}}$ (unnormalized) is the $i$-row cofactor
$\left(C_{i,j=0}\right)$ of the matrix $\mathbf{V}$ containing $\mathbf{v}_k$ in its columns:
\begin{equation}
  n_{i} = \left(-1\right)^{i+1}M_{i,j=0}\left(
\begin{bmatrix}
        |         &        & |              \\
  \mathbf{v}_{1}  & \ldots & \mathbf{v}_{k} \\
        |         &        & |              \\
\end{bmatrix}
\right)
\label{eq:art146:hyperplane_normal}
\end{equation}
Here, $M_{i,j=0}\left(\mathbf{V}\right)$ denotes the
$i$-row minor of $\mathbf{V}$,
\ie, the determinant of the submatrix formed by removing the $i$-row.

The algorithm then enters a loop over the facets of the convex hull until no points are declared ``outside'',
defined in the hyperplane description by the signed point-plane distance (Figure~\ref{fig:art146:hull_workflow}(b)).
Each point outside of the hull is singularly assigned to the outside set of a facet (\textcolor{pranab_red}{{\bf red}}
in Figure~\ref{fig:art146:hull_workflow}(c)).
The furthest point from each facet --- by standard point-plane distance --- is selected from the outside set
(marked with a triangle in Figure~\ref{fig:art146:hull_workflow}(d)).
Each neighboring facet is visited to determine whether the furthest point is also outside of it, defining
the set of visible planes (\textcolor{pranab_green}{{\bf green}}) and its boundary,
the horizon ridges (\textcolor{pranab_red}{{\bf red}}) (Figure~\ref{fig:art146:hull_workflow}(d)).
The furthest point is combined with each ridge of the horizon to form new facets (Figure~\ref{fig:art146:hull_workflow}(e)).
The visible planes --- the dotted line in Figure~\ref{fig:art146:hull_workflow}(e) --- are then removed from the
convex hull (Figure~\ref{fig:art146:hull_workflow}(f)).
The fully constructed convex hull --- with all points on the hull or inside of it --- is
summarized in Figure~\ref{fig:art146:hull_workflow}(g).

A challenge arises with lower dimensional data in higher dimensional convex hull constructions.
For example, binary phases composed of the same species all exist on the same (vertical) plane in three dimensions.
A half-space partitioning scheme can make no ``inside'' \vs\ ``outside'' differentiation between such points.
These ambiguously-defined facets\nocite{qhull}\footnote{Ambiguously-defined facets occur when a set of $d+1$ points (or more) define a $(d-1)$-flat~\cite{qhull}.}
constitute a hull outside the scope of the \QHULL\ algorithm~\cite{qhull}.
In the case of three dimensions, the creation of ill-defined facets with collinear edges can result.
Hyper-collinearity --- planes defined with collinear edges, tetrahedra defined with coplanar faces, \etc\ ---
is prescribed by the content (hyper-volume) of the facet.
The quantity resolves the length of the line ($1$-simplex), the area of a triangle ($2$-simplex),
the volume of a tetrahedron ($3$-simplex), \etc,
and is calculated for a simplex of $N$-dimensions via the Cayley-Menger determinant~\cite{sommerville_1929_n_dimensional_geometry}.
Both vertical and content-less facets are problematic for thermodynamic characterizations,
particularly when calculating hull distances, which require facets within finite energetic distances
and well-defined normals.

A dimensionally-iterative scheme is implemented in \AFLOWHULL\ to solve the issue.
It calculates the convex hull for each dimension consecutively
(Figure~\ref{fig:art146:dimensions}).
In the case of a ternary hull, the three binary hulls are calculated first, and the relevant
thermodynamic data is extracted and then propagated forward.
Though vertical and content-less facets are still created in higher dimensions, no thermodynamic
descriptors are extracted from them.
To optimize the calculation, only stable binary structures are propagated forward to
the ternary hull calculation, and this approach is generalized for $N$-dimensions.
The scheme is the default for thermodynamic hulls, resorting back to
the general convex hull algorithm otherwise.

\fig
\includegraphics[width=0.75\linewidth]{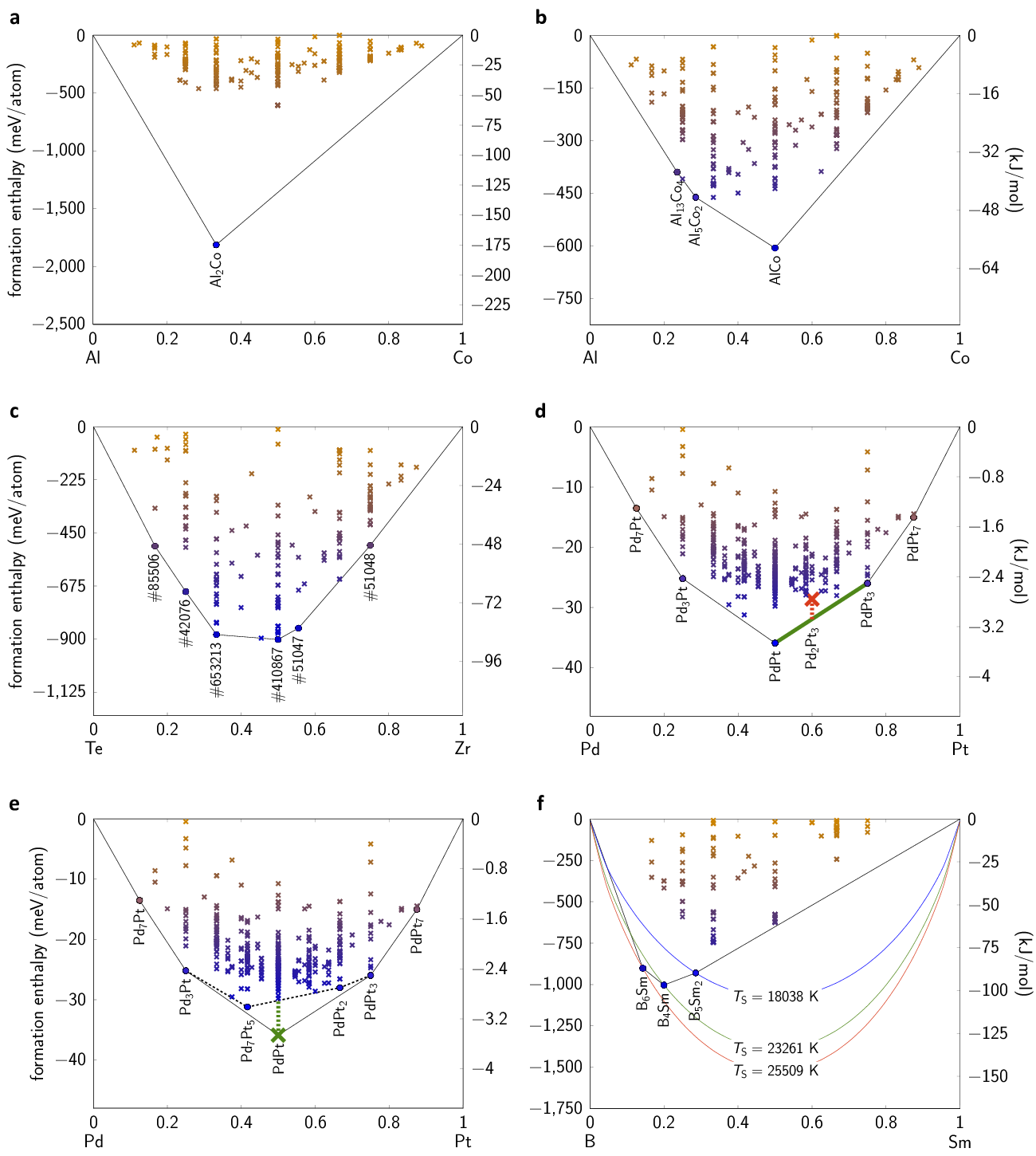}
\mycaption[Illustrations of various automated convex hull analyses in \AFLOWHULL.]
{(\textbf{a}) A plot showing an egregious outlier in the Al-Co convex hull.
(\textbf{b}) The corrected Al-Co convex hull with the outlier removed.
(\textbf{c}) The Te-Zr convex hull with the traditional compound labels replaced
with the corresponding \ICSD\ number designations as determined by a structure
comparison analysis.
If multiple \ICSD\ entries are found for the same stoichiometry, the lowest number
\ICSD\ entry is chosen (chronologically reported, usually).
(\textbf{d}) The Pd-Pt convex hull. The decomposition energy of Pd$_{2}$Pt$_{3}$ is plotted in
\textcolor{pranab_red}{{\bf red}}, and highlighted in \textcolor{pranab_green}{{\bf green}} is
the equilibrium facet directly below it.
The facet is defined by ground-state phases PdPt$_{3}$ and PdPt.
(\textbf{e}) The Pd-Pt convex hull. The stability criterion $\delta_{\mathrm{sc}}$ of PdPt is plotted
in \textcolor{pranab_green}{{\bf green}}, with the pseudo-hull plotted with dashed lines.
(\textbf{f}) The B-Sm convex hull plotted with the
ideal ``{\it iso-max-latent-heat}'' lines of the grand-canonical ensemble~\cite{monsterPGM,curtarolo:art98}
for the ground-state structures.}
\label{fig:art146:hull_analyses}
\efig

\boldsection{Thermodynamic data.}
Structural and energetic data employed to construct the convex hull
is retrieved from the \AFLOWorg~\citeAFLOWLIB{} repository, which contains more than 2 million compounds and
200 million calculated properties.
The database is generated by the autonomous, \abinitio\ framework \AFLOW~\citeAFLOW{}
following the \AFLOW\ Standard for high-throughput materials science
calculations~\cite{curtarolo:art104}.
In particular, calculations are performed
with \VASP\ (\underline{V}ienna \textit{\underline{A}b initio} \underline{S}imulation \underline{P}ackage)~\citeVASP.
Wavefunctions are represented by a large basis set, including
all terms with kinetic energy up to a threshold 1.4 times larger than the recommended defaults.
\AFLOW\ also leverages a large $\mathbf{k}$-point mesh --- as standardized by
a $\mathbf{k}$-points-per-reciprocal-atom scheme~\cite{curtarolo:art104} ---
which is critical for convergence and reliability of calculated properties.
Investigations show that the \AFLOW\ Standard of at least $6,000$ $\mathbf{k}$-points-per-reciprocal-atom
for structural relaxations and $10,000$ for the static calculations ensures
robust convergence of the energies to within one meV/atom in more than 95\% of systems
(including metals which suffer from the discontinuity in the occupancy function at zero temperature),
and within three meV/atom otherwise~\cite{Wisesa_Kgrids_PRB_2016}.

Special consideration is taken for the calculation of $H_{\mathrm{f}}$.
The reference energies for the elemental phases are calculated and stored in the
\LIBONE\ catalog for unary phases in the \AFLOWorg\ repository, and include variations for different
functionals and pseudopotentials.
For consistency, \AFLOWHULL\ only employs data calculated with the \underline{P}erdew-\underline{B}urke-\underline{E}rnzerhof
Generalized Gradient Approximation functional~\cite{PBE}
and pseudopotentials calculated with the
\underline{p}rojector \underline{a}ugmented \underline{w}ave method~\cite{PAW} (\PAW-\PBE).
{Calculations employing \DFT$+U$ corrections to rectify self-interaction errors and energy-gap issues for
electronic properties~\cite{curtarolo:art104} are neglected.
In general, these corrections are parameterized
and material-specific~\cite{curtarolo:art93}.
They artificially augment the energy of the system affecting the reliability of thermodynamic properties.}
It is possible to encounter stable (lowest energy) elemental phases with energy differences from the reference
of order meV/atom, which is the result of duplicate entries (by relaxation or otherwise)
as well as reruns with new parameters, \eg, a denser $\mathbf{k}$-point mesh.
To avoid any issues with the convex hull calculation, the algorithm fixes
the half-space plane at zero.
However, a ``warning'' is prompted in the event that the stable elemental phase differs from
the reference energy by more than 15 meV/atom, yielding a ``skewed'' hull.

Data is retrieved via the \AFLUX\ Search-\API~\cite{aflux}, designed for accessing
property-specific datasets efficiently.
The following is an example of a relevant request:
\begin{center}
\noindent{\sf http://aflowlib.duke.edu/search/API/?species(Mn,Pd),nspecies(2),*,paging(0)}
\end{center}
where {\sf http://aflowlib.duke.edu/search/API/} is the \URL\ for the \AFLUX\ server and
{\sf species(Mn,Pd),nspecies(2),*,paging(0)} is the query.
{\sf species(Mn,Pd)} queries for any entry containing the elements
Mn or Pd, {\sf nspecies(2)} limits the search to binaries only, {\sf *} returns the data
for all available fields, and {\sf paging(0)} amalgamates all data into a single response
without paginating (warning, this can be a large quantity of data).
Such queries are constructed combinatorially for each dimension, \eg,
a general ternary hull $ABC$ constructs the following seven queries:
{\sf species($A$)},
{\sf species($B$)}, and
{\sf species($C$)} with {\sf nspecies(1)},
{\sf species($A$,$B$)},
{\sf species($A$,$C$)}, and
{\sf species($B$,$C$)} with {\sf nspecies(2)}, and
{\sf species($A$,$B$,$C$)} with {\sf nspecies(3)}.

\boldsection{Validation schemes.}
Various statistical analyses and data curation procedures are employed
by \AFLOWHULL\ to maximize fidelity.
At a minimum, each binary hull must contain 200 structures to ensure
a sufficient sampling size for inference.
There is never any guarantee that all stable structures have been identified~\cite{curtarolo:art54,monsterPGM},
but convergence is approached with larger datasets.
With continued growth of \LIBTHREE\ (ternary phases) and beyond, higher dimensional parameters will be incorporated,
though it is expected that the parameters are best defined along tie-lines (\vs\ tie-surfaces).
A comprehensive list of available alloys and structure counts are included in the
Supporting Information of Reference~\cite{curtarolo:art146}.

\boldsection{Outlier detection.} In addition to having been calculated with a standard set of parameters~\cite{curtarolo:art104},
database entries
should also be well-converged.
Prior to the injection of new entries into the \AFLOWorg\ database,
various verification tests are employed to ensure convergence, including an analysis of the
relaxed structure's stress tensor~\cite{aflux}.
Issues stemming from poor convergence and failures in the functional parameterization~\cite{curtarolo:art54,curtarolo:art113}
can change the topology of the convex hull,
resulting in contradictions with experiments.
Hence, an outlier detection algorithm is applied before the hull is constructed:
structures are classified as outliers and discarded if
they have energies that fall well below the first
quartile by a multiple of the interquartile range (conservatively set to 3.25 by default)~\cite{Miller_QJEPSA_1991}.
Only points existing in the lower half-space (phases stable against end-members)
are considered for the outlier analysis, and hence systems need to show
some miscibility, \ie, at least four points for a proper interquartile range determination.
Despite its simplicity, the interquartile range is the preferred estimate of scale
over other measures such as the standard deviation or the median absolute deviation,
which require knowledge of the underlying distribution (normal or otherwise)~\cite{Leys_JESP_2013}.
An example hull (Al-Co) showing an outlier is plotted in Figure~\ref{fig:art146:hull_analyses}(a)
and the corrected hull with the outlier removed is presented in Figure~\ref{fig:art146:hull_analyses}(b).

\boldsection{Duplicate detection.} A procedure for identifying duplicate entries is also employed.
By database construction, near-exact duplicates of elemental phases exist in \LIBTWO,
which is created spanning the full range of compositions for each alloy system (including elemental phases).
These degenerate entries are detected and removed by comparing composition, prototype,
and formation enthalpy.
Other structures may have been created distinctly, but converge to duplicates
via structural relaxation.
These equivalent structures are detected via \AFLOWXTALMATCH\
(\AFLOW\ crys\underline{tal} \underline{match})~\cite{aflow_compare_2018},
which determines structural/material uniqueness via the Burzlaff criteria~\cite{Burzlaff_ActaCrystA_1997}.
To compare two crystals, a commensurate representation between structures is resolved by
\textbf{i.} identifying common unit cells,
\textbf{ii.} exploring cell orientations and origin choices,
and \textbf{iii.} matching atomic positions.
For each description, the structural similarity is measured by
a composite misfit quantity based on the lattice deviations and mismatch of the mapped atomic positions,
with a match occurring for sufficiently small misfit values ($<0.1$).
Depending on the size of the structures, the procedure can be quite expensive,
and only applied to find duplicate stable structures.
Candidates are first screened by composition, space group, and
formation enthalpies (must be within 15~meV/atom of the relevant stable configuration).
{By identifying duplicate stable phases, \AFLOWHULL\ can
cross-reference the \AFLOWorg\ \ICSD\ (\underline{I}norganic \underline{C}rystal \underline{S}tructure \underline{D}atabase)
catalog~\citeICSD{} to reveal whether the structure has already been observed.}
The analysis is depicted in Figure~\ref{fig:art146:hull_analyses}(c), where
the Te-Zr convex hull is plotted with the \verb|compound| labels replaced with the
corresponding \ICSD\ number designation.

\boldsection{Thermodynamic descriptors.}
A wealth of properties can be extracted from the convex hull construction beyond
a simple determination of stable/unstable phases.
For unstable structures, the energetic vertical-distance to the hull $\Delta H_{\mathrm{f}}$,
depicted in Figure~\ref{fig:art146:hull_analyses}(d), serves as a useful metric for quasi-stability.
$\Delta H_{\mathrm{f}}$ is the magnitude of the energy driving the decomposition reaction.
Without the temperature and pressure contributions to the energy,
near-stable structures should also be considered \text{(meta-)stable} candidates,
\eg, those within $k_{\mathrm{B}}T=25$~meV (room temperature) of the hull.
Highly disordered systems can be realized with even larger distances~\cite{Sato_Science_2006,curtarolo:art113}.

\fig
\includegraphics[width=1.00\linewidth]{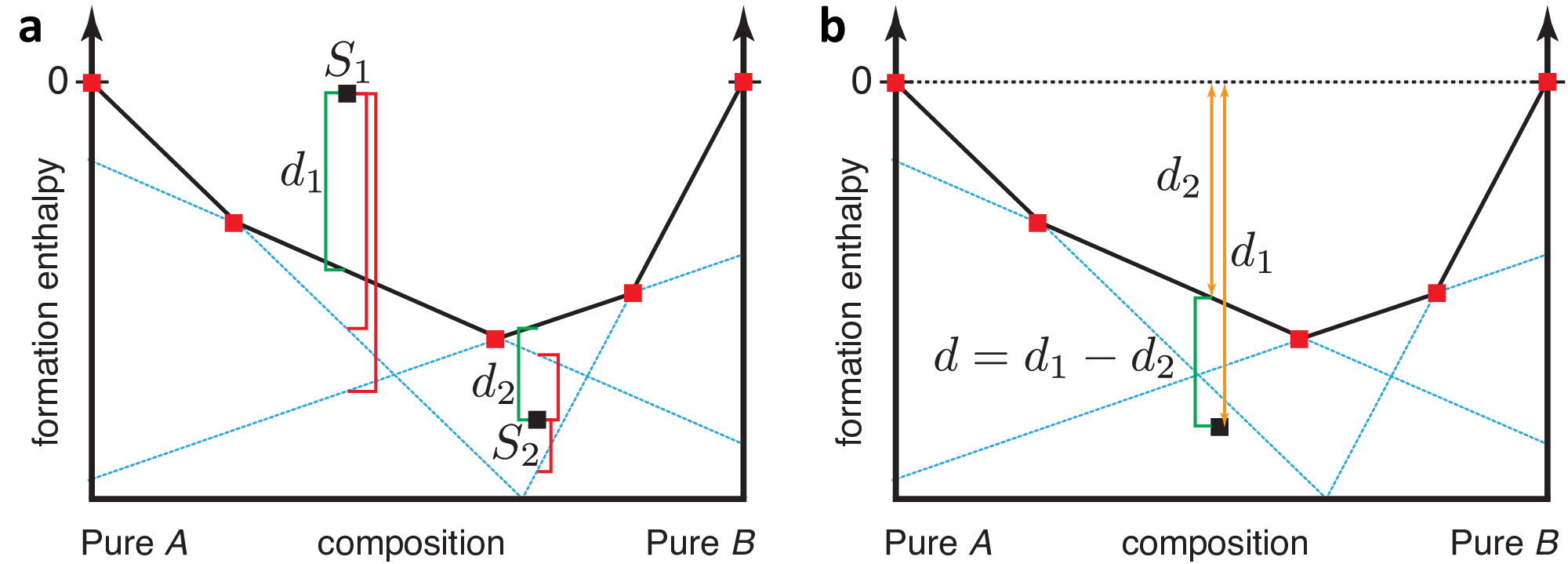}
\mycaption[Distance to the hull algorithm.]
{(\textbf{a}) The correct distance (shown in \textcolor{pranab_green}{{\bf green}}) for $d_1$ is the minimum distance of structure $S_1$
to all hyperplanes defining the convex hull.
In case of structure $S_2$, the minimum distance is not $d_2$ (\textcolor{pranab_green}{{\bf green}}) line), an artifact of the hyperplane
description for hull facets.
(\textbf{b}) Projecting the points to the zero energy line guarantees that all points will lie within the hull,
thus enabling the use of minimization algorithm to calculate the correct distance.
The distance to the hull $d$ is given as the difference of the projected distance $d_2$ from the distance to the zero energy line $d_1$.
The image is adapted from Figure A10 in Reference~\cite{curtarolo:art113}.}
\label{fig:art146:hyperplane_confusion}
\efig

To calculate $\Delta H_{\mathrm{f}}$ of phase $\mathbf{p}$ (Equation~\ref{eq:art146:point}),
\AFLOWHULL\ first resolves the energy of the hull $H_{\mathrm{hull}}$ at
stoichiometric coordinates $\mathbf{x}$, and then
subtracts it from
the phase's formation enthalpy $H_{\mathrm{f}}$:
\begin{equation}
\Delta H_{\mathrm{f}}[\mathbf{p}]=\left|H_{\mathrm{f}}-H_{\mathrm{hull}}[\mathbf{x}]\right|.
\label{eq:art146:dist2hull}
\end{equation}
The procedure is depicted in Figure~\ref{fig:art146:hull_analyses}(d), which involves
identifying the facet (highlighted in \textcolor{pranab_green}{{\bf green}}) that encloses $\mathbf{x}$ and thus defines
$H_{\mathrm{hull}}(\mathbf{x})$.
Here, the hyperplane description can be misleading (Equations~\ref{eq:art146:plane_eq}~and~\ref{eq:art146:hyperplane_normal}) as
it lacks information about facet boundaries (Figure~\ref{fig:art146:hyperplane_confusion}).
The enclosing facet is identified as that which
minimizes the distance to the zero $H_{\mathrm{f}}$ tie-line at $\mathbf{x}$:
\begin{equation}
H_{\mathrm{hull}}[\mathbf{x}]=-\min_{\mathrm{facets}\in \mathrm{hull}}\left|n_N^{-1} \left(D + \sum_{i=1}^{N-1} n_i x_i\right)\right|.
\label{eq:art146:energy_hull}
\end{equation}
Vertical facets and those showing hyper-collinearity (having no content) are excluded from the calculation.

With the appropriate facet identified, the $l$ coefficients of the balanced decomposition reaction
are derived to yield the full equation.
The decomposition of an $N$-ary phase into $l-1$ stable phases
defines an $\left(l \times N\right)$-dimensional chemical composition matrix $\mathbf{C}$,
where $C_{j,i}$ is the atomic concentration of the $i$-species
of the $j$-phase (the first of which is the unstable mixed phase).
Take, for example, the decomposition of $\mathrm{Pd}_{2}\mathrm{Pt}_{3}$
to $\mathrm{PdPt}$ and $\mathrm{PdPt}_{3}$ as presented in Figure~\ref{fig:art146:hull_analyses}(d):
\begin{equation}
N_{1}~\mathrm{Pd}_{0.4}\mathrm{Pt}_{0.6} \to N_{2}~\mathrm{Pd}_{0.5}\mathrm{Pt}_{0.5} + N_{3}~\mathrm{Pd}_{0.25}\mathrm{Pt}_{0.75},
\label{eq:art146:decomp_reaction}
\end{equation}
where $N_{j}$ is the balanced chemical coefficient for the $j$-phase.
In this case, $\mathbf{C}$ is defined as:
\begin{equation}
\begin{bmatrix}
x_{\mathrm{Pd}} \in \mathrm{Pd}_{2}\mathrm{Pt}_{3} & x_{\mathrm{Pt}} \in \mathrm{Pd}_{2}\mathrm{Pt}_{3} \\
-x_{\mathrm{Pd}} \in \mathrm{PdPt} & -x_{\mathrm{Pt}} \in \mathrm{PdPt} \\
-x_{\mathrm{Pd}} \in \mathrm{PdPt}_{3} & -x_{\mathrm{Pt}} \in \mathrm{PdPt}_{3} \\
\end{bmatrix}
=
\begin{bmatrix}
0.4 & 0.6 \\
-0.5 & -0.5 \\
-0.25 & -0.75 \\
\end{bmatrix},
\end{equation}
where a negative sign differentiates the right hand side of the equation from the left.
Reference~\onlinecite{Thorne_ARXIV_2011} shows that $N_{j}$ can be extracted from the null space of $\mathbf{C}$.
\AFLOWHULL\ accesses the null space via a full $\mathbf{QR}$ decomposition of $\mathbf{C}$, specifically employing a general
Householder algorithm~\cite{trefethen1997numerical}.
The last column of the $\left(l \times l\right)$-dimensional $\mathbf{Q}$ orthogonal matrix spans the null space $\mathbf{N}$:
\begin{equation}
  \mathbf{Q} =
\begin{bmatrix}
        |         &       |        & 0.8111 \\
  \mathbf{q}_{1}  & \mathbf{q}_{2} & 0.4867 \\
        |         &       |        & 0.3244 \\
\end{bmatrix}.
\end{equation}
By normalizing $\mathbf{N}$ such that the first element $N_{1}=1$, the approach yields $N_{2}=0.6$ and $N_{3}=0.4$,
which indeed balances Equation~\ref{eq:art146:decomp_reaction}.
These coefficients can be used to verify
the decomposition energy
observed in Figure~\ref{fig:art146:hull_analyses}(d).
The formation enthalpies of Pd$_{2}$Pt$_{3}$, PdPt, and PdPt$_{3}$ are
\mbox{-286~meV/(10~atoms)}, \mbox{-72~meV/(2~atoms)}, and \mbox{-104~meV/(4~atoms)}, respectively.
The decomposition energy is calculated as:
\begin{equation}
0.6 H_{\mathrm{f}}\left[\mathrm{PdPt}\right] + 0.4 H_{\mathrm{f}}\left[\mathrm{PdPt}_{3}\right] - H_{\mathrm{f}}\left[\mathrm{Pd}_{2}\mathrm{Pt}_{3}\right]
= -3~\mathrm{meV/atom},
\end{equation}

For a given stable structure, \AFLOWHULL\ determines the phases with which it is in equilibrium.
For instance, PdPt is in two-phase equilibria with Pd$_{3}$Pt as well as
with PdPt$_{3}$ (Figure~\ref{fig:art146:hull_analyses}(d)).
Phase coexistence plays a key role in defining a descriptor for precipitate-hardened superalloys.
Candidates are chosen if a relevant composition is in two-phase equilibrium with the host matrix,
suggesting that the formation of coherent precipitates in the matrix is feasible~\cite{Kirklin_ActaMat_2016,curtarolo:art113}.

An analysis similar to that quantifying instability $\left(\Delta H_{\mathrm{f}}\right)$
determines the robustness of stable structures.
The stability criterion $\delta_{\mathrm{sc}}$ is defined as the distance of a stable
structure to the pseudo-hull constructed without it
(Figure~\ref{fig:art146:hull_analyses}(e)).
Its calculation is identical to that of $\Delta H_{\mathrm{f}}$ for the pseudo-hull (Equations~\ref{eq:art146:dist2hull}~and~\ref{eq:art146:energy_hull}).
This descriptor quantifies the effect of the structure on the minimum energy surface, as
well as the structure's susceptibility to destabilization by a new phase that has yet to be explored.
As with the decomposition analysis, $\delta_{\mathrm{sc}}$ also serves to anticipate
the effects of temperature and pressure on the minimum energy surface.
The descriptor played a pivotal role in screening Heusler structures for new magnetic systems~\cite{curtarolo:art109}.
$\delta_{\mathrm{sc}}$ calls for the recalculation of facets local to the structure
and all relevant duplicates as well, thus employing the results of the structure comparison
protocol.

\AFLOWHULL\ can also plot the entropic temperature envelopes characterizing nucleation
in hyper-thermal synthesis methods for binary systems~\cite{curtarolo:art98}.
The entropic temperature is the ratio of the formation enthalpy to the mixing entropy for an ideal solution ---
a simple quantification for the resilience against disorder~\cite{monsterPGM}.
The ideal ``{\it iso-max-latent-heat}'' lines
shown in Figure~\ref{fig:art146:hull_analyses}(f)
try to reproduce the phase's capability to absorb latent heat, which can
promote its nucleation over more stable phases when starting from large
Q reservoirs/feedstock.
The descriptor successfully predicts the synthesis of SmB$_{6}$ over SmB$_{4}$
with hyper-thermal plasma co-sputtering~\cite{monsterPGM,curtarolo:art98}.

\subsection{Results}

\fig
\includegraphics[width=1.00\linewidth]{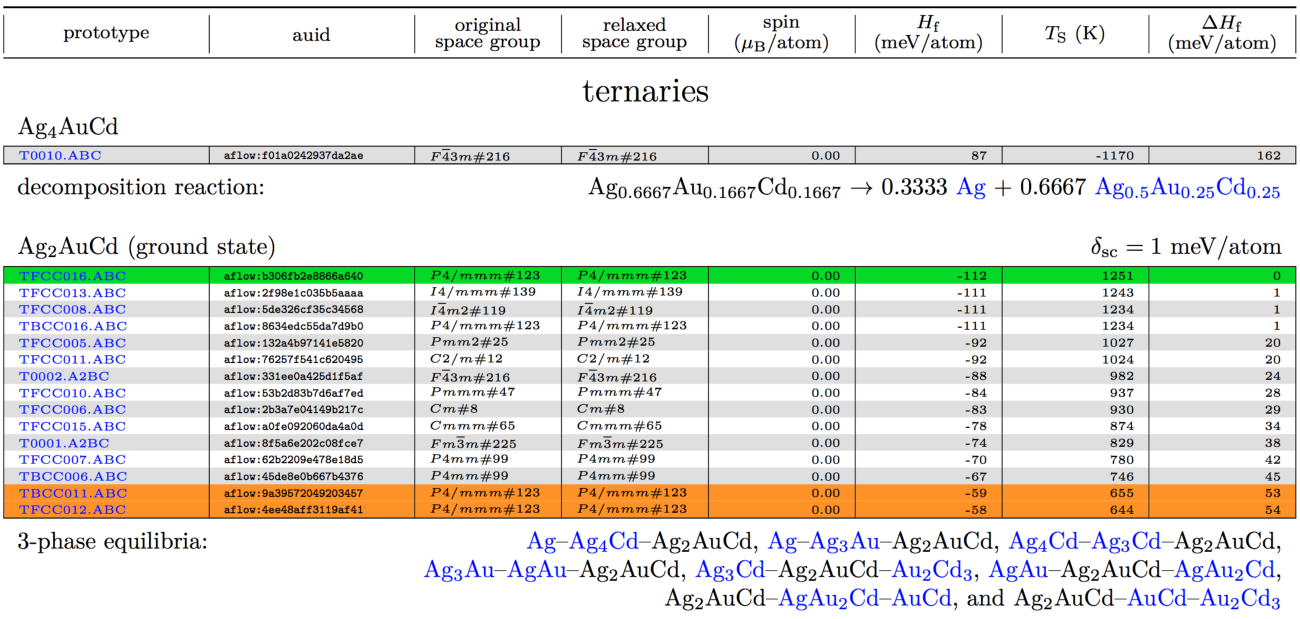}
\mycaption[Excerpt from the Ag-Au-Cd thermodynamic analysis report.]
{The document is generated by \AFLOWHULL\ and showcases
entry-specific data from the \AFLOWorg\ database as well as calculated thermodynamic descriptors.
Structures highlighted in \textcolor{pranab_green}{{\bf green}} are structurally equivalent stable structures,
and those in \textcolor{orange}{{\bf orange}} are structurally similar (same relaxed space group).
The working document includes a variety of links,
including hyperlinks to the entry page of each phase (see prototypes)
and links to relevant parts of the report (see decomposition reaction and
$N$-phase equilibria).}
\label{fig:art146:report}
\efig

\boldsection{Analysis output.}
Following the calculation of the convex hull and relevant thermodynamic descriptors,
\AFLOWHULL\ generates a \PDF\ file summarizing the results.
Included in the \PDF\ are \textbf{i.} an illustration of the convex hull as shown in
Figure~\ref{fig:art146:hull_examples} (for binary and ternary systems)~\cite{pgfplots_manual} and
\textbf{ii.} a report with the aforementioned calculated
thermodynamic descriptors --- an excerpt is shown in Figure~\ref{fig:art146:report}.

In the illustrations, color is used to differentiate points with different enthalpies
and indicate depth of the facets (3-dimensions).
The report includes entry-specific data from the \AFLOWorg\ database (prototype, \AUID,
original and relaxed space groups, spin, formation enthalpy $H_{\mathrm{f}}$, and entropic temperature $T_{\mathrm{S}}$)
as well as calculated thermodynamic data (distance to the hull $\Delta H_{\mathrm{f}}$,
the balanced decomposition reaction for unstable phases, the
stability criterion $\delta_{\mathrm{sc}}$ for stable phases, and
phases in coexistence).
Stable phases (and those that are structurally equivalent) are highlighted in \textcolor{pranab_green}{{\bf green}},
and similar phases (comparing relaxed space groups) are highlighted in \textcolor{orange}{{\bf orange}}.
Links are also incorporated in the report, including external
hyperlinks to entry pages on \AFLOWorg\ (see prototypes) and internal
links to relevant parts of the report (see decomposition reaction and $N$-phase equilibria).
Internal links are also included on the convex hull illustration (see Supporting Information of Reference~\cite{curtarolo:art146}).
The information is provided in the form of plain text and \JSON\ files.
Keys and format are explained in the Supporting Information.

\fig
\includegraphics[width=0.95\linewidth]{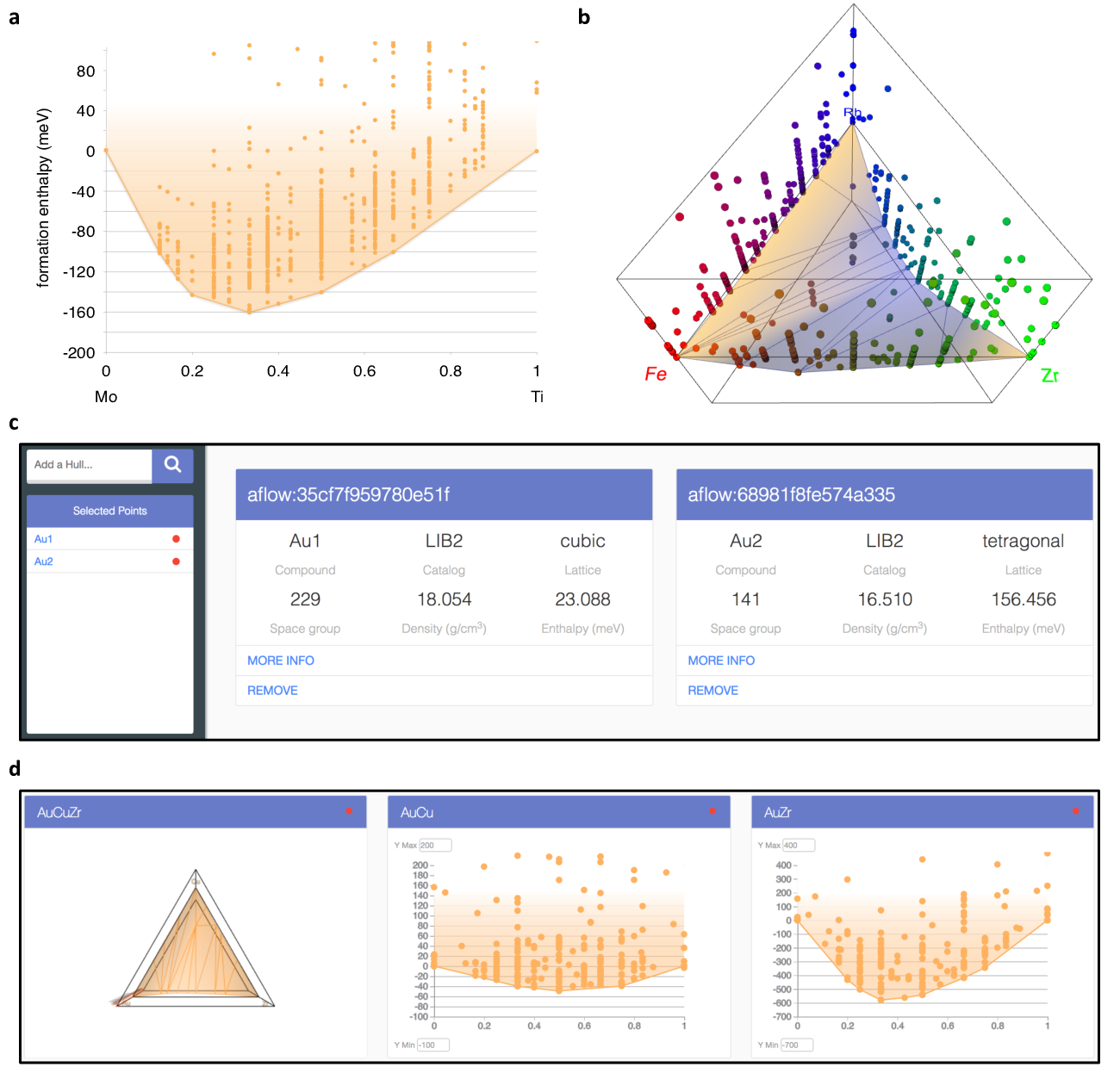}
\mycaption[The convex hull web application powered by \AFLOWHULL.]
{(\textbf{a}) An example 2-dimensional convex hull illustration (Mo-Ti).
(\textbf{b}) An example 3-dimensional convex hull illustration (Fe-Rh-Zr).
(\textbf{c}) The information component of the hull application.
Pertinent thermodynamic data for selected points is displayed within the grid of cards.
Each card includes a link to the \AFLOWorg\ entry page and the option to remove a point.
As points are selected within the visualization, more cards will be added to the grid.
(\textbf{d}) The comparison component of the hull application.
Each hull visualization is displayed as part of a grid of cards.
From this page, new hulls can be added to the store by typing a query in the search box (sidebar).}
\label{fig:art146:hull_app}
\efig

\boldsection{Web application.}
A modern web application has been developed to provide an enhanced, command-line-free platform for \AFLOWHULL.
The project includes a rich feature set consisting of binary and ternary convex
hull visualizations, \AFLOWorg\ entry data retrieval, and a convex hull comparison interface.
The application is divided into four components: the periodic table, the visualization viewport,
the selected entries list, and the comparison page.

The periodic table component is initially displayed.
Hulls can be queried by selecting/typing in the elemental combination.
As elements are added to the search, the periodic table reacts to the query depending on the
reliability of the hull:
\textcolor{pranab_green}{{\bf green}} (fully reliable, $N_{\mathrm{entries}} \geq 200$),
\textcolor{orange}{{\bf orange}} (potentially reliable, $100 \leq N_{\mathrm{entries}} < 200$),
\textcolor{pranab_red}{{\bf red}} (unreliable, $N_{\mathrm{entries}} < 100$), and
\textcolor{gray}{\bf gray} (unavailable, $N_{\mathrm{entries}} =0$).
Each new hull request triggers a fresh data download and analysis,
offering the most up-to-date results given that new calculations are injected into the
\AFLOWorg\ repository daily.
Once the analysis is performed and results are retrieved,
the application loads the visualization viewport
prompting a redirect to the \URL\ endpoint of the selected hull, \eg, {\sf /hull/AlHfNi}.
The \URL\ is ubiquitous and can be shared/cited.

When a binary convex hull is selected, the viewport reveals a traditional
2-dimensional plot (Figure~\ref{fig:art146:hull_app}(a)),
while a ternary hull yields a 3-dimensional visualization (Figure~\ref{fig:art146:hull_app}(b)).
The scales of both are tunable, and the 3-dimensional visualization offers
mouse-enabled pan and zoom.

Common to both types is the ability to select and highlight points.
When a point is selected, its name will appear within the sidebar.
The information component is populated with a grid of cards containing properties of each
selected point (entry), including a link to the \AFLOWorg\ entry page (Figure~\ref{fig:art146:hull_app}(c)).

The application environment stores all previously selected hulls,
which are retrievable via the hull comparison component (Figure~\ref{fig:art146:hull_app}(d)).
On this page each hull visualization is displayed as a card on a grid.
This grid serves as both a history and a means to compare hulls.

\tab
\mycaption[The 25 binary phases predicted to be most stable by \AFLOWHULL.]
{Phases with equivalent structures in the \AFLOW\ \ICSD\ catalog are excluded.
The list is sorted by the absolute value ratio between the stability criterion $\left(\delta_{\mathrm{sc}}\right)$
and the formation enthalpy $\left(H_{\mathrm{f}}\right)$ (shown as a percentage).
${}^{\dagger}$ indicates no binary phase diagram is available on the
\ASM\ Alloy Phase Diagram database~\cite{ASMAlloyInternational}.
\POCC\ denotes a \underline{p}artially-\underline{occ}upied (disordered) structure~\cite{curtarolo:art110}.
Comparisons with the \ASM\ database include phases that are observed at high temperatures and pressures.}
\tabvspace
\resizebox{\linewidth}{!}{
\begin{tabular}{l|l|r|r|r|R{4.75in}}
compound                                                                                       & \AUID                         &             relaxed space group &     $\left|\delta_{\mathrm{sc}}/H_{\mathrm{f}}\right|$ & Figure & comparison with \ASM\ Alloy Phase Diagrams~\cite{ASMAlloyInternational} \\
\hline
\href{http://aflow.org/material.php?id=aflow:38ecc639e4504b9d}{Hf$_{5}$Pb}$^{\dagger}$         & \texttt{aflow:38ecc639e4504b9d}       &                  $P4/mmm~\#123$ &                                               78\% & \ref{fig:art146:HfPb_binary_hull_supp} & no diagram \\
\href{http://aflow.org/material.php?id=aflow:11ba11a3ee157f2e}{AgIn$_{3}$}                     & \texttt{aflow:11ba11a3ee157f2e}       &              $P6_{3}/mmc~\#194$ &                                               54\% & \ref{fig:art146:AgIn_binary_hull_supp} & composition not found, nearest are \href{http://aflow.org/material.php?id=aflow:b60c1f9a1528ba5b}{AgIn$_{2}$} (space group $I4/mcm$, $\Delta H_{\mathrm{f}}$ = 53 meV/atom) and \href{http://aflow.org/material.php?id=aflow:d30bd203dd3b4049}{In} (space group $I4/mmm$)  \\
\href{http://aflow.org/material.php?id=aflow:1da75eb5f31b6dd5}{Hf$_{3}$In$_{4}$}$^{\dagger}$   & \texttt{aflow:1da75eb5f31b6dd5}       &                  $P4/mbm~\#127$ &                                               45\% & \ref{fig:art146:HfIn_binary_hull_supp} & no diagram \\
\href{http://aflow.org/material.php?id=aflow:66dda41a34fe3ad6}{AsTc$_{2}$}$^{\dagger}$         & \texttt{aflow:66dda41a34fe3ad6}       &                     $C2/m~\#12$ &                                               41\% & \ref{fig:art146:AsTc_binary_hull_supp} & no diagram \\
\href{http://aflow.org/material.php?id=aflow:57e1a1246f813f27}{MoPd$_{8}$}                     & \texttt{aflow:57e1a1246f813f27}       &                  $I4/mmm~\#139$ &                                               40\% & \ref{fig:art146:MoPd_binary_hull_supp} & composition not found, nearest are Mo$_{0.257}$Pd$_{0.743}$ (space group $Fm\overline{3}m$, \POCC\ structure) and \href{http://aflow.org/material.php?id=aflow:53b1a8ec286d7fe5}{Pd} (space group $Fm\overline{3}m$) \\
\href{http://aflow.org/material.php?id=aflow:32051219452f8e0f}{Ga$_{4}$Tc}$^{\dagger}$         & \texttt{aflow:32051219452f8e0f}       &         $Im\overline{3}m~\#229$ &                                               39\% & \ref{fig:art146:GaTc_binary_hull_supp} & no diagram \\
\href{http://aflow.org/material.php?id=aflow:7bd140d7b4c65bc1}{Pd$_{8}$V}                      & \texttt{aflow:7bd140d7b4c65bc1}       &                  $I4/mmm~\#139$ &                                               36\% & \ref{fig:art146:PdV_binary_hull_supp} & composition not found, nearest are V$_{0.1}$Pd$_{0.9}$ (space group $Fm\overline{3}m$, \POCC\ structure) and \href{http://aflow.org/material.php?id=aflow:4c0207df2fbbd51e}{VPd$_{3}$} (space group $I4/mmm$, $\Delta H_{\mathrm{f}}$ = 5 meV/atom) \\
\href{http://aflow.org/material.php?id=aflow:e7ed70c4711eb718}{InSr$_{3}$}                     & \texttt{aflow:e7ed70c4711eb718}       &                  $P4/mmm~\#123$ &                                               35\% & \ref{fig:art146:InSr_binary_hull_supp} & composition not found, nearest are Sr$_{28}$In$_{11}$ (space group $Imm2$) and \href{http://aflow.org/material.php?id=aflow:cb9aeb10d6379029}{Sr} (space group $Fm\overline{3}m$) \\
\href{http://aflow.org/material.php?id=aflow:f5cc5eaf65e692a9}{CoNb$_{2}$}                     & \texttt{aflow:f5cc5eaf65e692a9}       &                  $I4/mcm~\#140$ &                                               35\% & \ref{fig:art146:CoNb_binary_hull_supp} & composition not found, nearest are Nb$_{6.7}$Co$_{6.3}$ (space group $R\overline{3}m$, \POCC\ structure) and Nb$_{0.77}$Co$_{0.23}$ (space group $Fm\overline{3}m$, \POCC\ structure) \\
\href{http://aflow.org/material.php?id=aflow:6ee057decaf093d0}{Ag$_{3}$In$_{2}$}               & \texttt{aflow:6ee057decaf093d0}       &                     $Fdd2~\#43$ &                                               34\% & \ref{fig:art146:AgIn_binary_hull_supp} & composition not found, nearest are \href{http://aflow.org/material.php?id=aflow:89453842555b9d95}{Ag$_{9}$In$_{4}$} (space group $P\overline{4}3m$, $\Delta H_{\mathrm{f}}$ = 21 meV/atom) and \href{http://aflow.org/material.php?id=aflow:b60c1f9a1528ba5b}{AgIn$_{2}$} (space group $I4/mcm$, $\Delta H_{\mathrm{f}}$ = 53 meV/atom) \\
\href{http://aflow.org/material.php?id=aflow:360240dae753fec6}{AgPt}                           & \texttt{aflow:360240dae753fec6}       &         $P\overline{6}m2~\#187$ &                                               34\% & \ref{fig:art146:AgPt_binary_hull_supp} & polymorph found (space group $Fm\overline{3}m$, \POCC\ structure) \\
\href{http://aflow.org/material.php?id=aflow:bd3056780447faf0}{OsY$_{3}$}                      & \texttt{aflow:bd3056780447faf0}       &                     $Pnma~\#62$ &                                               34\% & \ref{fig:art146:OsY_binary_hull_supp} & composition found, one-to-one match \\
\href{http://aflow.org/material.php?id=aflow:96142e32718a5ee0}{RuZn$_{6}$}                     & \texttt{aflow:96142e32718a5ee0}       &                $P4_{1}32~\#213$ &                                               33\% & \ref{fig:art146:RuZn_binary_hull_supp} & composition found, one-to-one match \\
\href{http://aflow.org/material.php?id=aflow:1ba6b4b5c0ed9788}{Ag$_{2}$Zn}                     & \texttt{aflow:1ba6b4b5c0ed9788}       &         $P\overline{6}2m~\#189$ &                                               33\% & \ref{fig:art146:AgZn_binary_hull_supp} & composition not found, nearest are \href{http://aflow.org/material.php?id=aflow:46dec61deb1ed379}{Ag} (space group $Fm\overline{3}m$) and Ag$_{4.5}$Zn$_{4.5}$ (space group $P\overline{3}$, \POCC\ structure) \\
\href{http://aflow.org/material.php?id=aflow:87d6637b32224f7b}{MnRh}                           & \texttt{aflow:87d6637b32224f7b}       &         $Pm\overline{3}m~\#221$ &                                               32\% & \ref{fig:art146:MnRh_binary_hull_supp} & \href{http://aflow.org/material.php?id=aflow:19c39238f5d3feb5}{polymorph} found (space group $P4/mmm$, $\Delta H_{\mathrm{f}}$ = 156 meV/atom) \\
\href{http://aflow.org/material.php?id=aflow:f08f2f61de18aa61}{AgNa$_{2}$}                     & \texttt{aflow:f08f2f61de18aa61}       &                  $I4/mcm~\#140$ &                                               32\% & \ref{fig:art146:AgNa_binary_hull_supp} & composition not found, nearest are \href{http://aflow.org/material.php?id=aflow:a174f130a5b9b61f}{NaAg$_{2}$} (space group $Fd\overline{3}m$, $\Delta H_{\mathrm{f}}$ = 208 meV/atom) and \href{http://aflow.org/material.php?id=aflow:95da3ef7fcc58eea}{Na} (space group $R\overline{3}m$) \\
\href{http://aflow.org/material.php?id=aflow:7ce4fcc3660c16cf}{BeRe$_{2}$}                     & \texttt{aflow:7ce4fcc3660c16cf}       &                  $I4/mcm~\#140$ &                                               31\% & \ref{fig:art146:BeRe_binary_hull_supp} & composition not found, nearest are \href{http://aflow.org/material.php?id=aflow:2bb092148157834d}{Be$_{2}$Re} (space group $P6_{3}/mmc$) and \href{http://aflow.org/material.php?id=aflow:47d6720be60b12f3}{Re} (space group $P6_{3}/mmc$) \\
\href{http://aflow.org/material.php?id=aflow:e94ab366799a008c}{As$_{2}$Tc}$^{\dagger}$         & \texttt{aflow:e94ab366799a008c}       &                     $C2/m~\#12$ &                                               30\% & \ref{fig:art146:AsTc_binary_hull_supp} & no diagram \\
\href{http://aflow.org/material.php?id=aflow:eec0d7b6b0d1dfa0}{Be$_{2}$Mn}$^{\dagger}$         & \texttt{aflow:eec0d7b6b0d1dfa0}       &              $P6_{3}/mmc~\#194$ &                                               30\% & \ref{fig:art146:BeMn_binary_hull_supp} & no diagram \\
\href{http://aflow.org/material.php?id=aflow:6f3f5b696f5aa391}{AgAu}                           & \texttt{aflow:6f3f5b696f5aa391}       &                  $P4/mmm~\#123$ &                                               29\% & \ref{fig:art146:AgAu_binary_hull_supp} & polymorph found (space group $Fm\overline{3}m$, \POCC\ structure) \\
\href{http://aflow.org/material.php?id=aflow:ca051dbe25c55b92}{Nb$_{5}$Re$_{24}$}              & \texttt{aflow:ca051dbe25c55b92}       &         $I\overline{4}3m~\#217$ &                                               29\% & \ref{fig:art146:NbRe_binary_hull_supp} & composition not found, nearest are Nb$_{0.25}$Re$_{0.75}$ (space group $I\overline{4}3m$, \POCC\ structure) and Nb$_{0.01}$Re$_{0.99}$ (space group $P6_{3}/mmc$, \POCC\ structure) \\
\href{http://aflow.org/material.php?id=aflow:a9daa69940d3a59a}{La$_{3}$Os}$^{\dagger}$         & \texttt{aflow:a9daa69940d3a59a}       &                     $Pnma~\#62$ &                                               28\% & \ref{fig:art146:LaOs_binary_hull_supp} & no diagram \\
\href{http://aflow.org/material.php?id=aflow:8ce84acfd6f9ea44}{Be$_{5}$Pt}                     & \texttt{aflow:8ce84acfd6f9ea44}       &         $F\overline{4}3m~\#216$ &                                               28\% & \ref{fig:art146:BePt_binary_hull_supp} & composition found, one-to-one match \\
\href{http://aflow.org/material.php?id=aflow:487f7cf6c3fb13f0}{Ir$_{8}$Ru}                     & \texttt{aflow:487f7cf6c3fb13f0}       &                  $I4/mmm~\#139$ &                                               27\% & \ref{fig:art146:IrRu_binary_hull_supp} & composition not found, nearest are \href{http://aflow.org/material.php?id=aflow:1513b1faeafa2d61}{Ir} (space group $Fm\overline{3}m$) and Ru$_{0.3}$Ir$_{0.7}$ (space group $Fm\overline{3}m$, \POCC\ structure) \\
\href{http://aflow.org/material.php?id=aflow:66af8171e22dc212}{InK}                            & \texttt{aflow:66af8171e22dc212}       &          $R\overline{3}m~\#166$ &                                               27\% & \ref{fig:art146:InK_binary_hull_supp} & composition not found, nearest are K$_{8}$In$_{11}$ (space group $R\overline{3}c$) and \href{http://aflow.org/material.php?id=aflow:a9c9107790b0344c}{K} (space group $Im\overline{3}m$) \\
\end{tabular}}
\label{tab:art146:stable_binaries}
\etab

\tab
\mycaption[The 25 ternary phases predicted to be most stable by \AFLOWHULL.]
{Phases with equivalent structures in the \AFLOW\ \ICSD\ catalog are excluded.
The list is sorted by the absolute value ratio between the stability criterion $\left(\delta_{\mathrm{sc}}\right)$
and the formation enthalpy $\left(H_{\mathrm{f}}\right)$ (shown as a percentage).
${}^{\dagger}$ indicates no ternary phase diagram is available on the
\ASM\ Alloy Phase Diagram database~\cite{ASMAlloyInternational},
while ${}^{\ddagger}$ indicates all three relevant binaries are available.
\POCC\ denotes a \underline{p}artially-\underline{occ}upied (disordered) structure~\cite{curtarolo:art110}.
Comparisons with the \ASM\ database include phases that are observed at high temperatures and pressures.}
\tabvspace
\resizebox{\linewidth}{!}{
\begin{tabular}{l|l|r|r|r|R{4.75in}}
compound                                                                                               & \AUID                             &             relaxed space group &     $\left|\delta_{\mathrm{sc}}/H_{\mathrm{f}}\right|$ & Figure & comparison with \ASM\ Alloy Phase Diagrams~\cite{ASMAlloyInternational} \\
\hline
\href{http://aflow.org/material.php?id=aflow:df0cdf0f1ad3110d}{MgSe$_{2}$Zn$_{2}$}$^{\dagger}$         & \texttt{aflow:df0cdf0f1ad3110d}           &                     $Fmmm~\#69$ &                                               58\% & \ref{fig:art146:MgSeZn_ternary_hull_supp} & no diagram, two of three binary phase diagrams found (no Mg-Se) \\
\href{http://aflow.org/material.php?id=aflow:8c51c7ab71f25d11}{Be$_{4}$OsTi}$^{\dagger}$               & \texttt{aflow:8c51c7ab71f25d11}           &         $F\overline{4}3m~\#216$ &                                               38\% & \ref{fig:art146:BeOsTi_ternary_hull_supp} & no diagram, two of three binary phase diagrams found (no Be-Os) \\
\href{http://aflow.org/material.php?id=aflow:4e5711451dc4b601}{Be$_{4}$OsV}$^{\dagger}$                & \texttt{aflow:4e5711451dc4b601}           &         $F\overline{4}3m~\#216$ &                                               38\% & \ref{fig:art146:BeOsV_ternary_hull_supp} & no diagram, two of three binary phase diagrams found (no Be-Os) \\
\href{http://aflow.org/material.php?id=aflow:1684c02e75b0d950}{Ag$_{2}$InZr}                           & \texttt{aflow:1684c02e75b0d950}           &         $Fm\overline{3}m~\#225$ &                                               35\% & \ref{fig:art146:AgInZr_ternary_hull_supp} & composition not found, nearest are Ag$_{0.8}$In$_{0.2}$ (space group $Fm\overline{3}m$, \POCC\ structure), Zr$_{0.5}$In$_{0.5}$ (space group $Fm\overline{3}m$, \POCC\ structure), and AgZr$_{5}$In$_{3}$ (space group $P6_{3}/mcm$) \\
\href{http://aflow.org/material.php?id=aflow:b85addbb42c47ae9}{Be$_{4}$RuTi}$^{\dagger \ddagger}$      & \texttt{aflow:b85addbb42c47ae9}           &         $F\overline{4}3m~\#216$ &                                               32\% & \ref{fig:art146:BeRuTi_ternary_hull_supp} & no diagram, all three binary phase diagrams found \\
\href{http://aflow.org/material.php?id=aflow:cabd6decf5b6c991}{Be$_{4}$FeTi}$^{\dagger \ddagger}$      & \texttt{aflow:cabd6decf5b6c991}           &         $F\overline{4}3m~\#216$ &                                               29\% & \ref{fig:art146:BeFeTi_ternary_hull_supp} & no diagram, all three binary phase diagrams found \\
\href{http://aflow.org/material.php?id=aflow:7010472778d429f7}{Be$_{4}$ReV}$^{\dagger \ddagger}$       & \texttt{aflow:7010472778d429f7}           &         $F\overline{4}3m~\#216$ &                                               29\% & \ref{fig:art146:BeReV_ternary_hull_supp} & no diagram, all three binary phase diagrams found \\
\href{http://aflow.org/material.php?id=aflow:e4cc9eea02d9d303}{Ba$_{2}$RhZn}$^{\dagger}$               & \texttt{aflow:e4cc9eea02d9d303}           &                        $Cm~\#8$ &                                               29\% & \ref{fig:art146:BaRhZn_ternary_hull_supp} & no diagram, two of three binary phase diagrams found (no Ba-Rh) \\
\href{http://aflow.org/material.php?id=aflow:2ace5c5383f8ea10}{Be$_{4}$HfOs}$^{\dagger}$               & \texttt{aflow:2ace5c5383f8ea10}           &         $F\overline{4}3m~\#216$ &                                               27\% & \ref{fig:art146:BeHfOs_ternary_hull_supp} & no diagram, two of three binary phase diagrams found (no Be-Os) \\
\href{http://aflow.org/material.php?id=aflow:de79192a0c4e751f}{Be$_{4}$ReTi}$^{\dagger \ddagger}$      & \texttt{aflow:de79192a0c4e751f}           &         $F\overline{4}3m~\#216$ &                                               27\% & \ref{fig:art146:BeReTi_ternary_hull_supp} & no diagram, all three binary phase diagrams found \\
\href{http://aflow.org/material.php?id=aflow:d484b95ba623f9f7}{Be$_{4}$TcV}$^{\dagger}$                & \texttt{aflow:d484b95ba623f9f7}           &         $F\overline{4}3m~\#216$ &                                               27\% & \ref{fig:art146:BeTcV_ternary_hull_supp} & no diagram, two of three binary phase diagrams found (no Be-Tc) \\
\href{http://aflow.org/material.php?id=aflow:c13660b990eb9570}{Be$_{4}$TcTi}$^{\dagger}$               & \texttt{aflow:c13660b990eb9570}           &         $F\overline{4}3m~\#216$ &                                               27\% & \ref{fig:art146:BeTcTi_ternary_hull_supp} & no diagram, two of three binary phase diagrams found (no Be-Tc) \\
\href{http://aflow.org/material.php?id=aflow:07840d9e13694f7e}{Be$_{4}$RuV}$^{\dagger \ddagger}$       & \texttt{aflow:07840d9e13694f7e}           &         $F\overline{4}3m~\#216$ &                                               27\% & \ref{fig:art146:BeRuV_ternary_hull_supp} & no diagram, all three binary phase diagrams found \\
\href{http://aflow.org/material.php?id=aflow:5778f3b725d5f850}{AsCoTi}$^{\dagger \ddagger}$            & \texttt{aflow:5778f3b725d5f850}           &         $F\overline{4}3m~\#216$ &                                               26\% & \ref{fig:art146:AsCoTi_ternary_hull_supp} & no diagram, all three binary phase diagrams found \\
\href{http://aflow.org/material.php?id=aflow:9a10dd8a8224e158}{Be$_{4}$MnTi}$^{\dagger}$               & \texttt{aflow:9a10dd8a8224e158}           &         $F\overline{4}3m~\#216$ &                                               26\% & \ref{fig:art146:BeMnTi_ternary_hull_supp} & no diagram, two of three binary phase diagrams found (no Be-Mn) \\
\href{http://aflow.org/material.php?id=aflow:de412213bdefbd14}{Be$_{4}$OsZr}$^{\dagger}$               & \texttt{aflow:de412213bdefbd14}           &         $F\overline{4}3m~\#216$ &                                               26\% & \ref{fig:art146:BeOsZr_ternary_hull_supp} & no diagram, two of three binary phase diagrams found (no Be-Os) \\
\href{http://aflow.org/material.php?id=aflow:07bcc161f57da109}{Be$_{4}$IrTi}$^{\dagger}$               & \texttt{aflow:07bcc161f57da109}           &         $F\overline{4}3m~\#216$ &                                               26\% & \ref{fig:art146:BeIrTi_ternary_hull_supp} & no diagram, two of three binary phase diagrams found (no Be-Ir) \\
\href{http://aflow.org/material.php?id=aflow:90b98cdcd6eea146}{Mg$_{2}$ScTl}$^{\dagger}$               & \texttt{aflow:90b98cdcd6eea146}           &                  $P4/mmm~\#123$ &                                               25\% & \ref{fig:art146:MgScTl_ternary_hull_supp} & no diagram, two of three binary phase diagrams found (no Sc-Tl) \\
\href{http://aflow.org/material.php?id=aflow:086b4a89f8d62804}{Be$_{4}$MnV}$^{\dagger}$                & \texttt{aflow:086b4a89f8d62804}           &         $F\overline{4}3m~\#216$ &                                               25\% & \ref{fig:art146:BeMnV_ternary_hull_supp} & no diagram, two of three binary phase diagrams found (no Be-Mn) \\
\href{http://aflow.org/material.php?id=aflow:0595e3d45678a85c}{AuBe$_{4}$Cu}$^{\dagger \ddagger}$      & \texttt{aflow:0595e3d45678a85c}           &         $F\overline{4}3m~\#216$ &                                               25\% & \ref{fig:art146:AuBeCu_ternary_hull_supp} & no diagram, all three binary phase diagrams found \\
\href{http://aflow.org/material.php?id=aflow:d7fed8d4996290f4}{BiRhZr}$^{\dagger \ddagger}$            & \texttt{aflow:d7fed8d4996290f4}           &         $F\overline{4}3m~\#216$ &                                               24\% & \ref{fig:art146:BiRhZr_ternary_hull_supp} & no diagram, all three binary phase diagrams found \\
\href{http://aflow.org/material.php?id=aflow:80bf8ad33a5bb33b}{LiMg$_{2}$Zn}                           & \texttt{aflow:80bf8ad33a5bb33b}           &         $Fm\overline{3}m~\#225$ &                                               21\% & \ref{fig:art146:LiMgZn_ternary_hull_supp} & composition not found, nearest are \href{http://aflow.org/material.php?id=aflow:a66c0917c0faf13f}{Li} (space group $Im\overline{3}m$, $\Delta H_{\mathrm{f}}$ = 2 meV/atom), \href{http://aflow.org/material.php?id=aflow:b83b8ffef10abaa0}{Mg} (space group $P6_{3}/mmc$), and Li$_{0.77}$MgZn$_{1.23}$ (space group $Fd\overline{3}m$, \POCC\ structure) \\
\href{http://aflow.org/material.php?id=aflow:faa814b1222e8aea}{Be$_{4}$RhTi}$^{\dagger}$               & \texttt{aflow:faa814b1222e8aea}           &         $F\overline{4}3m~\#216$ &                                               21\% & \ref{fig:art146:BeRhTi_ternary_hull_supp} & no diagram, two of three binary phase diagrams found (no Be-Rh) \\
\href{http://aflow.org/material.php?id=aflow:26cc4fc55644b0d8}{AuCu$_{4}$Hf}$^{\dagger \ddagger}$      & \texttt{aflow:26cc4fc55644b0d8}           &         $F\overline{4}3m~\#216$ &                                               21\% & \ref{fig:art146:AuCuHf_ternary_hull_supp} & no diagram, all three binary phase diagrams found \\
\href{http://aflow.org/material.php?id=aflow:ab57b1ae74f4c6d4}{Mg$_{2}$SeZn$_{2}$}$^{\dagger}$         & \texttt{aflow:ab57b1ae74f4c6d4}           &                     $Fmmm~\#69$ &                                               21\% & \ref{fig:art146:MgSeZn_ternary_hull_supp} & no diagram, two of three binary phase diagrams found (no Mg-Se) \\
\end{tabular}}
\label{tab:art146:stable_ternaries}
\etab

\boldsection{Candidates for synthesis.}
To demonstrate the capability of \AFLOWHULL, all binary and ternary systems in the \AFLOWorg\ repository
are explored for ones yielding well-converged thermodynamic properties.
Since reliability constraints are built-in, {no pre-filtering is required and}
all potential elemental combinations {are attempted.}
Across all catalogs present in the database, there exist materials composed of
86 elements, including:
H, He, Li, Be, B, C, N, O, F, Ne, Na, Mg, Al, Si, P, S, Cl,
Ar, K, Ca, Sc, Ti, V, Cr, Mn, Fe, Co, Ni, Cu, Zn, Ga, Ge, As, Se, Br, Kr, Rb,
Sr, Y, Zr, Nb, Mo, Tc, Ru, Rh, Pd, Ag, Cd, In, Sn, Sb, Te, I, Xe, Cs, Ba, La,
Ce, Pr, Nd, Pm, Sm, Eu, Gd, Tb, Dy, Ho, Er, Tm, Yb, Lu, Hf, Ta, W, Re, Os, Ir,
Pt, Au, Hg, Tl, Pb, Bi, Ac, Th, and Pa.
Hulls are eliminated if systems
\textbf{i.} are unreliable based on count (fewer than 200 entries among binary combinations), and
\textbf{ii.} show significant immiscibility (fewer than 50 points below the zero $H_{\mathrm{f}}$ tie-line).
Ternary systems are further screened for those containing ternary ground-state structures.
The analysis resulted in the full thermodynamic characterization of \CHULLCountBinaryHulls\ binary and \CHULLCountTernaryHulls\ ternary systems.
The results are provided in the Supporting Information of Reference~\cite{curtarolo:art146}.

Leveraging the \JSON\ outputs, reliable hulls are further explored for new stable phases.
Phases are first screened (eliminated) if an equivalent structure exists in the \AFLOWorg\ \ICSD\
catalog, and candidates are sorted by their relative stability criterion,
\ie, $\left|\delta_{\mathrm{sc}}/H_{\mathrm{f}}\right|$.
This dimensionless quantity captures the effect of the phase on the minimum energy
surface relative to its depth, enabling comparisons across hulls.
{An example Python script that performs this analysis is provided in the Supporting Information.}

\fig
\includegraphics[width=1.0\linewidth]{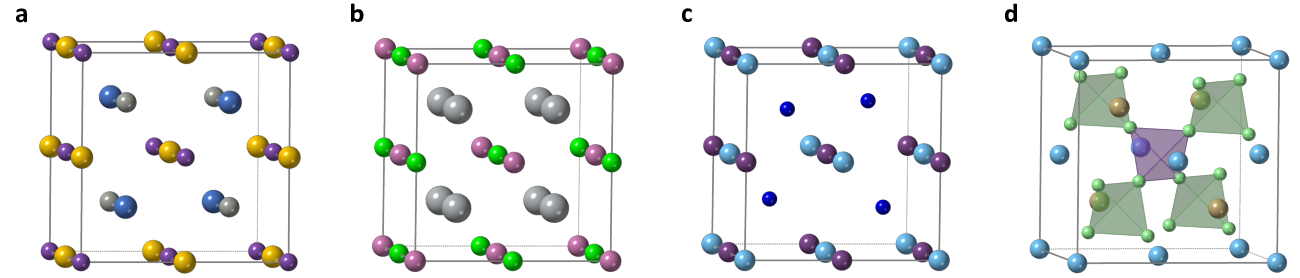}
\mycaption[Illustration of the most prevalent stable ternary structures.]
{(\textbf{a}) The conventional cubic cell of the ``quaternary-Heusler'' structure, LiMgPdSn~\cite{Eberz_ZfNaturfB_35_1341_1980,anrl_pt2_2018}.
Each species occupies a Wyckoff site of space group $F\overline{4}3m~\#216$:
Sn (purple) (4a),
Mg (yellow) (4b),
Pd (gray) (4c), and
Li (blue) (4d).
(\textbf{b}) The conventional cubic cell of the Heusler structure, here represented by
\href{http://aflow.org/material.php?id=aflow:1684c02e75b0d950}{Ag$_{2}$InZr}.
Each species occupies a Wyckoff site of space group $Fm\overline{3}m~\#225$:
In (pink) (4a), Zr (green) (4b), Ag (light gray) (8c).
(\textbf{c}) The conventional cubic cell of the half-Heusler $C1_{b}$ structure, here represented by
\href{http://aflow.org/material.php?id=aflow:5778f3b725d5f850}{AsCoTi}.
Each species occupies a Wyckoff site of space group $F\overline{4}3m~\#216$:
Ti (light blue) (4a), As (purple) (4b), Co (dark blue) (4c).
The (4d) site is empty.
(\textbf{d}) The conventional cubic cell of the $C15_{b}$-type crystal, here represented by
\href{http://aflow.org/material.php?id=aflow:8c51c7ab71f25d11}{Be$_{4}$OsTi}.
Each species occupies a Wyckoff site of space group $F\overline{4}3m~\#216$:
Ti (light blue) (4a),
Os (brown) (4c), and
Be (light green) (8e).
The (4d) site is empty, and the Be atoms form a tetrahedron centered around the (4b) site of (\textbf{a}).}
\label{fig:art146:heuslers}
\efig

The top 25 most stable binary and ternary phases are presented in Tables~\ref{tab:art146:stable_binaries}
and \ref{tab:art146:stable_ternaries}, respectively, for which extended analysis is performed
based on information stored in the \ASM\ (\underline{A}merican \underline{S}ociety for \underline{M}etals)
Alloy Phase Diagram database~\cite{ASMAlloyInternational}.
The \ASM\ database is the largest of its kind, aggregating a wealth of experimental phase diagram information:
40,300 binary and ternary alloy phase diagrams from over 9,000 systems.
Upon searching the \ASM\ website, many binary systems from Table~\ref{tab:art146:stable_binaries}
are unavailable and denoted by the symbol ${}^{\dagger}$.
Among those that are available, some stable phases have already been observed,
including
\href{http://aflow.org/material.php?id=aflow:bd3056780447faf0}{OsY$_{3}$},
\href{http://aflow.org/material.php?id=aflow:96142e32718a5ee0}{RuZn$_{6}$},
and
\href{http://aflow.org/material.php?id=aflow:8ce84acfd6f9ea44}{Be$_{5}$Pt}.
For
\href{http://aflow.org/material.php?id=aflow:360240dae753fec6}{AgPt},
\href{http://aflow.org/material.php?id=aflow:87d6637b32224f7b}{MnRh},
and
\href{http://aflow.org/material.php?id=aflow:6f3f5b696f5aa391}{AgAu},
the composition is successfully predicted,
but polymorphs (structurally distinct phases) are observed instead.
For all other phases on the list, the composition has not been observed.
The discrepancy may be isolated to the phase, or indicative of a more extreme
contradiction in the topology of the hull, and thus, nearby phases are also analyzed.
For the Be-Re system, though \href{http://aflow.org/material.php?id=aflow:7ce4fcc3660c16cf}{BeRe$_{2}$}
has not been observed,
both \href{http://aflow.org/material.php?id=aflow:2bb092148157834d}{Be$_{2}$Re}
and \href{http://aflow.org/material.php?id=aflow:47d6720be60b12f3}{Re}
are successfully identified.
Most of the remaining phases show the nearest phase to be a disordered (partially
occupied) structure, which are excluded from the \AFLOWorg\ repository.
Addressing disorder is a particularly challenging task in \abinitio\ studies.
However, recent high-throughput techniques~\cite{curtarolo:art110} show promise for future investigations
and will be integrated in future releases of the code.

Among the most stable ternary phases, only two systems have available
phase diagrams in the \ASM\ database, Ag-In-Zr and Li-Mg-Zn.
For the Ag-In-Zr system, the composition of
\href{http://aflow.org/material.php?id=aflow:1684c02e75b0d950}{Ag$_{2}$InZr}
is not observed and the nearest stable phases include disordered structures and
AgZr$_{5}$In$_{3}$, which has not yet been included the \AFLOWorg\ repository.
For Li-Mg-Zn, the composition of
\href{http://aflow.org/material.php?id=aflow:80bf8ad33a5bb33b}{LiMg$_{2}$Zn}
is also not observed and the nearest stable phases include unaries
\href{http://aflow.org/material.php?id=aflow:a66c0917c0faf13f}{Li},
\href{http://aflow.org/material.php?id=aflow:b83b8ffef10abaa0}{Mg},
and a disordered structure.
All other ternary systems are entirely unexplored.
Ternary phases with all three binary phase diagrams available
are denoted with the symbol ${}^{\ddagger}$, suggesting experimental feasibility.

A striking feature of Table~\ref{tab:art146:stable_ternaries}
is that most of the stable structures are
found to be in space group $F\overline{4}3m~\#216$.
This structure has a face-centered cubic lattice with symmetry operations that
include a four-fold rotation about the ${<}001{>}$ axes, a three-fold rotation
about the ${<}111{>}$ axes, and no inversion.
Further study reveals that these phases, as well as $Fm\overline{3}m~\#225$
\href{http://aflow.org/material.php?id=aflow:1684c02e75b0d950}{Ag$_{2}$InZr}
and
\href{http://aflow.org/material.php?id=aflow:80bf8ad33a5bb33b}{LiMg$_{2}$Zn},
can be obtained from the ``quaternary-Heusler'' structure,
LiMgPdSn~\cite{Eberz_ZfNaturfB_35_1341_1980,anrl_pt2_2018} (Figure~\ref{fig:art146:heuslers}(a)).
The prototype can be considered a $2\times2\times2$ supercell of the body-centered cubic structure.
The Sn, Mg, Au and Li atoms all occupy different Wyckoff positions of space group
$F\overline{4}3m$ and each atom has two sets of nearest neighbors, each four-fold coordinated.
Various decorations of these Wyckoff positions generate the other structures:
\begin{itemize}
\item By decorating two second-neighbor atom sites identically, a Heusler alloy forms
({\em Strukturbericht} symbol $L2_{1}$)~\cite{Bradley_PRSL_A144_340_1934,aflowANRL}.
For example, the following substitutions generate
\href{http://aflow.org/material.php?id=aflow:1684c02e75b0d950}{Ag$_{2}$InZr}
(Figure~\ref{fig:art146:heuslers}(b)):
Pd $\rightarrow$ Ag, Li $\rightarrow$ Ag,
Sn $\rightarrow$ In, and Mg $\rightarrow$ Zr.
Since the crystal now has an inversion center, the space group becomes
$Fm\overline{3}m~\#225$.
As in LiMgPdSn, each atom has two sets of four-fold coordinated nearest neighbors,
each arranged as a tetrahedron.
Now, however, one species (Ag) has second-neighbors of the same type.
\item By removing the Li atom completely, a half-Heusler forms
($C1_{b}$)~\cite{Nowotny_Z_f_Metallk_33_391_1941,aflowANRL}.
There are two half-Heusler systems in Table~\ref{tab:art146:stable_ternaries}:
\href{http://aflow.org/material.php?id=aflow:5778f3b725d5f850}{AsCoTi}
(Figure~\ref{fig:art146:heuslers}(c)) and
\href{http://aflow.org/material.php?id=aflow:d7fed8d4996290f4}{BiRhZr}.
The structure does differ from that of LiMgPdSn and $L2_{1}$,
as the Ag and Ti atoms are four-fold coordinated, with only Co having the
coordination seen in the previous structures.
\item The majority of structures in Table~\ref{tab:art146:stable_ternaries}
are type $C15_{b}$, prototype AuBe$_{5}$~\cite{Batchelder_Acta_Crist_11_122_1958,aflowANRL}
(\AFLOW\ prototype: \verb|AB5_cF24_216_a_ce|~\cite{AB5_cF24_216_a_ce}),
represented by
\href{http://aflow.org/material.php?id=aflow:8c51c7ab71f25d11}{Be$_{4}$OsTi}
shown in Figure~\ref{fig:art146:heuslers}(d).
Compared to the $C1_{b}$, $C15_{b}$ contains an (8e) Wyckoff position
forming a tetrahedra centered around the (4b) Wyckoff position.
Replacing the tetrahedra with a single atom returns the $C1_{b}$ structure.
\end{itemize}
Hence, of the 25 most stable ternary structures, 21 are of related structure.

Sampling bias likely plays a role in the high prominence of space group $F\overline{4}3m~\#216$
structures in Table~\ref{tab:art146:stable_ternaries}, but cannot fully account for the anomaly.
Space group $F\overline{4}3m~\#216$ constitutes about 17\% of the \LIBTHREE\ catalog,
containing the bulk of the \AFLOWorg\ repository (at over 1.5 million ternary systems)
generated largely by small structure prototypes.
For context, space group $F\overline{4}3m~\#216$ is ranked about twentieth of the
most common space groups in the \ICSD~\cite{Urusov_JSC_2009},
appearing in about 1\% of all entries.
Further exploration of larger structure ternary prototypes covering the full range
of space groups is needed to fully elucidate the nature of this structure's stability.

The \mbox{regular-}, inverse-, and half-Heusler prototypes were added to \LIBTHREE\
for the exploration of new magnets, of which two were discovered~\cite{curtarolo:art109}.
The Heusler set includes more 236,000 structures, most of which remains unexplored.
The fully sorted lists of stable binary and ternary phases are presented in the
Supporting Information of Reference~\cite{curtarolo:art146}.

\clearpage

\subsection{Convex hulls of most stable candidates}

\figsec
\includegraphics[width=0.75\linewidth]{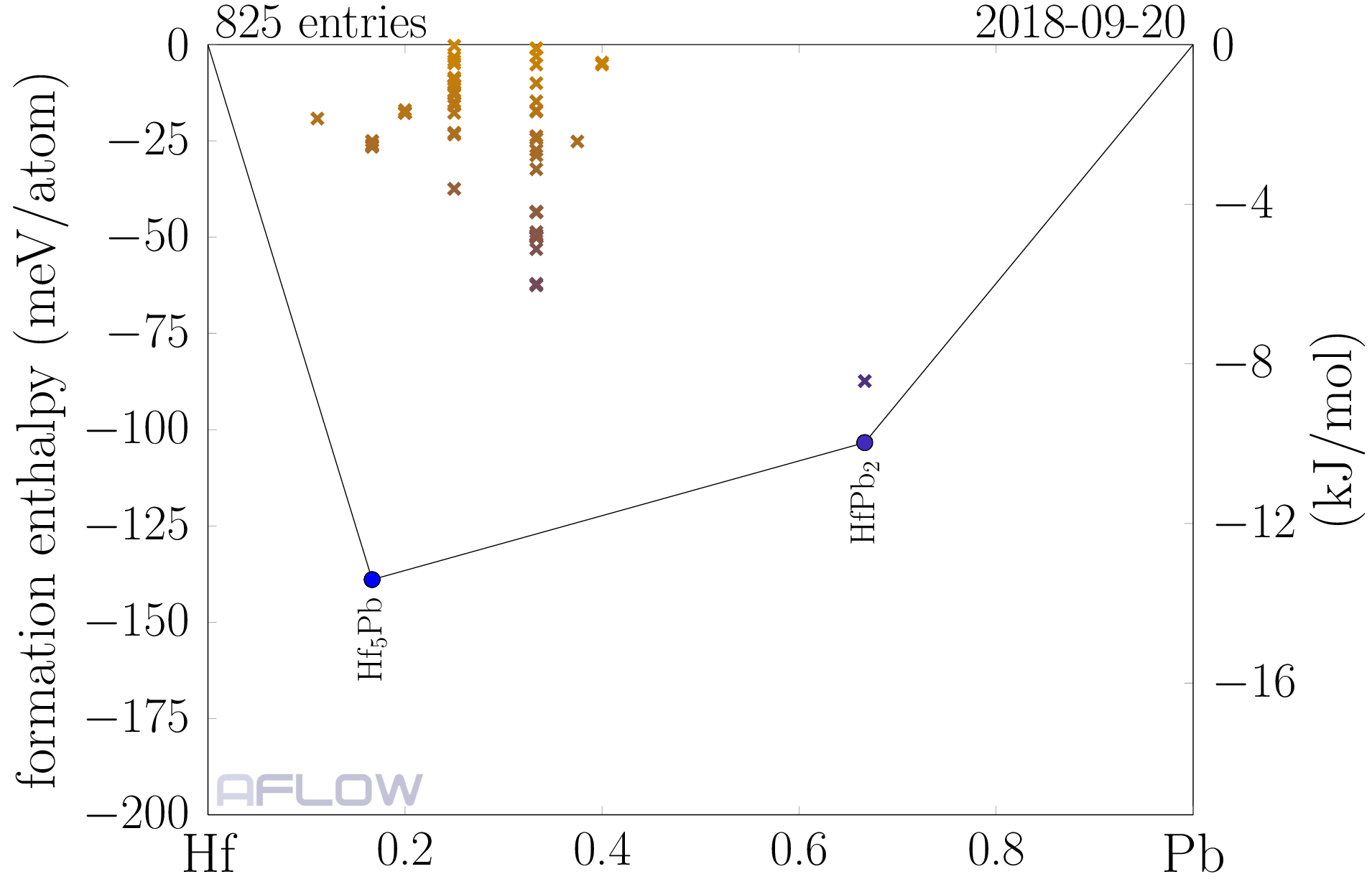}
\mycaption{Hf-Pb binary convex hull as plotted by \AFLOWHULL.}
\label{fig:art146:HfPb_binary_hull_supp}
\efig

\vspace{\fill}

\figsec
\includegraphics[width=0.75\linewidth]{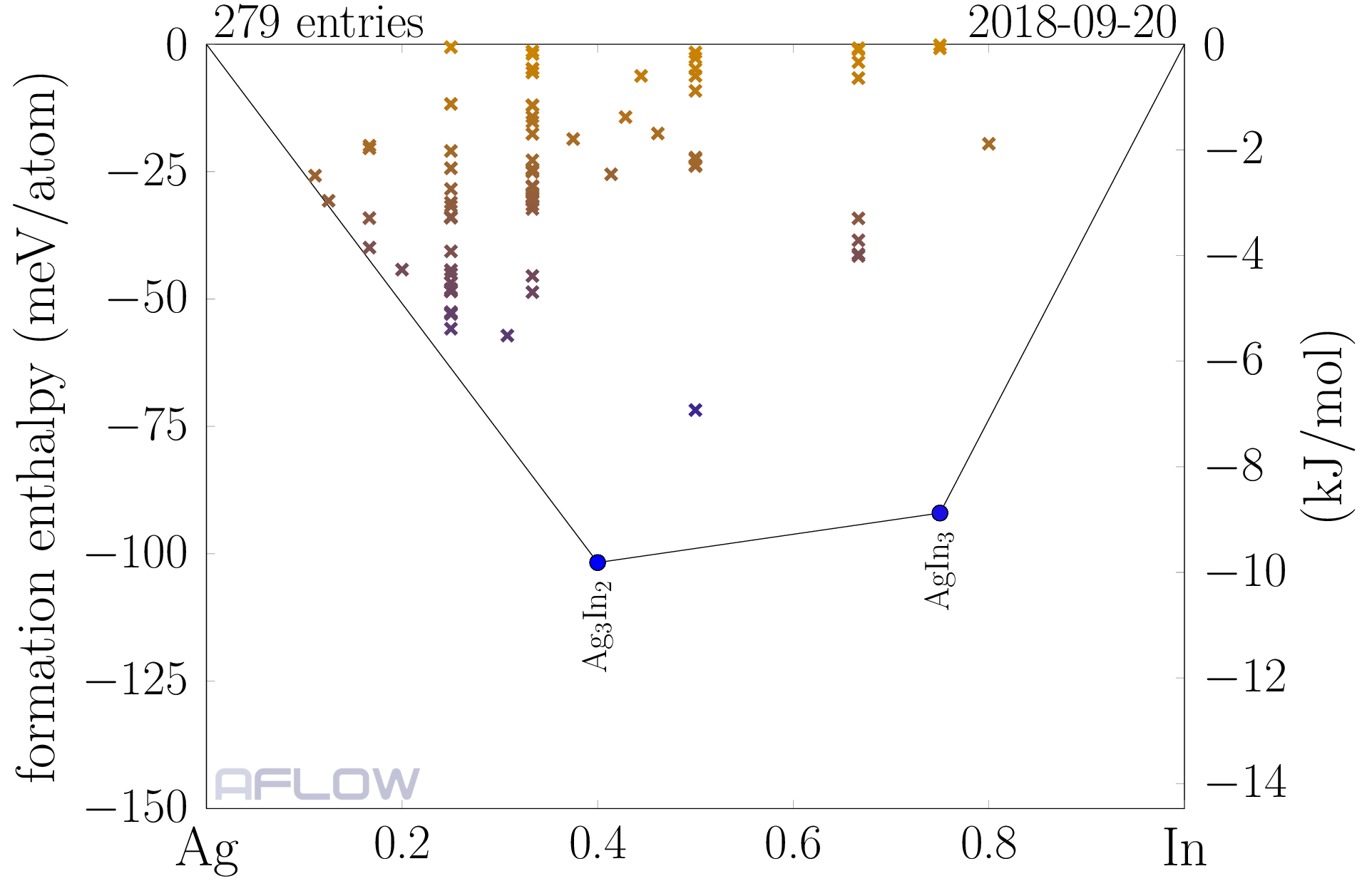}
\mycaption{Ag-In binary convex hull as plotted by \AFLOWHULL.}
\label{fig:art146:AgIn_binary_hull_supp}
\efig

\clearpage

\figsec
\includegraphics[width=0.75\linewidth]{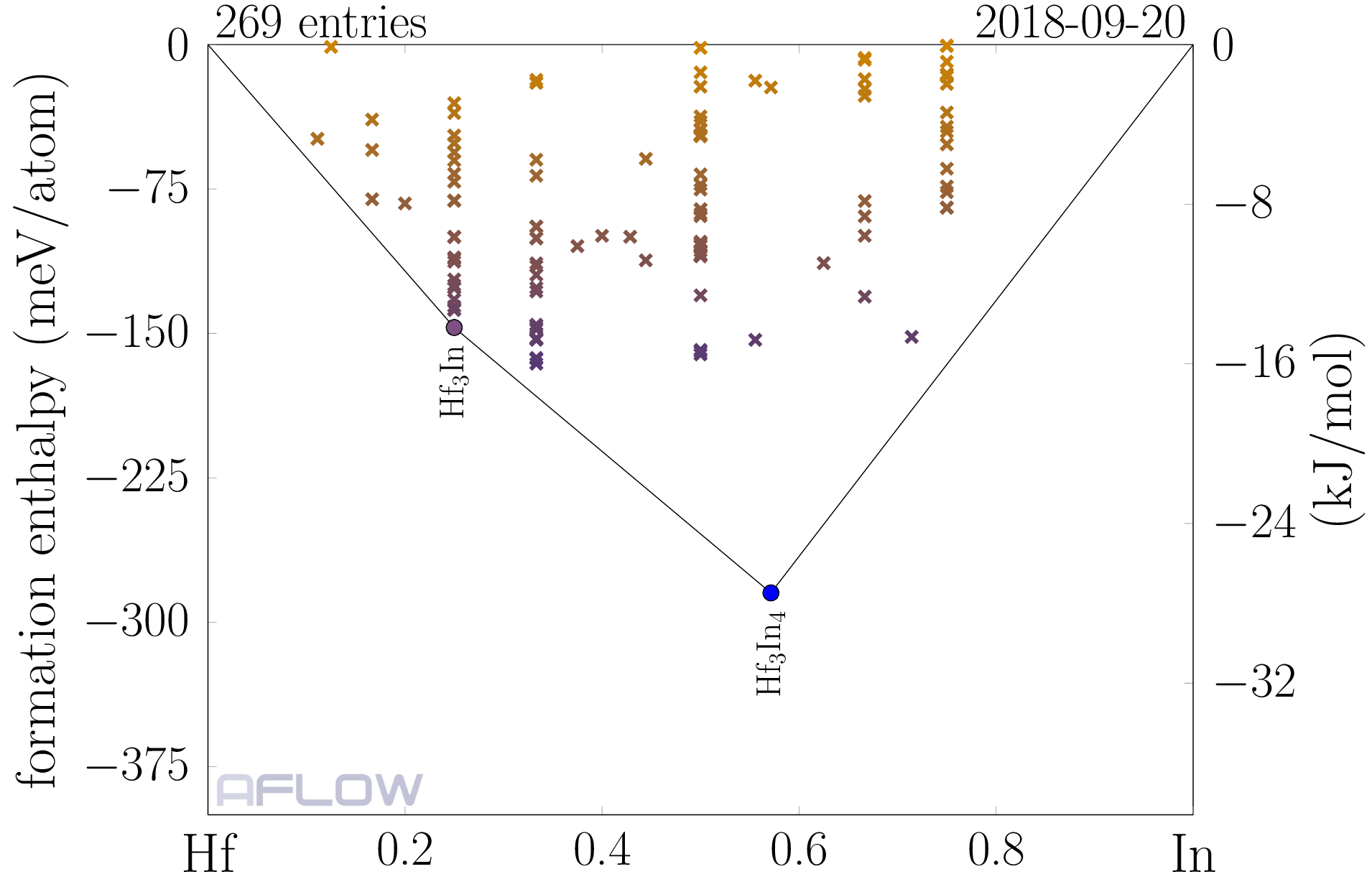}
\mycaption{Hf-In binary convex hull as plotted by \AFLOWHULL.}
\label{fig:art146:HfIn_binary_hull_supp}
\efig

\vspace{\fill}

\figsec
\includegraphics[width=0.75\linewidth]{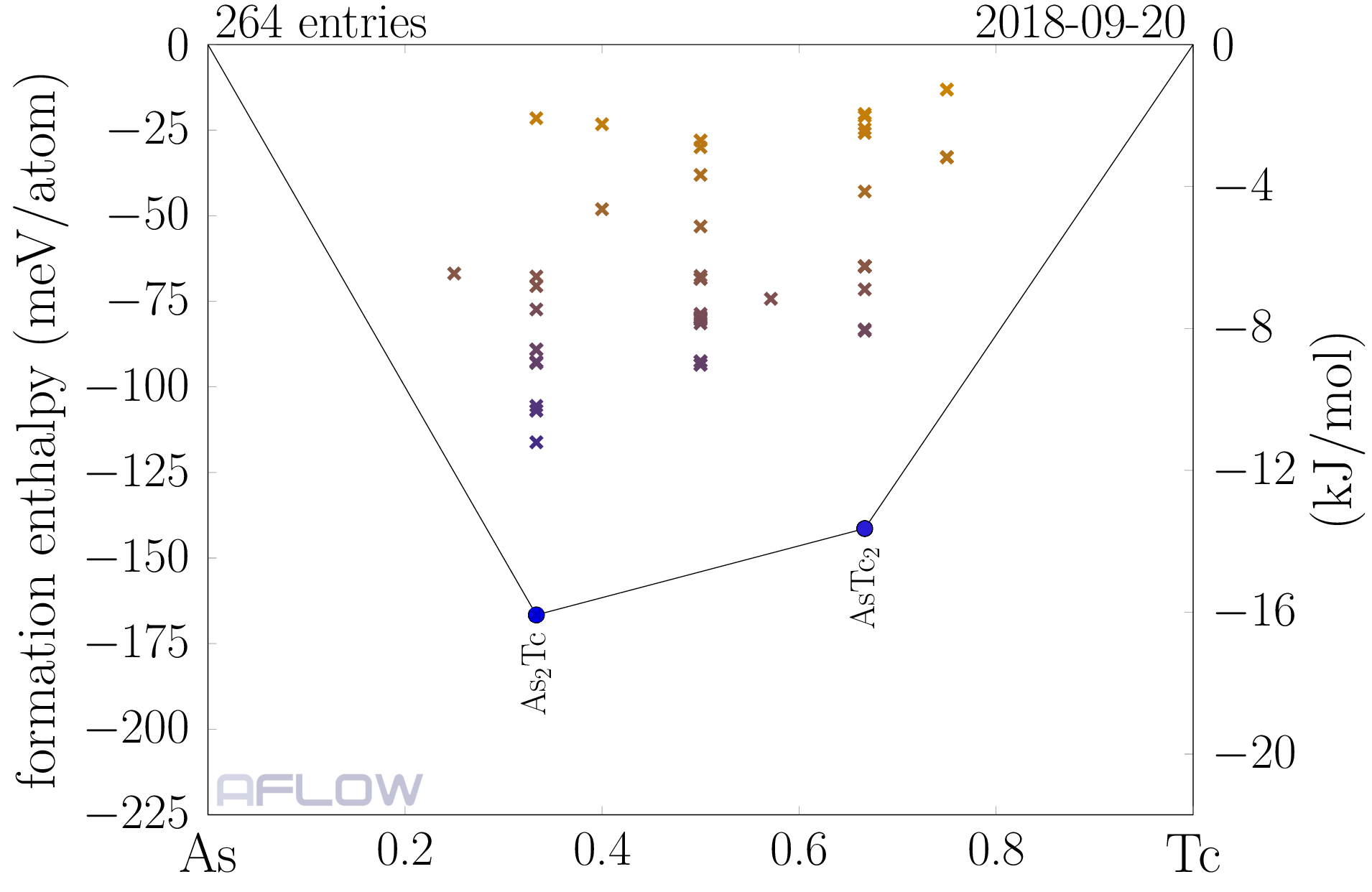}
\mycaption{As-Tc binary convex hull as plotted by \AFLOWHULL.}
\label{fig:art146:AsTc_binary_hull_supp}
\efig

\clearpage

\figsec
\includegraphics[width=0.75\linewidth]{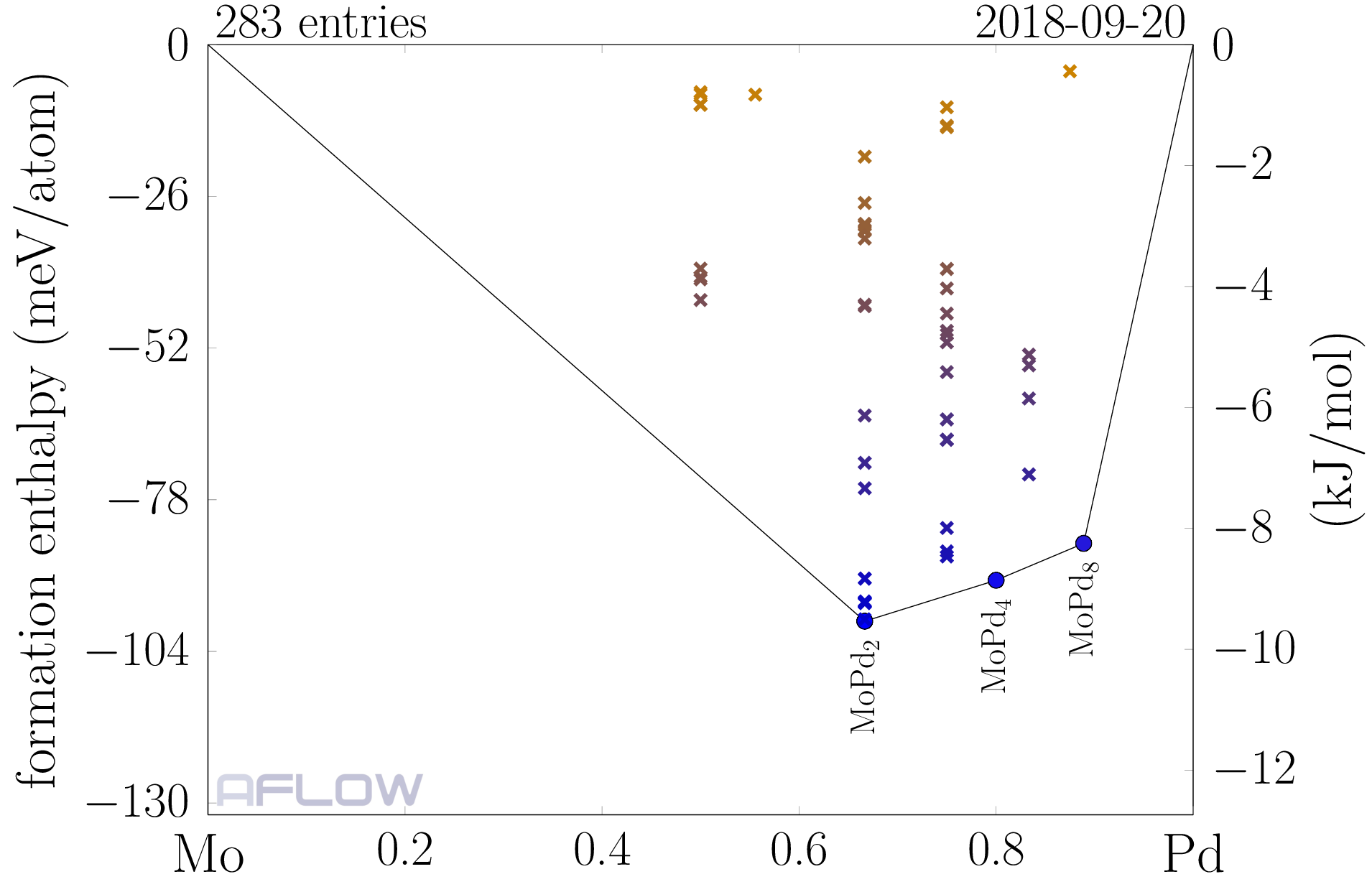}
\mycaption{Mo-Pd binary convex hull as plotted by \AFLOWHULL.}
\label{fig:art146:MoPd_binary_hull_supp}
\efig

\vspace{\fill}

\figsec
\includegraphics[width=0.75\linewidth]{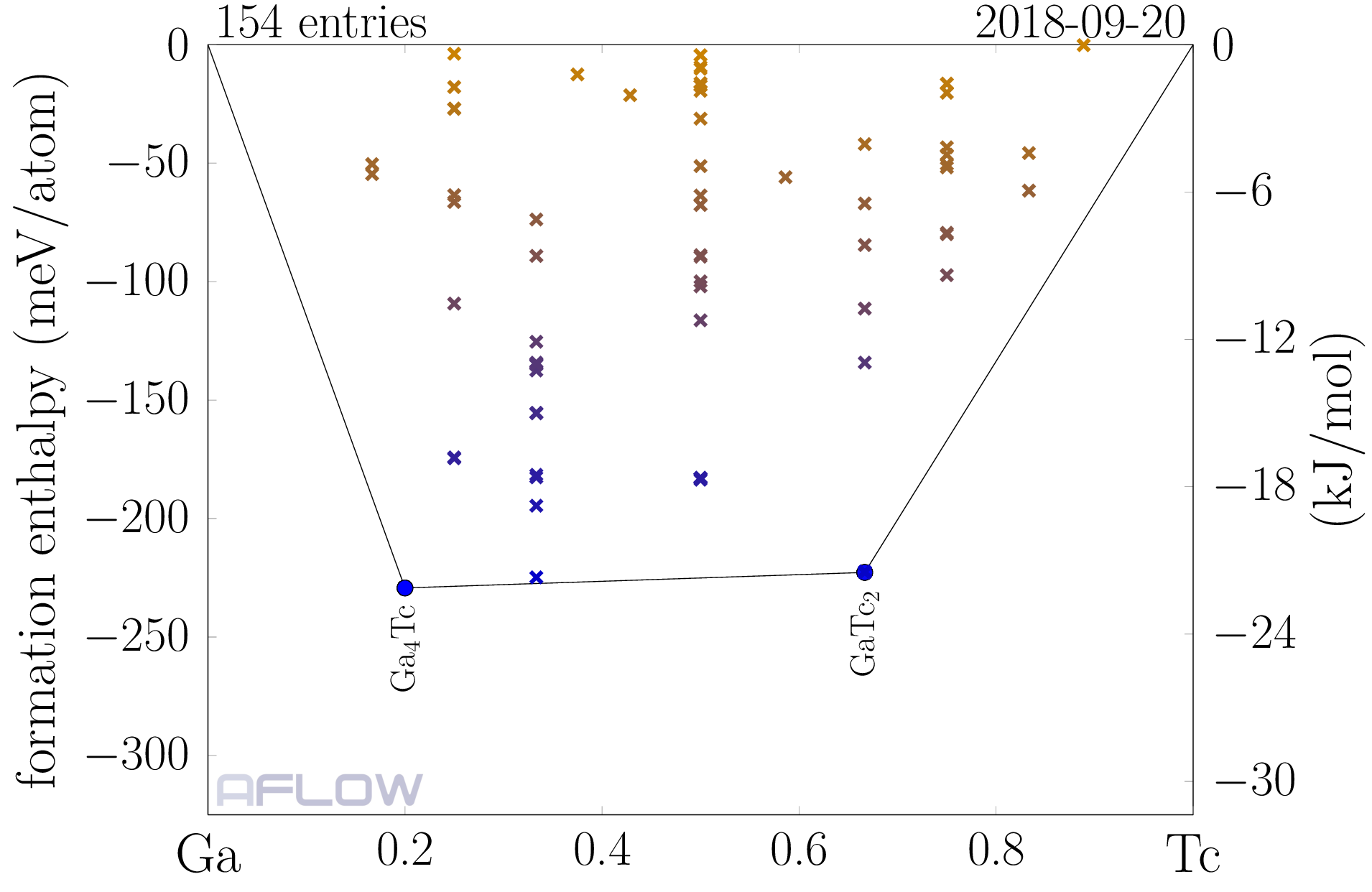}
\mycaption{Ga-Tc binary convex hull as plotted by \AFLOWHULL.}
\label{fig:art146:GaTc_binary_hull_supp}
\efig

\clearpage

\figsec
\includegraphics[width=0.75\linewidth]{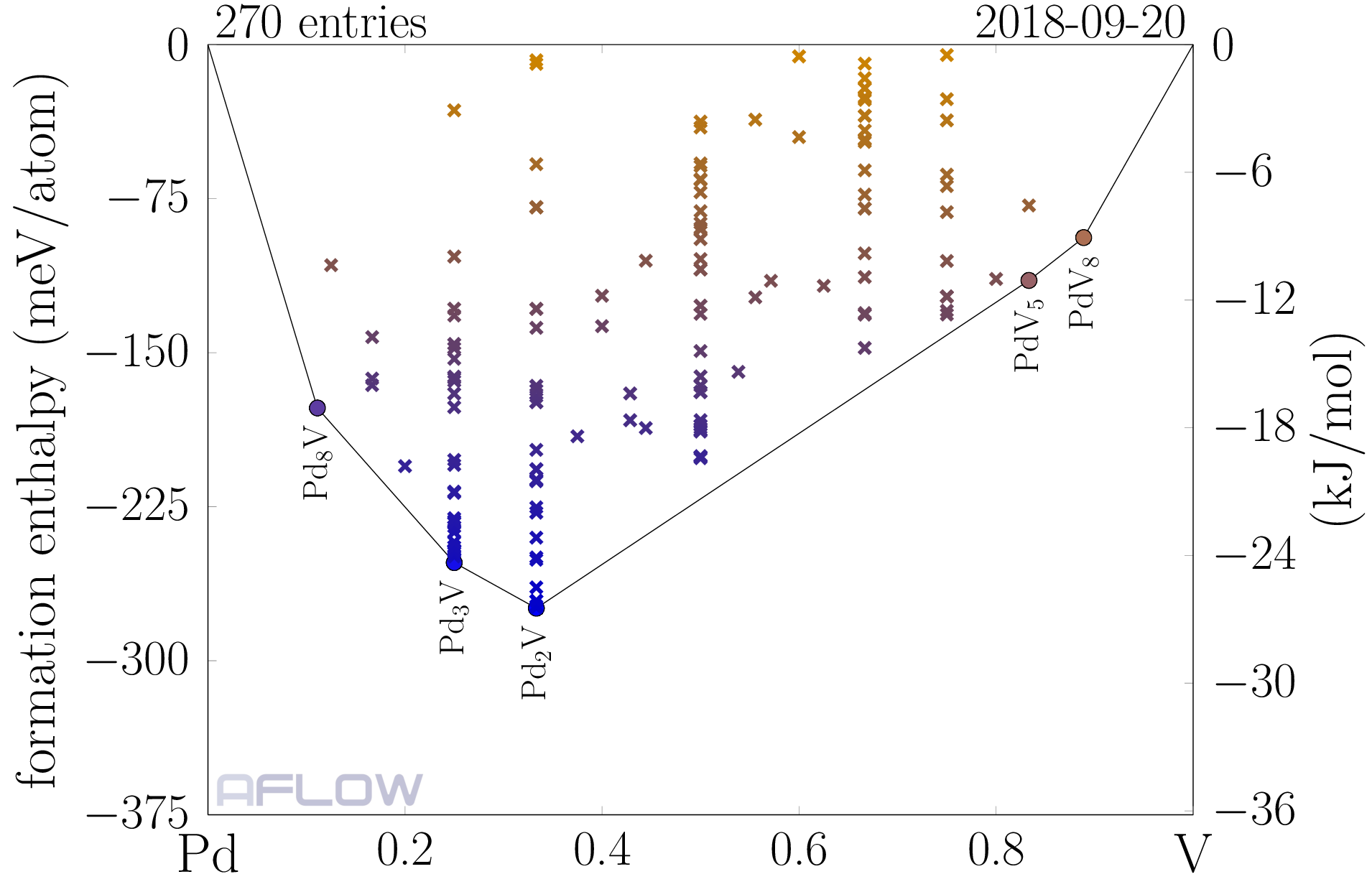}
\mycaption{Pd-V binary convex hull as plotted by \AFLOWHULL.}
\label{fig:art146:PdV_binary_hull_supp}
\efig

\vspace{\fill}

\figsec
\includegraphics[width=0.75\linewidth]{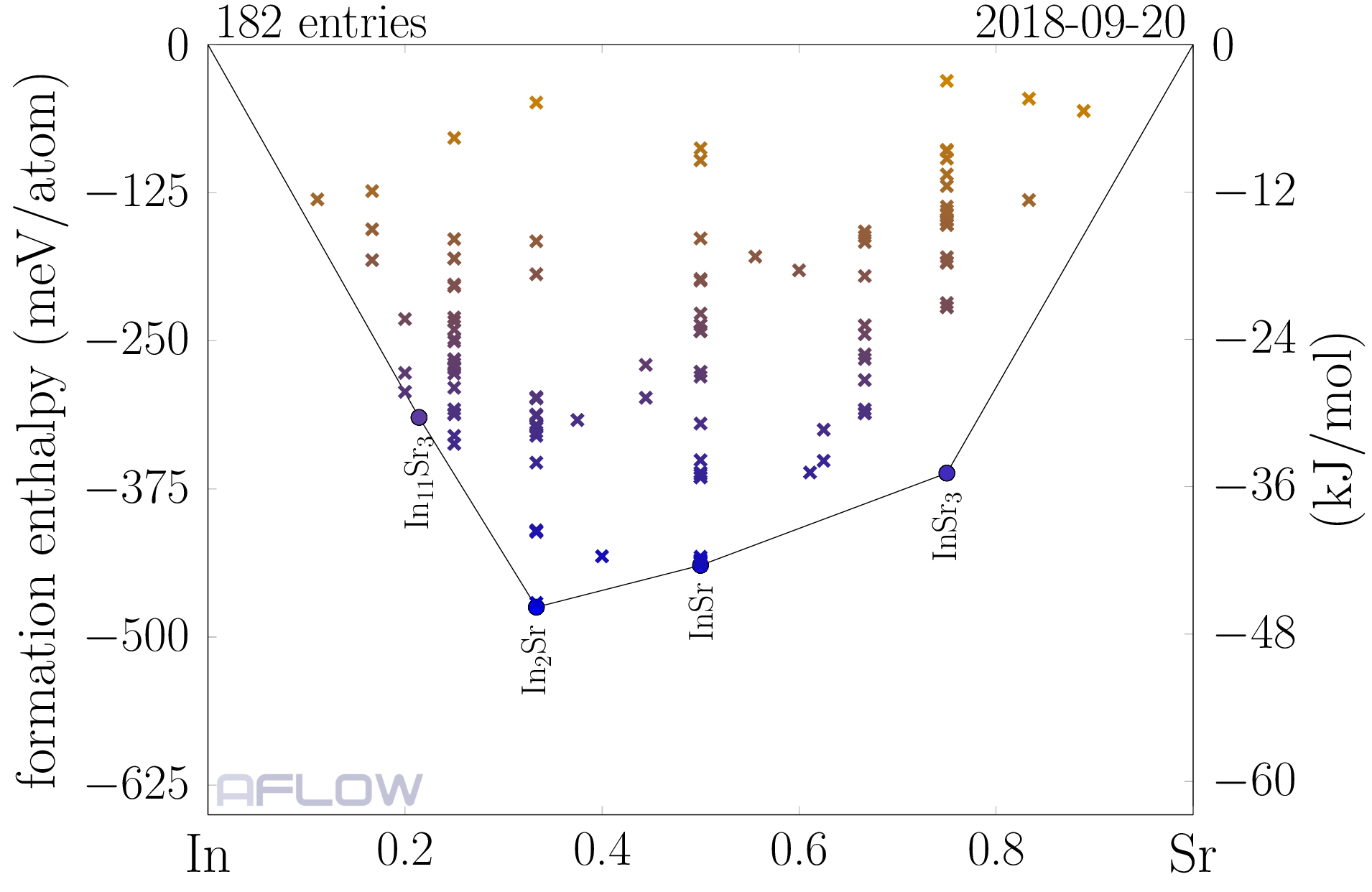}
\mycaption{In-Sr binary convex hull as plotted by \AFLOWHULL.}
\label{fig:art146:InSr_binary_hull_supp}
\efig

\clearpage

\figsec
\includegraphics[width=0.75\linewidth]{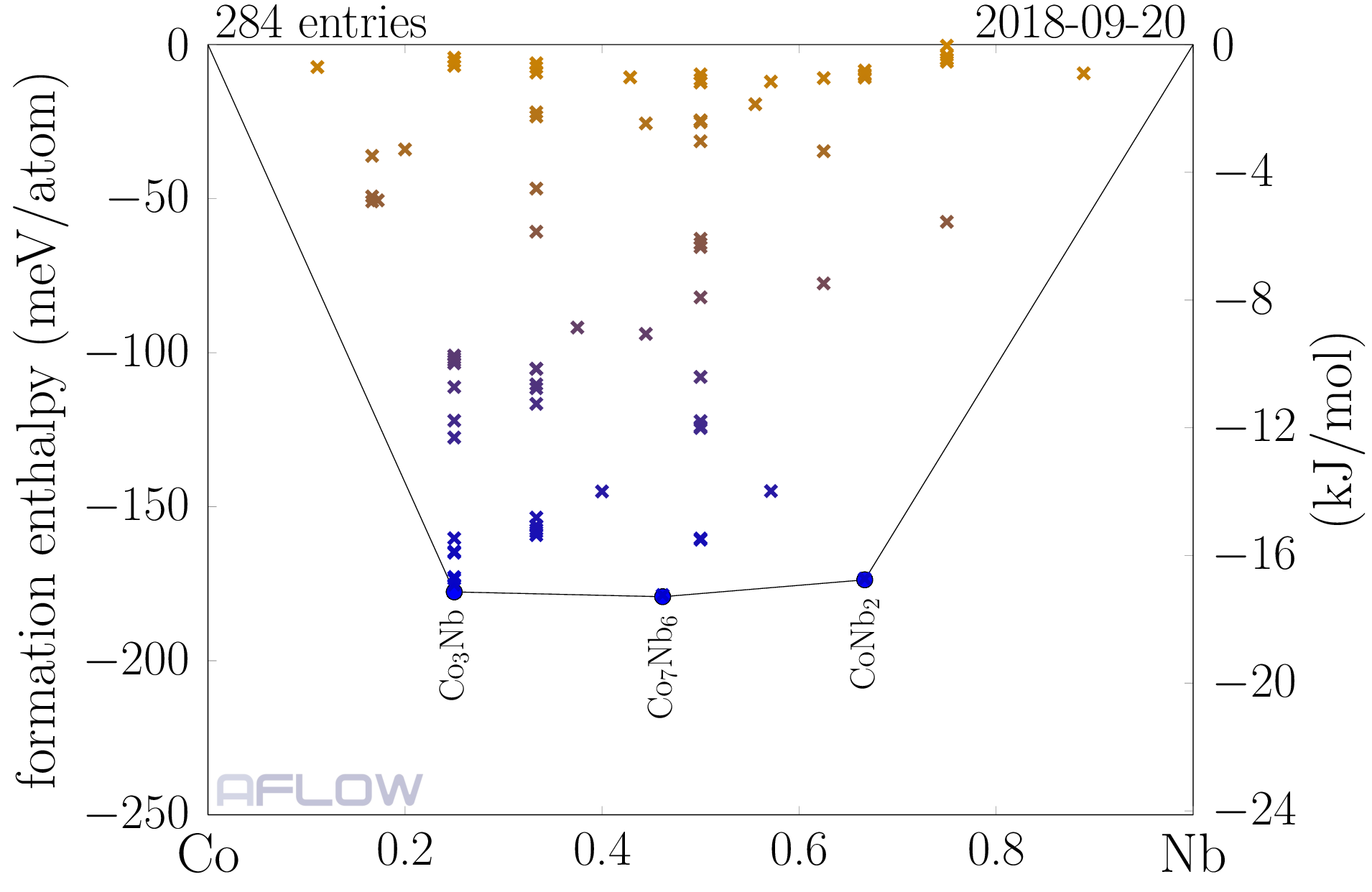}
\mycaption{Co-Nb binary convex hull as plotted by \AFLOWHULL.}
\label{fig:art146:CoNb_binary_hull_supp}
\efig

\vspace{\fill}

\figsec
\includegraphics[width=0.75\linewidth]{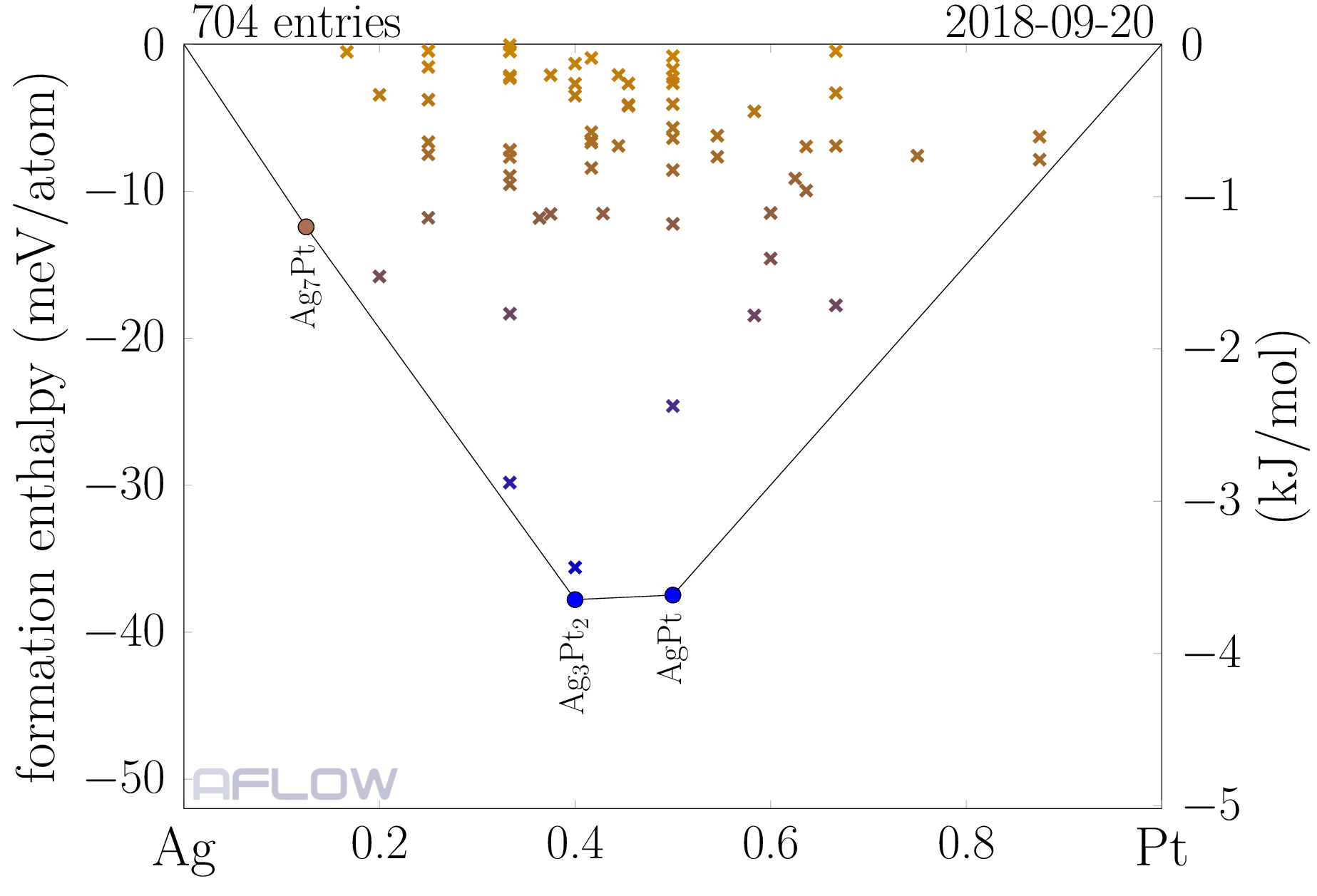}
\mycaption{Ag-Pt binary convex hull as plotted by \AFLOWHULL.}
\label{fig:art146:AgPt_binary_hull_supp}
\efig

\clearpage

\figsec
\includegraphics[width=0.75\linewidth]{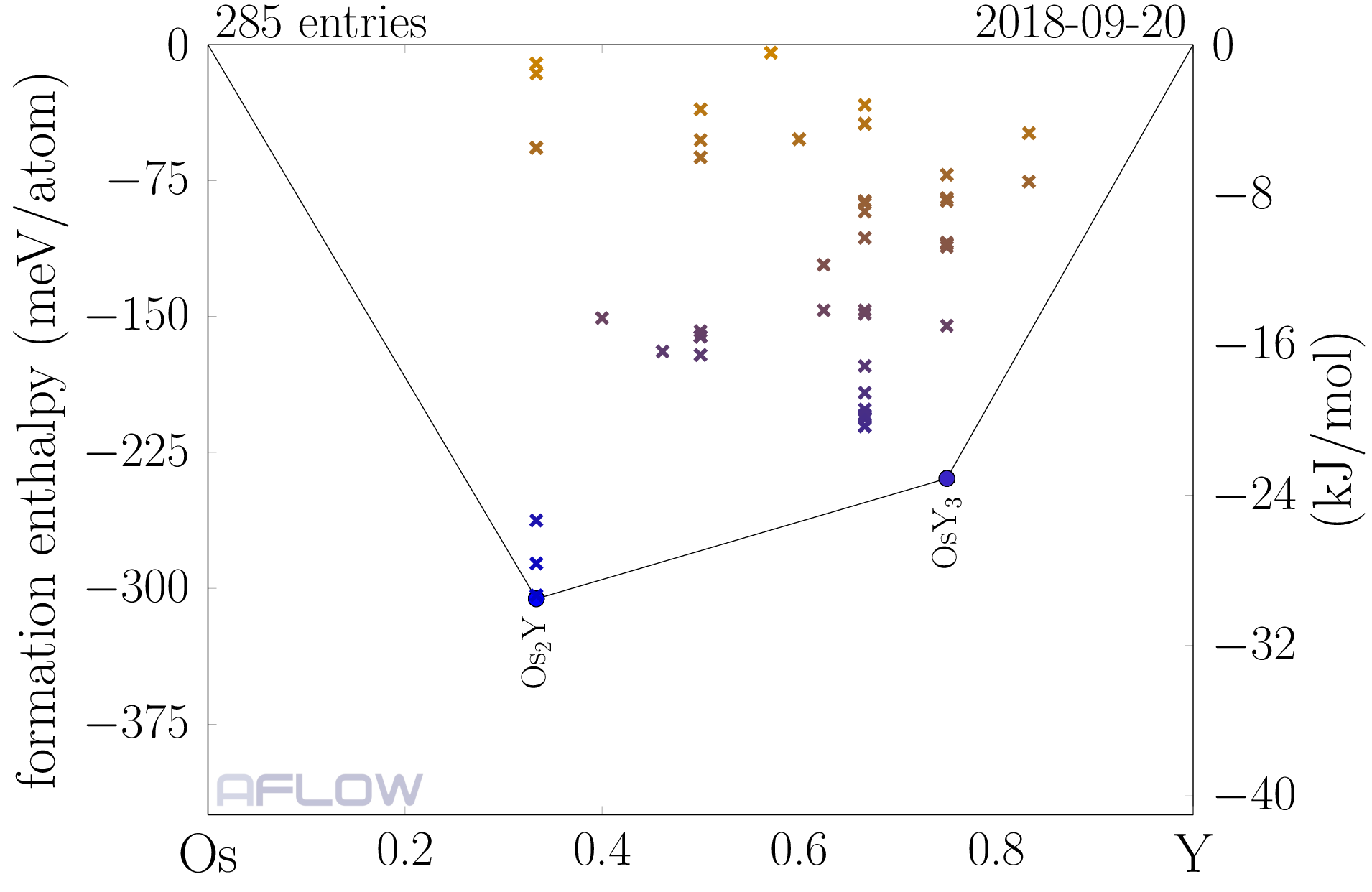}
\mycaption{Os-Y binary convex hull as plotted by \AFLOWHULL.}
\label{fig:art146:OsY_binary_hull_supp}
\efig

\vspace{\fill}

\figsec
\includegraphics[width=0.75\linewidth]{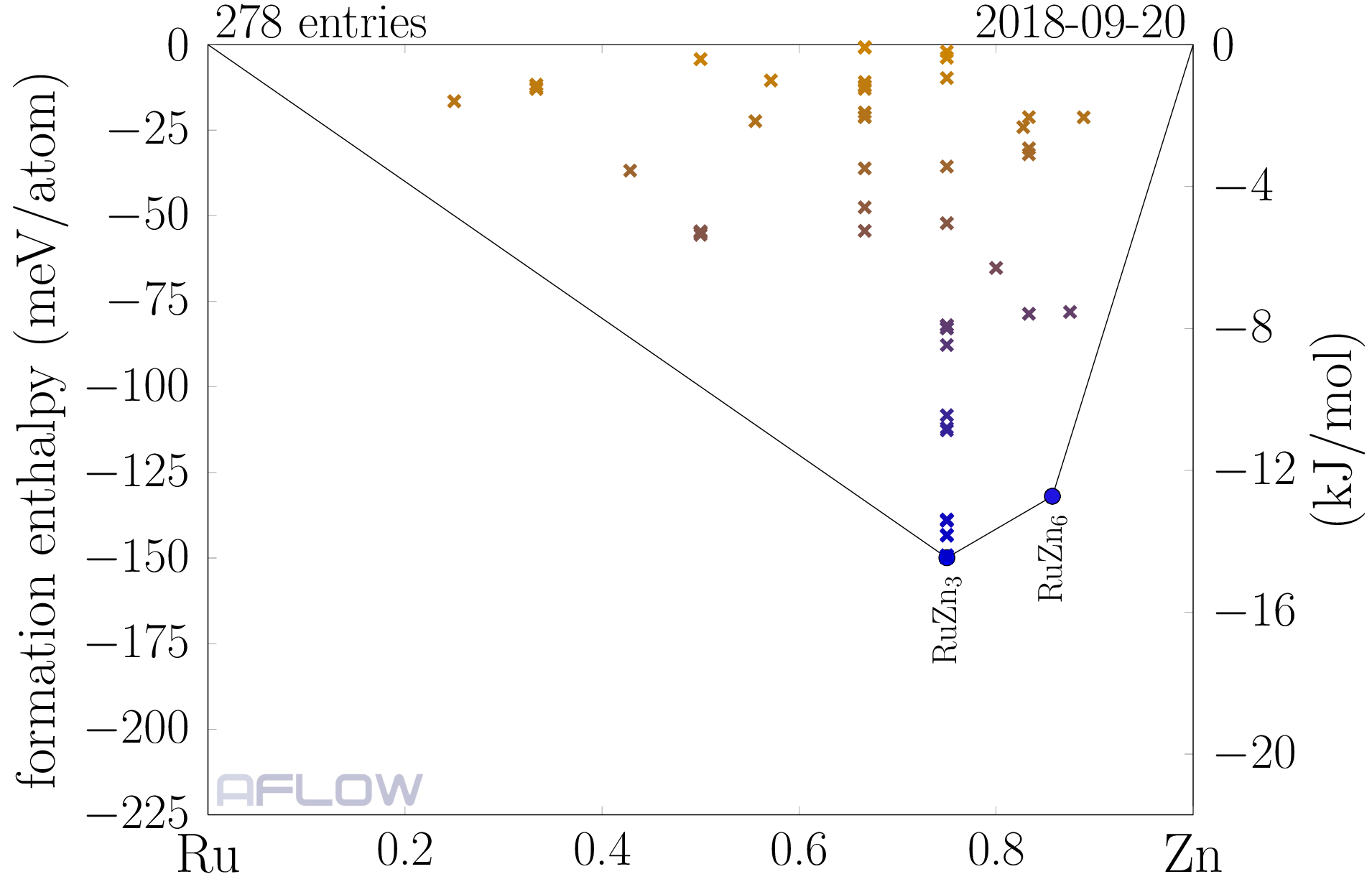}
\mycaption{Ru-Zn binary convex hull as plotted by \AFLOWHULL.}
\label{fig:art146:RuZn_binary_hull_supp}
\efig

\clearpage

\figsec
\includegraphics[width=0.75\linewidth]{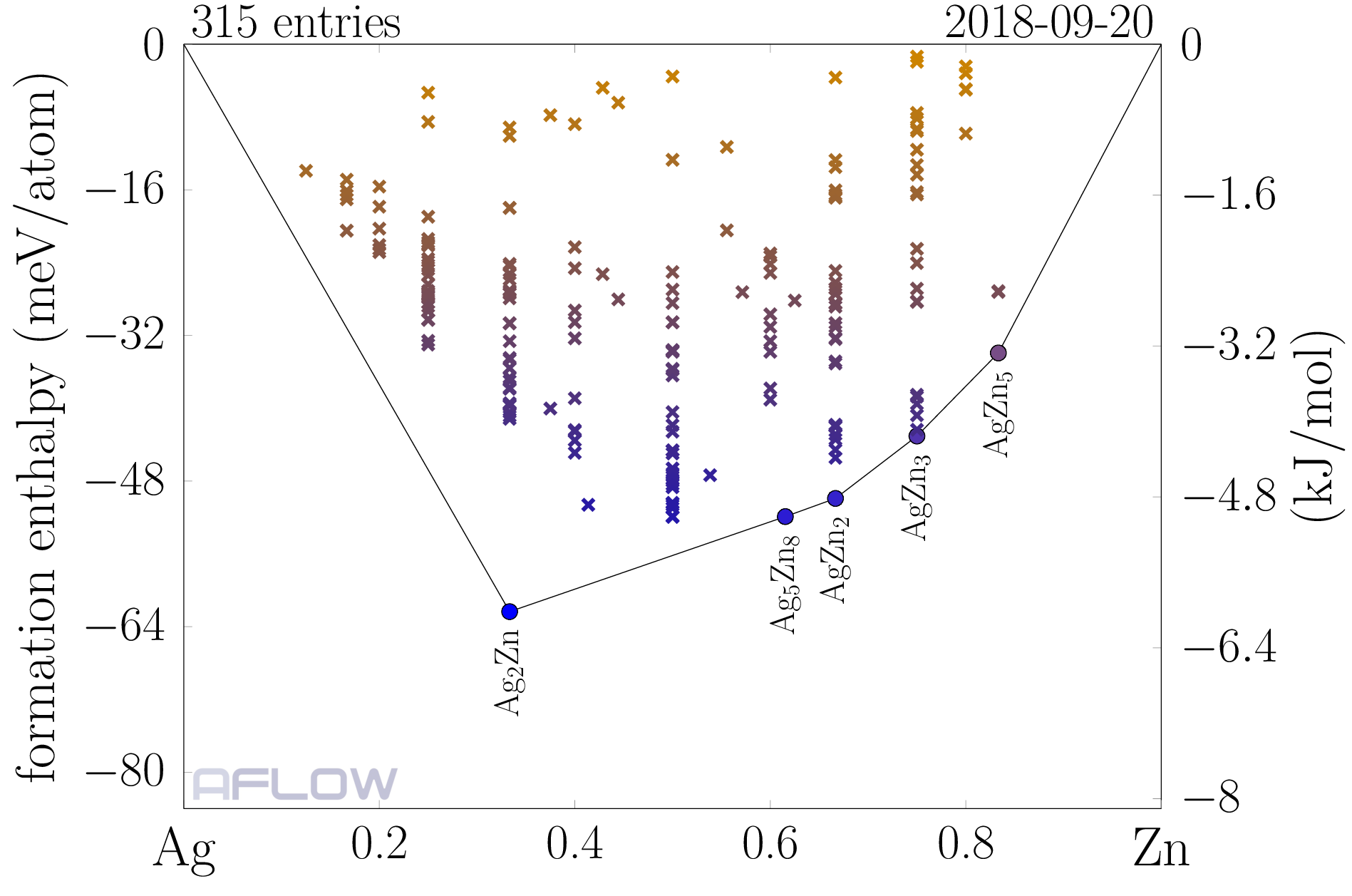}
\mycaption{Ag-Zn binary convex hull as plotted by \AFLOWHULL.}
\label{fig:art146:AgZn_binary_hull_supp}
\efig

\vspace{\fill}

\figsec
\includegraphics[width=0.75\linewidth]{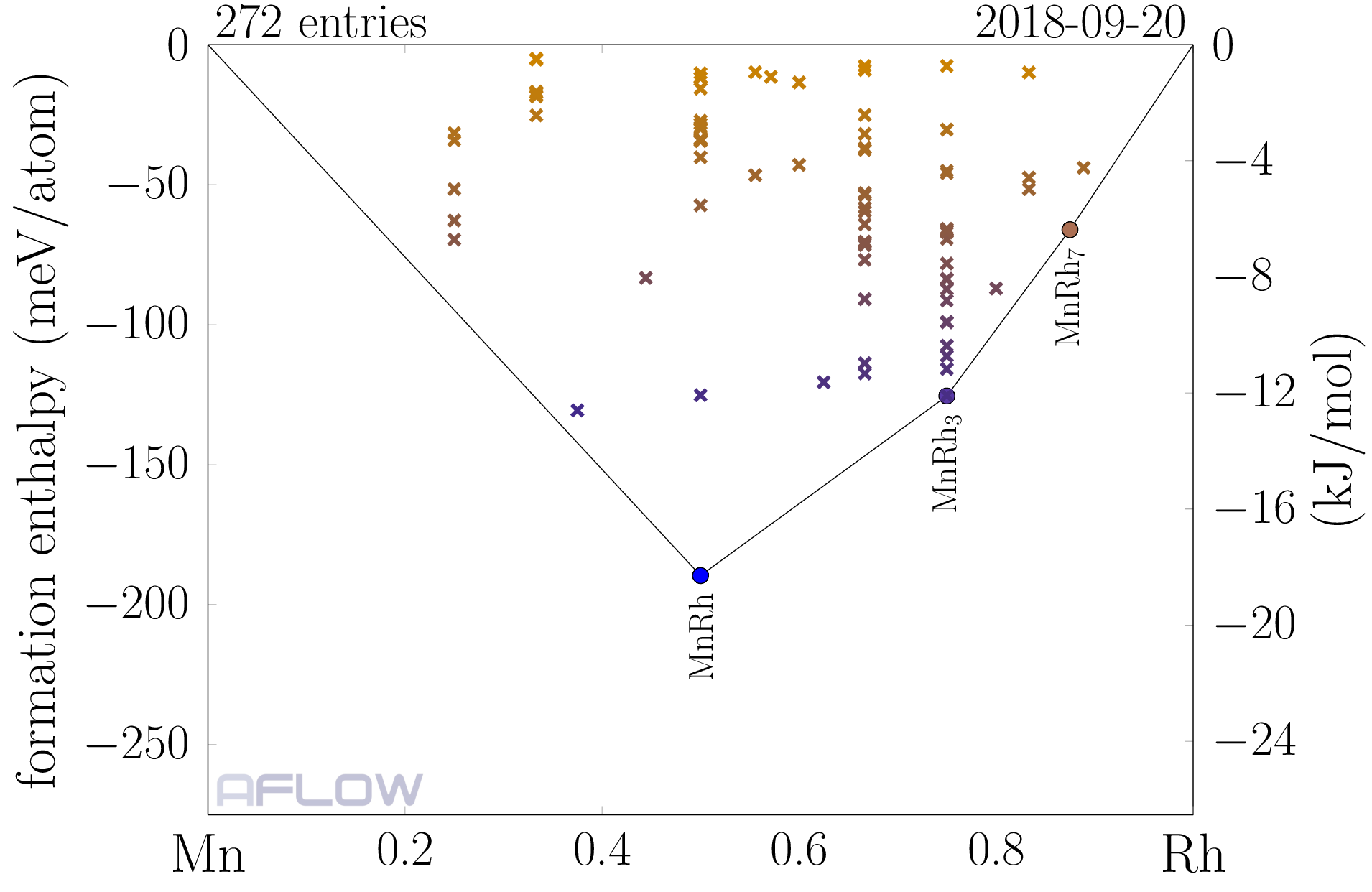}
\mycaption{Mn-Rh binary convex hull as plotted by \AFLOWHULL.}
\label{fig:art146:MnRh_binary_hull_supp}
\efig

\clearpage

\figsec
\includegraphics[width=0.75\linewidth]{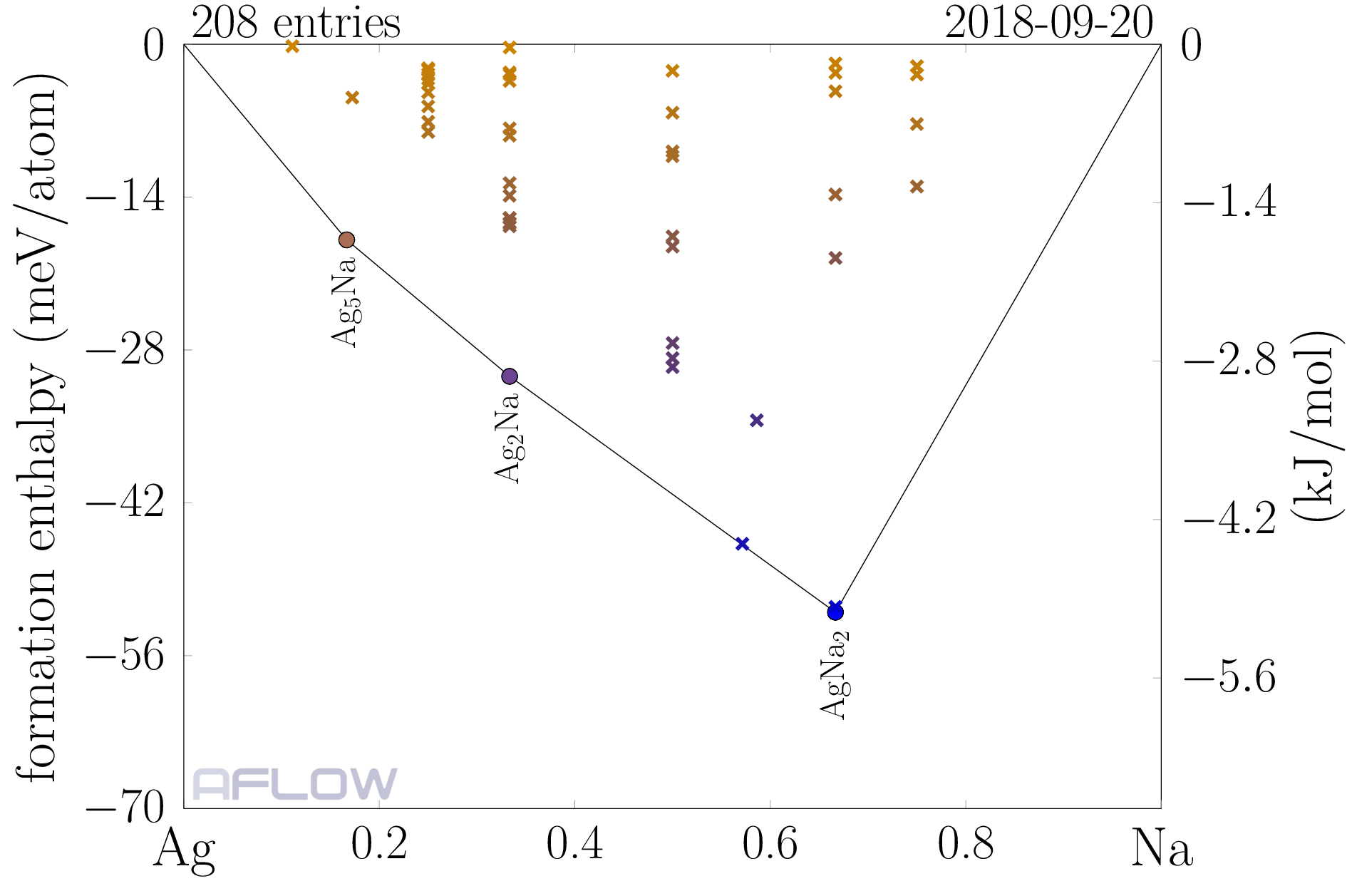}
\mycaption{Ag-Na binary convex hull as plotted by \AFLOWHULL.}
\label{fig:art146:AgNa_binary_hull_supp}
\efig

\vspace{\fill}

\figsec
\includegraphics[width=0.75\linewidth]{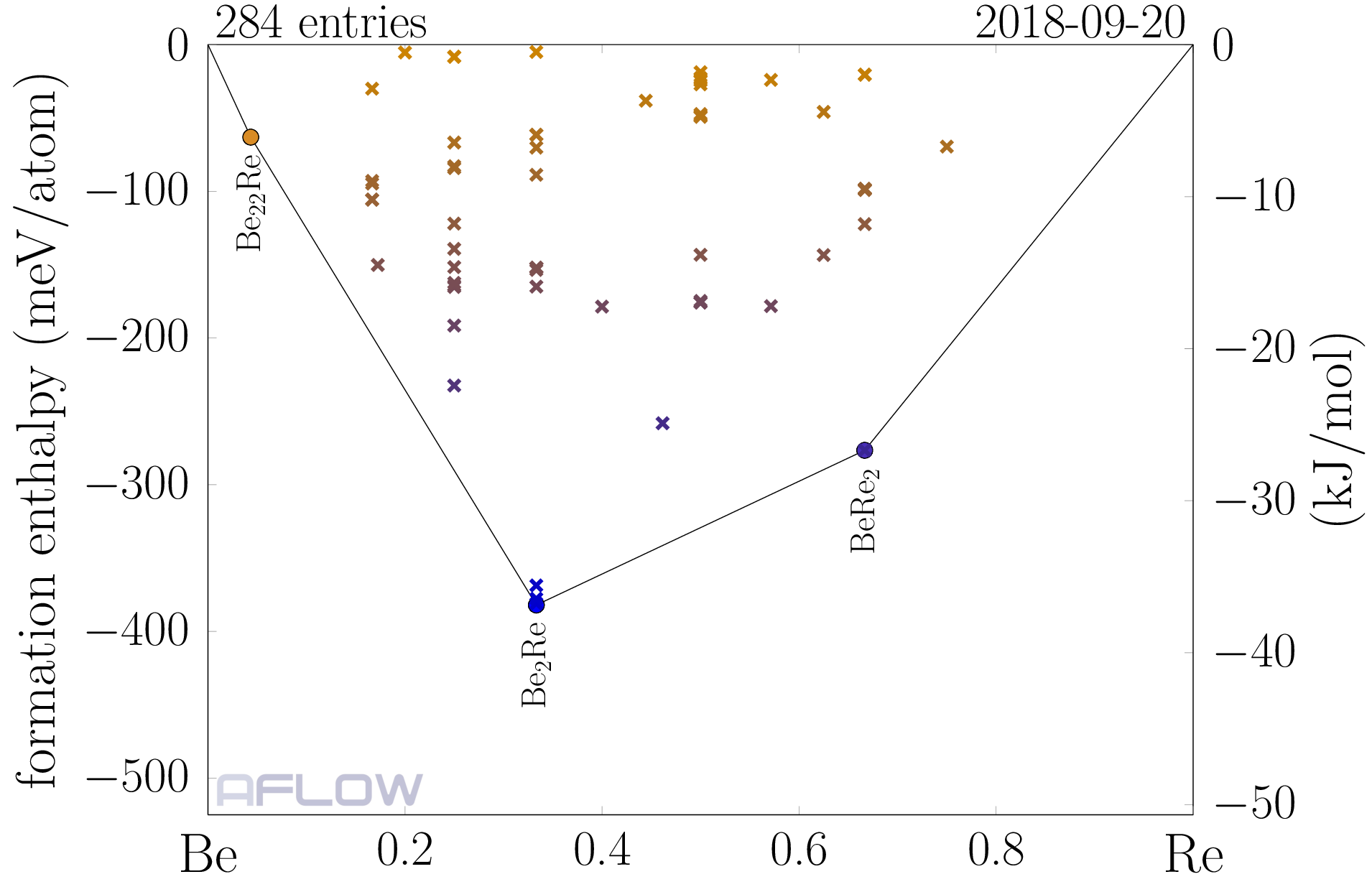}
\mycaption{Be-Re binary convex hull as plotted by \AFLOWHULL.}
\label{fig:art146:BeRe_binary_hull_supp}
\efig

\clearpage

\figsec
\includegraphics[width=0.75\linewidth]{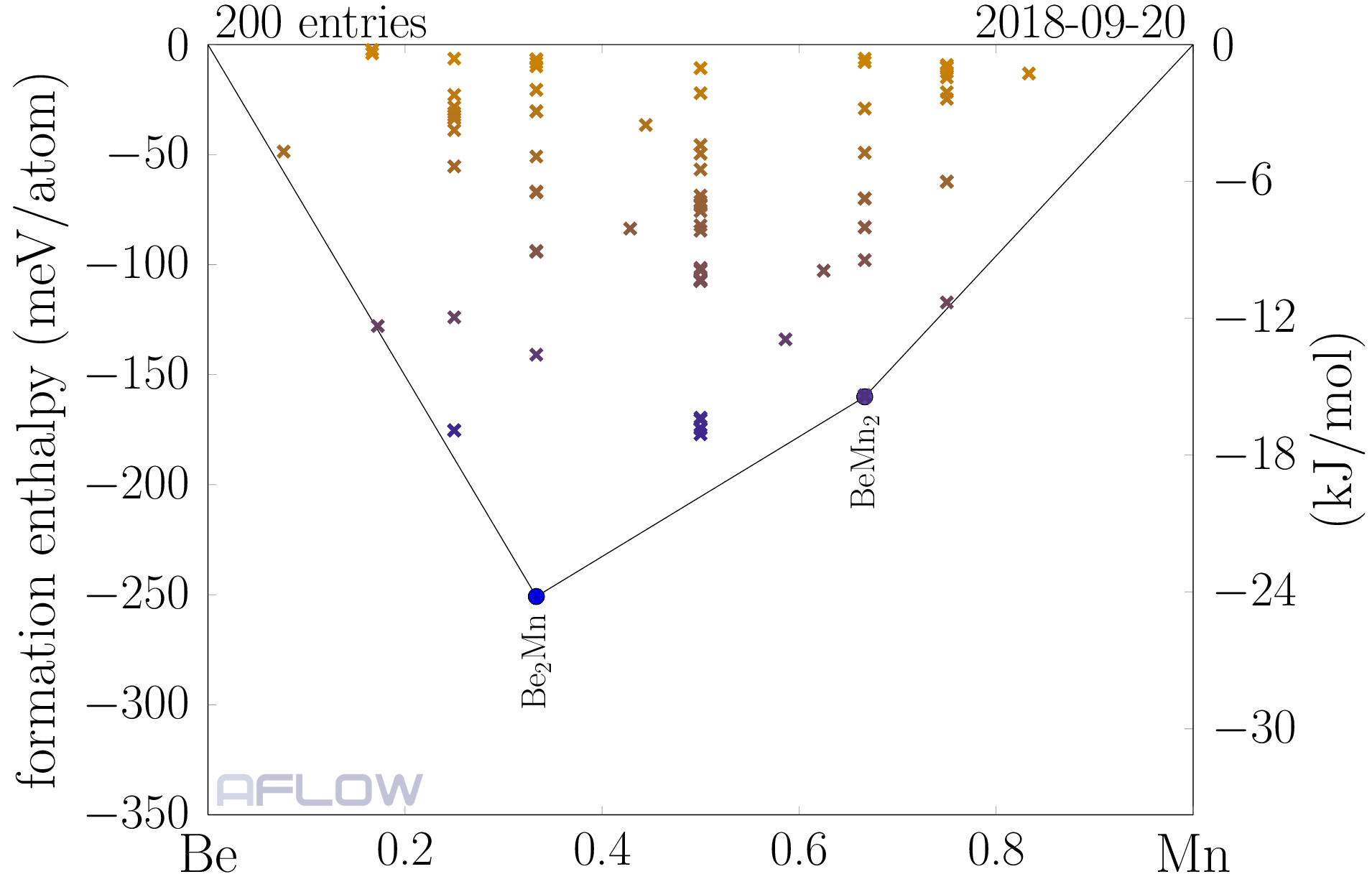}
\mycaption{Be-Mn binary convex hull as plotted by \AFLOWHULL.}
\label{fig:art146:BeMn_binary_hull_supp}
\efig

\vspace{\fill}

\figsec
\includegraphics[width=0.75\linewidth]{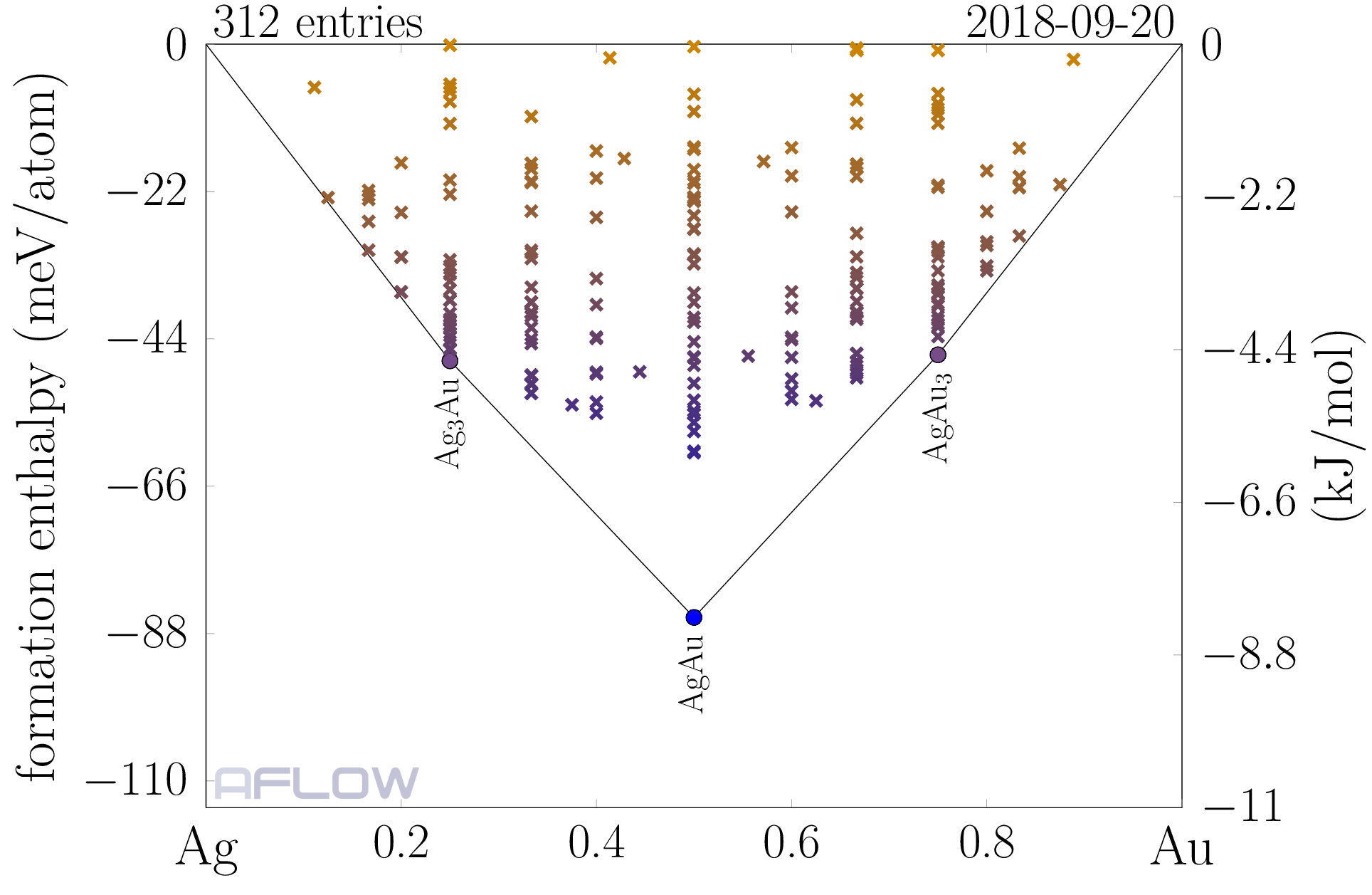}
\mycaption{Ag-Au binary convex hull as plotted by \AFLOWHULL.}
\label{fig:art146:AgAu_binary_hull_supp}
\efig

\clearpage

\figsec
\includegraphics[width=0.75\linewidth]{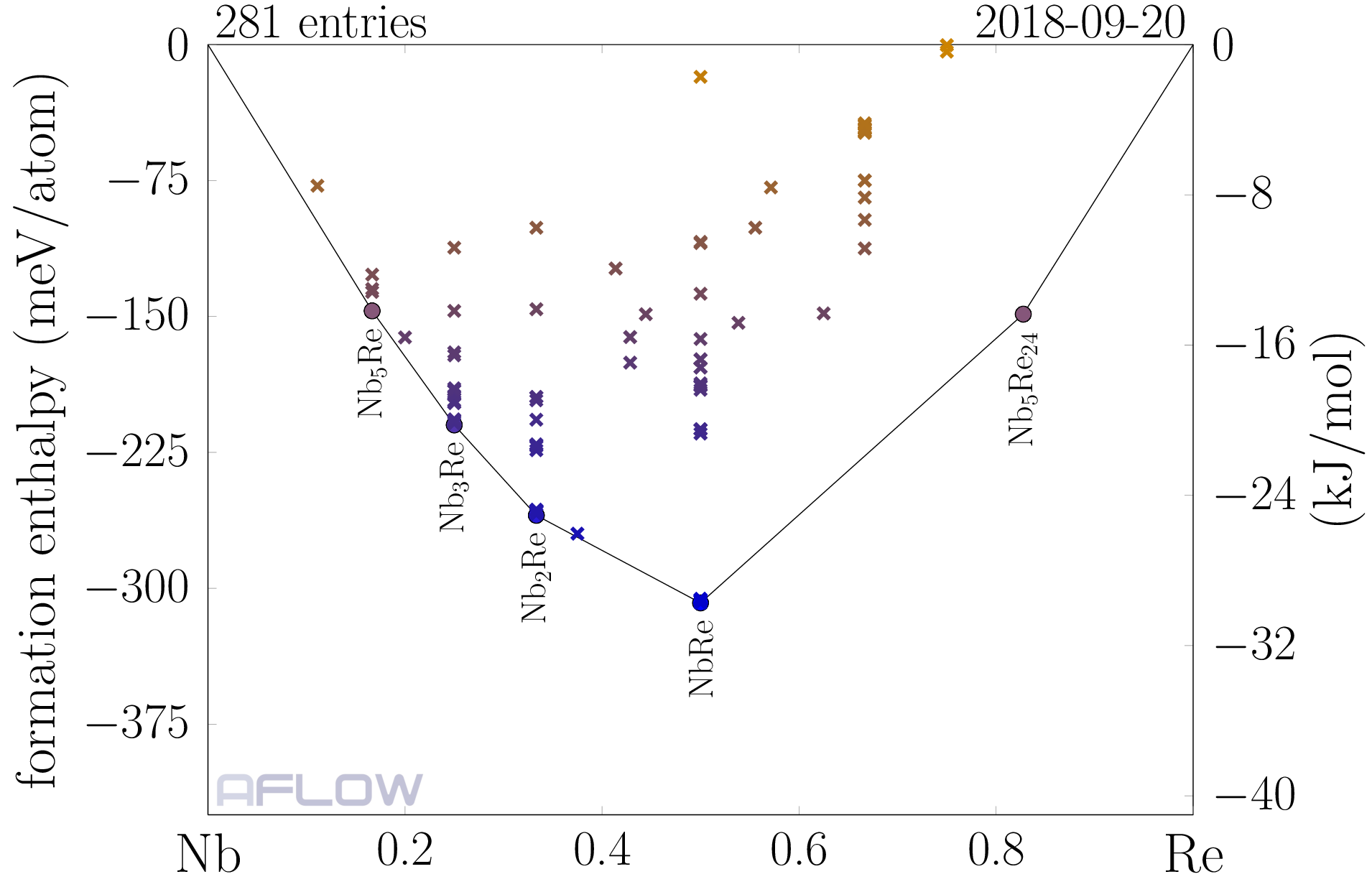}
\mycaption{Nb-Re binary convex hull as plotted by \AFLOWHULL.}
\label{fig:art146:NbRe_binary_hull_supp}
\efig

\vspace{\fill}

\figsec
\includegraphics[width=0.75\linewidth]{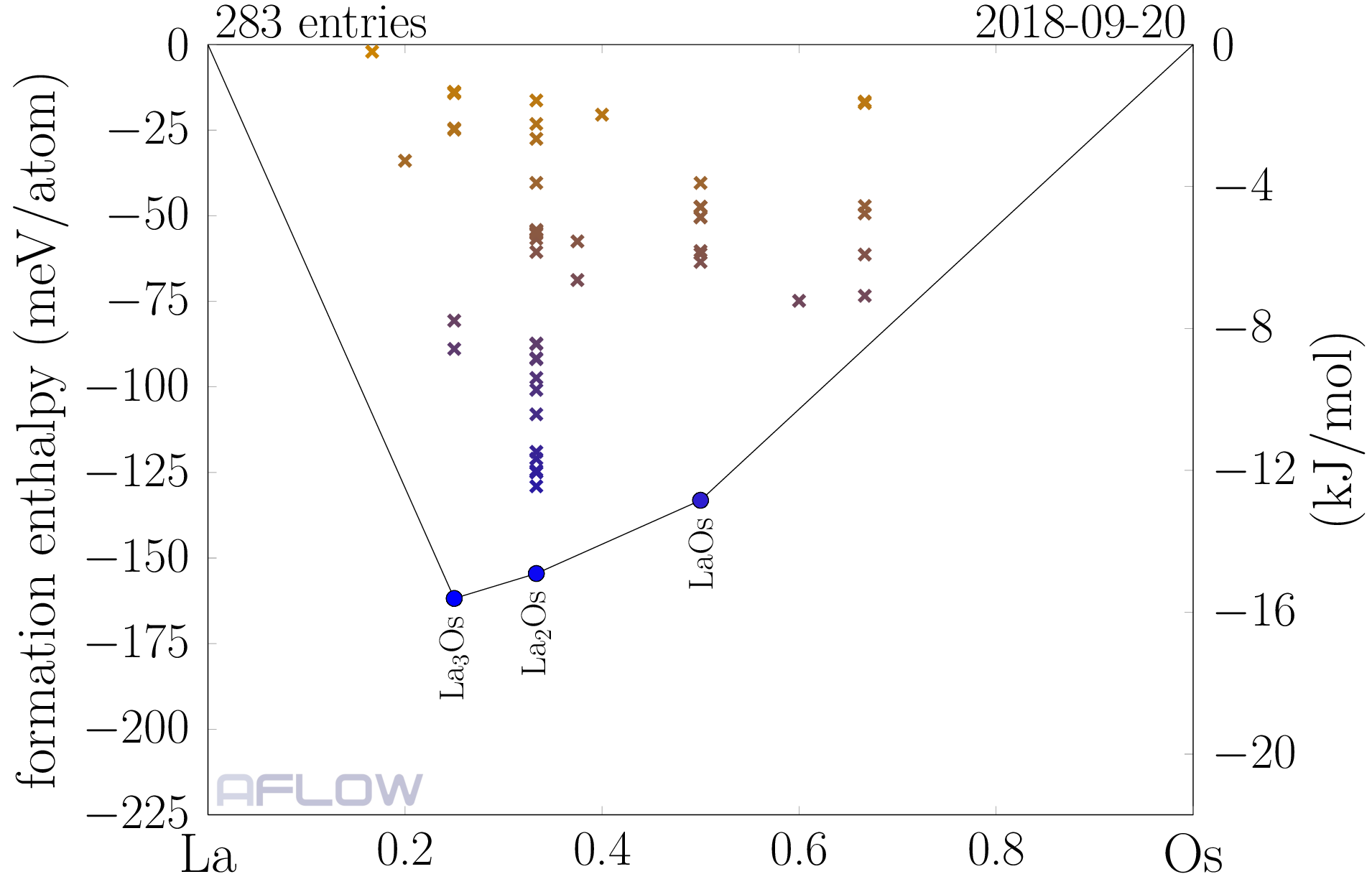}
\mycaption{La-Os binary convex hull as plotted by \AFLOWHULL.}
\label{fig:art146:LaOs_binary_hull_supp}
\efig

\clearpage

\figsec
\includegraphics[width=0.75\linewidth]{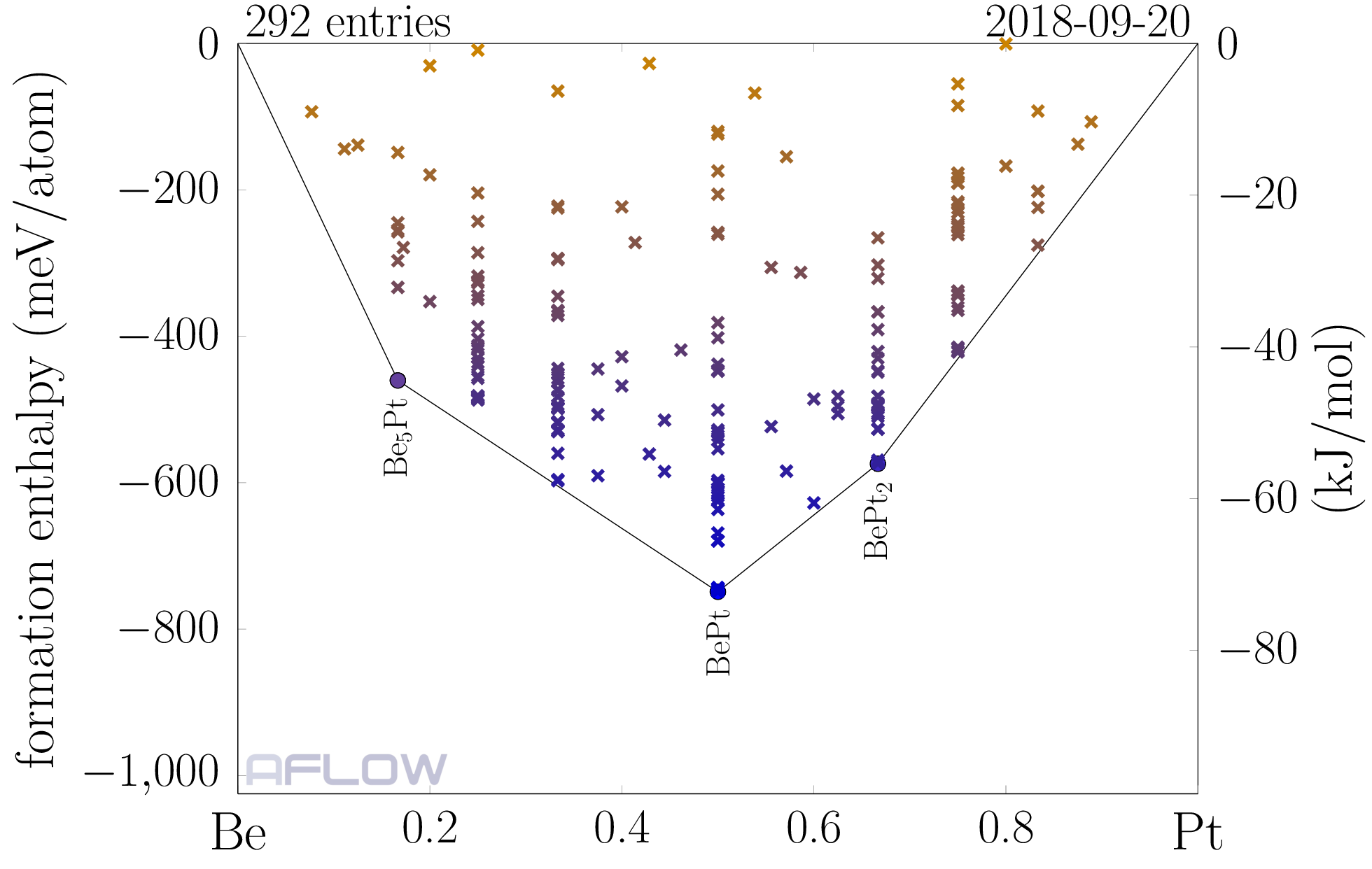}
\mycaption{Be-Pt binary convex hull as plotted by \AFLOWHULL.}
\label{fig:art146:BePt_binary_hull_supp}
\efig

\vspace{\fill}

\figsec
\includegraphics[width=0.75\linewidth]{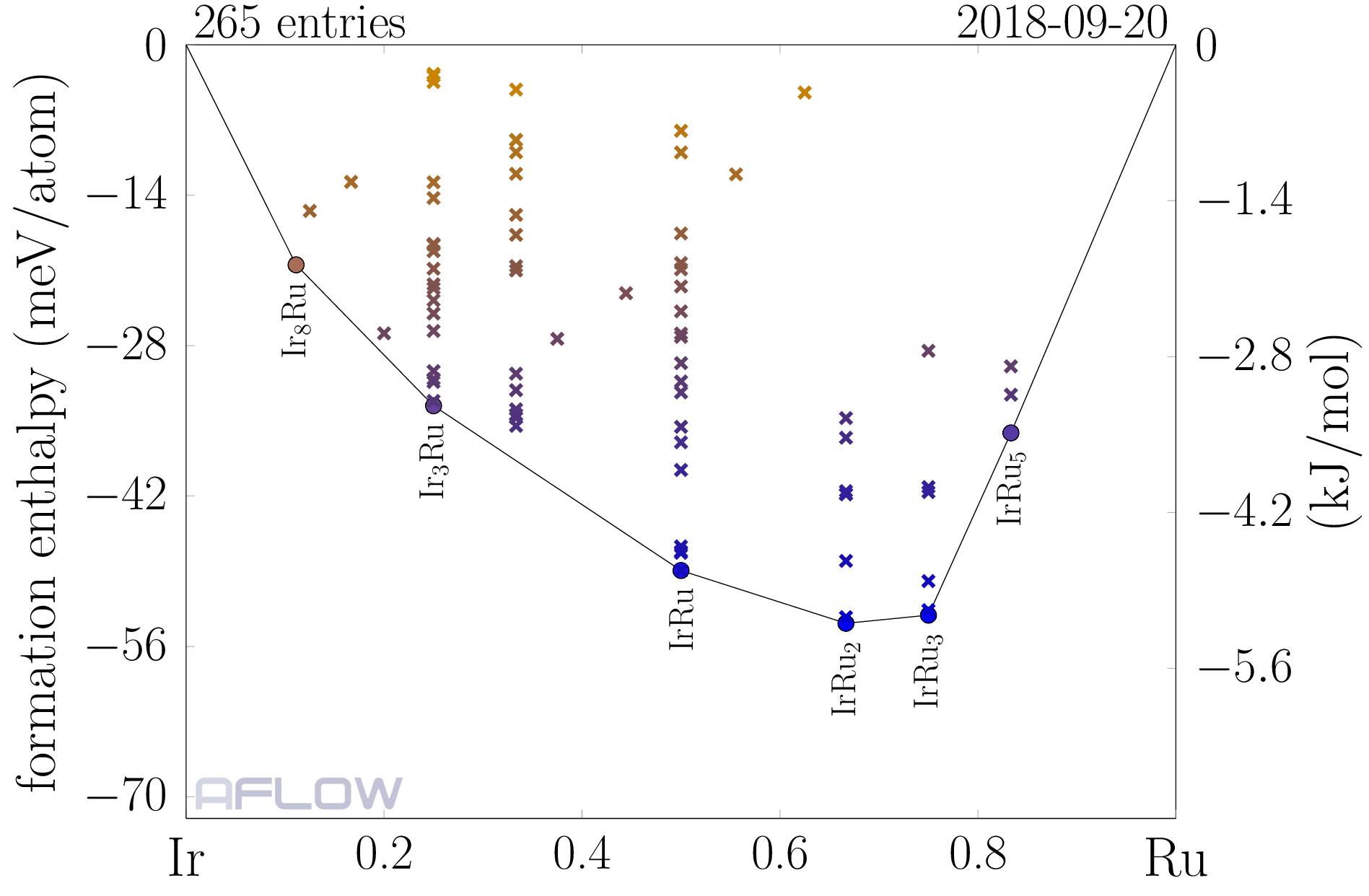}
\mycaption{Ir-Ru binary convex hull as plotted by \AFLOWHULL.}
\label{fig:art146:IrRu_binary_hull_supp}
\efig

\clearpage

\figsec
\includegraphics[width=0.75\linewidth]{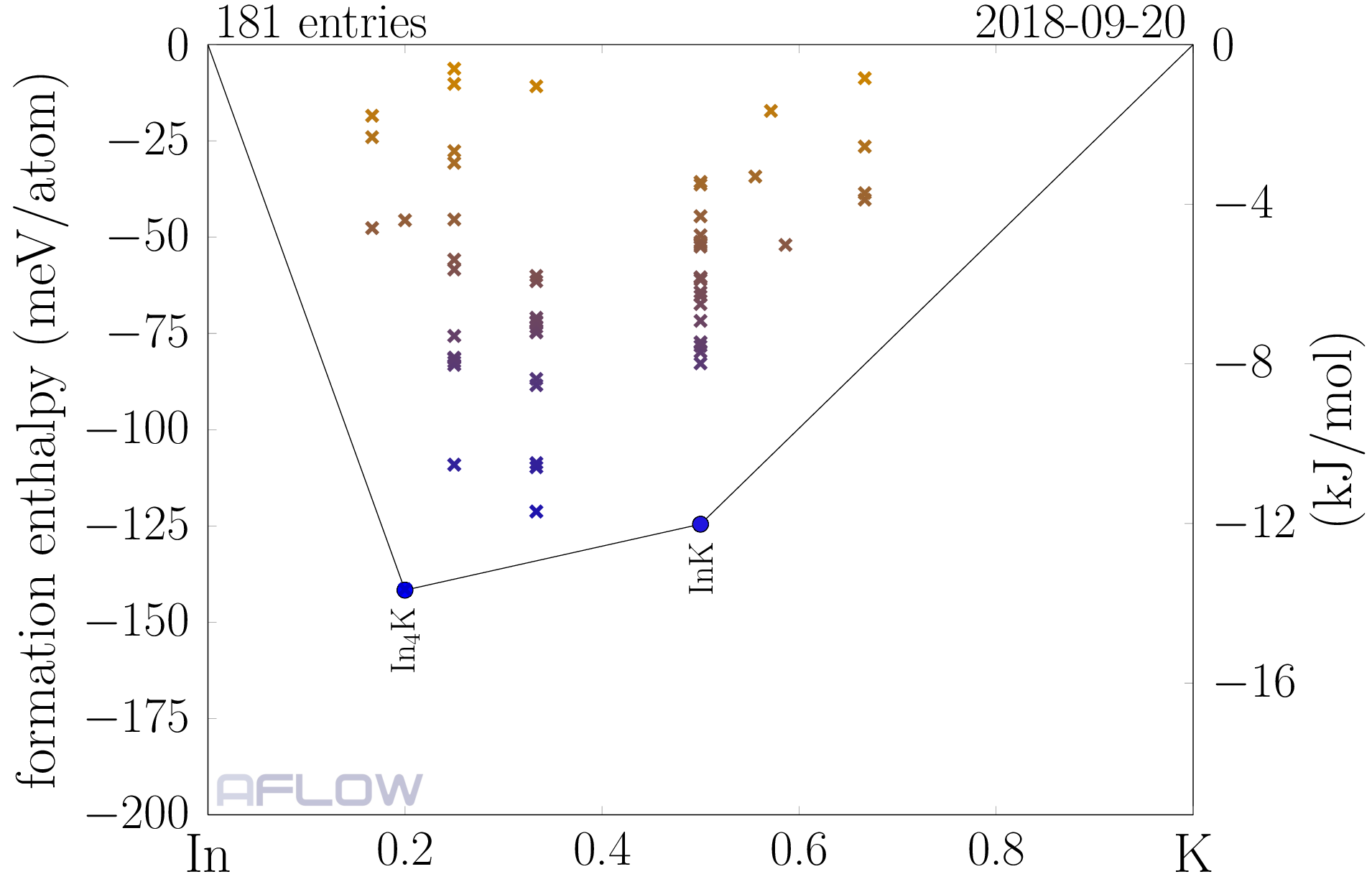}
\mycaption{In-K binary convex hull as plotted by \AFLOWHULL.}
\label{fig:art146:InK_binary_hull_supp}
\efig

\vspace{\fill}

\figsec
\includegraphics[width=0.75\linewidth]{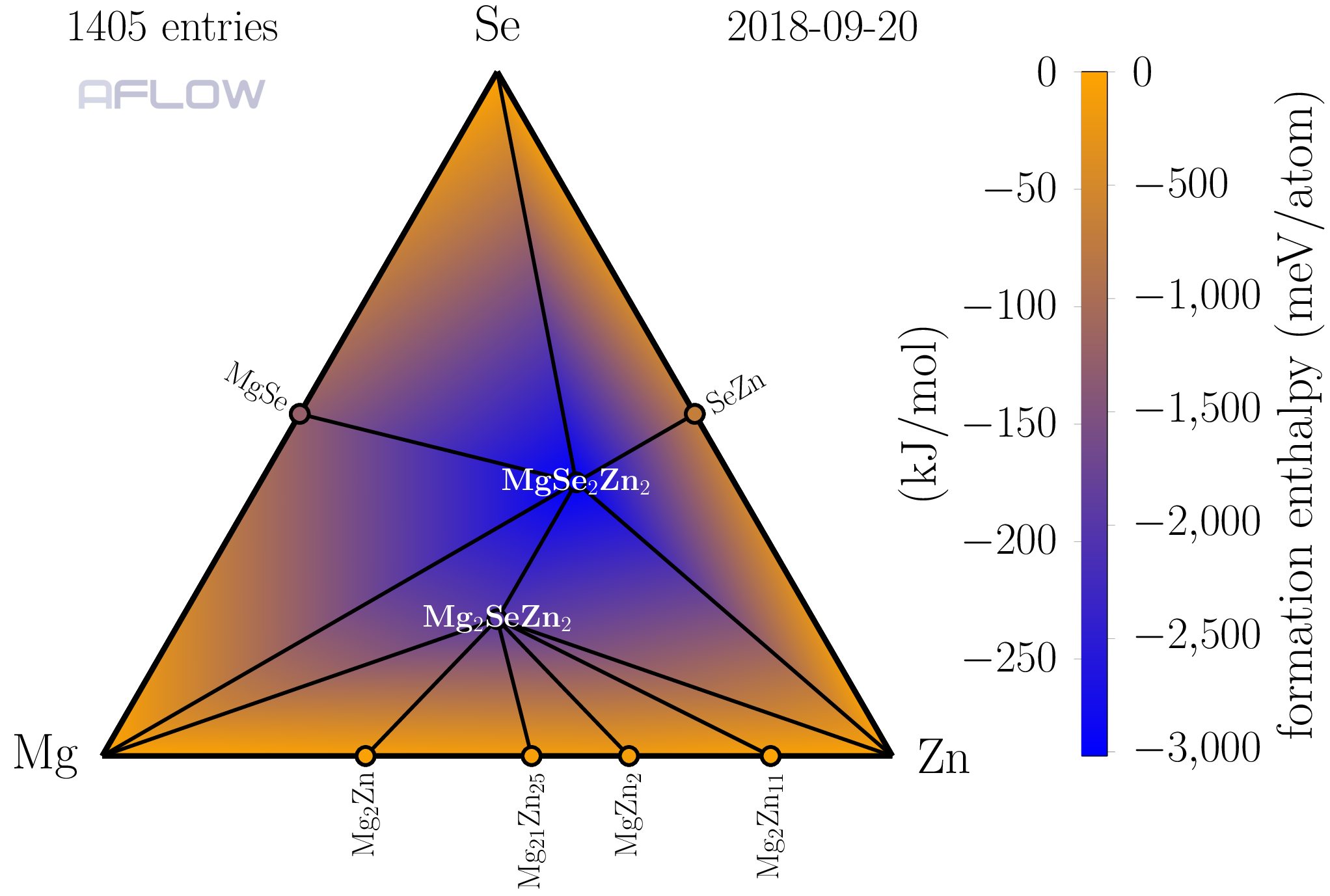}
\mycaption{Mg-Se-Zn ternary convex hull as plotted by \AFLOWHULL.}
\label{fig:art146:MgSeZn_ternary_hull_supp}
\efig

\clearpage

\figsec
\includegraphics[width=0.75\linewidth]{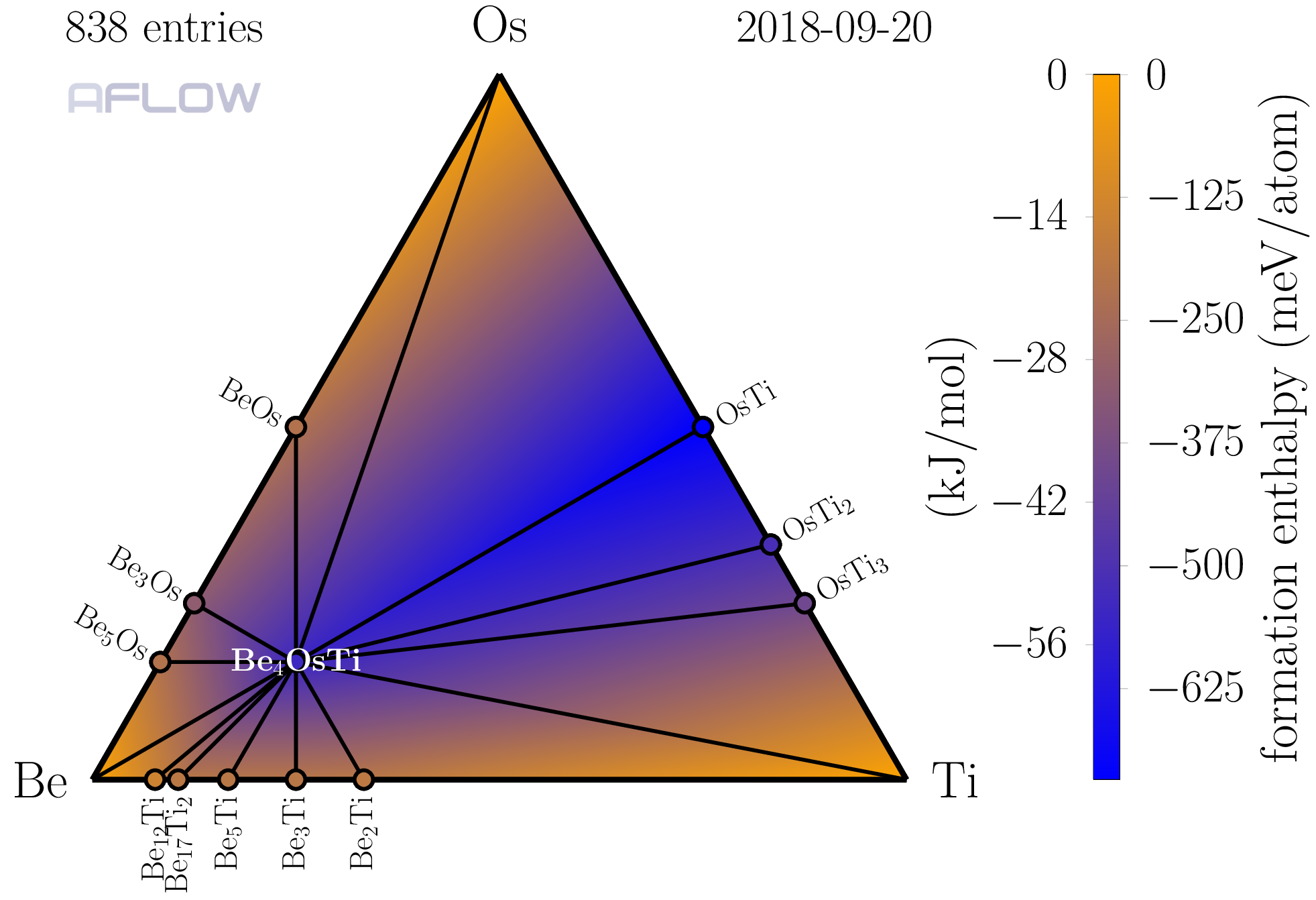}
\mycaption{Be-Os-Ti ternary convex hull as plotted by \AFLOWHULL.}
\label{fig:art146:BeOsTi_ternary_hull_supp}
\efig

\vspace{\fill}

\figsec
\includegraphics[width=0.75\linewidth]{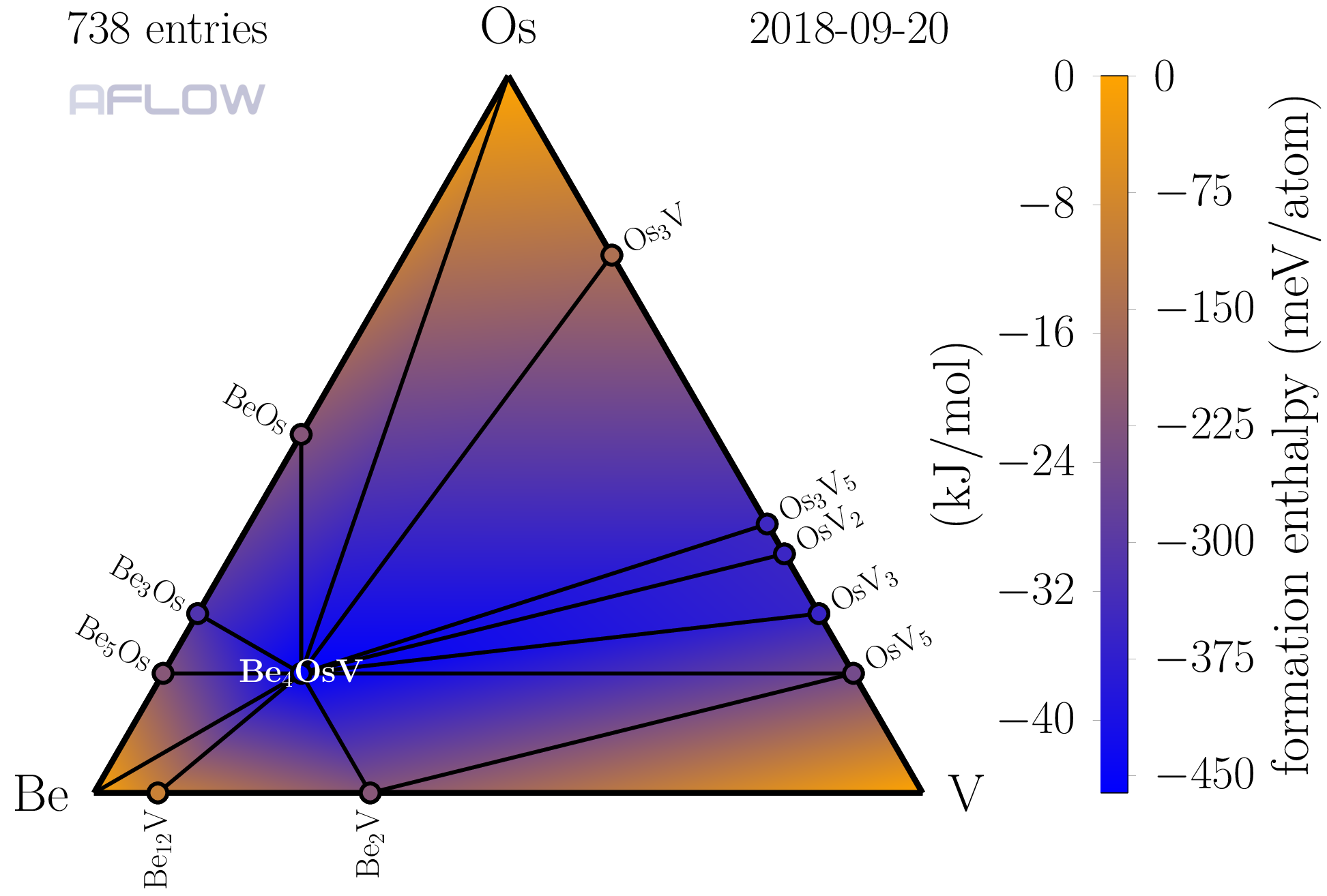}
\mycaption{Be-Os-V ternary convex hull as plotted by \AFLOWHULL.}
\label{fig:art146:BeOsV_ternary_hull_supp}
\efig

\clearpage

\figsec
\includegraphics[width=0.75\linewidth]{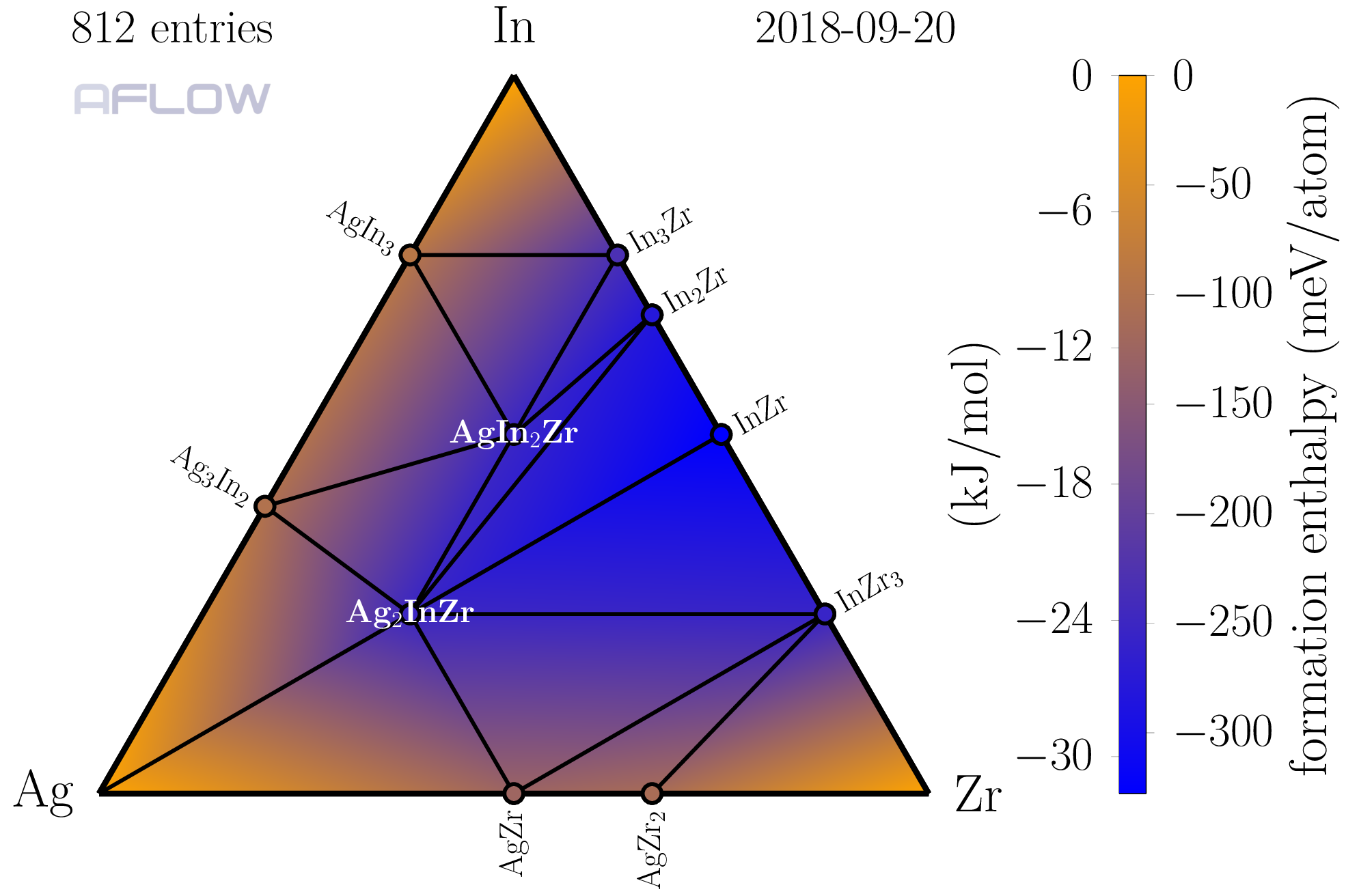}
\mycaption{Ag-In-Zr ternary convex hull as plotted by \AFLOWHULL.}
\label{fig:art146:AgInZr_ternary_hull_supp}
\efig

\vspace{\fill}

\figsec
\includegraphics[width=0.75\linewidth]{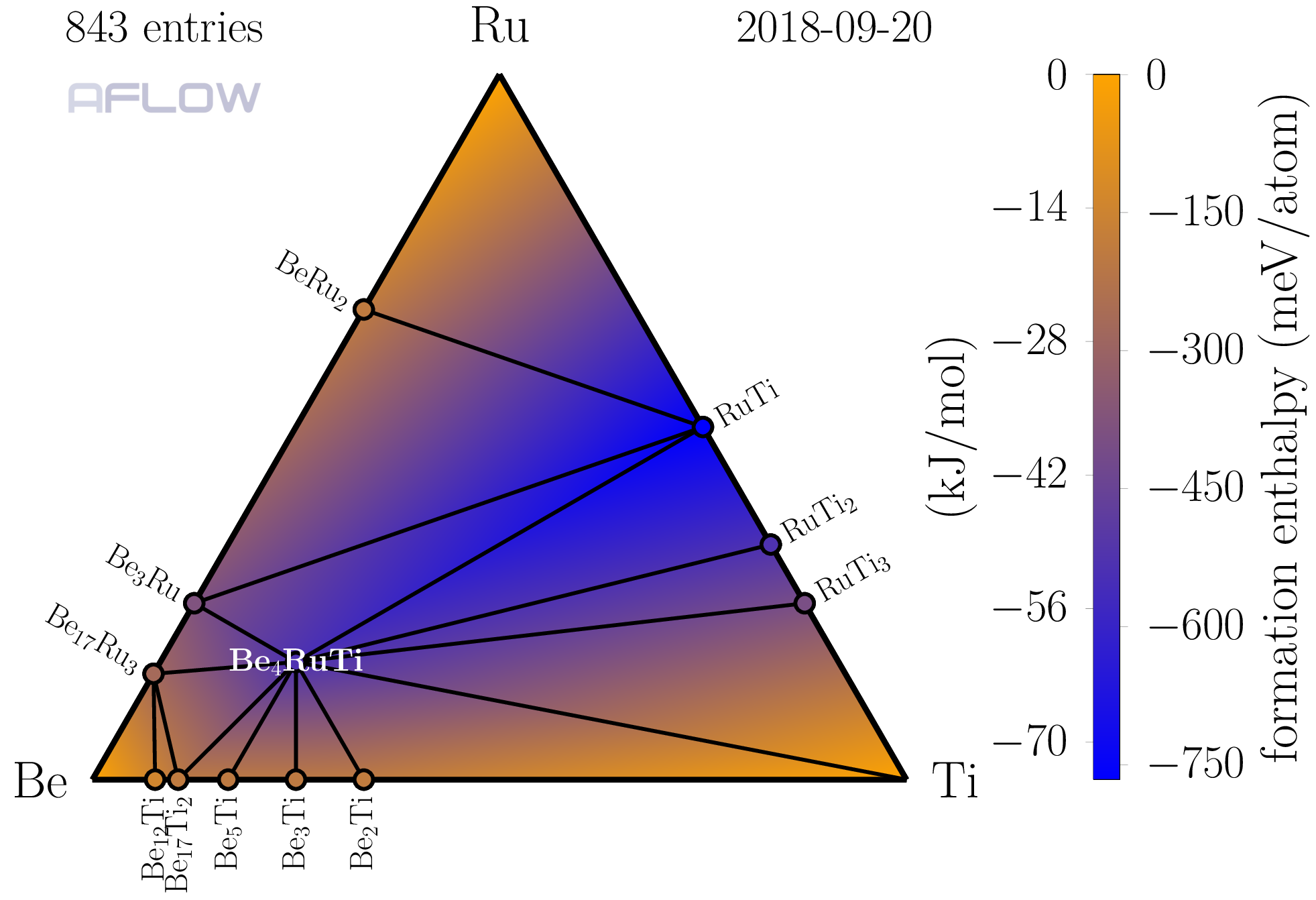}
\mycaption{Be-Ru-Ti ternary convex hull as plotted by \AFLOWHULL.}
\label{fig:art146:BeRuTi_ternary_hull_supp}
\efig

\clearpage

\figsec
\includegraphics[width=0.75\linewidth]{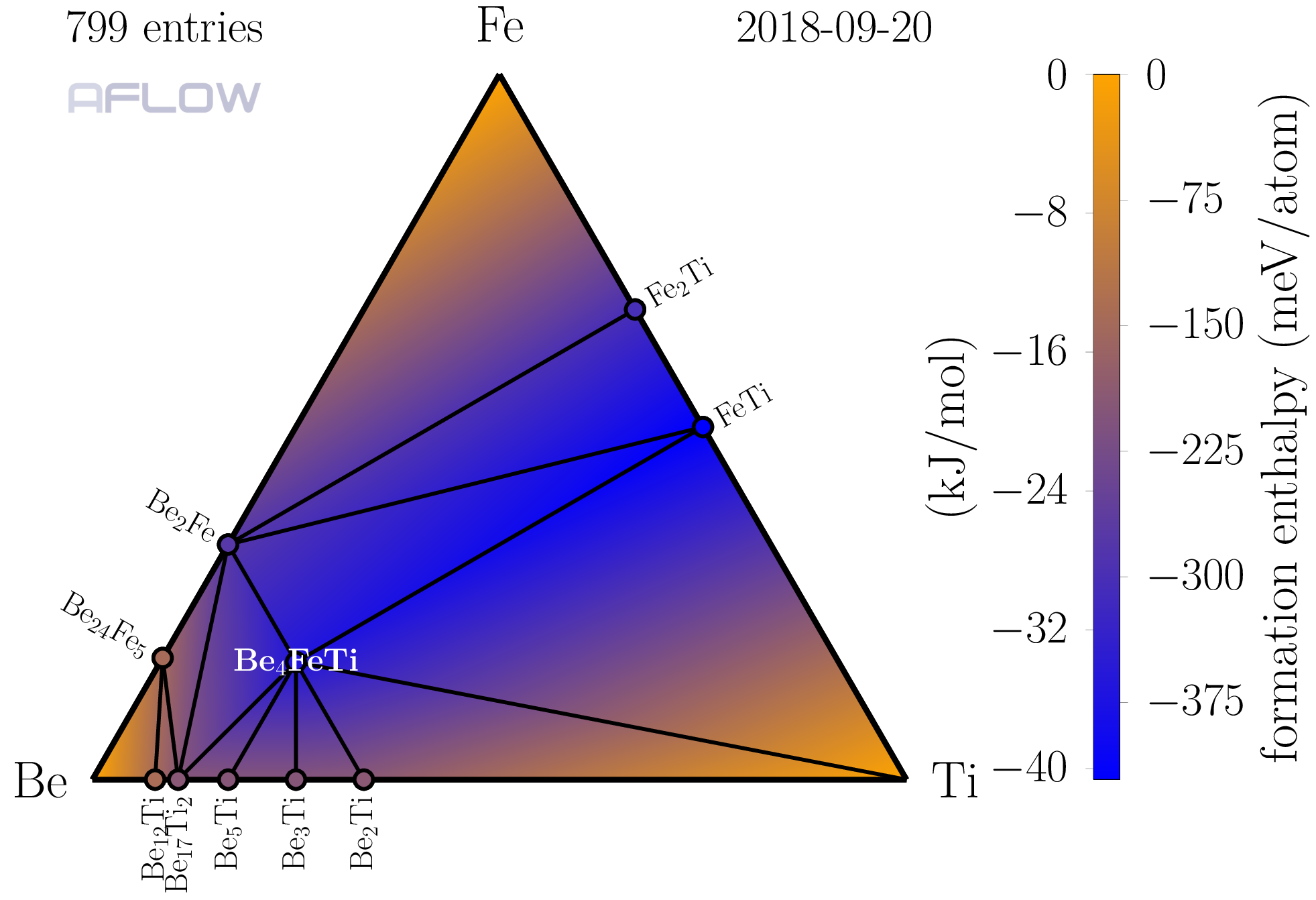}
\mycaption{Be-Fe-Ti ternary convex hull as plotted by \AFLOWHULL.}
\label{fig:art146:BeFeTi_ternary_hull_supp}
\efig

\vspace{\fill}

\figsec
\includegraphics[width=0.75\linewidth]{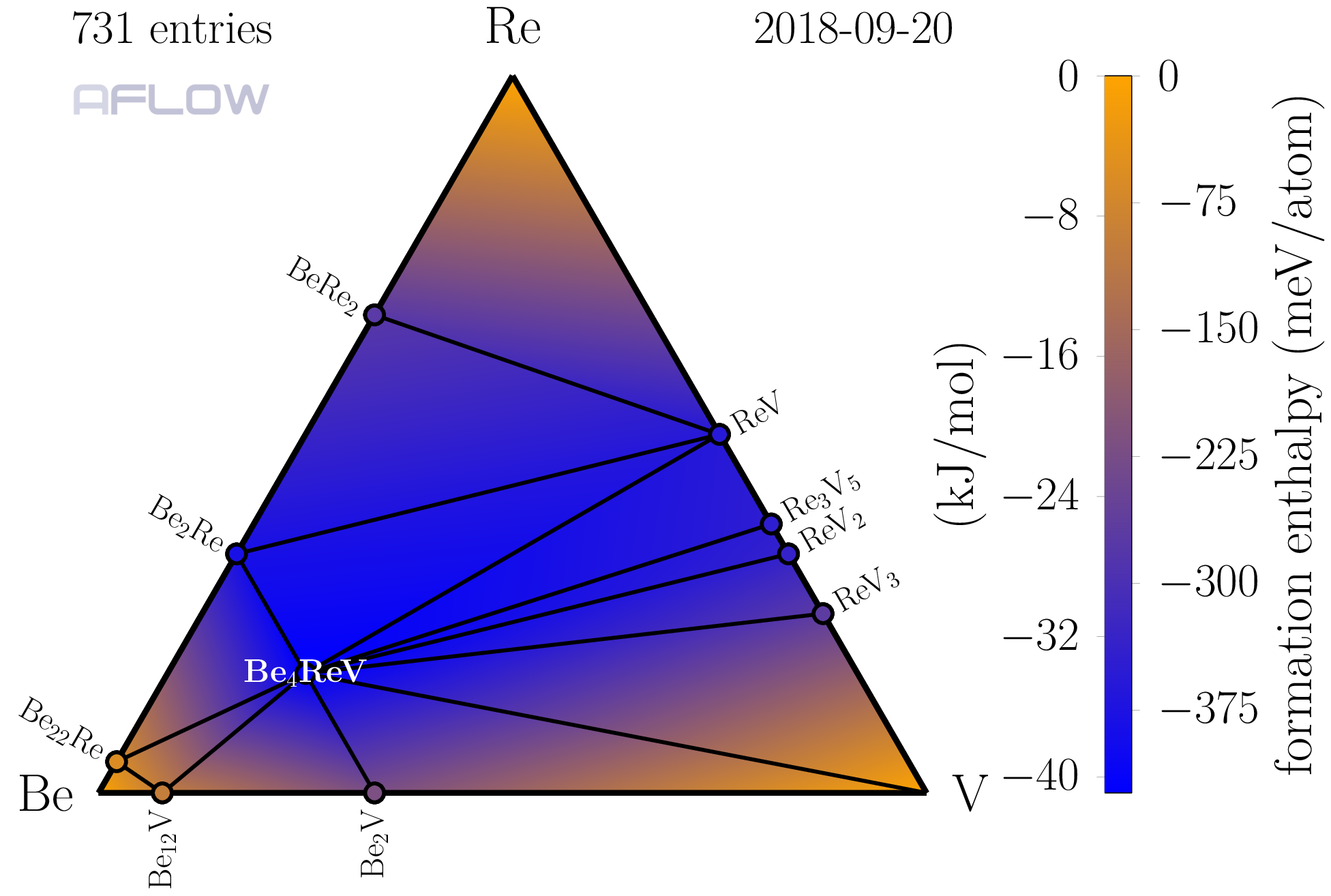}
\mycaption{Be-Re-V ternary convex hull as plotted by \AFLOWHULL.}
\label{fig:art146:BeReV_ternary_hull_supp}
\efig

\clearpage

\figsec
\includegraphics[width=0.75\linewidth]{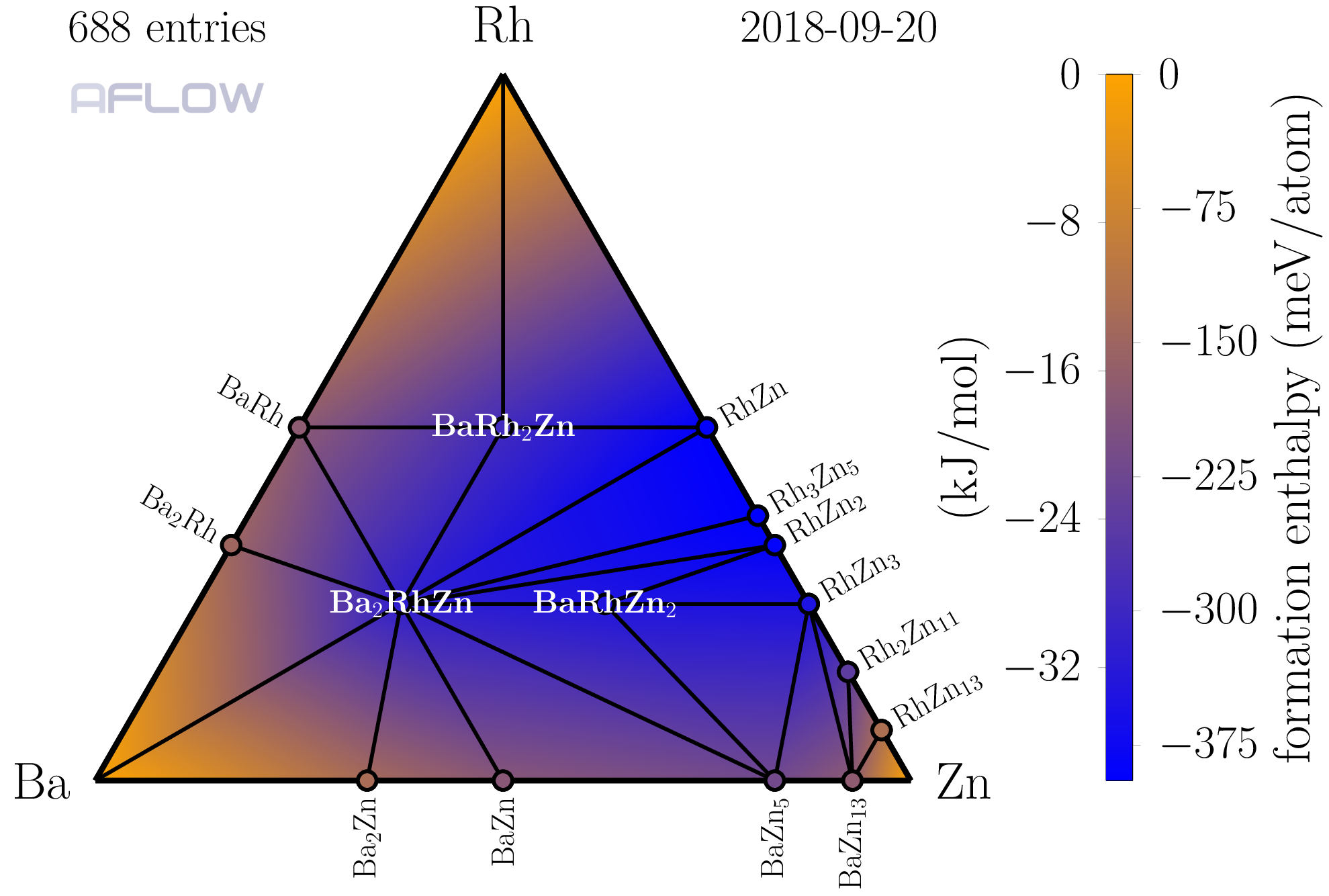}
\mycaption{Ba-Rh-Zn ternary convex hull as plotted by \AFLOWHULL.}
\label{fig:art146:BaRhZn_ternary_hull_supp}
\efig

\vspace{\fill}

\figsec
\includegraphics[width=0.75\linewidth]{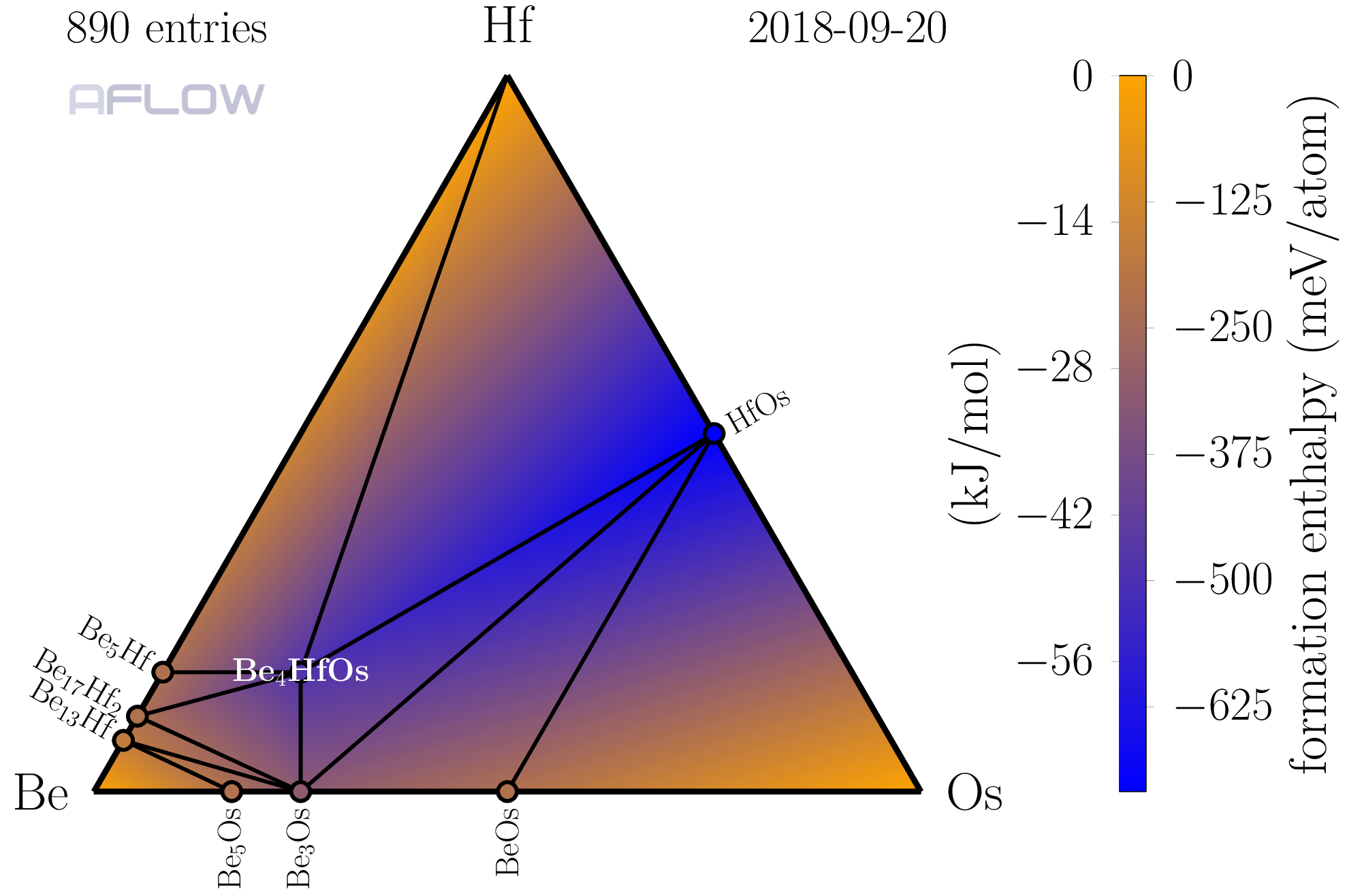}
\mycaption{Be-Hf-Os ternary convex hull as plotted by \AFLOWHULL.}
\label{fig:art146:BeHfOs_ternary_hull_supp}
\efig

\clearpage

\figsec
\includegraphics[width=0.75\linewidth]{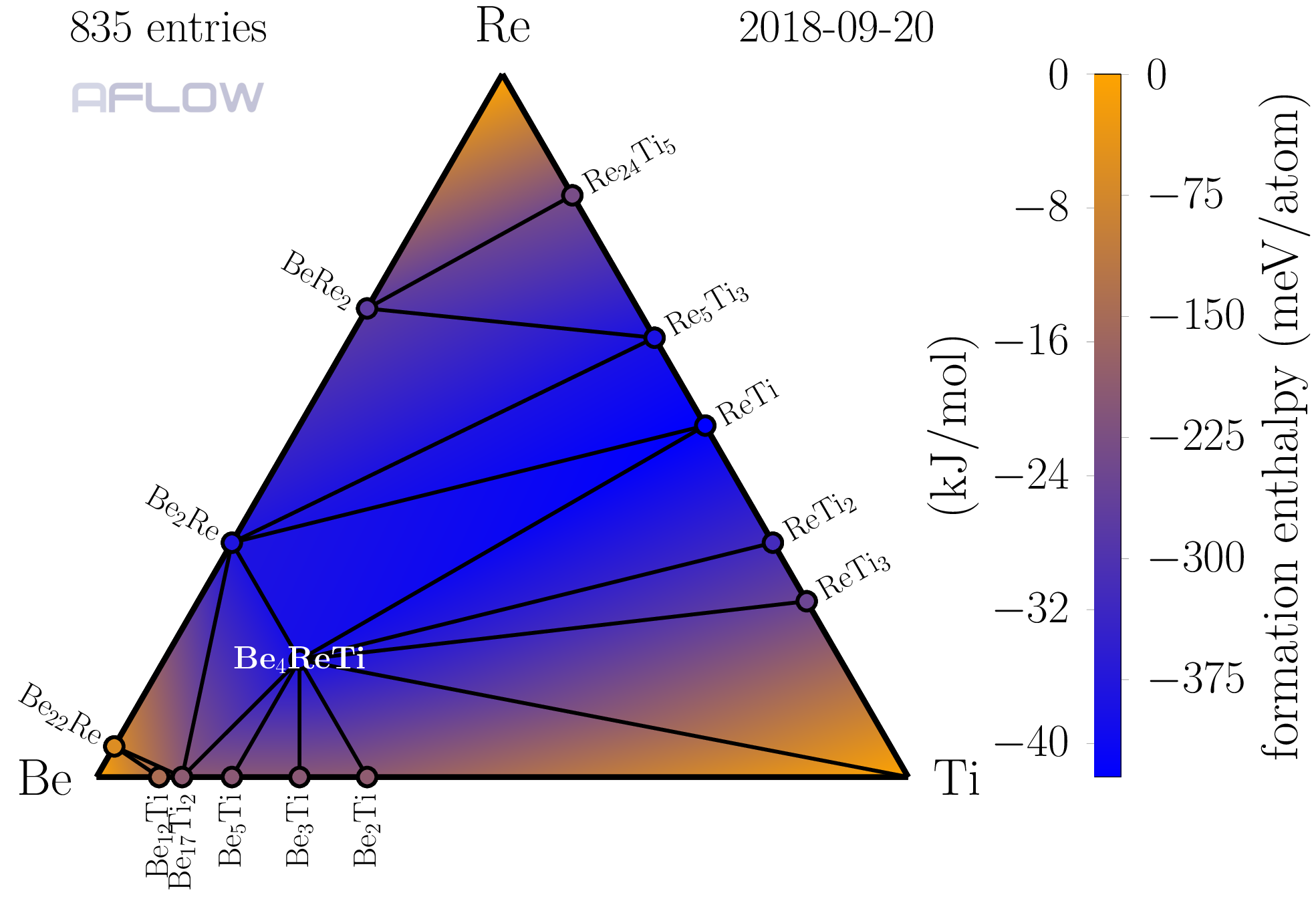}
\mycaption{Be-Re-Ti ternary convex hull as plotted by \AFLOWHULL.}
\label{fig:art146:BeReTi_ternary_hull_supp}
\efig

\vspace{\fill}

\figsec
\includegraphics[width=0.75\linewidth]{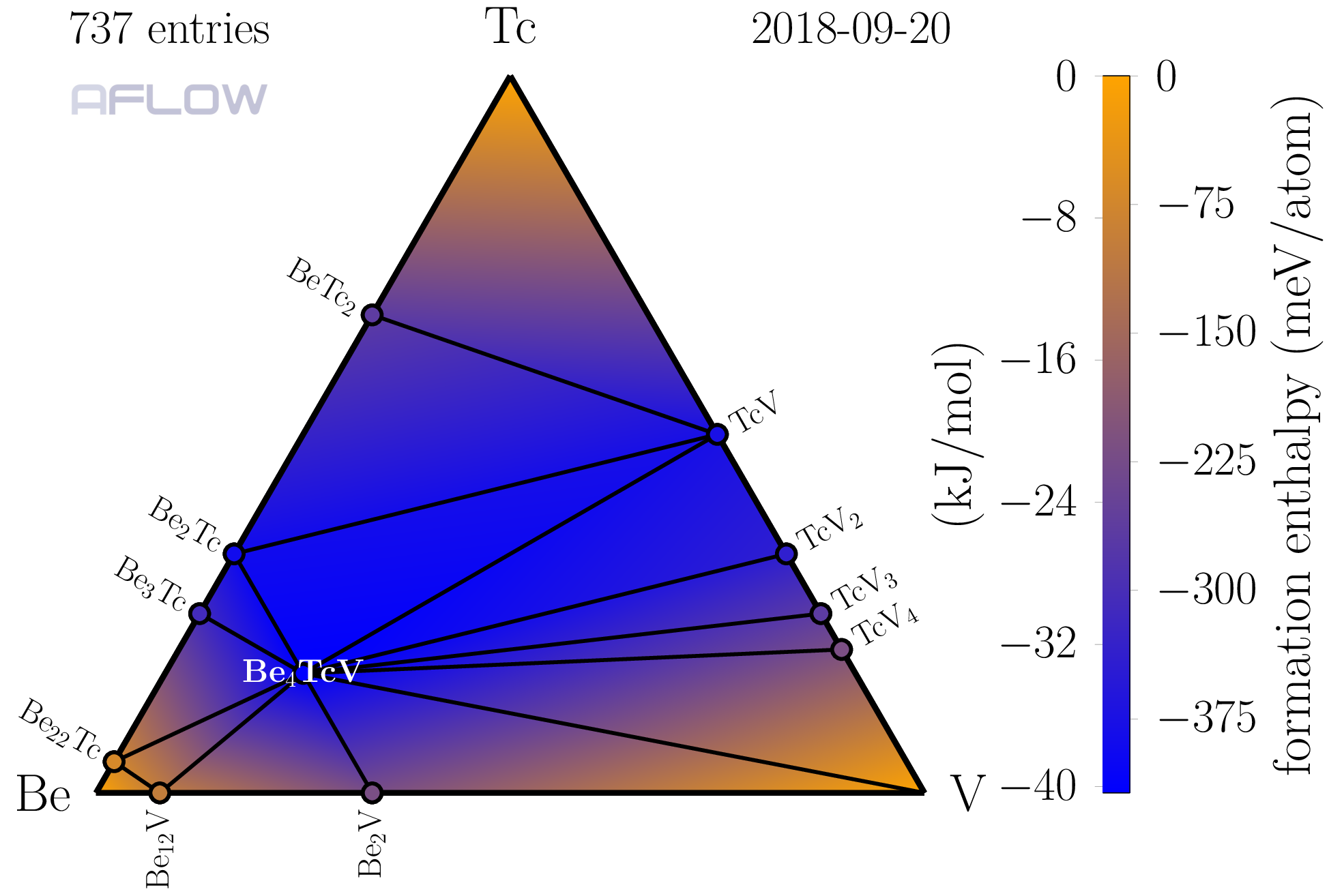}
\mycaption{Be-Tc-V ternary convex hull as plotted by \AFLOWHULL.}
\label{fig:art146:BeTcV_ternary_hull_supp}
\efig

\clearpage

\figsec
\includegraphics[width=0.75\linewidth]{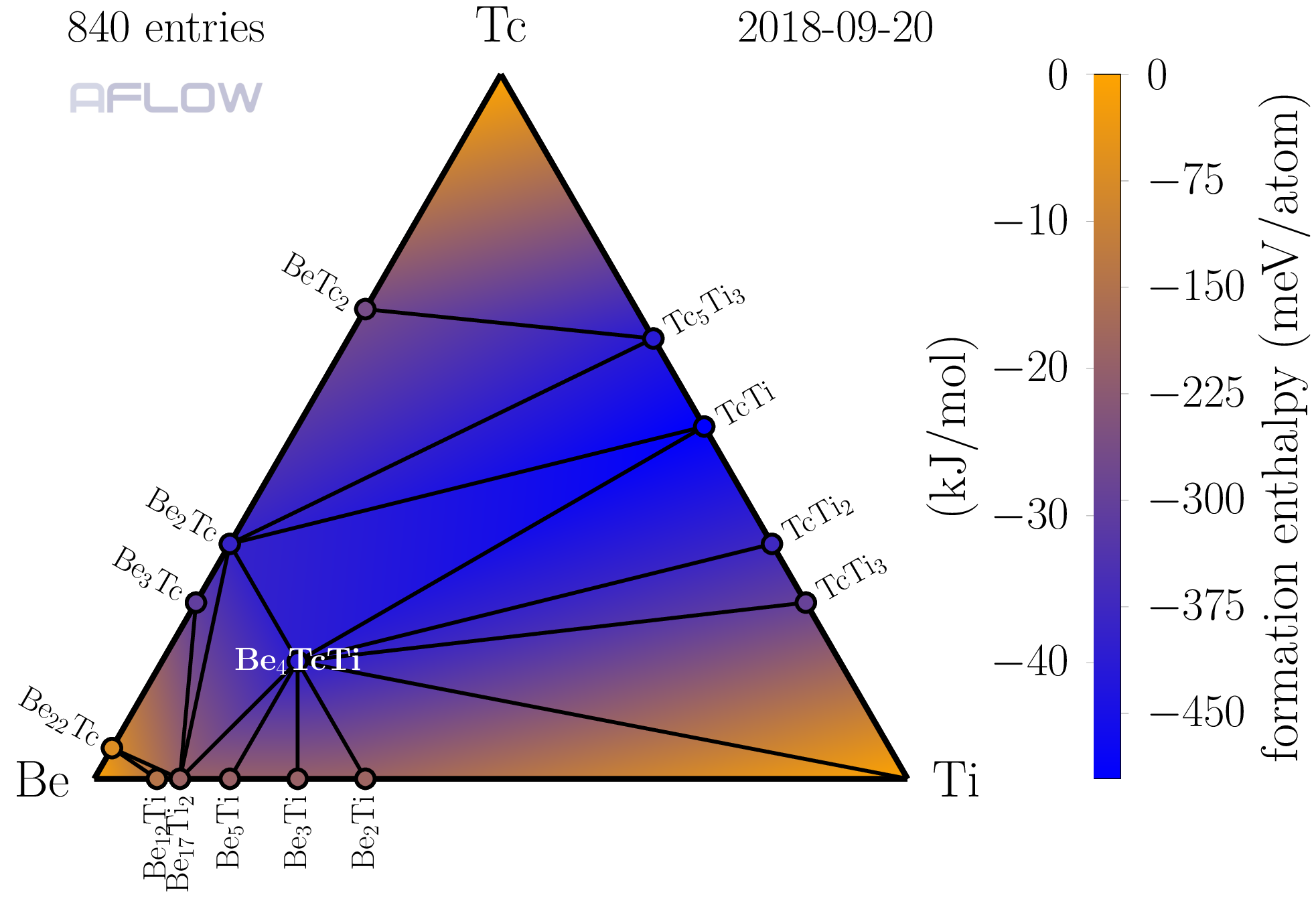}
\mycaption{Be-Tc-Ti ternary convex hull as plotted by \AFLOWHULL.}
\label{fig:art146:BeTcTi_ternary_hull_supp}
\efig

\vspace{\fill}

\figsec
\includegraphics[width=0.75\linewidth]{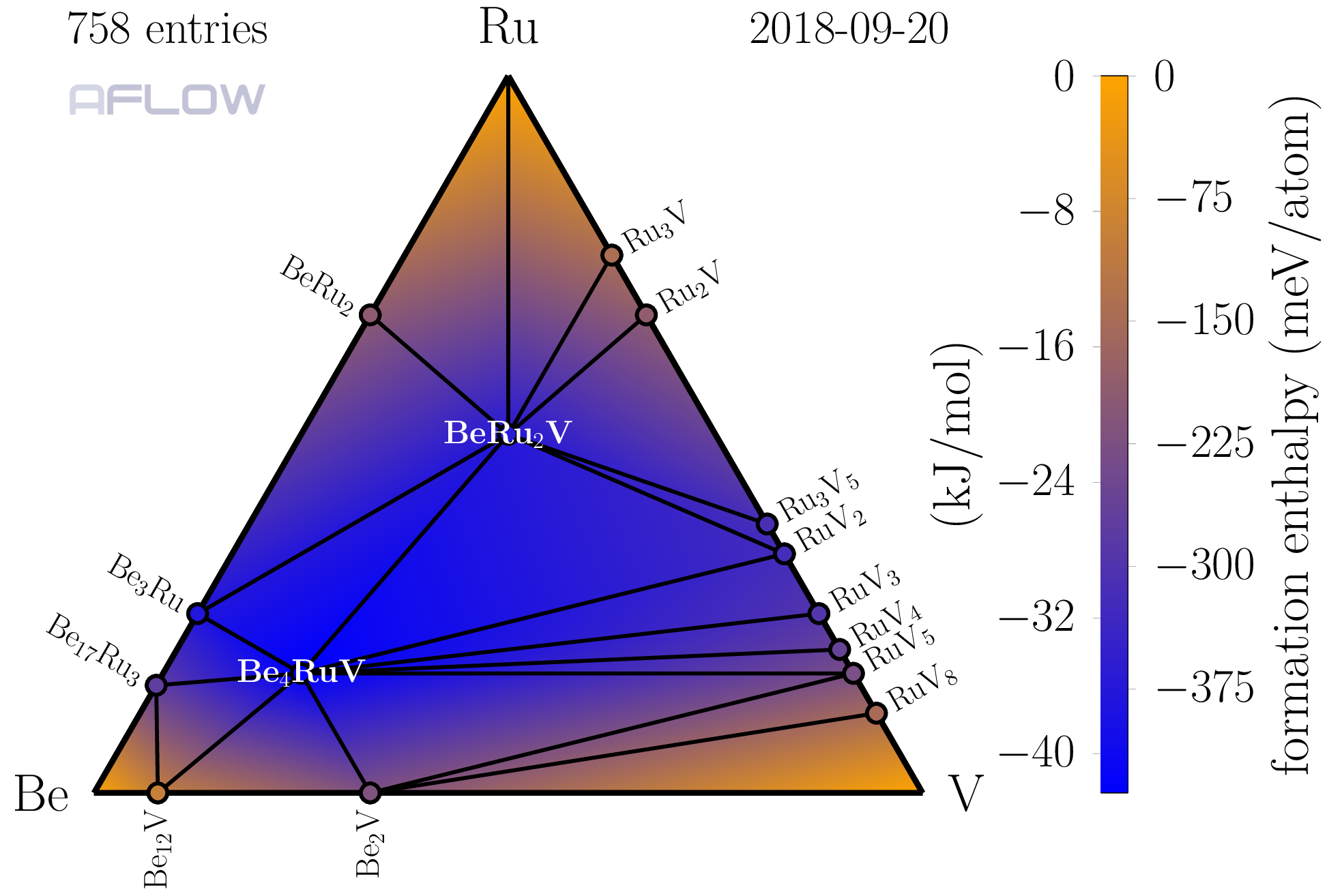}
\mycaption{Be-Ru-V ternary convex hull as plotted by \AFLOWHULL.}
\label{fig:art146:BeRuV_ternary_hull_supp}
\efig

\clearpage

\figsec
\includegraphics[width=0.75\linewidth]{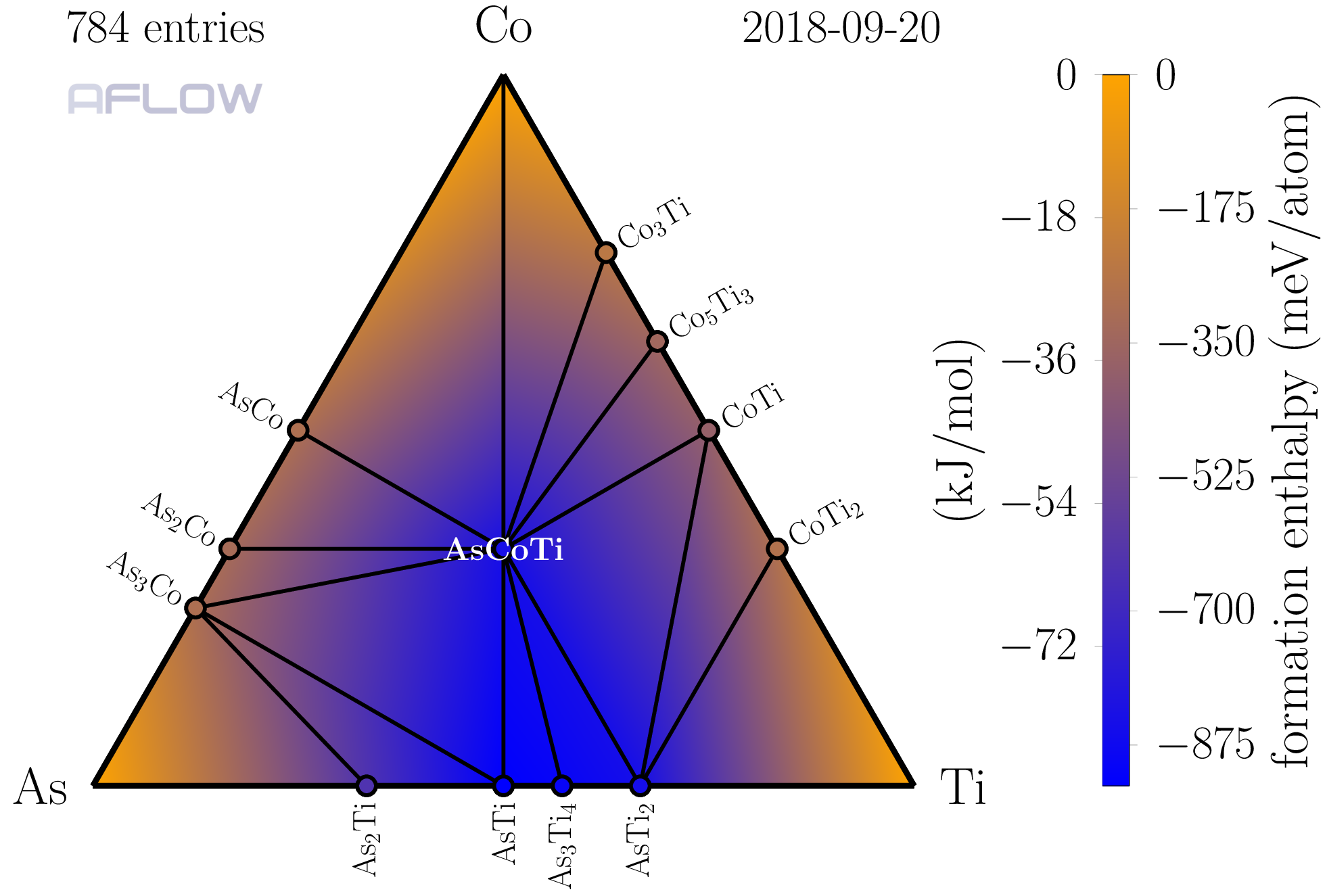}
\mycaption{As-Co-Ti ternary convex hull as plotted by \AFLOWHULL.}
\label{fig:art146:AsCoTi_ternary_hull_supp}
\efig

\vspace{\fill}

\figsec
\includegraphics[width=0.75\linewidth]{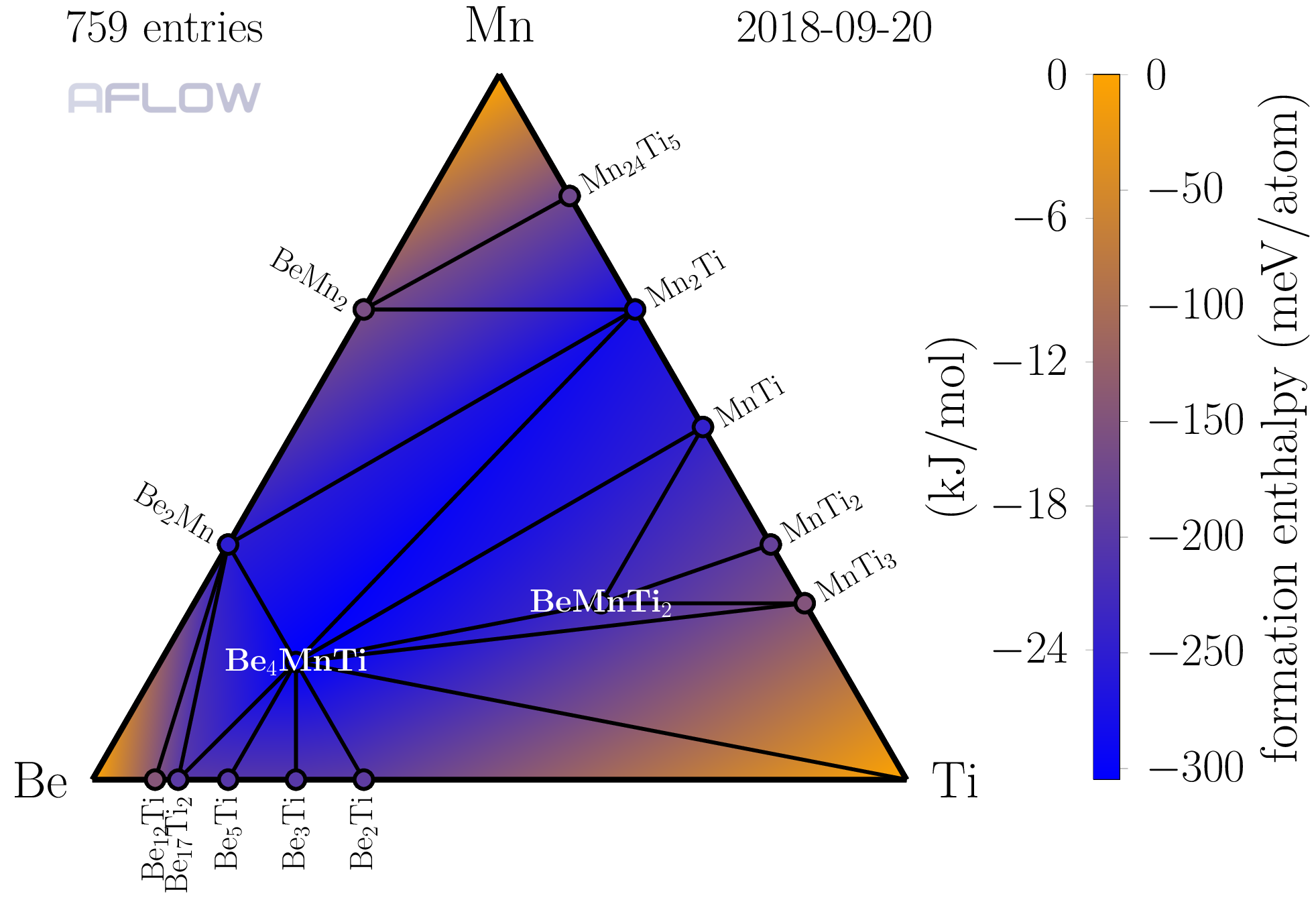}
\mycaption{Be-Mn-Ti ternary convex hull as plotted by \AFLOWHULL.}
\label{fig:art146:BeMnTi_ternary_hull_supp}
\efig

\clearpage

\figsec
\includegraphics[width=0.75\linewidth]{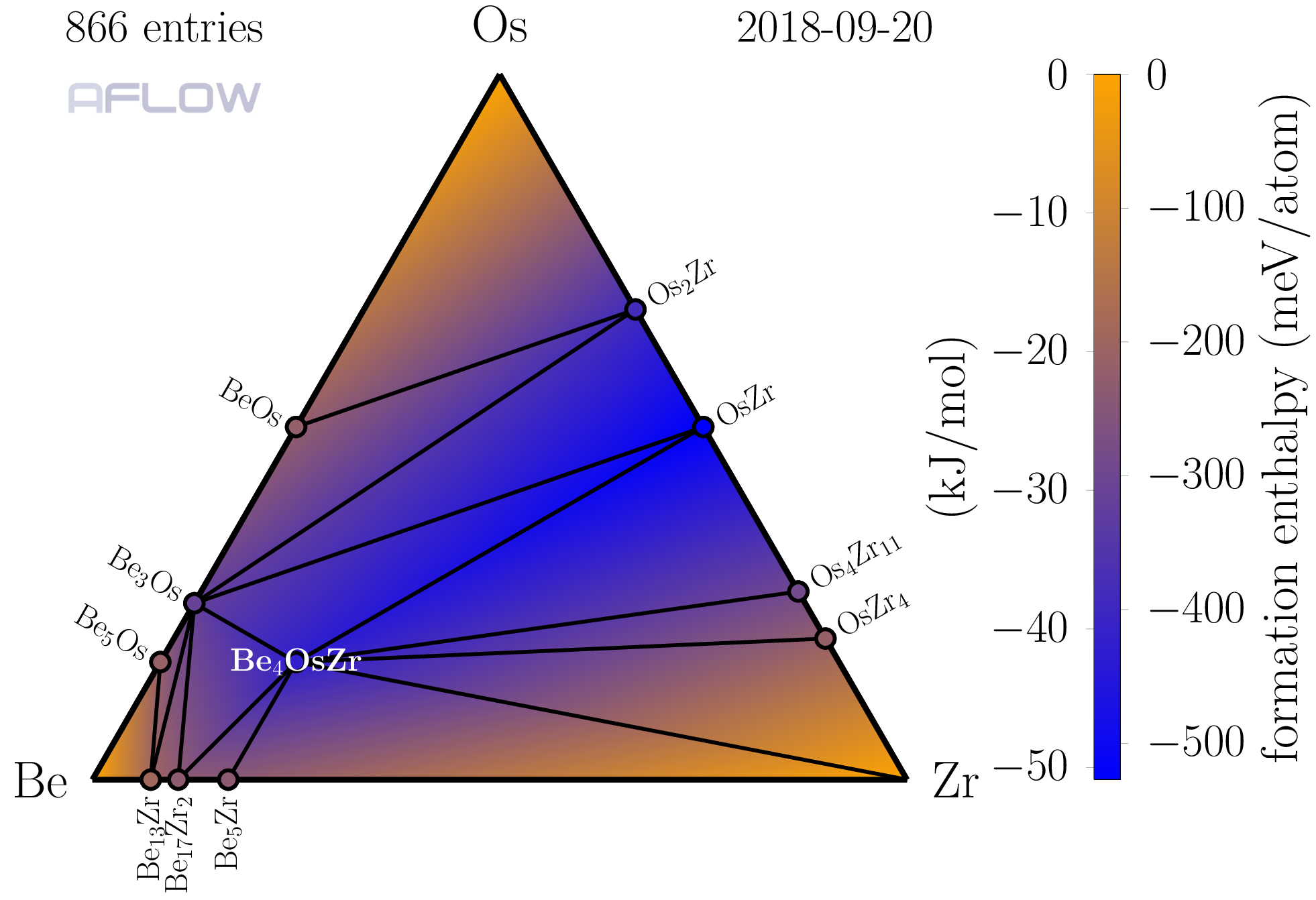}
\mycaption{Be-Os-Zr ternary convex hull as plotted by \AFLOWHULL.}
\label{fig:art146:BeOsZr_ternary_hull_supp}
\efig

\vspace{\fill}

\figsec
\includegraphics[width=0.75\linewidth]{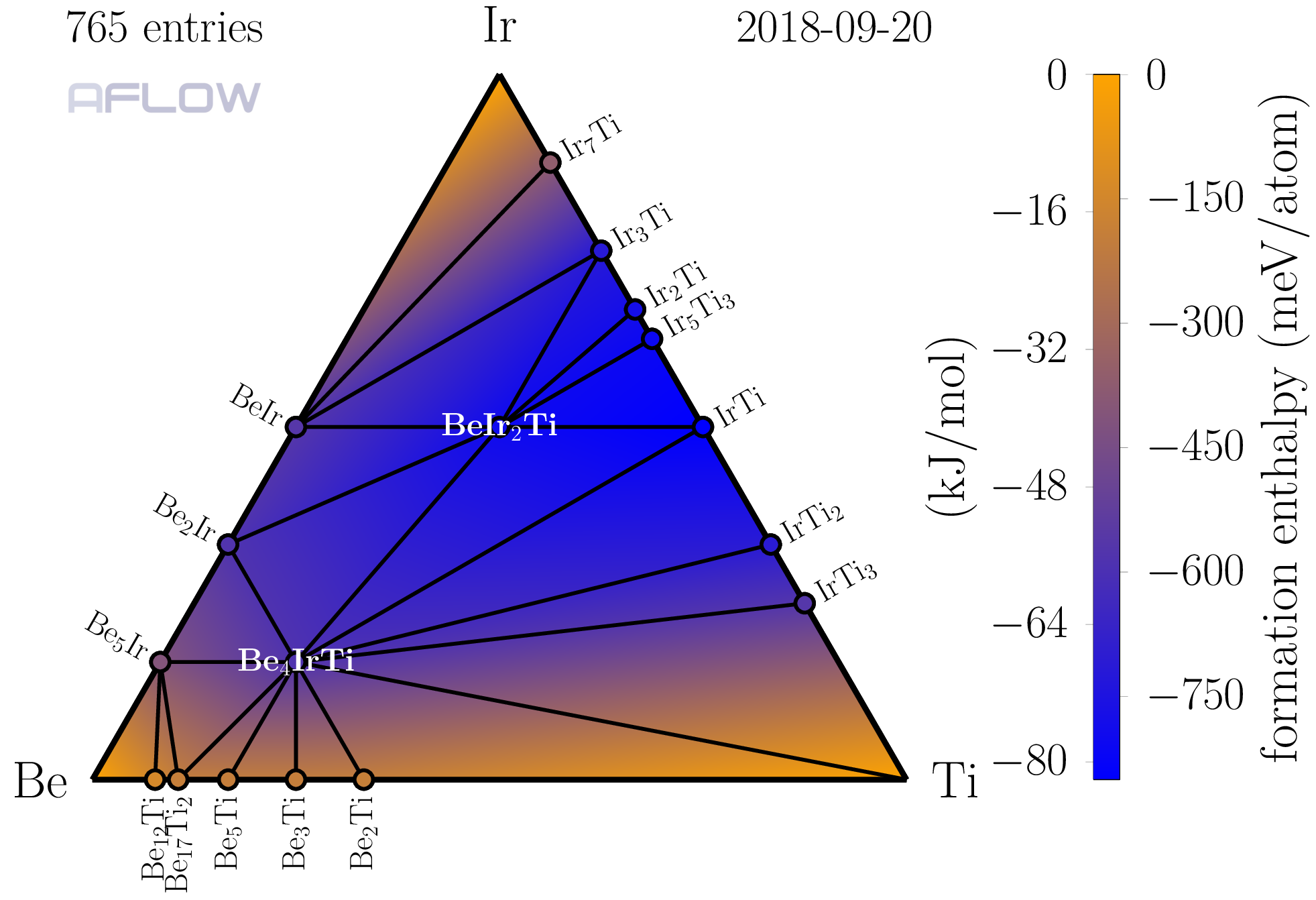}
\mycaption{Be-Ir-Ti ternary convex hull as plotted by \AFLOWHULL.}
\label{fig:art146:BeIrTi_ternary_hull_supp}
\efig

\clearpage

\figsec
\includegraphics[width=0.75\linewidth]{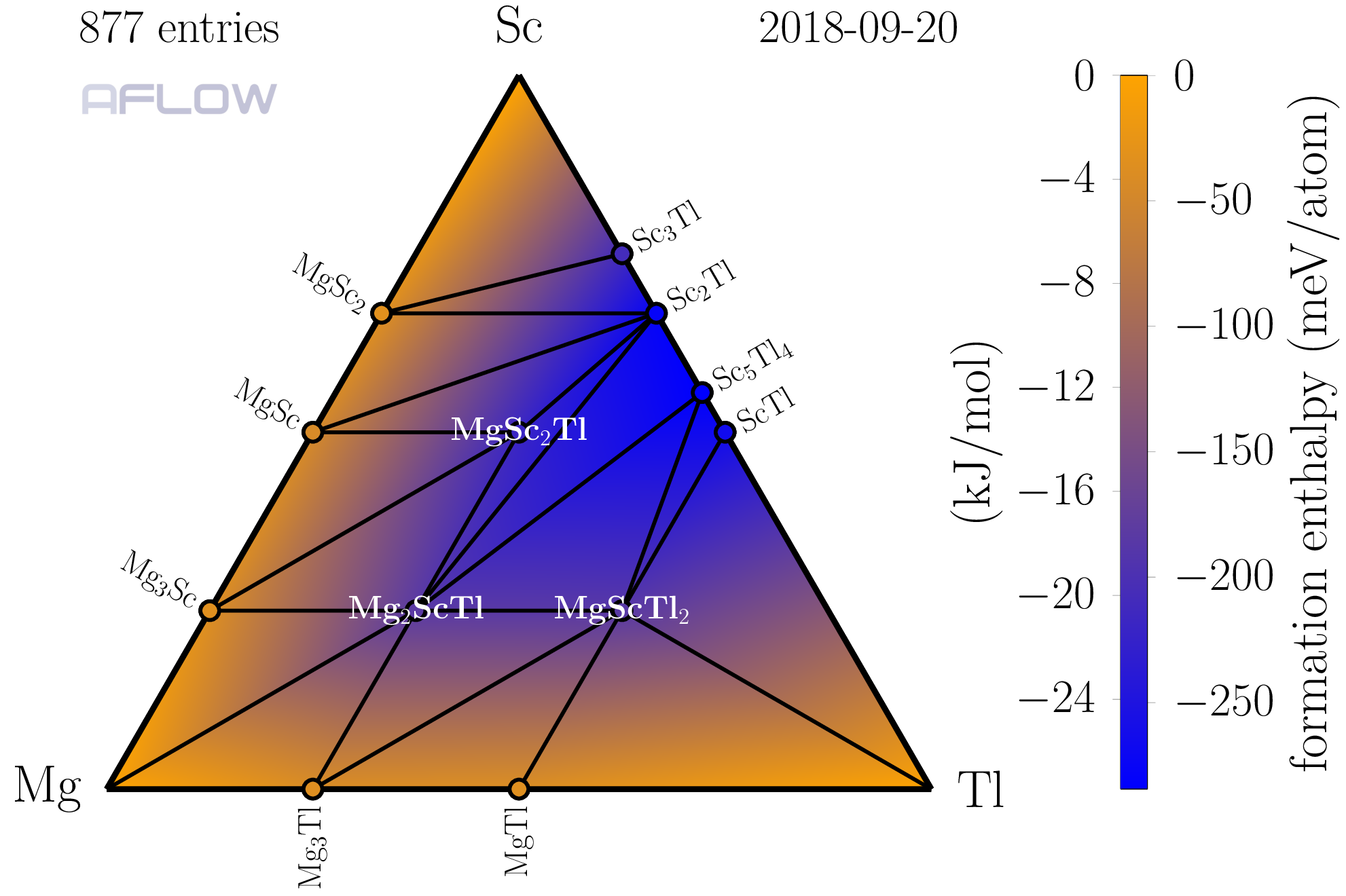}
\mycaption{Mg-Sc-Tl ternary convex hull as plotted by \AFLOWHULL.}
\label{fig:art146:MgScTl_ternary_hull_supp}
\efig

\vspace{\fill}

\figsec
\includegraphics[width=0.75\linewidth]{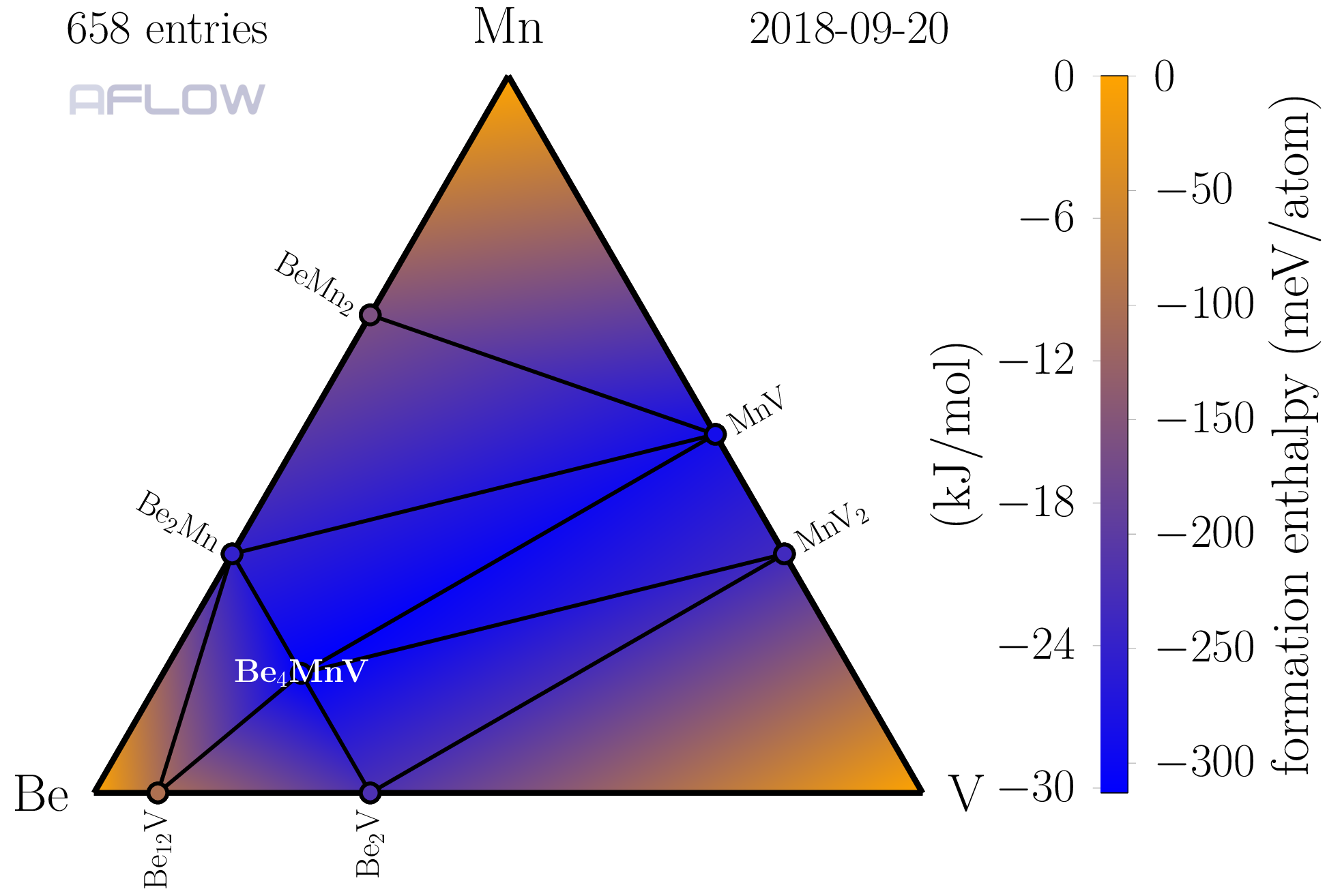}
\mycaption{Be-Mn-V ternary convex hull as plotted by \AFLOWHULL.}
\label{fig:art146:BeMnV_ternary_hull_supp}
\efig

\clearpage

\figsec
\includegraphics[width=0.75\linewidth]{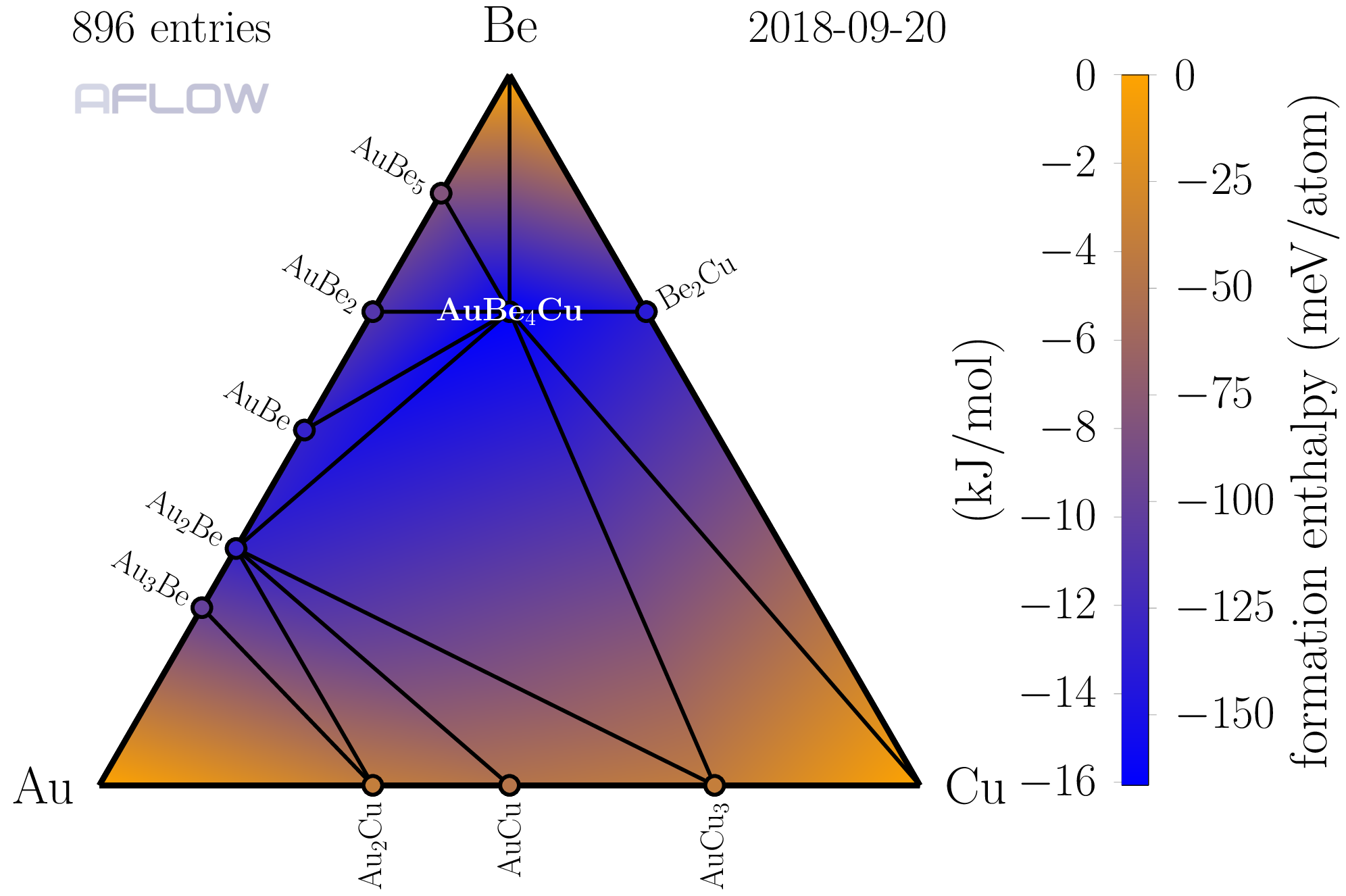}
\mycaption{Au-Be-Cu ternary convex hull as plotted by \AFLOWHULL.}
\label{fig:art146:AuBeCu_ternary_hull_supp}
\efig

\vspace{\fill}

\figsec
\includegraphics[width=0.75\linewidth]{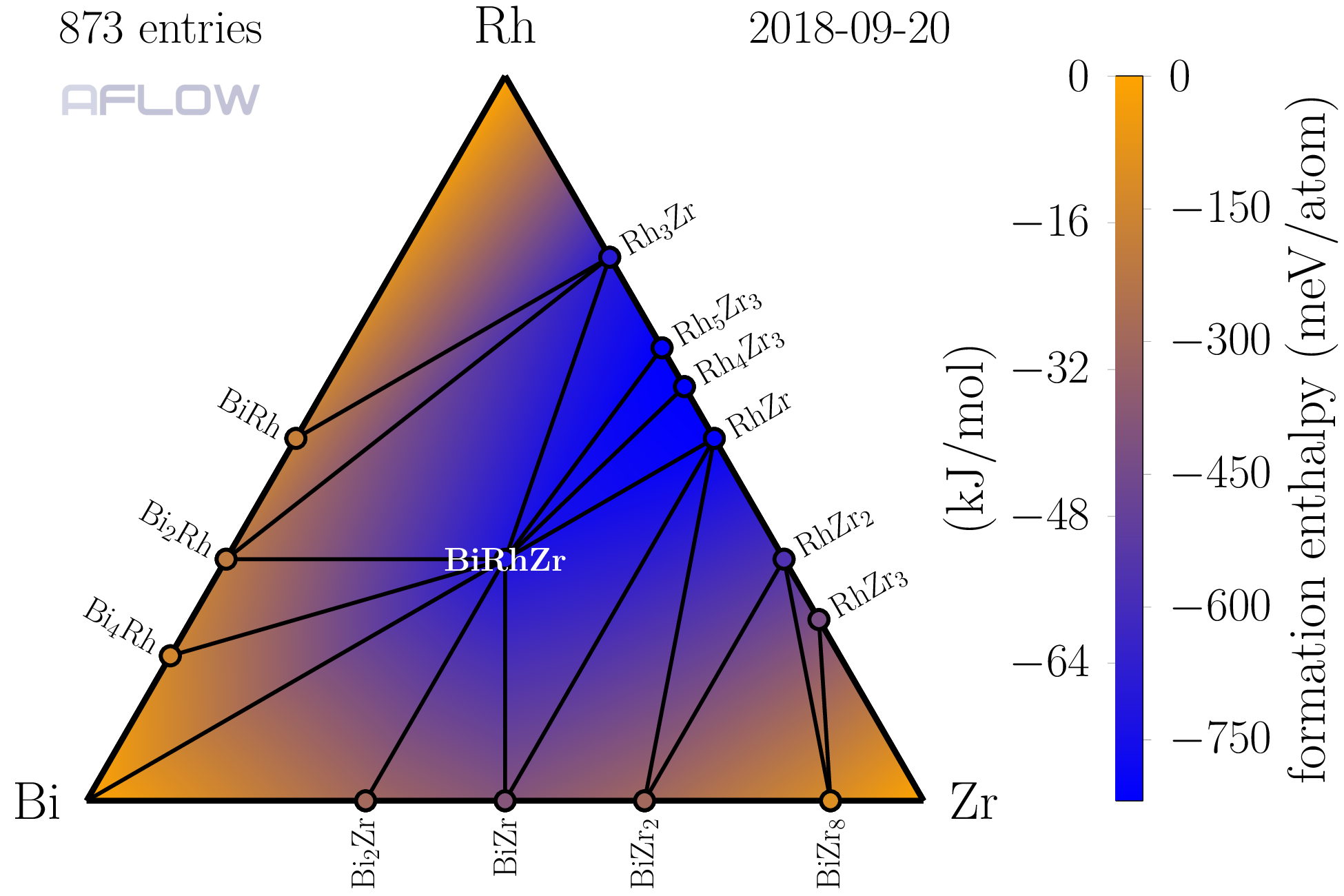}
\mycaption{Bi-Rh-Zr ternary convex hull as plotted by \AFLOWHULL.}
\label{fig:art146:BiRhZr_ternary_hull_supp}
\efig

\clearpage

\figsec
\includegraphics[width=0.75\linewidth]{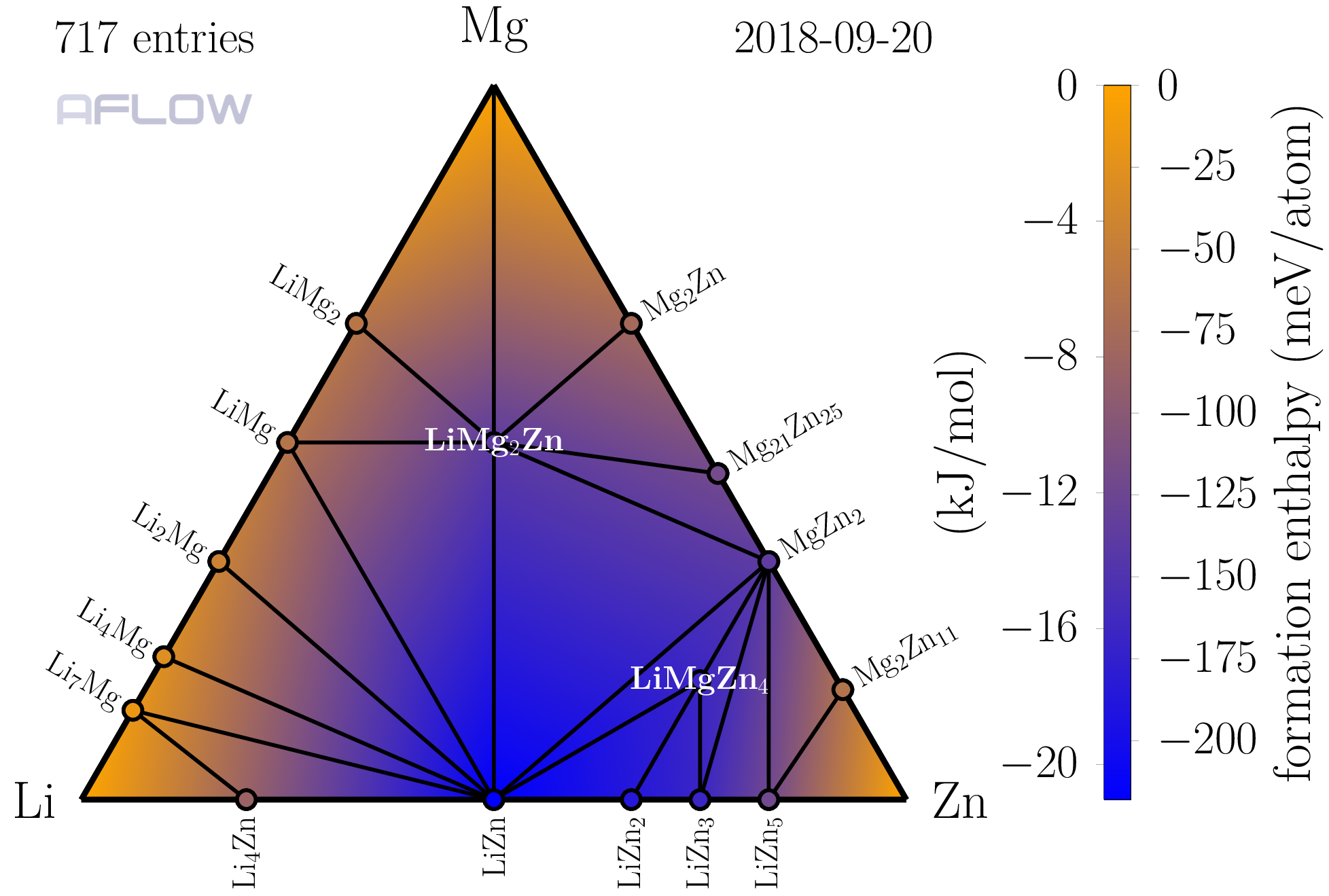}
\mycaption{Li-Mg-Zn ternary convex hull as plotted by \AFLOWHULL.}
\label{fig:art146:LiMgZn_ternary_hull_supp}
\efig

\vspace{\fill}

\figsec
\includegraphics[width=0.75\linewidth]{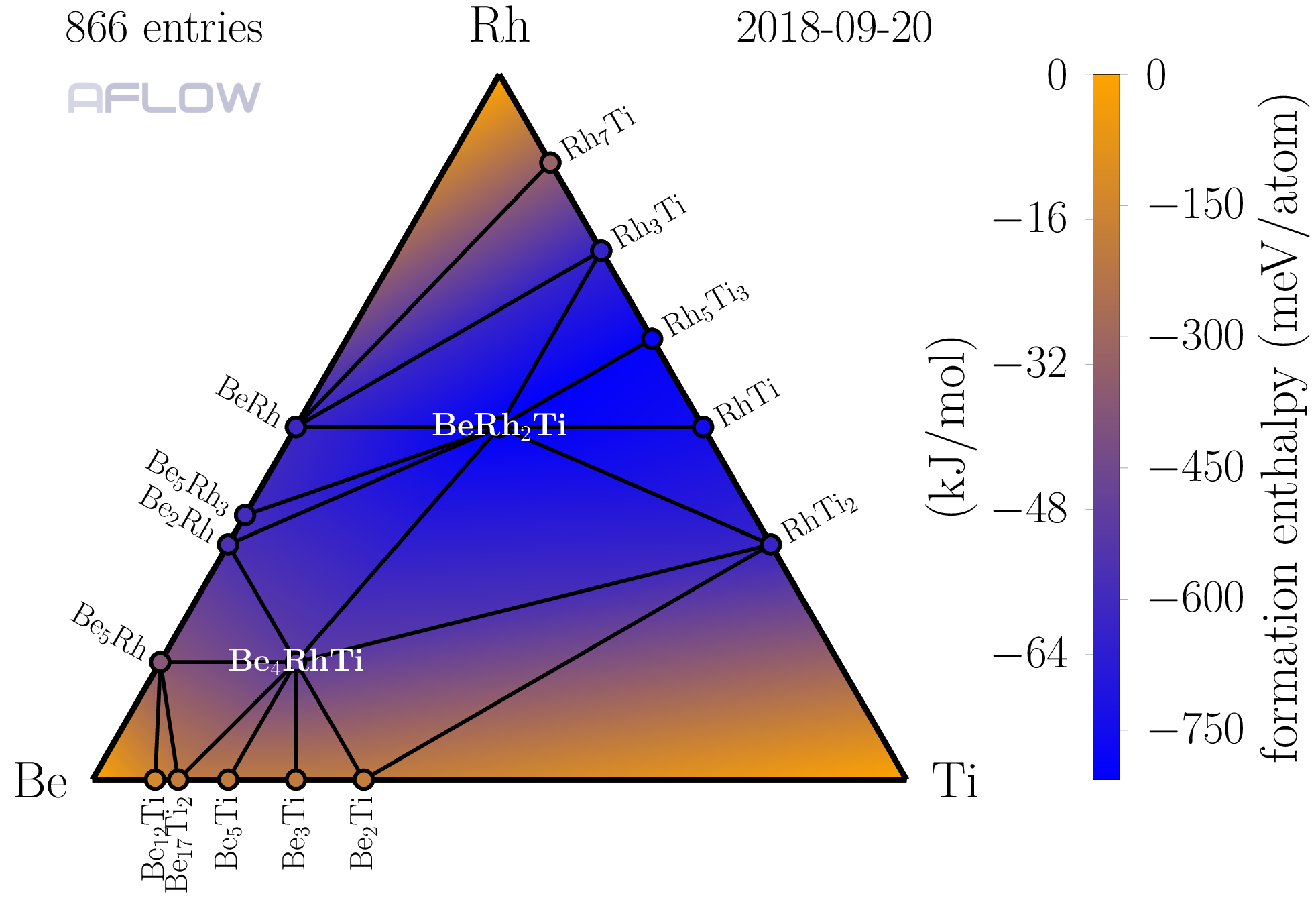}
\mycaption{Be-Rh-Ti ternary convex hull as plotted by \AFLOWHULL.}
\label{fig:art146:BeRhTi_ternary_hull_supp}
\efig

\clearpage

\figsec
\includegraphics[width=0.75\linewidth]{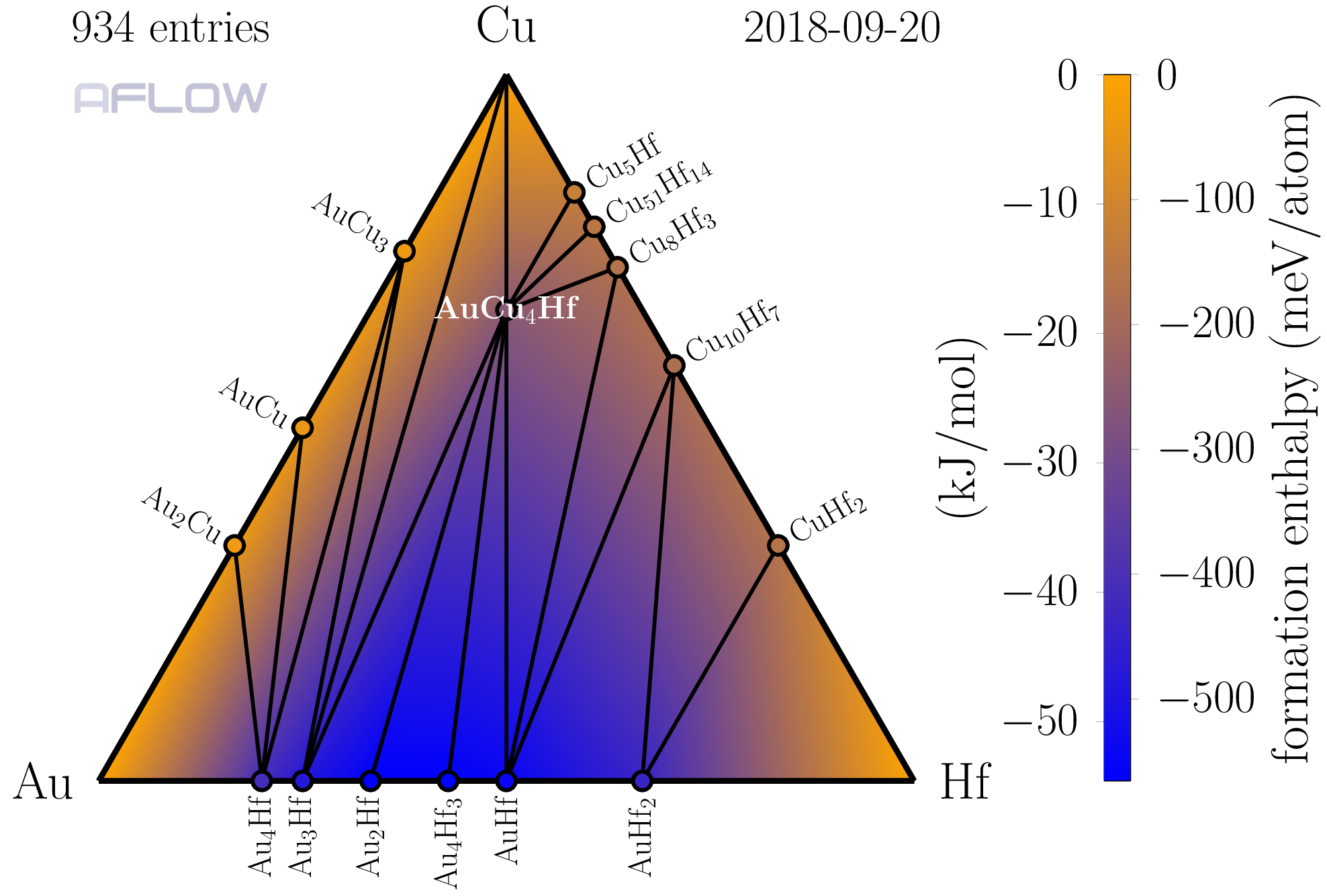}
\mycaption{Au-Cu-Hf ternary convex hull as plotted by \AFLOWHULL.}
\label{fig:art146:AuCuHf_ternary_hull_supp}
\efig

\def\description{\item {{\it Description:}\ }}
\def\type{\item {{\it Type:}\ }}
\def\units{\item {{\it Units:}\ }}
\def\similarto{\item {{\it Similar to:}\ }}
\def\bluedescription{{\color{blue}{\item {{\it Description:}\ }}}}
\def\bluetype{{\color{blue}{\item {{\it Type:}\ }}}}
\def\bluesimilarto{{\color{blue}{\item {{\it Similar to:}\ }}}}

\subsection{AFLOW-CHULL manual}

\boldsection{Command-line options.}
\AFLOWHULL\ is an integrated module of the \AFLOW\ \abinitio\ framework
which runs on any \UNIX-like computer, including those running macOS.
The most up-to-date binary can be downloaded from {\sf aflow.org/src/aflow}:
current version \AFLOWVERSION.
\AFLOWHULL\ depends on the compiled binary executable and an internet connection,
as all data is retrieved and analyzed \textit{in-situ}.
The default output option also requires the \LaTeX\ package.
The results (graphics and \PDF\ reports) presented herein are
compiled using pdf\TeX, Version 3.14159265-2.6-1.40.18 (\TeX\ Live 2017).

\vspace{0.5cm}

\noindent Primary commands:
\begin{itemize}
  \item{\verb!aflow --chull --alloy=InNiY!}
  \begin{itemize}
    \item{Calculates and returns the convex hull for system In-Ni-Y.}
  \end{itemize}
  \item{\verb!aflow --chull --alloy=InNiY! \\ \verb!--distance_to_hull=aflow:375066afdfb5a93f!}
  \begin{itemize}
    \item{Calculates and returns the distance to the hull $\Delta H_{\mathrm{f}}$ for \href{http://aflow.org/material.php?id=aflow:375066afdfb5a93f}{InNiY$_{4}$}.}
  \end{itemize}
  \item{\verb!aflow --chull --alloy=InNiY! \\ \verb!--stability_criterion=aflow:60a36639191c0af8!}
  \begin{itemize}
    \item{Calculates and returns the stability criterion $\delta_{\mathrm{sc}}$ for \href{http://aflow.org/material.php?id=aflow:60a36639191c0af8}{InNi$_{4}$Y}.
      The structure and relevant duplicates (if any) are removed to create the pseudo-hull.}
  \end{itemize}
  \item{\verb!aflow --chull --alloy=InNiY --hull_formation_enthalpy=0.25,0.25!}
  \begin{itemize}
    \item{Calculates and returns the formation enthalpy of the minimum energy surface at In$_{0.25}$Ni$_{0.25}$Y$_{0.5}$.
      The input composition is specified by implicit coordinates (refer to Equation~\ref{eq:art146:point}), where the last coordinate
      offers an optional energetic shift.
      }
  \end{itemize}
  \item{\verb!aflow --chull --usage!}
  \begin{itemize}
    \item{Prints full set of commands to the screen.}
  \end{itemize}
  \item{\verb!aflow --readme=chull!}
  \begin{itemize}
    \item{Prints a verbose manual (commands and descriptions) to the screen.}
  \end{itemize}
\end{itemize}

\vspace{0.5cm}

\noindent General options:
\begin{myitemize}
  \item{\verb!--output=pdf!}
  \begin{myitemize}
    \item{Selects the output format. Options include: \verb|pdf|, \verb|png|, \verb|json|, \verb|txt|, and \verb|full|. For multiple output, provide a comma-separated value list. A file with the corresponding extension is created, \eg, {\sf aflow\_InNiY\_hull.pdf}.}
  \end{myitemize}
\item{\verb!--destination=$HOME/!}
  \begin{myitemize}
    \item{Sets the output path to {\sf \${\small HOME}}. All output will be redirected to this destination.}
  \end{myitemize}
\item \verb!--keep=log!
  \begin{myitemize}
    \item{Creates a log file with verbose output of the calculation, \eg, {\sf aflow\_InNiY\_hull.log}.}
  \end{myitemize}
\end{myitemize}

\vspace{0.5cm}

\noindent Loading options:
\begin{myitemize}
\item \verb!--load_library=icsd!
  \begin{myitemize}
    \item{Limits the catalogs from which entries are loaded. Options include: \verb!icsd!, \verb!lib1!, \verb!lib2!, and \verb!lib3!. For multiple catalogs, provide a comma-separated value list.}
  \end{myitemize}
\item \verb!--load_entries_entry_output!
  \begin{myitemize}
    \item{Prints verbose output of the entries loaded.  This output is included in the log file by default.}
  \end{myitemize}
\item \verb!--neglect=aflow:60a36639191c0af8,aflow:3f24d2be765237f1!
  \begin{myitemize}
    \item{Excludes individual points from the convex hull calculation.}
  \end{myitemize}
\item \verb!--see_neglect!
  \begin{myitemize}
    \item{Prints verbose output of the entries neglected from the calculation, including ill-calculated entries, duplicates, outliers, and those requested via \verb!--neglect!.}
  \end{myitemize}
\item \verb!--remove_extreme_points=-1000!
  \begin{myitemize}
    \item{Excludes all points with formation enthalpies below -1000 meV/atom.}
  \end{myitemize}
\item \verb!--include_paw_gga!
  \begin{myitemize}
    \item{Includes all entries calculated with \PAW-\GGA\ (in addition to those calculated with \PAW-\PBE).
    \PAW-\GGA\ refers to the \underline{G}eneralized \underline{G}radient \underline{A}pproximation functional~\cite{PBE}
    with pseudopotentials calculated with the
    \underline{p}rojector \underline{a}ugmented \underline{w}ave method~\cite{PAW}.
    This flag is needed to generate Figure~\ref{fig:art146:hull_analyses}(f).}
  \end{myitemize}
\end{myitemize}

\vspace{0.5cm}

\noindent Analysis options:
\begin{myitemize}
\item \verb!--skip_structure_comparison!
  \begin{myitemize}
    \item{Avoids determination of structures equivalent to stable phases (speed).}
  \end{myitemize}
\item \verb!--skip_stability_criterion_analysis!
  \begin{myitemize}
    \item{Avoids determination of the stability criterion of stable phases (speed).}
  \end{myitemize}
\item \verb!--include_skewed_hulls!
  \begin{myitemize}
    \item{Proceeds to calculate the hull in the event that it is determined ``skewed'', \ie,
      the stable elemental phase differs from the reference energy by more than 15~meV/atom.
      This flag is needed to generate Figure~\ref{fig:art146:hull_analyses}(f).}
  \end{myitemize}
\item \verb!--include_unreliable_hulls!
  \begin{myitemize}
    \item{Proceeds to calculate the hull in the event that it is determined unreliable (fewer than 200 entries along the binary hulls).}
  \end{myitemize}
\item \verb!--include_outliers!
  \begin{myitemize}
    \item{Includes outliers in the calculation.}
  \end{myitemize}
\item \verb!--force!
  \begin{myitemize}
    \item{Forces an output, ignoring all warnings.}
  \end{myitemize}
\end{myitemize}

\vspace{0.5cm}

\noindent \PDF/\LaTeX\ options:
\begin{myitemize}
\item \verb!--image_only!
  \begin{myitemize}
    \item{Creates a \PDF\ with the hull illustration only.}
  \end{myitemize}
\item \verb!--document_only!
  \begin{myitemize}
    \item{Creates a \PDF\ with the thermodynamic report only. Default for dimensions $N>3$.}
  \end{myitemize}
\item \verb!--keep=tex!
  \begin{myitemize}
    \item{Saves the \LaTeX\ input file (deleted by default), allowing for customization of the resulting \PDF, \eg, {\sf aflow\_InNiY\_hull.tex}.}
  \end{myitemize}
\item \verb!--latex_interactive!
  \begin{myitemize}
    \item{Displays the \LaTeX\ compilation output and enables interaction with the program.}
  \end{myitemize}
\item \verb!--plot_iso_max_latent_heat!
  \begin{myitemize}
    \item{Plots the entropic temperature envelopes shown in Figure~\ref{fig:art146:hull_analyses}(f). Limited to binary systems only.}
  \end{myitemize}
\end{myitemize}

\vspace{0.5cm}

\boldsection{{\AFLOW}rc options.}
The {\sf .aflow.rc} file is a new protocol for specifying \AFLOW\ default options.
The file emulates the {\sf .bashrc} script that runs when initializing an interactive environment in
Bash (\underline{B}ourne \underline{a}gain \underline{sh}ell).
A fresh {\sf .aflow.rc} file is created in {\sf \${\small HOME}} if one is not already
present.

\noindent Relevant \AFLOWHULL\ options include:
\begin{myitemize}
\item \verb!DEFAULT_CHULL_ALLOWED_DFT_TYPES="PAW_PBE"!
  \begin{myitemize}
      \description Defines the allowed entries based on \underline{d}ensity \underline{f}unctional \underline{t}heory (\DFT) calculation type (comma-separated value).
      Options include: \verb|US|, \verb|GGA|, \verb|PAW_LDA|, \verb|PAW_GGA|, \verb|PAW_PBE|, \verb|GW|, and \verb|HSE06|~\cite{aflowAPI}.
    \type \verb|string|
  \end{myitemize}
\item \verb!DEFAULT_CHULL_ALLOW_ALL_FORMATION_ENERGIES=0!
  \begin{myitemize}
    \description Allows all entries independent of \DFT\ calculation type~\cite{aflowAPI}.
    \type \verb|0 (false) or 1 (true)|
  \end{myitemize}
\item \verb!DEFAULT_CHULL_COUNT_THRESHOLD_BINARIES=200!
  \begin{myitemize}
    \description Defines the minimum number of entries for a reliable binary hull.
    \type \verb|integer|
  \end{myitemize}
\item \verb!DEFAULT_CHULL_PERFORM_OUTLIER_ANALYSIS=1!
  \begin{myitemize}
    \description Enables determination of outliers.
    \type \verb|0 (false) or 1 (true)|
  \end{myitemize}
\item \verb!DEFAULT_CHULL_OUTLIER_ANALYSIS_COUNT_THRESHOLD_BINARIES=50!
  \begin{myitemize}
    \description Defines the minimum number of entries for a reliable outlier analysis.
      Only phases stable with respect to their end-members are considered for the outlier analysis (below the zero $H_{\mathrm{f}}$ tie-line).
    \type \verb|integer|
  \end{myitemize}
\item \verb!DEFAULT_CHULL_OUTLIER_MULTIPLIER=3.25!
  \begin{myitemize}
    \description Defines the bounds beyond the interquartile range for which points are considered outliers~\cite{Miller_QJEPSA_1991}.
    \type \verb|double|
  \end{myitemize}
\item \verb!DEFAULT_CHULL_IGNORE_KNOWN_ILL_CONVERGED=1!
  \begin{myitemize}
      \description \AFLOW\ maintains a list of (older) prototypes known to have converged poorly.
      These entries are likely outliers, \eg, see prototype $549$ in Figure~\ref{fig:art146:hull_analyses}(a).
      If this flag is on (\verb|1|), then the entries are removed before the analysis.
      Turning this flag off (\verb|0|) is not recommended.
    \type \verb|0 (false) or 1 (true)|
  \end{myitemize}
\item \verb!DEFAULT_CHULL_LATEX_PLOT_UNARIES=0!
  \begin{myitemize}
    \description Incorporates the end-members in the convex hull illustration.
    \type \verb|0 (false) or 1 (true)|
  \end{myitemize}
\item \verb!DEFAULT_CHULL_LATEX_PLOT_OFF_HULL=-1!
  \begin{myitemize}
    \description Incorporates off-hull phases in the convex hull illustration, but excludes phases unstable with respect to their end-members (above the zero $H_{\mathrm{f}}$ tie-line).
      Only three values are accepted: \verb|-1| (default: true for 2-dimensional systems, false for 3-dimensional systems), \verb|0| (false), \verb|1| (true).
      \type \verb|-1 (default), 0 (false), or 1 (true)|
  \end{myitemize}
\item \verb!DEFAULT_CHULL_LATEX_PLOT_UNSTABLE=0!
  \begin{myitemize}
    \description Incorporates all unstable phases in the convex hull illustration.
    \type \verb|0 (false) or 1 (true)|
  \end{myitemize}
\item \verb!DEFAULT_CHULL_LATEX_FILTER_SCHEME="energy-axis"!
  \begin{myitemize}
    \description Defines the exclusion scheme for the convex hull illustration.
    In contrast to \verb!--neglect!, this scheme is limited to the illustration, and points are still included in the analysis/report.
    The following strings are accepted: \verb!energy-axis!, \verb!distance!, and an empty string.
    \verb!energy-axis! refers to a scheme that eliminates structures from the illustration based on their formation enthalpies.
    On the other hand, \verb!distance! refers to a scheme that eliminates structures from the illustration based on their distances to the hull.
    An empty string signifies no exclusion scheme.
    The criteria (value) for elimination is defined by \\ \verb!DEFAULT_CHULL_LATEX_FILTER_VALUE!.
    \type \verb|string|
  \end{myitemize}
\item \verb!DEFAULT_CHULL_LATEX_FILTER_VALUE=50!
  \begin{myitemize}
    \description Defines the value beyond which points are excluded per the scheme defined with \verb!DEFAULT_CHULL_LATEX_FILTER_SCHEME!.
      In this case, \\ \AFLOWHULL\ would filter points with formation enthalpies greater than 50 meV.
    \type \verb|double|
  \end{myitemize}
\item \verb!DEFAULT_CHULL_LATEX_COLOR_BAR=1!
  \begin{myitemize}
    \description Defines whether to show the color bar graphic (3-dimensional illustration only). Colors can still be incorporated without the color bar graphic.
    \type \verb|0 (false) or 1 (true)|
  \end{myitemize}
\item \verb!DEFAULT_CHULL_LATEX_HEAT_MAP=1!
  \begin{myitemize}
      \description Defines whether to color facets with heat maps illustrating their depth (3-dimensional illustration only).
    \type \verb|0 (false) or 1 (true)|
  \end{myitemize}
\item \verb!DEFAULT_CHULL_LATEX_COLOR_GRADIENT=1!
  \begin{myitemize}
    \description Defines whether to incorporate a color scheme at all in the illustration.
      Turning this flag off will also turn off \verb!DEFAULT_CHULL_LATEX_COLOR_BAR! and \verb!DEFAULT_CHULL_LATEX_HEAT_MAP!.
    \type \verb|0 (false) or 1 (true)|
  \end{myitemize}
\item \verb!DEFAULT_CHULL_LATEX_COLOR_MAP=""!
  \begin{myitemize}
      \description Defines the color map, options are presented in Reference~\onlinecite{pgfplots_manual}.
      Default is \\ \verb!rgb(0pt)=(0.035,0.270,0.809); rgb(63pt)=(1,0.644,0)!.
    \type \verb|string|
  \end{myitemize}
\item \verb!DEFAULT_CHULL_LATEX_LINKS=1!
  \begin{myitemize}
      \description Defines the links scheme. True/false, \ie, \verb|0|/\verb|1|, will toggle all links on/off.
      \verb|2| enables external hyperlinks only (no links to other sections of the \PDF).
      \verb|3| enables internal links only (no links to external pages).
      \type \verb|0 (false), 1 (true), 2 (external-only), or 3 (internal-only)|
  \end{myitemize}
\item \verb!DEFAULT_CHULL_LATEX_LABEL_NAME=""!
  \begin{myitemize}
    \description Defines the labeling scheme for phases shown on the convex hull.
      By default, the \verb|compound| label is shown, while the \verb|prototype| label can also be specified.
      \verb|icsd| shows the \ICSD\ entry number designation
      (lowest for multiple equivalent ground-state structures reflecting \verb|icsd_canonical_auid|) if appropriate,
      as shown in Figure~\ref{fig:art146:hull_analyses}(c).
      Also acceptable: \verb|both| (\verb|compound| and \verb|prototype|) and \verb|none|.
    \type \verb|string|
  \end{myitemize}
\item \verb!DEFAULT_CHULL_LATEX_META_LABELS=0!
  \begin{myitemize}
      \description Enables verbose labels, including \verb|compound|, \verb|prototype|, $H_{\mathrm{f}}$, $T_{\mathrm{S}}$,
      and $\Delta H_{\mathrm{f}}$. Warning, significant overlap of labels should be expected.
    \type \verb|0 (false) or 1 (true)|
  \end{myitemize}
\item \verb!DEFAULT_CHULL_LATEX_LABELS_OFF_HULL=0!
  \begin{myitemize}
    \description Enables labels for off-hull points.
    \type \verb|0 (false) or 1 (true)|
  \end{myitemize}
\item \verb!DEFAULT_CHULL_LATEX_HELVETICA_FONT=1!
  \begin{myitemize}
      \description Switches the font scheme from Computer Modern (default) to Helvetica.
    \type \verb|0 (false) or 1 (true)|
  \end{myitemize}
\item \verb!DEFAULT_CHULL_LATEX_FONT_SIZE=""!
  \begin{myitemize}
    \description Defines the font size of the labels on the convex hull illustration. Warning,
      other settings may override this default. Options include: \verb|tiny|, \verb|scriptsize|,
      \verb|footnotesize|, \verb|small|, \verb|normalsize|, \verb|large| (default), \verb|Large|,
      \verb|LARGE|, \verb|huge|, and \verb|Huge|.
    \type \verb|string|
  \end{myitemize}
\item \verb!DEFAULT_CHULL_LATEX_ROTATE_LABELS=1!
  \begin{myitemize}
    \description Toggles whether labels are rotated.
    \type \verb|0 (false) or 1 (true)|
  \end{myitemize}
\item \verb!DEFAULT_CHULL_LATEX_BOLD_LABELS=-1!
  \begin{myitemize}
    \description Toggles whether labels are bolded.
      Three values are accepted: \verb|-1| (default: false unless phase is a ternary), \verb|0| (false), \verb|1| (true).
      \type \verb|-1 (default), 0 (false), or 1 (true)|
  \end{myitemize}
\end{myitemize}

\vspace{0.5cm}

\boldsection{Image generation.}
Instructions for generating the images herein are provided below.
Many of these images can be generated automatically with \AFLOWHULL. \\
\noindent Figure~\ref{fig:art146:hull_examples}(a): run \verb|aflow --chull --alloy=CoTi --image_only|. \\
\noindent Figure~\ref{fig:art146:hull_examples}(b): run \verb|aflow --chull --alloy=MnPdPt --image_only|. \\
\noindent Figure~\ref{fig:art146:hull_workflow}: \textbf{i.} the Pd-Pt hull was first generated by running \\ \verb|aflow --chull --alloy=PdPt --image_only --keep=tex|,
\textbf{ii.} the resulting \LaTeX\ input file ({\sf aflow\_PdPt\_hull.tex}) was modified by hand and compiled to get the various hull illustrations,
\textbf{iii.} the overall flowchart was constructed with Microsoft PowerPoint. \\
\noindent Figure~\ref{fig:art146:dimensions}: \textbf{i.} the Al-Ni, Al-Ti, and Ni-Ti binary hulls were first generated by running \\ \verb|aflow --chull --alloy=AlNi,AlTi,NiTi --image_only --keep=tex|,
\textbf{ii.} the resulting \LaTeX\ input files ({\sf aflow\_AlNi\_hull.tex}, {\sf aflow\_AlTi\_hull.tex}, and {\sf aflow\_NiTi\_hull.tex}) were modified by hand and compiled to get the binary hull images,
\textbf{iii.} a snapshot of the Al-Ni-Ti ternary hull was taken from the web application at {\sf aflow.org/aflow-chull},
\textbf{iv.} the overall illustration was constructed with Adobe Illustrator. \\
\noindent Figure~\ref{fig:art146:hull_analyses}(a): set \verb|DEFAULT_CHULL_IGNORE_KNOWN_ILL_CONVERGED=0| in the {\sf .aflow.rc} and run \verb|aflow --chull --alloy=AlCo --image_only|. \\
\noindent Figure~\ref{fig:art146:hull_analyses}(b): set \verb|DEFAULT_CHULL_IGNORE_KNOWN_ILL_CONVERGED=1| in the {\sf .aflow.rc} and run \verb|aflow --chull --alloy=AlCo --image_only|. \\
\noindent Figure~\ref{fig:art146:hull_analyses}(c): set \verb|DEFAULT_CHULL_LATEX_LABEL_NAME=``icsd''| in the {\sf .aflow.rc} and run \verb|aflow --chull --alloy=TeZr --image_only|. \\
\noindent Figure~\ref{fig:art146:hull_analyses}(d): \textbf{i.} the Pd-Pt hull was first generated by running \\ \verb|aflow --chull --alloy=PdPt --image_only --keep=tex|,
\textbf{ii.} the resulting \LaTeX\ input file ({\sf aflow\_PdPt\_hull.tex}) was modified by hand and compiled to get the hull illustration.
$\Delta H_{\mathrm{f}}[\text{aflow:71bc1b15525ffa35}]$ can be calculated individually by running \verb|aflow --chull --alloy=PdPt --distance_to_hull=aflow:71bc1b15525ffa35|. \\
\noindent Figure~\ref{fig:art146:hull_analyses}(e): \textbf{i.} the Pd-Pt hull was first generated by running \\ \verb|aflow --chull --alloy=PdPt --image_only --keep=tex|,
\textbf{ii.} the resulting \LaTeX\ input file ({\sf aflow\_PdPt\_hull.tex}) was modified by hand and compiled to get the hull illustration.
$\delta_{\mathrm{sc}}[\text{aflow:f31b0e27897cd162}]$ can be calculated individually by running \verb|aflow --chull --alloy=PdPt| \\ \verb|--stability_criterion=aflow:f31b0e27897cd162|. \\
\noindent Figure~\ref{fig:art146:hull_analyses}(f): run \verb|aflow --chull --alloy=BSm --image_only| \\ \verb|--plot_iso_max_latent_heat --include_paw_gga --include_skewed_hulls|. \\
\noindent Figure~\ref{fig:art146:report}: run \verb|aflow --chull --alloy=AgAuCd|. \\
\noindent Figure~\ref{fig:art146:hull_app}(a): navigate to {\sf aflow.org/aflow-chull} and select the Mo-Ti hull. \\
\noindent Figure~\ref{fig:art146:hull_app}(b): navigate to {\sf aflow.org/aflow-chull} and select the Fe-Rh-Zr hull. \\
\noindent Figure~\ref{fig:art146:hull_app}(c): navigate to {\sf aflow.org/aflow-chull}, select the Au-Cu-Zr hull, click on several points in the 3-dimensional illustration to populate the ``Select Points'' table on the left side of the screen, then click on one of the points in the table. \\
\noindent Figure~\ref{fig:art146:hull_app}(d): navigate to {\sf aflow.org/aflow-chull}, select the Au-Cu-Zr, Au-Cu, and AuZr hulls by clicking ``Periodic Table'' from the navigation bar on the top right corner of the screen between selections, and click ``Hull History'' from the navigation bar on the top right corner of the screen. \\
\noindent Figures~\ref{fig:art146:heuslers}(a-d): the structures were visualized with the CrystalMaker X software.

\vspace{0.5cm}

\boldsection{Python environment.}
A module has been created that employs \AFLOWHULL\
within a Python environment.
The module and its description closely follow that of the \AFLOWSYM\ Python module~\cite{curtarolo:art135}.
It connects to a local \AFLOW\ installation and imports the \AFLOWHULL\ results into a
\verb|CHull| class.
A \verb|CHull| object is initialized with:

\begin{python}
from aflow_hull import CHull
from pprint import pprint

chull = CHull(aflow_executable = './aflow')
alloy = 'AlCuZr'
output = chull.get_hull(alloy)
pprint(output)
\end{python}

\noindent By default, the \verb|CHull| object searches for an \AFLOW\ executable in
the {\sf \${\small PATH}}.
However, the location of an \AFLOW\ executable can be specified as
follows:

\noindent \verb|CHull(aflow_executable=$HOME/bin/aflow)|.

\noindent The \verb|CHull| object contains built-in methods corresponding to the command line calls mentioned previously:

\begin{myitemize}
\item \verb|get_hull(`InNiY', options = `--keep=log')|
\item \verb|get_distance_to_hull(`InNiY', `aflow:375066afdfb5a93f',| \\ \verb|options = `--keep=log')|
\item \verb|get_stability_criterion(`InNiY', `aflow:60a36639191c0af8',| \\ \verb|options = `--keep=log')|
\item \verb|get_hull_energy(`InNiY', [0.25,0.25], options = `--keep=log')|
\end{myitemize}
Each method requires an input alloy string and allows an additional parameters/flags string to be passed via \verb|options|.
\verb|get_distance_to_hull| and \\ \verb|get_stability_criterion| require an additional string input for the
\AUID, while \verb|get_hull_energy| takes an array of doubles as its input
for the composition.

\vspace{0.5cm}

\boldsection{Python module.}
The module to run the aforementioned \AFLOWHULL\ commands
is provided below.
This module can be modified to incorporate additional/customized options.
\begin{python}
import json
import subprocess
import os

class CHull:

    def __init__(self, aflow_executable='aflow'):
        self.aflow_executable = aflow_executable

    def aflow_command(self, cmd):
        try:
            return subprocess.check_output(
                self.aflow_executable + cmd,
                shell=True
            )
        except subprocess.CalledProcessError:
            print('Error aflow executable not found at: ' + self.aflow_executable)

    def get_hull(self, alloy, options = None):
        command = ' --chull'
        if options:
            command += ' ' + options

        output = ''
        output = self.aflow_command(
            command + ' --print=json --screen_only --alloy=' + alloy
        )
        res_json = json.loads(output)
        return res_json

    def get_distance_to_hull(self, alloy, off_hull_point, options = None):
        command = ' --chull --distance_to_hull=' + off_hull_point
        if options:
            command += ' ' + options

        output = ''
        output = self.aflow_command(
            command + ' --print=json --screen_only --alloy=' + alloy
        )
        res_json = json.loads(output)
        return res_json

    def get_stability_criterion(self, alloy, hull_point, options = None):
        command = ' --chull --stability_criterion=' + hull_point
        if options:
            command += ' ' + options

        output = ''
        output = self.aflow_command(
            command + ' --print=json --screen_only --alloy=' + alloy
        )
        res_json = json.loads(output)
        return res_json

    def get_hull_energy(self, alloy, composition, options = None):
        command = ' --chull --hull_energy=' + ','.join([ str(comp) for comp in composition ])
        if options:
            command += ' ' + options

        output = ''
        output = self.aflow_command(
            command + ' --print=json --screen_only --alloy=' + alloy
        )
        res_json = json.loads(output)
        return res_json
\end{python}

\vspace{0.5cm}

\noindent{\textbf{Stability analysis.}
A Python script is provided below demonstrating
how to perform the stability analysis presented in the Results
section.
The script gathers the most stable binary compounds generated
from 2-element combinations of \texttt{elements}.
Compounds are filtered for binary ground-state structures not in the \ICSD.
Only unique compositions are saved.
The script writes the results to the \JSON\ file {\sf most\_stable\_binaries.json}
and prints them to screen.
The script can be adapted to incorporate the full set of elements and
for the calculation of ternary systems.
Considering the number of combinations, it is recommended that the script be adapted
to generate the hulls in parallel.
}
\begin{python}
from aflow_hull import CHull
import json
from pprint import pprint

elements = ['Mn', 'Pd', 'Pt'] #extend as needed
elements.sort()

most_stable_binaries = [] #final list
saved_points_rc = []      #easy way to avoid adding duplicate compositions

chull = CHull(aflow_executable = './aflow') #initialize hull object
for i in range(len(elements)):              #generate binary alloy combinations
    for j in range(i + 1, len(elements)):   #generate binary alloy combinations
        alloy = elements[i]+elements[j]     #generate binary alloy combinations
        output = chull.get_hull(alloy)      #get hull data
        points_data = output['points_data'] #grab points data
        for point in points_data:
            #filter for binary ground-state structures not in the \ICSD\
            if point['ground_state'] and not point['icsd_ground_state'] and point['nspecies'] == 2:
                #easy way to avoid adding duplicate compositions
                if point['reduced_compound'] not in saved_points_rc:
                    saved_points_rc.append(point['reduced_compound'])
                    #save only what is necessary
                    abridged_entry = {}
                    abridged_entry['compound'] = point['compound']
                    abridged_entry['prototype'] = point['prototype']
                    abridged_entry['auid'] = point['auid']
                    abridged_entry['aurl'] = point['aurl']
                    abridged_entry['relative_stability_criterion'] = point['relative_stability_criterion']
                    most_stable_binaries.append(abridged_entry)

most_stable_binaries = sorted(most_stable_binaries, key=lambda point: -point['relative_stability_criterion'])    #sort in descending order

#save data to JSON file
with open('most_stable_binaries.json', 'w') as fout:
    json.dump(most_stable_binaries, fout)

#also print output to screen
pprint(most_stable_binaries)
\end{python}

\vspace{0.5cm}

\boldsection{Output list.}
This section details the output fields for the thermodynamic analysis.
The lists describe the keywords as they appear in the \JSON\ format.
Similar keywords are used for the standard text output.

\boldsection{Points data} (\verb|points_data|).
\begin{myitemize}
\item \verb|auid|
  \begin{myitemize}
      \description {\small \underline{A}FLOW} \underline{u}nique \underline{ID} (\AUID)~\cite{aflowAPI}.
    \type \verb|string|
  \end{myitemize}
\item \verb|aurl|
  \begin{myitemize}
      \description {\small \underline{A}FLOW} \underline{u}niform \underline{r}esource \underline{l}ocator (\AURL)~\cite{aflowAPI}.
    \type \verb|string|
  \end{myitemize}
\item \verb|compound|
  \begin{myitemize}
    \description Compound name~\cite{aflowAPI}.
    \type \verb|string|
  \end{myitemize}
\item \verb|enthalpy_formation_atom|
  \begin{myitemize}
      \description Formation enthalpy per atom $\left(H_{\mathrm{f}}\right)$~\cite{aflowAPI}.
    \type \verb|double|
    \units meV/atom
  \end{myitemize}
\item \verb|enthalpy_formation_atom_difference|
  \begin{myitemize}
      \description The energetic vertical-distance to the hull $\left(\Delta H_{\mathrm{f}}\right)$, \ie, the magnitude of the energy driving the decomposition reaction.
    \type \verb|double|
    \units meV/atom
  \end{myitemize}
\item \verb|entropic_temperature|
  \begin{myitemize}
      \description The ratio of the formation enthalpy and the ideal mixing entropy $\left(T_{\mathrm{S}}\right)$~\cite{monsterPGM}.
      This term defines the ideal ``{\it iso-max-latent-heat}'' lines of the grand-canonical ensemble~\cite{monsterPGM,curtarolo:art98}. Refer to Figure~\ref{fig:art146:hull_analyses}(f).
    \type \verb|double|
    \units Kelvin
  \end{myitemize}
\item \verb|equivalent_structures_auid|
  \begin{myitemize}
    \description \AUID\ of structurally equivalent entries. This analysis is limited to stable phases only.
    \type \verb|array of strings|
  \end{myitemize}
\item \verb|ground_state|
  \begin{myitemize}
    \description True for stable phases, and false otherwise.
    \type \verb|boolean|
  \end{myitemize}
\item \verb|icsd_canonical_auid|
  \begin{myitemize}
    \description \AUID\ of an equivalent \ICSD\ entry. If there are multiple equivalent \ICSD\ entries, the one with the lowest number designation is chosen (original usually). This analysis is limited to stable phases only.
    \type \verb|string|
  \end{myitemize}
\item \verb|icsd_ground_state|
  \begin{myitemize}
    \description True for stable phases with an equivalent \ICSD\ entry, and false otherwise.
    \type \verb|boolean|
  \end{myitemize}
\item \verb|nspecies|
  \begin{myitemize}
    \description The number of species in the system (\eg, binary = 2 and ternary = 3).
    \type \verb|integer|
  \end{myitemize}
\item \verb|phases_decomposition_auid|
  \begin{myitemize}
      \description \AUID\ of the products of the decomposition reaction (stable phases). This analysis is limited to unstable phases only.
    \type \verb|array of strings|
  \end{myitemize}
\item \verb|phases_decomposition_coefficient|
  \begin{myitemize}
      \description Coefficients of the decomposition reaction normalized to reactant, \ie, $\textbf{N}$ from Equation~\ref{eq:art146:decomp_reaction}. Hence, the first entry is always 1. This analysis is limited to unstable phases only.
    \type \verb|array of doubles|
  \end{myitemize}
\item \verb|phases_decomposition_compound|
  \begin{myitemize}
      \description \verb|compound| of the products of the decomposition reaction (stable phases). This analysis is limited to unstable phases only.
    \type \verb|array of strings|
  \end{myitemize}
\item \verb|phases_equilibrium_auid|
  \begin{myitemize}
    \description \AUID\ of phases in coexistence. This analysis is limited stable phases only.
    \type \verb|array of strings|
  \end{myitemize}
\item \verb|phases_equilibrium_compound|
  \begin{myitemize}
    \description \verb|compound| of phases in coexistence. This analysis is limited stable phases only.
    \type \verb|array of strings|
  \end{myitemize}
\item \verb|prototype|
  \begin{myitemize}
    \description \AFLOW\ prototype designation~\cite{aflowAPI}.
    \type \verb|string|
  \end{myitemize}
\item \verb|relative_stability_criterion|
  \begin{myitemize}
      \description A dimensionless quantity capturing the effect of the phase on the minimum energy surface relative to its depth, \ie, $\left|\delta_{\mathrm{sc}}/H_{\mathrm{f}}\right|$.
    \type \verb|double|
  \end{myitemize}
\item \verb|space_group_orig|
  \begin{myitemize}
      \description The space group (symbol and number) of the structure pre-relaxation as determined by \AFLOWSYM~\cite{curtarolo:art135}.
    \type \verb|string|
  \end{myitemize}
\item \verb|space_group_relax|
  \begin{myitemize}
      \description The space group (symbol and number) of the structure post-relaxation as determined by \AFLOWSYM~\cite{curtarolo:art135}.
    \type \verb|string|
  \end{myitemize}
\item \verb|spin_atom|
  \begin{myitemize}
    \description The magnetization per atom for spin polarized calculations~\cite{aflowAPI}.
    \type \verb|double|
      \units $\mu_{\mathrm{B}}$/atom.
  \end{myitemize}
\item \verb|stability_criterion|
  \begin{myitemize}
      \description A metric for robustness of a stable phase $\left(\delta_{\mathrm{sc}}\right)$, \ie,
      the distance of a stable phase from the pseudo-hull constructed without it.
      This analysis is limited to stable phases only.
    \type \verb|double|
    \units meV/atom
  \end{myitemize}
\item \verb|url_entry_page|
  \begin{myitemize}
      \description The \URL\ to the entry page: \\ {\sf http://aflow.org/material.php?id=aflow:60a36639191c0af8}.
    \type \verb|string|
  \end{myitemize}
\end{myitemize}

\vspace{0.5cm}

\boldsection{Facets data} (\verb|facets_data|).
\begin{myitemize}
\item \verb|artificial|
  \begin{myitemize}
      \description True if the facet is artificial, \ie, defined solely by artificial end-points, and false otherwise.
    \type \verb|boolean|
  \end{myitemize}
\item \verb|centroid|
  \begin{myitemize}
    \description The centroid of the facet.
    \type \verb|array of doubles|
      \units Stoichiometric-energetic coordinates as defined by Equation~\ref{eq:art146:point}.
  \end{myitemize}
\item \verb|content|
  \begin{myitemize}
      \description The content (hyper-volume) of the facet.
    \type \verb|array of doubles|
      \units Stoichiometric-energetic coordinates as defined by Equation~\ref{eq:art146:point}.
  \end{myitemize}
\item \verb|hypercollinearity|
  \begin{myitemize}
      \description True if the facet has no content, \ie, exhibits hyper-collinearity, and false otherwise.
    \type \verb|boolean|
  \end{myitemize}
\item \verb|normal|
  \begin{myitemize}
      \description The normal vector characterizing the facet, \ie, $\mathbf{n}$ in Equation~\ref{eq:art146:plane_eq}.
    \type \verb|double|
    \units Stoichiometric-energetic coordinates as defined by Equation~\ref{eq:art146:point}.
  \end{myitemize}
\item \verb|offset|
  \begin{myitemize}
      \description The offset characterizing the facet, \ie, $D$ in Equation~\ref{eq:art146:plane_eq}.
    \type \verb|double|
    \units Stoichiometric-energetic coordinates as defined by Equation~\ref{eq:art146:point}.
  \end{myitemize}
\item \verb|vertical|
  \begin{myitemize}
    \description True if the facet is vertical along the energetic axis, and false otherwise.
    \type \verb|boolean|
  \end{myitemize}
\item \verb|vertices_auid|
  \begin{myitemize}
    \description \AUID\ of the phases that define the vertices of the facet.
    \type \verb|array of strings|
  \end{myitemize}
\item \verb|vertices_compound|
  \begin{myitemize}
    \description \verb|compound| of the phases that define the vertices of the facet.
    \type \verb|array of strings|
  \end{myitemize}
\item \verb|vertices_position|
  \begin{myitemize}
    \description Coordinates that define the vertices of the facet.
    \type \verb|array of arrays of doubles|
    \units Stoichiometric-energetic coordinates as defined by Equation~\ref{eq:art146:point}.
  \end{myitemize}
\end{myitemize}

\vspace{0.5cm}

\boldsection{\AFLOW\ forum.}
Updates about \AFLOWHULL\ are discussed in the \AFLOW\ forum ({\sf aflow.org/forum}):
``Thermodynamic analysis''.

\subsection{Conclusions}
Thermodynamic analysis is a critical step for any effective materials design workflow.
Being a collective characterization, thermodynamics requires comparisons between many configurations of the system.
The availability of large databases \cite{aflowlibPAPER,aflowAPI,curtarolo:art104,aflux,nomad,APL_Mater_Jain2013,Saal_JOM_2013,cmr_repository}
allows the construction of computationally-based phase diagrams.
\AFLOWHULL\ presents a complete software infrastructure, including
flexible protocols for data retrieval, analysis, and verification~\cite{aflowPI,nomad}.
The module is exhaustively applied to the \AFLOWorg\ repository and identified
several new candidate phases: \CHULLCountPromisingTernaries\ promising $C15_{b}$-type structures and two half-Heuslers.
The extension of \AFLOWHULL\ to repositories beyond \AFLOWorg\ {can}
be performed {by adapting} the open-source \texttt{C++} code and/or Python module.
Computational platforms such as \AFLOWHULL\ are valuable tools for guiding synthesis, including high-throughput and
even autonomous approaches~\cite{Xiang06231995,Takeuchi:2003fe,koinuma_nmat_review2004,nmatHT}.

\clearpage
\section{Modeling Off-Stoichiometry Materials with a High Throughput \Abinitio\ Approach}
\label{sec:art110}

This study follows from a collaborative effort described in Reference~\cite{curtarolo:art110}.
Author contributions are as follows:
Stefano Curtarolo designed the study.
Kesong Yang and Corey Oses implemented the \AFLOWPOCC\ framework and performed proof of concept studies.
All authors discussed the results and their implications and contributed to the paper.

\subsection{Introduction}

Crystals are characterized by their regular, repeating structures.
Such a description allows us to reduce our focus from the macroscopic material to a microscopic subset of
unique atoms and positions.
A full depiction of material properties, including mechanical, electronic, and magnetic features,
follows from an analysis of the primitive lattice.
First principles quantum mechanical calculations have been largely successful in reproducing ground state properties of
perfectly ordered crystals~\cite{DFT,Hohenberg_PR_1964,nmatHT}.
However, such perfection does not exist in nature.
Instead, crystals display a degree of randomness, or disorder, in their lattices.
There are several types of disorder; including topological, spin, substitutional, and vibrational~\cite{Elliott_PoAM_1990}.
This work focuses on substitutional disorder, in which equivalent sites of a crystal are not uniquely or fully occupied.
Rather, each site is characterized by a statistical, or partial, occupation.
Such disorder is intrinsic in many technologically significant systems, including those used in fuel cells~\cite{Xie_ACatB_2015},
solar cells~\cite{Kurian_JPCC_2013}, high-temperature superconductors~\cite{Bednorz_ZPBCM_1986,Maeno_Nature_1994}, low thermal conductivity
thermoelectrics~\cite{Winter_JACerS_2007}, imaging and communications devices~\cite{Patra_JAP_2012}, as well as promising
rare-earth free materials for use in sensors, actuators, energy-harvesters, and spintronic devices~\cite{Wang_SR_2013}.
Hence, a comprehensive computational study of substitutionally disordered materials at the atomic scale is of paramount importance for
optimizing key physical properties of materials in technological applications.

Unfortunately, structural parameters with partial occupancy cannot be used directly in first principles
calculations --- a significant hindrance for computational studies of disordered systems.
Therefore, additional efforts must be made to model disorder or aperiodic systems~\cite{curtarolo:art25,Mihalkovic_PHM_2006,
Nordheim_1931_AP_VCA,Vanderbilt_2000_PRB_VCA,Soven_PhysRev_1967,Korringa1947392,
Kohn_1954_PhysRev,Stocks_PRL_1978,zunger_sqs,Shan_PRL_1999,Popescu_PRL_2010,Faulkner_PMS_1982}.
A rigorous statistical treatment of substitutional disorder at the atomic scale requires utility of large ordered supercells
containing a composition consistent with the compound's stoichiometry~\cite{sod,Habgood_PCCP_2011,Haverkort_ArXiv_2011}.
However, the computational cost of such large supercell calculations has traditionally inhibited their use.
Fortunately, the emergence of high-throughput (HT) computational techniques \cite{nmatHT}
coupled with the exponential growth of computational power is
now allowing the study of disordered systems from first principles~\cite{MGI}.

Herein, we present an approach to perform such a treatment working within the HT computational framework
\AFLOW~\cite{curtarolo:art104,aflowPAPER}.
We highlight three novel and attractive features central to this method:  complete implementation into an automatic high throughput framework (optimizing speed without
mitigating accuracy), utility of a novel occupancy optimization algorithm, and use of the Universal Force Field method \cite{Rappe_1992_JCAS_UFF}
to reduce the number of \DFT\ calculations needed per system.
To illustrate the effectiveness of the approach, \AFLOWPOCC\ is applied to three disordered systems,
a zinc chalcogenide (ZnS$_{1-x}$Se$_x$), a wide-gap oxide semiconductor (Mg$_{x}$Zn$_{1-x}$O), and an iron alloy (Fe$_{1-x}$Cu$_{x}$).
Experimental observations are successfully reproduced and new phenomena are predicted:
\begin{itemize}
\item ZnS$_{1-x}$Se$_x$ shows a small, yet smooth optical bowing over the complete compositional space.
Additionally, the stoichiometrically-evolving ensemble average DOS demonstrates that
this system is of the amalgamation type and not of the persistence type.
\item Mg$_{x}$Zn$_{1-x}$O exhibits an abrupt transition in optical bowing consistent with a phase transition
over its compositional range.
\item The ferromagnetic behavior of Fe$_{1-x}$Cu$_{x}$ is predicted to be smoothly stifled as more
copper is introduced into the structure, even through a phase transition.
\end{itemize}
Overall, these systems exhibit highly-tunable properties already exploited in many technologies.
Through the approach, these features are not only recovered, but additional insight into the underlying physical mechanisms is
also revealed.

\subsection{Methodology}

This section details the technicalities of representing a partially occupied disordered system as a series of unique supercells.
Here is an outline of the approach:
\begin{enumerate}
\item For a given disordered material, optimize its partial occupancy values and determine the size of the derivative superlattice $n$.
\item
\begin{enumerate}
\item Use the superlattice size $n$ to generate a set of unique derivative superlattices and corresponding sets of
unique supercells with the required stoichiometry.
\item Import these non-equivalent supercells into the automatic computational framework \AFLOW\ for HT
first principles electronic structure calculations.
\end{enumerate}
\item Obtain and use the relative formation enthalpy to calculate the equilibrium probability of each
supercell as a function of temperature $T$ according to the Boltzmann distribution.
\item Determine the disordered system's material properties through ensemble averages of those calculated for each supercell.
Specifically, the following properties are resolved: the density of states (DOS), band gap energy $E_{\mathrm{gap}}$, and magnetic moment $M$.
\end{enumerate}

In the following sections, a model disordered system, Ag$_{8.733}$Cd$_{3.8}$Zr$_{3.267}$, is presented
to illustrate the technical procedures mentioned above.
This disordered system has two partially occupied sites:  one shared between silver and zirconium, and another shared between
cadmium and a vacancy.
Working within the \AFLOW\ framework~\cite{aflowBZ}, a simple structure file has been designed for partially occupied systems.
Adapted from \VASP{}'s \POSCAR~\cite{vasp_cms1996,vasp_prb1996}, the \PARTCAR\ contains within it a description of lattice parameters and
site coordinates/occupants, along
with a concentration tolerance (explained in the next section), and (partial) occupancy values for each site.
To see more details about this structure or its \PARTCAR, see Section~\ref{subsec:art110:PARCAR}.

\subsubsection{Determining superlattice size}
In order to fully account for the partial occupancy of the disordered system, the set of superlattices of
a size corresponding to the lowest common denominator of the fractional partial occupancy values should be generated.
With partial occupancy values of 0.733 (733/1000) and 0.267 (267/1000) in the disordered system Ag$_{8.733}$Cd$_{3.8}$Zr$_{3.267}$,
superlattices of size 1000 would need to be constructed.
Not only would this require working with correspondingly large supercells (16,000 atoms per supercell in this example),
but the number of unique supercells in the set would be substantial.
This extends well beyond the capabilities of first principles calculations, and thus, is not practical.
It is therefore necessary to optimize the partial occupancy values to produce an appropriate superlattice size.

\tab
\mycaption[Evolution of the algorithm used to optimize the partial occupancy values and superlattice size for the disordered system
Ag$_{8.733}$Cd$_{3.8}$Zr$_{3.267}$.]
{$f_i$ indicates the iteration's choice fraction for each partially occupied site, ($i$ = 1, 2, 3, \ldots);
$e_i$ indicates the error between the iteration's choice fraction and the actual partial occupancy value.
$e_{\mathrm{max}}$ is the maximum error of the system.}
\tabvspace
\begin{tabular}{llrlrlrrr}
\multirow{2}{*}{$n^{\prime}$} & \multicolumn{2}{c}{occup. 1 (Ag)} & \multicolumn{2}{c}{occup. 2 (Zr)} & \multicolumn{2}{c}{occup. 3 (Cd)} & \multirow{2}{*}{\textit{$e_{\mathrm{max}}$} } & \multirow{2}{*}{$n$} \\
\cline{2-7}
 & \textit{$f_{1}$}  & \textit{$e_{1}$}  & \textit{$f_{2}$}  & \textit{$e_{2}$}  & \textit{$f_{3}$}  & \textit{$e_{3}$}  &  &  \\
\hline
1	& 1/1   & 0.267  & 0/1  & 0.267  & 1/1   & 0.2    & 0.267  & 1 \\
\hline
2	& 1/2   & 0.233  & 1/2  & 0.233  & 2/2   & 0.2    & 0.233  & 2 \\
\hline
3	& 2/3   & 0.067  & 1/3  & 0.067  & 2/3   & 0.133  & 0.133  & 3 \\
\hline
4	& 3/4   & 0.017  & 1/4  & 0.017  & 3/4   & 0.05   & 0.05   & 4 \\
\hline
5	& 4/5   & 0.067  & 1/5  & 0.067  & 4/5   & 0      & 0.067  & 5 \\
\hline
6	& 4/6   & 0.067  & 2/6  & 0.067  & 5/6   & 0.033  & 0.067  & 6 \\
\hline
7	& 5/7   & 0.019  & 2/7  & 0.019  & 6/7   & 0.057  & 0.057  & 7 \\
\hline
8	& 6/8   & 0.017  & 2/8  & 0.017  & 6/8   & 0.05   & 0.05   & 4 \\
\hline
9	& 7/9   & 0.044  & 2/9  & 0.044  & 7/9   & 0.022  & 0.044  & 9 \\
\hline
10	& 7/10  & 0.033  & 3/10 & 0.033  & 8/10  & 0      & 0.033  & 10 \\
\hline
11	& 8/11  & 0.006  & 3/11 & 0.006  & 9/11  & 0.018  & 0.018  & 11 \\
\hline
12	& 9/12  & 0.017  & 3/12 & 0.017  & 10/12 & 0.033  & 0.033  & 12 \\
\hline
13	& 10/13 & 0.036  & 3/13 & 0.036  & 10/13 & 0.031  & 0.036  & 13 \\
\hline
14	& 10/14 & 0.019  & 4/14 & 0.019  & 11/14 & 0.014  & 0.019  & 14 \\
\hline
15	& 11/15 & 0.00003 & 4/15 & 0.00003 & 12/15 & 0      & 0.00003 & 15 \\
\end{tabular}
\label{tab:art110:pocc_algo}
\etab

An efficient algorithm is presented to calculate the optimized partial occupancy values and corresponding superlattice size
with the example disordered system Ag$_{8.733}$Cd$_{3.8}$Zr$_{3.267}$ in Table~\ref{tab:art110:pocc_algo}.
For convenience, the algorithm's iteration step is referred as $n^{\prime}$,
the superlattice index, and $n$ as the superlattice size.
Quite simply, the algorithm iterates, increasing the superlattice index from 1 to $n^{\prime}$ until the optimized partial occupancy values reach the required accuracy.
At each iteration, a fraction is generated for each partially occupied site, all of which have the common denominator $n^{\prime}$.
The numerator is determined to be the integer that reduces the overall fraction's error relative to the actual site's fractional partial occupancy value.
The superlattice size corresponds to the lowest common denominator of the irreducible fractions (\eg, see iteration step 8).
The maximum error among all of the sites is chosen to be the accuracy metric for the system.

For the disordered system Ag$_{8.733}$Cd$_{3.8}$Zr$_{3.267}$, given a tolerance of 0.01, the calculated superlattice size is 15
(240 atoms per supercell).
By choosing a superlattice with a nearly equivalent stoichiometry as the disordered system, the supercell size has been
reduced by over a factor of 60 and entered the realm of feasibility with this calculation.
Notice that the errors in partial occupancy values calculated for
silver and zirconium are the same, as they share the same site.
The same holds true for cadmium and its vacant counterpart (not shown).
Therefore, the algorithm only needs to determine one choice fraction per site, instead of per occupant (as shown).
Such an approach reduces computational costs by guaranteeing that only the smallest supercells (both in number and size)
with the lowest tolerable error in composition are funneled into the HT first principles calculation framework.

\fig
\includegraphics[width=\linewidth]{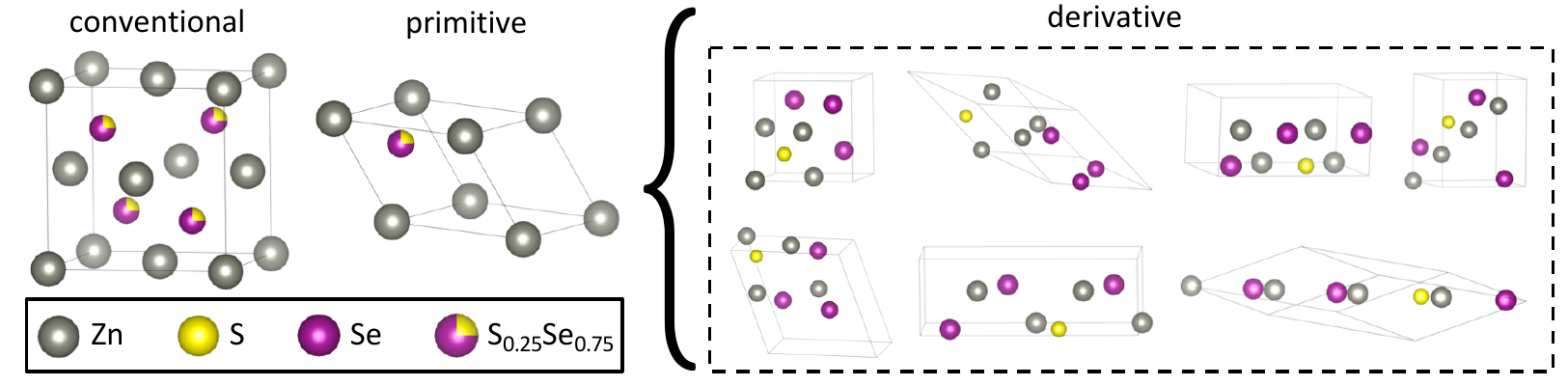}
\mycaption[Structure enumeration for off-stoichiometric materials modeling.]
{For the off-stoichiometric material ZnS$_{0.25}$Se$_{0.75}$, a superlattice of size $n=4$ accommodates the stoichiometry exactly.
By considering all possibilities of decorated supercells and eliminating duplicates by UFF energies, seven structures are identified as unique.
These representative structures are fully characterized by \AFLOW\ and \VASP, and are ensemble-averaged to resolve the system-wide properties.}
\label{fig:art110:pocc}
\efig

\subsubsection{Unique supercells generation}
With the optimal superlattice size $n$, the unique derivative superlattices of the disordered system can be generated using
Hermite Normal Form (HNF) matrices~\cite{enum1} as depicted in Figure~\ref{fig:art110:pocc}.
Each HNF matrix generates a superlattice of a size corresponding to its determinant, $n$.
There exists many HNF matrices with the same determinant, each creating a variant superlattice.
For each unique superlattice, a complete set of possible supercells is generated with the required stoichiometry by exploring all
possible occupations of partially occupied sites.
However, not all of these combinations are unique --- nominally warranting an involved structure comparison analysis that becomes
extremely time consuming for large supercells~\cite{enum1}.
Instead, duplicates are identified by estimating the total energy of each supercell in a HT manner based on the Universal Force Field (UFF)
method~\cite{Rappe_1992_JCAS_UFF}.
This classical molecular mechanics force field approximates the energy of a structure by considering its composition,
connectivity, and geometry, for which parameters have been tabulated.
Only supercells with the same total energy are structurally compared and potentially treated as duplicate structures to be discarded, if necessary.
The count of duplicate structures determines the degeneracy of the structure.
Only non-equivalent supercells are imported into the automatic computational framework \AFLOW\ for HT
quantum mechanics.

\subsubsection{Supercell equilibrium probability calculation}
The unique supercells representing a partially occupied disordered material are labeled as {$S_1$, $S_2$, $S_3$, \ldots,
$S_n$}.
Their formation enthalpies (per atom) are labeled as {$H_{\mathrm{f},1}$, $H_{\mathrm{f},2}$, $H_{\mathrm{f},3}$, \ldots, $H_{\mathrm{f},n}$}, respectively.
The formation enthalpy of each supercell is automatically calculated from HT first principles calculations using the \AFLOW\
framework~\cite{curtarolo:art104,aflowPAPER}.
The supercell with the lowest formation enthalpy is selected as a reference (ground state structure), and its formation enthalpy is denoted as $H_{\mathrm{f},0}$.
The relative formation enthalpy of the \emph{i}th supercell is calculated as $\Delta {H_{\mathrm{f},i}} = {H_{\mathrm{f},i}} - {H_{\mathrm{f},0}}$
and characterizes its disorder relative to the ground state.
The probability $P_i$ of the \emph{i}th supercell is determined by the Boltzmann factor:
\begin{equation}
{P_i} = \frac{{{g_ie^{ - \Delta {H_{\mathrm{f},i}}/{k_{\mathrm{B}}}T}}}}{{\sum\limits_{i = 1}^n {{g_ie^{ - \Delta {H_{\mathrm{f},i}}/{k_{\mathrm{B}}}T}}} }},
\end{equation}
where $g_i$ is the degeneracy of the \emph{i}th supercell,
$\Delta {H_{\mathrm{f},i}}$ is the relative formation enthalpy of the \emph{i}th supercell,
$k_{\mathrm{B}}$ is the Boltzmann constant,
and $T$ is a virtual ``roughness'' temperature.
$T$ is not a true temperature \textit{per se}, but instead a parameter describing how much disorder has been
statistically explored during synthesis.
To elaborate further, consider two extremes in the ensemble average (ignoring structural degeneracy):
\begin{enumerate}
\item $k_{\mathrm{B}} T \lesssim \max\left(\Delta {H_{\mathrm{f},i}}\right)$
neglecting highly disordered structures $(\Delta {H_{\mathrm{f},i}} \ggg 0)$
as $T\to 0$, and
\item $k_{\mathrm{B}} T\ggg\max\left(\Delta {H_{\mathrm{f},i}}\right)$
representing the annealed limit ($T\to \infty$) in which all structures are equiprobable.
\end{enumerate}
The probability $P_i$ describes the weight of the \emph{i}th supercell among the thermodynamically equivalent states of the disordered material
at equilibrium.

\subsubsection{Ensemble average density of states, band gap energy, and magnetic moment}
With the calculated material properties of each supercell and its equilibrium probability in hand, the overall system properties
can be determined by ensemble averages of those calculated for each supercell.
This work focuses on the calculation of the ensemble average density of states (DOS), band gap energy $E_{\mathrm{gap}}$, and magnetic moment $M$.
The DOS of the \emph{i}th supercell is labeled as $N_i(E)$ and indicates the number of electronic states per energy interval.
The ensemble average DOS of the system is then determined by the following formula:
\begin{equation}
N(E) = \sum\limits_{i = 1}^n {{P_i} \times {N_i}(E)}.
\end{equation}
Additionally, a band gap $E_{\mathrm{gap},i}$ can be extracted from the DOS of each supercell.
In this fashion, an ensemble average band gap $E_{\mathrm{gap}}$ can be calculated for the system.
It is important to note that standard density functional theory (\DFT) calculations are limited to a description of the ground
state~\cite{DFT,Hohenberg_PR_1964,nmatHT}.
As such, calculated excited state properties may contain substantial errors.
In particular, \DFT\ tends to underestimate the band gap~\cite{Perdew_IJQC_1985}.
Despite these known hindrances in the theory, the framework is capable of predicting significant trends
specific to the disordered systems.
As a bonus, the calculation of these results are performed in a high-throughput fashion.
It is expected that a more accurate, fine-grained description of the electronic structure in such systems will be obtained through a combination of
this software framework and more advanced first principles approaches~\cite{GW,Hedin_GW_1965,Heyd2003,Liechtenstein1995,curtarolo:art86,curtarolo:art93,curtarolo:art103}.

In the same spirit as the $N(E)$ and $E_{\mathrm{gap}}$, \AFLOWPOCC\ calculates the ensemble average magnetic moment $M$ of the system.
The magnetic moment of the \emph{i}th supercell is labeled as $M_i$.
If the ground state of the \emph{i}th structure is non-spin-polarized, then its magnetic moment is set to zero, \ie, $M_i=0$.
Taking into account the impact of signed spins on the ensemble average, this approach is limited only to ferromagnetic solutions.
Additionally, as an initialization for the self-consistent run, the same ferromagnetic alignment is assumed among all of the spins in the system
(an \AFLOW\ calculation standard)~\cite{curtarolo:art104}.
Finally, the ensemble average magnetic moment of the system is calculated with the following formula:
\begin{equation}
M = \sum\limits_{i = 1}^n {{P_i} \times } |{M_i}|.
\end{equation}

\subsection{Example applications}
Three disordered systems of technological importance are analyzed using \AFLOWPOCC:
a zinc chalcogenide, a wide-gap oxide semiconductor, and an iron alloy.
Unless otherwise stated, the supercells used in these calculations were generated with the lowest superlattice size $n_{\mathrm{xct}}$ needed to represent
the composition exactly.

\subsubsection{Zinc chalcogenides}

\fig
\includegraphics[width=\linewidth]{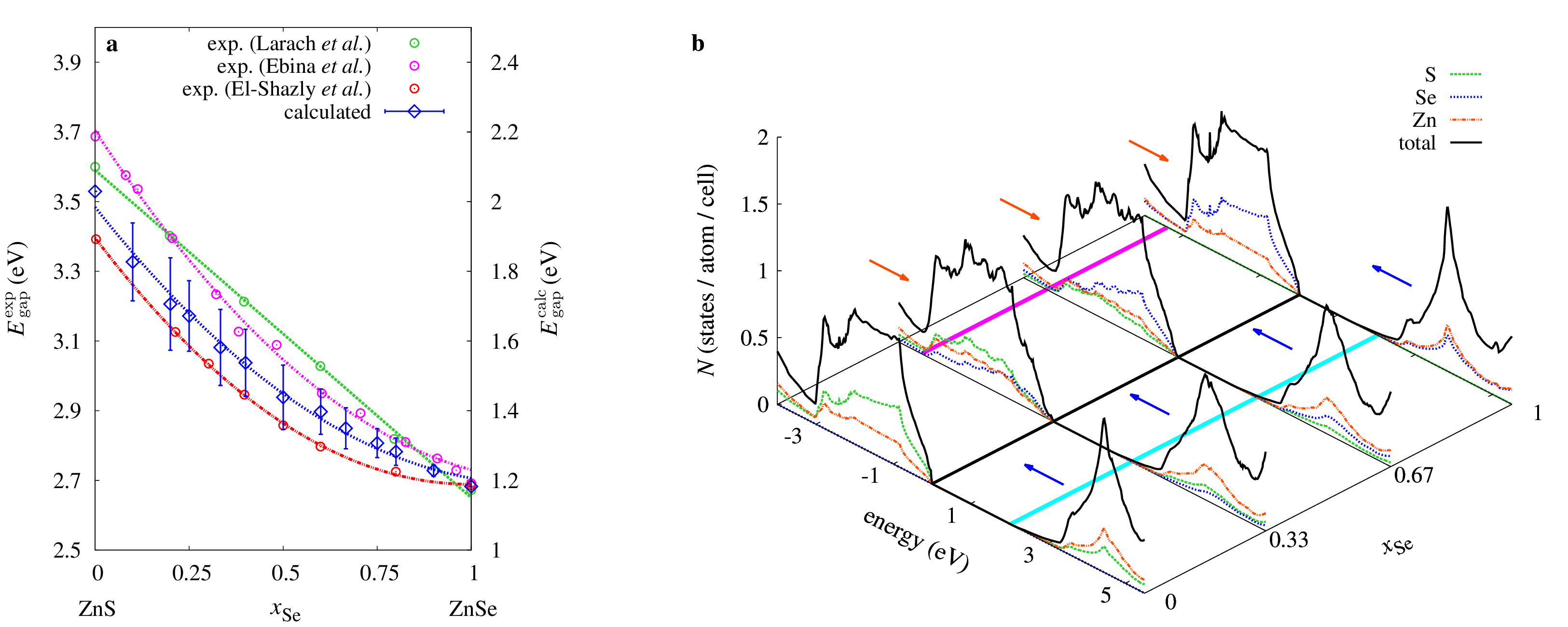}
\mycaption[Disordered ZnS$_{1-x}$Se$_x$.]
{(\textbf{a}) A comparison of the experimental \cite{Larach_PR_1957,Ebina_PRB_1974,El-Shazly_APA_1985} \vs\
calculated compositional dependence of the band gap energy $E_{\mathrm{gap}}$ at room temperature.
A rigid shift in the $E_{\mathrm{gap}}$ axis relative to the experimental results of ZnSe (second ordinate axis) accounts for the expected systematic
deviation in \DFT\ calculations~\cite{Perdew_IJQC_1985}.
Only the lowest empirical $E_{\mathrm{gap}}$ trends are shown.
Error bars indicate the weighted standard deviation of the ensemble average $E_{\mathrm{gap}}$.
(\textbf{b}) Calculated density of states plots for various compositions:
$x_{\mathrm{Se}}=0.00$ ($n=1$),
$0.33$ ($n=3$),
$0.67$ ($n=3$), and
$1.00$ ($n=1$).
The straight black line indicates the position of the valence band maximum,
while the straight magenta and cyan lines indicate the positions of the valence band minimum at $x_{\mathrm{Se}}=0.33$
and the conduction band minimum at $x_{\mathrm{Se}}=0.00$, respectively. }
\label{fig:art110:ZnSSe}
\efig

Over the years, zinc chalcogenides have garnered interest for a dynamic range of applications --- beginning with the creation
of the first blue-light emitting laser diodes~\cite{Haase_APL_1991}, and recently have been studied
as inorganic graphene analogues (IGAs) with potential applications in flexible and transparent nanodevices~\cite{Sun_NComm_2012}.
These wide-gap II-VI semiconductors have demonstrated a smoothly tunable band gap energy $E_{\mathrm{gap}}$ with respect to
composition~\cite{Larach_PR_1957,Ebina_PRB_1974,El-Shazly_APA_1985}.
Both linear and quadratic dependencies have been observed, with the latter phenomenon referred to as
\textit{optical bowing}~\cite{Bernard_PRB_1986}.
Specifically, given the pseudo-ternary system $A_{x}B_{1-x}C$,
\begin{equation}
E_{\mathrm{gap}}(x)=\left[x \epsilon_{AC}+(1-x)\epsilon_{BC}\right] - b x(1-x),
\end{equation}
with $b$ characterizing the bowing.
While Larach \etal\ reported a linear dependence ($b=0$)~\cite{Larach_PR_1957},
Ebina \etal\ \cite{Ebina_PRB_1974} and
El-Shazly \etal\ \cite{El-Shazly_APA_1985} reported similar bowing parameters of
$b=0.613\pm0.027$~eV and $b=0.457\pm0.044$~eV, respectively, averaged over the two observed direct transitions.

As a proof of concept, \AFLOWPOCC\ is employed to calculate the compositional dependence of the $E_{\mathrm{gap}}$ and
DOS for ZnS$_{1-x}$Se$_x$ at room temperature (annealed limit).
Overall, this system shows relatively low disorder ($\max\left(\Delta {H_{\mathrm{f},i}}\right)\sim 0.005$~eV),
exhibiting negligible variations in the ensemble average properties at higher temperatures.
These results are compared to experimental measurements~\cite{Larach_PR_1957,Ebina_PRB_1974,El-Shazly_APA_1985} in Figure~\ref{fig:art110:ZnSSe}.
Common among all three trends (Figure~\ref{fig:art110:ZnSSe}(a)) is the $E_{\mathrm{gap}}$ shrinkage with increasing $x_{\mathrm{Se}}$,
as well as a near 1~eV tunable $E_{\mathrm{gap}}$ range.
The calculated trend demonstrates a non-zero bowing similar to that observed by both Ebina \etal~\cite{Ebina_PRB_1974} and
El-Shazly \etal~\cite{El-Shazly_APA_1985}.
A fit shows a bowing parameter of $b=0.585\pm0.078$~eV, lying in the range between the two experimental bowing parameters.

The ensemble average DOS plots at room temperature are illustrated in Figure~\ref{fig:art110:ZnSSe}(b) for $x_{\mathrm{Se}}=0.00$ ($n=1$), $0.33$ ($n=3$), $0.67$ ($n=3$),
and $1.00$ ($n=1$).
The plots echo the negatively correlated band gap relationship illustrated in Figure~\ref{fig:art110:ZnSSe}(a), highlighting
that the replacement of sulfur with selenium atoms reduces the band gap.
Specifically, two phenomena are observed as the concentration of selenium increases:  (\textcolor{red}{\bf red arrows})
the reduction of the valence band width
(with the exception of $x_{\mathrm{Se}} = 0.00$ (ZnS) concentration), and (\textcolor{blue}{\bf blue arrows})
a shift of the conduction band peak back towards the Fermi energy.
The valence band of ZnS more closely resembles that of its extreme concentration counterpart at $x_{\mathrm{Se}} = 1.00$
(ZnSe) than the others.
The extreme concentration conduction peaks appear more defined than their intermediate concentration counterparts, which is likely an artifact
of the ensemble averaging calculation.

Finally, a partial-DOS analysis is performed in both species and orbitals (not shown).
In the valence band, sulfur and selenium account for the majority of the states, in agreement with their relative concentrations.
Meanwhile, zinc accounts for the majority of the states in the conduction band at all concentrations.
Correspondingly, at all concentrations, the $p$-orbitals make up the majority of the valence band,
whereas the conduction band consists primarily of $s$- and $p$-orbitals.
These observations are consistent with conclusions drawn from previous optical reflectivity measurements that optical transitions
are possible from sulfur or selenium valence bands to zinc conduction bands~\cite{Kirschfeld_PRL_1972}.

Overall, the concentration-evolving $E_{\mathrm{gap}}$ trend and DOS plots support a continuing line of
work~\cite{Larach_PR_1957,Ebina_PRB_1974,El-Shazly_APA_1985} corroborating that this system
is of the amalgamation type~\cite{Onodera_JPSJ_1968}
and not of the persistence type~\cite{Kirschfeld_PRL_1972}.
Notably, however, reflectivity spectra shows that the peak position in the $E_{\mathrm{gap}}$ for ZnS rich alloys may remain
stationary~\cite{Ebina_PRB_1974},
which may have manifested itself in the aforementioned anomaly observed in this structure's valence band width.

\subsubsection{Wide-gap oxide semiconductor alloys}

\fig
\includegraphics[width=\linewidth]{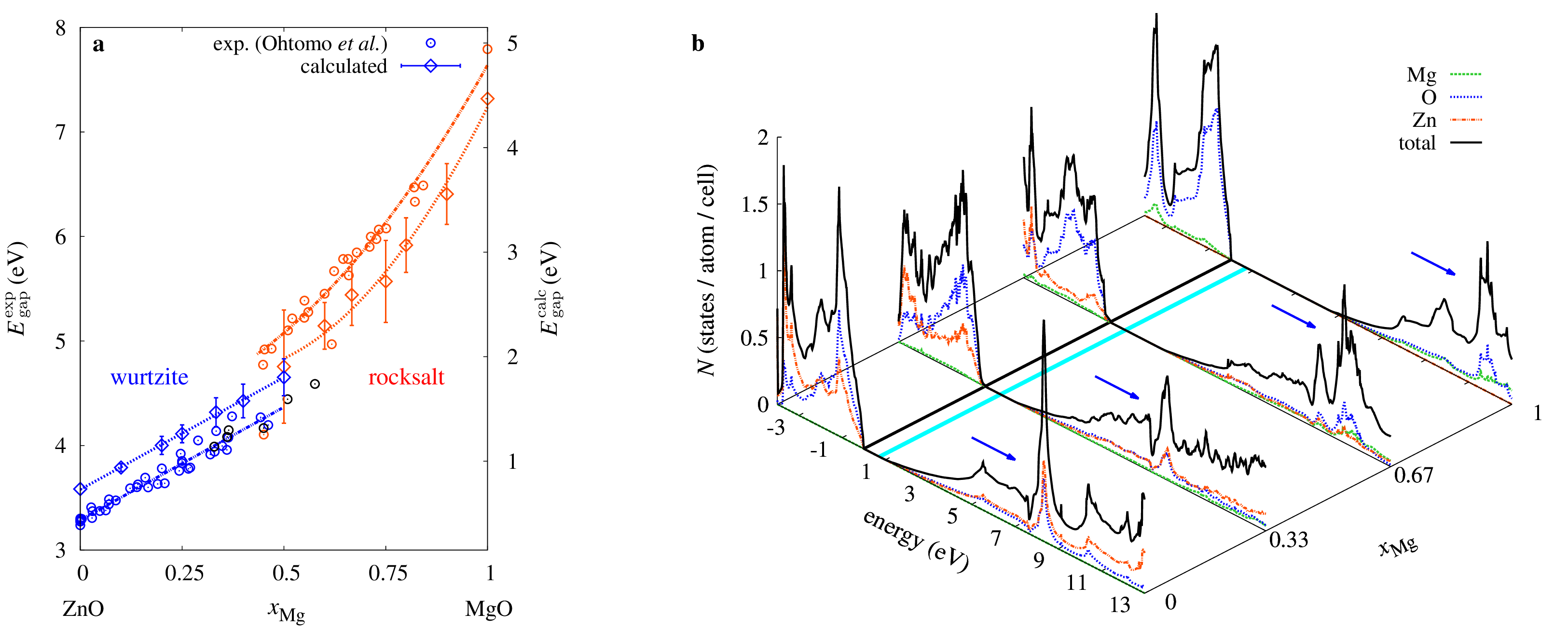}
\mycaption[Disordered Mg$_{x}$Zn$_{1-x}$O.]
{(\textbf{a}) A comparison of the experimental~\cite{Ohtomo_SST_2005,Takeuchi_JAP_2003,
Chen_JAPCM_2003,Takagi_JJAP_2003,Choopun_APL_2002,Minemoto_TSF_2000,Sharma_APL_1999,Ohtomo_APL_1998} \vs\ calculated compositional
dependence of the band gap energy $E_{\mathrm{gap}}$ at room temperature.
A rigid shift in the $E_{\mathrm{gap}}$ axis relative to the experimental results of MgO (second ordinate axis) accounts for the expected
systematic deviation in \DFT\ calculations~\cite{Perdew_IJQC_1985}.
The \textcolor{blue}{\bf wurtzite} and \textcolor{red}{\bf rocksalt} structures are highlighted in blue and red, respectively,
while the mixed phase structures are shown in black.
Error bars indicate the weighted standard deviation of the ensemble average $E_{\mathrm{gap}}$.
(\textbf{b}) Calculated density of states plots for various compositions:
$x_{\mathrm{Mg}}=0.00$ ($n=1$),
$0.33$ ($n=3$),
$0.67$ ($n=3$), and
$1.00$ ($n=1$).
The straight black line indicates the position of the valence band maximum,
while the straight cyan line indicates the position of the conduction band minimum at $x_{\mathrm{Mg}}=0.00$.}
\label{fig:art110:MgZnO}
\efig

Zinc oxide (ZnO) has proven to be a pervasive material, with far reaching applications such as paints, catalysts,
pharmaceuticals (sun creams), and optoelectronics~\cite{Takeuchi_MgZnO_Patent}.
It has long been investigated for its electronic properties, and falls into the class of transparent conducting
oxides~\cite{Ellmer_ZnO_2007}.
Just as with the previous zinc chalcogenide example, ZnO is a wide-gap II-VI semiconductor that has demonstrated
a tunable band gap energy $E_{\mathrm{gap}}$ with composition.
In particular, ZnO has been engineered to have an $E_{\mathrm{gap}}$ range as large as 5~eV by synthesizing it with magnesium.
This pairing has been intensively studied because of the likeness in ionic radius between zinc and magnesium
which results in mitigated misfit strain in the heterostructure~\cite{Yoo_TSF_2015}.
While the solubility of MgO and ZnO is small, synthesis has been made possible throughout the full compositional
spectrum~\cite{Ohtomo_SST_2005,Takeuchi_JAP_2003,Chen_JAPCM_2003,Takagi_JJAP_2003,Choopun_APL_2002,
Minemoto_TSF_2000,Sharma_APL_1999,Ohtomo_APL_1998}.

As another proof of concept, the compositional dependence of the $E_{\mathrm{gap}}$ and DOS for Mg$_{x}$Zn$_{1-x}$O
are modeled at room temperature (annealed limit).
In particular, this disordered system is chosen to illustrate the breath of materials which this framework can model.
Similar to ZnS$_{1-x}$Se$_x$, this system shows relatively low disorder ($\max\left(\Delta {H_{\mathrm{f},i}}\right)\sim 0.007$~eV),
exhibiting negligible variations in the ensemble average properties at higher temperatures.
The results are compared to that observed empirically~\cite{Ohtomo_SST_2005,Takeuchi_JAP_2003,Chen_JAPCM_2003,Takagi_JJAP_2003,Choopun_APL_2002,
Minemoto_TSF_2000,Sharma_APL_1999,Ohtomo_APL_1998} in Figure~\ref{fig:art110:MgZnO}.
As illustrated in Figure~\ref{fig:art110:MgZnO}(a), Ohtomo \etal\ observed a composition dependent phase transition
from a wurtzite to a rocksalt structure with increasing $x_{\mathrm{Mg}}$; the transition occurring around the mid concentrations.
This transition is enforced in the calculations.
Empirically, the overall trend in the wurtzite phase shows a negligible bowing in the $E_{\mathrm{gap}}$ trend,
contrasting the significant bowing observed in the rocksalt phase.
The wurtzite phase $E_{\mathrm{gap}}$ trend shows a slope of $2.160\pm0.080$~eV, while the rocksalt phase shows a bowing
parameter of $3.591\pm0.856$~eV.
Calculated trends are shown in Figure~\ref{fig:art110:MgZnO}(a).
Qualitatively, linear and non-linear $E_{\mathrm{gap}}$ trends are also observed in the wurtzite and rocksalt phases, respectively.
The fits are as follows:  a slope of $2.147\pm0.030$~eV in the wurtzite phase and a bowing parameter of
$5.971\pm1.835$~eV in the rocksalt phase.
These trends match experiment well within the margins of error.
A larger margin of error is detected in the rocksalt phase, particular in the phase separated region
($0.4\lesssim x_{\mathrm{Mg}} \lesssim 0.6$).
This may be indicative of the significant shear strain and complex nucleation behavior characterizing the region~\cite{Takeuchi_JAP_2003}.

The ensemble average DOS plots at room temperature are illustrated in Figure~\ref{fig:art110:MgZnO}(b) for $x_{\mathrm{Mg}}=0.00$ ($n=1$), $0.33$ ($n=3$), $0.67$ ($n=3$),
and $1.00$ ($n=1$)
The plots not only echo the positively correlated band gap relationship illustrated in Figure~\ref{fig:art110:MgZnO}(a),
but also exhibit the aforementioned change from a linear to non-linear trend.
This is most easily seen by observing the shift in the conduction band away from the Fermi energy,
highlighted by the \textcolor{blue}{\bf blue arrows}.
Contrasting ZnS$_{1-x}$Se$_x$, a significant change in width of the valence band is not observed over the range of the stoichiometry.

Finally, a partial-DOS analysis is performed in both species and orbitals (not shown).
Overall, the constant oxygen backbone plays a major role in defining the shape of both the valence and conduction bands,
particularly as $x_{\mathrm{Mg}}$ increases.
This resonates with the strong $p$-orbital presence in both bands throughout all concentrations.
Zinc and its $d$-orbitals play a particularly dominant role in the valence band in magnesium-poor structures.

\subsubsection{Iron alloys}

\fig
\includegraphics[width=\linewidth]{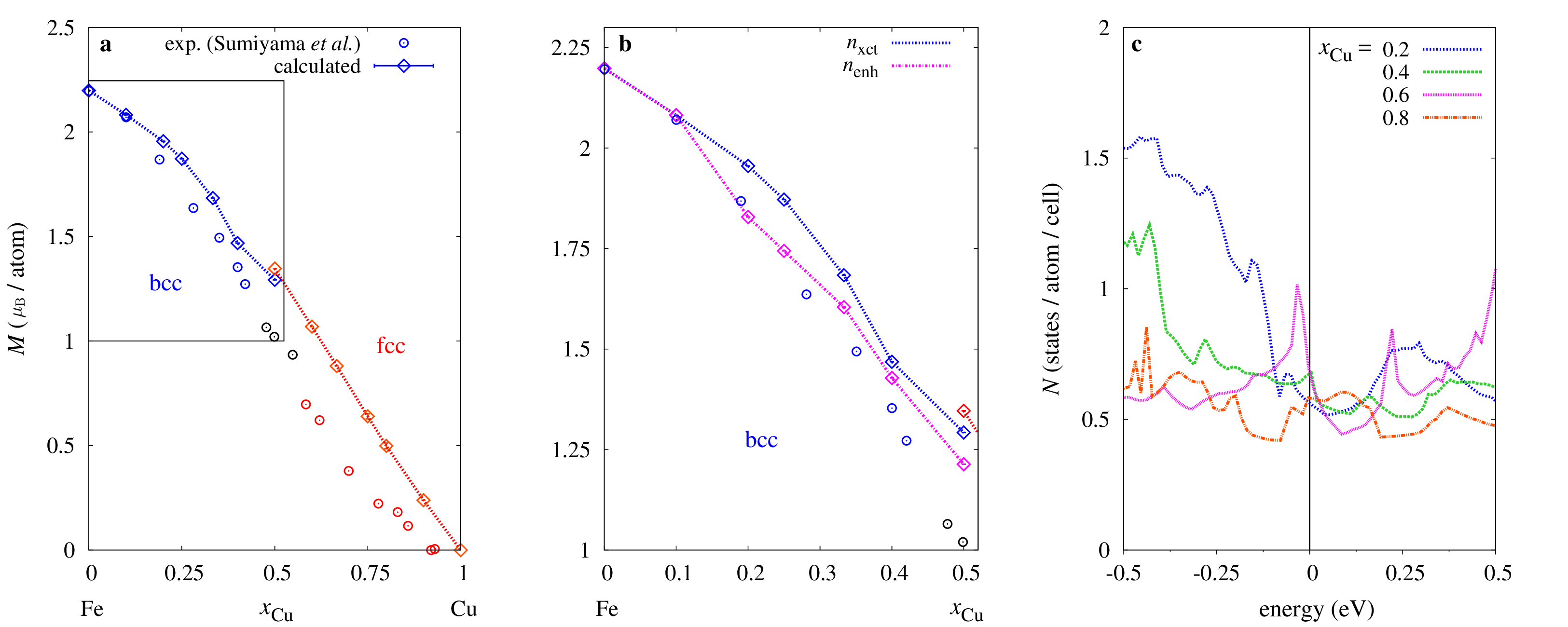}
\mycaption[Disordered Fe$_{1-x}$Cu$_{x}$.]
{(\textbf{a}) A comparison of the experimental~\cite{Sumiyama_JPSJ_1984} \vs\ calculated compositional
dependence of the magnetic moment $M$.
The calculations mimic the following phases observed at 4.2~K:
$x_{\mathrm{Cu}}\leq0.42$ \textcolor{blue}{\bf bcc} phase shown in blue,
$0.42<x_{\mathrm{Cu}}<0.58$ mixed bcc-fcc phases shown in black,
$x_{\mathrm{Cu}}\geq0.58$ \textcolor{red}{\bf fcc} phase shown in red.
Error bars indicate the weighted standard deviation of the ensemble average $E_{\mathrm{gap}}$.
(\textbf{b}) A comparison of the aforementioned trends with calculations performed with enhanced superlattice sizes $n_{\mathrm{enh}}$:
$x_{\mathrm{Cu}}=0.00$ ($n_{\mathrm{xct}}=1$),
$0.10$ ($n_{\mathrm{xct}}=10$),
$0.20$ ($n_{\mathrm{xct}}=5$ \vs\ $n_{\mathrm{enh}}=10$),
$0.25$ ($n_{\mathrm{xct}}=4$ \vs\ $n_{\mathrm{enh}}=12$),
$0.33$ ($n_{\mathrm{xct}}=3$),
$0.40$ ($n_{\mathrm{xct}}=5$ \vs\ $n_{\mathrm{enh}}=10$), and
$0.50$ ($n_{\mathrm{xct}}=2$ \vs\ $n_{\mathrm{enh}}=10$).
(\textbf{c}) Calculated unpolarized density of states (DOS) plots at $x_{\mathrm{Cu}}=0.2$ ($n=5$), $0.4$ ($n=5$), $0.6$ ($n=5$), $0.8$ ($n=5$).}
\label{fig:art110:CuFe}
\efig

Despite its ubiquity, iron remains at the focus of critical materials research.
Even as new phenomena are discovered with an ever-growing effort to explore extreme
conditions~\cite{Bonetti_PRL_1999,Shimizu_Nature_2001,Bose_PRB_2003},
there exist long-standing, interesting aspects that are not fully resolved.
This includes the magnetic character of the (fcc) $\gamma$-Fe phase at low
temperatures~\cite{Gorria_PRB_2004,Abrahams_PR_1962,Sandratskii_AP_1998}, among other complexities in its magnetic phase diagram~\cite{Pepperhoff_Fe_2013}.
One popular approach to studying the $\gamma$-Fe phase is through the Fe$_{1-x}$Cu$_{x}$ disordered
alloy~\cite{Gorria_PRB_2004,Gorria_PRB_2005,Orecchini_JAC_2006}.
Nominally, unary copper and iron metals with fcc structures are nonmagnetic, but together exhibit ferromagnetic ordering with very
high magnetic moments.
This observation has led to identification of Invar and anti-Invar behaviors, which may pave the way to enhanced
thermomechanical actuators~\cite{Gorria_PRB_2004,Gorria_PRB_2005}.
Fe$_{1-x}$Cu$_{x}$ is an interesting structure in its own right, as it has extremely low miscibility~\cite{Liu_PRB_2005}.
Overcoming the hurdle of developing metastable structures throughout the full compositional range has been the focus of much
research~\cite{Ma_PMS_2005}.
Such metastable structures have demonstrated novel properties like high thermal and electrical conductivity~\cite{Korn_ZPBCM_1976},
magnetoresistance, and coercivity~\cite{Hihara_JJap_1997}.

As a final proof of concept, \AFLOWPOCC\ is employed to calculate the compositional dependence of
the magnetic moment $M$ for Fe$_{1-x}$Cu$_{x}$ at $T=4.2$~K for direct comparison against experimental results~\cite{Sumiyama_JPSJ_1984}.
Considering both the sensitivity of magnetic properties to temperature as well as the significant disorder exhibited in
this system ($\max\left(\Delta {H_{\mathrm{f},i}}\right)\sim 1.63$~eV), the analysis is limited to low temperatures.
This is also where \AFLOWPOCC\ is expected to perform optimally, which considers structures relaxed at zero temperature and pressure~\cite{curtarolo:art104}.
The results are illustrated in Figure~\ref{fig:art110:CuFe}.
Sumiyama \etal\ show that the disordered system's phase is concentration dependent,
with a phase transition from bcc to fcc in the mid concentrations as $x_{\mathrm{Cu}}$ increases.
Just as with Mg$_{x}$Zn$_{1-x}$O, the transition is enforced in these calculations.
The overall decreasing trend in $M$ with reduced $x_{\mathrm{Fe}}$ in Figure~\ref{fig:art110:CuFe}(a) matches these expectations well.

With such a simple system, there is an opportunity to test whether an augmented superlattice size $n$ enhances the results.
While the concentration remains constant for $n$ above that which is needed for the desired concentration $n_{\mathrm{xct}}$,
more structures are introduced into the ensemble average.
The structures themselves also increase in size by a factor of $n$ relative to their parent structure.
For $x_{\mathrm{Cu}}=0.2,~0.4$, $n$ is doubled ($n_{\mathrm{enh}}=10$), and tripled for $x_{\mathrm{Cu}}=0.25$ ($n_{\mathrm{enh}}=12$).
With only two two-atom structures needed to describe $x_{\mathrm{Cu}}=0.5$ at $n_{\mathrm{xct}}$, $n$ can be increased by a factor of five ($n_{\mathrm{enh}}=10$)
without compromising the feasibility of the calculation.
A comparison of results calculated at $n_{\mathrm{enh}}$ is shown in Figure~\ref{fig:art110:CuFe}(b).
At most concentrations, substantial improvements are observed as the calculated trend more closely follows that which was observed
empirically.

Finally, this system's ensemble average DOS plots are illustrated in Figure~\ref{fig:art110:CuFe}(c) for $x_{\mathrm{Cu}}=0.2$, 0.4, 0.6, and 0.8.
In general, the DOS near the Fermi energy decreases with increasing $x_{\mathrm{Cu}}$, with some instability near the mixed phase regions.
This can be understood using the Stoner criterion model for transitional metals~\cite{Xie_CMS_2011,James_PRB_1999}.
Namely, ferromagnetism appears when the gain in exchange energy is larger than the loss in kinetic energy.
A larger DOS at the Fermi energy induces a higher exchange energy and favors a split into the ferromagnetic state.
The competition between ferromagnetic and paramagnetic phases can be inferred from the decreasing $M$
trend as depicted in Figure~\ref{fig:art110:CuFe}(a).

\subsection{\texorpdfstring{\PARTCAR}{PARTCAR}} \label{subsec:art110:PARCAR}
A universal file format is defined for detailing parameters of a disordered system recognizable by the \AFLOW\ framework.
Herein, \PARTCAR\ refers to the file describing the lattice geometry and partial occupancy values for a given structure.
These files will be formatted as follows:
\tabvspace

\noindent\begin{minipage}{\linewidth}
\begin{lstlisting}[numbers=none,language={POSCAR}]
PARTCAR of Ag8.733Cd3.8Zr3.267
-191.600 0.01
5.76 5.76 5.76 90 90 90
8*1+1*0.7330 3*1+1*0.8000 3*1+1*0.2670
Direct Partial
0.25 0.25 0.25 Ag
0.75 0.75 0.25 Ag
0.75 0.25 0.75 Ag
0.25 0.75 0.75 Ag
0.25 0.25 0.75 Ag
0.75 0.75 0.75 Ag
0.25 0.75 0.25 Ag
0.75 0.25 0.25 Ag
0.50 0.50 0.50 Ag
0.00 0.50 0.50 Cd
0.50 0.00 0.50 Cd
0.50 0.50 0.00 Cd
0.00 0.00 0.00 Cd
0.50 0.00 0.00 Zr
0.00 0.50 0.00 Zr
0.00 0.00 0.50 Zr
0.50 0.50 0.50 Zr
\end{lstlisting}
\end{minipage}

The first line is a comment line showing the name of the disordered system.
The atoms in the system are arranged in alphabetical order.
The second line provides a universal scaling factor for the lattice vectors and a tolerance for optimizing the partial occupancy values.
If the tolerance value is a negative integer number, then it is interpreted as the superlattice size $n$.
The third line defines the lattice parameters, namely axial lengths and interaxial angles~\cite{aflowBZ}.
The fourth line provides the number of sites occupied by each atom per unit cell.
For each atom (listed in alphabetical order), fully occupied sites are separated from partially occupied ones with a `+' sign.
Fully occupied sites are shown before partially occupied sites, although this can reversed as long as it matches the order of atomic
positions listed below.
Additionally, the degeneracy of a specified occupation value is denoted with a `*'.
Fully occupied sites have an occupation value of 1.
For example, in the above \PARTCAR\ for the disordered system Ag$_{8.733}$Cd$_{3.8}$Zr$_{3.267}$ (illustrated in Figure~\ref{fig:art110:AgCdZr}),
one silver atom and one zirconium atom share a
site at the fractional coordinate (0.5, 0.5, 0.5).
The partial occupancy values are 0.733 and 0.267 for the silver and zirconium atoms, respectively.
The \PARTCAR\ also shows one cadmium atom and one vacancy position share another site at coordinate (0, 0, 0).
The partial occupancy values are 0.8 and 0.2 for the cadmium atom and the vacancy, respectively.
The fifth line specifies that the atomic positions are given in `Direct' (fractional) coordinates
and the structure shows `Partial' occupation.
Only the first character of the two words, \ie, `D' and `P', are significant and recognized by \AFLOW.
The following (final) lines provide the fractional coordinates and occupation of each site, in the order matching that of the fourth line.

\fig
\includegraphics[width=0.6\linewidth]{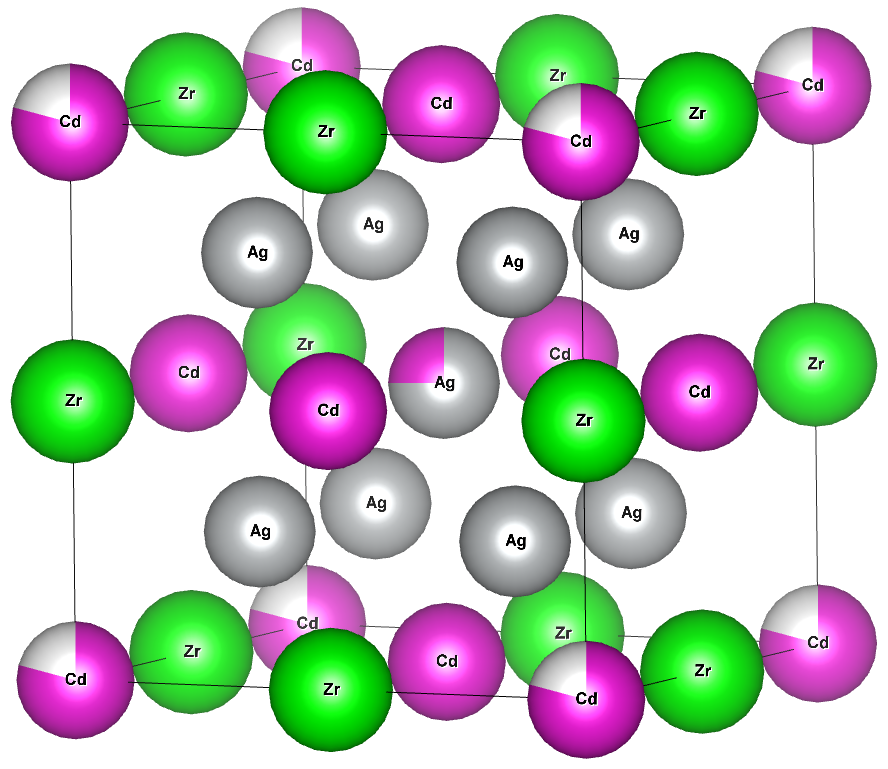}
\mycaption[Ag$_{8.733}$Cd$_{3.8}$Zr$_{3.267}$ structure showing two partially occupied sites.]
{The first disordered site is in the middle of the structure between Ag (73.3\%) and Zr (26.7\%),
and the second is on the corners between Cd (80\%) and a vacancy (20\%).
This image was created using {\small VESTA}~\cite{VESTA}.}
\label{fig:art110:AgCdZr}
\efig

\subsection{Variation of the band gap energy with superlattice size}

To illustrate the variation of the band gap energy $E_{\mathrm{gap}}$ with superlattice size $n$,
the root-mean-square error (RMSE) of the $E_{\mathrm{gap}}$ is plotted as a function of $n$ for
the ZnS$_{1-x}$Se$_{x}$ and Mg$_{x}$Zn$_{1-x}$O disordered systems in
Figure~\ref{fig:art110:Egapvsn}.
RMSE $E_{\mathrm{gap}}$ is defined as
\begin{equation}
\mathrm{RMSE}~E_{\mathrm{gap}} = \sqrt{\frac{\sum_{i}\left(E_{\mathrm{gap},i}^{\mathrm{calc}}-E_{\mathrm{gap}}^{\mathrm{exp}}\right)^{2}}
{\sum_{i}g_{i}}},
\end{equation}
where $E_{\mathrm{gap},i}^{\mathrm{calc}}$ and $g_{i}$ are the calculated band gap energy and degeneracy of the
\emph{i}$^{\mathrm{th}}$ enumerated structure, respectively, and
$E_{\mathrm{gap}}^{\mathrm{exp}}$ is the experimentally observed band gap energy.
The two limits of this variation are explored --- as $n$ approaches that having the exact stoichiometry $\left(n_{\mathrm{xct}}\right)$
and $n$ goes beyond $n_{\mathrm{xct}}$ $\left(n_{\mathrm{enh}}\right)$.
The former limit is of particular importance in the extreme concentration (dilute) limits, as $n_{\mathrm{xct}}$ can be large
even for simple systems.
RMSE~$E_{\mathrm{gap}}$ decreases with $n$ in this limit for
Mg$_{0.9}$Zn$_{0.1}$O,
Mg$_{0.1}$Zn$_{0.9}$O,
ZnS$_{0.9}$Se$_{0.1}$,
and
ZnS$_{0.1}$Se$_{0.9}$.
Conversely, the latter limit is of interest for reducing the high translational symmetry expected in smaller
supercells.
In the case of ZnS$_{0.5}$Se$_{0.5}$ and rocksalt Mg$_{0.5}$Zn$_{0.5}$O,
RMSE~$E_{\mathrm{gap}}$ also falls with $n$ in this limit, as expected.

\fig
\includegraphics[width=1.00\linewidth]{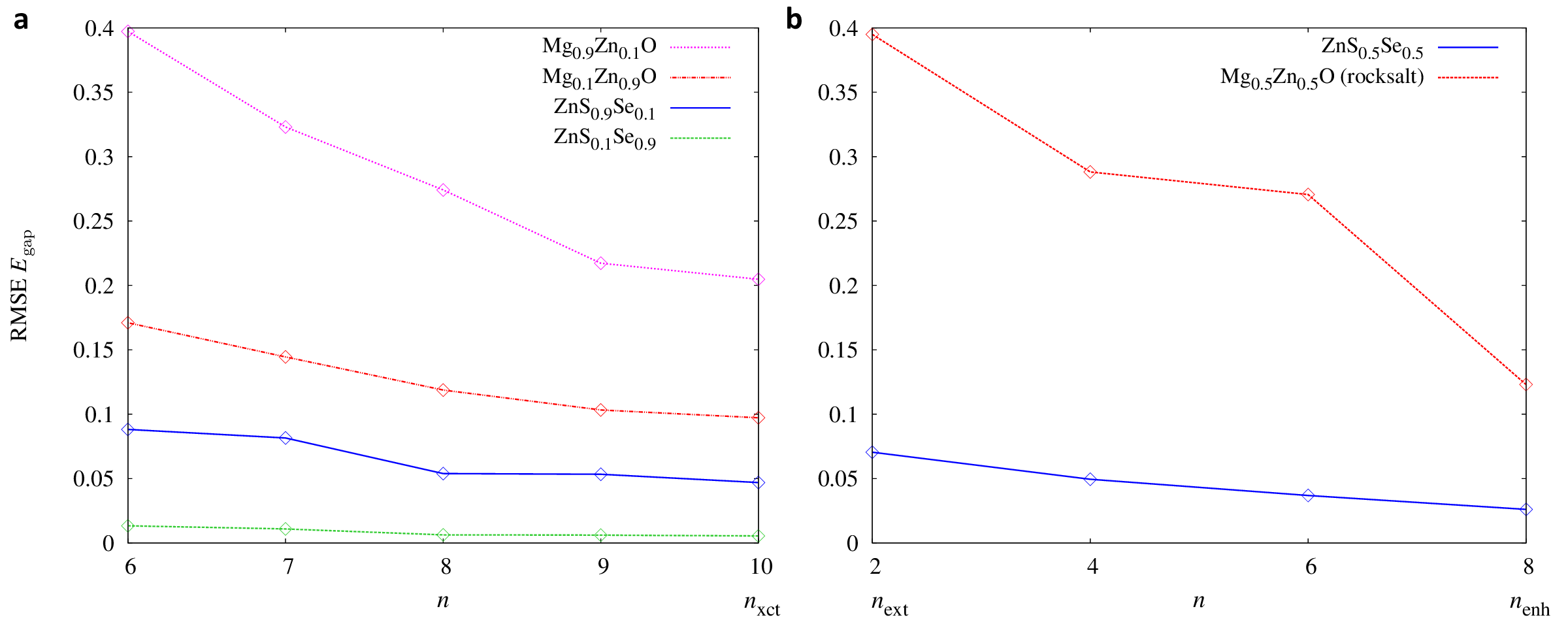}
\mycaption[Band gap energy $E_{\mathrm{gap}}$ variation with superlattice size $n$.]
{(\textbf{a}) In the extreme concentration limits, the $E_{\mathrm{gap}}$ variation is shown as
$n$ approaches that having exact stoichiometry $\left(n_{\mathrm{xct}}\right)$ for
Mg$_{0.9}$Zn$_{0.1}$O,
Mg$_{0.1}$Zn$_{0.9}$O,
ZnS$_{0.9}$Se$_{0.1}$,
and
ZnS$_{0.1}$Se$_{0.9}$.
(\textbf{b}) In the mid concentration limit, the $E_{\mathrm{gap}}$ variation is shown as $n$ increases
beyond that that having the exact stoichiometry $\left(n_{\mathrm{enh}}\right)$ for
ZnS$_{0.5}$Se$_{0.5}$ and rocksalt Mg$_{0.5}$Zn$_{0.5}$O.
The root-mean-square error (RMSE) of the $E_{\mathrm{gap}}$ is
calculated with respect to the trend observed experimentally, \ie,
the value at the specific concentration was interpolated if not provided exactly.}
\label{fig:art110:Egapvsn}
\efig

\clearpage

\subsection{Computational summary}

\tabsec
\cprotect\mycaption[Computational details for the ZnS$_{1-x}$Se$_x$ system.]
{For each composition, the parent structure, the superlattice size, the number of derivative structures enumerated by the framework,
the number of \DFT\ runs,
and the total calculation time for all \DFT\ runs are specified.
Each \DFT\ run consists of four stages (two \verb!RELAX! runs, a \verb!STATIC!, and a \verb!BANDS! calculation)
in accordance with the \AFLOW\ Standard~\cite{curtarolo:art104}.
hours* indicates the number of hours for a job parallelized across a 32-CPU node.}
\tabvspace
\resizebox{\linewidth}{!}{
\begin{tabular}{l|r|r|r|r|r}
$x$ & structure & $n$ & structure count & \DFT\ run count & total time (hours*) \\
\hline
$0.0$  & zincblende & 1                      & 1    & 1   & 0.05    \\
$0.1$  & zincblende & 10                     & 180  & 18  & 281.44  \\
$0.2$  & zincblende & 5                      & 25   & 5   & 25.79   \\
$0.25$ & zincblende & 4                      & 28   & 7   & 8.53    \\
$0.33$ & zincblende & 3                      & 9    & 3   & 3.51    \\
$0.4$  & zincblende & 5                      & 50   & 9   & 46.80   \\
$0.5$  & zincblende & 2                      & 4    & 2   & 3.60    \\
$0.6$  & zincblende & 5                      & 50   & 9   & 55.32   \\
$0.67$ & zincblende & 3                      & 9    & 3   & 2.05    \\
$0.75$ & zincblende & 4                      & 28   & 7   & 8.90    \\
$0.8$  & zincblende & 5                      & 25   & 5   & 26.79   \\
$0.9$  & zincblende & 10                     & 180  & 18  & 331.21  \\
$1.0$  & zincblende & 1                      & 1    & 1   & 0.16    \\
\hline
\multicolumn{3}{c|}{total} & 590  & 88  & 794.15  \\
\end{tabular}}
\etab

\tabsec
\cprotect\mycaption[Computational details for the Mg$_{x}$Zn$_{1-x}$O system.]
{For each composition, the parent structure, the superlattice size, the number of derivative structures enumerated by the framework,
the number of \DFT\ runs,
and the total calculation time for all \DFT\ runs are specified.
Each \DFT\ run consists of four stages (two \verb!RELAX! runs, a \verb!STATIC!, and a \verb!BANDS! calculation)
in accordance with the \AFLOW\ Standard~\cite{curtarolo:art104}.
hours* indicates the number of hours for a job parallelized across a 32-CPU node.}
\tabvspace
\resizebox{\linewidth}{!}{
\begin{tabular}{l|r|r|r|r|r}
$x$ & structure & $n$ & structure count & \DFT\ run count & total time (hours*) \\
\hline
$0.0$  & wurtzite & 1                      & 1    & 1   & 0.24    \\
$0.1$  & wurtzite & 10                     & 3300 & 300 & 3652.96 \\
$0.2$  & wurtzite & 5                      & 175  & 34  & 103.16  \\
$0.25$ & wurtzite & 4                      & 176  & 41  & 65.31   \\
$0.33$ & wurtzite & 3                      & 45   & 13  & 18.45   \\
$0.4$  & wurtzite & 5                      & 700  & 104 & 415.59  \\
$0.5$  & wurtzite & 2                      & 12   & 6   & 5.84    \\
$0.5$  & rocksalt & 2                      & 4    & 2   & 0.45    \\
$0.6$  & rocksalt & 5                      & 50   & 9   & 17.26   \\
$0.67$ & rocksalt & 3                      & 9    & 3   & 1.19    \\
$0.75$ & rocksalt & 4                      & 28   & 7   & 2.99    \\
$0.8$  & rocksalt & 5                      & 25   & 5   & 7.67    \\
$0.9$  & rocksalt & 10                     & 180  & 18  & 53.64   \\
$1.0$  & rocksalt & 1                      & 1    & 1   & 0.04    \\
\hline
\multicolumn{3}{c|}{total} & 590  & 88  & 794.15  \\
\end{tabular}}
\etab

\tabsec
\cprotect\mycaption[Computational details for the Fe$_{1-x}$Cu$_{x}$ system.]
{For each composition, the parent structure, the superlattice size, the number of derivative structures enumerated by the framework,
the number of \DFT\ runs,
and the total calculation time for all \DFT\ runs are specified.
Each \DFT\ run consists of four stages (two \verb!RELAX! runs, a \verb!STATIC!, and a \verb!BANDS! calculation)
in accordance with the \AFLOW\ Standard~\cite{curtarolo:art104}.
hours* indicates the number of hours for a job parallelized across a 32-CPU node.}
\tabvspace
\resizebox{\linewidth}{!}{
\begin{tabular}{l|r|r|r|r|r}
$x$ & structure & $n$ & structure count & \DFT\ run count & total time (hours*) \\
\hline
$0.0$  & bcc & 1                      & 1    & 1   & 0.08    \\
$0.1$  & bcc & 10                     & 180  & 18  & 82.22   \\
$0.2$  & bcc & 5                      & 25   & 5   & 136.76  \\
$0.25$ & bcc & 4                      & 28   & 7   & 4.99    \\
$0.33$ & bcc & 3                      & 9    & 3   & 91.70   \\
$0.4$  & bcc & 5                      & 50   & 9   & 658.56  \\
$0.5$  & bcc & 2                      & 4    & 2   & 0.75    \\
$0.5$  & fcc & 2                      & 4    & 2   & 0.78    \\
$0.6$  & fcc & 5                      & 50   & 9   & 26.48   \\
$0.67$ & fcc & 3                      & 9    & 3   & 2.44    \\
$0.75$ & fcc & 4                      & 28   & 7   & 6.12    \\
$0.8$  & fcc & 5                      & 25   & 5   & 14.56   \\
$0.9$  & fcc & 10                     & 180  & 18  & 119.52  \\
$1.0$  & fcc & 1                      & 1    & 1   & 0.07    \\
\hline
\multicolumn{3}{c|}{total} & 594  & 90  & 1145.03 \\
\end{tabular}}
\etab

\clearpage

\subsection{Conclusion}
In this work, the \AFLOWPOCC\ software framework is introduced capable of modeling substitutionally disordered materials.
Specifically, the framework delivers high value properties of disordered systems, including the density of states (DOS),
band gap energy $E_{\mathrm{gap}}$, and magnetic moment $M$,
as well as additional insight into underlying physical mechanisms.
Through a number of technologically significant examples, the prowess of this highly efficient and
convenient framework is illustrated.
Such materials that exhibit highly tunable properties are of critical importance toward the goal of rational materials design.
Without loss of feasibility or accuracy, the framework exploits highly successful high-throughput first principles approaches
in more complex, real-world systems.
\clearpage
\section{Universal Fragment Descriptors for Predicting Properties of Inorganic Crystals}
\label{sec:art124}

This study follows from a collaborative effort described in Reference~\cite{curtarolo:art124}.
Author contributions are as follows:
Olexandr Isayev developed and implemented the method.
Corey Oses and Cormac Toher prepared the data and worked with the \AFLOW\ database.
Eric Gossett developed the open-access online application available at
\url{aflow.org/aflow-ml} leveraging the \ML\ models.
Olexandr Isayev and Corey Oses contributed equally to the work.
All authors discussed the results and their implications and contributed
to the writing of the article.

\subsection{Introduction}
Advances in materials science are often slow and fortuitous~\cite{nmatHT}.
Coupling the field's combinatorial challenges with the demanding efforts required for materials characterization
makes progress uniquely difficult.
The number of materials currently characterized, either experimentally~\cite{ICSD,ICSD3} or
computationally~\cite{aflowlibPAPER},
pales in comparison to the anticipated potential diversity.
Only considering naturally occurring elements, 9,000 crystal structure prototypes~\cite{ICSD,ICSD3},
and stoichiometric compositions, there are roughly $3 \times 10^{11}$ potential quaternary compounds and
$10^{13}$ quinary combinations.
Indeed, it has been estimated that the total number of theoretical materials can be as large as
$10^{100}$~\cite{Walsh_NChem_2015}.
Exacerbating the issue, standard materials characterization practices, such as calculating the band structure,
can become quite expensive when considering
finite-size scaling, charge corrections~\cite{Castleton_PRB_2006},
and going beyond standard density functional theory (\DFT) with Green's function methods such as the fully
self-consistent \GW\ approximation~\cite{Lindgren_BetheSalpeter_2011,vanSchilfgaarde_PRL_2006}.
Ultimately, brute force exploration of this search space,
even in high-throughput fashion~\cite{nmatHT,koinuma_nmat_review2004}, is entirely impractical.

To circumvent the issue, many knowledge-based structure-property
relationships have been conjectured over the years to aid in the search for novel functional materials---ranging
from the simplest empirical relationships~\cite{Mizutani_HR_2011}
to complex advanced
models~\cite{curtarolo:art94,Ghiringhelli_PRL_2015,PyzerKnapp_AdFM_2015,Rajan_ARMR_2015,curtarolo:art85,curtarolo:art120,Furmanchuk_RSCA_2016}.
For instance, many (semi-)empirical rules have been developed to predict
band gap energies, such as those incorporating
(optical~\cite{Duffy_JCP_1977}) electronegativity~\cite{DiQuarto_JPCB_1997}.
More sophisticated Machine Learning (\ML) models were also developed for chalcopyrite
semiconductors~\cite{Zeng_ChemMat_2002}, perovskites~\cite{Pilania_SR_2016}, and binary
compounds~\cite{Gu_SSS_2006}.
Unfortunately, many of these models are limited to a single family of materials,
with narrow applicability outside of their training scope.

The development of such structure-property relationships has become an integral practice in the drug industry,
which faces a similar combinatorial challenge.
The number of potential organic molecules is estimated to be anywhere between $10^{13}$ to $10^{180}$~\cite{Gorse_CTMC_2006}.
In computational medicinal chemistry, Quantitative Structure-Activity
Relationship (\QSAR) modeling coupled with
virtual screening of chemical libraries have been largely successfully in the discovery of
novel bioactive compounds~\cite{Baskin_VirScreen_2008}.

Here, we introduce fragment descriptors of materials structure.
The combination of these descriptors with \ML\ approaches affords the development of universal models
capable of accurate prediction of properties for virtually any stoichiometric inorganic crystalline material.
First, the algorithm for descriptor generation is described, along with implementation of
\ML\ methods for Quantitative Materials Structure-Property
Relationship (\QMSPR) modeling.
Next, the effectiveness of this approach is assessed through prediction of eight critical electronic
and thermomechanical properties of materials,
including the metal/insulator classification, band gap energy, bulk and shear moduli,
Debye temperature, heat capacities (at constant pressure and volume), and thermal expansion coefficient.
The impact and interaction among the most significant descriptors as determined
by the \ML\ algorithms are highlighted.
As a proof-of-concept, the \QMSPR\ models are then employed to predict thermomechanical properties
for compounds previously uncharacterized, and the predictions are validated via
the \AEL-\AGL\ integrated framework~\cite{curtarolo:art96, curtarolo:art115}.
Such predictions are of particular value as proper calculation pathways for thermomechanical properties
in the most efficient scenarios still require analysis of multiple \DFT-runs, elevating the cost
of already expensive calculations.
Finally, \ML-predictions and calculations are both compared to experimental values which ultimately corroborate the validity of the approach.

Other investigations have predicted a subset of the target properties
discussed here
by building \ML\ approaches where computationally obtained quantities,
such as the cohesive energy, formation energy, and energy above the
convex hull,
are the part of the input data~\cite{deJong_SR_2016}.
The approach presented here is orthogonal.
Once trained, our proposed models achieve comparable accuracies without the need of
further \abinitio\ data.
All necessary input properties are either tabulated or derived directly
from the geometrical structures.
There are advantages:
\textit{(i)} \textit{a priori}, after the training, no further calculations need
to be performed,
\textit{(ii)} \textit{a posteriori}, the modeling framework becomes
independent of the source or nature of the training data,
\eg, calculated \vs\ experimental.
The latter allows for rapid extension of predictions to online
applications---given the geometry of a cell and the species involved, eight \ML\
predicted properties are returned
(\href{http://aflow.org/aflow-ml}{aflow.org/aflow-ml}).

\fig
\includegraphics[width=\linewidth]{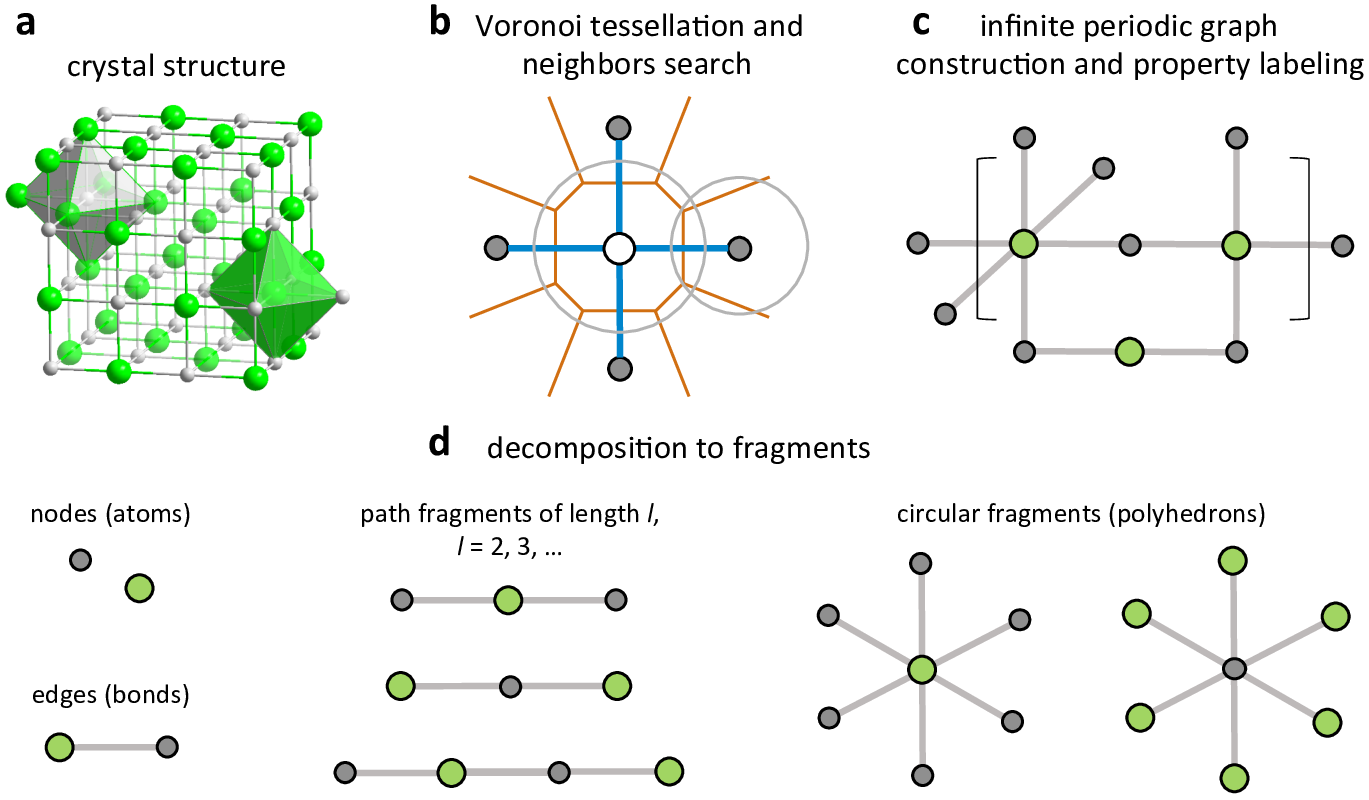}
\mycaption[Schematic representing the construction of the Property-Labeled
Materials Fragments (\PLMF).]
{The crystal structure (\textbf{a}) is analyzed for atomic neighbors (\textbf{b}) via Voronoi tessellation.
After property labeling, the resulting periodic graph (\textbf{c}) is decomposed into simple
subgraphs (\textbf{d}).}
\label{fig:art124:figure1}
\efig

\subsection{Results}
\boldsection{Universal property-labeled materials fragments.}
Many cheminformatics investigations have demonstrated the critical importance of
molecular descriptors, which are known to
influence model accuracy more than the choice of the
\ML\ algorithm~\cite{Young_MI_2012,Polishchuk_MI_2013}.
For the purposes of this investigation, fragment descriptors typically used for organic
molecules were adapted to serve for materials characterization~\cite{Ruggiu_MI_2010}.
Molecular systems can be described as graphs whose vertices correspond to atoms and edges to chemical bonds.
In this representation, fragment descriptors characterize subgraphs of the full 3D molecular network.
Any molecular graph invariant can be uniquely represented as a linear combination of fragment descriptors.
They offer several advantages over other types of chemical descriptors~\cite{Todeschini_MolecularDescriptors_2000},
including simplicity of calculation, storage, and interpretation~\cite{Varnek_JCAMD_2005}.
However, they also come with a few disadvantages.
Models built with fragment descriptors perform poorly when presented with new fragments for which they were not trained.
Additionally, typical fragments are constructed solely with information of the individual atomic symbols
(\eg, C, N, Na).
Such a limited context would be insufficient for modeling the complex chemical interactions within materials.

Mindful of these constraints, fragment descriptors for materials were conceptualized
by differentiating atoms not
by their symbols but by a plethora of well-tabulated chemical and physical properties~\cite{Lide_CRC_2004}.
Descriptor features comprise of various combinations of these atomic properties.
From this perspective, materials can be thought of as ``colored'' graphs, with vertices decorated according
to the nature of the atoms they represent~\cite{Varnek_CCADD_2008}.
Partitions of these graphs form Property-Labeled Materials Fragments (\PLMF).

Figure~\ref{fig:art124:figure1} shows the scheme for constructing {\PLMF}s.
Given a crystal structure, the first step is to determine the atomic connectivity within it.
In general, atomic connectivity is not a trivial property to determine within materials.
Not only must the potential bonding distances among atoms be considered, but also whether
the topology of nearby atoms allows for bonding.
Therefore, a computational geometry approach is employed to partition the crystal structure (Figure~\ref{fig:art124:figure1}(a))
into atom-centered Voronoi-Dirichlet
polyhedra~\cite{Blatov_CR_2004,Carlucci_ChemRev_2014} (Figure~\ref{fig:art124:figure1}(b)).
This partitioning scheme was found to be invaluable in the topological analysis of metal
organic frameworks (\MOF), molecules, and inorganic crystals~\cite{Zolotarev_CGD_2014}.
Connectivity between atoms is established by satisfying two criteria:
\textit{(i)} the atoms must share a Voronoi face (perpendicular bisector between neighboring atoms), and
\textit{(ii)} the interatomic distance must be shorter than the sum of the Cordero covalent
radii~\cite{Cordero_DT_2008} to within a 0.25 \text{\AA} tolerance.
Here, only strong interatomic interactions are modeled, such as covalent, ionic, and metallic
bonding, ignoring van der Waals interactions.
Due to the ambiguity within materials, the bond order (single/double/triple bond classification) is not considered.
Taken together, the Voronoi centers that share a Voronoi face and are within the sum of their
covalent radii form a three-dimensional graph defining the connectivity within the material.

In the final steps of the \PLMF\ construction, the full graph and
corresponding adjacency matrix (Figure~\ref{fig:art124:figure1}(c)) are constructed from the total list of connections.
The adjacency matrix $\mathbf{A}$ of a simple graph (material) with $n$ vertices (atoms) is a square
matrix $\left(n \times n\right)$ with entries $a_{ij}=1$ if atom $i$ is
connected to atom $j$, and $a_{ij}=0$ otherwise.
This adjacency matrix reflects the global topology for a given system, including interatomic bonds and contacts within the crystal.
The full graph is partitioned into smaller subgraphs, corresponding to individual fragments (Figure~\ref{fig:art124:figure1}(d)).
While there are several subgraphs to consider in general, the length $l$ is restricted to a maximum of three,
where $l$ is the largest number of consecutive, non-repetitive edges in the subgraph.
This restriction serves to curb the complexity of the final descriptor vector.
In particular, there are two types of fragments.
Path fragments are subgraphs of at most $l=3$ that encode any linear strand of up to four atoms.
Only the shortest paths between atoms are considered.
Circular fragments are subgraphs of $l=2$ that encode the first shell of nearest neighbor atoms.
In this context, circular fragments represent coordination polyhedra, or clusters of atoms with
anion/cation centers each surrounded by a set of its respective counter ion.
Coordination polyhedra are used extensively in crystallography and mineralogy~\cite{pauling_bond}.

The {\PLMF}s are differentiated by local (standard atomic/elemental) reference properties~\cite{Lide_CRC_2004},
which include:
\textit{(i)} general properties:
the Mendeleev group and period numbers $\left(g_\sP,~p_\sP\right)$,
number of valence electrons $\left(N_\sV\right)$;
\textit{(ii)} measured properties~\cite{Lide_CRC_2004}:
atomic mass $\left(m_\satom\right)$,
electron affinity $\left(EA\right)$,
thermal conductivity $\left(\lambda\right)$,
heat capacity $\left(C\right)$,
enthalpies of atomization $\left(\Delta H_{\mathrm{at}}\right)$,
fusion $\left(\Delta H_\sfusion\right)$, and
vaporization $\left(\Delta H_\svapor\right)$,
first three ionization potentials $\left(IP_{1,2,3}\right)$; and
\textit{(iii)} derived properties:
effective atomic charge $\left(Z_\seff\right)$,
molar volume $\left(V_\smolar\right)$,
chemical hardness $\left(\eta\right)$~\cite{Parr_JACS_1983,Lide_CRC_2004},
covalent $\left(r_\scov\right)$~\cite{Cordero_DT_2008},
absolute~\cite{Ghosh_IJMS_2002}, and
van der Waals radii~\cite{Lide_CRC_2004},
electronegativity $\left(\chi\right)$,
and
polarizability $\left(\alpha_\sP\right)$.
Pairs of properties are included in the form of their multiplication and ratio,
as well as the property value divided by the atomic connectivity
(number of neighbors in the adjacency matrix).
For every property scheme $\mathbf{q}$, the following quantities are also considered:
minimum $\left(\min(\mathbf{q})\right)$,
maximum $\left(\max(\mathbf{q})\right)$,
total sum $\left(\sum \mathbf{q}\right)$,
average $\left(\avg (\mathbf{q}) \right)$, and
standard deviation $\left(\std(\mathbf{q})\right)$ of
$\mathbf{q}$ among the atoms in the material.

To incorporate information about shape, size, and symmetry of the crystal unit cell,
the following crystal-wide properties are incorporated:
lattice parameters ($a$, $b$, $c$), their ratios ($a/b$, $b/c$, $a/c$), angles ($\alpha$, $\beta$, $\gamma$),
density, volume, volume per atom, number of atoms, number of species (atom types),
lattice type, point group, and space group.

All aforementioned descriptors (fragment-based and crystal-wide) can be concatenated together to represent each
material uniquely.
After filtering out low variance ($<0.001$) and highly correlated $\left(r^{2}\!>\!0.95\right)$ features,
the final feature vector captures 2,494 total descriptors.

Descriptor construction is inspired by the topological charge indices~\cite{Galvez_JCICS_1995}
and the Kier-Hall
electro-topological state indices~\cite{Kier_Electrotopological_1999}.
Let $\mathbf{M}$ be the matrix obtained by multiplying the adjacency
matrix $\mathbf{A}$ by the reciprocal square distance matrix $\mathbf{D}$ $\left(D_{ij}=1/r_{i,j}^{2}\right)$:
\begin{equation}
\mathbf{M}=\mathbf{A} \cdot \mathbf{D}.
\end{equation}
The matrix $\mathbf{M}$, called the Galvez matrix, is a square $n \times n$ matrix,
where $n$ is the number of atoms in the unit cell.
From $\mathbf{M}$, descriptors of reference property $\mathbf{q}$ are calculated as
\begin{equation}
T^{\mathrm{E}}=\sum_{i=1}^{n-1}\sum_{j=i+1}^{n}\left|q_{i}-q_{j}\right|M_{ij}
\end{equation}
and
\begin{equation}
T_{\sbond}^{\mathrm{E}}=\sum_{\{i,j\}\in\mathrm{bonds}}\left|q_{i}-q_{j}\right|M_{ij},
\end{equation}
where the first set of indices count over all pairs of atoms and the second
is restricted to all pairs $i,j$ of bonded atoms.

\fig
\includegraphics[width=\linewidth]{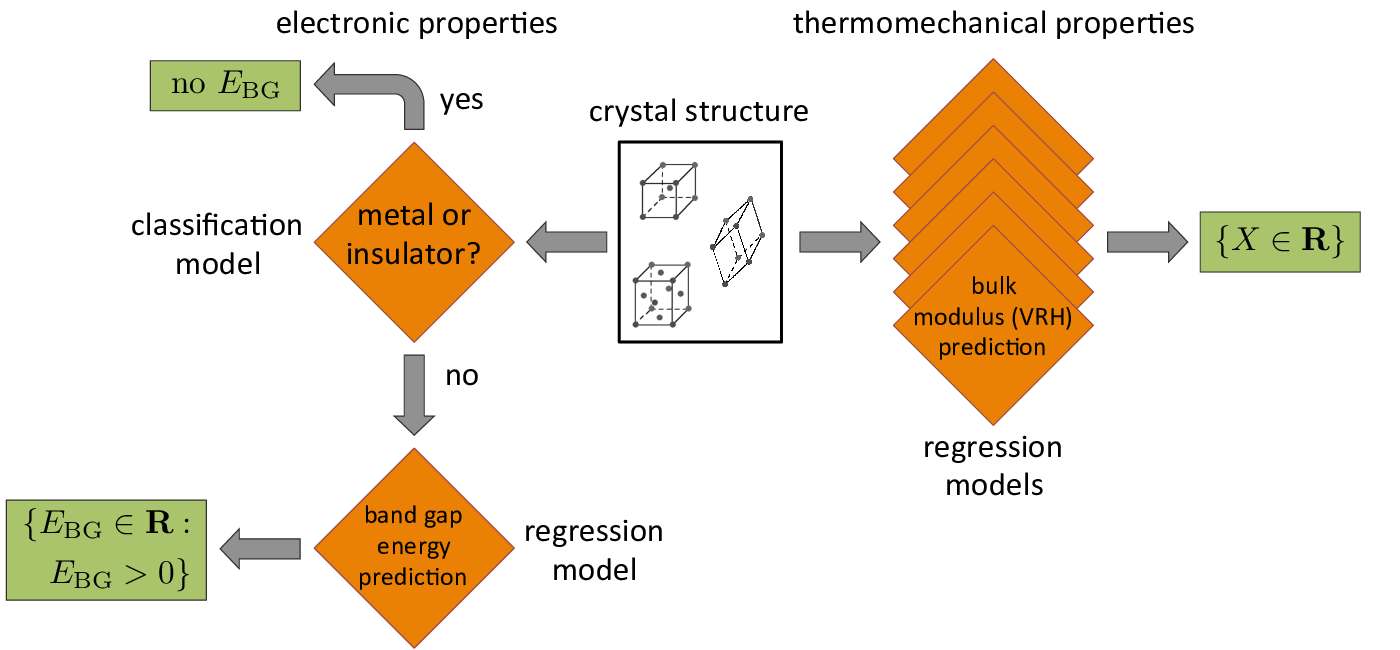}
\mycaption[Outline of the modeling work-flow.]
{\ML\ models are represented by orange diamonds. Target properties predicted by these models are highlighted in green.}
\label{fig:art124:figure2}
\efig

\boldsection{Quantitative materials structure-property relationship modeling.}
In training the models, the same \ML\ method and descriptors are employed without any hand tuning or variable selection.
Specifically, models are constructed using gradient boosting decision tree
(\GBDT) technique~\cite{Friedman_AnnStat_2001}.
All models were validated through $y$-randomization (label scrambling).
Five-fold cross validation is used to assess how well each model will generalize to an independent dataset.
Hyperparameters are determined with grid searches on the training set and 10-fold cross validation.

The gradient boosting decision trees (\GBDT) method~\cite{Friedman_AnnStat_2001}
evolved from the application of boosting
methods~\cite{gbm} to regression trees~\cite{Loh_ISR_2014}.
The boosting method is based on the observation that finding many weakly accurate
prediction rules can be a lot easier than finding a single, highly accurate rule~\cite{Schapire_ML_1990}.
The boosting algorithm calls this ``weak'' learner repeatedly, at each stage feeding it
a different subset of the training examples.
Each time it is called, the weak learner generates a new weak prediction rule.
After many iterations, the boosting algorithm combines these weak rules into
a single prediction rule aiming to be much more accurate than any single weak rule.

The \GBDT\ approach is an additive model of the following form:
\begin{equation}
F(\mathbf{x};\{\gamma_{m},\mathbf{a}\}_{1}^{M})=\sum_{m=1}^{M}\gamma_{m} h_{m}(\mathbf{x};\mathbf{a}_{m}),
\end{equation}
where $h_{m}(\mathbf{x};\mathbf{a}_{m})$ are the weak learners (decision trees in this case)
characterized by parameters
$\mathbf{a}_{m}$, and $M$ is the total
count of decision trees obtained through boosting.

It builds the additive model in a forward stage-wise fashion:
\begin{equation}
F_m(\mathbf{x})=F_{m-1}(\mathbf{x})+\gamma_{m} h_{m}(\mathbf{x};\mathbf{a}_{m}).
\end{equation}
At each stage $\left(m=1,2,\ldots,M\right)$, $\gamma_{m}$ and $\mathbf{a}_{m}$ are chosen to minimize the loss function
$f_L$ given the current model $F_{m-1}(x_{i})$ for all data points (count $N$),
\begin{equation}
\left(\gamma_{m},\mathbf{a}_{m}\right)=\argmin_{\gamma,\mathbf{a}} \sum_{i=1}^{N}
f_{L} \left[y_{i},F_{m-1}\left(\mathbf{x_{i}}\right)+\gamma h\left(\mathbf{x}_{i};\mathbf{a}\right)\right].
\end{equation}
Gradient boosting attempts to solve this minimization problem numerically via steepest descent.
The steepest descent direction is the negative gradient of the loss function
evaluated at the current model $F_{m-1}$, where the step length is chosen using line search.

An important practical task is to quantify variable importance.
Feature selection in decision tree ensembles cannot differentiate between primary
effects and effects caused by interactions between variables.
Therefore, unlike regression coefficients, a direct comparison of captured effects is prohibited.
For this purpose, variable influence is quantified in the
following way~\cite{Friedman_AnnStat_2001}.
Let us define the influence of variable $j$ in a single tree $h$.
Consider that the tree has $l$ splits and therefore $l-1$ levels.
This gives rise to the definition of the variable influence,
\begin{equation}
K_{j}^2(h)=\sum_{i=1}^{l-1} I_{i}^{2} \mathbbm{1}\left(x_{i}=j\right),
\end{equation}
where $I_{i}^{2}$ is the empirical squared improvement resulting from this split,
and $\mathbbm{1}$ is the indicator function.
Here, $\mathbbm{1}$ has a value of one if the split at node $x_{i}$ is on variable $j$, and
zero otherwise,
\ie, it measures the number of times a variable $j$ is selected for splitting.
To obtain the overall influence of variable $j$ in the ensemble of decision trees (count $M$),
it is averaged over all trees,
\begin{equation}
K_{j}^{2}={M}^{-1} \sum_{m=1}^{M} K_{j}^{2} (h_{m}).
\end{equation}
The influences $K_{j}^{2}$ are normalized so that they add to one.
Influences capture the importance of the variable, but
not the direction of the response (positive or negative).

\fig
\includegraphics[width=\linewidth]{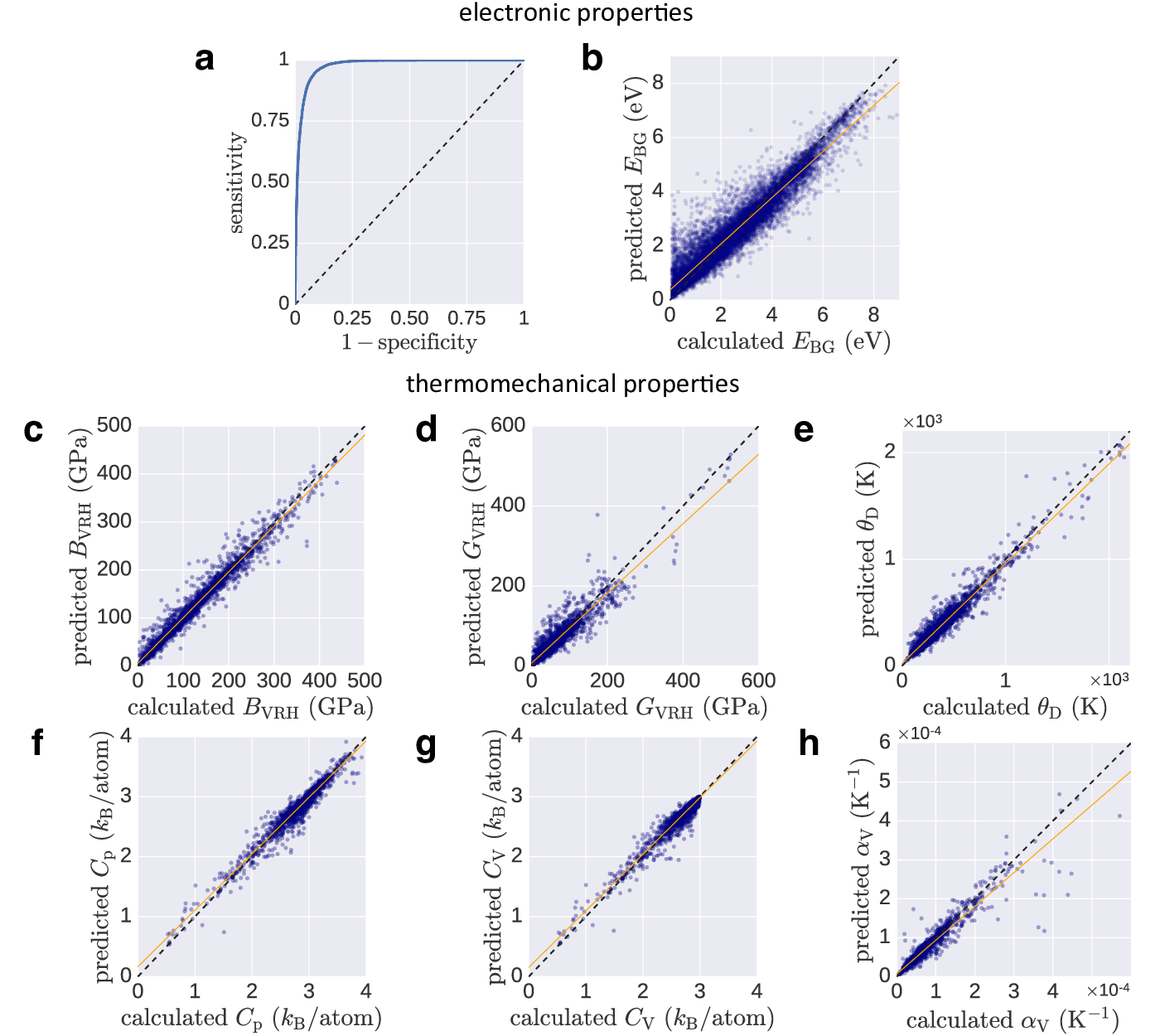}
\mycaption[Five-fold cross validation plots for the eight \ML\ models predicting electronic and thermomechanical properties.]
{\textbf{(a)} Receiver operating characteristic (\ROC) curve for the classification \ML\ model.
\textbf{(b)}-\textbf{(h)} Predicted \vs\ calculated values
for the regression \ML\ models:
\textbf{(b)} band gap energy $\left(E_{\scriptstyle \mathrm{BG}}\right)$,
\textbf{(c)} bulk modulus $\left(B_{\scriptstyle \mathrm{VRH}}\right)$,
\textbf{(d)} shear modulus $\left(G_{\scriptstyle \mathrm{VRH}}\right)$,
\textbf{(e)} Debye temperature $\left(\theta_{\scriptstyle \mathrm{D}}\right)$,
\textbf{(f)} heat capacity at constant pressure $\left(C_{\scriptstyle \mathrm{P}}\right)$,
\textbf{(g)} heat capacity at constant volume $\left(C_{\scriptstyle \mathrm{V}}\right)$, and
\textbf{(h)} thermal expansion coefficient $\left(\alpha_{\scriptstyle \mathrm{V}}\right)$.
}
\label{fig:art124:figure3}
\efig

\boldsection{Integrated modeling work-flow.}
Eight predictive models are developed in this work, including:
a binary classification model that predicts if a material is a metal or an insulator
and seven regression models that predict:
the band gap energy $\left(E_\sBG\right)$ for insulators,
bulk modulus $\left(B_\sVRH\right)$,
shear modulus $\left(G_\sVRH\right)$,
Debye temperature $\left(\theta_\sD\right)$,
heat capacity at constant pressure $\left(C_\sp\right)$,
heat capacity at constant volume $\left(C_\sV\right)$, and
thermal expansion coefficient $\left(\alpha_\sV\right)$.

Figure~\ref{fig:art124:figure2} shows the overall application work-flow.
A novel candidate material is first classified as a metal or an insulator.
If the material is classified as an insulator, $E_\sBG$ is predicted,
while classification as a metal implies that the material has no $E_\sBG$.
The six thermomechanical properties are then predicted independent of the material's metal/insulator classification.
The integrated modeling work-flow has been implemented as a web application at
\href{http://aflow.org/aflow-ml}{aflow.org/aflow-ml},
requiring only the atomic species and positions as input for predictions.

While all three models were trained independently, the accuracy of the
$E_\sBG$ regression model is inherently dependent on the accuracy of the metal/insulator classification model
in this work-flow.
However, the high accuracy of the metal/insulator classification model suggests this not to be a practical concern.

\boldsection{Model generalizability.}
One technique for assessing model quality is five-fold cross validation, which gauges how well
the model is expected to generalize to an independent dataset.
For each model, the scheme involves randomly partitioning the set into five groups and predicting the value of
each material in one subset while training the model on the other four subsets.
Hence, each subset has the opportunity to play the role of the ``test set''.
Furthermore, any observed deviations in the predictions are addressed.
For further analysis, all predicted and calculated results are available in
Supplementary Note 2 of Reference~\cite{curtarolo:art124}.

The accuracy of the metal/insulator classifier is reported as the
area under the curve (\AUC)
of the receiver operating characteristic (\ROC) plot (Figure~\ref{fig:art124:figure3}(a)).
The \ROC\ curve illustrates the model's ability to differentiate between metallic and insulating input materials.
It plots the prediction rate for insulators (correctly \vs\ incorrectly predicted) throughout the
full spectrum of possible prediction thresholds.
An area of 1.0 represents a perfect test, while an area of 0.5 characterizes a random guess (the dashed line).
The model shows excellent external predictive power with the \AUC\ at 0.98,
an insulator-prediction success rate (sensitivity) of 0.95,
a metal-prediction success rate (specificity) of 0.92,
and an overall classification rate (\CCR) of 0.93.
For the complete set of \PLMFelectronicTotal\ materials, this corresponds to
2,103 misclassified materials, including 1,359 misclassified metals and 744 misclassified insulators.
Evidently, the model exhibits positive bias toward predicting insulators, where bias refers to whether a
\ML\ model tends to over- or under-estimate the predicted property.
This low false-metal rate is fortunate as the model is unlikely to
misclassify a novel, potentially interesting semiconductor as a metal.
Overall, the metal classification model is robust enough to handle the full complexity of the periodic table.

\tab
\mycaption[Statistical summary of the five-fold cross-validated predictions for the seven regression models.]
{The summary corresponds with Figure~\ref{fig:art124:figure3}.}
\tabvspace
\begin{tabular}{l|r|r|r}
property & \RMSE\ & \MAE\ & $r^{2}$ \\
\hline
$E_\sBG$                                & 0.51~eV                                     & 0.35~eV                                 & 0.90 \\
$B_\sVRH$                               & 14.25~GPa                                   & 8.68~GPa                                & 0.97 \\
$G_\sVRH$                               & 18.43~GPa                                   & 10.62~GPa                               & 0.88 \\
$\theta_\sD$                            & 56.97~K                                     & 35.86~K                                 & 0.95 \\
$C_\sp$                                 & 0.09~$k_\sB$/atom                           & 0.05~$k_\sB$/atom                       & 0.95 \\
$C_\sV$                                 & 0.07~$k_\sB$/atom                           & 0.04~$k_\sB$/atom                       & 0.95 \\
$\alpha_\sV$                            & $1.47 \times 10^{-5}$~K$^{-1}$              & $5.69 \times 10^{-6}$~K$^{-1}$           & 0.91 \\
\end{tabular}
\label{tab:art124:table1}
\etab

The results of the five-fold cross validation analysis for the band gap energy $\left(E_\sBG\right)$ regression model
are plotted in Figure~\ref{fig:art124:figure3}(b).
Additionally, a statistical profile of these predictions, along with that of the six thermomechanical regression models,
is provided in Table~\ref{tab:art124:table1}, which includes metrics such as
the root-mean-square error (\RMSE),
mean absolute error (\MAE), and coefficient of determination $\left(r^2\right)$.
Similar to the classification model, the $E_\sBG$ model exhibits a positive predictive bias.
The biggest errors come from materials with narrow band gaps,
\ie, the scatter in the lower left corner in Figure~\ref{fig:art124:figure3}(b).
These materials predominantly include complex fluorides and nitrides.
N$_{2}$H$_{6}$Cl$_{2}$ (\ICSD\ \#23145)
exhibits the worst prediction accuracy with signed error SE = 3.78 eV~\cite{Donohue_JCP_1947}.
The most underestimated materials are HCN (\ICSD\ \#76419) and, respectively
N$_{2}$H$_{6}$Cl$_{2}$ (\ICSD\ \#240903) with SE = -2.67 and -3.19 eV~\cite{Dulmage_ActaCrist_1951,Kruszynski_ActaCristE_2007}, respectively.
This is not surprising considering that all three are molecular crystals.
Such systems are anomalies in the \ICSD, and fit better in other databases, such as
the Cambridge Structural Database~\cite{Groom_CSD_2016}.
Overall, 10,762 materials are predicted within 25\% accuracy of calculated values,
whereas 824 systems have errors over 1 eV.

Figures~\ref{fig:art124:figure3}(c-h) and Table~\ref{tab:art124:table1} showcase the results of the five-fold cross validation analysis
for the six thermomechanical regression models.
For both bulk $\left(B_\sVRH\right)$ and shear $\left(G_\sVRH\right)$ moduli,
over 85\% of materials are predicted within 20~GPa of their calculated values.
The remaining models also demonstrate high accuracy, with
at least 90\% of the full training set $\left(>2,546~\mathrm{systems}\right)$
predicted to within 25\% of the calculated values.
Significant outliers in predictions of the bulk modulus include
graphite (\ICSD\ \#187640, SE = 100 GPa, likely
due to extreme anisotropy) and two theoretical high-pressure boron nitrides (\ICSD\ \#162873 and \#162874,
under-predicted by over 110 GPa)~\cite{Lian_JCP_2013,Doll_PRB_2008}.
Other theoretical systems are ill-predicted throughout the six properties, including
ZN (\ICSD\ \#161885), CN$_{2}$ (\ICSD\ \#247676), C$_{3}$N$_{4}$ (\ICSD\ \#151782),
and CH (\ICSD\ \#187642)~\cite{EscorciaSalas_MJ_2008,Li_PCCP_2012,Marques_PRB_2004,Lian_JCP_2013}.
Predictions for the $G_\sVRH$, Debye temperature $\left(\theta_\sD\right)$, and thermal expansion coefficient
$\left(\alpha_\sV\right)$
tend to be slightly underestimated, particularly for higher calculated values.
Additionally, mild scattering can be seen for $\theta_\sD$ and
$\alpha_\sV$, but not enough to have a significant
impact on the error or correlation metrics.

Despite minimal deviations, both \RMSE\ and \MAE\ are within 4\% of the ranges covered for each property,
and the predictions demonstrate excellent correlation with the calculated properties.
Note the tight clustering of points just below 3 $k_\sB$/atom for the heat
capacity at constant volume $\left(C_\sV\right)$.
This is due to $C_\sV$ saturation in accordance with the Dulong-Petit law occurring at or below
300 K for many compounds.

\fig
\includegraphics[width=0.65\linewidth]{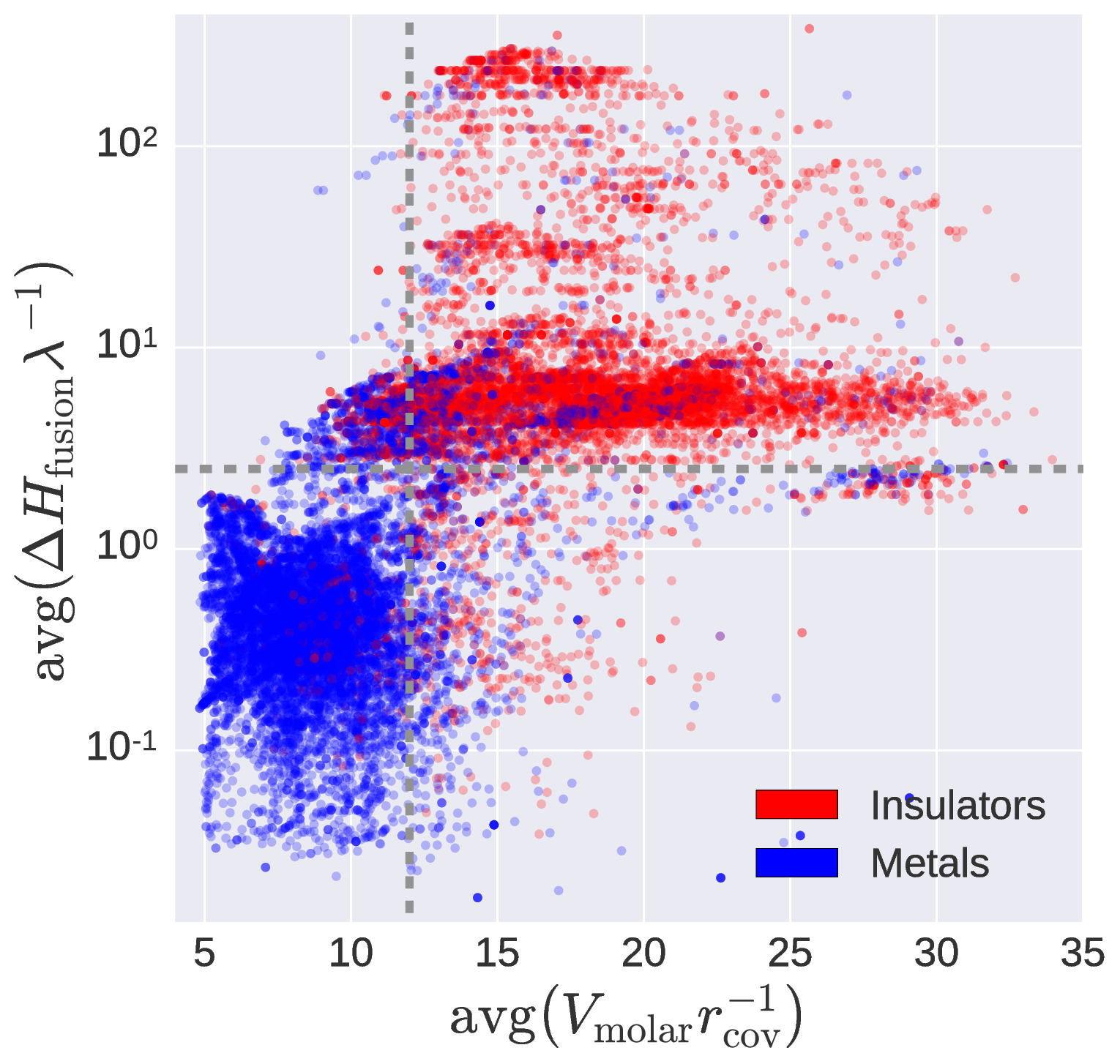}
\mycaption[Semi-log scatter plot of the full dataset (\PLMFelectronicTotal\ unique materials) in a dual-descriptor space.]
{$\avg\left(\Delta H_{\scriptstyle \mathrm{fusion}}\lambda^{-1}\right)$ \vs\
$\avg\left(V_{\scriptstyle \mathrm{molar}}r_{\scriptstyle \mathrm{cov}}^{-1}\right)$.
Insulators and metals are colored in red and blue, respectively.}
\label{fig:art124:figure4}
\efig

\fig
\includegraphics[width=\linewidth]{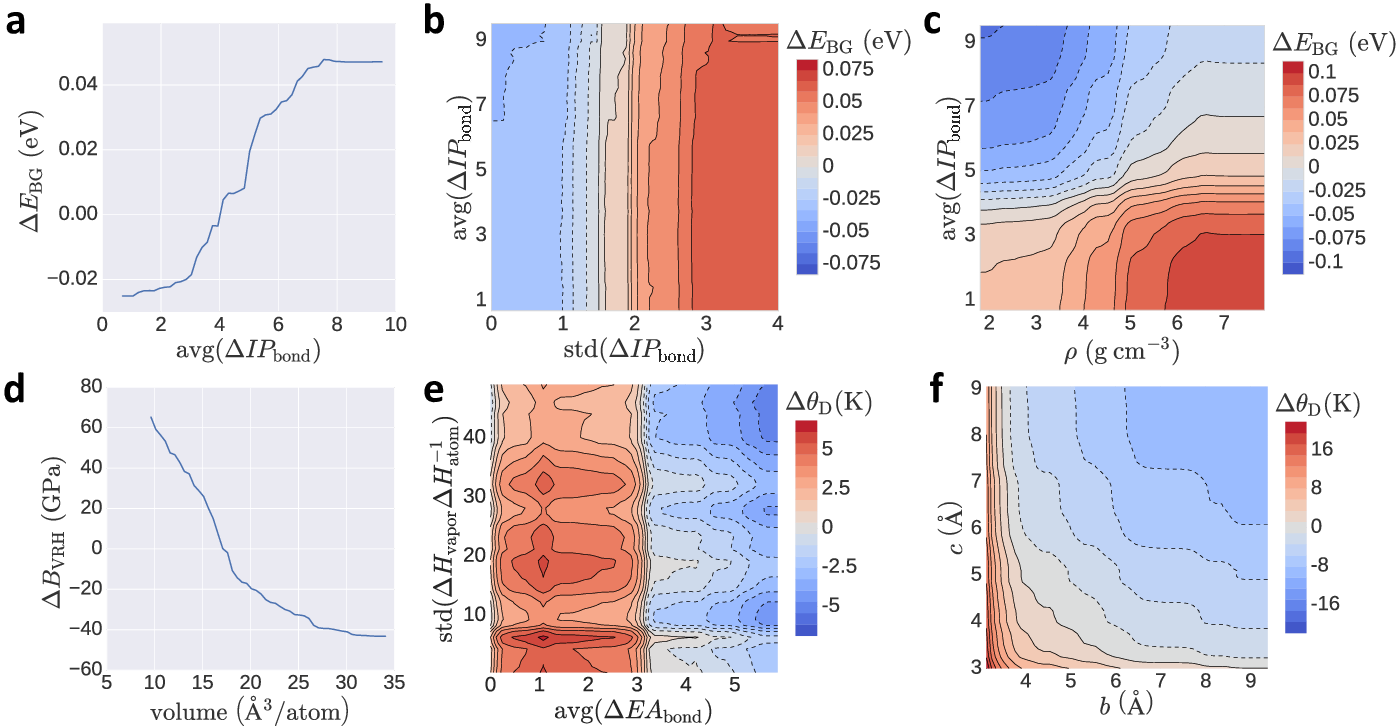}
\mycaption[Partial dependence plots of the $E_{\scriptstyle \mathrm{BG}}$, $B_{\scriptstyle \mathrm{VRH}}$, and
$\theta_{\scriptstyle \mathrm{D}}$ models.]
{\textbf{(a)} Partial dependence of $E_{\scriptstyle \mathrm{BG}}$ on the $\avg\left(\Delta IP_{\mathrm{bond}}\right)$
descriptor.
For $E_{\scriptstyle \mathrm{BG}}$,
the 2D interaction between $\std\left(\Delta IP_{\mathrm{bond}}\right)$ and $\avg\left(\Delta IP_{\mathrm{bond}}\right)$
and between $\rho$ (density) and $\avg\left(\Delta IP_{\mathrm{bond}}\right)$ are illustrated in panels
\textbf{(b)} and \textbf{(c)}, respectively.
\textbf{(d)} Partial dependence of the $B_{\scriptstyle \mathrm{VRH}}$ on the crystal volume per atom descriptor.
For $\theta_{\scriptstyle \mathrm{D}}$,
the 2D interaction between
$\avg\left(\Delta EA_{\scriptstyle \mathrm{bond}}\right)$ and
$\std\left(\Delta H_{\scriptstyle \mathrm{vapor}} \Delta H_{\scriptstyle \mathrm{atom}}^{-1}\right)$
and between
crystal lattice parameters $b$ and $c$ are illustrated
in panels \textbf{(e)} and \textbf{(f)}, respectively.}
\label{fig:art124:figure5}
\efig

\boldsection{Model interpretation.}
Model interpretation is of paramount importance in any \ML\ study.
The significance of each descriptor is determined in order to gain insight into
structural features that impact molecular properties of interest.
Interpretability is a strong advantage of decision tree methods, particularly with the \GBDT\ approach.
One can quantify the predictive power of a specific descriptor by analyzing the reduction
of the \RMSE\ at each node of the tree.

Partial dependence plots offer yet another opportunity for \GBDT\ model interpretation.
Similar to the descriptor significance analysis, partial dependence resolves the
effect of a variable (descriptor) on a property, but only after marginalizing over all other
explanatory variables~\cite{Hastie_StatLearn_2001}.
The effect is quantified by the change of that property as relevant descriptors are varied.
The plots themselves highlight the most important interactions among relevant descriptors
as well as between properties and their corresponding descriptors.
While only the most important descriptors are highlighted and discussed,
an exhaustive list of relevant descriptors and their relative contributions
can be found in
Supplementary Note 1 of Reference~\cite{curtarolo:art124}.

For the metal/insulator classification model, the descriptor significance analysis
shows that two descriptors have the highest importance (equally), namely
$\avg\left(\Delta H_\sfusion\lambda^{-1}\right)$ and
$\avg\left(V_\smolar r_\scov^{-1}\right)$.
$\avg\left(\Delta H_\sfusion\lambda^{-1}\right)$ is the ratio between the
fusion enthalpy $\left(\Delta H_\sfusion\right)$
and the thermal conductivity $\left(\lambda\right)$ averaged over all atoms in the material, and
$\avg\left(V_\smolar r_\scov^{-1}\right)$ is the ratio between the
molar volume $\left(V_\smolar\right)$
and the covalent radius $\left(r_\scov\right)$ averaged over all atoms in the material.
Both descriptors are simple node-specific features.
The presence of these two prominent descriptors accounts for the high accuracy of the classification model.

Figure~\ref{fig:art124:figure4} shows the projection of the full dataset onto the dual-descriptor space of
$\avg\left(\Delta H_\sfusion\lambda^{-1}\right)$ and $\avg\left(V_\smolar r_\scov^{-1}\right)$.
In this 2D space, metals and insulators are substantially partitioned.
To further resolve this separation, the plot is split into four quadrants
(see dashed lines) with an origin approximately at
$\avg\left(V_\smolar r_\scov^{-1}\right)=11$,~$\avg\left(\Delta H_\sfusion\lambda^{-1}\right)=2$.
Insulators are predominately located in quadrant I.
There are several clusters (one large and several small) parallel to the $x$-axis.
Metals occupy a compact square block in quadrant III within intervals
$5<\avg\left(V_\smolar r_\scov^{-1}\right)<12$ and $0.02<\avg\left(\Delta H_\sfusion\lambda^{-1}\right)<2$.
Quadrant II is mostly empty with a few materials scattered about the origin.
In the remaining quadrant (IV), materials have mixed character.

Analysis of the projection shown in Figure~\ref{fig:art124:figure4} suggests a simple heuristic rule:
all materials within quadrant I are classified as insulators $\left(E_\sBG\!>\!0\right)$,
and all materials outside of this quadrant are metals.
Remarkably, this unsupervised projection approach achieves a very high
classification accuracy of 86\% for the entire dataset of \PLMFelectronicTotal\ materials.
The model misclassifies only 3,621 materials:
2,414 are incorrectly predicted as insulators and 1,207 are incorrectly predicted as metals.
This example illustrates how careful model analysis of the most significant descriptors
can yield simple heuristic rules for materials design.

The regression model for the band gap energy $\left(E_\sBG\right)$ is more complex.
There are a number of descriptors in the model with comparable contributions,
and thus, all individual contributions are small.
This is expected as a number of conditions can affect $E_\sBG$.
The most important are $\avg\left(\chi Z_{\mathrm{eff}}^{-1}\right)$ and $\avg\left(C \lambda^{-1}\right)$ with
significance scores of 0.075 and 0.071,
respectively, where $\chi$ is the electronegativity, $Z_{\mathrm{eff}}$ is the effective nuclear charge,
$C$ is the specific heat capacity, and $\lambda$ is the thermal conductivity of each atom.

Figure~\ref{fig:art124:figure5} shows partial dependence plots focusing on $\avg\left(\Delta IP_{\sbond}\right)$ as an example.
It is derived from edge fragments of bonded atoms $\left(l=1\right)$ and defined as an absolute difference in
ionization potentials averaged over the material.
In other words, it is a measure of bond polarity, similar to electronegativity.
Figure~\ref{fig:art124:figure5}(a) shows a steady monotonic increase in $\Delta E_\sBG$ for larger
values of $\avg\left(\Delta IP_{\sbond}\right)$.
The effect is small, but captures an expected physical principle:
polar inorganic materials (\eg, oxides, fluorides) tend to have larger $E_\sBG$.

\fig
\includegraphics[width=\linewidth]{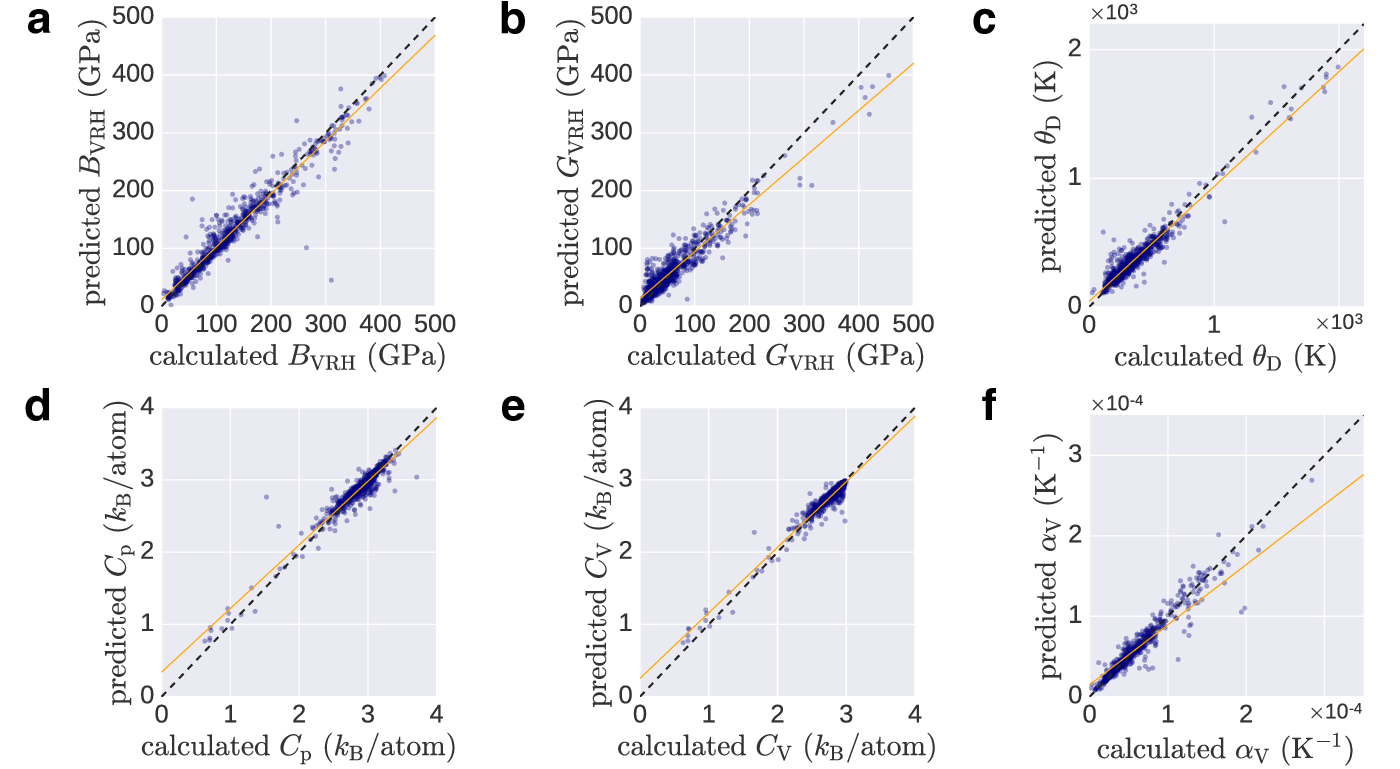}
\mycaption[Model performance evaluation for the six \ML\ models predicting thermomechanical properties
of \PLMFthermoTestTotal\ newly characterized materials.]
{Predicted \vs\ calculated values for the regression \ML\ models:
\textbf{(a)} bulk modulus $\left(B_{\scriptstyle \mathrm{VRH}}\right)$,
\textbf{(b)} shear modulus $\left(G_{\scriptstyle \mathrm{VRH}}\right)$,
\textbf{(c)} Debye temperature $\left(\theta_{\scriptstyle \mathrm{D}}\right)$,
\textbf{(d)} heat capacity at constant pressure $\left(C_{\scriptstyle \mathrm{P}}\right)$,
\textbf{(e)} heat capacity at constant volume $\left(C_{\scriptstyle \mathrm{V}}\right)$, and
\textbf{(f)} thermal expansion coefficient $\left(\alpha_{\scriptstyle \mathrm{V}}\right)$.}
\label{fig:art124:figure6}
\efig

Given the number of significant interactions involved with this phenomenon,
tailoring $E_\sBG$ involves the
optimization of a highly non-convex, multidimensional object.
Figure~\ref{fig:art124:figure5}(b) illustrates a 2D slice of this object as
$\std\left(\Delta IP_{\sbond}\right)$ and $\avg\left(\Delta IP_{\sbond}\right)$ vary simultaneously.
Like $\avg\left(\Delta IP_{\sbond}\right)$,
$\std\left(\Delta IP_{\sbond}\right)$ is the standard deviation of the set of absolute differences in $IP$ among
all bonded atoms.
In the context of these two variables, $E_\sBG$ responds to deviations in $\Delta IP_{\sbond}$
among the set of bonded atoms, but remains constant across shifts in $\avg\left(\Delta IP_{\sbond}\right)$.
This suggests an opportunity to tune $E_\sBG$ by considering another composition that varies the deviations among bond polarities.
Alternatively, a desired $E_\sBG$ can be maintained
by considering another composition that preserves the deviations among bond polarities, even as the overall average
shifts.
Similarly, Figure~\ref{fig:art124:figure5}(c) shows the partial dependence on both
the density $\left(\rho\right)$ and $\avg\left(\Delta IP_{\sbond}\right)$.
Contrary to the previous trend, larger $\avg\left(\Delta IP_{\sbond}\right)$
values correlate with smaller $E_\sBG$, particularly for low density structures.
Materials with higher density and lower $\avg\left(\Delta IP_{\sbond}\right)$ tend to have higher $E_\sBG$.
Considering the elevated response (compared to Figure~\ref{fig:art124:figure5}(b)), the inverse correlation of $E_\sBG$ with the average
bond polarity in the context of density suggests an even more effective means of tuning $E_\sBG$.

A descriptor analysis of the thermomechanical property models reveals the importance of
one descriptor in particular, the volume per atom of the crystal.
This conclusion certainly resonates with the nature of these properties, as they generally correlate
with bond strength~\cite{curtarolo:art115}.
Figure~\ref{fig:art124:figure5}(d) exemplifies such a relationship, which shows
the partial dependence plot of the bulk modulus $\left(B_\sVRH\right)$ on the volume per atom.
Tightly bound atoms are generally indicative of stronger bonds.
As the interatomic distance increases, properties like $B_\sVRH$ generally reduce.

\tab
\mycaption[Statistical summary of the new predictions for the six thermomechanical regression models.]
{The summary corresponds with Figure~\ref{fig:art124:figure6}.}
\tabvspace
\begin{tabular}{l|r|r|r}
property & \RMSE\ & \MAE\ & $r^{2}$ \\
\hline
$B_\sVRH$                           & 21.13~GPa                               & 12.00~GPa                               & 0.93 \\
$G_\sVRH$                           & 18.94~GPa                               & 13.31~GPa                               & 0.90 \\
$\theta_\sD$                        & 64.04~K                                 & 42.92~K                                 & 0.93 \\
$C_\sp$                             & 0.10~$k_\sB$/atom                       & 0.06~$k_\sB$/atom                       & 0.92 \\
$C_\sV$                             & 0.07~$k_\sB$/atom                       & 0.05~$k_\sB$/atom                       & 0.95 \\
$\alpha_\sV$                        & $1.95 \times 10^{-5}$~K$^{-1}$          & $5.77 \times 10^{-6}$~K$^{-1}$          & 0.76 \\
\end{tabular}
\label{tab:art124:table2}
\etab

Two of the more interesting dependence plots are also shown in Figure~\ref{fig:art124:figure5}(e-f),
both of which offer opportunities for tuning the Debye temperature ($\theta_\sD$).
Figure~\ref{fig:art124:figure5}(e) illustrates the interactions among two descriptors,
the absolute difference in electron affinities among bonded atoms
averaged over the material
$\left(\avg\left(\Delta EA_\sbond\right)\right)$, and
the standard deviation of the set of ratios of the enthalpies of vaporization $\left(\Delta H_\svapor\right)$
and atomization $\left(\Delta H_\satom\right)$ for all atoms in the material
$\left(\std\left(\Delta H_\svapor \Delta H_\satom^{-1}\right)\right)$.
Within these dimensions, two distinct regions emerge of increasing/decreasing $\theta_\sD$ separated by a
sharp division
at about $\avg\left(\Delta EA_\satom\right) = 3$.
Within these partitions, there are clusters of maximum gradient in $\theta_\sD$---peaks within the left
partition and troughs within the right.
The peaks and troughs alternate with varying $\std\left(\Delta H_\svapor \Delta H_\satom^{-1}\right)$.
Although $\std\left(\Delta H_\svapor \Delta H_\satom^{-1}\right)$
is not an immediately intuitive descriptor, the alternating clusters may be a manifestation
of the periodic nature of $\Delta H_\svapor$ and $\Delta H_\satom$~\cite{webelements_periodicity}.
As for the partitions themselves,
the extremes of $\avg\left(\Delta EA_\satom\right)$ characterize covalent and ionic materials, as
bonded atoms with similar $EA$ are likely to share electrons, while those
with varying $EA$ prefer to donate/accept electrons.
Considering that $EA$ is also periodic, various opportunities for carefully tuning $\theta_\sD$
should be available.

Finally, Figure~\ref{fig:art124:figure5}(f) shows the partial dependence of $\theta_\sD$ on the lattice parameters $b$ and $c$.
It resolves two notable correlations:
\textit{(i)} uniformly increasing the cell size of the system decreases $\theta_\sD$, but
\textit{(ii)} elongating the cell ($c/b \gg 1$) increases it.
Again, \textit{(i)} can be attributed to the
inverse relationship between volume per atom and bond strength,
but does little to address \textit{(ii)}.
Nevertheless, the connection between elongated, or layered, systems and the Debye temperature is certainly not
surprising---anisotropy can be leveraged to enhance phonon-related interactions associated with
thermal conductivity~\cite{Minnich_PRB_2015}
and superconductivity~\cite{Shimahara_PRB_2002,Jha_PT_1989,Klein_SSC_1980}.
While the domain of interest is quite narrow,
the impact is substantial, particularly in comparison to that shown in Figure~\ref{fig:art124:figure5}(e).

\boldsection{Model validation.}
While the expected performances of the \ML\ models can be projected through five-fold cross validation,
there is no substitute for validation against an independent dataset.
The \ML\ models for the thermomechanical properties are leveraged to make predictions
for materials previously uncharacterized, and subsequently validated
these predictions via the \AEL-\AGL\ integrated framework~\cite{curtarolo:art96, curtarolo:art115}.
Figure~\ref{fig:art124:figure6} illustrates the models' performance on the set of \PLMFthermoTestTotal\ additional materials,
with relevant statistics displayed in Table~\ref{tab:art124:table2}.
For further analysis, all predicted and calculated results are available in
Supplementary Note 3 of Reference~\cite{curtarolo:art124}.

Comparing with the results of the generalizability analysis shown in Figure~\ref{fig:art124:figure3} and Table~\ref{tab:art124:table1},
the overall errors are consistent with five-fold cross validation.
Five out of six models have $r^2$ of 0.9 or higher.
However, the $r^2$ value for the thermal expansion coefficient
$\left(\alpha_\sV\right)$ is lower than forecasted.
The presence of scattering suggests the need for a larger training set---as new,
much more diverse materials were likely introduced in the test set.
This is not surprising considering the number of variables that can affect thermal expansion~\cite{Figge_APL_2009}.
Otherwise, the accuracy of these predictions confirm the effectiveness of the \PLMF\ representation,
which is particularly compelling considering:
\textit{(i)} the limited diversity training dataset (only about 11\% as large as that available for
predicting the electronic properties), and
\textit{(ii)} the relative size of the test set (over a quarter the size of the training set).

In the case of the bulk modulus $\left(B_\sVRH\right)$, 665 systems (86\% of test set) are predicted within 25\%
of calculated values.
Only the predictions of four materials, Bi (\ICSD\ \#51674), PrN (\ICSD\ \#168643),
Mg$_{3}$Sm (\ICSD\ \#104868), and ZrN (\ICSD\ \#161885), deviate beyond 100~GPa from calculated values.
Bi is a high-pressure phase (Bi-III) with a caged, zeolite-like structure~\cite{McMahon_BiIII_2001}.
The structures of zirconium nitride (wurtzite phase) and praseodymium nitride (B3 phase) were hypothesized and
investigated via \DFT\ calculations~\cite{EscorciaSalas_MJ_2008,Kocak_PSCB_2010} and have yet to be observed
experimentally.

For the shear modulus $\left(G_\sVRH\right)$, 482 materials (63\% of the test set) are predicted within 25\%
of calculated values.
Just one system, C$_{3}$N$_{4}$ (\ICSD\ \#151781), deviates beyond 100~GPa from its calculated value.
The Debye temperature $\left(\theta_\sD\right)$ is predicted to within 50 K accuracy for 540 systems (70\% of the test set).
BeF$_{2}$ (\ICSD\ \#173557), yet another cage (sodalite) structure~\cite{Zwijnenburg_JACS_2008}, has among the largest errors
in three models including $\theta_\sD$ (SE = -423 K) and both heat capacities
($C_\sp$: SE = 0.65 $k_\sB$/atom; $C_\sV$: SE = 0.61 $k_\sB$/atom).
Similar to other ill-predicted structures, this polymorph is theoretical, and has yet to be synthesized.

\fig
\includegraphics[width=\linewidth]{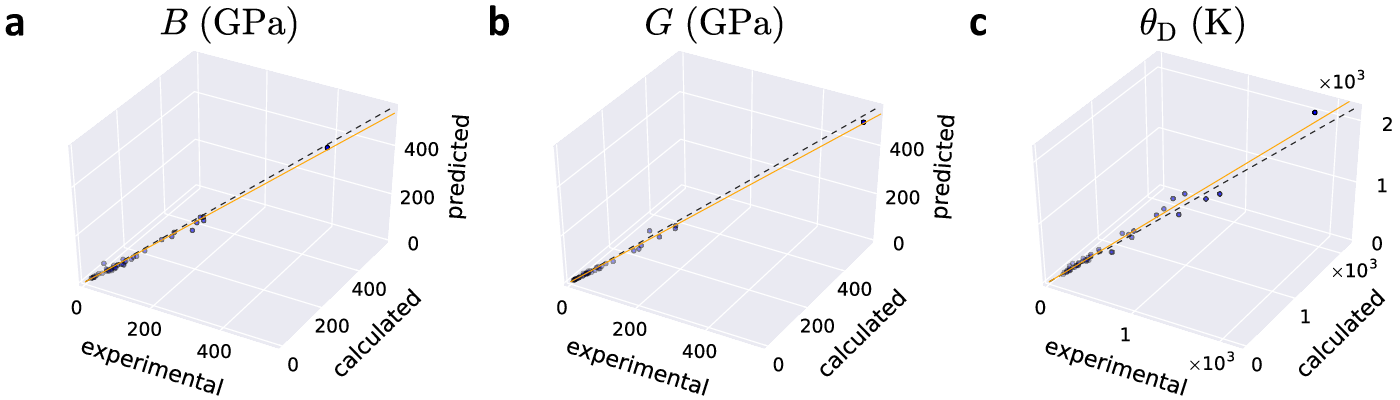}
\mycaption[Comparison of the \AEL-\AGL\ calculations and \ML\ predictions with experimental values for three thermomechanical properties.]
{\textbf{(a)} bulk modulus $\left(B\right)$,
\textbf{(b)} shear modulus $\left(G\right)$,
and
\textbf{(c)} Debye temperature $\left(\theta_{\scriptstyle \mathrm{D}}\right)$.
}
\label{fig:art124:figure7}
\efig

\tab
\mycaption[Statistical summary of the \AEL-\AGL\ calculations and
\ML\ predictions \vs\ experimental values for three thermomechanical properties.]
{The summary corresponds with Figure~\ref{fig:art124:figure7}.}
\tabvspace
\resizebox{\linewidth}{!}{
\begin{tabular}{l|r|r|r|r|r|r}
\multirow{2}{*}{property} & \multicolumn{2}{c|}{\RMSE} & \multicolumn{2}{c|}{\MAE} & \multicolumn{2}{c}{$r^{2}$} \\
\cline{2-7}
 & exp. \vs\ calc. & exp. \vs\ pred. & exp. \vs\ calc. & exp. \vs\ pred. & exp. \vs\ calc. & exp. \vs\ pred. \\
\hline
$B$               & 8.90~GPa        & 10.77~GPa   & 6.36~GPa    & 8.12~GPa    & 0.99  & 0.99 \\
$G$               & 7.29~GPa        & 9.15~GPa    & 4.76~GPa    & 6.09~GPa    & 0.99  & 0.99 \\
$\theta_\sD$      & 76.13~K         & 65.38~K     & 49.63~K     & 42.92~K     & 0.97  & 0.97 \\
\end{tabular}}
\label{tab:art124:table3}
\etab

\boldsection{Comparison with experiments.}
A comparison between calculated, predicted, and experimental results is presented in
Figure~\ref{fig:art124:figure7}, with relevant statistics summarized in Table~\ref{tab:art124:table3}.
Data is considered for the bulk modulus $B$, shear modulus $G$, and (acoustic) Debye temperature $\theta_\sa$
for 45 well-characterized materials with
diamond (SG\# 227, \AFLOW\ prototype \texttt{A\_cF8\_227\_a}),
zincblende (SG\# 216, \texttt{AB\_cF8\_216\_c\_a}),
rocksalt (SG\# 225, \texttt{AB\_cF8\_225\_a\_b}),
and wurtzite (SG\# 186, \texttt{AB\_hP4\_186\_b\_b})
structures~\cite{Morelli_Slack_2006,Semiconductors_BasicData_Springer}.
Experimental $B$ and $G$ are compared to the $B_\sVRH$ and $G_\sVRH$ values predicted here, and
$\theta_\sa$ is converted to the traditional Debye temperature $\theta_\sD=\theta_\sa n^{1/3}$,
where $n$ is the number of atoms in the unit cell.
All relevant values are listed in
Supplementary Note 4 of Reference~\cite{curtarolo:art124}.

Excellent agreement is found between experimental and calculated values,
but more importantly, between experimental and predicted results.
With error metrics close to or under expected tolerances from the generalizability analysis,
the comparison highlights effective experimental confidence in the approach.
The experiments/prediction validation is clearly the ultimate objective of the research presented here.

\subsection{Discussion}
Traditional trial-and-error approaches have proven ineffective in discovering practical materials.
Computational models developed with \ML\ techniques may provide
a truly rational approach to materials design.
Typical high-throughput \DFT\ screenings involve exhaustive
calculations of all materials in the database, often without
consideration of previously calculated results.
Even at high-throughput rates, an average \DFT\ calculation of a medium
size structure (about 50 atoms per unit cell) takes about 1,170 CPU-hours of
calculations or about 37 hours on a 32-CPU cores node.
However, in many cases, the desired range of values for the target property is known.
For instance,
the optimal band gap energy and thermal conductivity for optoelectronic applications
will depend on the power and voltage conditions of the device~\cite{Figge_APL_2009,Zhou_JACerS_2016}.
Such cases offer an opportunity to leverage previous results and savvy \ML\ models,
such as those developed in this work, for rapid pre-screening of potential materials.
Researchers can quickly narrow the list of candidate materials and avoid many extraneous
\DFT\ calculations---saving money, time, and computational resources.
This approach takes full advantage of previously calculated results,
continuously accelerating materials discovery.
With prediction rates of about 0.1 seconds per material, the same 32-CPU cores node can screen
over 28 million material candidates per day with this framework.

Furthermore, interaction diagrams as depicted in Figure~\ref{fig:art124:figure5} offer a pathway to design
materials that meet certain constraints and requirements.
For example, substantial differences in thermal expansion coefficients among the materials used
in high-power, high-frequency optoelectronic applications leads to bending and cracking of the structure
during the growth process~\cite{Figge_APL_2009,Zhou_JACerS_2016}.
Not only would this work-flow facilitate the search for semiconductors with large band gap energies,
high Debye temperatures (thermal conductivity),
but also materials with similar thermal expansion coefficients.

While the models themselves demonstrate excellent predictive power with minor deviations, outlier analysis reveals
theoretical structures to be among the worst offenders.
This is not surprising, as the true stability conditions (\eg, high-pressure/high-temperature) have yet
to be determined, if they exist at all.
The \ICSD\ estimates that structures for over 7,000 materials (or roughly 4\%) come
from calculations rather than actual experiment.
Such discoveries exemplify yet another application for \ML\ modeling, rapid/robust curation of large datasets.

To improve large-scale high-throughput computational screening for the identification
of materials with desired properties, fast and accurate data mining approaches
should be incorporated into the standard work-flow.
In this work, we developed a universal \QMSPR\ framework for predicting electronic
properties of inorganic materials.
Its effectiveness is validated through the prediction of eight key materials properties
for stoichiometric inorganic crystalline materials, including
the metal/insulator classification,
band gap energy, bulk and shear moduli, Debye temperature, heat capacity (at constant
pressure and volume), and thermal expansion coefficient.
Its applicability extends to all 230 space groups and the vast majority of
elements in the periodic table.
All models are freely available at \href{http://aflow.org/aflow-ml}{aflow.org/aflow-ml}.

\subsection{Methods}
\boldsection{Data preparation.}
Two independent datasets were prepared for the creation and validation of the \ML\ models.
The training set includes
electronic~\cite{aflowlibPAPER,aflowPAPER,aflowBZ,curtarolo:art67,monsterPGM,curtarolo:art49}
and thermomechanical properties~\cite{curtarolo:art96, curtarolo:art115} for a broad diversity of
compounds already characterized in the \AFLOW\ database.
This set is used to build and analyze the \ML\ models, one model per property.
The constructed thermomechanical models are then employed to make predictions of previously uncharacterized compounds in the \AFLOW\ database.
Based on these predictions and consideration of computational cost, several compounds are selected to validate the models' predictive
power.
These compounds and their newly computed properties define the test set.
The compounds used in both datasets are specified in
Supplementary Notes 2 and 3 of Reference~\cite{curtarolo:art124}, respectively.

\boldsection{Training set.}
{\bf I.}
Band gap energy data for 49,934 materials were extracted from the \AFLOW\
repository~\cite{aflowlibPAPER,aflowPAPER,aflowBZ,curtarolo:art67,monsterPGM,curtarolo:art49}, representing approximately
60\% of the known stoichiometric inorganic crystalline materials listed in the
Inorganic Crystal Structure Database (\ICSD)~\cite{ICSD,ICSD3}.
While these band gap energies are generally underestimated with respect to experimental
values~\cite{Perdew_IJQC_1985}, \DFT+$U$ is robust enough to
differentiate between metallic (no $E_\sBG$) and insulating $\left(E_\sBG\!>\!0\right)$ systems~\cite{curtarolo:art104}.
Additionally, errors in band gap energy prediction are typically systematic.
Therefore, the band gap energy values can be corrected \textit{ad-hoc} with fitting
schemes~\cite{Yazyev_PRB_2012,Zheng_PRL_2011}.
Prior to model development, both \ICSD\ and \AFLOW\ data were curated:
duplicate entries, erroneous structures, and ill-converged calculations were corrected or removed.
Noble gases crystals are not considered.
The final dataset consists of \PLMFelectronicTotal\ unique materials (\PLMFmetalTotal\ with no $E_\sBG$
and \PLMFinsulatorTotal\ with $E_\sBG\!>\!0$),
covering the seven lattice systems, 230 space groups, and 83 elements
(H-Pu, excluding noble gases, Fr, Ra, Np, At, and Po).
All referenced \DFT\ calculations were performed with the Generalized Gradient Approximation
(\GGA) \PBE~\cite{PBE}
exchange-correlation functional and projector-augmented wavefunction (\PAW)
potentials~\cite{PAW,kresse_vasp_paw} according to the
\AFLOW\ Standard for High-Throughput (HT) Computing~\cite{curtarolo:art104}.
The Standard ensures reproducibility of the data, and provides visibility/reasoning for any parameters
set in the calculation, such as accuracy thresholds, calculation
pathways, and mesh dimensions.
{\bf II.}
Thermomechanical properties data for just over 3,000 materials were extracted from the \AFLOW\
repository~\cite{curtarolo:art115}.
These properties include the bulk modulus, shear modulus, Debye temperature, heat capacity at constant pressure,
heat capacity at constant volume, and thermal expansion coefficient, and were
calculated using the \AEL-\AGL\ integrated framework~\cite{curtarolo:art96, curtarolo:art115}.
The \AEL\ (\AFLOW\ Elasticity Library)
method~\cite{curtarolo:art115} applies a set of independent normal and shear strains to the structure, and then fits the calculated stress
tensors to obtain the elastic constants~\cite{curtarolo:art100}.
These can then be used to calculate the elastic moduli in
the Voigt and Reuss approximations, as well as the Voigt-Reuss-Hill (\VRH) averages which are the values of the bulk and
shear moduli modeled in this work.
The \AGL\ (\AFLOW\ \GIBBS\ Library) method~\cite{curtarolo:art96}
fits the energies from a set of isotropically
compressed and expanded volumes of a structure to a quasiharmonic Debye-Gr{\"u}neisen model~\cite{Blanco_CPC_GIBBS_2004}
to obtain thermomechanical
properties, including the bulk modulus, Debye temperature, heat capacity, and thermal expansion coefficient.
\AGL\ has been
combined with \AEL\ in a single workflow, so that it can utilize the Poisson ratios obtained from \AEL\ to improve the
accuracy of the thermal properties predictions~\cite{curtarolo:art115}.
After a similar curation of ill-converged calculations, the final dataset consists of
\PLMFthermoTrainingTotal\ materials.
It covers the seven lattice systems, includes unary, binary, and ternary compounds, and
spans broad ranges of each thermomechanical property, including
high thermal conductivity systems such as C (\ICSD\ \#182729), BN (\ICSD\ \#162874), BC$_{5}$ (\ICSD\ \#166554),
CN$_{2}$ (\ICSD\ \#247678), MnB$_{2}$ (\ICSD\ \#187733), and SiC (\ICSD\ \#164973), as well as
low thermal conductivity systems such as Hg$_{33}$(Rb,K)$_{3}$ (\ICSD\ \#410567 and \#410566),
Cs$_{6}$Hg$_{40}$ (\ICSD\ \#240038), Ca$_{16}$Hg$_{36}$ (\ICSD\ \#107690), CrTe (\ICSD\ \#181056),
and Cs (\ICSD\ \#426937).
Many of these systems additionally exhibit extreme values of the bulk and shear moduli,
such as C (high bulk and shear moduli) and Cs (low bulk and shear moduli).
Interesting systems such as
RuC (\ICSD\ \#183169) and NbC (\ICSD\ \#189090)
with a high bulk modulus ($B_\sVRH$ = 317.92 GPa, 263.75 GPa) but
low shear modulus ($G_\sVRH$ = 16.11 GPa, 31.86 GPa)
also populate the set.

\boldsection{Test set.}
While nearly all \ICSD\ compounds are characterized electronically within the \AFLOW\ database,
most have not been characterized thermomechanically due to the added computational cost.
This presented an opportunity to validate the \ML\ models.
Of the remaining compounds, several were prioritized for immediate characterization via
the \AEL-\AGL\ integrated framework~\cite{curtarolo:art96, curtarolo:art115}.
In particular, focus was placed on systems predicted to have a large bulk modulus, as this property
is expected to scale well with the other aforementioned thermomechanical
properties~\cite{curtarolo:art96, curtarolo:art115}.
The set also includes various other small cell, high symmetry systems expected to span the full
applicability domains of the models.
This effort resulted in the characterization of \PLMFthermoTestTotal\ additional compounds.

\boldsection{Data availability.}
All the \abinitio\ data are freely available to the public as
part of the \AFLOW\ online repository and can be accessed through \AFLOWorg\
following the \RESTAPI\ interface~\cite{aflowPAPER}.
\clearpage
\chapter{Applications}
\section{Materials Cartography: Representing and Mining Materials Space Using Structural and Electronic Fingerprints}
\label{sec:art094}

This study follows from a collaborative effort described in Reference~\cite{curtarolo:art94},
which was awarded with ACS Editors' Choice.
Author contributions are as follows:
Stefano Curtarolo and Alexander Tropsha designed the study.
Olexandr Isayev and Denis Fourches developed the fingerprinting and cartography methods.
Eugene N. Muratov adapted the SiRMS method for materials.
Corey Oses and Kevin M. Rasch prepared the data and worked with the \AFLOWorg\ database.
All authors discussed the results and their implications and contributed to the paper.

\subsection{Introduction}
Designing materials with desired physical and chemical properties is recognized as an
outstanding challenge in materials research~\cite{Rajan_materialstoday_2005,nmatHT,Potyrailo_ACSCombSci_2011}.
Material properties directly depend on a large number of key variables, often making the property prediction complex.
These variables include constitutive elements, crystal forms, and geometrical and electronic characteristics; among others.
The rapid growth of materials research has led to the accumulation of vast amounts of data.
For example, the Inorganic Crystal Structure Database (\ICSD) includes more than 170,000 entries~\cite{ICSD}.
Experimental data are also included in other databases, such as MatWeb~\cite{MatWeb} and MatBase~\cite{Matbase}.
In addition, there are several large databases such as the \AFLOWorg\ repository~\cite{aflowBZ,aflowSCINT},
the Materials Project~\cite{APL_Mater_Jain2013},
and the Harvard Clean Energy Project~\cite{Hachmann_JPCL_2011,Hachmann_EES_2014}
that contain thousands of unique materials and their theoretically calculated properties.
These properties include electronic structure profiles estimated with quantum mechanical methods.
The latter databases have great potential to serve as a source of novel functional materials.
Promising candidates from these databases may in turn be selected for experimental
confirmation using rational design approaches~\cite{MGI}.

The rapidly growing compendium of experimental and theoretical materials data offers
a unique opportunity for scientific discovery.
Specialized data mining and data visualization methods are being developed within
the nascent field of materials
informatics~\cite{Rajan_materialstoday_2005,Suh_MST_2009,Olivares-Amaya_EES_2011,Potyrailo_ACSCombSci_2011,nmatHT,Schuett_PRB_2014,Seko_PRB_2014}.
Similar approaches have been used extensively in cheminformatics with resounding success.
For example, in many cases, these approaches have served to help identify and design
small organic molecules with desired biological activity and acceptable
environmental/human-health safety profiles~\cite{Laggner_NCB_2012,Besnard_Nature_2012,Cherkasov_JMC_2013,Lusci_JCIM_2013}.
Application of cheminformatics approaches to materials science would allow researchers to
{\bf i.} define, visualize, and navigate through materials space,
{\bf ii.} analyze and model structural and electronic characteristics of materials
with regard to a particular physical or chemical property, and
{\bf iii.} employ predictive materials informatics models to forecast the experimental properties of
{\it de novo} designed or untested materials.
Such rational design approaches in materials science constitute a rapidly growing
field~\cite{Olivares-Amaya_EES_2011,Balachandran_PRSA_2011,Kong_JCIM_2012,Balachandran_ActaCristB_2012,Srinivasan_MAT_2013,Schuett_PRB_2014,Seko_PRB_2014,Broderick_APL_2014,Dey_CMS_2014}.

Herein, we introduce a novel materials fingerprinting approach.
We combine this with graph theory, similarity searches, and machine learning algorithms.
This enables the unique characterization, comparison, visualization, and design of materials.
We introduce the concept and describe the development of materials fingerprints that encode
materials' band structures, density of states (\DOS), crystallographic, and constitutional information.
We employ materials fingerprints to visualize this territory via advancing the new concept of ``{\it materials cartography}''.
We show this technology identifies clusters of materials with similar properties.
Finally, we develop Quantitative Materials Structure-Property Relationship (\QMSPR) models
that rely on these materials fingerprints.
We then employ these models to discover novel materials with desired properties that
lurk within the materials databases.

\subsection{Methods}
\label{subsec:art094:methods}

\subsubsection{\AFLOWorg\ repository and data}
The \AFLOWorg\ repository of density functional theory (\DFT) calculations is managed
by the software package \AFLOW~\cite{aflowPAPER,aflowlibPAPER}.
At the time of the study, the \AFLOWorg\ database included the results of calculations
characterizing over 20,000 crystals, but has since grown to include 50,000 entries ---
representing about a third of the contents of the \ICSD~\cite{ICSD}.
Of the characterized systems, roughly half are metallic and half are insulating.
\AFLOW\ leverages the \VASP\ Package~\cite{vasp_cms1996} to calculate the total energy
of a given crystal structure with \PAW\ pseudopotentials~\cite{PAW} and the \PBE~\cite{PBE} exchange-correlation functional.
The entries of the repositories have been described previously~\cite{aflowBZ,aflowlibPAPER,aflowAPI}.

\subsubsection{Data set of superconducting materials}
We have compiled experimental data for superconductivity critical temperatures,
$T_{\mathrm{c}}$, for more than 700 records from the Handbook of Superconductivity~\cite{Poole_Superconductivity_2000} and the
CRC Handbook of Chemistry and Physics~\cite{Lide_CRC_2004}, as well as the SuperCon Database~\cite{SuperCon}.
As we have shown recently~\cite{Fourches_JCIM_2010}, data curation is a necessary
step for any Quantitative Structure-Property Relationship (\QSAR) modeling.
In the compiled data set, several $T_{\mathrm{c}}$ values have been measured under strained conditions,
such as different pressures and magnetic fields.
We have only kept records taken under standard pressure and  with no external magnetic fields.
For materials with variations in reported $T_{\mathrm{c}}$ values in excess of 4~K,
original references were revisited and records have been discarded when no reliable information was available.
$T_{\mathrm{c}}$ values with a variation of less than 3~K have been averaged.
Of the remaining 465 materials ($T_{\mathrm{c}}$ range of 0.1-133~K), most records show
a variability in $T_{\mathrm{c}}$ of $\pm$1~K between different sources.
Such a level of variability would be extremely influential in materials with
low $T_{\mathrm{c}}$ ($T_{\mathrm{c}}\!<\!1$~K) because we have used the decimal
logarithm of the experimentally measured critical temperature ($\log(T_{\mathrm{c}})$) as our target property.

To appropriately capture information inherent to materials over the full range of
$T_{\mathrm{c}}$, we have constructed two data sets for the development of three models.
The {\bf continuous model} serves to predict $T_{\mathrm{c}}$ and utilizes
records excluding materials with $T_{\mathrm{c}}$ values less than 2~K.
This data set consists of 295 unique materials with a $\log(T_{\mathrm{c}})$ range of 0.30-2.12.
The {\bf classification model} serves to predict the position of $T_{\mathrm{c}}$
(above/below) with respect to the threshold $T_{\mathrm{thr}}$
(unbiasedly set to 20~K as observed in Figure~\ref{fig:art094:bands}(e), see the \nameref{subsec:art094:results} section).
It utilizes records incorporating the aforementioned excluded materials,
as well as lanthanum cuprate (La$_2$CuO$_4$, \ICSD\ \#19003).
Lanthanum cuprate had been previously discarded for high variability
($T_{\mathrm{c}}$ = 21-39~K), but now satisfies the classification criteria.
This data set consists of 464 materials (29 with $T_{\mathrm{c}}\!>\!T_{\mathrm{thr}}$
and 435 with $T_{\mathrm{c}}\!\leq\!T_{\mathrm{thr}}$).
Finally, the {\bf structural model} serves to identify geometrical components that most
influence $T_{\mathrm{c}}$. It utilizes the same data set as the continuous model.

\subsubsection{Materials fingerprints}
Following the central paradigms of structure-property relationships, we assume that
{\bf i.} properties of materials are a direct function of their structure and
{\bf ii.} materials with similar structures (as determined by constitutional,
topological, spatial, and electronic characteristics) are likely to have similar physical and chemical properties.

Thus, encoding material characteristics in the form of numerical arrays,
namely descriptors~\cite{nmatHT,Schuett_PRB_2014} or
``{\it fingerprints}''~\cite{Valle_ActaCristA_2010}, enables the use of classical cheminformatics and
machine-learning approaches to mine, visualize, and model any set of materials.
We have encoded the electronic structure diagram for each material as two distinct types of arrays
(Figure~\ref{fig:art094:fingerprints_construction}):
a {\it symmetry-dependent fingerprint} (band structure based ``B-fingerprint'') and a
{\it symmetry-independent fingerprint} (\DOS\ based ``D-fingerprint'').

\fig
\includegraphics[width=1.00\linewidth]{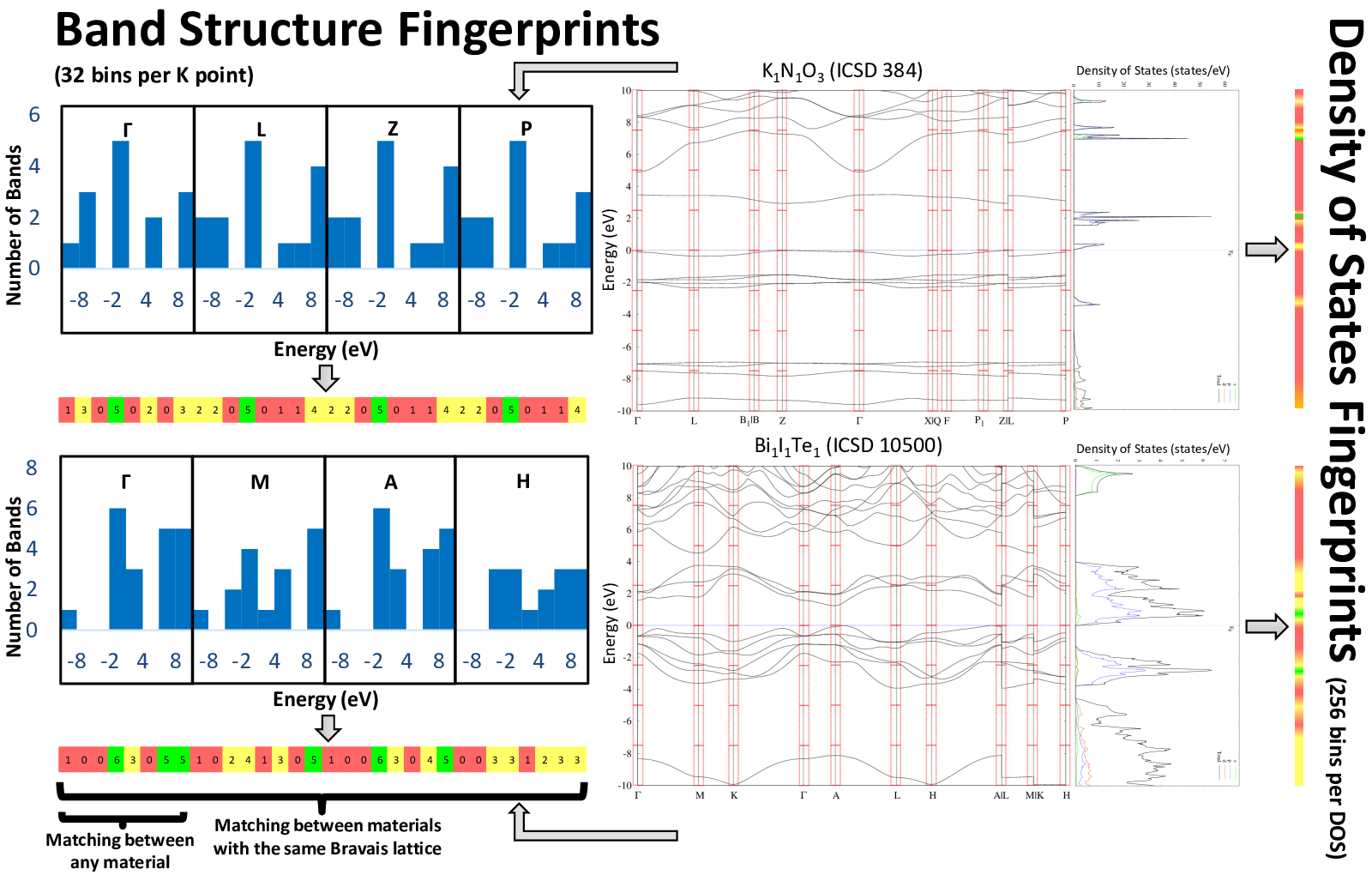}
\mycaption[Construction of materials fingerprints from the band structure and \DOS.]
{For simplicity, we illustrate the idea of B-fingerprints with only 8 bins.}
\label{fig:art094:fingerprints_construction}
\efig

\boldsection{B-fingerprint.} Along every special high-symmetry point of the Brillouin zone (\BZ),
the energy diagram has been discretized into 32 bins to serve as our fingerprint array.
Each \BZ\ has a unique set of high-symmetry points~\cite{aflowBZ}.
The comparison set of high-symmetry points belonging to a single \BZ\ type is considered symmetry-dependent.
To name a few examples, the Brillouin zone path of a cubic lattice
($\Gamma–X–M–\Gamma–R–X\!\!\mid\!\! M–R$) is encoded with just four points ($\Gamma, M, R, X$),
giving rise to a fingerprint array of length 128.
The body-centered orthorhombic lattice is more complex~\cite{aflowBZ,aflowSCINT}
($\Gamma–X–L–T–W–R–X_1–Z–\Gamma–Y–S–W\!\mid\!L_1–Y\!\mid\!Y_1–Z$)
and is represented by 13 points ($\Gamma, L, L_1, L_2, R, S, T, W, X, X_1, Y, Y_1, Z)$,
giving a fingerprint array of length 416.
Conversely, the comparison of identical {\bf k}-points not specifically belonging to any \BZ\
is always possible when only restricted to $\Gamma$.
Consequently, we limit our models to the $\Gamma$ point B-fingerprint in the present work.

\boldsection{D-fingerprint.} A similar approach can be taken for the \DOS\ diagrams,
which are sampled in 256 bins (from min to max) and the magnitude of each bin is discretized in 32 bits.
Therefore, the D-fingerprint is a total of 1024 bytes.
Owing to the complexity and limitations of the symmetry-dependent B-fingerprints,
we have only generated symmetry-independent D-fingerprints.
The length of these fingerprints is tunable depending on the objects, applications, and other factors.
We have carefully designed the domain space and length of these fingerprints to avoid
the issues of enhancing boundary effects or discarding important features.

\boldsection{SiRMS descriptors for materials.}
To characterize the structure of materials from several different perspectives,
we have developed descriptors similar to those used for small organic molecules
that can reflect their compositional, topological, and spatial (stereochemical) characteristics.
Classical cheminformatics tools can only handle small organic molecules.
Therefore, we have modified the Simplex (SiRMS) approach~\cite{Kuzmin_JCAMD_2008}
based on our experience with mixtures~\cite{Muratov_SC_2013,Muratov_MI_2012}
in order to make this method suitable for computing descriptors for materials.

The SiRMS approach~\cite{Kuzmin_JCAMD_2008} characterizes small organic molecules
by splitting them into multiple molecular fragments called simplexes.
Simplexes are tetratomic fragments of fixed composition (1D), topology (2D), and chirality and symmetry (3D).
The occurrences of each of these fragments in a given compound are then counted.
As a result, each molecule of a given data set can be characterized by its SiRMS fragment profiles.
These profiles take into account atom types, connectivity, \etc~\cite{Kuzmin_JCAMD_2008}.
Here, we have adapted the SiRMS approach to describe materials with their fragmental compositions.

Every material is represented according to the structure of its crystal unit cell (Figure~\ref{fig:art094:sirms_generation}).
Computing SiRMS descriptors for materials is equivalent to the computation of
SiRMS fragments for nonbonded molecular mixtures.
Bounded simplexes describe only a single component of the mixture.
Unbounded simplexes could either belong to a single component, or could span up to four components of the unit cell.
A special label is used during descriptor generation to distinguish ``mixture''
simplexes (belonging to different molecular moieties) from those incorporating elements from a single compound~\cite{Muratov_MI_2012}.

\fig
\includegraphics[width=1.00\linewidth]{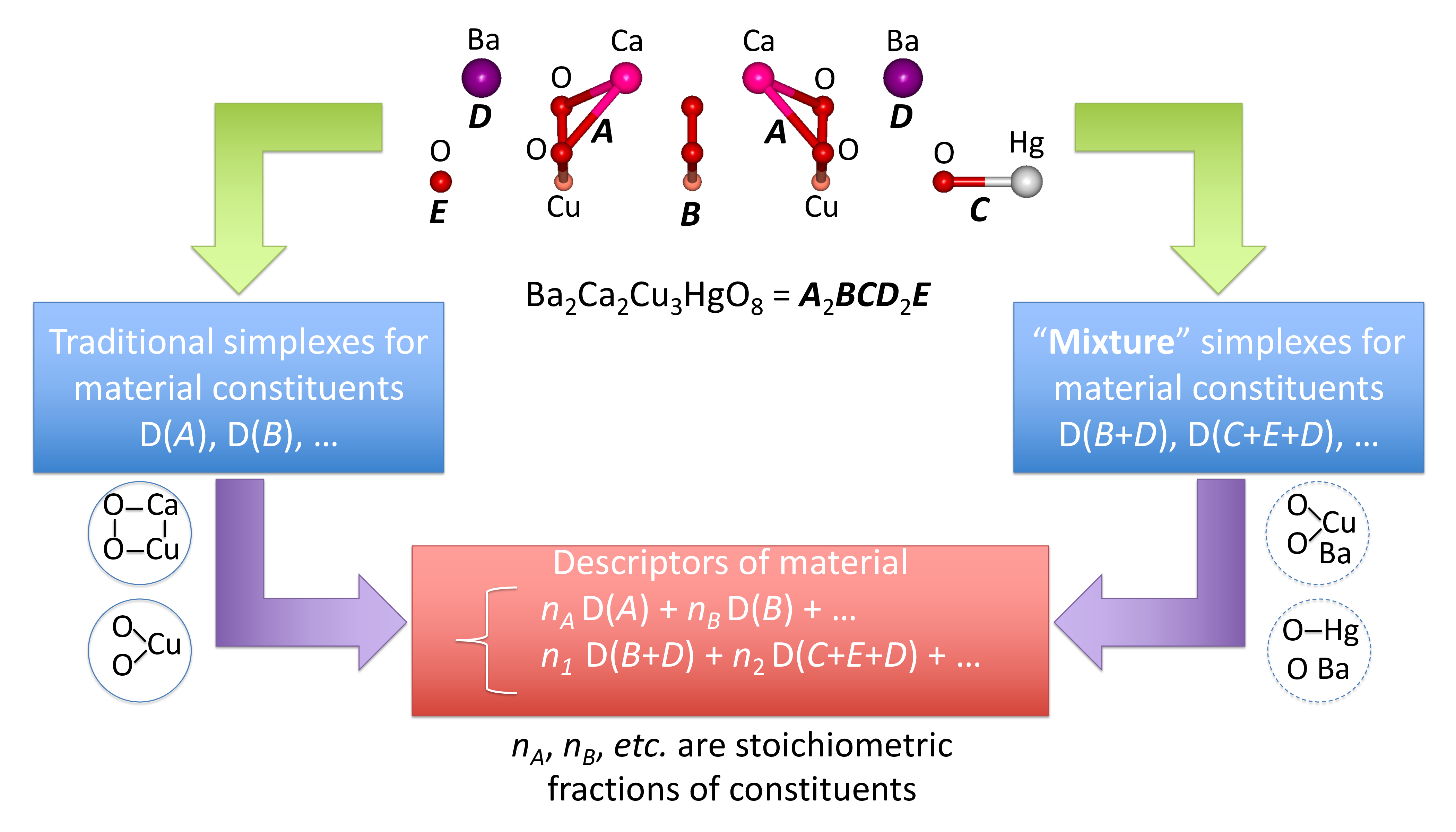}
\mycaption{Generation of SiRMS descriptors for materials.}
\label{fig:art094:sirms_generation}
\efig

Thus, the structure of every material is characterized by both bounded and unbounded
SiRMS descriptors as illustrated in Figure~\ref{fig:art094:sirms_generation}.
The descriptor value of a given simplex fragment is equal to the number of its occurrences in the system.
In the case of materials, this value has been summed throughout all the constituents of a system;
taking into account their stoichiometric ratios and crystal lattices (see Figure~\ref{fig:art094:sirms_generation}).
``Mixture'' descriptors are weighted according to the smallest stoichiometric
ratio of constituents within this mixture, and added throughout all the mixtures in a system.
Atoms in simplexes are differentiated according to their type (element) and partial charge.
For the latter, atoms are divided into six groups corresponding to their partial charge:
$A\!\leq\!-2\!<\!B\!\leq\!-1\!<\!C\!\leq\!0\!<\!D\!\leq\!1\!<\!E\!\leq\!2\!<\!F$.
In addition, we have developed a special differentiation of atoms in simplexes to account for their groups on the periodic table.
That is, all elements belonging to the same group are encoded by the same symbol.

\subsubsection{Network representation (materials cartograms)}
To represent the library of materials as a network, we considered each material, encoded by its fingerprints, as a node.
Edges exist between nodes with similarities greater than or equal to certain thresholds.
In this study, we use fingerprint-based Tanimoto similarity and a threshold $S=0.7$.
This network representation of materials is defined as the graph $G(V,E)$, where $V=\left\{\nu_1|\nu_2\in L\right\}$ and
$E\!=\!\left\{(\nu_1,\nu_2)\mid\mathrm{sim}(\nu_1,\nu_2)\geq T\right\}$.
Here, $L$ denotes a materials library, $\mathrm{sim}(\nu_1,\nu_2)$
denotes a similarity between materials $\nu_1$ and $\nu_2$, and $T$ denotes a similarity threshold.

To examine if the materials networks are scale-free, we analyzed the degree distributions of the networks.
Networks are considered scale-free if the distribution of vertex degrees of the nodes follows the power law:
$p(x)=kx^{-\alpha}$ where $k$ is the normalization constant, and $\alpha$ is the exponent.
The materials networks have been visualized using the Gephi package~\cite{Bastian_ICWSM_2009}.
The ForceAtlas 2 algorithm~\cite{Jacomy_PLoS_2014}, a type of force-directed layout
algorithm, has been used for the graph layout.
A force-directed layout algorithm considers a force between any two nodes,
and minimizes the ``energy'' of the system by moving the nodes and changing the forces between them.
The algorithm guarantees that the topological similarity among nodes determines their vicinity, leading to accurate and
visually-informative representations of materials space.

\subsection{Results and discussion}
\label{subsec:art094:results}

\subsubsection{Similarity search in materials space}
In the first phase of this study, the optimized geometries, symmetries,
band structures, and \DOS{}s available in the \AFLOWorg\ repository were converted
into fingerprints, or arrays of numbers.

We encoded the electronic structure diagram for each material as two distinct types of
fingerprints (Figure~\ref{fig:art094:fingerprints_construction}):
band structure symmetry-dependent fingerprints (B-fingerprints) and
\DOS\ symmetry-independent fingerprints (D-fingerprints).
The B-fingerprint is defined as a collated digitalized histogram of energy eigenvalues
sampled at the high-symmetry reciprocal points with 32 bins.
The D-fingerprint is a string containing 256 4-byte real numbers,
each characterizing the strength of the \DOS\ in one of the 256 bins dividing the [-10, 10]~eV interval.
More details are in the \nameref{subsec:art094:methods} section.

This unique, condensed representation of materials enabled the use of cheminformatics methods,
such as similarity searches, to retrieve materials with similar properties but different compositions from the \AFLOWorg\ database.
As an added benefit, our similarity search can also quickly find duplicate records.
For example, we have identified several barium titanate (BaTiO$_3$) records with identical fingerprints
(\ICSD\ \#15453, \#27970, \#6102, and \#27965 in the \AFLOWorg\ database).
Thus, fingerprint representation afforded rapid identification of duplicates,
which is the standard first step in our cheminformatics data curation workflow~\cite{Fourches_JCIM_2010}.
It is well known that standard \DFT\ has severe limitations in the description of excited states, and needs to be substituted
with more advanced approaches to characterize semiconductors and
insulators~\cite{Hedin_GW_1965,GW,Heyd2003,Liechtenstein1995,Cococcioni_reviewLDAU_2014}.
However, there is a general trend of \DFT\ errors being comparable in similar classes of systems.
These errors may thus be considered ``systematic'', and are irrelevant when one seeks only similarities between  materials.

The first test case is gallium arsenide, GaAs (\ICSD\ \#41674),
a very important material for electronics~\cite{INSPEC_PGA_1986} in the \AFLOWorg\ database.
GaAs is taken as the reference material, and the remaining 20,000+ materials from the
\AFLOWorg\ database are taken as the virtual screening library.
The pairwise similarity between GaAs and any of the materials represented by our D-fingerprints
is computed using the Tanimoto similarity coefficient ($S$)~\cite{Maggiora_JMC_2014}.
The top five materials (GaP, Si, SnP, GeAs, InTe) retrieved show very high similarity ($S\!>\!0.8$)
to GaAs, and all five are known to be semiconductor materials~\cite{Lide_CRC_2004,Littlewood_CRSSMS_1983,Madelung_Semiconductors_2004}.

In addition, we have searched the \AFLOWorg\ database for materials similar to BaTiO$_3$
with the perovskite structure (\ICSD\ \#15453) using B-fingerprints.
BaTiO$_3$ is widely used as a ferroelectric ceramic or piezoelectric~\cite{Bhalla_MRI_2000}.
Out of the six most similar materials with $S>0.8$, five (BiOBr, SrZrO$_3$, BaZrO$_3$, KTaO$_3$ and KNbO$_3$)
are well known for their optical properties~\cite{Rabe_Ferroelectrics_2010}.
The remaining material, cubic YbSe (\ICSD\ \#33675), is largely unexplored.
One can therefore formulate a testable hypothesis suggesting that this material may be ferroelectric or piezoelectric.

We also investigated the challenging case of topological insulators.
They form a rare group of insulating materials with conducting surface-segregated states (or interfaces)~\cite{nmatTI}
arising from a combination of spin-orbit coupling and time-reversal symmetry~\cite{RevModPhys.82.3045}.
Although \DFT\ calculations conducted for materials in the \AFLOWorg\ repository do not
incorporate spin-orbit coupling for the most part~\cite{nmatTI}, various topological insulators show exceptionally
high band-structure similarities --- validating the B-fingerprints scheme.
The two materials most similar to Sb$_2$Te$_3$~\cite{RevModPhys.82.3045} (based on B-fingerprints)
with $S\!>\!0.9$ are Bi$_2$Te$_3$~\cite{Chen09science,zhang_PRL_2009} and Sb$_2$Te$_2$Se~\cite{Xu2010arxiv1007}.
Five out of six materials most similar to Bi$_2$Te$_2$Se~\cite{Xu2010arxiv1007,Arakane2010NC}
are also known topological insulators: Bi$_2$Te$_2$S, Bi$_2$Te$_3$, Sb$_2$Te$_2$Se,
GeBi$_2$Te$_4$~\cite{Xu2010arxiv1007}, and Sb$_2$Se$_2$Te~\cite{nmatTI,Zhang_Nat.Phys._2009}.

These examples demonstrate proof of concept and illustrate the power of simple yet uncommon
fingerprint-based similarity searches for rapid and effective identification of
materials with similar properties in large databases.
They also illuminate the intricate link between structures and properties of materials by demonstrating
that similar materials (as defined by their fingerprint similarity) have similar
properties (such as being ferroelectric or insulating).
This observation sets the stage for building and exploring \QMSPR\ models; as discussed in the following sections.

\subsubsection{Visualizing and exploring materials space}

\fig
\includegraphics[width=1.00\linewidth]{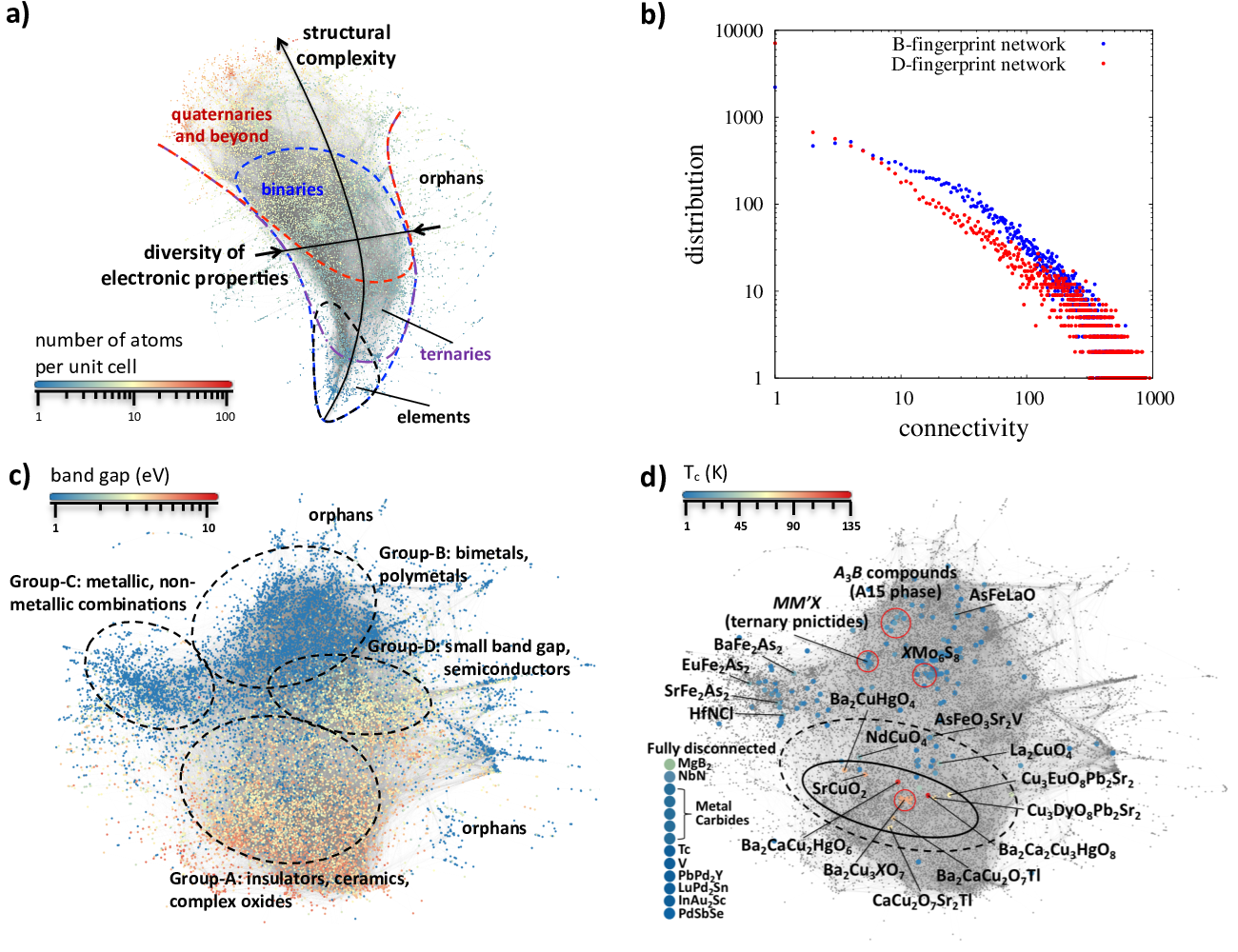}
\mycaption[Materials cartograms with D- (top) and B-fingerprint network representations (bottom).]
{({\bf a}) D-fingerprint network representation of materials.  Materials are color-coded
according to the number of atoms per unit cell.
Regions corresponding to pure elements, binary, ternary and quaternary compounds are outlined.
({\bf b}) Distribution of connectivity within the network.
({\bf c}) Mapping band gaps of materials. Points colored in deep blue are metals;
insulators are color-coded according to the band gap value. Four large communities are outlined.
({\bf d}) Mapping the superconductivity critical temperature, $T_{\mathrm{c}}$, with relevant regions outlined.}
\label{fig:art094:cartograms}
\efig

The use of fingerprint representation and similarity concepts led us to develop the materials network.
Compounds are mapped as nodes.
We use the ``{\it force directed graph drawing}'' algorithm~\cite{Herman_IEEEtvcg_2000}
in which positions of the compounds are initially taken randomly.
There is a force between the nodes: a repulsive Coulomb component and an optional
attractive contribution with a spring constant equal to the Tanimoto coefficient
between D-fingerprints (effective when $S\ge0.7$).
Two nodes are connected only when the coefficient is greater than or equal to the threshold.
The model is equilibrated through a series of heating and quenching steps.
Figure~\ref{fig:art094:cartograms}(a) shows the result in which we add
Bezier-curved lines depicting regions of accumulation.
We shall refer to this approach to visualizing and analyzing materials and their properties as ``{\it materials cartography}''.

The network shown in Figure~\ref{fig:art094:cartograms}(a) is color-coded according to overall complexity.
Pure systems, 79\% of the total 246 unary nodes, are confined in a small, enclosed region.
Binary nodes cover more configurational space, with 82\% of the 3700+ binaries lying in a compact region.
Ternaries are scattered. They mostly populate the center of the space (91\% of the 5300+ ternaries).
Quaternaries and beyond are located at the top part of the network (92\% of the 1080 nodes).
This region is the most distant from that of the unary nodes, which tends to be disconnected from the others.
Indeed, overlap between binaries and ternaries is substantial.
The diversification of electronic properties and thickness of the compact envelope grows with structural complexity.
Orphans are defined as nodes with a very low degree of connectivity: only the vertices (materials)
connected by edges are shown ($\sim$39\% of the database).
Interestingly, of the 200 materials with connectivity smaller than 12,
most are La-based (36 bimetallic and 126 polymetallic) or Ce-based (10 nodes).

\tab
\mycaption[Topological properties for constructed materials cartograms.]
{In network theory, a ``component'' is a group of nodes that are all connected to each other.
A ``giant component'' is a connected component of a given random graph that contains a constant
fraction of the entire graph's vertices~\cite{Chung_Complex_2006}.
Figures in parenthesis are calculated by fitting only the asymptotic portion of the curve in Figure~\ref{fig:art094:cartograms}(b).
}
\tabvspace
\begin{tabular}{l | r r}
 & D-fingerprints network & B-fingerprints network \\
\hline
total number of cases & 17420 & 17420 \\
giant component & 10521 (60.4\%) & 15535 (89.2\%)\\
edges & 466,000 & 564,000 \\
average degree & 88.60 & 72.59 \\
network diameter (edges) & 27 & 23 \\
power law $\gamma$ & 2.745 & 0.916 (2.04) \\
\end{tabular}
\label{tab:art094:cartograms}
\etab

The degree of connectivity is illustrated in Figure~\ref{fig:art094:cartograms}(b).
The panel indicates the log-log distribution of connectivity across the sample set.
The red and blue points measure the D-fingerprints (Figure~\ref{fig:art094:cartograms}(a))
and B-fingerprints connectivity (Figure~\ref{fig:art094:cartograms}(c)), respectively.
Table~\ref{tab:art094:cartograms} contains relevant statistical information about the cartograms.
Although the power law distribution of Figure~\ref{fig:art094:cartograms}(b) is typical of
scale-free networks and similar to many networks examined in cheminformatics and
bioinformatics~\cite{Girvan_PNAS_2002,Newman_SiRev_2003,Yildirim_NB_2007}, in our case, connectivity differs.
In previous examples~\cite{Girvan_PNAS_2002,Newman_SiRev_2003,Yildirim_NB_2007},
most of the nodes have only a few connections; with a small minority being highly
connected to a small set of ``hubs''~\cite{Jeong_Nature_2000,Barabasi_Science_1999}.
In contrast, the \AFLOWorg\ database is highly heterogeneous:
most of the hubs' materials are concentrated along the long, narrow belt along the middle of the network.
The top 200 nodes (ranked by connectivity) are represented by 83 polymetallics
(CoCrSi, Al$_2$Fe$_3$Si$_3$, Al$_8$Cr$_4$Y, \etc),
102 bimetallics (Al$_3$Mo, As$_3$W$_2$, FeZn$_{13}$, \etc),
14 common binary compounds (GeS, AsIn, \etc), and boron (\ICSD\ \#165132).
This is not entirely surprising, since these materials are well studied
and represent the lion's share of the \ICSD\ database.
Al$_3$FeSi$_2$ (\ICSD\ \#79710), an uncommonly used material, has the highest connectivity of 946.
Meanwhile, complex ceramics and exotic materials are relatively disconnected.

A second network, built with B-fingerprints, is illustrated in Figure~\ref{fig:art094:cartograms}(c).
While this network preserves most of the topological features described
in the D-fingerprint case (Figure~\ref{fig:art094:cartograms}(a)), critical distinctions appear.
The B-fingerprint network separates metals from insulators.
Clustering and subsequent community analyses show four large groups of materials.
Group-A ($\sim$3000 materials) consists predominately of insulating compounds (63\%) and semiconductors (10\%).
Group-B distinctly consists of compounds with polymetallic character (70\% of $\sim$2500 materials).
In contrast, Group-C includes $\sim$500 zero band gap materials with nonmetal atoms,
including halogenides, carbides, silicides, \etc\
Lastly, Group-D has a mixed character with $\sim$300 small band gap materials (below 1.5~eV);
and $\sim$500 semimetals and semiconductors.

Lithium scandium diphosphate, LiScP$_2$O$_7$ (\ICSD\ \#91496), has the highest connectivity
of 746 in the B-fingerprint network.
Very highly connected materials are nearly evenly distributed between Groups-A and -B,
forming dense clusters within their centers.
As in the case of the D-fingerprint network, the connectivity distribution follows a power law
(Figure~\ref{fig:art094:cartograms}(b), see Table~\ref{tab:art094:cartograms} for
additional statistics); indicating that this is a scale-free network.

To illustrate one possible application of the materials networks, we chose superconductivity ---
one of the most elusive challenges in solid-state physics.
We have compiled experimental data for 295 stoichiometric superconductors that
are also available in the \AFLOWorg\ repository.
All materials in the data set are characterized with the fingerprints specified in
the \nameref{subsec:art094:methods} section.
The data set includes both prominently high temperature superconducting materials
such as layered cuprates, ferropnictides, iron arsenides-122, MgB$_2$; as well as more
conventional compounds such as A15, ternary pnictides, \etc\
Our model does not consider the effect of phonons, which play a
dominant role in many superconductors~\cite{tinkham_superconductivity}.
High-throughput parameterization of phonon spectra is still in its infancy~\cite{curtarolo:Ru},
and only recently have vibrational descriptors been
adapted to large databases~\cite{curtarolo:art96}.
We envision that future development of vibrational fingerprints
following these guidelines will capture similarities between
known, predicted, and verified superconductors (\ie,
MgB$_2$ \vs\ LiB$_2$~\cite{curtarolo:art21,curtarolo:art26} and MgB$_2$ \vs\
Fe-B compounds~\cite{Kolmogorov_FeB_PRL2010,Gou_PRL_2013_FeB_superconductor}).

\fig
\includegraphics[width=1.00\linewidth]{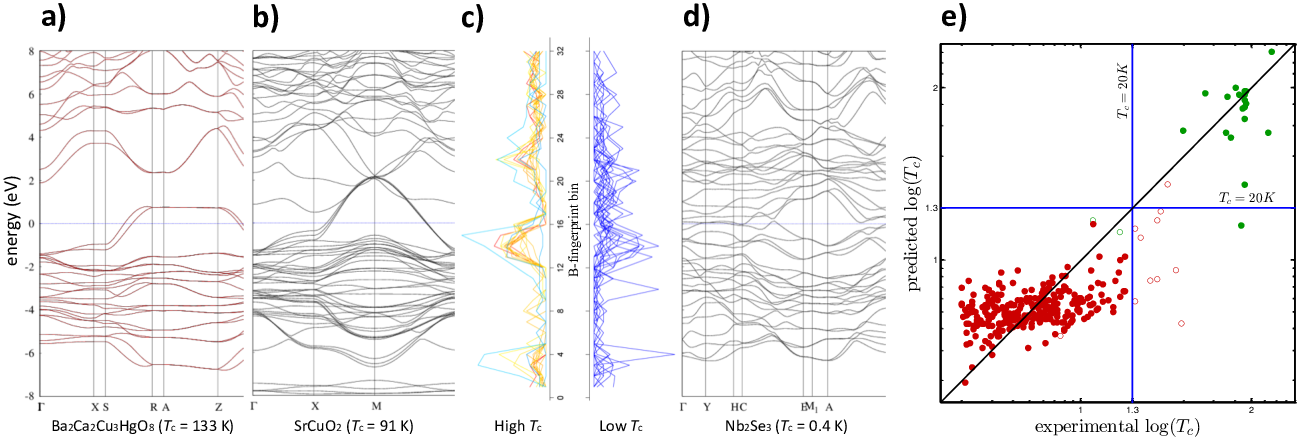}
\mycaption[Comparison high-low $T_{\mathrm{c}}$ aligned band structures and $T_{\mathrm{c}}$ predictions.]
{({\bf a}) Band structure of Ba$_2$Ca$_2$Cu$_3$HgO$_8$ ($T_{\mathrm{c}}=$133~K).
({\bf b}) Band structure of SrCuO$_2$ (\ICSD\ \#16217, $T_{\mathrm{c}}=$91~K~\cite{Takahashi_PSCC_1994_SrCuO2_superconductor}).
({\bf c}) Aligned B-fingerprints for the 15 materials with the highest and lowest $T_{\mathrm{c}}$.
({\bf d}) Band structure of Nb$_2$Se$_3$ (\ICSD\ \#42981, $T_{\mathrm{c}}=$0.4~K).
({\bf e}) Plot of the predicted \vs\ experimental critical temperatures for the continuous model.
Materials are color-coded according to the classification model: solid/open green (red) circles indicate
correct/incorrect predictions in $T_{\mathrm{c}}\!>\!T_{\mathrm{thr}}$
($T_{\mathrm{c}}\!\leq\!T_{\mathrm{thr}}$), respectively.}
\label{fig:art094:bands}
\efig

All materials are identified and marked on the B-fingerprint network, and are
color-coded according to their critical temperature, $T_{\mathrm{c}}$ (Figure~\ref{fig:art094:cartograms}(d)).
All high-$T_{\mathrm{c}}$ superconductors are localized in a relatively compact region.
The distribution is centered on a tight group of Ba$_2$Cu$_3X$O$_7$ compounds
(the so-called Y123, where $X$= lanthanides).
The materials with the two highest $T_{\mathrm{c}}$ values in our set are
Ba$_2$Ca$_2$Cu$_3$HgO$_8$ (\ICSD\ \#75730, $T_{\mathrm{c}}=$133~K) and
Ba$_2$CaCu$_2$HgO$_6$ (\ICSD\ \#75725, $T_{\mathrm{c}}=$125~K).
Their close grouping manifests a significant superconductivity hot-spot of materials with similar fingerprints.
We aligned the B-fingerprints for the 15 superconductors with the highest $T_{\mathrm{c}}$
values in Figure~\ref{fig:art094:bands}(c).

All the top 15 high $T_{\mathrm{c}}$ superconductors are layered cuprates,
which have dominated high $T_{\mathrm{c}}$ superconductor research since 1986~\cite{Bednorz_ZPBCM_1986}.
These compounds are categorized as Charge-Transfer Mott Insulators (CTMI)~\cite{Zaanen_PRL_1985}.
There are three distinct bands that are conserved for these structures around -6, -1, and 4~eV
relative to the Fermi energy at $\Gamma$ (within the simple \DFT+$U$ description available in the \AFLOWorg\ repository,
Figure~\ref{fig:art094:bands}(c)).
These features are consistent with the three-band Hubbard-like picture characteristic of
CTMIs~\cite{Manske_Superconductors_2004,Emery_PRL_1987}.

Meanwhile, the fingerprint distribution for the 15 materials with the lowest
$T_{\mathrm{c}}$ is random (Figure~\ref{fig:art094:bands}(c)).
The importance of band structure features in superconductivity has long
been recognized~\cite{Zaanen_NPhys_2006,Micnas_RMP_1990,Orenstein_Science_2000}.
Thus, materials cartography based on the B-fingerprint network allows us to visualize this phenomenon concisely.

\subsubsection{Predictive QMSPR modeling}
We developed \QMSPR\ models (continuous~\cite{Bramer_PDM_2007}, classification, and structural)
to compute superconducting properties of materials from their structural characteristics.
To achieve this objective, we compiled two superconductivity data sets consisting of
{\bf i.} 295 materials with continuous $T_{\mathrm{c}}$ values ranging from 2~K to 133~K; and
{\bf ii.} 464 materials with binary $T_{\mathrm{c}}$ values.
The models were generated with Random Forest (RF)~\cite{Breiman_ML_2001} and
Partial Least Squares (PLS)~\cite{Wold_CILS_2001} techniques.
These used both B- and D-fingerprints, as well as Simplex (SiRMS)~\cite{Kuzmin_JCAMD_2008} descriptors.
These fingerprints were adapted for materials modeling for the first time in this study
(see the \nameref{subsec:art094:methods} section).
Additionally, we incorporated atomic descriptors that differentiate
by element, charge, and group within the periodic table.
Statistical characteristics for all 464 materials used for the \QMSPR\ analysis
are reported in Tables~\ref{tab:art094:continuous}-\ref{tab:art094:fragments}.

Attempts to develop \QMSPR\ models using B- and D-fingerprints for both data sets were not satisfactory,
indicating that our fingerprints, while effective in qualitative clustering,
do not contain enough information for quantitatively predicting target properties
(\QMSPR\ model acceptance criteria has been discussed previously~\cite{Tropsha_MI_2010}).
Thus, we employed more sophisticated chemical fragment descriptors,
such as SiRMS~\cite{Kuzmin_JCAMD_2008}, and adapted them for
materials modeling (see the \nameref{subsec:art094:methods} section).

\boldsection{Continuous model.}
We constructed a continuous model which serves to predict the value of
$T_{\mathrm{c}}$ with a consensus RF- and PLS-SiRMS approach.
It has a cross-validation determination coefficient of $Q^2=0.66$ (five-fold external CV;
see Table~\ref{tab:art094:continuous}).
Figure~\ref{fig:art094:bands}(e) shows predicted \vs\ experimental $T_{\mathrm{c}}$
values for the continuous model: all materials having
$\log(T_{\mathrm{c}})$$\leq$1.3 are scattered, but within the correct range.
Interestingly, we notice that systems with
$\log(T_{\mathrm{c}})$$\geq$1.3 received higher accuracy, with the exceptions of
MgB$_{2}$ (\ICSD\ \#26675), Nb$_{3}$Ge (\ICSD\ \#26573), Cu$_{1}$Nd$_{2}$O$_{4}$ (\ICSD\ \#4203),
As$_{2}$Fe$_{2}$Sr (\ICSD\ \#163208), Ba$_{2}$CuHgO$_{4}$ (\ICSD\ \#75720), and
ClHfN (\ICSD\ \#87795) (all highly underestimated).
Not surprisingly MgB$_2$~\cite{Buzea_SST_2001_MgB2} is an outlier in our statistics.
This is in agreement with the fact that to date
no superconductor with an electronic structure similar to MgB$_2$ has been found.

\tab
\mycaption[Statistical characteristics of the continuous \QMSPR\ models for superconductivity.]
{$Q^{2}$(ext) refers to the leave-one-out five-fold external cross-validation coefficient,
RMSE refers to root-mean-square error,
MAE refers to the mean absolute error,
RF-SiRMS refers to the application of the Random Forest technique with Simplex descriptors,
PLS-SiRMS refers to the application of the Partial Least Squares regression technique with Simplex descriptors,
and consensus refers to the average of the RF-SiRMS and PLS-SiRMS results.}
\tabvspace
\begin{tabular}{l | r r r r}
model & $N$ & $Q^{2}$(ext) & RMSE & MAE \\
\hline
RF-SiRMS & 295 & 0.64 & 0.24 & 0.18\% \\
PLS-SiRMS & 295 & 0.61 & 0.25 & 0.20\% \\
consensus & 295 & 0.66 & 0.23 & 0.18\% \\
\end{tabular}
\label{tab:art094:continuous}
\etab

\boldsection{Classification model.}
By observing the existence of the threshold $T_{\mathrm{thr}}$=20~K ($\log(T_{\mathrm{thr}})$=1.3),
we developed a classification model.
It is based on the same RF-SiRMS technique, but it is strictly used to predict the position of
$T_{\mathrm{c}}$ with respect to the threshold, above or below.
The classification model has a balanced accuracy of 0.97 with five-fold external CV analysis.
The type of points in Figure~\ref{fig:art094:bands}(e) illustrates the classification model outcome:
solid/open green (red) circles for correct/incorrect
predictions in $T_{\mathrm{c}}\!>\!T_{\mathrm{thr}}$ ($T_{\mathrm{c}}\leq T_{\mathrm{thr}}$), respectively.

\tab
\mycaption[Statistical characteristics of the classification \QMSPR\ models for superconductivity.]
{AD refers to applicability domain~\cite{Tropsha_CPD_2007}.
Accuracy is determined by the ratio of correct predictions to the total number of predictions,
sensitivity is determined by the ratio of correctly predicted $T_{\mathrm{c}}\!>\!T_{\mathrm{thr}}$
to the number of empirical $T_{\mathrm{c}}\!>\!T_{\mathrm{thr}}$,
specificity is determined by the ratio of correctly predicted $T_{\mathrm{c}}\!\leq\!T_{\mathrm{thr}}$
to the number of empirical $T_{\mathrm{c}}\!\leq\!T_{\mathrm{thr}}$,
CCR (correct classification rate) is the average of the sensitivity and the specificity,
and coverage is determined by the ratio of the total number of predictions to the total number of cases.}
\tabvspace
\begin{tabular}{l | r r}
 & no AD & with AD \\
\hline
total number of cases & 464 & 464\\
total number of predictions & 464 & 451\\
number of correct predictions & 452 & 446\\
number of wrong predictions & 12 & 5\\
number of empirical $T_{\mathrm{c}}\!>\!T_{\mathrm{thr}}$ & 29 & 22\\
number of empirical $T_{\mathrm{c}}\!\leq\!T_{\mathrm{thr}}$ & 435 & 429\\
number of correctly predicted $T_{\mathrm{c}}\!>\!T_{\mathrm{thr}}$ & 19 & 17\\
number of correctly predicted $T_{\mathrm{c}}\!\leq\!T_{\mathrm{thr}}$ & 433 & 429\\
number of incorrectly predicted $T_{\mathrm{c}}\!>\!T_{\mathrm{thr}}$ & 2 & 0\\
number of incorrectly predicted $T_{\mathrm{c}}\!\leq\!T_{\mathrm{thr}}$ & 10 & 5\\
$T_{\mathrm{c}}\!>\!T_{\mathrm{thr}}$ prediction value & 0.90 & 1.00\\
$T_{\mathrm{c}}\!\leq\!T_{\mathrm{thr}}$ prediction value & 0.98 & 0.99\\
accuracy & 0.97 & 0.99\\
sensitivity & 0.66 & 0.77\\
specificity & 1.00 & 1.00\\
CCR & 0.83 & 0.89\\
coverage & 1.00 & 0.97\\
\end{tabular}
\label{tab:art094:classification}
\etab

For $T_{\mathrm{c}}\!\leq\!T_{\mathrm{thr}}$ and $T_{\mathrm{c}}\!>\!T_{\mathrm{thr}}$,
accuracies of prediction are 98\% and 90\% (cumulative 94\%).
(Figure~\ref{fig:art094:bands}(e), see Table~\ref{tab:art094:classification} for additional statistics).
Among the 464 materials, ten systems with experimental $T_{\mathrm{c}}\!>\!T_{\mathrm{thr}}$ are predicted to have
$T_{\mathrm{c}}\!\leq\!T_{\mathrm{thr}}$)
[FeLaAsO (\ICSD\ \#163496), AsFeO$_{3}$Sr$_{2}$V (\ICSD\ \#165984), As$_{2}$EuFe$_{2}$ (\ICSD\ \#163210),
As$_{2}$Fe$_{2}$Sr, CuNd$_{2}$O$_{4}$ (\ICSD\ \#86754), As$_{2}$BaFe$_{2}$
(\ICSD\ \#166018), MgB$_{2}$, ClHfN, La$_{2}$CuO$_{4}$, and Nb$_{3}$Ge].
Only two with experimental $T_{\mathrm{c}}\!\leq\!T_{\mathrm{thr}}$
are predicted with $T_{\mathrm{c}}\!>\!T_{\mathrm{thr}}$
(AsFeLi (\ICSD\ \#168206), As$_{2}$CaFe$_{2}$ (\ICSD\ \#166016)).
Owing to the spread around the threshold, additional information about
borates and Fe-As compounds is required for proper training of the learning algorithm.

In the past, it has been shown that \QSAR\ approaches can be used for the
detection of mis-annotated chemical compounds, a critical step in data curation~\cite{Fourches_JCIM_2010}.
We have employed a similar approach here.
In our models, three materials, ReB$_2$ (\ICSD\ \#23871),
Li$_2$Pd$_3$B (\ICSD\ \#84931), and La$_2$CuO$_4$, were significantly mis-predicted.
More careful examination of the data revealed that the $T_{\mathrm{c}}$'s of
ReB$_2$ and Li$_2$Pd$_3$B were incorrectly extracted from literature.
We also found that La$_2$CuO$_4$ has the largest variation of reported values within the data set.
Therefore, it was excluded from the regression.
This approach illustrates that \QMSPR\ modeling should be automatically implemented to reduce and correct erroneous entries.

\boldsection{Structural model.}
We also developed a structural model meant to capture the geometrical
features that most influence $T_{\mathrm{c}}$.
It employs SiRMS descriptors, PLS approaches, and five-fold external cross-validation.
The predictive performance of this model ($Q^2=0.61$) is comparable to that
of the SiRMS-based RF model (see Table~\ref{tab:art094:continuous} for additional statistics).
The top 10 statistically significant geometrical fragments and their contributions to
$T_{\mathrm{c}}$ variations are shown in Table~\ref{tab:art094:classification}.
All descriptor contributions were converted to atomic contributions
(details discussed previously~\cite{Muratov_FMC_2010}) and related to material structures.
Examples of unit cell structures for pairs of similar materials
with different $T_{\mathrm{c}}$ values were color-coded according to
atomic contributions to $T_{\mathrm{c}}$, and are shown in Figure~\ref{fig:art094:structure_fragments}
(green for $T_{\mathrm{c}}\!\uparrow$, red for $T_{\mathrm{c}}\!\downarrow$, and gray for neutral).

\tab
\mycaption[Top statistically significant fragments and their contributions to $T_{\mathrm{c}}$ variation.]
{``-'' demonstrates that the collection is bonded, while ``and'' demonstrates that the collection is not bonded.}
\tabvspace
\begin{tabular}{l | r}
fragment name & contribution to $log(T_{\mathrm{c}})$ score\\
\hline
O-Cu-O & 18\%\\
periodic groups IB-IIB-IVA & 14\%\\
periodic groups IIA and IB & 12\%\\
As, As, Fe fragment count & 5\%\\
periodic groups IIB-IVA & 5\%\\
periodic groups IIA and IVA & 5\%\\
charges~\cite{bader_atoms_1994} (-1.5)(-1.5)(+2.5) & 3\%\\
O element count & 2\%\\
Cu element count & 2\%\\
O, O, O fragment count & 2\%\\
charge~\cite{bader_atoms_1994} (+2.5) & 2\%\\
Nb element count & 2\%\\
charge~\cite{bader_atoms_1994} (-1.5) & 2\%\\
\end{tabular}
\label{tab:art094:fragments}
\etab

\fig
\includegraphics[width=1.00\linewidth]{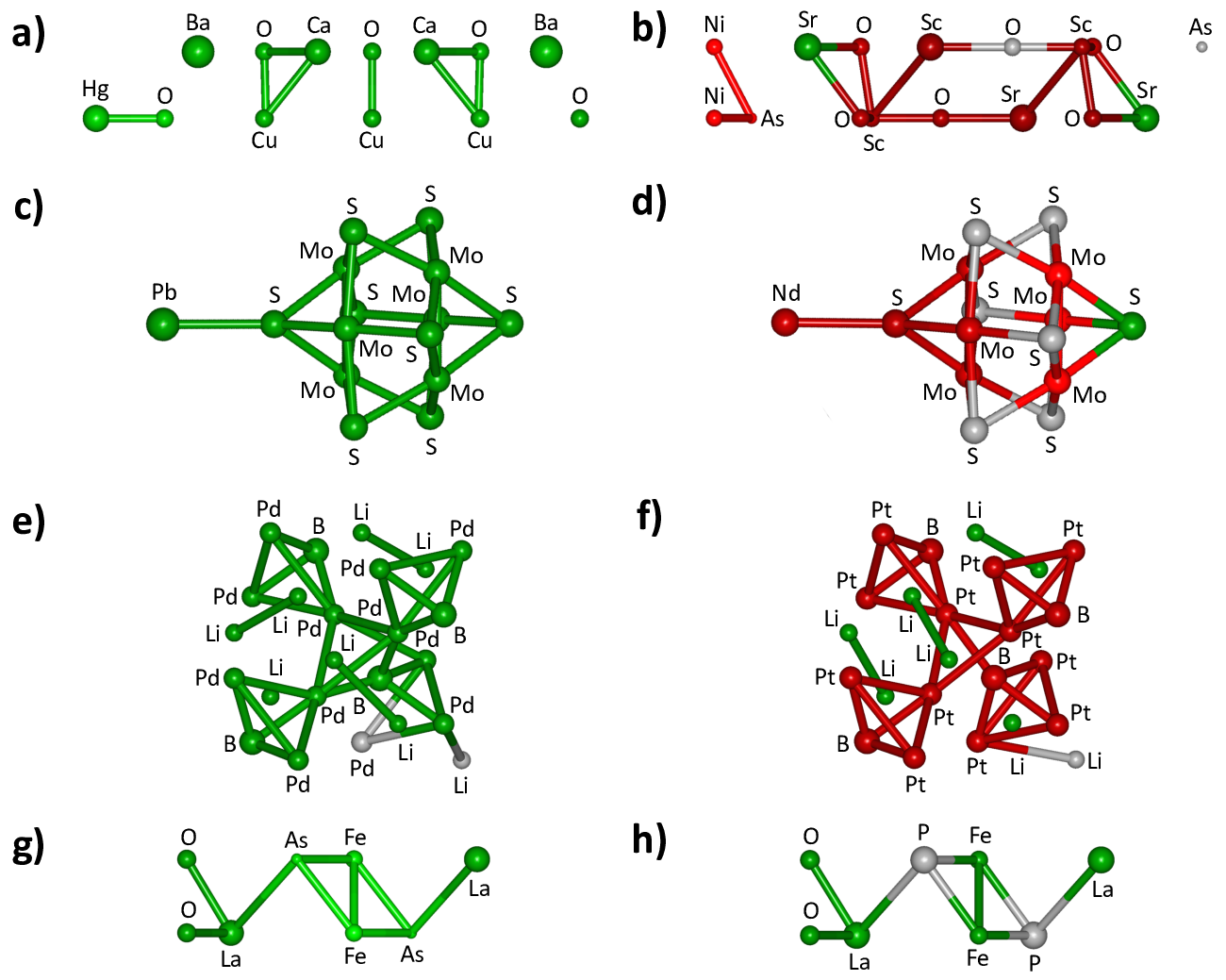}
\mycaption[Materials color-coded according to atom contributions to $\log(T_{\mathrm{c}})$.]
{Atoms and structural fragments that decrease superconductivity critical temperatures
are colored in red and those enhancing $T_{\mathrm{c}}$ are shown in green.
Non-influential fragments are in gray.
({\bf a}) Ba$_2$Ca$_2$Cu$_3$HgO$_8$,
({\bf b}) As$_2$Ni$_2$O$_6$Sc$_2$Sr$_4$,
({\bf c}) Mo$_6$PbS$_8$,
({\bf d}) Mo$_6$NdS$_8$,
({\bf e}) Li$_2$Pd$_3$B,
({\bf f}) Li$_2$Pt$_3$B,
({\bf g}) FeLaAsO, and
({\bf h}) FeLaPO. }
\label{fig:art094:structure_fragments}
\efig

Examples of fragments for materials having
$T_{\mathrm{c}}\!>\!T_{\mathrm{thr}}$ [Ba$_2$Ca$_2$Cu$_3$HgO$_8$, \ICSD\ \#75730,
$\log(T_{\mathrm{c}})$=2.12]
and $T_{\mathrm{c}}\!\!\leq T_{\mathrm{thr}}$
[As$_2$Ni$_2$O$_6$Sc$_2$Sr$_4$, \ICSD\ \#180270, $\log(T_{\mathrm{c}})$=0.44]
are shown in Figures~\ref{fig:art094:structure_fragments}(a) and~\ref{fig:art094:structure_fragments}(b), respectively.
They indicate that individual atom contributions are nonlocal as they strongly
depend upon the atomic environment (Figures~\ref{fig:art094:structure_fragments}(c)-\ref{fig:art094:structure_fragments}(h)).
For example, Mo$_6$PbS$_8$ [\ICSD\ \#644102, $\log(T_{\mathrm{c}})$=1.13] and
Mo$_6$NdS$_8$ [\ICSD\ \#603458, $\log(T_{\mathrm{c}})$=0.54]
differ by a substitution --- yet the difference in $T_{\mathrm{c}}$ is substantial.
Furthermore, substitution of Nd for Pb affects contributions to the
target property from all the remaining atoms in the unit cell
(Figure~\ref{fig:art094:structure_fragments}(c) and~\ref{fig:art094:structure_fragments}(d)).
The same observation holds for Li$_2$Pd$_3$B [\ICSD\ \#84931,
$\log(T_{\mathrm{c}})$=0.89] and Li$_2$Pt$_3$B [\ICSD\ \#84932,
$\log(T_{\mathrm{c}})$=0.49] Figure~\ref{fig:art094:structure_fragments}(e) and
~\ref{fig:art094:structure_fragments}(f); as well as FeLaAsO [\ICSD\ \#163496, $\log(T_{\mathrm{c}})$=1.32]
and FeLaPO [\ICSD\ \#162724, $\log(T_{\mathrm{c}})$=0.82]
Figure~\ref{fig:art094:structure_fragments}(g) and~\ref{fig:art094:structure_fragments}(h).

\subsection{Conclusion}
With high-throughput approaches in materials science
increasing the data-driven content of the field,
the gap between accumulated-information and derived knowledge widens.
The issue can be overcome by adapting the data-analysis approaches
developed during the past decade for chem- and bioinformatics.

Our study gives an example of this.
We introduce novel materials fingerprint descriptors that lead to the generation of
networks called ``{\it materials cartograms}'': nodes represent compounds;
connections represent similarities.
The representation can identify regions with distinct physical and chemical properties,
the key step in searching for interesting, yet unknown compounds.

Starting from atomic-compositions, bond-topologies, structure-geometries,
and electronic properties of materials publicly available in the \AFLOWorg\ repository,
we have introduced cheminformatics models leveraging novel materials fingerprints.
Within our formalism, simple band-structure and \DOS\ fingerprints are adequate to
locate metals, semiconductors, topological insulators, piezoelectrics, and superconductors.
More complex \QMSPR\ modeling~\cite{Kuzmin_JCAMD_2008} are used to tackle
qualitative and quantitative values of superconducting critical temperature
and geometrical features helping/hindering criticality~\cite{Kuzmin_JCAMD_2008}.

In summary, the fingerprinting cartography introduced in this work
has demonstrated its utility in an initial set of problems.
This shows the possibility of designing new materials and gaining
insight into the relationship between the structure
and physical properties of materials.
Further advances in the analysis and exploration of databases may become the
foundation for rationally designing novel compounds with desired properties.
\clearpage
\section{Machine Learning Modeling of Superconducting Critical Temperature}
\label{sec:art137}

This study follows from a collaborative effort described in Reference~\cite{curtarolo:art137}.
Author contributions are as follows:
Valentin Stanev, Ichiro Takeuchi and Aaron Gilad Kusne designed the research.
Valentin Stanev worked on the model.
Corey Oses and Stefano Curtarolo performed the \AFLOW\ calculations.
Valentin Stanev, Ichiro Takeuchi, Efrain Rodriguez and Johnpierre Paglione analyzed the results.
Valentin Stanev, Corey Oses, Ichiro Takeuchi and Efrain Rodriguez wrote the text of the manuscript.
All authors discussed the results and commented on the manuscript.

\subsection{Introduction}

Superconductivity, despite being the subject of intense physics,
chemistry and materials science research for more than a century,
remains among one of the most puzzling scientific topics~\cite{SCSpecial_PSCC_2015}.
It is an intrinsically quantum phenomenon caused by a finite attraction between paired electrons,
with unique properties including zero DC resistivity, Meissner and Josephson effects, and
with an ever-growing list of current and potential applications.
There is even a profound connection between phenomena in the
superconducting state and the Higgs mechanism in particle physics~\cite{PWAnderson_PR_1963}.
However, understanding the relationship between superconductivity and materials'
chemistry and structure presents significant theoretical and experimental challenges.
In particular, despite focused research efforts in the last 30 years,
the mechanisms responsible for high-temperature superconductivity in
cuprate and iron-based families remain elusive~\cite{Chu_PSCC_2015,Paglione_NatPhys_2010}.

Recent developments, however, allow a different approach to investigate what ultimately determines the superconducting
critical temperatures $\left(T_{\mathrm{c}}\right)$ of materials.
Extensive databases covering various measured and calculated materials properties have been created over
the years~\cite{ICSD,aflowPAPER,cmr_repository,Saal_JOM_2013,APL_Mater_Jain2013}.
The shear quantity of accessible information also makes possible,
and even necessary,
the use of data-driven approaches, \eg, statistical and machine learning (ML)
methods~\cite{Agrawal_APLM_2016,Lookman_MatInf_2016,Jain_JMR_2016,Mueller_MLMS_2016}.
Such algorithms can be
developed/trained on the variables collected in these databases,
and employed to predict macroscopic properties
such as the melting temperatures of binary compounds~\cite{Seko_PRB_2014},
the likely crystal structure at a given composition~\cite{Balachandran_SR_2015},
band gap energies~\cite{Pilania_SR_2016,curtarolo:art124} and
density of states~\cite{Pilania_SR_2016} of certain classes of materials.

Taking advantage of this immense increase of readily accessible and potentially relevant information, we develop several
ML methods modeling $T_{\mathrm{c}}$  from
the complete list of reported (inorganic) superconductors~\cite{SuperCon}.
In their simplest form, these methods take as input a number of predictors
generated from the elemental composition of each material.
Models developed with these basic features are surprisingly accurate, despite
lacking information of relevant properties, such as space group, electronic structure,
and phonon energies.
To further improve the predictive power of the models,
as well as the ability to extract useful information out of them,
another set of features are constructed based on crystallographic and electronic information
taken from the \AFLOW\ Online Repositories~\citeAFLOWLIB.

Application of statistical methods in the context of superconductivity began in the early
eighties with simple clustering methods~\cite{Villars_PRB_1988,Rabe_PRB_1992}.
In particular, three ``golden'' descriptors confine the sixty known (at the time) superconductors with
$T_{\mathrm{c}} > 10$~K to three small islands in space:
the averaged valence-electron numbers, orbital radii differences, and metallic
electronegativity differences.
Conversely, about $600$ other superconductors with $T_{\mathrm{c}} < 10$~K appear randomly dispersed
in the same space.
These descriptors were selected heuristically due to their
success in classifying binary/ternary structures and predicting stable/metastable ternary quasicrystals.
Recently, an investigation stumbled on this clustering problem again by observing a
threshold $T_{\mathrm{c}}$ closer to $\log\left(T_{\mathrm{c}}^{\mathrm{thres}}\right)\approx1.3$
$\left(T_{\mathrm{c}}^{\mathrm{thres}}=20~\text{K}\right)$~\cite{curtarolo:art94}.
Instead of a heuristic approach, random forests and simplex fragments were
leveraged on the structural/electronic properties
data from the \AFLOW\ Online Repositories to find the optimum clustering descriptors.
A classification model was developed showing good performance.
Separately, a sequential learning framework was evaluated on superconducting materials,
exposing the limitations of relying on random-guess (trial-and-error) approaches for
breakthrough discoveries~\cite{Ling_IMMI_2017}.
Subsequently, this study also highlights the impact machine learning can have
on this particular field.
In another early work,
statistical methods were used to find correlations between normal state properties and $T_{\mathrm{c}}$
of the metallic elements in the first six rows of the periodic table~\cite{Hirsch_PRB_1997}.
Other contemporary works hone in on specific materials~\cite{Owolabi_JSNM_2015,Ziatdinov_NanoTech_2016}
and families of superconductors~\cite{Klintenberg_CMS_2013,Owolabi_APTA_2014} (see also Reference~\cite{Norman_RPP_2016}).

Whereas previous investigations explored several hundred compounds at most,
this work considers more than $16,000$ different compositions.
These are extracted from the SuperCon database, which contains an exhaustive
list of superconductors, including many closely-related materials varying only by small changes in stoichiometry (doping plays a significant role in optimizing $T_{\mathrm{c}}$).
The order-of-magnitude increase in training data
\textbf{i.} presents crucial subtleties in chemical composition among related compounds,
\textbf{ii.} affords family-specific modeling exposing different superconducting mechanisms, and
\textbf{iii.} enhances model performance overall.
It also enables the optimization of several model construction procedures.
Large sets of independent variables can be constructed and rigorously filtered
by predictive power (rather than selecting them by intuition alone).
These advances are crucial to uncovering insights into the
emergence/suppression of superconductivity with composition.

As a demonstration of the potential of ML methods in looking for novel superconductors, we combined and
applied several models to search for candidates among the roughly
$110,000$ different compositions contained in the Inorganic Crystallographic Structure Database (\ICSD),
a large fraction of which have not been tested for superconductivity.
The framework highlights 35  compounds with predicted $T_{\mathrm{c}}$'s
above 20~K for experimental validation.
Of these, some exhibit interesting chemical and structural similarities to cuprate superconductors, demonstrating the ability of the ML models to identify meaningful patterns in the data.
In addition, most materials from the list share a peculiar feature in their electronic band structure:
one (or more) flat/nearly-flat bands just below the energy of the highest occupied electronic state.
The associated large peak in the density of states (infinitely large in the limit of truly flat bands)
can lead to strong electronic instability, and has been discussed recently as one possible way to
high-temperature superconductivity~\cite{Kopnin_PRB_2011,Peotta_NComm_2015}.

\subsection{Results}
\boldsection{Data and predictors.}
The success of any ML method ultimately depends on access to reliable and plentiful data.
Superconductivity data used in this work is extracted from the SuperCon database~\cite{SuperCon},
created and maintained by the Japanese National Institute for Materials Science.
It houses information such as the $T_{\mathrm{c}}$
and reporting journal publication for superconducting materials known from experiment.
Assembled within it is a uniquely exhaustive list of all reported superconductors,
as well as related non-superconducting compounds.
As such, SuperCon is the largest database of its kind, and has never before been employed
{\it en masse} for machine learning modeling.

From SuperCon, we have extracted a list of approximately $16,400$ compounds,
of which $4,000$ have no $T_{\mathrm{c}}$ reported (see Methods for details).
Of these, roughly $5,700$ compounds are cuprates and $1,500$ are iron-based
(about 35\% and 9\%, respectively), reflecting the significant research efforts invested in these two families.
The remaining set of about $8,000$ is a mix of various materials, including conventional phonon-driven superconductors
(\eg, elemental superconductors, A15 compounds), known unconventional superconductors like the
layered nitrides and heavy fermions, and many materials for which the mechanism of superconductivity
is still under debate (such as bismuthates and borocarbides).
The distribution of materials by $T_{\mathrm{c}}$ for the three groups is shown in Figure~\ref{fig:art137:Class_score}(a).

Use of this data for the purpose of creating ML models can be problematic.
ML models have an intrinsic applicability domain, \ie, predictions
are limited to the patterns/trends encountered in the training set.
As such, training a model only on superconductors can lead to significant selection bias
that may render it ineffective when applied to new
materials\footnote{\textit{N.B.}, a model suffering from selection bias
can still provide valuable statistical information about known superconductors.}.
Even if the model learns to correctly recognize factors promoting superconductivity,
it may miss effects that strongly inhibit it.
To mitigate the effect, we incorporate about $300$ materials found by H.
Hosono's group not to display superconductivity~\cite{Hosono_STAM_2015}.
However, the presence of non-superconducting materials,
along with those without $T_{\mathrm{c}}$ reported in SuperCon, leads to a conceptual problem.
Some of these compounds emerge as non-superconducting ``end-members'' from
doping/pressure studies, indicating no superconducting transition was observed despite some efforts to find one.
However, the transition may still exist,
albeit at experimentally difficult to reach or altogether inaccessible temperatures
(for most practical purposes below $10$~mK)\footnote{There are theoretical arguments for this --- according
to the Kohn-Luttinger theorem, a
superconducting instability should be present as $T \rightarrow 0$ in any fermionic metallic system
with Coulomb interactions~\cite{Kohn_PRL_1965}.}.~\nocite{Kohn_PRL_1965}
This presents a conundrum:
ignoring compounds with no reported $T_{\mathrm{c}}$ disregards a potentially important
part of the dataset, while assuming $T_{\mathrm{c}} = 0$~K prescribes an inadequate description
for (at least some of) these compounds.
To circumvent the problem,
materials are first partitioned in two groups by their $T_{\mathrm{c}}$,
above and below a threshold temperature $\left(T_{\mathrm{sep}}\right)$,
for the creation of a classification model.
Compounds with no reported critical temperature can be classified in the ``below-$T_{\mathrm{sep}}$'' group
without the need to specify a $T_{\mathrm{c}}$ value (or assume it is zero).
The ``above-$T_{\mathrm{sep}}$'' bin also enables the development of a regression model
for $\ln{(T_{\mathrm{c}})}$, without problems arising in the $T_{\mathrm{c}}\to0$ limit.

For most materials, the SuperCon database provides
only the chemical composition and $T_{\mathrm{c}}$.
To convert this information into meaningful features/predictors (used interchangeably),
we employ the Materials Agnostic Platform for Informatics and Exploration (Magpie)~\cite{Ward_ML_GFA_NPGCompMat_2016}.
Magpie computes a set of attributes for each material, including elemental property
statistics like the mean and the standard deviation of 22 different elemental properties
(\eg, period/group on the periodic table,
atomic number, atomic radii, melting temperature), as well as electronic structure attributes, such as the average
fraction of electrons from the $s$, $p$, $d$ and $f$ valence shells among all
elements present.

The application of Magpie predictors, though appearing to lack \textit{a priori} justification,
expands upon past clustering approaches by Villars and Rabe~\cite{Villars_PRB_1988,Rabe_PRB_1992}.
They show that, in the space of a few judiciously chosen
heuristic predictors, materials separate and cluster according to their
crystal structure and even complex properties such as high-temperature
ferroelectricity and superconductivity.
Similar to these features, Magpie predictors capture significant chemical information, which
plays a decisive role in determining
structural and physical properties of materials.

Despite the success of Magpie predictors in modeling materials properties~\cite{Ward_ML_GFA_NPGCompMat_2016},
interpreting their connection to superconductivity presents a serious challenge.
They do not encode (at least directly) many important properties, particularly those
pertinent to superconductivity.
Incorporating features
like lattice type and density of states would undoubtedly lead to significantly more powerful and interpretable models.
Since such information is not generally available in SuperCon,
we employ data from the \AFLOW\ Online Repositories~\citeAFLOWLIB.
The materials database houses more than 200 million properties calculated with
the software package \AFLOW~\citeAFLOW.
It contains information for the vast majority of compounds in the \ICSD~\cite{ICSD}.
Although the \AFLOW\ Online Repositories contain calculated properties,
the density functional theory (\DFT) results have been extensively validated with
\ICSD\ records~\cite{curtarolo:art94,curtarolo:art96,curtarolo:art112,curtarolo:art115,curtarolo:art120,curtarolo:art124}.

\fig
\includegraphics[width=\linewidth]{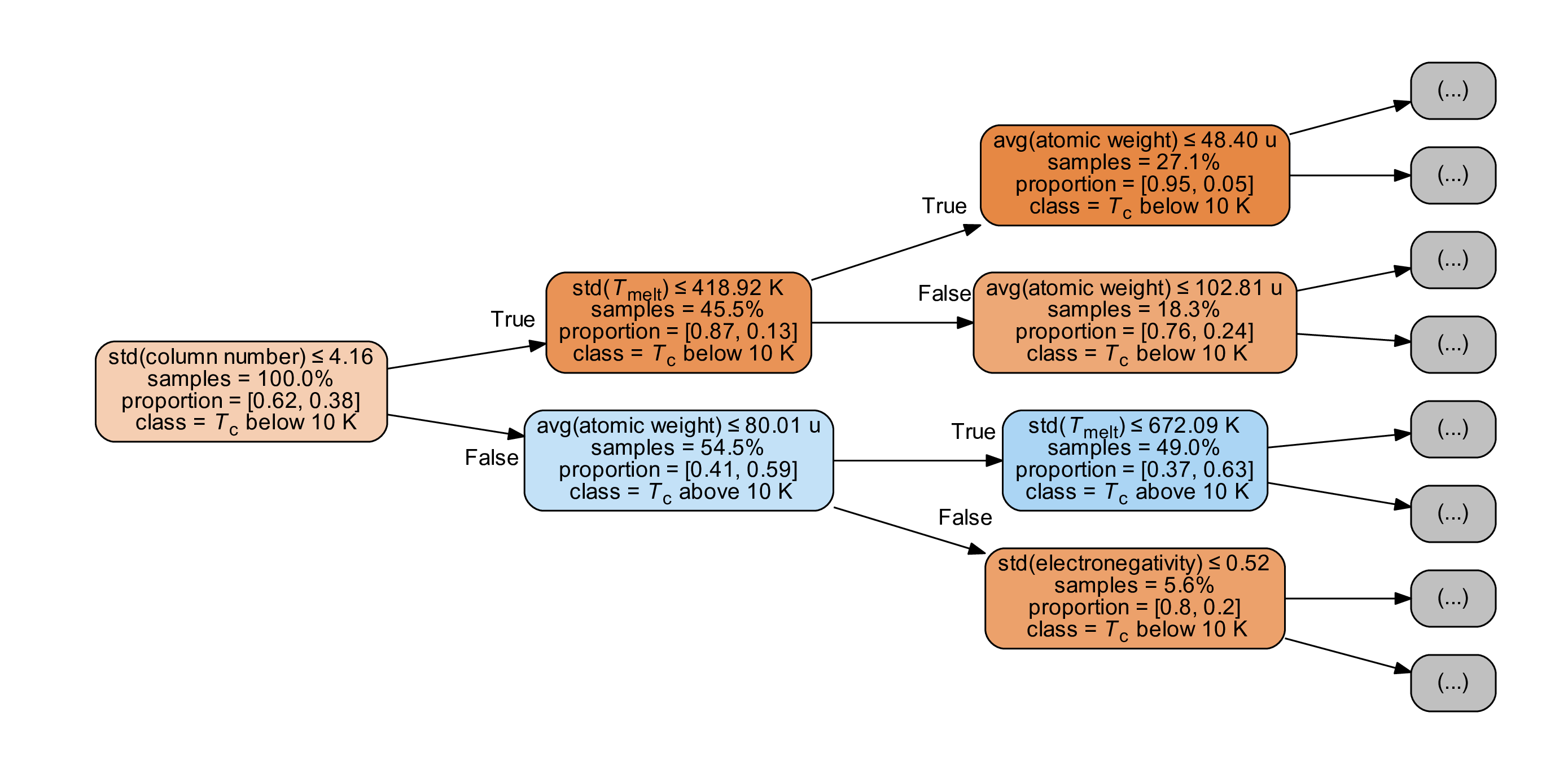}
\mycaption[Schematic of the random forest ML approach.]
{Example of a single decision tree used to classify materials depending on whether
$T_{\mathrm{c}}$ is above or below $10$~K.
A tree can have many levels, but only the three top are shown.
The decision rules leading to each subset are written inside individual rectangles.
The subset population percentage
is given by ``samples'', and the node color/shade
represents the degree of separation,
\ie, dark blue/orange illustrates a high proportion of
$T_{\mathrm{c}} >10$~K/$T_{\mathrm{c}} < 10$~K materials
(the exact value is given by ``proportion'').
A random forest consists of a large number --- could be hundreds or thousands --- of such individual trees.}
\label{fig:art137:tree_example}
\efig

Unfortunately, only a small subset of materials in SuperCon overlaps with those in the \ICSD:
about $800$ with finite $T_{\mathrm{c}}$ and less than $600$ are contained within \AFLOW.
For these, a set of 26 predictors are incorporated
from the \AFLOW\ Online Repositories, including structural/chemical information like the lattice type, space group,
volume of the unit cell, density, ratios of the lattice parameters,
Bader charges and volumes, and formation energy (see Methods for details).
In addition, electronic properties are considered, including the
density of states near the Fermi level as calculated by \AFLOW.
Previous investigations exposed limitations in applying ML methods to a similar dataset
in isolation~\cite{curtarolo:art94}.
Instead, a framework is presented here for combining models built on Magpie descriptors
(large sampling, but features limited to compositional data) and \AFLOW\ features
(small sampling, but diverse and pertinent features).

Once we have a list of relevant predictors, various ML models can be applied to the
data~\cite{Bishop_ML_2006,Hastie_StatLearn_2001}.
All ML algorithms in this work are
variants of the random forest method~\cite{randomforests}.
Fundamentally, this approach combines many individual decision trees, where
each tree is a non-parametric supervised learning method used for
modeling either categorical or numerical variables (\ie,
classification or regression modeling).
A tree predicts the value of a target variable by learning simple decision rules
inferred from the available features (see Figure~\ref{fig:art137:tree_example} for an example).

Random forest is one of the most powerful, versatile, and widely-used ML methods~\cite{Caruana_2006}.
There are several advantages that make it especially suitable for this problem.
First, it can learn complicated non-linear dependencies from the data.
Unlike many other methods (\eg, linear regression),
it does not make assumptions about the functional form of the relationship between the predictors
and the target variable (\eg, linear, exponential or some other {\it a priori} fixed function).
Second, random forests are quite tolerant to heterogeneity in the training data.
It can handle both numerical and categorical data which, furthermore, does not
need extensive and potentially dangerous preprocessing, such as scaling or normalization.
Even the presence of strongly correlated predictors is not a problem for model
construction (unlike many other ML algorithms).
Another significant advantage of this method is that, by combining information from
individual trees, it can estimate the importance of each predictor, thus making the model more interpretable.
However, unlike model construction, determination of predictor importance is complicated by the presence of
correlated features.
To avoid this, standard feature selection procedures are employed along with
a rigorous predictor elimination scheme (based on their strength and correlation with others).
Overall, these methods
reduce the complexity of the models and improve our
ability to interpret them.

\boldsection{Classification models.}
As a first step in applying ML methods to the dataset, a sequence of classification models
are created, each designed to separate materials into two distinct groups depending on whether
$T_{\mathrm{c}}$ is above or below some predetermined value.
The temperature that separates the two groups ($T_{\mathrm{sep}}$)
is treated as an adjustable parameter of the model, though some physical
considerations should guide its choice as well.
Classification ultimately allows compounds with no reported $T_{\mathrm{c}}$ to be used
in the training set by including them in the below-$T_{\mathrm{sep}}$ bin.
Although discretizing continuous variables is not generally recommended, in this case
the benefits of including compounds without $T_{\mathrm{c}}$ outweigh
the potential information loss.

In order to choose the optimal value of $T_{\mathrm{sep}}$, a series of random forest models
are trained with different threshold temperatures separating the two classes.
Since setting $T_{\mathrm{sep}}$ too low or too high creates strongly imbalanced classes
(with many more instances in one group), it is important to compare the models using several different metrics.
Focusing only on the accuracy (count of correctly-classified instances)
can lead to deceptive results.
Hypothetically, if $95\%$ of the observations in the dataset are in the below-$T_{\mathrm{sep}}$ group,
simply classifying all materials as such would
yield a high accuracy ($95\%$), while being trivial in any other sense\footnote{There are more sophisticated techniques to deal with severely imbalanced datasets, like undersampling the majority class or generating synthetic data points for the minority class (see, for example, Reference~\cite{SMOTE}.}~\nocite{SMOTE}.
To avoid this potential pitfall, three other standard metrics
for classification are considered: precision, recall, and $F_{\mathrm{1}}$ score.
They are defined using the values $tp$, $tn$, $fp$, and $fn$ for
the count of true/false positive/negative
predictions of the model:
\begin{eqnarray}
\text{accuracy} \equiv \frac{tp+tn}{tp+tn+fp+fn},
\label{eq:art137:accur}
\end{eqnarray}
\begin{eqnarray}
\text{precision}\equiv\frac{tp}{tp+fp},
\label{eq:art137:precision}
\end{eqnarray}
\begin{eqnarray}
\text{recall} \equiv\frac{tp}{tp+fn},
\label{eq:art137:recall}
\end{eqnarray}
\begin{eqnarray}
F_{\mathrm{1}}\equiv2*\frac{\text{precision}*\text{recall}}{\text{precision}+\text{recall}},
\label{eq:art137:f1}
\end{eqnarray}
where positive/negative refers to above-$T_{\mathrm{sep}}$/below-$T_{\mathrm{sep}}$.
The accuracy of a classifier is the total proportion of correctly-classified materials,
while precision measures the proportion of correctly-classified
above-$T_{\mathrm{sep}}$ superconductors out of all predicted above-$T_{\mathrm{sep}}$.
The recall is the proportion of correctly-classified above-$T_{\mathrm{sep}}$
materials out of all truly above-$T_{\mathrm{sep}}$ compounds.
While the precision measures the probability that a
material selected by the model actually has $T_{\mathrm{c}} > T_{\mathrm{sep}}$,
the recall reports how sensitive the model is to above-$T_{\mathrm{sep}}$ materials.
Maximizing the precision or recall would require some compromise with
the other, \ie, a model that labels all materials as above-$T_{\mathrm{sep}}$ would have perfect recall but dismal precision.
To quantify the trade-off between recall and precision, their harmonic mean ($F_{\mathrm{1}}$ score) is
widely used to measure the performance of a classification model.
With the exception of accuracy, these metrics are not symmetric with respect to the exchange of positive and negative labels.

\fig
\includegraphics[width=0.90\linewidth]{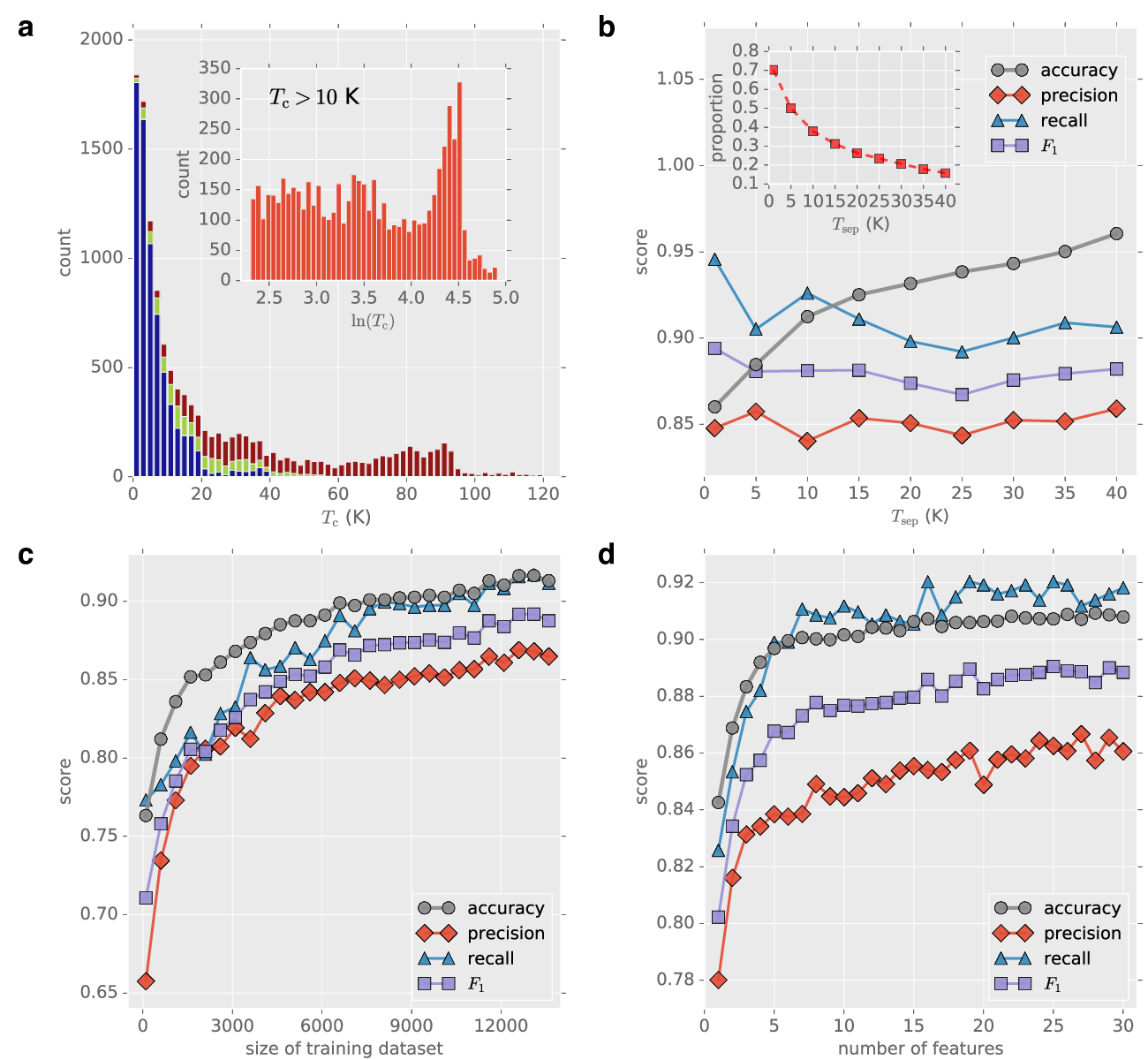}
\mycaption[SuperCon dataset and classification model performance.]
{(\textbf{a}) Histogram of materials categorized by
$T_{\mathrm{c}}$ (bin size is $2$~K, only those with finite $T_{\mathrm{c}}$ are counted).
Blue, green, and red denote low-$T_{\mathrm{c}}$, iron-based, and cuprate superconductors, respectively.
In the inset: histogram of materials categorized by $\ln{(T_{\mathrm{c}})}$
restricted to those with $T_{\mathrm{c}} >10$~K.
(\textbf{b}) Performance of different classification models as a function of the threshold temperature
$\left(T_{\mathrm{sep}}\right)$ that separates materials in two classes by $T_{\mathrm{c}}$.
Performance is measured by accuracy (gray), precision (red), recall (blue), and $F_{\mathrm{1}}$ score (purple).
The scores are calculated from predictions on an independent test set, \ie, one separate
from the dataset used to train the model.
In the inset: the dashed red curve gives the proportion of materials in the above-$T_{\mathrm{sep}}$ set.
(\textbf{c}) Accuracy, precision, recall, and $F_{\mathrm{1}}$
score as a function of the size of the training set with a fixed test set.
(\textbf{d}) Accuracy, precision, recall, and $F_{\mathrm{1}}$ as a function of the number of
predictors.}
\label{fig:art137:Class_score}
\efig

For a realistic estimate of the performance of each model,
the dataset is randomly split ($85\%/15\%$) into training and test subsets.
The training set is employed to fit the model, which is then applied to the test set for subsequent benchmarking.
The aforementioned metrics (Equations~\ref{eq:art137:accur}-\ref{eq:art137:f1}) calculated on the test set provide
an unbiased estimate of how well the model is expected to generalize to a new (but similar) dataset.
With the random forest method, similar estimates can be obtained intrinsically at the training stage.
Since each tree is trained only on a bootstrapped subset of the data,
the remaining subset can be used as an internal test set.
These two methods for quantifying model performance usually yield very similar results.

\fig
\includegraphics[width=\linewidth]{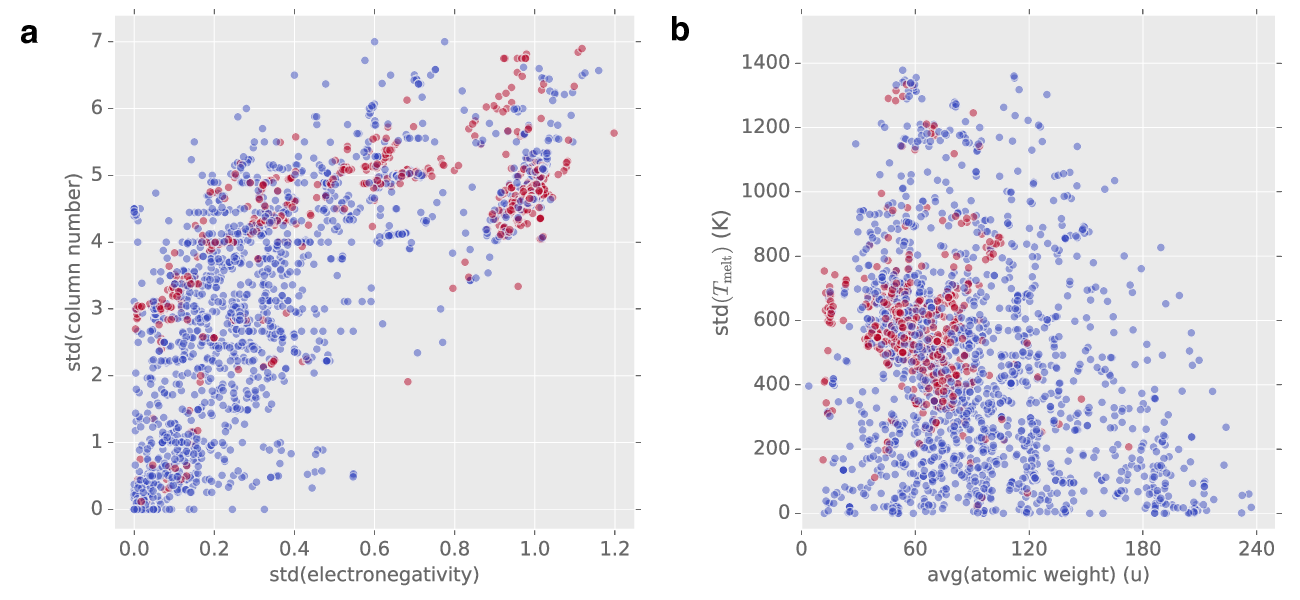}
\mycaption[Scatter plots of $3,000$ superconductors in the space of the four most important classification predictors.]
{Blue/red represent below-$T_{\mathrm{sep}}$/above-$T_{\mathrm{sep}}$ materials, where $T_{\mathrm{sep}} = 10$~K.
(\textbf{a}) Feature space of the first and second most important predictors:
standard deviations of the column numbers and electronegativities (calculated over the values for the constituent elements in each compound).
(\textbf{b}) Feature space of the third and fourth most important predictors:
standard deviation of the elemental melting temperatures and average of the
atomic weights.}
\label{fig:art137:Class_features}
\efig

With the procedure in place, the models' metrics are evaluated for a range of $T_{\mathrm{sep}}$ and illustrated
in Figure~\ref{fig:art137:Class_score}(b).
The accuracy increases as $T_{\mathrm{sep}}$ goes from $1$~K to $40$~K,
and the proportion of above-$T_{\mathrm{sep}}$ compounds drops from above $70\%$ to about $15\%$,
while the recall and $F_{\mathrm{1}}$ score generally decrease.
The region between $5-15$~K is especially appealing in (nearly) maximizing
all benchmarking metrics while balancing the sizes of the bins.
In fact, setting $T_{\mathrm{sep}}=10$~K is a particularly convenient choice.
It is also the temperature used in References~\cite{Villars_PRB_1988,Rabe_PRB_1992}
to separate the two classes, as it is just above the
highest $T_{\mathrm{c}}$ of all elements and pseudoelemental
materials (solid solution whose range of composition includes a pure element).
Here, the proportion of above-$T_{\mathrm{sep}}$ materials is approximately $38\%$ and
the accuracy is about $92\%$, \ie, the model can
correctly classify nine out of ten materials --- much better than random guessing.
The recall --- quantifying how well all above-$T_{\mathrm{sep}}$ compounds are labeled and,
thus, the most important metric when searching for new superconducting materials --- is even higher.
(Note that the models' metrics also depend on random factors such as the composition of the
training and test sets, and their exact values can vary.)

\tab
\mycaption[The most relevant predictors and their importances for the classification and general regression models.]
{$\avg(x)$ and $\std(x)$ denote the composition-weighted average and
standard deviation, respectively,
calculated over the vector of elemental values for each compound~\cite{Ward_ML_GFA_NPGCompMat_2016}.
For the classification model, all predictor importances are quite close.}
\tabvspace
\resizebox{\linewidth}{!}{
\begin{tabular}{l | l r| l r}
predictor & 	\multicolumn{4}{c}{model}
\\
\cline{2-5}
rank& \multicolumn{2}{l|}{classification} & \multicolumn{2}{l}{regression (general; $T_{\mathrm{c}}>10$ K)} \\
\hline
1 & $\std ($column number$)$ & \setlength{\tabcolsep}{4pt} 0.26 & $\avg ($number of unfilled orbitals$)$ & \setlength{\tabcolsep}{4pt} 0.26\\
2 & $\std ($electronegativity$)$ & \setlength{\tabcolsep}{4pt} 0.26 & $\std ($ground state volume$)$ & \setlength{\tabcolsep}{4pt} 0.18\\
3 & $\std ($melting temperature$)$ & \setlength{\tabcolsep}{4pt} 0.23 & $\std ($space group number$)$ & \setlength{\tabcolsep}{4pt} 0.17\\
4 & $\avg ($atomic weight$)$ & \setlength{\tabcolsep}{4pt} 0.24 & $\avg ($number of $d$ unfilled orbitals$)$ & \setlength{\tabcolsep}{4pt} 0.17\\
5 & & \setlength{\tabcolsep}{4pt} - & $\std ($number of $d$ valence electrons$)$ & \setlength{\tabcolsep}{4pt} 0.12\\
6 & & \setlength{\tabcolsep}{4pt} - & $\avg ($melting temperature$)$ & \setlength{\tabcolsep}{4pt} 0.1\\
\end{tabular}}
\label{tab:art137:Table1}
\etab

The most important factors that determine the model's performance are the size of the available
dataset and the number of meaningful predictors.
As can be seen in Figure~\ref{fig:art137:Class_score}(c), all
metrics improve significantly with the increase of the training set size. The effect is most dramatic for sizes between several hundred and few thousands instances, but there is no obvious saturation even for the largest available datasets. This validates efforts herein to incorporate as much relevant data as possible into model training.
The number of predictors is another very important model parameter.
In Figure~\ref{fig:art137:Class_score}(d),
the accuracy is calculated at each step of the backward feature elimination process.
It quickly saturates when the number of predictors reaches $10$.
In fact, a model using only the five most informative predictors, selected out of the full list of 145 ones,
achieves almost $90\%$ accuracy.

For an understanding of what the model has learned, an analysis of the chosen predictors is needed.
In the random forest method, features can be ordered by their importance quantified via
the so-called Gini importance or
``mean decrease in impurity''~\cite{Bishop_ML_2006,Hastie_StatLearn_2001}.
For a given feature, it is the sum of the Gini impurity\footnote{Gini impurity is calculated as
\unexpanded{$\sum_i p_i \left(1-p_i\right)$},
where \unexpanded{$p_i$} is the probability of randomly chosen data point
from a given decision tree leaf to be in class
\unexpanded{$i$}~\cite{Bishop_ML_2006,Hastie_StatLearn_2001}.}~\nocite{Bishop_ML_2006,Hastie_StatLearn_2001}
over the number of splits that include the feature, weighted by the number of samples
it splits, and averaged over the entire forest.
Due to the nature of the algorithm, the closer to the top of the tree a predictor is used,
the greater number of predictions it impacts.

Although correlations between predictors do not affect the model's ability to learn, it can distort importance estimates.
For example, a material property with a strong effect on $T_{\mathrm{c}}$ can be shared
among several correlated predictors.
Since the model can access the same information through any of these variables,
their relative importances are diluted across the group.
To reduce the effect and limit the list of predictors to a manageable size,
the backward feature elimination method is employed.
The process begins with a model constructed with the full list of predictors,
and iteratively removes the least significant one, rebuilding the model and recalculating importances
with every iteration.
(This iterative procedure is necessary since the ordering of the predictors by importance can change at each step.)
Predictors are removed until the overall accuracy of the model drops by $2\%$, at which point there are only five left.
Furthermore, two of these predictors are strongly correlated with each other, and we remove the less important one. This
has a negligible impact on the model performance,
yielding four predictors total (see Table~\ref{tab:art137:Table1})
with an above $90\%$ accuracy score --- only slightly worse than the full model.
Scatter plots of the pairs of the most important predictors are shown in Figure~\ref{fig:art137:Class_features}, where
blue/red denotes whether the material is in the below-$T_{\mathrm{sep}}$/above-$T_{\mathrm{sep}}$ class.
Figure~\ref{fig:art137:Class_features}(a) shows a scatter plot of $3,000$ compounds
in the space spanned by the standard deviations of the column numbers and electronegativities
calculated over the elemental values.
Superconductors with $T_{\mathrm{c}} > 10$~K tend to
cluster in the upper-right corner of the plot and in a relatively thin elongated region extending to the left of it.
In fact, the points in the upper-right corner represent mostly cuprate materials,
which with their complicated compositions and large number of elements are likely
to have high standard deviations in these variables.
Figure~\ref{fig:art137:Class_features}(b) shows
the same compounds projected in the space of the standard deviations of
the melting temperatures and the averages of the atomic weights of the elements forming each compound.
The above-$T_{\mathrm{sep}}$ materials tend to cluster in areas with lower mean atomic weights --- not
a surprising result given the role of phonons in conventional superconductivity.

For comparison, we create another classifier based on the average number of valence electrons,
metallic electronegativity differences, and orbital radii differences, \ie, the predictors used
in References~\cite{Villars_PRB_1988,Rabe_PRB_1992} to cluster materials with $T_{\mathrm{c}} > 10$ K.
A classifier built only with these three predictors
is less accurate than both the full and the truncated models presented herein,
but comes quite close: the full model has about $3\%$ higher
accuracy and $F_{\mathrm{1}}$ score, while the truncated model with four predictors is less that $2\%$ more accurate.
The rather small (albeit not insignificant) differences demonstrates that even on the scale of the
entire SuperCon dataset, the predictors used by Villars and Rabe~\cite{Villars_PRB_1988,Rabe_PRB_1992}
capture much of the relevant chemical information for superconductivity.

\tab
\mycaption[The most significant predictors and their importances for the three material-specific regression models.]
{$\avg(x)$, $\std(x)$, $\max(x)$ and $\fraction(x)$ denote the composition-weighted average,
standard deviation, maximum, and fraction, respectively,
taken over the elemental values for each compound.
$l^2$-norm of a composition is calculated by $||x||_{2} = \sqrt{\sum_i x_i^2}$, where $x_i$ is the proportion of each element $i$ in the compound.}
\tabvspace
\resizebox{\linewidth}{!}{
\begin{tabular}{l | l r| l r |l r}
pred. & 	\multicolumn{6}{c}{model}
\\
\cline{2-7}
rank & \multicolumn{2}{l|}{regression (low-$T_{\mathrm{c}}$)} & \multicolumn{2}{l}{regression (cuprates)}& \multicolumn{2}{|l} {regression (Fe-based)} \\
\hline
1 & $\fraction (d$ valence electrons$)$ &\setlength{\tabcolsep}{4pt} 0.18 & $\avg ($number of unfilled orbitals$)$ &\setlength{\tabcolsep}{4pt} 0.22 & $\std ($column number$)$ &\setlength{\tabcolsep}{4pt} 0.17\\
2 & $\avg ($number of $d$ unfilled orbitals$)$ &\setlength{\tabcolsep}{4pt} 0.14 & $\std ($number of $d$ valence electrons$)$ &\setlength{\tabcolsep}{4pt} 0.13 & $\avg ($ionic character$)$ &\setlength{\tabcolsep}{4pt} 0.15\\
3 & $\avg ($number of valence electrons$)$ &\setlength{\tabcolsep}{4pt} 0.13 & $\fraction (d$ valence electrons$)$ &\setlength{\tabcolsep}{4pt} 0.13 & $\std ($Mendeleev number$)$ &\setlength{\tabcolsep}{4pt} 0.14\\
4 & $\fraction (s$ valence electrons$)$ &\setlength{\tabcolsep}{4pt} 0.11 & $\std ($ground state volume$)$ &\setlength{\tabcolsep}{4pt} 0.13 & $\std ($covalent radius$)$ &\setlength{\tabcolsep}{4pt} 0.14\\
5 & $\avg ($number of $d$ valence electrons$)$ &\setlength{\tabcolsep}{4pt} 0.09 & $\std ($number of valence electrons$)$ &\setlength{\tabcolsep}{4pt} 0.1 & $\max ($melting temperature$)$ &\setlength{\tabcolsep}{4pt} 0.14\\
6 & $\avg ($covalent radius$)$ &\setlength{\tabcolsep}{4pt} 0.09 & $\std ($row number$)$ &\setlength{\tabcolsep}{4pt} 0.08 & $\avg ($Mendeleev number$)$ &\setlength{\tabcolsep}{4pt} 0.14\\
7 & $\avg ($atomic weight$)$ &\setlength{\tabcolsep}{4pt} 0.08 & $||$composition$||_{2}$ &\setlength{\tabcolsep}{4pt} 0.07 & $||$composition$||_{2}$ &\setlength{\tabcolsep}{4pt} 0.11\\
8 & $\avg ($Mendeleev number$)$ &\setlength{\tabcolsep}{4pt} 0.07 & $\std ($number of $s$ valence electrons$)$ &\setlength{\tabcolsep}{4pt} 0.07 & &\setlength{\tabcolsep}{4pt} -\\
9 & $\avg ($space group number$)$ &\setlength{\tabcolsep}{4pt} 0.07 & $\std ($melting temperature$)$ &\setlength{\tabcolsep}{4pt} 0.07 & &\setlength{\tabcolsep}{4pt} -\\
10 & $\avg ($number of unfilled orbitals$)$ &\setlength{\tabcolsep}{4pt} 0.06 & &\setlength{\tabcolsep}{4pt} - & &\setlength{\tabcolsep}{4pt} -\\
\end{tabular}}
\label{tab:art137:Table2}
\etab

\boldsection{Regression models.}
After constructing a successful classification model, we now move to the more difficult challenge of predicting $T_{\mathrm{c}}$.
Creating a regression model may enable better understanding of the factors controlling
$T_{\mathrm{c}}$ of known superconductors,
while also serving as an organic part of a system for identifying potential new ones.
Leveraging the same set of elemental predictors as the classification model, several regression models are presented
focusing on materials with $T_{\mathrm{c}} > 10$~K.
This approach avoids the problem of materials with no reported $T_{\mathrm{c}}$ with the assumption that,
if they were to exhibit superconductivity at all, their critical temperature would be below $10$~K.
It also enables the substitution of $T_{\mathrm{c}}$ with $\ln{(T_{\mathrm{c}})}$ as the target variable
(which is problematic as $T_{\mathrm{c}}\to0$), and thus addresses the problem of the uneven distribution
of materials along the $T_{\mathrm{c}}$ axis (see Figure~\ref{fig:art137:Class_score}(a)).
Using $\ln{(T_{\mathrm{c}})}$ creates a more uniform distribution (Figure~\ref{fig:art137:Class_score}(a) inset),
and is also considered a best practice when the range of a target variable covers more than one
order of magnitude (as in the case of $T_{\mathrm{c}}$).
Following this transformation, the dataset is parsed randomly ($85\%$/$15\%$) into training
and test subsets (similarly performed for the classification model).

Present within the dataset are distinct families of superconductors with different driving
mechanisms for superconductivity, including cuprate and iron-based high-temperature superconductors,
with all others denoted ``low-$T_{\mathrm{c}}$'' for brevity (no specific mechanism in this group).
Surprisingly, a single regression model does reasonably well among the
different families -- benchmarked on the test set,
the model achieves $R^2 \approx 0.88$ (Figure~\ref{fig:art137:Rerg_r2}(a)).
It suggests that the random forest algorithm is flexible and powerful enough
to automatically separate the compounds into groups
and create group-specific branches with distinct predictors (no explicit group labels were used during training and testing).
As validation, three separate models are constructed, each trained only on a specific family, namely the
low-$T_{\mathrm{c}}$, cuprate, and iron-based superconductors, respectively.
Benchmarking on mixed-family test sets, the models performed well on compounds belonging
to their training set family while demonstrating no predictive power on the others.
Figures~\ref{fig:art137:Rerg_r2}(b)-(d) illustrate a cross-section of this comparison.
Specifically, the model trained on low-$T_{\mathrm{c}}$ compounds dramatically underestimates
the $T_{\mathrm{c}}$ of both high-temperature superconducting families (Figures~\ref{fig:art137:Rerg_r2}(b) and (c)),
even though this test set only contains compounds with $T_{\mathrm{c}} < 40$~K.
Conversely, the model trained on the cuprates tends to overestimate the $T_{\mathrm{c}}$
of low-$T_{\mathrm{c}}$ (Figure~\ref{fig:art137:Rerg_r2}(d)) and iron-based (Figure~\ref{fig:art137:Rerg_r2}(e)) superconductors.
This is a clear indication that superconductors from these groups have different factors determining their $T_{\mathrm{c}}$.
Interestingly, the family-specific models do not perform better than the general regression containing
all the data points: $R^2$ for the low-$T_{\mathrm{c}}$ materials is about $0.85$, for cuprates is just below $0.8$,
and for iron-based compounds is about $0.74$.
In fact, it is a purely geometric effect that
the combined model has the highest $R^2$.
Each group of superconductors contributes mostly to a distinct $T_{\mathrm{c}}$ range, and, as a result, the combined regression is better determined over longer temperature interval.

In order to reduce the number of predictors and increase the interpretability of these models without
significant detriment to their performance, a backward feature elimination process is again employed.
The procedure is very similar to the one described previously for the classification model,
with the only difference being that the reduction is guided by $R^2$ of the model, rather than the accuracy
(the procedure stops when $R^2$ drops by $3\%$).

The most important predictors for the four models (one general and three family-specific) together with
their importances are shown in Tables~\ref{tab:art137:Table1} and \ref{tab:art137:Table2}.
Differences in important predictors across the family-specific models reflect the fact that
distinct mechanisms are responsible for driving superconductivity among these groups.
The list is longest for the low-$T_{\mathrm{c}}$ superconductors, reflecting the eclectic nature of
this group.
Similar to the general regression model,
different branches are likely created for distinct sub-groups.
Nevertheless, some important predictors have straightforward interpretation.
As illustrated in Figure~\ref{fig:art137:Tc_atomWeigth}(a),
low average atomic weight is a necessary (albeit not sufficient) condition for
achieving high $T_{\mathrm{c}}$ among the low-$T_{\mathrm{c}}$ group.
In fact, the maximum $T_{\mathrm{c}}$ for a given weight roughly follows $1/\sqrt{m_A}$.
Mass plays a significant role in conventional superconductors
through the Debye frequency of phonons, leading to the well-known formula $T_{\mathrm{c}} \sim 1/\sqrt{m}$,
where $m$ is the ionic mass\footnote{See, for example, References~\cite{Maxwell_PR_1950,Reynolds_PR_1950,Reynolds_PR_1951}.}~\nocite{Maxwell_PR_1950,Reynolds_PR_1950,Reynolds_PR_1951}.
Other factors like density of states are also important,
which explains the spread in $T_{\mathrm{c}}$ for a given $m_A$.
Outlier materials clearly lying above the $\sim 1/\sqrt{m_A}$ line include
bismuthates and chloronitrates, suggesting the conventional electron-phonon mechanism is not driving
superconductivity in these materials.
Indeed, chloronitrates exhibit a very weak isotope effect~\cite{Kasahara_PSCC_2015}, though
some unconventional electron-phonon coupling could still be relevant for superconductivity~\cite{Yin_PRX_2013}.
Another important feature for low-$T_{\mathrm{c}}$ materials
is the average number of valence electrons.
This recovers the empirical relation first discovered by Matthias more than sixty years ago~\cite{Matthias_PR_1955}.
Such findings validate the ability of ML approaches
to discover meaningful patterns that encode true physical phenomena.

Similar $T_{\mathrm{c}}$-\vs-predictor plots reveal more interesting and subtle features.
A narrow cluster of materials with $T_{\mathrm{c}} > 20$~K emerges in the context of the mean covalent radii of compounds
(Figure ~\ref{fig:art137:Tc_atomWeigth}(b)) --- another
important predictor for low-$T_{\mathrm{c}}$ superconductors.
The cluster includes (left-to-right) alkali-doped C$_{60}$, MgB$_2$-related compounds, and bismuthates.
The sector likely characterizes a region of strong covalent bonding and corresponding high-frequency phonon modes
that enhance $T_{\mathrm{c}}$ (however, frequencies that are too high become irrelevant for superconductivity).
Another interesting relation appears in the context of the average number of $d$ valence electrons.
Figure~\ref{fig:art137:Tc_atomWeigth}(c) illustrates a fundamental bound on
$T_{\mathrm{c}}$ of all non-cuprate and non-iron-based superconductors.

A similar limit exists for cuprates based on the average number of unfilled orbitals (Figure ~\ref{fig:art137:Tc_atomWeigth}(d)).
It appears to be quite rigid --- several data points found above it on inspection are actually
incorrectly recorded entries in the database and were subsequently removed.
The connection between $T_{\mathrm{c}}$ and the average number of unfilled orbitals\footnote{The
number of unfilled orbitals refers to the
electron configuration of the substituent elements before combining to form oxides.
For example, Cu has one unfilled orbital ([Ar]$4s^23d^9$) and Bi has
three ([Xe]$4f^{14}6s^25d^{10}6p^3$).
These values are averaged per formula unit.}
may offer new insight into the mechanism for superconductivity in this family.
Known trends include higher $T_{\mathrm{c}}$'s for structures that
\textbf{i.} stabilize more than one superconducting Cu-O plane per unit cell
and \textbf{ii.} add more polarizable cations such as Tl$^{3+}$ and Hg$^{2+}$ between these planes.
The connection reflects these observations,
since more copper and oxygen per formula unit
leads to lower average number of unfilled orbitals (one for copper, two for oxygen).
Further, the lower-$T_{\mathrm{c}}$ cuprates typically consist of Cu$^{2-}$/Cu$^{3-}$-containing
layers stabilized by the addition/substitution of hard cations,
such as Ba$^{2+}$ and La$^{3+}$, respectively.
These cations have a large number of unfilled orbitals, thus increasing the compound's average.
Therefore, the ability of between-sheet cations
to contribute charge to the Cu-O planes may be indeed quite important.
The more polarizable the $A$ cation, the more electron density it can contribute
to the already strongly covalent Cu$^{2+}$--O bond.

\fig
\includegraphics[width=\linewidth]{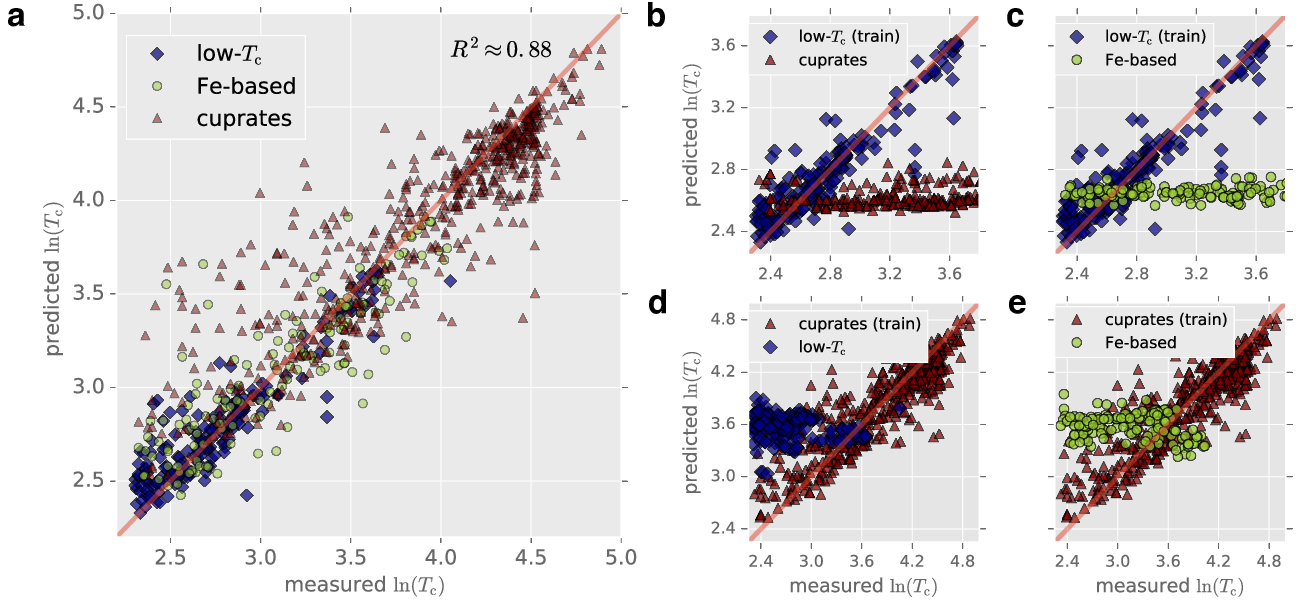}
\mycaption[Benchmarking of regression models predicting $\ln(T_{\mathrm{c}})$.]
{(\textbf{a}) Predicted
\vs\ measured $\ln(T_{\mathrm{c}})$ for the general regression model.
The test set comprises of a mix of low-$T_{\mathrm{c}}$, iron-based, and cuprate superconductors
with $T_{\mathrm{c}}>10$~K.
With an $R^2$ of about $0.88$, this one model can accurately predict
$T_{\mathrm{c}}$ for materials in different superconducting groups.
(\textbf{b} and \textbf{c}) Predictions of the regression model
trained solely on low-$T_{\mathrm{c}}$ compounds
for test sets containing cuprate and iron-based materials.
(\textbf{d} and \textbf{e}) Predictions of the regression model
trained solely on cuprates for test sets containing low-$T_{\mathrm{c}}$ and iron-based superconductors.
Models trained on a single group have no predictive power for materials from other groups.}
\label{fig:art137:Rerg_r2}
\efig

\fig
\includegraphics[width=\linewidth]{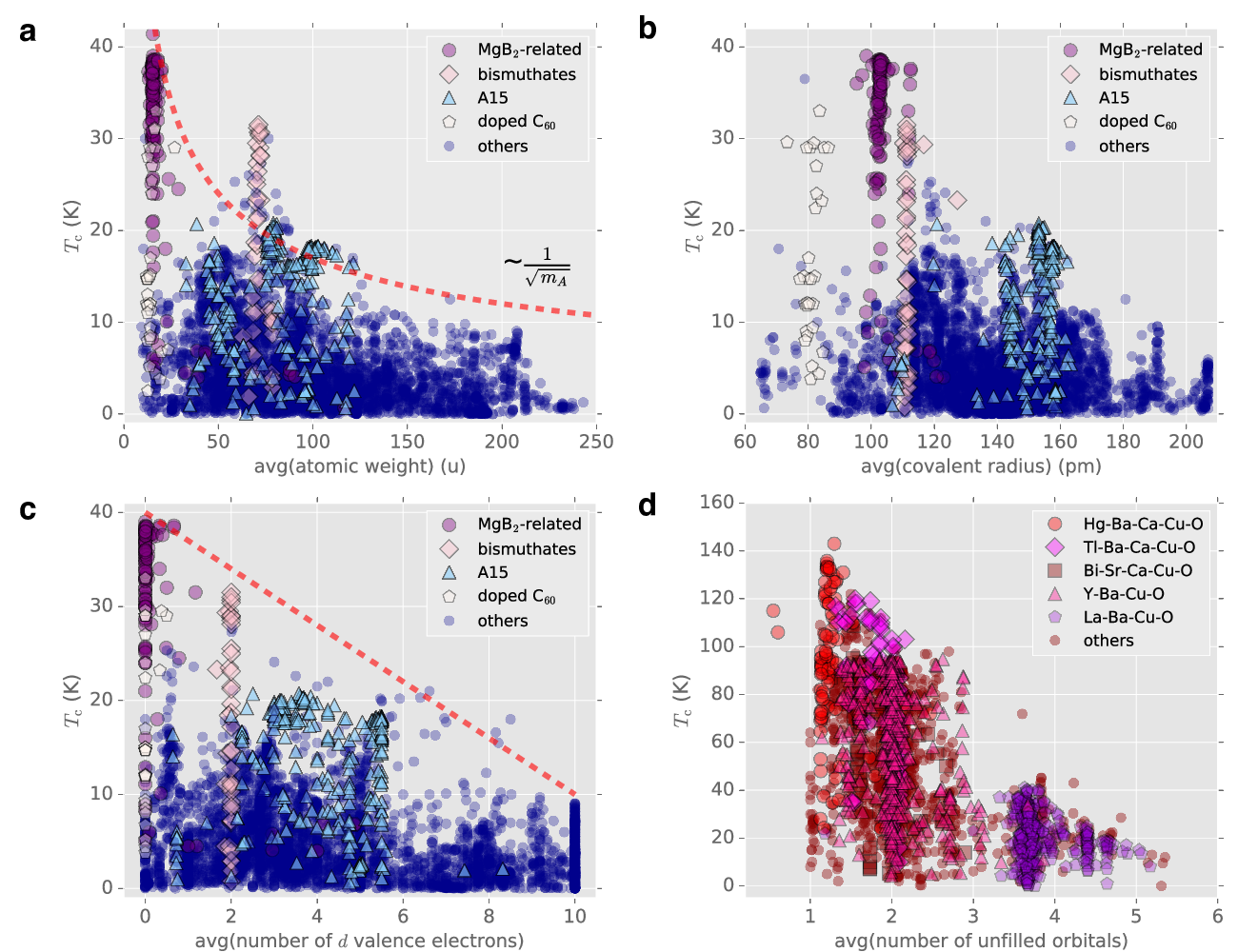}
\mycaption[Scatter plots of $T_{\mathrm{c}}$ for superconducting materials in the space of significant,
family-specific regression predictors.]
{For $4,000$ ``low-$T_{\mathrm{c}}$'' superconductors (\ie, non-cuprate and non-iron-based),
$T_{\mathrm{c}}$ is plotted
\vs\ the
(\textbf{a}) average atomic weight,
(\textbf{b}) average covalent radius, and
(\textbf{c}) average number of $d$ valence electrons.
The dashed red line in (\textbf{a}) is $\sim 1/\sqrt{m_A}$.
Having low average atomic weight and low average number of $d$ valence
electrons are necessary (but not sufficient) conditions for achieving high $T_{\mathrm{c}}$
in this group.
(\textbf{d}) Scatter plot of $T_{\mathrm{c}}$ for all known superconducting cuprates \vs\ the mean number of unfilled orbitals.
(\textbf{c} and \textbf{d}) suggest that the values of these predictors lead to
hard limits on the maximum achievable $T_{\mathrm{c}}$.}
\label{fig:art137:Tc_atomWeigth}
\efig

\boldsection{Including AFLOW.}
The models described previously demonstrate
surprising accuracy and predictive power, especially considering the difference between the
relevant energy scales of most Magpie predictors (typically in the range of eV) and superconductivity (meV scale).
This disparity, however, hinders the interpretability of the models,
\ie, the ability to extract meaningful physical correlations.
Thus, it is highly desirable to create accurate ML models with features based on
measurable macroscopic properties of the actual compounds
(\eg, crystallographic and electronic properties)
rather than composite elemental predictors.
Unfortunately, only a small subset of materials in SuperCon
is also included in the \ICSD:
about $1,500$ compounds in total, only about $800$ with finite $T_{\mathrm{c}}$,
and even fewer are characterized with \abinitiospace\ calculations\footnote{Most of the superconductors in
\protect\ICSD\ but not in \protect\AFLOW\ are non-stoichiometric/doped compounds, and thus not amenable to conventional \DFT\ methods.
For the others, \protect\AFLOW\ calculations were attempted but did not converge to a reasonable solution.}.
In fact, a good portion of known superconductors are disordered (off-stoichiometric) materials and
notoriously challenging to address with \DFT\ calculations.
Currently, much faster and efficient methods are becoming available~\cite{curtarolo:art110}
for future applications.

To extract suitable features, data is incorporated from
the \AFLOW\ Online Repositories --- a database of
\DFT\ calculations managed by the software package \AFLOW.
It contains information for the vast majority of compounds
in the \ICSD\ and about 550 superconducting materials.
In Reference~\onlinecite{curtarolo:art94}, several
ML models using a similar set of materials are presented.
Though a classifier shows good accuracy, attempts to create a
regression model for $T_{\mathrm{c}}$ led to disappointing results.
We verify that using Magpie predictors for the superconducting compounds in the \ICSD\
also yields an unsatisfactory regression model.
The issue is not the lack of compounds \textit{per se}, as
models created with randomly drawn subsets from SuperCon with
similar counts of compounds perform much better.
In fact, the
problem is the chemical sparsity of superconductors in the \ICSD, \ie,
the dearth of closely-related compounds (usually created by chemical substitution).
This translates to compound scatter in predictor space --- a challenging learning environment for the model.

The chemical sparsity in \ICSD\ superconductors is a significant hurdle, even when both sets of predictors
(\ie, Magpie and \AFLOW\ features) are combined via feature fusion.
Additionally, this approach alone neglects the majority of the $16,000$ compounds available via SuperCon.
Instead, we constructed separate models employing
Magpie and \AFLOW\ features, and then judiciously combined the results
to improve model metrics --- known as late or decision-level fusion.
Specifically, two independent classification models are developed,
one using the full SuperCon dataset and Magpie predictors, and another based on
superconductors in the \ICSD\ and \AFLOW\ predictors.
Such an approach can improve the recall, for example, in the case where we classify ``high-$T_{\mathrm{c}}$''
superconductors as those predicted by either model to be above-$T_{\mathrm{sep}}$.
Indeed, this is the case here where, separately, the models obtain a recall of $40\%$ and $ 66\%$, respectively, and
together achieve a recall of about $76\%$\footnote{These numbers are based on (a relatively small) test set benchmarking and their uncertainty is roughly $3\%$.}.
In this way, the models' predictions complement each other in a constructive way such that
above-$T_{\mathrm{sep}}$ materials missed by one model (but not the other) are now accurately classified.

\fig
\includegraphics[width=\linewidth]{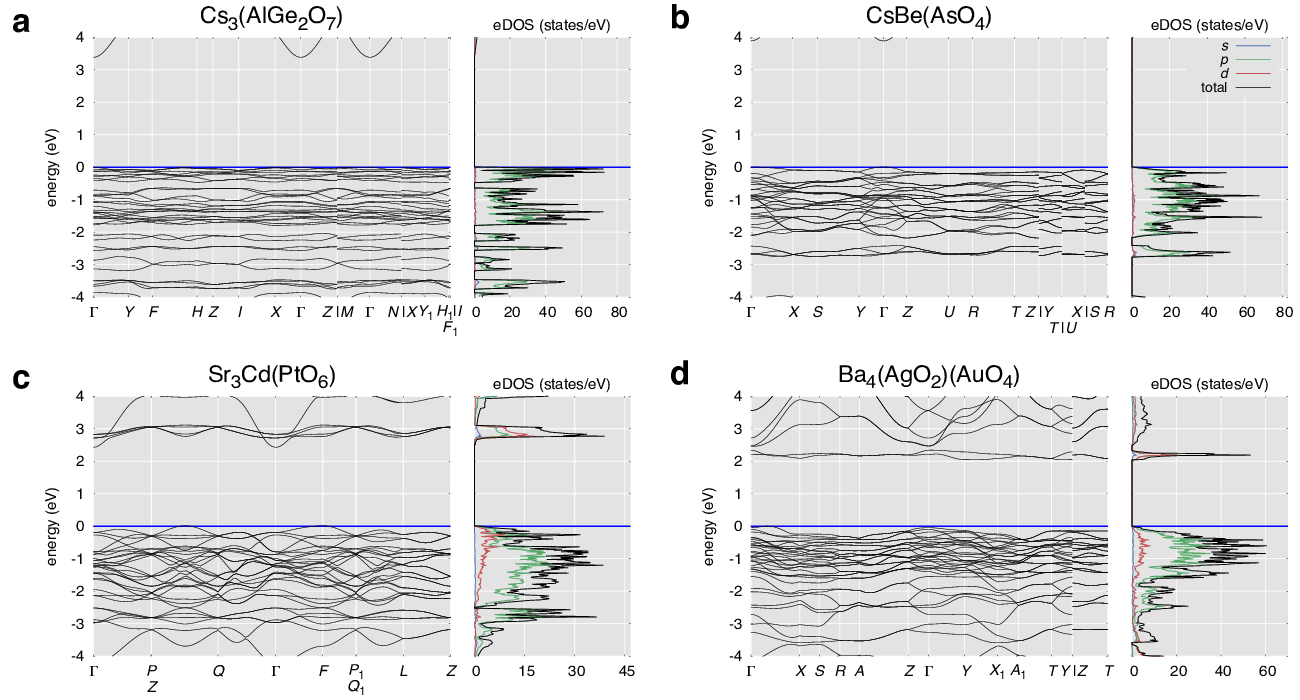}
\mycaption[DOS of four compounds identified by the ML algorithm as potential materials with $T_{\mathrm{c}} > 20$~K.]
{The partial DOS contributions from $s$, $p$ and $d$ electrons and total DOS are shown in blue, green, red, and black, respectively.
The large peak just below $E_F$ is a direct consequence of the flat band(s) present in all these materials.
These images were generated automatically via \AFLOW~\cite{curtarolo:art53}.
In the case of substantial overlap among \textbf{k}-point labels, the right-most label is offset below.}
\label{fig:art137:flat_bands}
\efig

\boldsection{Searching for new superconductors in the ICSD.}
As a final proof of concept demonstration,
the classification and regression models
described previously are integrated in one pipeline
and employed to screen the entire \ICSD\ database for candidate ``high-$T_{\mathrm{c}}$'' superconductors.
(Note that ``high-$T_{\mathrm{c}}$'' is a simple label,
the precise meaning of which can be adjusted.)
Similar tools power high-throughput screening workflows for materials with desired
thermal conductivity and magnetocaloric properties~\cite{curtarolo:art120,Bocarsly_ChemMat_2017}.
As a first step, the full set of Magpie predictors are generated for all
compounds in SuperCon.
A classification model similar to the one presented above is constructed,
but trained only on materials in SuperCon and not in the \ICSD\ (used
as an independent test set).
The model is then applied on the \ICSD\ set
to create a list of materials with predicted $T_{\mathrm{c}}$ above $10$~K.
Opportunities for model benchmarking are limited to those
materials both in the SuperCon and \ICSD\ datasets, though this test
set is shown to be problematic.
The set includes about 1,500 compounds, with $T_{\mathrm{c}}$ reported for only about half of them.
The model achieves an impressive accuracy of $0.98$, which is overshadowed by the fact that
$96.6\%$ of these compounds belong to the $T_{\mathrm{c}} < 10$~K class.
The precision, recall, and $F_{\mathrm{1}}$ scores are about $0.74$,
$0.66$, and $0.70$, respectively.
These metrics are lower than the estimates
calculated for the general classification model,
which is expected given that this set cannot
be considered randomly selected.
Nevertheless, the performance suggests a good opportunity to identify new candidate superconductors.

\tab
\mycaption[List of potential superconductors identified by the pipeline.]
{Also shown are their \ICSD\ numbers and symmetries.
Note that for some compounds there are several entries.
All of the materials contain oxygen.}
\tabvspace
{\small
\begin{tabular}{l|r|r}
compound & \ICSD\ & SYM  \\
\hline
CsBe(AsO$_4$) & 074027 & orthorhombic \\
RbAsO$_2$ & 413150 & orthorhombic \\
KSbO$_2$ & 411214 & monoclinic \\
RbSbO$_2$ & 411216 & monoclinic \\
CsSbO$_2$ & 059329 & monoclinic \\
\hline
AgCrO$_2$ & 004149/025624 & hexagonal \\
K$_{0.8}$(Li$_{0.2}$Sn$_{0.76}$)O$_2$ & 262638 & hexagonal \\
\hline
Cs(MoZn)(O$_3$F$_3$)& 018082 & cubic \\
\hline
Na$_3$Cd$_2$(IrO$_6$) & 404507 & monoclinic \\
Sr$_3$Cd(PtO$_6$) & 280518 & hexagonal \\
Sr$_3$Zn(PtO$_6$) & 280519 & hexagonal \\
\hline
(Ba$_5$Br$_2)$Ru$_2$O$_9$ & 245668 & hexagonal \\
\hline
Ba$_4$(AgO$_2$)(AuO$_4)$ & 072329 & orthorhombic \\
Sr$_5$(AuO$_4$)$_2$ & 071965 & orthorhombic \\
\hline
RbSeO$_2$F & 078399 & cubic \\
CsSeO$_2$F & 078400 & cubic \\
KTeO$_2$F & 411068 & monoclinic \\
\hline
Na$_2$K$_4$(Tl$_2$O$_6$) & 074956 & monoclinic \\
\hline
Na$_3$Ni$_2$BiO$_6$ & 237391 & monoclinic \\
Na$_3$Ca$_2$BiO$_6$ & 240975 & orthorhombic\\
\hline

CsCd(BO$_3$) & 189199 & cubic \\
\hline
K$_2$Cd(SiO$_4)$ & 083229/086917 & orthorhombic \\
Rb$_2$Cd(SiO$_4$) & 093879 & orthorhombic \\
K$_2$Zn(SiO$_4$) & 083227 & orthorhombic \\
K$_2$Zn(Si$_2$O$_6$) & 079705 & orthorhombic \\
\hline

K$_2$Zn(GeO$_4$) & 069018/085006/085007 & orthorhombic \\
(K$_{0.6}$Na$_{1.4})$Zn(GeO$_4)$ & 069166 & orthorhombic \\
K$_2$Zn(Ge$_2$O$_6$) & 065740 & orthorhombic \\
Na$_6$Ca$_3$(Ge$_2$O$_6$)$_3$ & 067315 & hexagonal \\
Cs$_3$(AlGe$_2$O$_7$) & 412140 & monoclinic \\
K$_4$Ba(Ge$_3$O$_9$) & 100203 & monoclinic \\
K$_{16}$Sr$_4$(Ge$_3$O$_9$)$_{4}$ & 100202 & cubic \\
K$_3$Tb[Ge$_3$O$_8$(OH)$_2$] & 193585 & orthorhombic \\
K$_3$Eu[Ge$_3$O$_8$(OH)$_2$] & 262677 & orthorhombic \\
\hline
KBa$_6$Zn$_4$(Ga$_7$O$_{21}$) & 040856 & trigonal \\
\end{tabular}}
\label{tab:art137:Table3}
\etab

Next in the pipeline, the list is fed into a random forest regression
model (trained on the entire SuperCon database)
to predict $T_{\mathrm{c}}$.
Filtering on the materials with $T_{\mathrm{c}} > 20$~K,
the list is further reduced to about 2,000 compounds.
This count may appear daunting, but should
be compared with the total number of compounds in the database --- about 110,000.
Thus, the method selects less than two percent of all materials,
which in the context of the training set (containing more than $20\%$ with ``high-$T_{\mathrm{c}}$''),
suggests that the model is not overly biased toward predicting high critical temperatures.

The vast majority of the compounds identified as
candidate superconductors are cuprates,
or at least compounds that contain copper and oxygen.
There are also some materials clearly related to the iron-based superconductors.
The remaining set has 35 members, and is composed of materials that are not obviously
connected to any high-temperature superconducting families (see Table~\ref{tab:art137:Table3})\footnote{For at least one compound
from the list --- Na$_3$Ni$_2$BiO$_6$ --- low-temperature measurements have been performed and no signs
of superconductivity were observed~\cite{Seibel_InChem_2013}.}~\nocite{Seibel_InChem_2013}.
None of them is predicted to have
$T_{\mathrm{c}}$ in excess of $40$~K, which is not surprising, given that no such instances exist in the training dataset. All contain oxygen --- also not a surprising result, since the group of
known superconductors with $T_{\mathrm{c}} > 20$~K is dominated by oxides.

The list comprises several distinct groups.
Most of the materials are insulators, similar to stoichiometric (and underdoped) cuprates that
generally require charge doping and/or pressure to drive these materials into a superconducting state.
Especially interesting are the compounds containing heavy metals (such as Au, Ir, Ru), metalloids (Se, Te),
and heavier post-transition metals (Bi, Tl), which are or could be pushed into interesting/unstable oxidation states.
The most surprising and non-intuitive of the compounds in the list are the silicates and the germanates.
These materials form corner-sharing SiO$_4$ or GeO$_4$ polyhedra, similar to quartz glass,
and also have counter cations with full or empty shells such as Cd$_2$$^+$ or K$^+$.
Converting these insulators to metals (and possibly superconductors) likely requires
significant charge doping. However, the similarity between these compounds and cuprates is meaningful.
In compounds like K$_2$CdSiO$_4$ or K$_2$ZnSiO$_4$,  K$_2$Cd (or K$_2$Zn) unit carries
a 4+ charge that offsets the (SiO$_4$)$^{4-}$ (or (GeO$_4$)$^{4-}$) charges.
This is reminiscent of the way Sr$_2$ balances the (CuO$_4$)$^{4-}$ unit in Sr$_2$CuO$_4$.
Such chemical similarities based on charge balancing and stoichiometry were likely identified and exploited by the ML algorithms.

The electronic properties calculated by \AFLOW\ offer additional insight into the results of the search, and suggest a possible connection among these candidate.
Plotting the electronic structure of the potential superconductors exposes a rather unusual feature shared
by almost all --- one or several (nearly) flat bands just below the energy of the highest occupied electronic state\footnote{The
flat band attribute is unusual for a superconducting material: the average DOS of the known superconductors in the \protect\ICSD\
(at least those available in the \protect\AFLOW\ Online Repositories) has no distinct features, demonstrating roughly uniform distribution of electronic states.
In contrast, the average DOS of the potential superconductors in Table~\ref{tab:art137:Table3} shows a sharp peak just below $E_{\mathrm{F}}$.
Also, most of the flat bands in the potential superconductors we discuss have a notable contribution from the oxygen $p$-orbitals.
Accessing/exploiting the potential strong instability this electronic structure feature creates can require significant charge doping.}.
Such bands lead to a large peak in the DOS (see Figure~\ref{fig:art137:flat_bands}) and
can cause a significant enhancement in $T_{\mathrm{c}}$.
Peaks in the DOS elicited by van Hove singularities can enhance $T_{\mathrm{c}}$
if sufficiently close to $E_{\mathrm{F}}$~\cite{Labbe_PRL_1967,Hirsch_PRL_1986,Dzyaloshinskii_JETPLett_1987}.
However, note that unlike typical van Hove points, a true flat band creates divergence
in the DOS (as opposed to its derivatives), which in turn leads to a critical temperature
dependence that is linear in the pairing interaction strength, rather than the usual exponential relationship
yielding lower $T_{\mathrm{c}}$~\cite{Kopnin_PRB_2011}.
Additionally, there is significant similarity
with the band structure and DOS of layered
BiS$_2$-based superconductors~\cite{Yazici_PSCC_2015}.

This band structure feature came as the surprising
result of applying the ML model.
It was not sought for, and, moreover,
no explicit information about the electronic band structure has been
included in these predictors.
This is in contrast to the algorithm presented in Reference~\onlinecite{Klintenberg_CMS_2013},
which was specifically designed to filter \ICSD\ compounds based on several preselected electronic structure features.

While at the moment it is not clear if some (or indeed any) of these compounds are really superconducting,
let alone with $T_{\mathrm{c}}$'s above 20~K,
the presence of this highly unusual electronic structure feature is encouraging.
Attempts to synthesize several of these compounds are already underway.

\subsection{Discussion}
Herein, several machine learning tools are developed to study the critical temperature of superconductors.
Based on information from the SuperCon database, initial coarse-grained
chemical features are generated using the Magpie software.
As a first application of ML methods, materials are divided into two classes depending on
whether $T_{\mathrm{c}}$ is above or below $10$~K.
A non-parametric random forest classification model is constructed
to predict the class of superconductors.
The classifier shows excellent performance, with out-of-sample accuracy and $F_{\mathrm{1}}$
score of about $92\%$.
Next,
several successful random forest regression models are created to predict the value of $T_{\mathrm{c}}$,
including separate models for three material sub-groups, \ie,
cuprate, iron-based, and low-$T_{\mathrm{c}}$ compounds.
By studying the importance of predictors for each family of superconductors,
insights are obtained about the
physical mechanisms driving superconductivity among the different groups.
With the incorporation of crystallographic-/electronic-based features
from the \AFLOW\ Online Repositories, the ML models are further improved.
Finally, we combined these models into one integrated pipeline, which is employed to search the entire
\ICSD\ database for new inorganic superconductors.
The model identified 35 oxides as candidate materials.
Some of these are chemically and structurally similar  to cuprates (even though no explicit structural information was provided during training of the model). Another feature that unites almost all of these materials is the presence of flat or nearly-flat bands just below the energy of the highest occupied electronic state.

In conclusion, this work demonstrates the important role
ML models can play in superconductivity research.
Records collected over several decades in SuperCon and other relevant databases can be consumed by ML models,
generating insights and promoting better understanding of the connection
between materials' chemistry/structure and superconductivity.
Application of sophisticated ML algorithms has the potential to dramatically accelerate
the search for candidate high-temperature superconductors.

\subsection{Methods}

\boldsection{Superconductivity data.}
The SuperCon database consists of two separate subsets: ``Oxide \& Metallic''
(inorganic materials containing metals, alloys, cuprate high-temperature superconductors, \etc)
and ``Organic'' (organic superconductors).
Downloading the entire inorganic materials dataset and removing compounds with
incompletely-specified chemical compositions leaves about $22,000$ entries.
If a single $T_{\mathrm{c}}$ record exists for a given material, it is taken to accurately reflect the critical temperature of this material.
In the case of multiple records for the same compound,
the reported material's $T_{\mathrm{c}}$'s are averaged, but only if
their standard deviation is less than $5$~K, and discarded otherwise.
This brings the total down to about $16,400$ compounds,
of which around $4,000$ have no critical temperature reported. Each entry in the set contains fields for the chemical composition,
$T_{\mathrm{c}}$, structure, and a journal reference to the information source.
Here, structural information is ignored as it is not always available.

There are occasional problems with the validity and consistency of some of the data.
For example, the database includes some reports based on tenuous experimental evidence and
only indirect signatures of superconductivity, as well as reports of inhomogeneous (surface, interfacial)
and nonequilibrium phases.
Even in cases of \textit{bona fide} bulk superconducting phases, important relevant variables
like pressure are not recorded.
Though some of the obviously erroneous records were removed from the data,
these issues were largely ignored
assuming their effect on the entire dataset to be relatively modest. The data cleaning and processing is carried out using the Python Pandas package for data analysis~\cite{Mckinney_Pandas_2012}.

\boldsection{Chemical and structural features.}
The predictors are calculated using the Magpie software \cite{magpie_software}.
It computes a set of 145 attributes
for each material, including:
\textbf{i.} stoichiometric features (depends only on the ratio of elements and
not the specific species);
\textbf{ii.} elemental property statistics: the mean, mean absolute deviation, range, minimum,
maximum, and mode of 22 different elemental properties
(\eg, period/group on the periodic table,
atomic number, atomic radii, melting temperature);
\textbf{iii.} electronic structure attributes: the average
fraction of electrons from the $s$, $p$, $d$ and $f$ valence shells among all
elements present; and
\textbf{iv.} ionic compound features that include whether it is possible to form an ionic
compound assuming all elements exhibit a single oxidation state.

ML models are also constructed with the
superconducting materials in the \AFLOW\ Online Repositories.
\AFLOW\ is a high-throughput \abinitiospace\ framework that manages density functional theory (\DFT)
calculations in accordance with the \AFLOW\ Standard~\cite{curtarolo:art104}.
The Standard ensures that the calculations and derived properties are empirical (reproducible), reasonably
well-converged, and above all, consistent (fixed set of parameters), a particularly attractive feature for ML modeling.
Many materials properties important for superconductivity have been calculated within the \AFLOW\ framework,
and are easily accessible through the \AFLOW\ Online Repositories.
The features are built with the following properties:
number of atoms, space group, density, volume, energy per atom, electronic entropy per atom, valence of the cell,
scintillation attenuation length, the ratios of the unit cell's dimensions, and Bader charges and volumes.
For the Bader charges and volumes (vectors), the following statistics
are calculated and incorporated:
the maximum, minimum, average, standard deviation, and range.

\boldsection{Machine learning algorithms.}
Once we have a list of relevant predictors, various ML models can be applied to the
data~\cite{Bishop_ML_2006,Hastie_StatLearn_2001}.
All ML algorithms in this work are
variants of the random forest method~\cite{randomforests}.
It is based on creating a set of individual decision trees (hence the ``forest''),
each built to solve the same classification/regression problem.
The model then combines their results, either by voting or averaging depending on the problem.
The deeper individual tree are, the more complex the relationships the model can learn,
but also the greater the danger of overfitting, \ie, learning
some irrelevant information or just ``noise''.
To make the forest more robust to overfitting, individual trees in the ensemble are
built from samples drawn with replacement (a bootstrap sample) from the training set.
In addition, when splitting a node during the construction of a tree, the model chooses the best split
of the data only considering a random subset of the features.

The random forest models above are developed using scikit-learn --- a powerful and efficient machine
learning Python library \cite{Pedregosa_JMLR_2011}.
Hyperparameters of these models include the number of trees in the forest,
the maximum depth of each tree, the minimum number of samples required to split an internal node,
and the number of features to consider when looking for the best split.
To optimize the classifier and the combined/family-specific regressors, the
GridSearch function in scikit-learn is employed, which generates and compares candidate models from a grid of parameter values.
To reduce computational expense, models are not optimized at each step of the backward feature selection process.

To test the influence of using log-transformed target variable $\ln(T_{\mathrm{c}})$,
a general regression model is trained and tested on raw $T_{\mathrm{c}}$ data.
This model is very similar to the one described in section ``Results'', and its $R^2$ value is fairly similar as well
(although comparing $R^2$ scores of models built using different target data can be misleading).
However, note the relative sparsity of data points in some $T_{\mathrm{c}}$ ranges, which makes the model susceptible to outliers.

\fig
\includegraphics[width=0.6\linewidth]{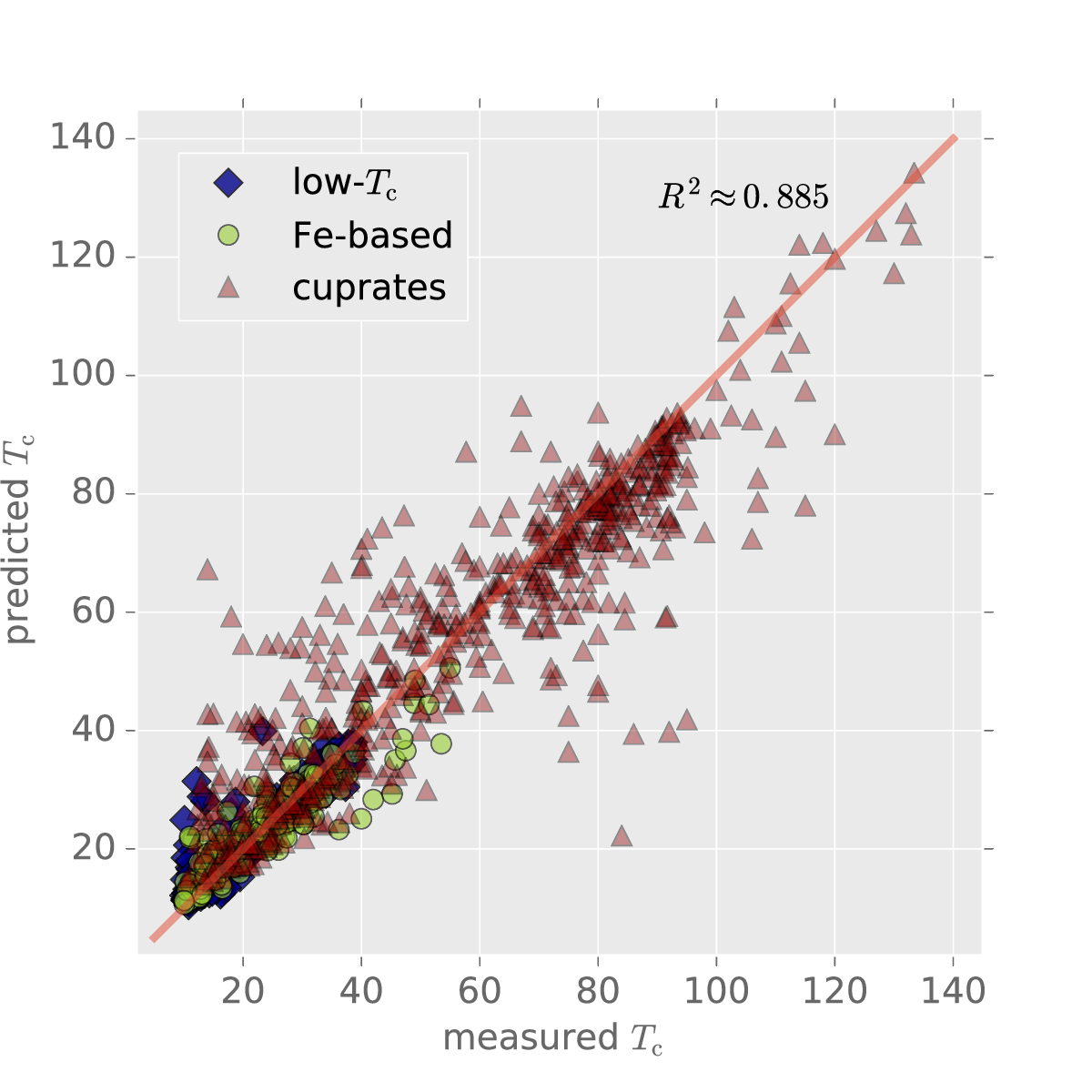}
\mycaption[Regression model predictions of $T_{\mathrm{c}}$.]
{Predicted
\vs\ measured $T_{\mathrm{c}}$ for  general regression model.
$R^2$ score is comparable to the one obtained testing regression modeling $\ln(T_{\mathrm{c}})$.}
\label{fig:art137:Regr_non_loc}
\efig

\fig
\includegraphics[width=\linewidth]{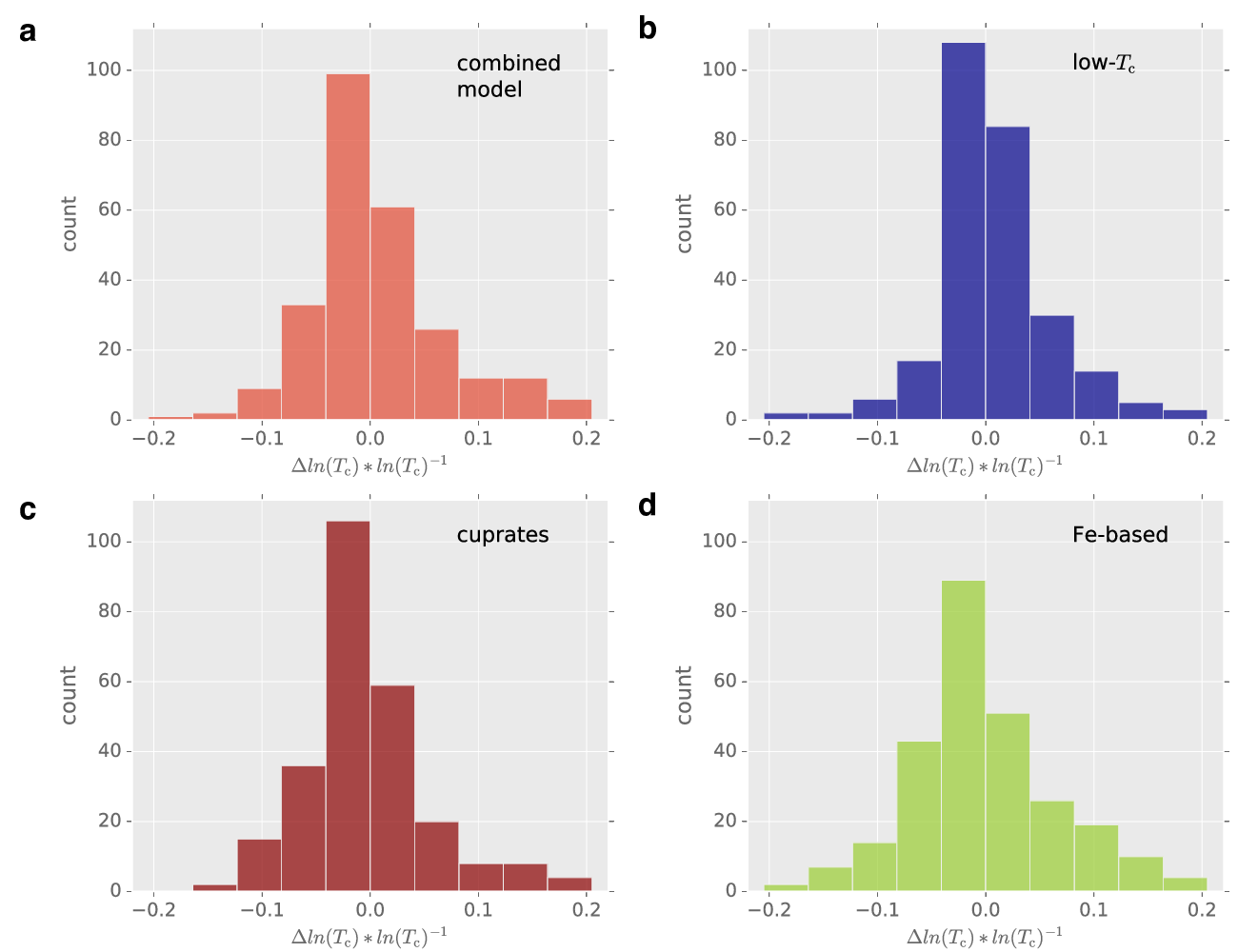}
\mycaption[Histograms of $\Delta\ln(T_{\mathrm{c}}) * \ln(T_{\mathrm{c}})^{-1}$ for the four regression models.]
{$\Delta\ln(T_{\mathrm{c}}) \equiv (\ln(T^{\mathrm{meas}}_{\mathrm{c}}) - \ln(T^{\mathrm{pred}}_{\mathrm{c}}))$
and $\ln(T_{\mathrm{c}}) \equiv \ln(T^{\mathrm{meas}}_{\mathrm{c}})$.}
\label{fig:art137:Regr_err}
\efig

\boldsection{Prediction errors of the regression models.}
Previously, several regression models were described,
each one designed to predict the critical temperatures of materials from different superconducting groups.
These models achieved an impressive $R^{2}$ score, demonstrating
good predictive power for each group.
However, it is also important to consider the accuracy of the predictions
for individual compounds (rather than on the aggregate set),
especially in the context of searching for new materials.
To do this, we calculate the prediction errors for about 300 materials from a test set.
Specifically, we consider the difference between the logarithm of the predicted and measured
critical temperature $[\ln(T^{\mathrm{meas}}_{\mathrm{c}})- \ln(T^{\mathrm{pred}}_{\mathrm{c}})]$
normalized by the value of $\ln(T^{\mathrm{meas}}_{\mathrm{c}})$
(normalization compensates the  different $T_{\mathrm{c}}$ ranges of different groups).
The models show comparable spread of errors.
The histograms of errors for the four models
(combined and three group-specific) are shown in Fig.~\ref{fig:art137:Regr_err}.
The errors approximately follow a normal distribution,
centered not at zero but at a small negative value.
This suggests the models are marginally biased, and on average tend to slightly underestimate $T_{\mathrm{c}}$.
The variance is comparable for all models, but largest for the model trained
and tested on iron-based materials, which also shows the smallest $R^2$.
Performance of this model is expected to benefit from a larger training set.

\boldsection{Data availability.} The superconductivity data used to generate the results
in this work can be downloaded from \url{https://github.com/vstanev1/Supercon}.
\clearpage
\section{High Throughput Thermal Conductivity of High Temperature Solid Phases: The Case of Oxide and Fluoride Perovskites}
\label{sec:art120}

This study follows from a collaborative effort described in Reference~\cite{curtarolo:art120}.

\subsection{Introduction}
\label{subsec:art120:Introduction}

High throughput \abinitio\ screening of materials is a new and rapidly
growing discipline~\cite{nmatHT}. Amongst the basic properties
of materials, thermal conductivity is a particularly relevant one.
Thermal management is a crucial factor to a vast range of technologies,
including power electronics, CMOS interconnects, thermoelectric energy
conversion, phase change memories, turbine thermal coatings and many
others~\cite{Cahill_APR_2014}. Thus, rapid determination
of thermal conductivity for large pools of compounds is a desirable
goal in itself, which may enable the identification of suitable compounds
for targeted applications. A few recent works have investigated thermal
conductivity in a high throughput fashion~\cite{aflowKAPPA,Seko_PRL_2015}.
A drawback of these studies is that they were restricted to use the
zero Kelvin phonon dispersions. This is often fine when the room temperature
phase is mechanically stable at 0~K. It however poses a problem
for materials whose room or high temperature phase is not the 0~K
structure: when dealing with structures exhibiting displacive distortions,
including temperature effects in the phonon spectrum is a crucial
necessity.

Such a phenomenon often happens for perovskites. Indeed, the perovskite
structure can exhibit several distortions from the ideal cubic lattice,
which is often responsible for rich phase diagrams. When the structure
is not stable at low temperatures, a simple computation of the phonon
spectrum using forces obtained from density functional theory and
the finite displacement method yields imaginary eigenvalues. This
prevents us from assessing the mechanical stability of those compounds
at high temperatures or calculating their thermal conductivity. Moreover,
taking into account finite-temperature effects in phonon calculations
is currently a very demanding task, especially for a high-throughput
investigation.

In this study, we are interested in the \textit{high-temperature}
properties of perovskites, notably for thermoelectric applications.
For this reason, we focus on perovskites with the highest symmetry
cubic structure, which are most likely to exist at high temperatures
\cite{Landau_CTP5_SP_1969,Howard_ActaCrisA_2005,Thomas_PRL_1968,Cochran_PSSB_1968,Angel_PRL_2005}.
We include the effects of anharmonicity in our \abinitio\ calculations
of mechanical and thermal properties.

\subsection{Finite-temperature calculations of mechanical stability and thermal properties}
\label{subsec:art120:Finite-T_calculations}

Recently, several methods have been developed to deal with anharmonic
effects at finite temperatures in solids~\cite{Souvatzis_PRL_2008,Hellman_PRB_2011,Hellman_PRB_2013,Errea_PRL_2013,Tadano_PRB_2015,VanRoekeghem_ARXIV_2016}.
In this study, we use the method presented in Reference~\cite{VanRoekeghem_ARXIV_2016}
to compute the temperature-dependent interatomic force constants,
which uses a regression analysis of forces from density functional
theory coupled with a harmonic model of the quantum canonical ensemble.
This is done in an iterative way to achieve self-consistency of the
phonon spectrum.
The workflow is summarized in Figure~\ref{fig:art120:finite-T-phonon}.
In the following (in particular Section~\ref{subsec:art120:PCA-regression}),
it will be referred as ``SCFCS'' -- standing for self-consistent
force constants. As a trade-off between accuracy and throughput, we
choose a 3x3x3 supercell and a cutoff of 5~\AA\ for the third order
force constants. Special attention is paid to the computation of the
thermal displacement matrix~\cite{VanRoekeghem_ARXIV_2016}, due to the imaginary
frequencies that can appear during the convergence process, as well
as the size of the supercell that normally prevents us from sampling
the usual soft modes at the corners of the Brillouin zone (see Supplementary Material of Reference~\cite{curtarolo:art120}).
This allows us to assess the stability at 1000~K of the
391 hypothetical compounds mentioned in Section~\ref{subsec:art120:Introduction}.
Among this set, we identify 92 mechanically stable compounds, for
which we also check the stability at 300~K. The phonon spectra of
the stable compounds are provided in the Supplementary Material of Reference~\cite{curtarolo:art120}.
Furthermore, we compute the thermal conductivity using the finite temperature force
constants and the full solution of the Boltzmann transport equation
as implemented in the ShengBTE code~\cite{Li_ShengBTE_CPC_2014}.

\fig
\includegraphics[width=0.6\linewidth]{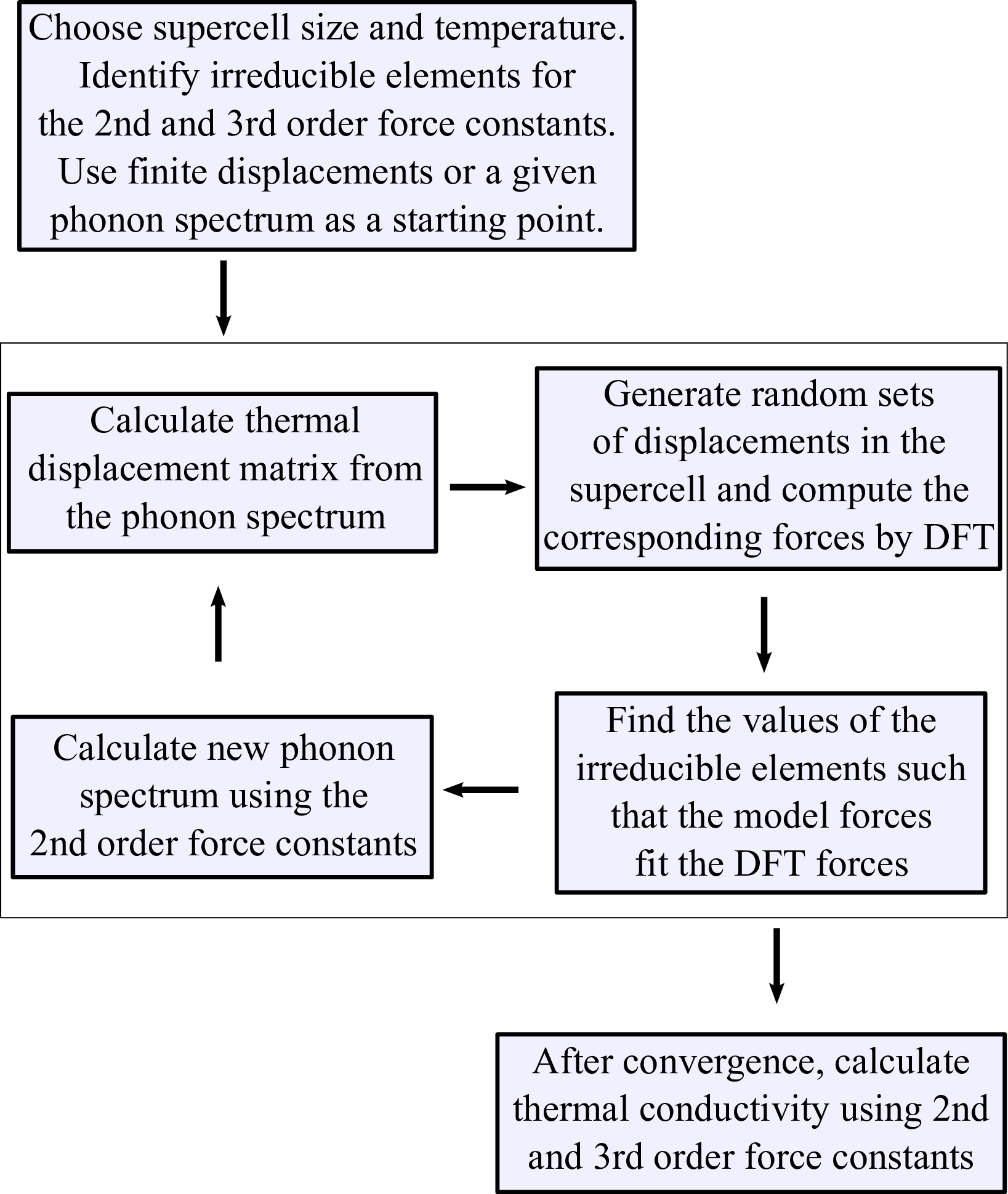}
\mycaption{Workflow of the method used to calculate the phonon spectrum and thermal
conductivity including finite-temperature anharmonic effects.}
\label{fig:art120:finite-T-phonon}
\efig

We list the stable compounds and their thermal conductivities in Table
\ref{tab:art120:List-of-perovskites}.
Remarkably, this list contains 37
perovskites that have been reported experimentally in the ideal cubic
structure (see References in Table~\ref{tab:art120:List-of-perovskites}),
which lends support to our screening method.
On the other hand, we
also find that 11 compounds are reported only in a non-perovskite
form. This is not necessarily indicative of mechanical instability,
but instead suggests thermodynamical stability may be an issue for
these compounds, at least near this temperature and pressure. 36 compounds
remain unreported experimentally in the literature to our knowledge.
Thus, by screening only for mechanical stability at high-temperatures,
we reduce the number of potential new perovskites by a factor of 10.
Furthermore, we find that 50 of them are mechanically stable in the
cubic form close to room temperature.

Of the full list of perovskites, only a few measurements of thermal
conductivity are available in the literature. They are displayed in
parentheses in Table~\ref{tab:art120:List-of-perovskites} along with their
calculated values. Our method tends to slightly underestimate the
value of the thermal conductivity, due to the compromises we made
to limit the computational cost of the study (see Supplementary Material of Reference~\cite{curtarolo:art120}).
This discrepancy could also be partially related to the electronic
thermal conductivity, which was not subtracted in the measurements.
Still, we expect the order of magnitude of the thermal conductivity
and the relative classification of different materials to be consistent.
More importantly, this large dataset allows us to analyze the global
trends driving thermal conductivity. These trends are discussed in
Section~\ref{subsec:art120:Descriptors}.

\newcommand{\fluoridestabfootone}{
AuMgF$_{3}$ was mentioned theoretically in Reference~\cite{Uetsuji_TJSME_2006}.}
\newcommand{\fluoridestabfoottwo}{
The thermal diffusivity of BaLiF$_{3}$ was measured at 300~K in
Reference~\cite{Duarte_MSEB_1994} as $\alpha$=0.037~cm$^{2}$s$^{-1}$.}

\clearpage

\tab
\mycaption[List of cubic perovskites found to be mechanically stable at 1000~K
and their corresponding computed lattice thermal conductivity (in
W/m/K).]
{We also report the computed lattice thermal conductivity at
300~K (in W/m/K) when we obtain stability at that temperature. We
highlight in blue the compounds that are experimentally reported in
the ideal cubic perovskite structure, and in red those that are reported
only in non-perovskite structures (references provided in the table).
When no reference is provided, no mention of the compound in this
stoichiometry has been found in the experimental literature. Experimental
measurements of the thermal conductivity are reported in parentheses,
and in italics when the structure is not cubic.}
\tabvspace
\resizebox{\linewidth}{!}{
\begin{tabular}{l|r|r|r|r|rcl|r|r|r|r|rcl|r|r|r|r|r}
 & $\kappa_{1000}$ &  & $\kappa_{300}$ &  & References &  &  & $\kappa_{1000}$ &  & $\kappa_{300}$ &  & References &  &  & $\kappa_{1000}$ &  & $\kappa_{300}$ &  & References\tabularnewline
\cline{1-6} \cline{8-13} \cline{15-20}
\textcolor{blue}{CaSiO$_{3}$} & 4.89 &  &  &  & \cite{Komabayashi_EPSL_2007} &  & CdYF$_{3}$ & 1.29 &  & 3.51 &  &  &  & TlOsF$_{3}$ & 0.62 &  & 0.95 &  & \tabularnewline
\textcolor{blue}{RbTaO$_{3}$} & 3.61 &  &  &  & \cite{Lebedev_PhysSolStat_2015} &  & \textcolor{blue}{RbCaF$_{3}$} & 1.15 &  & 2.46 & (3.2) & \cite{Ludekens_ActaCrist_1952,Ridou_Ferroelectrics_1976,Martin_Phonons_1976} &  & InZnF$_{3}$ & 0.61 &  & 1.86 &  & \tabularnewline
\textcolor{blue}{NaTaO$_{3}$} & 3.45 &  &  &  & \cite{Kennedy_JPCM_1999} &  & HgInF$_{3}$ & 1.15 &  & 3.85 &  &  &  & \textcolor{blue}{CsCdF$_{3}$} & 0.59 &  & 1.73 &  & \cite{Rousseau_PRB_1975}\tabularnewline
\textcolor{red}{CuCF$_{3}$} & 3.32 &  & 8.79 &  & \cite{Zanardi_JACS_2011} &  & AlFeF$_{3}$ & 1.14 &  &  &  &  &  & AlMgF$_{3}$ & 0.56 &  &  &  & \tabularnewline
\textcolor{blue}{SrSiO$_{3}$} & 3.23 &  & 10.10 &  & \cite{Xiao_AM_2013} &  & \textcolor{blue}{PbHfO$_{3}$} & 1.12 &  &  &  & \cite{Kwapulinski_JPCM_1994} &  & AuZnF$_{3}$ & 0.53 &  &  &  & \tabularnewline
\textcolor{blue}{NaNbO$_{3}$} & 3.05 &  &  & (\textit{1.5}) & \cite{Shirane_PR_1954,Mishra_PRB_2011,Tachibana_APL_2008} &  & \textcolor{blue}{AgMgF$_{3}$} & 1.11 &  &  &  & \cite{Portier_CRASC_1970} &  & InOsF$_{3}$ & 0.52 &  &  &  & \tabularnewline
\textcolor{blue}{BaHfO$_{3}$} & 3.04 &  (4.5) & 8.26 & (10.4) & \cite{Maekawa_BaHfO3_SrHfO3_JAC_2006} &  & ZnScF$_{3}$ & 1.10 &  & 3.66 &  &  &  & \textcolor{blue}{RbSrF$_{3}$} & 0.51 &  &  &  & \cite{Pies_Landolt_Bornstein_1973}\tabularnewline
\textcolor{blue}{KNbO$_{3}$} & 2.94 &  &  & (\textit{10}) & \cite{Shirane_PR_1954,Tachibana_APL_2008} &  & \textcolor{blue}{RbFeF$_{3}$} & 1.09 &  & 4.62 &  & \cite{Kestigian_IC_1966} &  & \textcolor{blue}{CsSrF$_{3}$} & 0.50 &  & 1.13 &  & \cite{Pies_Landolt_Bornstein_1973}\tabularnewline
\textcolor{red}{TlTaO$_{3}$} & 2.86 &  &  &  & \cite{Ramadass_SSC_1975} &  & \textcolor{black}{TlMgF$_{3}$} & 1.06 &  & 3.42 &  & \cite{Arakawa_JPCM_2006} &  & BeYF$_{3}$ & 0.48 &  & 2.34 &  & \tabularnewline
\textcolor{blue}{AgTaO$_{3}$} & 2.77 &  &  &  & \cite{Kania_PT_1981,Pawelczyk_PT_1987} &  & \textcolor{blue}{KCaF$_{3}$} & 1.06 &  &  &  & \cite{Demetriou_SSI_2005} &  & BeScF$_{3}$ & 0.48 &  & 1.59 &  & \tabularnewline
\textcolor{blue}{KMgF$_{3}$} & 2.74 &  & 8.25 & (10) & \cite{Wood_JACryst_2002,Martin_Phonons_1976} &  & HgScF$_{3}$ & 1.01 &  & 5.42 &  &  &  & \textcolor{blue}{TlCdF$_{3}$} & 0.44 &  &  &  & \cite{Rousseau_PRB_1975}\tabularnewline
\textcolor{red}{GaTaO$_{3}$} & 2.63 &  &  &  & \cite{Xu_Thesis_2000,Armiento_PRB_2011,Castelli_EES_2012} &  & \textcolor{blue}{CsCaF$_{3}$} & 0.98 &  & 3.03 &  & \cite{Rousseau_SSC_1981} &  & \textcolor{blue}{RbHgF$_{3}$} & 0.43 &  &  &  & \cite{Hoppe_ZAAC_1969}\tabularnewline
\textcolor{blue}{BaTiO$_{3}$} & 2.51 &  & 4.99 & (\textit{4-5}) & \cite{Tachibana_APL_2008,Strukov_JPCM_2003} &  & AuMgF$_{3}$ & 0.96 &  &  &  & \tablefootnote{\fluoridestabfootone} &  & PdYF$_{3}$ & 0.43 &  & 0.99 &  & \tabularnewline
\textcolor{blue}{PbTiO$_{3}$} & 2.42 &  &  & (\textit{5}) & \cite{Tachibana_APL_2008} &  & InMgF$_{3}$ & 0.96 &  & 3.53 &  &  &  & AlZnF$_{3}$ & 0.39 &  &  &  & \tabularnewline
\textcolor{blue}{SrTiO$_{3}$} & 2.36 & (4) & 6.44 & (10.5) & \cite{Muta_JAC_2005,Popuri_RSCA_2014,Yamanaka_JSSC_2004} &  & \textcolor{blue}{RbZnF$_{3}$} & 0.91 &  & 2.64 &  & \cite{Daniel_PRB_1995} &  & \textcolor{black}{KHgF$_{3}$} & 0.37 &  &  &  & \cite{Hoppe_ZAAC_1969}\tabularnewline
\textcolor{blue}{SrHfO$_{3}$} & 2.20 &  (\textit{2.7}) &  & (\textit{5.2}) & \cite{Kennedy_PRB_1999,Yamanaka_JSSC_2004} &  & ZnInF$_{3}$ & 0.88 &  & 1.89 &  &  &  & \textcolor{red}{RbSnF$_{3}$} & 0.37 &  & 0.82 &  & \cite{Tran_JSSC_2014}\tabularnewline
\textcolor{blue}{BaZrO$_{3}$} & 2.13 &  (2.9) & 5.61 &  (5.2) & \cite{Yamanaka_JAC_2003} &  & \textcolor{black}{BaSiO$_{3}$} & 0.87 &  &  &  & \cite{Yusa_AM_2007} &  & ZnBiF$_{3}$ & 0.37 &  & 1.29 &  & \tabularnewline
XeScF$_{3}$ & 1.87 &  & 4.40 &  &  &  & TlCaF$_{3}$ & 0.86 &  &  &  &  &  & \textcolor{blue}{CsHgF$_{3}$} & 0.37 &  & 1.00 &  & \cite{Hoppe_ZAAC_1969}\tabularnewline
HgYF$_{3}$ & 1.84 &  & 5.37 &  &  &  & CdScF$_{3}$ & 0.85 &  & 2.37 &  &  &  & \textcolor{red}{KSnF$_{3}$} & 0.35 &  &  &  & \cite{Tran_JSSC_2014}\tabularnewline
\textcolor{blue}{AgNbO$_{3}$} & 1.79 &  &  &  & \cite{Lukaszewski_PT_1983,Sciau_JPCM_2004} &  & XeBiF$_{3}$ & 0.82 &  & 2.13 &  &  &  & CdBiF$_{3}$ & 0.33 &  & 0.98 &  & \tabularnewline
\textcolor{red}{TlNbO$_{3}$} & 1.75 &  &  &  & \cite{Ramadass_SSC_1975} &  & \textcolor{blue}{AgZnF$_{3}$} & 0.80 &  &  &  & \cite{Portier_CRASC_1970} &  & \textcolor{black}{RbPbF$_{3}$} & 0.32 &  &  &  & \cite{Yamane_SSI_2008}\tabularnewline
\textcolor{blue}{KFeF$_{3}$} & 1.72 &  & 6.37 & (3.0) & \cite{Okazaki_JPSJ_1961,Suemune_kappa_JPSJ_1964} &  & PdScF$_{3}$ & 0.79 &  & 1.63 &  &  &  & BeAlF$_{3}$ & 0.30 &  & 1.70 &  & \tabularnewline
SnSiO$_{3}$ & 1.66 &  & 4.22 &  & \cite{Clark_IC_2001,Armiento_PRB_2014} &  & \textcolor{blue}{KCdF$_{3}$} & 0.75 &  &  &  & \cite{Hidaka_SSC_1977,Hidaka_PT_1990} &  & \textcolor{red}{KPbF$_{3}$} & 0.30 &  &  &  & \cite{Hull_JPCM_1999}\tabularnewline
\textcolor{red}{PbSiO$_{3}$} & 1.66 &  & 3.69 &  & \cite{Mackay_MM_1952,Xiao_AM_2012} &  & \textcolor{blue}{BaLiF$_{3}$} & 0.73 &  & 2.21 & \tablefootnote{\fluoridestabfoottwo} & \cite{Mortier_SSC_1994,Duarte_MSEB_1994} &  & CsBaF$_{3}$ & 0.29 &  &  &  & \tabularnewline
\textcolor{black}{AuNbO$_{3}$} & 1.56 &  &  &  & \cite{Wu_AngChemInt_2013} &  & HgBiF$_{3}$ & 0.72 &  & 2.37 &  &  &  & InCdF$_{3}$ & 0.29 &  &  &  & \tabularnewline
\textcolor{red}{CaSeO$_{3}$} & 1.42 &  &  &  & \cite{Wildner_NJMA_2007} &  & ZnAlF$_{3}$ & 0.72 &  & 1.92 &  &  &  & BaCuF$_{3}$ & 0.28 &  &  &  & \tabularnewline
\textcolor{red}{NaBeF$_{3}$} & 1.40 &  & 2.53 &  & \cite{ODaniel_NJMMAA_1945,Roy_JACerS_1953} &  & GaZnF$_{3}$ & 0.69 &  &  &  &  &  & \textcolor{red}{TlSnF$_{3}$} & 0.27 &  & 0.63 &  & \cite{Foulon_EJSSIC_1993}\tabularnewline
\textcolor{blue}{RbMgF$_{3}$} & 1.37 &  & 4.54 &  & \cite{Shafer_JoPACoS_1969} &  & \textcolor{blue}{RbCdF$_{3}$} & 0.68 &  & 1.46 &  & \cite{Rousseau_PRB_1975} &  & \textcolor{blue}{TlHgF$_{3}$} & 0.26 &  &  &  & \cite{Hebecker_Naturwissenschaften_1973}\tabularnewline
GaMgF$_{3}$ & 1.34 &  & 2.11 &  &  &  & GaRuF$_{3}$ & 0.67 &  &  &  &  &  & CdSbF$_{3}$ & 0.26 &  &  &  & \tabularnewline
\textcolor{blue}{KZnF$_{3}$} & 1.33 &  & 4.15 &  (5.5) & \cite{Suemune_JPSJ_1964,Martin_Phonons_1976} &  & \textcolor{black}{CsZnF$_{3}$} & 0.67 &  & 1.12 &  & \cite{Longo_JSSC_1969} &  & \textcolor{blue}{TlPbF$_{3}$} & 0.22 &  &  &  & \cite{Buchinskaya_RCR_2004}\tabularnewline
ZnYF$_{3}$ & 1.32 &  & 3.72 &  &  &  & \textcolor{black}{TlZnF$_{3}$} & 0.64 &  & 1.96 &  & \cite{Babel_TlZnF3_1967} &  &  &  &  &  &  & \tabularnewline
\end{tabular}}
\label{tab:art120:List-of-perovskites}
\etab

\clearpage

We also investigate the (potentially) negative thermal expansion of
these compounds. Indeed, the sign of the coefficient of thermal expansion
$\alpha_{\text{V}}$ is the same as the sign of the weighted Gr\"{u}neisen
parameter $\gamma$, following $\alpha_{\text{V}}=\frac{\gamma c_{\text{V}}\rho}{K_{\text{T}}}$,
where $K_{\text{T}}$ is the isothermal bulk modulus, $c_{\text{V}}$ is the isochoric
heat capacity and $\rho$ is the density~\cite{Gruneisen_AnnPhys_1912,ashcroft_mermin}.
The weighted Gr\"{u}neisen parameter is obtained by summing the contributions
of the mode-dependent Gr\"{u}neisen parameters: $\gamma=\sum\gamma_{i}c_{Vi}/\sum c_{Vi}$.
Finally the mode-dependent parameters are related to the volume variation
of the mode frequency $\omega_{i}$ via $\gamma_{i}=-(V/\omega_{i})(\partial\omega_{i}/\partial V)$.
In our case, we calculate those parameters directly using the second
and third order force constants at a given temperature~\cite{Fabian_PRL_1997,Broido_PRB_2005,Hellman_PRB_2013}:
\begin{equation}
\gamma_{m}=-\frac{1}{6\omega_{m}^{2}}\sum_{ijk\alpha\beta\gamma}\frac{\epsilon_{mi\alpha}^{*}\epsilon_{mj\beta}}{\sqrt{M_{i}M_{j}}}r_{k}^{\gamma}\Psi_{ijk}^{\alpha\beta\gamma}e^{i\mathbf{q}\cdot\mathbf{r}_{j}}
\end{equation}

This approach has been very successful in predicting the thermal expansion
behavior in the empty perovskite ScF$_{3}$~\cite{VanRoekeghem_ARXIV_2016},
which switches from negative to positive around 1100~K~\cite{Greve_JACS_2010}.
In our list of filled perovskites, we have found only two candidates
with negative thermal expansion around room temperature: TlOsF$_{3}$
and BeYF$_{3}$, and none at 1000~K.
This shows that filling the perovskite structure is probably detrimental to the negative thermal
expansion.

We also examine the evolution of the thermal conductivity as a function
of temperature, for the compounds that are mechanically stable at
300~K and 1000~K. There is substantial evidence that the thermal
conductivity in cubic perovskites generally decreases more slowly
than the model $\kappa\propto T^{-1}$ behavior~\cite{Peierls_AnnPhys_1929,Roufosse_JGPR_1974}
at high temperatures, in contrast to the thermal conductivity of \eg\
Si or Ge that decreases faster than $\kappa\propto T^{-1}$~\cite{Glassbrenner_PR_1964}.
This happens for instance in SrTiO$_{3}$~\cite{Muta_JAC_2005,Popuri_RSCA_2014},
KZnF$_{3}$~\cite{Suemune_JPSJ_1964,Martin_Phonons_1976},
KMgF$_{3}$~\cite{Martin_Phonons_1976}, KFeF$_{3}$~\cite{Suemune_kappa_JPSJ_1964},
RbCaF$_{3}$~\cite{Martin_Phonons_1976}, BaHfO$_{3}$
\cite{Maekawa_BaHfO3_SrHfO3_JAC_2006}, BaSnO$_{3}$~\cite{Maekawa_BaSnO3_JAC_2006}
and BaZrO$_{3}$~\cite{Yamanaka_JAC_2003}.
We also predicted an anomalous behavior in ScF$_{3}$ using \abinitio\ calculations,
tracing its origin to the important anharmonicity of the soft modes
\cite{VanRoekeghem_ARXIV_2016}.
Figure~\ref{fig:art120:Kappa} displays several experimentally
measured thermal conductivities from the literature on a logarithmic
scale, along with the results of our high-throughput calculations.
As discussed above, the absolute values of the calculated thermal
conductivities are generally underestimated, but their relative magnitude
and the overall temperature dependence are generally consistent. Although
the behavior of the thermal conductivity $\kappa(T)$ is in general
more complex than a simple power-law behavior, we model the deviation
to the $\kappa\propto T^{-1}$ law by using a parameter $\alpha$
that describes approximately the temperature-dependence of $\kappa$
between 300~K and 1000~K as $\kappa\propto T^{-\alpha}$. For
instance, in Figure~\ref{fig:art120:Kappa}, KMgF$_{3}$ appears to have
the fastest decreasing thermal conductivity with $\alpha=0.9$ both
from experiment and calculations, while SrTiO$_{3}$ is closer to
$\alpha=0.6$. At present, there are too few experimental measurements
of the thermal conductivities in cubic perovskites to state that the
$\kappa\propto T^{-\alpha}$ behavior with $\alpha<1$ is the general
rule in this family. However, the large number of theoretical predictions
provides a way to assess this trend. Of the 50 compounds that we found
to be mechanically stable at room temperature, we find a mean $\alpha\simeq0.85$,
suggesting that this behavior is likely general and correlated to
structural characteristics of the perovskites.

\fig
\includegraphics[width=8.5cm]{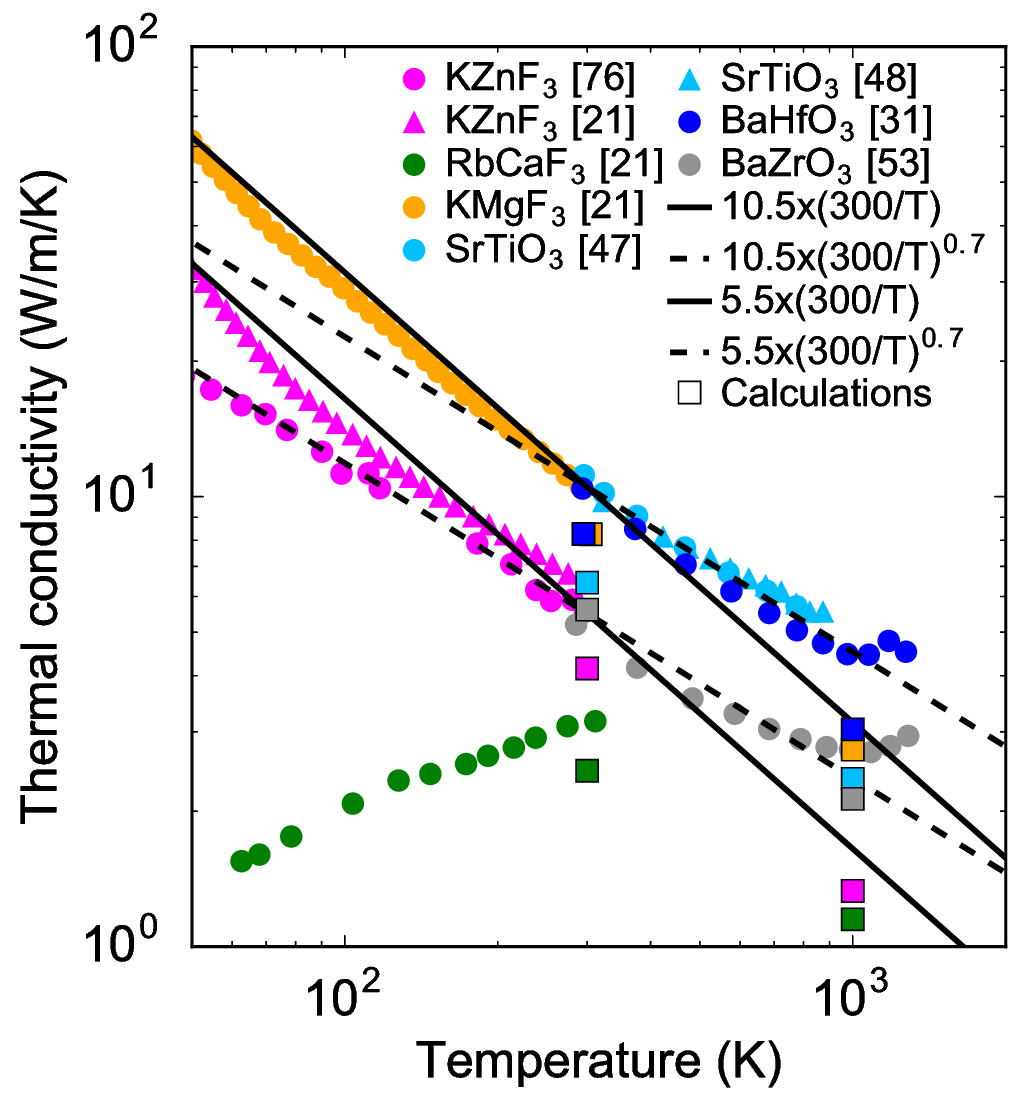}
\mycaption{A comparison between total thermal conductivities from References~\cite{Suemune_JPSJ_1964,Martin_Phonons_1976,Muta_JAC_2005,Popuri_RSCA_2014,Maekawa_BaHfO3_SrHfO3_JAC_2006,Yamanaka_JAC_2003},
high-throughput calculations of the lattice thermal conductivity at
300~K and 1000~K, and model behaviors in $\kappa\propto T^{-1}$
and $\kappa\propto T^{-0.7}$.}
\label{fig:art120:Kappa}
\efig

\subsection{Accelerating the discovery of stable compounds at high temperature}
\label{subsec:art120:PCA-regression}

Through brute-force calculations of the initial list of 391 compounds,
we extracted 92 that are mechanically stable at 1000~K. However,
this type of calculation is computationally expensive. Thus, it is
desirable for future high-throughput searches of other material classes
to define a strategy for exploring specific parts of the full combinatorial
space. In this section, we propose and test such a strategy based
on an iterative machine-learning scheme using principal component
analysis and regression.

We begin by calculating the second order force constants $\Phi_{0\text{~K}}$
of all compounds using the finite displacement method, which is more
than an order of magnitude faster than finite-temperature calculations.
This gives us a list of 29 perovskites that are mechanically stable
in the cubic phase at 0~K. Since this is the highest symmetry phase,
they are likely also mechanically stable at high-temperatures\footnote{However, we note that transitions to other structures can take place,
in particular with one of hexagonal symmetry, such as in BaTiO$_{3}$
\cite{Glaister_PPS_1960}, RbZnF$_{3}$\cite{Daniel_PRB_1995}
or RbMgF$_{3}$~\cite{Shafer_JoPACoS_1969}.
This phase transition is of
first order, in contrast to displacive transitions that are of second
order.}.
We calculate their self-consistent finite-temperature force constants
$\Phi{}_{1000\text{~K}}^{\text{SCFCS}}$ as described in Section~\ref{subsec:art120:Finite-T_calculations}.
This initial set allows us to perform principal component analysis
of the 0~K force constants so that we obtain a transformation that
retains the 10 most important components. In a second step, we use
regression analysis to find a relation between the principal components
at 0~K and at 1000~K. This finally gives us a model that extracts
the principal components of the force constants at 0~K, interpolate
their values at 1000~K, and reconstruct the full force constants
matrix at 1000~K: $\Phi_{1000\text{~K}}^{\text{model}}$. We say that this
model has been ``trained'' on the particular set of compounds described
above. Applying it to the previously calculated $\Phi_{0\text{~K}}$
for all compounds, we can efficiently span the full combinatorial
space to search for new perovskites with a phonon spectrum that is
unstable at 0~K but stable at 1000~K. For materials determined
mechanically stable with $\Phi_{1000\text{~K}}^{\text{model}}$, we calculate
$\Phi_{1000\text{~K}}^{\text{SCFCS}}$. If the mechanical stability is confirmed,
we add the new compound to the initial set and subsequently train
the model again with the enlarged set. When no new compounds with
confirmed mechanical stability at high temperatures are found, we
stop the search. This process is summarized in Figure~\ref{fig:art120:PCA-regression}.
Following this strategy, we find 79 perovskites that are stable according
to the model, 68 of which are confirmed to be stable by the full calculation.
This means that we have reduced the total number of finite-temperature
calculations by a factor of 5, and that we have retrieved mechanically
stable compounds with a precision of 86\% and a recall of 74\% \footnote{Precision is defined as the fraction of true positives in all positives
reported by the model and recall as the fraction of true positives
found using the model with respect to all true positives.}. It allows us to obtain approximate phonon spectra for unstable compounds,
which is not possible with our finite-temperature calculations scheme
(see Supplementary Material of Reference~\cite{curtarolo:art120}). It also allows us to find compounds
that had not been identified as mechanically stable by the first exhaustive
search due to failures in the workflow. Considering the generality
of the approach, we expect this method to be applicable to other families
of compounds as well. Most importantly, it reduces the computational
requirements, particularly if the total combinatorial space is much
larger than the space of interest.

\fig
\includegraphics[width=0.6\linewidth]{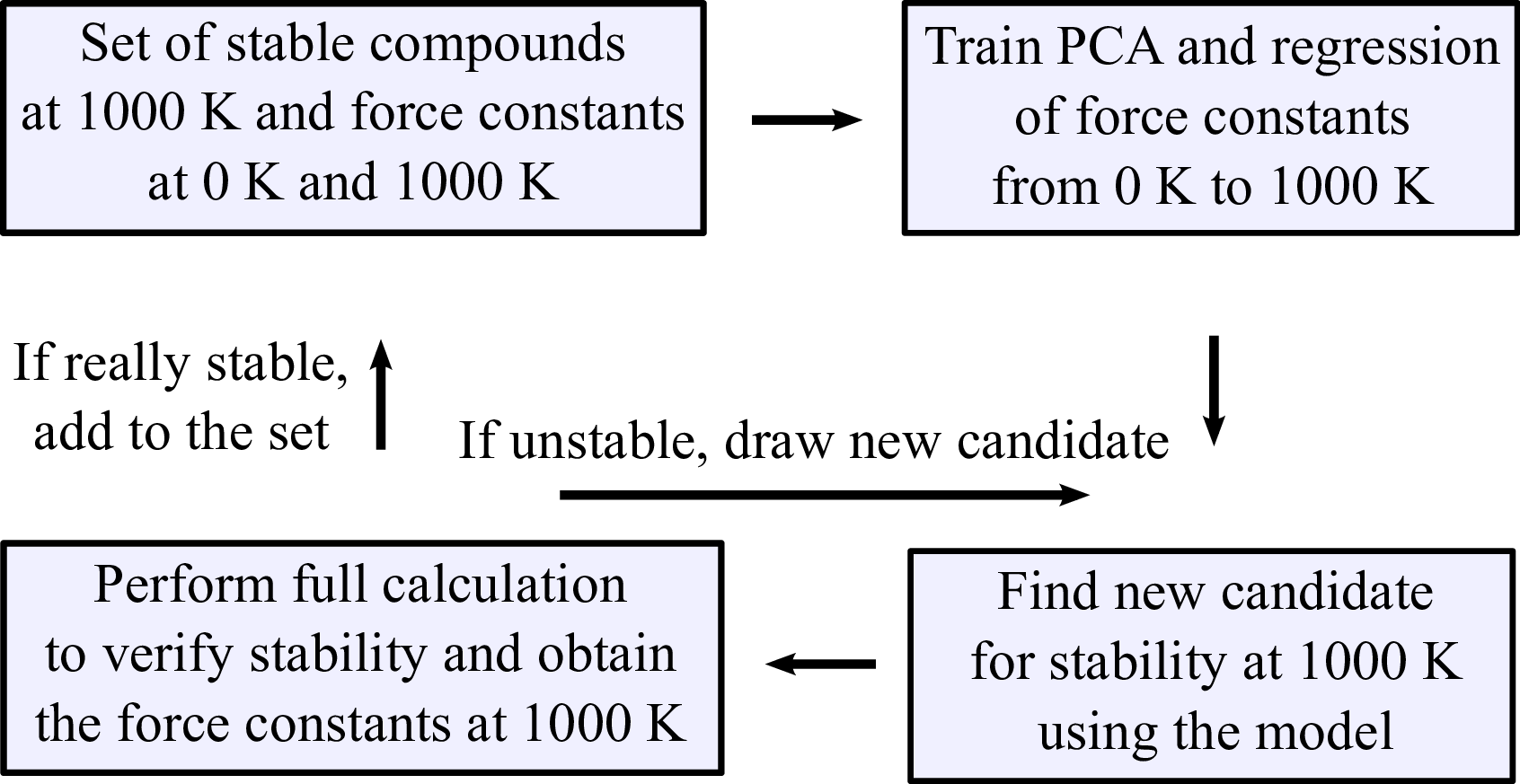}
\mycaption{Depiction of strategy for exploring the relevant combinatorial space
of compounds that are mechanically stable at high temperature.}
\label{fig:art120:PCA-regression}
\efig

\subsection{Simple descriptors of the thermal conductivity}
\label{subsec:art120:Descriptors}

We now focus on the analysis of the thermal conductivity data provided
in Table~\ref{tab:art120:List-of-perovskites}. We note that this set contains
about two times more fluorides than oxides. This was already the case
after the first screening in which we kept only the semiconductors,
and it can be explained by the strong electronegativity of fluorine,
which generally forms ionic solids with the alkali and alkaline earth
metals easily, as well as with elements from groups 12, 13 and 14.
This is shown on Figure~\ref{fig:art120:Columns}, in which we display histograms
of the columns of elements at sites \textit{A} and \textit{B} of the
perovskite in our initial list of paramagnetic semiconductors and
after screening for mechanical stability.

We can also see that the oxides tend to display a higher thermal conductivity
than the fluorides, as shown on the density plot of Figure~\ref{fig:art120:Fluorides_vs_oxides}.
This is once again due to the charge of the fluorine ion, which is
half that of the oxygen ion. In a model of a purely ionic solid, this
would cause the interatomic forces created by electrostatic interactions
to be divided by two in fluorides as compared to oxides. This is roughly
what we observe in our calculations of the second order force constants.
It translates into smaller phonon frequencies and mean group velocities
in fluorides as compared to oxides. Fluorides also have smaller heat
capacities, due to their larger lattice parameters (see Supplementary material of Reference~\cite{curtarolo:art120}).
Those two factors mainly drive the important discrepancy
of the thermal conductivity between fluorides and oxides. Following
the same reasoning, it means that halide perovskites in general should
have a very low thermal conductivity.

\fig
\includegraphics[width=8.5cm]{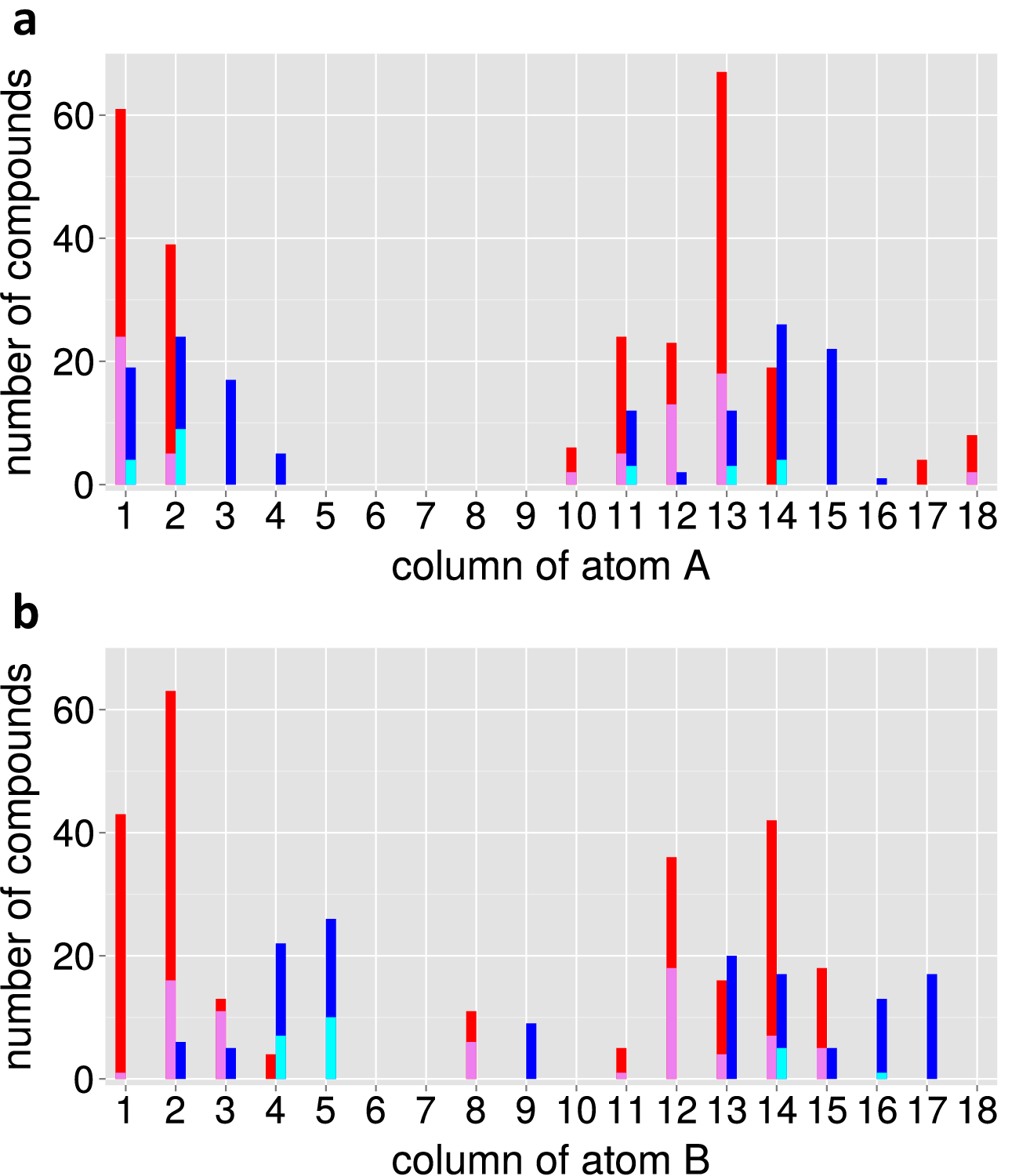}
\mycaption[Column number of the element at site (\textbf{a}) \textit{A} and (\textbf{b}) \textit{B}
of the perovskite \textit{ABX}$_{3}$.]
{Counts in the initial list of fluorides (red) and oxides (blue) paramagnetic semiconductors and
after screening for mechanical stability are shown in violet and cyan, respectively.}
\label{fig:art120:Columns}
\efig

\fig
\includegraphics[width=8.5cm]{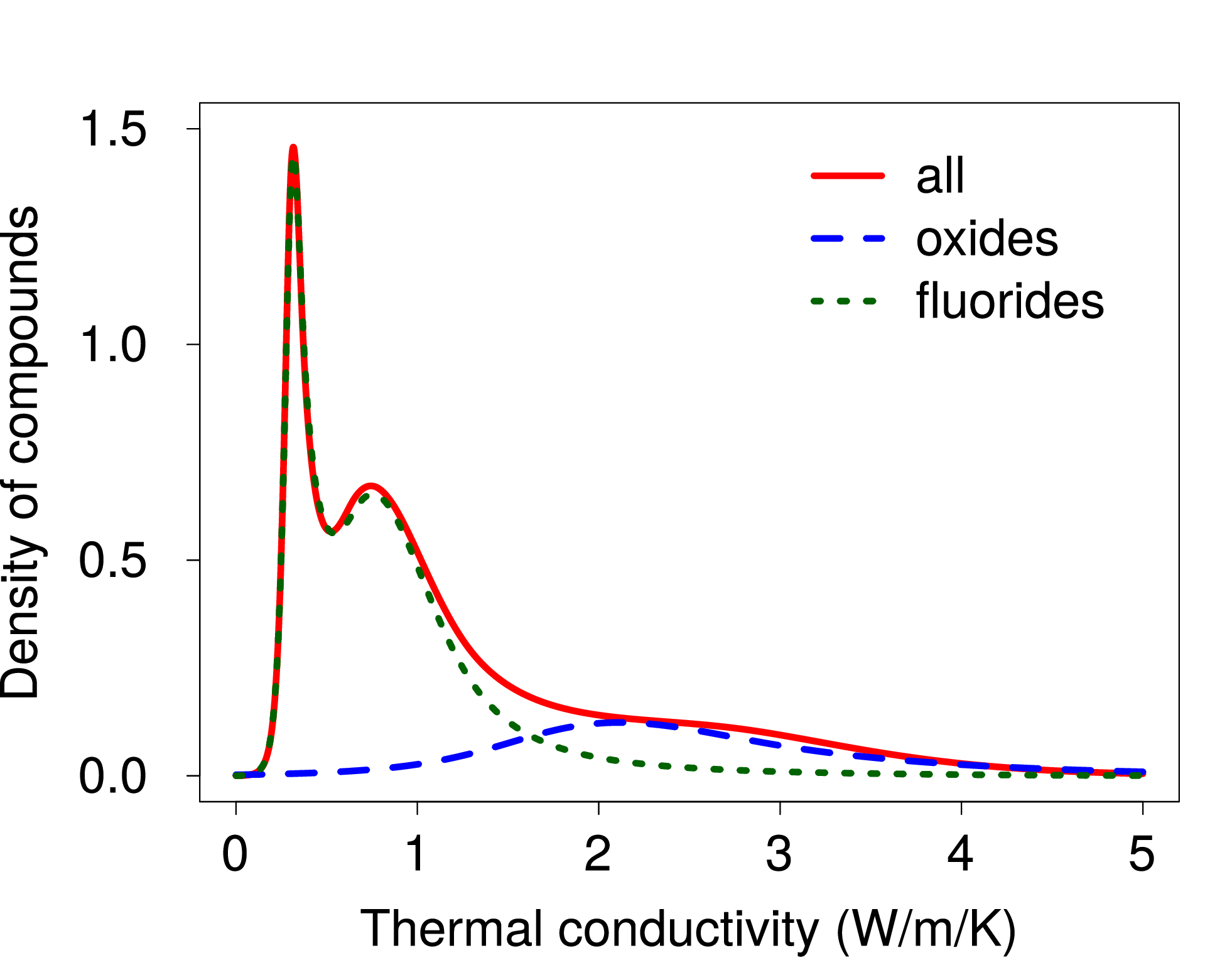}
\mycaption[Distribution of compounds as a function of the lattice thermal conductivity
at 1000~K.]
{The red curve corresponds to the distribution for all
mechanically stable compounds. The blue curve corresponds to the distribution
for fluorides only. The green curve corresponds to the distribution
for oxides only.}
\label{fig:art120:Fluorides_vs_oxides}
\efig

Finally, we analyze the correlations between the thermal conductivity
and different simple structural descriptors. Figure~\ref{fig:art120:Correlograms}
displays the correlograms for fluorides and oxides between the following
variables: the thermal conductivity $\kappa$, the thermal conductivity
in the small grain limit $\kappa_{\text{sg}}$~\cite{curtarolo:art85,aflowKAPPA},
the mean phonon group velocity v$_{\text{g}}$, the heat capacity
c$_{\text{V}}$, the root mean square Gr\"{u}neisen parameter $\gamma_{\text{rms}}$
\cite{Madsen_PRB_2014,Madsen_PSSA_2016}, the masses of atoms
at sites \textit{A} and \textit{B} of the perovskite \textit{ABX}$_{3}$,
their electronegativity, their Pettifor number~\cite{pettifor:1984},
their ionic radius, the lattice parameter of the compound and its
electronic gap. Remarkably, sites \textit{A} and \textit{B} play very
different roles in fluorides and oxides. In particular, the thermal
conductivity of fluorides is mostly influenced by substitutions of
the atom inside the fluorine octahedron (site \textit{B}), while the
interstitial atom at site \textit{A} has a negligible impact. The
opposite is true for the oxides. This means that when searching for
new compounds with a low lattice thermal conductivity, substitutions
at the \textit{A} site of fluorides can be performed to optimize cost
or other considerations without impacting thermal transport. It is
also interesting to note that the gap is largely correlated with the
electronegativity of atom \textit{B}, suggesting the first electronic
excitations likely involve electron transfer from the anion to the
\textit{B} atom.

Common to both fluorides and oxides, the lattice parameter is mostly
correlated with the ionic radius of atom \textit{B} rather than atom
\textit{A}. Interestingly, the lattice parameter is larger for fluorides,
although the ionic radius of fluorine is smaller than for oxygen.
This is presumably due to partially covalent bonding in oxides (see
\eg, Reference~\onlinecite{Kolezynski_Ferroelectrics_2005}). In contrast, fluorides
are more ionic: the mean degree of ionicity of the \textit{X-B} bond
calculated from Pauling's electronegativities~\cite{Pauling_JACS_1932}
$e_{X}$ and $e_{B}$ as $I{}_{XB}=100\left(1-e^{\left(e_{X}-e_{B}\right)/4}\right)$
yields a value of 56\% for oxides \vs\ 74\% for fluorides. Ionicity
is also reflected by the band structure, as can be seen from the weak
dispersion and hybridization of the F-2$p$ bands \footnote{See for instance the band structure of SrTiO$_{3}$~\cite{vanBenthem_JAP_2001}
compared to the one of KCaF$_{3}$~\cite{Ghebouli_SSS_2015}. In those
two compounds, the degree of ionicity of the \textit{X}-\textit{B}
bond calculated from Pauling's electronegativity is 59\% and 89\%,
respectively.}. This may explain why the role of atoms at site \textit{A} and \textit{B}
is so different between the two types of perovskites. We think that
the more ionic character combined to the small nominal charge in fluorides
makes the octahedron cage enclosing the atom \textit{B} less rigid,
such that the influence of the atom \textit{B} on the thermal conductivity
becomes more significant.

\fig
\includegraphics[width=1.0\linewidth]{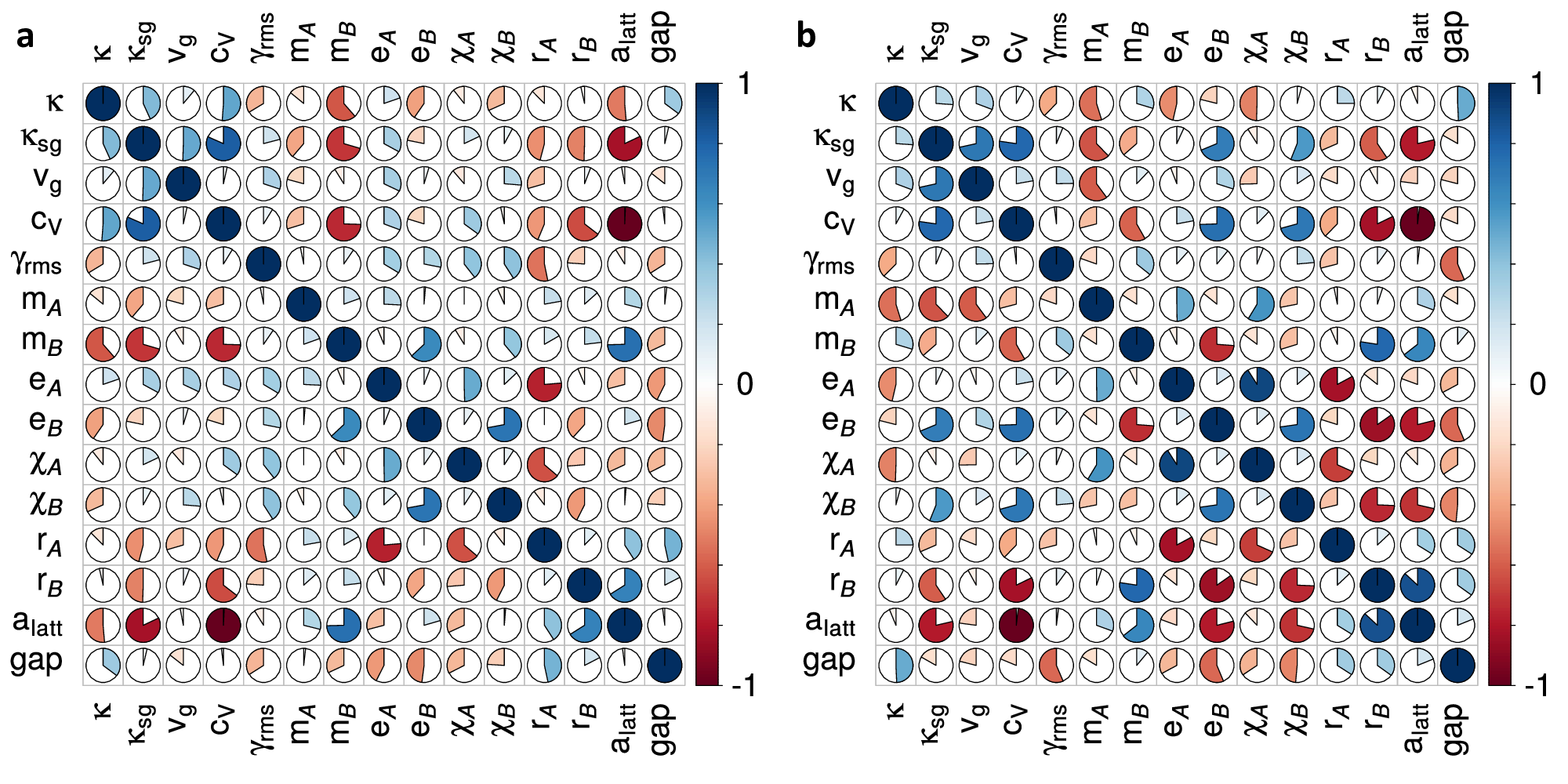}
\mycaption[Correlograms among properties of
mechanically stable (\textbf{a}) fluorides and
(\textbf{b}) oxides at 1000~K.]
{Properties compared include the thermal conductivity $\kappa$, the thermal
conductivity in the small grain limit $\kappa_{\text{sg}}$, the mean
phonon group velocity v$_{\text{g}}$, the heat capacity c$_{\text{V}}$,
the root mean square Gr\"{u}neisen parameter $\gamma_{\text{rms}}$,
the masses m$_{A}$ and m$_{B}$
of atoms at sites \textit{A} and \textit{B} of the perovskite \textit{ABX}$_{3}$,
their electronegativity e$_{A}$, e$_{B}$,
their Pettifor scale $\chi_{A}$, $\chi_{B}$,
their ionic radius
r$_{A}$, r$_{B}$,
the lattice parameter of the compound a$_{\text{latt}}$
and its electronic gap.}
\label{fig:art120:Correlograms}
\efig

\subsection{Conclusion}

Employing finite-temperature \abinitio\ calculations of force
constants in combination with machine learning techniques, we have
assessed the mechanical stability and thermal conductivity of hundreds
of oxides and fluorides with cubic perovskite structures at high temperatures.
We have shown that the thermal conductivities of fluorides are generally
much smaller than those of oxides, and we found new potentially stable
perovskite compounds. We have also shown that the thermal conductivity
of cubic perovskites generally decreases more slowly than the inverse
of temperature. Finally, we provide simple ways of tuning the thermal
properties of oxides and fluorides by contrasting the effects of substitutions
at the \textit{A} and \textit{B} sites. We hope that this work will
trigger further interest in halide perovskites for applications that
require a low thermal conductivity.
\clearpage
\section{Accelerated Discovery of New Magnets in the Heusler Alloy Family}
\label{sec:art109}

This study follows from a collaborative effort described in Reference~\cite{curtarolo:art109}.
Author contributions are as follows:
The initial idea for the project was developed by Stefano Sanvito and Stefano Curtarolo.
Junkai Xue and Thomas Archer constructed the Heusler database.
Corey Oses and Mario \v{Z}ic performed additional \DFT\ calculations for tetragonally distorted Heusler alloys.
Anurag Tiwari performed the regression analysis for the $T_\mathrm{C}$.
Corey Oses also performed the convex hull calculations.
Crystal growth and experimental characterization has been performed by Pelin Tozman under the supervision of
Munuswamy Venkatesan and J. Michael D. Coey.
The project was supervised by Stefano Sanvito, Stefano Curtarolo and J. Michael D. Coey, who also produced the manuscript.

\subsection{Introduction}
Very few types of macroscopic order in condensed matter are as sensitive to details as magnetism. The magnetic
interaction is usually based on the $m$-$J$ paradigm, where localized magnetic moments, $m$, are magnetically
coupled through the exchange interaction, $J$. Only a few elements in the periodic table can provide localized moments
in the solid state, namely 3$d$ transition metals, 4$f$ rare earths and some 4$d$ ions. Lighter 2$p$
elements are prone to form close shells, while in heavier ones the Hund's coupling is not strong enough to
sustain a high-spin configuration~\cite{Janak_PRB_1977}. The magnetic coupling then depends on how the wave-functions
of the magnetic ions overlap with each other, either directly, through other ions or via delocalized electrons.
This generates a multitude of mechanisms for magnetic coupling, operating at both sides of the
metal/insulator transition boundary, and specific to the details of the chemical environment. In general $J$ is sensitive
to the bond length, the bond angle, the magnetic ion valence. It is then not surprising
that among the $\sim$100,000 unique inorganic compounds known to mankind~\cite{ICSD4}, only about 2,000
show magnetic order of any kind~\cite{CoeyBook}.

When one focuses on the magnets that are useful for consumer applications, then the choice becomes even more restricted
with no more than two dozen compounds taking practically the entire global market. A useful magnet, regardless of the particular
technology, should operate in the -50$^\circ$C to +120$^\circ$C range, imposing the ordering temperature, $T_\mathrm{C}$,
to be at least 300$^\circ$C. Specific technologies then impose additional constraints. Permanent magnets should display a
large magnetization and hysteresis~\cite{CoeyBook}. Magnetic electrodes in high-performance magnetic tunnel junctions should
grow epitaxially on a convenient insulator and have a band-structure suitable for spin-filtering~\cite{Handbook_Spin_Mag_2011}. If the same tunnel
junction is used as spin-transfer torque magnetic random access memory element, the magnet should also have a low Gilbert
damping coefficient and a high Fermi-level spin polarization~\cite{Handbook_Spin_Mag_2011}. Indeed, there are not many magnets matching all the
criteria, hence the design of a new one suitable for a target application is a complex and multifaceted task.

The search for a new magnet usually proceeds by trial and error, but the path may hide surprises. For instance,
chemical intuition suggests that SrTcO$_3$ should be a poor magnet, since all Sr$X$O$_3$ perovskites with $X$ in
the chemical neighborhood of Tc are either low-temperature magnetic ($X$ = Ru, Cr, Mn, Fe) or do not present any
magnetic order ($X$ = Mo). Yet, SrTcO$_3$ is a G-type antiferromagnet~\cite{Rodriguez_STO2} with a remarkably
high N\'{e}el temperature, 750$^\circ$C, originating from a subtle interplay between $p$-$d$ hybridization and Jahn-Teller
distortion~\cite{Franchini_STOus}. This illustrates that often a high-performance magnet may represent a singularity
in physical/chemical trends and that its search can defy intuition. For this reason we take a completely different
approach to the discovery process and demonstrate that a combination of advanced electronic structure theory and
massive database creation and search, the high-throughput computational materials design approach~\cite{nmatHT},
can provide a formidable tool for finding new magnetic materials.

Our computational strategy consists of three main steps. Firstly, we construct an extensive database containing the
computed electronic structures of potential novel magnetic materials. Here we consider Heusler alloys (HAs), a prototypical
family of ternary compounds populated with several high-performance magnets~\cite{Graf_PSSC_2011}. A rough stability analysis,
based on evaluating the enthalpy of formation against reference single-phase compounds provides a first
screening of the database. This, however, is not a precise measure of the thermodynamic stability of a material, since
it does not consider decomposition into competing phases (single-element, binary, and ternary compounds). Such analysis
requires the computation of the electronic structure of all possible decomposition members associated with the given Heusler compounds.
This is our second step and it is carried out here only for intermetallic HAs, for which an extensive binary database is
available~\cite{aflowlibPAPER}. Finally, we analyze the magnetic order of the predicted stable magnetic intermetallic HAs and, via
a regression trained on available magnetic data, estimate their $T_\mathrm{C}$. The theoretical screening is then validated by
experimental synthesis of a few of the predicted compounds.

\subsection{Construction of the database}

The prototypical HA, $X_2YZ$ (Cu$_2$MnAl-type), crystallizes in the {\it Fm$\overline{3}$m} cubic
space group, with the $X$ atoms occupying the 8$c$ Wyckoff position (1/4, 1/4, 1/4) and the $Y$ and $Z$ atoms
being respectively at 4$a$ (0, 0, 0) and 4$b$ (1/2, 1/2, 1/2). The crystal can be described as four interpenetrating
{\it fcc} lattices with $Y$ and $Z$ forming an octahedral-coordinated rock-salt structure, while the $X$ atoms occupy the
tetrahedral voids [see Figure~\ref{fig:art109:Fig1}(a)]. Two alternative structures also exist. In the inverse Heusler $(XY)XZ$
(Hg$_2$CuTi-type), now $X$ and $Z$ form the rock-salt lattice, while the remaining $X$ and the $Y$ atoms fill the
tetrahedral sites [Figure~\ref{fig:art109:Fig1}(b)], so that one $X$ atom presents sixfold octahedral coordination, while the other
fourfold tetrahedral coordination. The second structure, the half-Heusler $XYZ$ (MgCuSb-type), is obtained by removing one of
the $X$ atoms, thus leaving a vacancy at one of the tetrahedral site [Figure~\ref{fig:art109:Fig1}(c)].
The minimal unit cell describing all three types can be constructed as a tetrahedral {\it F$\overline{4}$3m} cell, containing
4 (3 for the case of the half Heusler) atoms [Figure~\ref{fig:art109:Fig1}(d)]. Such a cell allows for a ferromagnetic spin configuration
and for a limited number of antiferromagnetic ones.

\fig
\includegraphics[width=\linewidth]{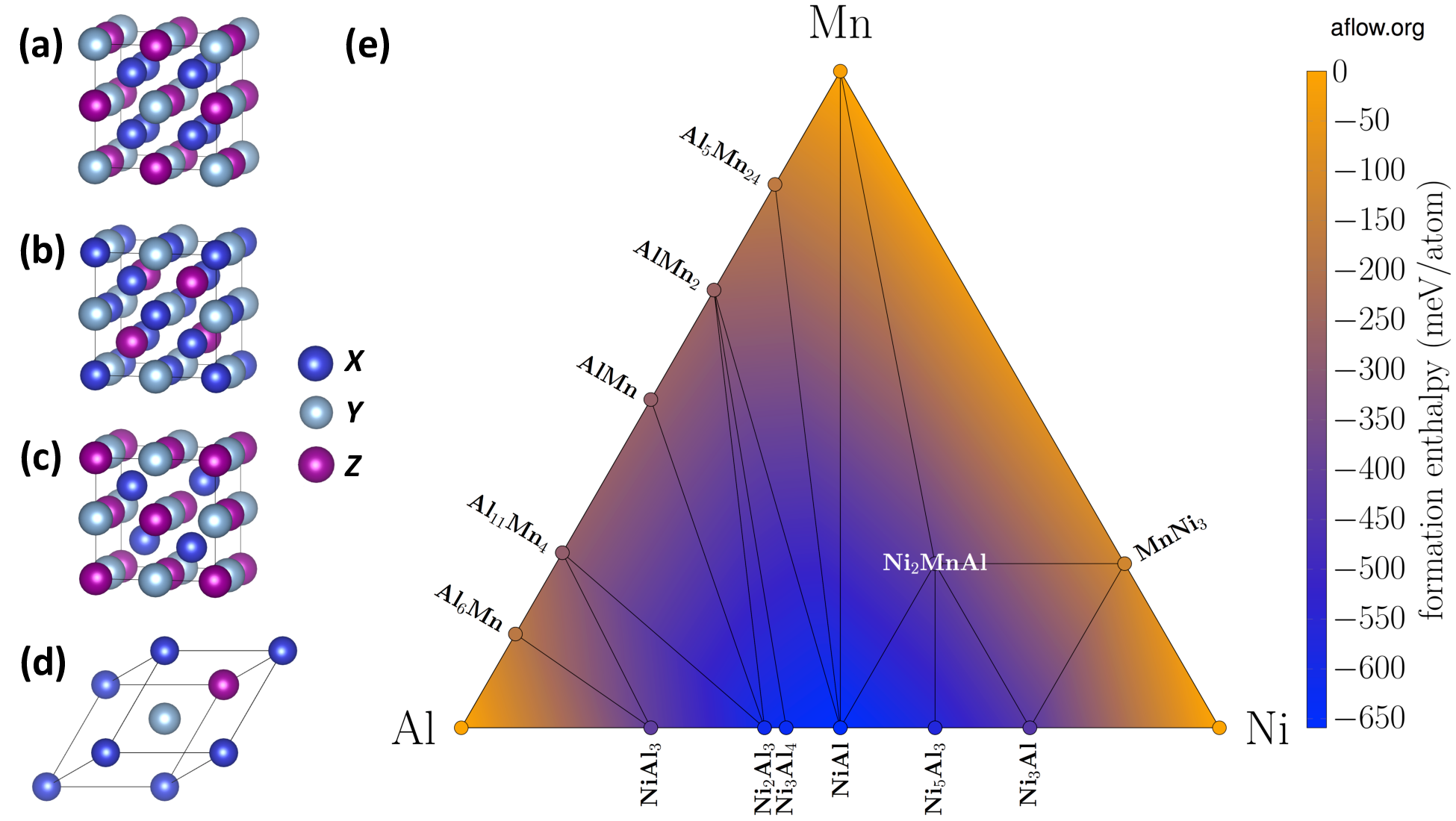}
\mycaption[Heusler structures.]
{(\textbf{a}) regular Heusler,
(\textbf{b}) inverse Heusler, and
(\textbf{c}) half Heusler.
(\textbf{d}) the tetrahedral {\it F$\overline{4}$3m} cell used to construct the electronic structure database.
(\textbf{e}) Ternary convex hull diagram for Al-Mn-Ni.
Note the presence of the stable HA, Ni$_2$MnAl.}
\label{fig:art109:Fig1}
\efig

\subsection{Results}

We construct the HAs database by considering all possible three-element combinations made of atoms from
the 3$d$, 4$d$ and 5$d$ periods and some elements from group III, IV, V and VI. In particular we use Ag, Al, As, Au,
B, Ba, Be, Bi, Br, Ca, Cd, Cl, Co, Cr, Cu, Fe, Ga, Ge, Hf, Hg, In, Ir, K, La, Li, Mg, Mn, Mo, Na, Nb, Ni, Os, P, Pb, Pd,
Pt, Re, Rh, Ru, Sb, Sc, Se, Si, Sn, Sr, Ta, Tc, Te, Ti, Tl, V, W, Y, Zn and Zr. Note that we have deliberately excluded
rare earths, responding to the global need to design new magnets with a reduced rare earth content.
Furthermore, we have not imposed constraints on the total number of valence electrons~\cite{zhang_sorting_2012,Yan_NComm_2015},
since magnetism is found for a broad range of electron counts.
For each combination of three elements ($X$, $Y$, $Z$) all the possible regular, inverse and half HAs are constructed.
These total to 236,115 decorations. The electronic structure of all the structures is computed by density functional theory
(\DFT) in the generalized gradient approximation (\GGA) of the exchange correlation functional as parameterized by
Perdew-Burke-Ernzerhof~\cite{PBE}. Our \DFT\ platform is the \VASP\ code~\cite{vasp_cms1996} and each structure is fully relaxed.
The typical convergence tolerance is 1~meV/atom and this is usually achieved by sampling the Brillouin zone
over a dense grid of 3000-4000 $k$-points per reciprocal atom. A much denser grid of 10,000 $k$-points is employed
for the static run to obtain accurate charge densities and density of states. The large volume of data is managed by the
AFLOW code~\cite{aflowPAPER}, which creates the appropriate entries for the AFLOW database~\cite{aflowlibPAPER}.
More details about the computational method are in Reference~\cite{curtarolo:art104}.

Let us begin our analysis by providing a broad overview of the database. Among the 236,115 decorations only 104,940
are unique, meaning that only a single structure is likely to form for a given stoichiometry. Strictly speaking, this is not true
since there are many examples of HAs presenting various degrees of site occupation disorder, and the
estimate gives an initial idea on how many compounds one may expect. Then a minimal criterion of stability is that the
enthalpy of formation of the $X_2YZ$ structure, $H_{X_2YZ}$, is lower than the sum of the enthalpies of formation
of its elementary constituents, namely $H_{\mathrm{f}}=H_{X_2YZ}-(2H_{X}+H_{Y}+H_{Z})<0$.
Such criterion returns us 35,602 compounds, with 6,778 presenting a magnetic moment. Note that this number can be
slightly underestimated as our unit cell can describe only a handful of possible anti-ferromagnetic configurations, meaning that
compounds where the magnetic cell is larger than the unit cell may then converge to a diamagnetic solution.
In any case, such a number is certainly significantly larger than the actual number of stable magnetic HAs. This can
only be established by computing the entire phase diagram of each ternary compound, \ie, by assessing the stability of
any given $X_2YZ$ structure against decomposition over all the possible alternative binary and ternary prototypes (for example
$X_2YZ$ can decompose into $XY$+$XZ$, $X_2Y$+$Z$, $XYZ$+$X$, \etc). Such a calculation is extremely intensive. An
informative phase diagram for a binary alloy needs to be constructed over approximately 10,000 prototypes~\cite{monsterPGM}, which
means that at least 30,000 calculations are needed for every ternary. As a consequence mapping the stability of every calculated HA
will require the calculation of approximately 15,000,000 prototypes, quite a challenging task.

When the electronic structure and the enthalpy of formation of the relevant binaries are available, then one can
construct the convex hull diagram for the associated ternary compounds~\cite{Lukas_CALPHAD_2007}. An example of such convex
hull diagram for Al-Mn-Ni is presented in Figure~\ref{fig:art109:Fig1}(e). The figure shows that there is a stable phase, Ni$_2$MnAl,
with a formation energy of -404~meV/atom. In this case, there are also three other unstable ternary structures with
$H_{\mathrm{f}}<0$, namely Mn$_2$NiAl, NiMnAl and Al$_2$MnNi. The enthalpy of formation of Mn$_2$NiAl is
$H_{\mathrm{f}}=-209$~meV/atom and it is 121~meV/atom higher than the tie-plane, that of NiMnAl is
-39~meV/atom (400~meV/atom above the tie plane), and that of Al$_2$MnNi is
-379~meV/atom (100~meV above the tie plane). This illustrates that $H_{\mathrm{f}}<0$
alone is not a stringent criterion for stability and that a full analysis needs to be performed before making the call on
a given ternary. Notably, Ni$_2$MnAl has been synthesized in a mixture of B2 and L2$_1$
phases~\cite{Ziebeck_JPFMP_1975} and it is a well-established magnetic shape memory alloy.

Given the enormous computational effort of mapping the stability of the entire database we have limited
further analysis to intermetallic HAs made only with elements of the 3$d$, 4$d$ and 5$d$
periods. These define 36,540 structures, for which the corresponding binaries are available in the \AFLOWorg\
database~\cite{aflowlibPAPER}. Our convex hull analysis then returns 248 thermodynamically stable compounds (full
list provided in Tables~\ref{fig:art109:magnetic_heusler_1}-\ref{fig:art109:magnetic_heusler_8}), of which only 22 possess a magnetic ground state in the tetrahedral
{\it F$\overline{4}$3m} unit cell. The details of their electronic structure are presented in Table~\ref{tab:art109:BLtab}.
Note that in the last column of the table we include an estimate of the robustness of a particular compound against
decomposition, $\delta_{\mathrm{sc}}^{30}$. A material is deemed as decomposable (`Y' in the table) if its enthalpy of formation
is negative but less than 30~meV/atom lower than the most stable balanced decomposition. In contrast a material is
deemed robust (`N' in the table) when $H_{\mathrm{f}}$ is more than 30~meV/atom away from that of the closest balanced
decomposition. When such a criterion is applied we find that 14 of the predicted HAs can potentially decompose,
while the other 8 are robust.

We have further checked whether such magnetic ground states are stable against tetragonal distortion, which may
occur in HAs in particular with the Mn$_2YZ$ composition. Indeed we find that the ground state of five structures, namely
Co$_2$NbZn, Co$_2$TaZn, Pd$_2$MnAu, Pd$_2$MnZn and Pt$_2$MnZn, is tetragonally distorted. Furthermore for
two of them, Co$_2$NbZn and Co$_2$TaZn, the tetragonal distortion suppresses the magnetic order indicating that
the competition between the Stoner and band Jahn-Teller instability~\cite{Labbe_JPFrance_1966} favors a distorted non-magnetic ground
state. The analysis so far tells us that the incidence of stable magnetic HAs among the possible intermetallics is
about 0.057\%. When this is extrapolated to the entire database we can forecast a total of about 140 stable magnetic
alloys, of which about 60 are already known. In the same way we can estimate approximately 1,450 stable non-magnetic
HAs, although this is just a crude forecast, since regions of strong chemical stability may be present in the complete
database and absent in the intermetallic subset.

\tab
\mycaption[Electronic structure parameters of the 22 magnetic HAs found among all possible
intermetallics.]
{The table lists the unit cell volume of the {\it F$\overline{4}$3m} cell, the $c/a$ ratio for tetragonal
cells, $a$, the Mn-Mn distance for Mn-containing alloys, $d_\mathrm{Mn-Mn}$, the magnetic moment per formula
unit, $m$, the spin polarization at the Fermi level, $P_\mathrm{F}$, the enthalpy of formation $H_{\mathrm{f}}$, the entropic
temperature, $T_\mathrm{S}$, and the magnetic ordering temperature, $T_\mathrm{C}$. Compounds labeled with
$*$ are not stable against tetragonal distortion (Co$_2$NbZn and Co$_2$TaZn become diamagnetic after distortion).
Note that $T_\mathrm{C}$ is evaluated only for Co$_2YZ$ and $X_2$Mn$Z$ compounds for which a sufficiently large
number of experimental data are available for other chemical compositions. In the case of Mn$_2YZ$ compounds we
report the magnetic moment of the ground state and in brackets that of the ferromagnetic solution. The last column
provides a more stringent criterion of stability. $\delta_{\mathrm{sc}}^{30}=$~Y if the given compound has an enthalpy within 30
meV/atom from that of its most favorable balanced decomposition (potentially decomposable), and $\delta_{\mathrm{sc}}^{30}=$~N if
such enthalpy is more than 30~meV/atom lower (robust).}
\tabvspace
\resizebox{\linewidth}{!}{
\begin{tabular}{l|r|r|r|r|r|r|r|r|r|r}
alloy & volume (\AA$^3$) & $c/a$ & $a$ (\AA) & $d_\mathrm{Mn-Mn}$ (\AA) & $m$ ($\mu_\mathrm{B}$/f.u.) & $P_\mathrm{F}$
& $H_{\mathrm{f}}$ (eV/atom) & $T_\mathrm{S}$~\K\ & $T_\mathrm{C}$~\K\ & $\delta_{\mathrm{sc}}^{30}$\\
\hline
Mn$_2$PtRh & 58.56 & & 6.16 & 3.08 & 0.00 (9.05) & 0.00 (0.86) &    -0.29 & 3247 & -- & N \\
Mn$_2$PtCo  & 54.28 && 6.00 & 3.00 & 1.13 (9.04) & 0.00 (0.86) & -0.17 & 1918 & -- & Y \\
Mn$_2$PtPd  & 60.75 && 6.24 & 3.12 & 0.00 (8.86) & 0.00 (0.38) & -0.29 & 3218 & -- & N \\
Mn$_2$PtV    & 55.73 && 6.06 & 3.03 & 4.87 (4.87) & 0.67 & -0.30 & 3353 & -- & Y \\
Mn$_2$CoCr  & 47.19 && 5.73 & 2.87 & 4.84 (4.84) & 0.016 & -0.05 & 529 & -- & N \\
Co$_2$MnTi & 49.68 && 5.84 & & 4.92 & 0.58 & -0.28 & 3122 & 940 & N \\
Co$_2$VZn & 46.87 && 5.73 & & 1.01 & 0.93 & -0.15 & 1653 & 228 & Y \\
Co$_2$NbZn$^*$ & 51.87 &1.0& 5.9 &&1.00 & 0.95 & -0.18 & 2034 & 212 & Y \\
Co$_2$NbZn  & 51.52 &     1.15 & 5.63 &&        0.0 &                  0.0 & -0.20& 2034 & 0 & Y \\
Co$_2$TaZn$^*$ & 51.80&1.0& 5.92 && 0.98& 0.63 & -0.22 & 2502 & 125 & N \\
Co$_2$TaZn  & 51.55 &     1.12 & 5.70 &&        0.0 &                  0.0 & -0.23& 2502 & 0 & N \\
Rh$_2$MnTi & 58.08 && 6.15 & 4.35 & 4.80 & 0.51 & -0.58 & 6500 & 417 & Y\\
Rh$_2$MnZr & 64.50 && 6.37 &4.50&4.75 & 0.34 & -0.58 & 6518 & 338 & Y \\
Rh$_2$MnHf & 63.22 && 6.32 & 4.47&4.74 &    0.34 & -0.67 & 7474 & 364 & Y \\
Rh$_2$MnSc & 61.62 && 6.27& 4.43&4.31 & 0.77 & -0.63 & 7031 & 429 & N \\
Rh$_2$MnZn & 54.95 && 6.03&4.27&3.37 & 0.63 & -0.31 & 3444 & 372 & Y \\
Pd$_2$MnAu$^*$ & 64.21 &1.0& 6.36& 4.49&4.60 & 0.06 & -0.20 & 2203 & 853 & Y \\
Pd$_2$MnAu  & 63.50 &     1.35 & 5.75&4.07 &    4.28 &      0.28 & -0.33 & 2203 & 331 & Y \\
Pd$_2$MnCu & 57.63 && 6.13&4.34&4.53 & 0.06 & -0.22 & 2492 & 415 & Y \\
Pd$_2$MnZn$^*$ & 58.88 & 1.0 &6.17&4.37& 4.33 & 0.38 & -0.39& 4399 & 894 & Y \\
Pd$_2$MnZn  & 58.74 &     1.18 &5.84&4.13&      4.22 &      0.16 & -0.47& 4399 & 402 & Y \\
Pt$_2$MnZn$^*$ & 59.23 &1.0&6.19&4.37&4.34 & 0.34 & -0.45 & 5035 & 694 & Y \\
Pt$_2$MnZn & 58.95 &     1.22 &5.79&4.10&      4.13 &       0.017 & -0.65& 5035 & 381 & Y \\
Ru$_2$MnNb & 59.64 &&6.20&4.39& 4.07 & 0.85 & -0.19 & 2068 & 276 & Y \\
Ru$_2$MnTa & 59.72 &&6.20&4.39& 4.06 & 0.86 & -0.26 & 2912 & 305 & N \\
Ru$_2$MnV & 54.38 &&6.01&4.25& 4.00 & 0.707 & -0.16 & 1832 & 342 & Y \\
Rh$_2$FeZn & 54.60 &&6.02&& 4.24 & 0.49 & -0.28 & 3150 & -- & N\\
\end{tabular}}
\label{tab:art109:BLtab}
\etab

In Table~\ref{tab:art109:BLtab}, together with structural details, the magnetic moment per formula unit, $m$, and the enthalpy
of formation we report a few additional quantities that help us in understanding the potential of a given alloy as
high-performance magnet. The spin polarization of the density of states at the Fermi level, $n_\mathrm{F}^\sigma$
($\sigma=\uparrow, \downarrow$) is calculated as~\cite{Mazin_PRL1999}
\begin{equation}
P_\mathrm{F}=\frac{n_\mathrm{F}^\uparrow-n_\mathrm{F}^\downarrow}{n_\mathrm{F}^\uparrow+n_\mathrm{F}^\downarrow}\:,
\end{equation}
and expresses the ability of a metal to sustain spin-polarized currents~\cite{Coey_JPDAP_2004}. We find a broad distribution of
$P_\mathrm{F}$s with values ranging from 0.93 (Co$_2$VZn) to 0.06 (Pd$_2$MnCu). None of the HAs display
half-metallicity, and in general their spin-polarization is similar to those of the elementary 3$d$ magnets
(Fe, Co and Ni).

We then calculate the entropic temperature~\cite{nmatHT,monsterPGM,curtarolo:art98}, $T_\mathrm{S}$. For simplicity we give
the definition for a $XY$ binary alloy, although all our calculations are performed for its ternary equivalent,
\begin{equation}
T_\mathrm{S}=\max_i\left[\frac{H_{\mathrm{f}}(X_{x_i}Y_{1-x_i})}{k_\mathrm{B}[x_i\log x_i+(1-x_i)\log(1-x_i)]}
\right]\:,
\end{equation}
where $k_\mathrm{B}$ is the Boltzmann constant and $i$ counts all the stable compounds in the $XY$ binary system.
Effectively $T_\mathrm{S}$ is a concentration-maximized formation enthalpy weighted by the inverse of its ideal entropic
contribution (random alloy). It measures the ability of an ordered phase to resist deterioration into a temperature-driven,
entropically-promoted, disordered mixture. The sign of $T_\mathrm{S}$ is chosen such that a positive temperature
is needed for competing against the compound stability (note that $T_\mathrm{S}<0$ if $H_{\mathrm{f}}>0$), and one expects
$T_\mathrm{S}\rightarrow0$ for a compound spontaneously decomposing into a disordered mixture.
If we analyze the $T_\mathrm{S}$ distribution for all the intermetallic HAs with $H_{\mathrm{f}}<0$
(8776 compounds) we find the behavior to closely follow that of a two-parameter Weibull distribution with a shape of
1.13 and a scale of 2585.63 (see histogram in Figure~\ref{fig:art109:histogram_full}). The same distribution for the 248 stable intermetallic HAs is
rather uniform in the range 1,000-10,000~K and presents a maximum at around 3,500~K. A similar trend is observed
for the 20 stable magnetic HAs, suggesting that several of them may be highly disordered.

Finally, Table~\ref{tab:art109:BLtab} includes an estimate of the magnetic ordering temperatures, $T_\mathrm{C}$. These
have been calculated based on available experimental data. Namely we have collected the experimental
$T_\mathrm{C}$'s of approximately 40 known magnetic Heusler compounds (see Section~\ref{subsec:art109:tc_known_heuslers}) and performed a linear
regression correlating the experimental $T_\mathrm{C}$'s with a range of calculated electronic structure properties,
namely equilibrium volume, magnetic moment per formula unit, spin-decomposition and number of valence
electrons. The regression is possible only for those compounds for which the set of available experimental data
is large enough, namely for Co$_2YZ$ and $X_2$Mn$Z$ HAs. We have trained the regression over the existing
data and found that for the two classes Co$_2YZ$ and $X_2$Mn$Z$ the typical error in the $T_\mathrm{C}$ estimate
is in the range of 50~K, which is taken as our uncertainty.

\subsection{Discussion}

We have found three different classes of stable magnetic HAs, namely Co$_2YZ$, $X_2$Mn$Z$ and Mn$_2YZ$. In
addition we have predicted also Rh$_2$FeZn to be stable. This is rather unique since there are no other HAs with Fe
in octahedral coordination and no magnetic ions at the tetrahedral positions.

\fig
\includegraphics[width=0.65\linewidth]{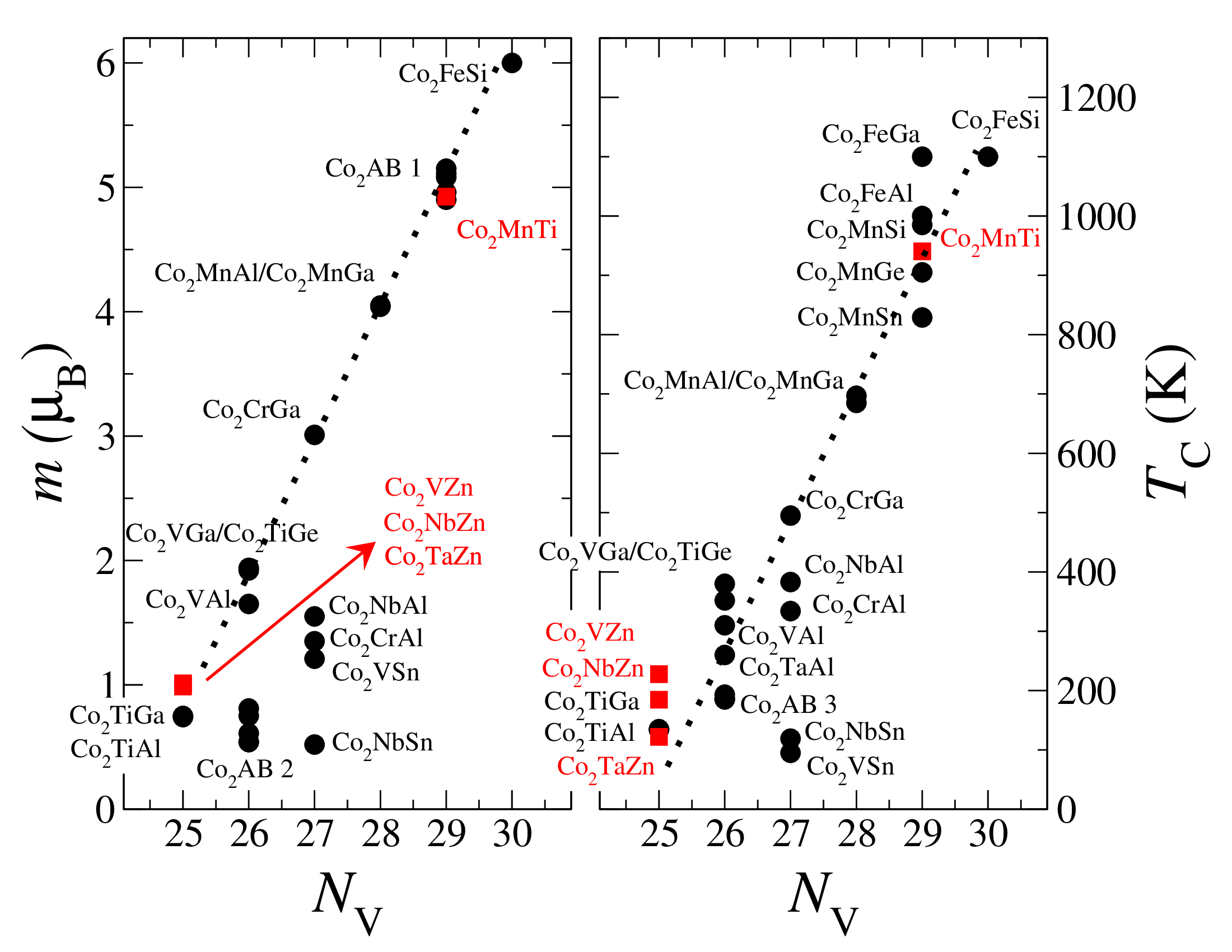}
\mycaption[Slater-Pauling curve for magnetic HAs of the form Co$_2YZ$.]
{The magnetic moment per formula unit, $m$,
is plotted against the number of valence electron, $N_\mathrm{V}$, in the left panel, while $T_\mathrm{C}$ is displayed
on the right. Red symbols corresponds to predicted HAs, while the black ones to existing materials. For the sake of clarity
several compounds have been named collectively on the picture. Co$_2AB$ 1: Co$_2$FeGa, Co$_2$FeAl, Co$_2$MnSi,
Co$_2$MnGe, Co$_2$MnSn; Co$_2AB$ 2: Co$_2$TaAl, Co$_2$ZrAl, Co$_2$HfGa, Co$_2$HfAl, Co$_2$TaGa;
Co$_2AB$ 3: Co$_2$ZrAl, Co$_2$HfAl, Co$_2$HfGa, Co$_2$TaGa.}
\label{fig:art109:Co2XY}
\efig

The first class is Co$_2YZ$, a class which is already populated by about 25 known compounds all lying on the
Slater-Pauling curve~\cite{Graf_PSSC_2011}. Our analysis reveals four new stable alloys, three of them with the low valence
electron counts of 25 (Co$_2$VZn, Co$_2$NbZn, Co$_2$TaZn) and one, Co$_2$MnTi, presenting the large count
of 29. The regression correctly places these four on the Slater-Pauling curve (see Figure~\ref{fig:art109:Co2XY}) and predicts
for Co$_2$MnTi the remarkably high $T_\mathrm{C}$ of 940~K. This is a rather interesting since only about
two dozen magnets are known to have a $T_\mathrm{C}$ in that range~\cite{CoeyBook}. Therefore, the discovery
of Co$_2$MnTi has to be considered as exceptional. The other three new compounds in this class are all predicted to
have a $T_\mathrm{C}$ around 200~K, but two of them become non-magnetic upon tetragonal distortion leaving
only Co$_2$VZn magnetic ($T_\mathrm{C}\sim228$~K).

The second class is $X_2$Mn$Z$ in which we find 13 new stable magnets, most of them including a 4$d$ ion (Ru, Rh
and Pd) in the tetrahedral $X$ position. In general, these compounds have a magnetic moment per formula unit ranging between
4~$\mu_\mathrm{B}$ and 5~$\mu_\mathrm{B}$, consistent with the nominal 2+ valence of Mn in octahedral coordination.
The regression, run against 18 existing compounds of which 13 are with $X$ = Ru, Rh or Pd, establishes a correlation
between the Mn-Mn nearest neighbors distance, $d_\mathrm{Mn-Mn}$, and $T_\mathrm{C}$ as shown in
Figure~\ref{fig:art109:X2MnY}.

\fig
\includegraphics[width=0.65\linewidth]{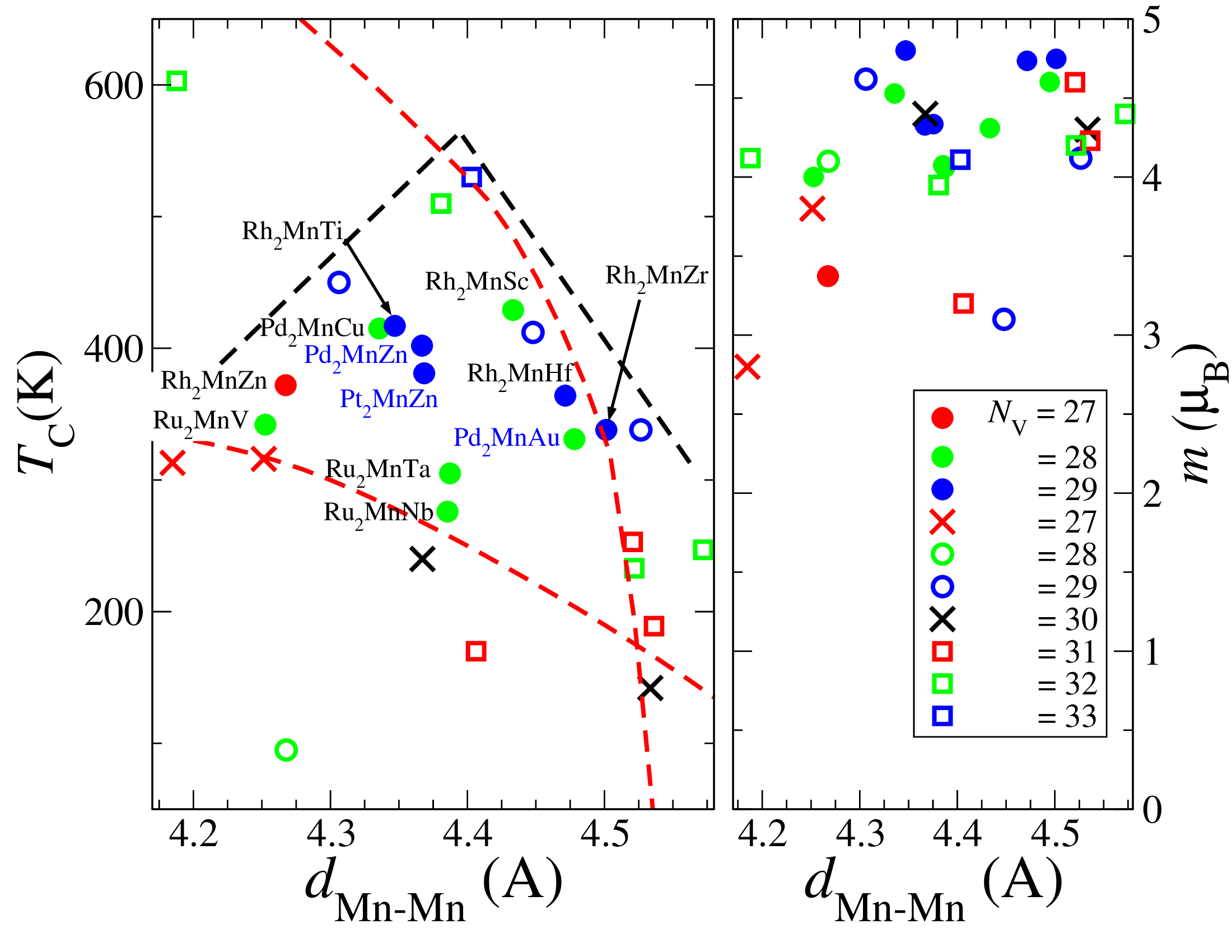}
\mycaption[Magnetic data for $X_2$Mn$Z$ magnets.]
{$T_\mathrm{C}$ (left) and magnetic moment per formula unit (right)
as a function of the Mn-Mn distance, $d_\mathrm{Mn-Mn}$.
Note that the $T_\mathrm{C}$ is limited to about 550~K and
peaks at a volume of about 60~\AA$^3$. In contrast the magnetic moment is approximately constant with values in between
4~$\mu_\mathrm{B}$ and 5~$\mu_\mathrm{B}$. Close circles (with associated chemical compositions) correspond to the
predicted compounds, while the other symbols correspond to experimental data. Different colors correspond to different
number of valence electrons, $N_\mathrm{V}$. Blue chemical formulas correspond to compound displaying tetragonal
distortion. The two red lines are Castelliz-Konamata curves, while the black one is to guide the eye.}
\label{fig:art109:X2MnY}
\efig

We find that $T_\mathrm{C}$ is a non-monotonic function of $d_\mathrm{Mn-Mn}$ with a single maximum at
$d_0$$\sim$4.4~\AA\ corresponding to a temperature of 550~K (the maximum coincides approximately with
Cu$_2$MnSn). The only apparent exception to such trend is the prototypical Cu$_2$MnAl, which displays a large
$T_\mathrm{C}$ and relatively small $d_\mathrm{Mn-Mn}$~\cite{Oxley1963}. A strong sensitivity of the $T_\mathrm{C}$
of Mn-containing compounds to $d_\mathrm{Mn-Mn}$ was observed long time ago and rationalized in an empirical
$T_\mathrm{C}$-$d_\mathrm{Mn-Mn}$ curve by Castelliz~\cite{Castelliz_ZM_1955}. This predicts that $T_\mathrm{C}$ is not
monotonically dependent on $d_\mathrm{Mn-Mn}$ and has a maximum at around $d_\mathrm{Mn-Mn}=3.6$. The curve
has been validated for a number of HAs and it has been used to explain the positive pressure coefficient of
$T_\mathrm{C}$, $(1/T_\mathrm{C})(\mathrm{d}T_\mathrm{C}/\mathrm{d}P)$, found, for instance, in
Rh$_2$MnSn~\cite{Adachi200437}. Refinements of the Castelliz curve predict that the rate of change of $T_\mathrm{C}$
with $d_\mathrm{Mn-Mn}$ in HAs is related to the valence count~\cite{Kanomata_JMMM_1987}, although the position
of the maximum is not. In general the results of Figure~\ref{fig:art109:X2MnY}, including several experimental data, seems to contradict
the picture since a monotonically decreasing $T_\mathrm{C}$ is expected for any $d_\mathrm{Mn-Mn}>3.6$~\AA, \ie,
practically for any HAs of the form $X_2$Mn$Z$. There are a few possible reasons for such disagreement. Firstly, the
Castelliz curve assumes that only Mn presents a magnetic moment, which is unlikely since many of the $X_2$Mn$Z$
compounds of Figure~\ref{fig:art109:X2MnY} have Rh or Pd in the X position, two highly spin-polarizable ions. Secondly,
many HAs in Figure~\ref{fig:art109:X2MnY} present various levels of disorder, meaning that Mn-Mn pairs separated
by less than the nominal $d_\mathrm{Mn-Mn}$ are likely to be present in actual samples. We then propose that the
trend of Figure~\ref{fig:art109:X2MnY} (see dashed black lines) represents a new empirical curve, valid for $X_2$Mn$Z$ HAs,
and taking into account such effects.

The last class of predicted magnetic HAs is populated by Mn$_2YZ$ compounds. These have recently
received significant attention because of their high $T_\mathrm{C}$ and the possibility of displaying tetragonal
distortion and hence large magneto-crystalline anisotropy~\cite{Kreiner2014}. Experimentally when the 4$c$ position
is occupied by an element from group III, IV or V one finds the regular Heusler structure if the atomic number of
the $Y$ ion is smaller than that of Mn, $Z$($Y$)$<$$Z$(Mn), and the inverse one for $Z$($Y$)$>$$Z$(Mn). To date only
Mn$_2$VAl and Mn$_2$VGa have been grown with a $Y$ element lighter than Mn, so that except those two all other
Mn$_2YZ$ HAs crystallize with the inverse structure (see Figure~\ref{fig:art109:Mn2YZ}). In the case of the two regular
HAs, Mn$_2$VAl and Mn$_2$VGa, the magnetic order is ferrimagnetic with the two Mn ions at the tetrahedral
sites being anti-ferromagnetically coupled to V~\cite{Nakamichi1983,Itoh1983,Kumar2008}. In contrast for
the inverse Mn$_2$-based HAs the antiferromagnetic alignment is between the two Mn ions and the magnetic
ground state then depends on whether there are other magnetic ions in the compound. In general, however, site disorder is
not uncommon (see Section~\ref{subsec:art109:tet_disorder_Mn2PtPd}) and so is tetragonal distortion, so that the picture becomes more complicated. There are
also some complex cases, such as that of Mn$_3$Ga, presenting a ground state with a non-collinear arrangement of
both the spin and angular momentum~\cite{Rode_PRB_2013}.

\fig
\includegraphics[width=0.65\linewidth]{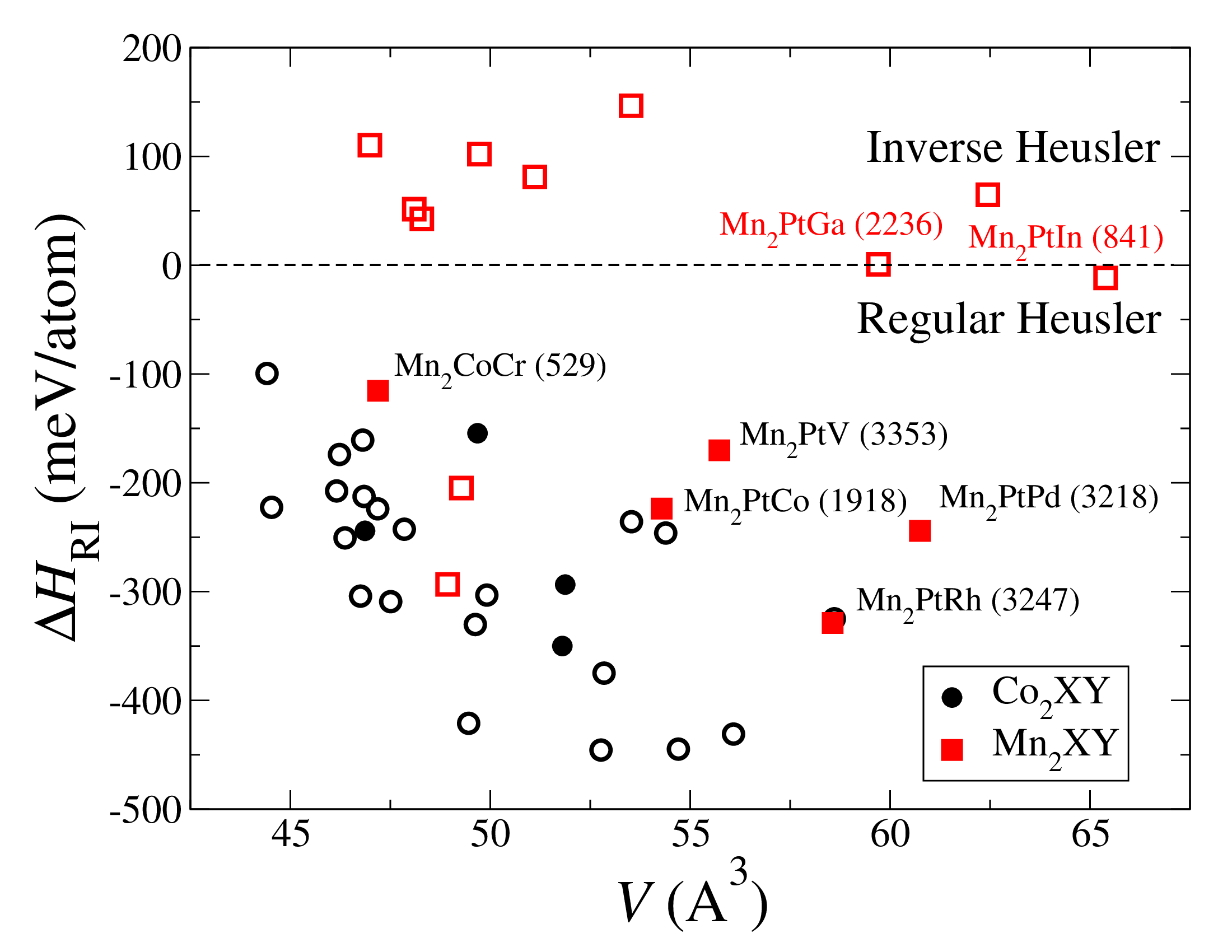}
\mycaption[Enthalpy of formation difference between the regular and inverse Heusler structure, $\Delta H_\mathrm{RI}$,
for Mn$_2$-containing compounds as a function of the cell volume.]
{The solid red squares (with chemical formulas) are
the predicted stable intermetallic materials, while the open red squares are existing compounds. For completeness we
also include data for Co$_2$-based HAs, again with open symbols for existing compounds and solid one for predicted.
In brackets beside the chemical formulas we report the value for the entropic temperature, $T_\mathrm{S}$, in \K.}
\label{fig:art109:Mn2YZ}
\efig

If we now turn our attention to the predicted compounds we find five stable compositions of which three match the
$\delta_{\mathrm{sc}}^{30}$ robustness criterion. Most intriguingly the regular {\it Fm$\overline{3}$m} structure
appears to be the ground state for all the compounds, regardless of their chemical composition. This sets Mn$_2$-based
intermetallic compounds aside from those with elements from the main groups. In Figure~\ref{fig:art109:Mn2YZ} we present the
enthalpy of formation difference between the regular and the inverse structure, $\Delta H_\mathrm{RI}=H_\mathrm{f,R}-H_\mathrm{f,I}$,
for the computed and the experimentally known Mn$_2$-based HAs, together with their $T_\mathrm{S}$ and reference
data for Co$_2$-based alloys. In general we find that $\Delta H_\mathrm{RI}$ for the Mn$_2YZ$ class is significantly smaller
than for the Co$_2YZ$ one. In fact there are cases, \eg, Mn$_2$PtGa and Mn$_2$PtIn, in which the two phases are almost
degenerate and different magnetic configurations can favor one over the other. Overall, one then expects such compounds
to be highly disordered. Finally, we take a look at the magnetic ground state. In all cases the compounds present some
degree of antiferromagnetic coupling, which results in either a zero-moment ground state when Mn is the only magnetic ion,
and in a ferrimagnetic configuration when other magnetic ions are present.

The last step in our approach consists in validating the theoretical predictions by experiments. We have attempted the
synthesis of four HAs, namely Co$_2$MnTi, Mn$_2$PtPd, Mn$_2$PtCo and Mn$_2$PtV. Co$_2$MnTi
is chosen because of its high Curie temperature, while among the Mn$_2$-based alloys we have selected two
presenting ferrimagnetic ground state (Mn$_2$PtCo and Mn$_2$PtV) and one meeting the stringent
$\delta_{\mathrm{sc}}^{30}$ robustness criterion (Mn$_2$PtPd). The alloys have been prepared by arc melting in high-purity Ar, with
the ingots being remelted four times to ensure homogeneity. An excess of 3 \% wt. Mn is added in order to
compensate for Mn losses during arc melting (see Section~\ref{subsec:art109:exp_data_Mn2_based} for details). Structural characterization has been carried
out by powder X-ray diffraction (XRD), while magnetic measurements were made using a superconducting magnetometer
in a field of up to 5~T. Furthermore, the microstructure has been analyzed by scanning electron microscopy
of the polished bulk samples, while the compositions are determined by Energy Dispersive X-ray (EDX)
spectroscopy.

Two of the four HAs have been successfully synthesized, Co$_2$MnTi and Mn$_2$PtPd, while the other two,
Mn$_2$PtCo and Mn$_2$PtV, decompose into binary compounds (see Section~\ref{subsec:art109:exp_data_Mn2_based} for details).

\fig
\includegraphics[width=\linewidth]{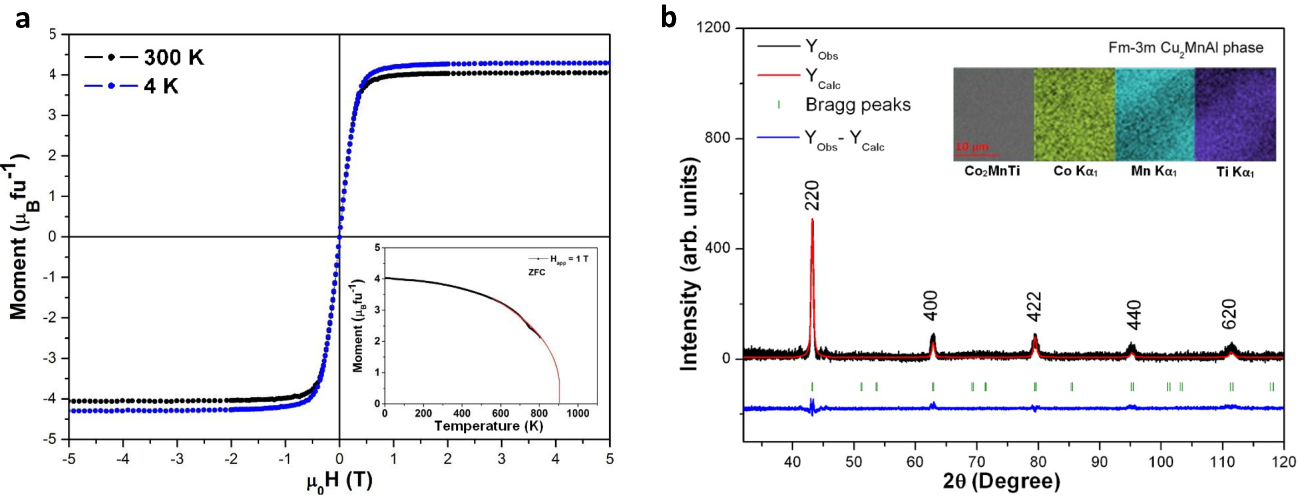}
\mycaption[Experimental magnetic characterization of Co$_2$MnTi.]
{(\textbf{a}) magnetization curve at 4~K and 300~K (inset: zero-field cooled magnetization
curve as a function of temperature in magnetic field of 1~T);
(\textbf{b}) XRD spectrum (inset: EDX chemical composition
analysis).
Co$_2$MnTi crystallizes in a single {\it Fm$\overline{3}$m} phase corresponding to a regular Heusler.
The $T_\mathrm{C}$ extrapolated from the magnetization curve is around 900~K.}
\label{fig:art109:Co2MnTi}
\efig

In Figure~\ref{fig:art109:Co2MnTi} we present the structural and magnetic characterization of Co$_2$MnTi. It crystallizes in
the regular {\it Fm$\overline{3}$m} Heusler structure with no evidence of secondary phases and a lattice parameter of
$a=5.89$~\AA\, in close agreement with theory, $a=5.84$~\AA. The magnetization curve displays little temperature
dependence and a saturation moment of 4.29~$\mu_\mathrm{B}$/f.u. at 4~K, fully consistent with the calculated
ferromagnetic ground state (see Table~\ref{tab:art109:BLtab}). Most notably, the $T_\mathrm{C}$ extrapolated from the
zero-field cooled magnetization curve in a field of 1~T is found to be 938~K, essentially identical that predicted
by our regression, 940~K.
This is a remarkable result, since it is the first time that a new high-temperature ferromagnet
has been discovered by HT means.

\fig
\includegraphics[width=\linewidth]{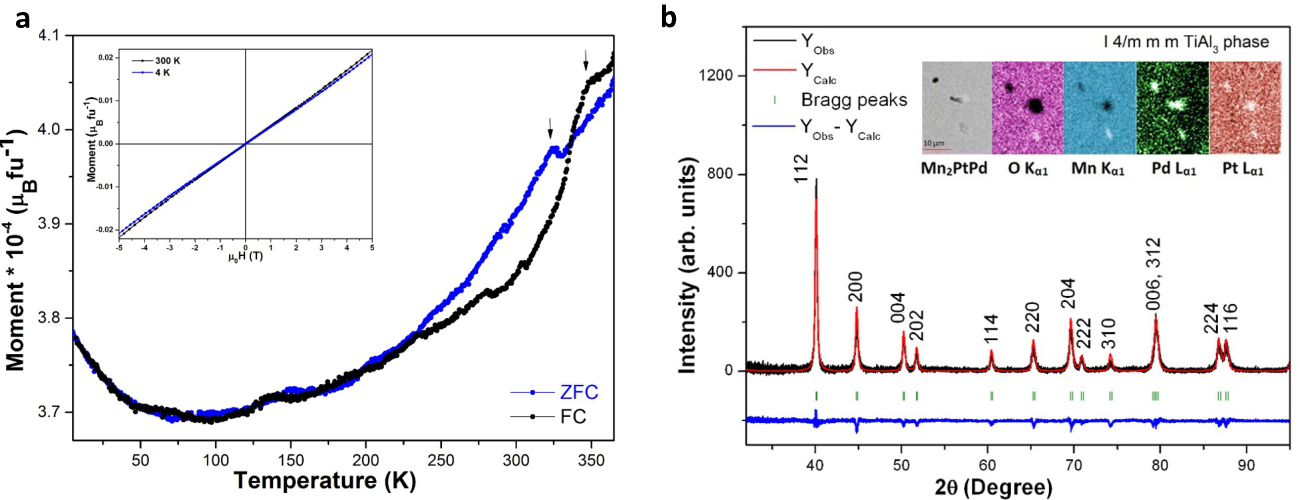}
\mycaption[Experimental magnetic characterization of Mn$_2$PtPd.]
{(\textbf{a}) field cooled and zero-field cooled magnetization curve as a function of temperature in
a magnetic field of 0.1~T (inset: magnetization curve at 4~K and 300~K);
(\textbf{b}) XRD spectrum (inset: EDX chemical
composition analysis). Mn$_2$PtPd crystallizes in a single {\it I}4{\it/mmm} (TiAl$_3$-type) phase corresponding
to a regular tetragonal distorted Heusler.
SEM images confirm that the bulk sample is mainly of Mn$_2$PtPd
composition (gray color) with a small amount of a secondary Mn-O inclusions, which have spherical shape of
diameter 400-900~nm and do not appear in the XRD spectrum.}
\label{fig:art109:Mn2PtPd}
\efig

Also in the case of Mn$_2$PtPd a single phase is found without evidence of decomposition. The XRD pattern
[Figure~\ref{fig:art109:Mn2PtPd}(b)] corresponds to a tetragonally-distorted regular Heusler with space group {\it I}4{\it/mmm}
(TiAl$_3$-type) and lattice parameters $a=4.03$~\AA\ and $c=7.24$~\AA. Our magnetic data show a magnetic
transition at $\sim$320~K, which shifts to a slightly higher temperature upon field cooling [Figure~\ref{fig:art109:Mn2PtPd}(a)].
Magnetization curves at room temperature and 4~K show no hysteresis or spontaneous magnetization indicating
that the compound is antiferromagnetic at low temperature.
From Table~\ref{tab:art109:BLtab} it will appear that the only difference between the calculated and experimental
data for Mn$_2$PtPd concerns the tetragonal distortion. However, the search for tetragonal distortion reported in the
table was performed only for the ferromagnetic state. Further analysis for the antiferromagnetic ground state (see Section~\ref{subsec:art109:tet_disorder_Mn2PtPd})
reveals that indeed Mn$_2$PtPd is antiferromagnetic and tetragonal distorted with a $c/a$ ratio of around 1.3, in good
agreement with experiments.

\subsection{Table of \texorpdfstring{$T_\mathrm{C}$}{Tc} of known Heusler alloys} \label{subsec:art109:tc_known_heuslers}

Here we present experimental data, collected from the literature, for known magnetic Heusler alloys.
These data have been used to perform the regression used to extract the $T_\mathrm{C}$ of the
new predicted compounds.

\tab
\mycaption[Summary Table for the magnetic Heusler alloys of the type Co$_2XY$.]
{Here are reported the compound,
the magnetic moment per formula unit, $m$, and the experimental $T_\mathrm{C}$, together with the
appropriate reference. The quantities labeled with a `*' have been used to run the
regression.}
\tabvspace
\begin{tabular}{l|r|r|r|r|r|r}
material & $m$/f.u. ($\mu_\mathrm{B}$) & $m^*$/f.u. ($\mu_\mathrm{B}$) & $T_\mathrm{C}$~\K\ & $T_\mathrm{C}^*$~\K\ & source & reference \\
\hline
Co$_2$TiAl & 0.74 & 0.74 & 134 & 134 & Exp. & \onlinecitesq{PhysRevB.76.024414} \\
Co$_2$TiGa & 0.82 & 0.82 & 128 & 128 & Exp. & \onlinecitesq{LandBorn1,Sasaki2001406} \\
Co$_2$TiSi & 1.96 & 1.96 & 380 & 380 & Exp. & \onlinecitesq{LandBorn1,Barth28092011} \\
Co$_2$TiGe & 1.94 & 1.94 & 380 & 380 & Exp. & \onlinecitesq{LandBorn1,Barth28092011} \\
Co$_2$TiSn & 1.97 & 1.97 & 355 & 355 & Exp. & \onlinecitesq{LandBorn1,Barth28092011} \\
Co$_2$ZrSn & 1.56 & 1.56 & 448 & 448 & Exp. & \onlinecitesq{Zhang2006255} \\
Co$_2$VGa & 2.04 & 2.04 & 357 & 357 & Exp. & \onlinecitesq{PhysRevB.76.024414,PhysRevB.82.144415} \\
Co$_2$VSn & 1.21 & 1.21 & 95 & 95 & Exp. & \onlinecitesq{PhysRevB.76.024414,0022-3727-40-6-S01} \\
Co$_2$VAl & 1.86 & 1.86 & 342 & 342 & Exp. & \onlinecitesq{LandBorn1,PhysRevB.82.144415} \\
Co$_2$ZrAl & 0.74 & 0.74 & 185 & 185 & Exp. & \onlinecitesq{LandBorn1,Kanomata200526} \\
Co$_2$ZrSn & 1.51 & 1.51 & 444 & 444 & Exp. & \onlinecitesq{LandBorn1} \\
Co$_2$NbAl & 1.35 & 1.35 & 383 & 383 & Exp. & \onlinecitesq{LandBorn1,0022-3727-40-6-S01} \\
Co$_2$NbSn & 0.52 & 0.52 & 119 & 119 & Exp. & \onlinecitesq{LandBorn1,PhysRevB.66.174428} \\
Co$_2$HfAl & 0.81 & 0.81 & 193 & 193 & Exp. & \onlinecitesq{LandBorn1} \\
Co$_2$HfGa & 0.54 & 0.54 & 186 & 186 & Exp. & \onlinecitesq{LandBorn1} \\
Co$_2$HfSn & 1.55 & 1.55 & 394 & 394 & Exp. & \onlinecitesq{LandBorn1} \\
Co$_2$CrGa & 3.01 & 3.01 & 495 & 495 & Exp. & \onlinecitesq{PhysRevB.76.024414} \\
Co$_2$CrAl & 1.55 & 1.55 & 334 & 334 & Exp. & \onlinecitesq{PhysRevB.76.024414,Hakimi20103443,JEPT2013Svyazhin} \\
Co$_2$MnAl & 4.01-4.04 & 4.04 & 693-697 & 697 & Exp. & \onlinecitesq{LandBorn1,PhysRevB.76.024414} \\
Co$_2$MnGa & 4.05 & 4.05 & 694 & 694 & Exp. & \onlinecitesq{LandBorn1} \\
Co$_2$MnGe & 5.11 & 5.11 & 905 & 905 & Exp. & \onlinecitesq{LandBorn1} \\
Co$_2$MnSi & 4.90 & 4.90 & 985 & 985 & Exp. & \onlinecitesq{PhysRevB.76.024414,LandBorn1} \\
Co$_2$MnSn & 5.08 & 5.08 & 829 & 829 & Exp. & \onlinecitesq{PhysRevB.76.024414,LandBorn1} \\
Co$_2$FeSi & 6.00 & 6.00 & 1100 & 1100 & Exp. & \onlinecitesq{PhysRevB.76.024414} \\
Co$_2$FeAl & 4.96 & 4.96 & 1000 & 1000 & Exp. & \onlinecitesq{Trudel:2013p323} \\
Co$_2$FeGa & 5.15 & 5.15 & $>$1100 & 1100 & Exp. & \onlinecitesq{Trudel:2013p323} \\
Co$_2$TaAl & 0.75 & 0.75 & 260 & 260 & Exp. & \onlinecitesq{Carbonari1996} \\
\end{tabular}
\etab

\tab
\mycaption[Summary Table magnetic Heuslers of the type $X_2$Mn$Y$.]
{Here are reported the compound,
the magnetic moment per formula unit, $m$, the experimental $T_\mathrm{C}$, the volume of
the $F\overline{4}3m$ cell, and the number of valence electrons per formula unit, $N_\mathrm{V}$,
together with the appropriate reference. The quantity labeled with a `*' are those, which have
been used to run the regression.}
\tabvspace
\resizebox{\linewidth}{!}{
\begin{tabular}{l|r|r|r|r|r|r|r|r}
material & $m$/f.u. ($\mu_\mathrm{B}$) & $m^*$/f.u. ($\mu_\mathrm{B}$) & $T_\mathrm{C}$~\K\ & $T_\mathrm{C}^*$~\K\ & volume (\AA$^3$) & $N_\mathrm{V}$ & order & reference \\
\hline
Rh$_2$MnGe & 4.17-4.62 & 4.62 & 400-470 & 450 & 56.46 & 29 & FM & \onlinecitesq{LandBorn1,Klaer2009,Suits1976,Adachi200437,Hames1971}  \\
Rh$_2$MnSn & 3.10-3.93 & 3.10 & 412-431 & 412 & 62.22 & 29 & FM & \onlinecitesq{LandBorn1,Suits1976,Adachi200437} \\
Rh$_2$MnPb & 4.12 & 4.12 & 338 & 338 & 65.58 & 29 & FM & \onlinecitesq{LandBorn1,Suits1976} \\
Rh$_2$MnAl & 4.1 & 4.1 & 85-105 & 95 & 54.96 & 28 & FM & \onlinecitesq{Wijn1,Suits1976} \\
Cu$_2$MnSn & 4.11 & 4.11 & 530 & 530 & 60.36 & 33 & FM & \onlinecitesq{LandBorn1,Wijn1,Oxley1963} \\
Cu$_2$MnAl & 3.73-4.12 & 4.12 & 603 & 603 & 51.93 & 32 & FM & \onlinecitesq{LandBorn1,Wijn1,Oxley1963} \\
Cu$_2$MnIn & 3.95 & 3.95 & 510 & 510 & 59.45 & 32& FM & \onlinecitesq{Oxley1963,Coles1949} \\
Pd$_2$MnAl & 4.4 & 4.4 & 240 & 240 & 58.89 & 30 & AFM & \onlinecitesq{Wijn1,Webster1968} \\
Pd$_2$MnSn & 4.23 & 4.23 & 189 & 189 & 66.00 & 31 & FM & \onlinecitesq{LandBorn1,Wijn1,Campbell1977,Webster1967} \\
Pd$_2$MnSb & 4.40 & 4.40 & 247 & 247 & 67.58 & 32 & FM & \onlinecitesq{LandBorn1,Wijn1,Webster1967} \\
Pd$_2$MnGe & 3.2 & 3.2 & 170 & 170 & 60.49 & 31 & FM & \onlinecitesq{Wijn1} \\
Pd$_2$MnIn & 4.3 & 4.3 & 142 & 142 & 65.88 & 30 & AFM & \onlinecitesq{Wijn1,Webster1967} \\
Au$_2$MnAl & 4.2 & 4.2 & 233 & 233 & 65.37 & 32 & FM & \onlinecitesq{Wijn1,Bacon1967} \\
Au$_2$MnZn & 4.6 & 4.6 & 253 & 253 & 65.32 & 31 & FM & \onlinecitesq{Wijn1,Bacon1973} \\
Ru$_2$MnGe & 3.2-3.8 & 3.8 & 316 & 316 & 54.33 & 27 & AFMII & \onlinecitesq{Kanomata2006,Gotoh1995} \\
Ru$_2$MnSi & 2.8 & 2.8 & 313 & 313 & 51.82 & 27 & AFMII & \onlinecitesq{Kanomata2006} \\
Ru$_2$MnSb & 3.9-4.4 & 4.4 & 195 & 195 & 58.98 & 28 & AFMII & \onlinecitesq{Kanomata2006,Gotoh1995} \\
Ru$_2$MnSn & 2.8 & 2.8 & 296 & 296 & 58.92 & 27 & AFMII & \onlinecitesq{Kanomata2006} \\
\end{tabular}}
\etab

\newcommand{\heuslerstabfootone}{
Note that the tetragonal phase is obtained when annealing at $400^\circ$C.
A higher annealing temperature of $800^\circ$C results in a disorder pseudo-cubic phase.
No magnetic data are available for this second phase.}
\newcommand{\heuslerstabfoottwo}{
Note that Mn$_2$NiGa is a shape memory alloy, displaying a martensitic transformation at a critical temperature
$T_{\mathrm{m}}=270$~K.
The structure is cubic for $T>T_{\mathrm{m}}$ and tetragonal for $T<T_{\mathrm{m}}$.}
\newcommand{\heuslerstabfootthree}{
The ground state magnetic configuration is non-collinear with a canting angle between the two inequivalent
magnetic ions of $180\pm55^\circ$.}
\newcommand{\heuslerstabfootfour}{
The $T_{\mathrm{C}}$ is evaluated from theory of Reference~\onlinecite{Meinert2011}.}

\clearpage

\tab
\mycaption[Summary Table magnetic Heuslers of the type Mn$_2$$YZ$.]
{Here are reported the compound,
the Heusler type (RE = regular, IN = inverse), the chemical order, the magnetic moment per
formula unit, $m$, the experimental $T_\mathrm{C}$, the magnetic order (FM = ferromagnetic,
FI = ferrimagnetic), together with the appropriate reference. For the chemical order we refer
to the conventional notation (see for instance Reference~\onlinecite{Graf_PSSC_2011}).}
\tabvspace
\resizebox{\linewidth}{!}{
\begin{tabular}{l|r|r|r|r|r|r}
material & Heusler & chemical order & $m$/f.u. ($\mu_\mathrm{B}$) & $T_\mathrm{C}$~\K\ & order & reference \\
\hline
Mn$_2$VAl & RE (cubic) & $L2_{1}$ & 1.2-1.81 (Mn); 0.7-0.9 (V) & 760 & FI & \onlinecitesq{Buschow1981,Nakamichi1983,Itoh1983,Jiang2001} \\
Mn$_2$FeGa & IN (tet. $c/a=1.89$)\tablefootnote{\heuslerstabfootone} & Disorder & 1.5 & 650 & FI & \onlinecitesq{Gasi2013} \\
Mn$_2$NiGa & IN (cubic)\tablefootnote{\heuslerstabfoottwo} & $L2_{1}B$ & 1.44-1.5 & 588 & FI & \onlinecitesq{Liu2005,Liu2006,Brown2010} \\
Mn$_2$PtGa & IN (tet.) & Disorder & 1.0-1.6 & 230 & FI & \onlinecitesq{Nayak2013,Nayak2013b,Nayak2015} \\
Mn$_2$CoGa & IN (cubic) & Order $L2_{1}$ & 1.95-2.02 & 718 & FI & \onlinecitesq{Alijani2011,Meinert_PRB_2011,Meinert2011}    \\
Mn$_2$RuGa & IN (cubic) & Order & 0.5-1.15 & 460 & FI & \onlinecitesq{Hori2002,Kurt2014}  \\
Mn$_2$RhGa & IN (cubic) & Disorder $L2_{1}B$ & &  & & \onlinecitesq{Kreiner2014} \\
Mn$_2$VGa & RE (cubic) & $L2_{1}$ & 1.88 & 783 & FI & \onlinecitesq{Kreiner2014,Kumar2008} \\
Mn$_2$RuGe & IN (cubic)  & Disorder $L2_{1}B$ & 1.55 & 303 & FI & \onlinecitesq{Kreiner2014,Yang2015}  \\
Mn$_2$PtIn & IN (tet. $c/a=1.57$) & Disorder & 1.6 & 350 & FI & \onlinecitesq{Nayak2012,Luo2013} \\
Mn$_2$RhSn & IN (tet. $c/a=1.54$) &  & 1.97 & 270 & FI\tablefootnote{\heuslerstabfootthree} & \onlinecitesq{Alijani2013Rh,Meshcheriakova2014} \\
Mn$_2$RuSi & IN (cubic)  & Disorder $L2_{1}B$ &  & 50 & Glass & \onlinecitesq{Endo2012} \\
Mn$_2$CoAl & IN (cubic)  & Disorder & 2 & 670-720 & FI & \onlinecitesq{Ouardi2013,Jamer2013,Zhang2013} \\
Mn$_2$CoGe & IN (cubic)  &  & 2.99 & 579\tablefootnote{\heuslerstabfootfour} & FI & \onlinecitesq{Liu2008} \\
Mn$_2$CoSn & IN (cubic)  & Disorder $L2_{1}B$/$B1$ & 2.98 & 598-610 & FI & \onlinecitesq{Kreiner2014,Liu2008,Lakshmi2002,Winterlik2011}  \\
Mn$_2$CoSb & IN (cubic)  &   & 3.92 & 485 & FI & \onlinecitesq{Dai2006} \\
Mn$_2$NiSn & IN (cubic)  & Disorder $B1$  & 2.95 & 530-565  & FI & \onlinecitesq{Lakshmi2002,Helmholdt1987,Luo2009} \\
Mn$_2$NiSb & IN (cubic)  &   & 4.20 & 647 & FM & \onlinecitesq{Luo2009,Luo2009a} \\
Mn$_3$Ga & (tet. $c/a=1.816$) & $DO_{22}$ & 0.26 & 770 & FI & \onlinecitesq{Winterlik2008} \\
\end{tabular}}
\etab

\clearpage

\subsection{Ternary phase diagrams}

Here we present the ternary phase diagrams (convex hull diagrams) for the Mn$_2$-based Heusler alloys for
which we have attempted the experimental growth.

\clearpage

\subsubsection{Mn-Pt-Co}

\tabsec
\mycaption[The most energetically favorable binary decomposition among structures in the \AFLOWorg\ database
and all competing phases found in the \ICSD\ for the Mn-Pt-Co system.]
{Whenever the experimental critical temperature (either Curie or N\'{e}el) of a given compound is known, it is reported in the
tables (FM = ferromagnet, AF = antiferromagnet).}
\tabvspace
\begin{tabular}{l|l|r|r|r}
Heusler & reactant & SG & lattice (\AA) & $T_\mathrm{C}$~\K\ \\
\hline
Mn$_2$CoPt 			  & Mn$_3$Pt          & $Pm\overline{3}m$     & 3.64                    & 475 (AF) \\
						      & Mn$_3$Pt$_5$      & $Cmmm$                & 5.51, 5.51, 4.01        & --- \\
 							    & Co                & $P6_{3}/mmc$          & 2.51, 2.51, 4.07        & 1388 (FM) \\
\hline
Mn$_2$CoPt 			  & MnPt              & $P4/mmm$              & 2.65, 2.65, 3.77        & 973 (AF) \\
                  & Mn$_3$Pt          & $Pm\overline{3}m$     & 3.64                    & 475 (AF) \\
                  & MnPt$_3$          & $Pm\overline{3}m$     & 3.93                    & 380 (FM) \\
				 			    & CoPt              & $P4/mmm$              & 2.69, 2.69, 3.70        & 813  (FM) \\
                  & CoPt$_3$          & $Pm\overline{3}m$     & 3.89                    & 288 (FM)  \\
                  & Mn$_{1-x}$Co$_x$  & $Fm\overline{3}m$     & 3.62                    & 143 (FM) 65 (AFM) \\
\end{tabular}
\etab

\figsec
\includegraphics[width=0.65\linewidth]{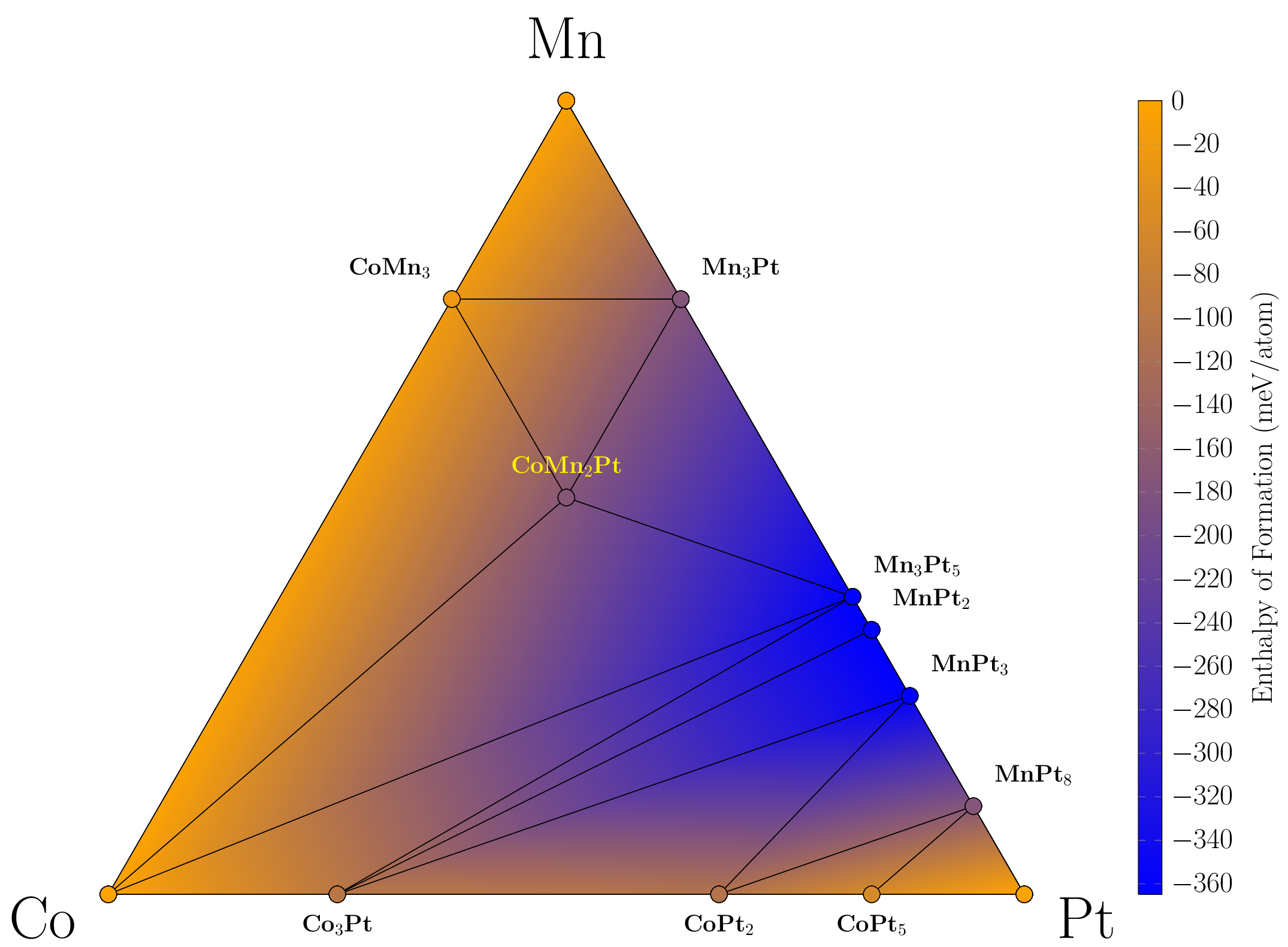}
\mycaption{Mn-Pt-Co ternary convex hull.}
\efig

\clearpage

\subsubsection{Mn-Pt-V}

\tabsec
\mycaption[The most energetically favorable binary decomposition among structures in the \AFLOWorg\ database
and all competing phases found in the \ICSD\ for the Mn-Pt-V system.]
{Whenever the experimental critical temperature (either Curie or N\'{e}el) of a given compound is known, it is reported in the
tables (FM = ferromagnet, AF = antiferromagnet).}
\tabvspace
\begin{tabular}{l|l|r|r|r}
Heusler & reactant & SG & lattice (\AA) & $T_\mathrm{C}$~\K\ \\
\hline
Mn$_2$PtV 			  & PtV           & $P4/mmm$              & 2.70, 2.70, 3.90        & 0  \\
			     			  & Mn$_3$Pt      & $Pm\overline{3}m$     & 3.64                    & 475 (AF) \\
						      & Mn$_3$Pt$_5$  & $Cmmm$                & 5.51, 5.51, 4.01        & --- \\
\hline
Mn$_2$PtV 			  & PtV           & $P4/mmm$              & 2.70, 2.70, 3.90        & 0  \\
			     			  & Pt$_8$V       & $I4/mmm$              & 6.24                    & 0  \\
			     			  & Pt$_3$V       & $I4/mmm$              & 4.81                    & 290 (FM) \\
			     			  & Pt$_2$V       & $Immm$                & 4.83                    & 0  \\
                  & PtV$_3$       & $Pm\overline{3}n$     & 4.81                    & 0 \\
				 			    & MnPt          & $P4/mmm$              & 2.65, 2.65, 3.77        & 973 (AF) \\
			     			  & Mn$_3$Pt      & $Pm\overline{3}m$     & 3.64                    & 475 (AF) \\
			     			  & MnPt$_3$      & $Pm\overline{3}m$     & 3.93                    & 380 (FM) \\
				 			    & MnV           & $Pm\overline{3}m$     & 2.87				            & 0  \\
\end{tabular}
\etab

\vspace{-0.5cm}

\figsec
\includegraphics[width=0.65\linewidth]{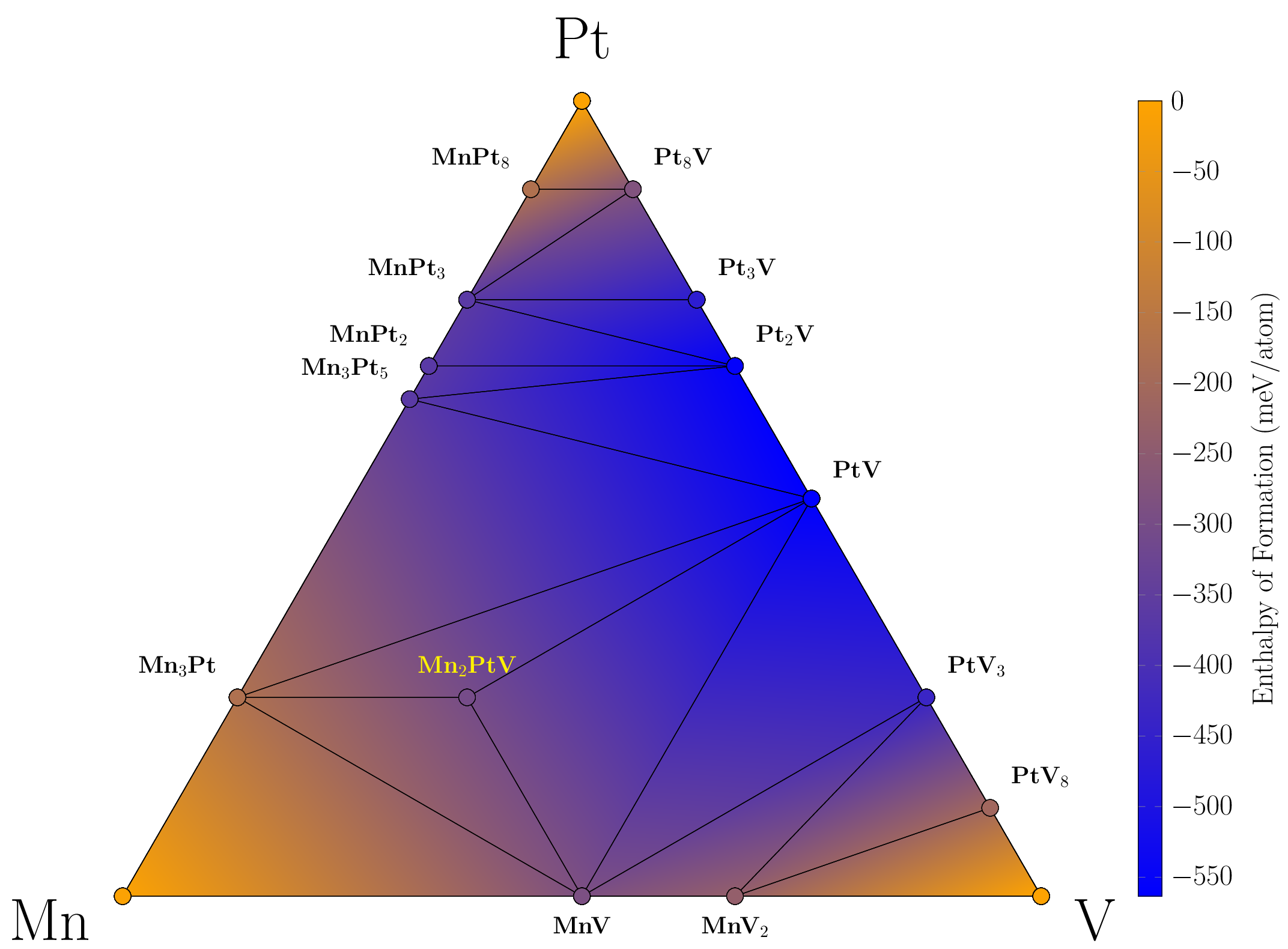}
\mycaption{Mn-Pt-V ternary convex hull.}
\efig

\clearpage

\subsubsection{Mn-Pt-Pd}

\tabsec
\mycaption[The most energetically favorable binary decomposition among structures in the \AFLOWorg\ database
and all competing phases found in the \ICSD\ for the Mn-Pt-Pd system.]
{Whenever the experimental critical temperature (either Curie or N\'{e}el) of a given compound is known, it is reported in the
tables (FM = ferromagnet, AF = antiferromagnet).}
\tabvspace
\begin{tabular}{l|l|r|r|r}
Heusler & reactant & SG & lattice (\AA) & $T_\mathrm{C}$~\K\ \\
\hline
Mn$_2$PtPd 			  & MnPd$_3$        & $I4/mmm$            & 8.30              & 170  \\
			     			  & MnPd$_2$        & $Pnma$              & 5.45, 4.11, 8.10  &  (AF) \\
			     			  & MnPt$_3$        & $Pm\overline{3}m$   & 3.93              & 380 (FM) \\
\hline
Mn$_2$PtPd 			  & MnPd$_3$        & $I4/mmm$            & 8.30              & 170  \\
				 			    & MnPd            & $P4/mmm$            & 2.62, 2.62, 3.81  & 813 (AF) \\
				 			    & MnPd            & $Pm\overline{3}m$   & 2.99              & 813 (AF) \\
				 			    & MnPt            & $P4/mmm$            & 2.65, 2.65, 3.77  & 973 (AF) \\
			     			  & Mn$_3$Pt        & $Pm\overline{3}m$   & 3.64              & 475 (AF) \\
			     			  & MnPt$_3$        & $Pm\overline{3}m$   & 3.93              & 380 (FM) \\
\end{tabular}
\etab

\figsec
\includegraphics[width=0.65\linewidth]{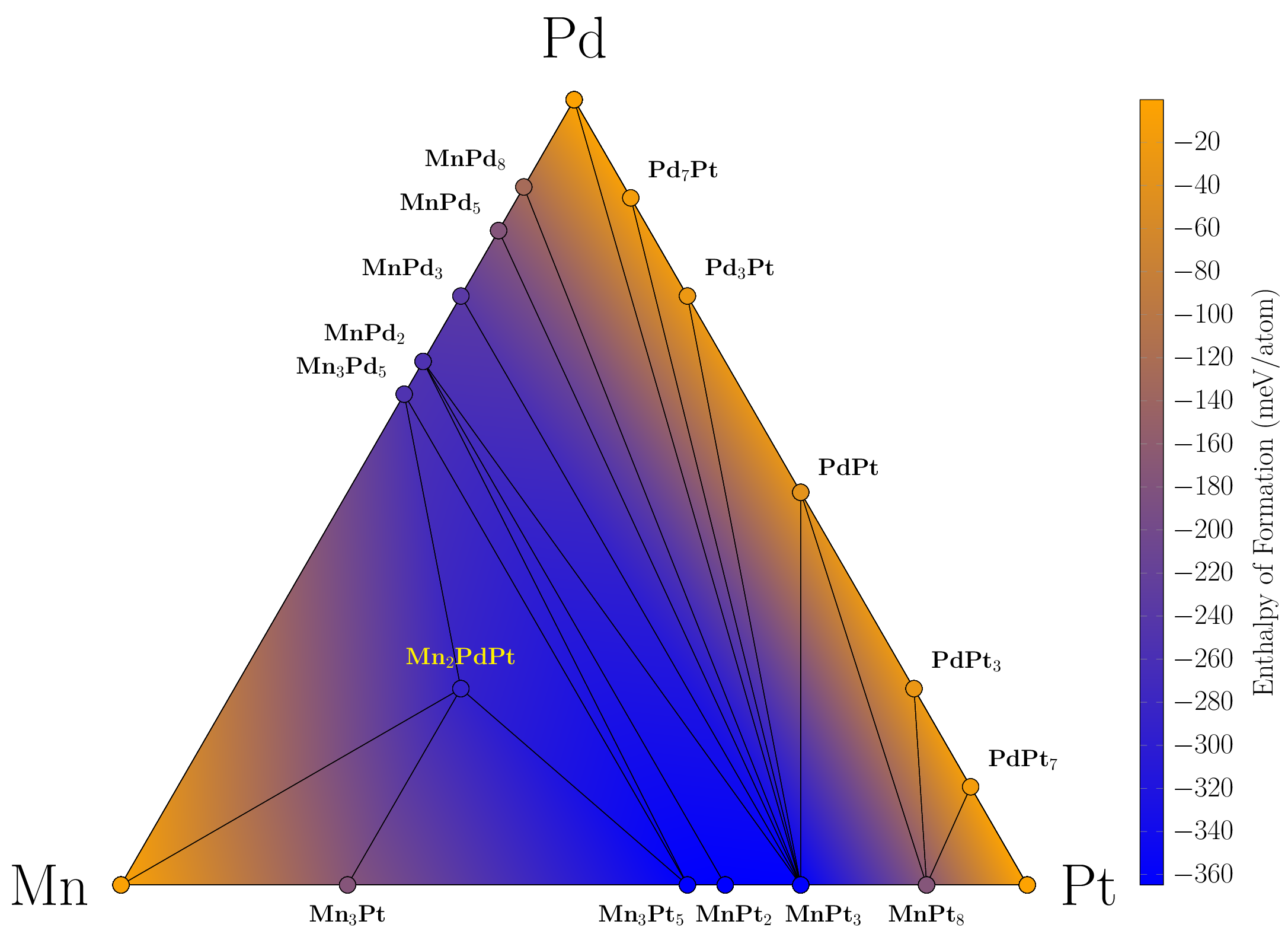}
\mycaption{Mn-Pt-Pd ternary convex hull.}
\efig

\clearpage

\subsection{\texorpdfstring{$T_\mathrm{C}$}{Tc} prediction:  regression analysis}
A standard generalized regression model with a Poisson link
function is used to predict $T_{\mathrm{C}}$ for type Co$_2YZ$ and $X_2$Mn$Z$ Heuslers~\cite{GLR}.
The link function was chosen because it performs the best under maximum likelihood fitting of the model parameters~\cite{Poisson}.

To determine the optimal set of regressors, we perform a correlation analysis to cluster predictor variables appropriately.
Variables within clusters should have high correlations among themselves and low correlations with variables of other clusters.
As expected, the following significant clusters are realized:  cluster one ($a$ and volume),
cluster two ($m$ and spin decomposition), cluster three ($N_{\mathrm{V}}$),
cluster 4 (H and $T_{\mathrm{S}}$), and cluster 5 ($P_{\mathrm{F}}$).
A check of the variance inflation factors (values less than 1.5) suggests sufficiently low correlation of variables.

In the regression, we perform a 10-fold cross validation (8:2 split).
Relevant parameters in the final model include volume, spin decomposition, $m$, and $N_{\mathrm{V}}$.

For Co$_2YZ$ Heuslers, the regression is improved by training only on experimental data, as $T_{\mathrm{C}}$
shows to be insensitive to volume and spin decomposition.
Therefore, we expect $T_{\mathrm{C}}$ prediction to be closely associated with the Slater-Pauling curve.

In general for the $X_2$Mn$Z$ Heuslers, the only active magnetic ions are Mn in the octahedral
positions with a $m\sim4\mu_{\text{B}}$.
This, along with low correlation values, led us to remove $m$ as a significant regressor.
Expecting short-range magnetic interaction, focus is placed on $a$ and $N_{\mathrm{V}}$ as major independent variables.
In general, we observe a negative correlation between $T_{\mathrm{C}}$ and $a$.

\fig
\includegraphics[width=\linewidth]{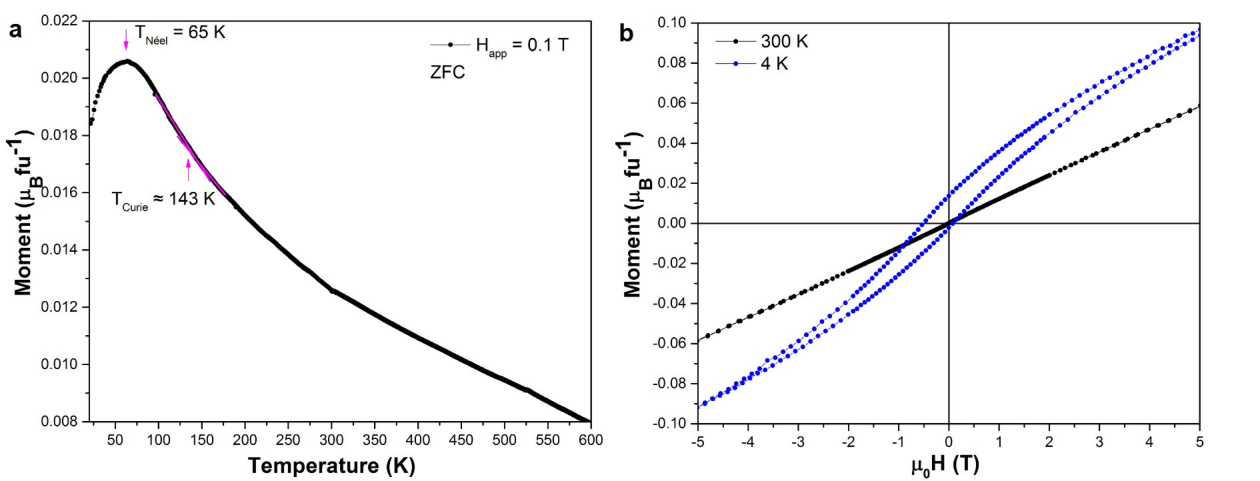}
\mycaption{(\textbf{a}) Zero field cooled magnetization curve as a function of temperature
and (\textbf{b}) magnetization curves at 300K and 4K of $\text{Mn}_{2}\text{PtCo}$.}
\label{fig:art109:Zero-field-cooled}
\efig

\fig
\includegraphics[width=\linewidth]{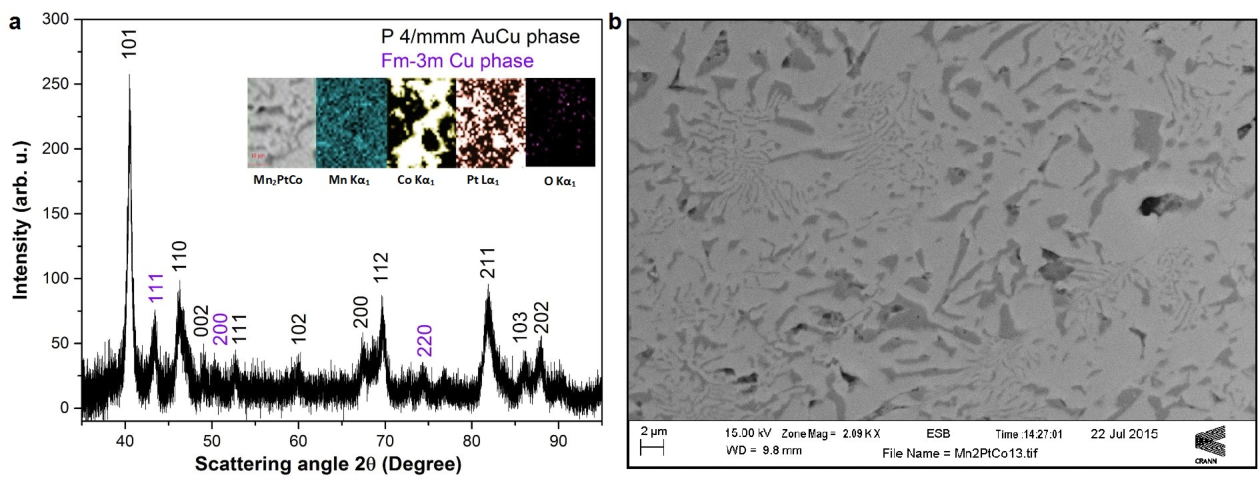}
\mycaption[(\textbf{a}) XRD pattern with main AuCu structure and EDX map analysis
and (\textbf{b}) SEM image of $\text{Mn}_{2}\text{PtCo}$ (\textbf{b}).]
{In the EDX map analysis, white/black indicates the absence/maximum amount of the element.}
\label{fig:art109:XRDMn2PtCo}
\efig

\subsection{Experimental structural and magnetic analysis \texorpdfstring{Mn$_2$}{Mn2}-based compounds} \label{subsec:art109:exp_data_Mn2_based}

We we provide information on the two Heusler alloys, namely Mn$_2$PtCo and Mn$_2$PtV
for which the synthesis was not successful and resulted in phase-segregated phases.

The Heusler alloys were prepared by arc melting in high-purity argon. The ingots were remelted four
times to ensure homogeneity. An excess of 3 \% wt. Mn was added in order to compensate for Mn losses
during arc melting. Ingots were sealed under vacuum in quartz tube ($10^{-6}$ Torr) slowly heated at
5~C/min up to 850~C and soaked at 850~C for 1 week, then slowly cooled down to room temperature at
2~C/min. Structural characterization was carried out by powder X-ray diffraction (XRD) with a PANalytical
X'Pert Pro diffractometer with Cu-K$_{\alpha}$ radiation.The bulk pieces were held in a gel cap and the
magnetic measurements were carried out using a Quantum Design superconducting quantum interference
device magnetometer in a field of up to 5 T. The microstructure and composition were analyzed with a scanning
electron microscope (SEM Carl Zeiss Evo) for the polished bulk samples. The compositions are determined by
Energy Dispersive X-ray Spectroscopy (EDX).

\subsubsection{\texorpdfstring{Mn$_{2}$}{Mn2}PtCo}

Mn$_2$PtCo is unstable and decomposes into two phases: MnPt with the tetragonal AuCu-type structure
($P4/mmm$) and Mn$_{1-x}$Co$_x$, with $x$ = 0.34-0.37 with a face centered cubic Cu type structure
($Fm\overline{3}m$). MnPt is antiferromagnetic with a high N\'{e}el temperature~\cite{Kren1968},
$\text{T}_{\text{N\'{e}el}}=$975~K. $\text{Mn}_{1-x}\text{Co}_{x}$ is expected to consist both ferromagnetic
and antiferromagnetic phases when the composition is $x$ = 0.34 - 0.37, with the Curie and N\'{e}el temperature
around 140~K and 60~K respectively \cite{KouvelCoMn,M.MatsuiT.IdoK.Sato1970,CoMnRussian}.

Our data reveal a N\'{e}el temperature around $\sim$65~K and a Curie temperature around $\sim$148~K
(Figure~\ref{fig:art109:Zero-field-cooled}a). We saw no trace of $\text{MnPt}_{3}$ ($T_{\text{C}}=$ 380K)
or $\text{Mn}_{3}\text{Pt}$ ($T_{\text{N}}=$475K). Our experimental magnetization data and XRD
measurements confirm the presence of $\text{Mn}_{{1-x}}\text{Co}_{{x}}$ with $x$ = 0.34-0.37 and
MnPt, which have cubic and tetragonal structures respectively (Figure~\ref{fig:art109:XRDMn2PtCo}). The room
temperature magnetization curve is linear due to the antiferromagnetic MnPt, with a paramagnetic
contribution from $\text{Mn}_{{1-x}}\text{Co}_{{x}}$. At 4~K, the hysteresis loop exhibits exchange bias,
due to the coexistence intergrown of AFM and FM phases in Mn-Co alloy as originally reported
by Kouvel~\cite{KouvelCoMn} .

XRD results show that the main phase is MnPt ({\it P4/mmm}) with lattice parameters $a=278$~pm
and $c=372$~pm and the secondary phase is Mn$_{1-x}$Co$_x$ with $x=0.34-0.37$
($Fm\overline{3}m$) with lattice parameter $a=362$~pm (Figure \ref{fig:art109:XRDMn2PtCo}). SEM images
demonstrate the decomposition of $\text{Mn}_{2}\text{PtCo}$ into Mn-Co and Mn-Pt phases
(see Figure~\ref{fig:art109:XRDMn2PtCo}). The EDX analysis confirms the absence of any Co-Pt phase, and
the existence of Mn-Pt and Mn-Co phases. EDX map analysis, line and point spectra show that white
or light grey parts belong to Mn-Pt, while the dark grey part belongs to Co and Mn-rich phases and the
black spots belong to Co-rich material. Thin grey line features may be $\text{Mn}_{1-x}\text{Co}_{x}$
with $x=0.34-0.37$. Elemental maps confirms that Co-rich areas show no sign of any Pt, and the Pt and
Mn coexists.

\subsubsection{\texorpdfstring{Mn$_{2}$}{Mn2}PtV}

\fig
\includegraphics[width=0.975\linewidth]{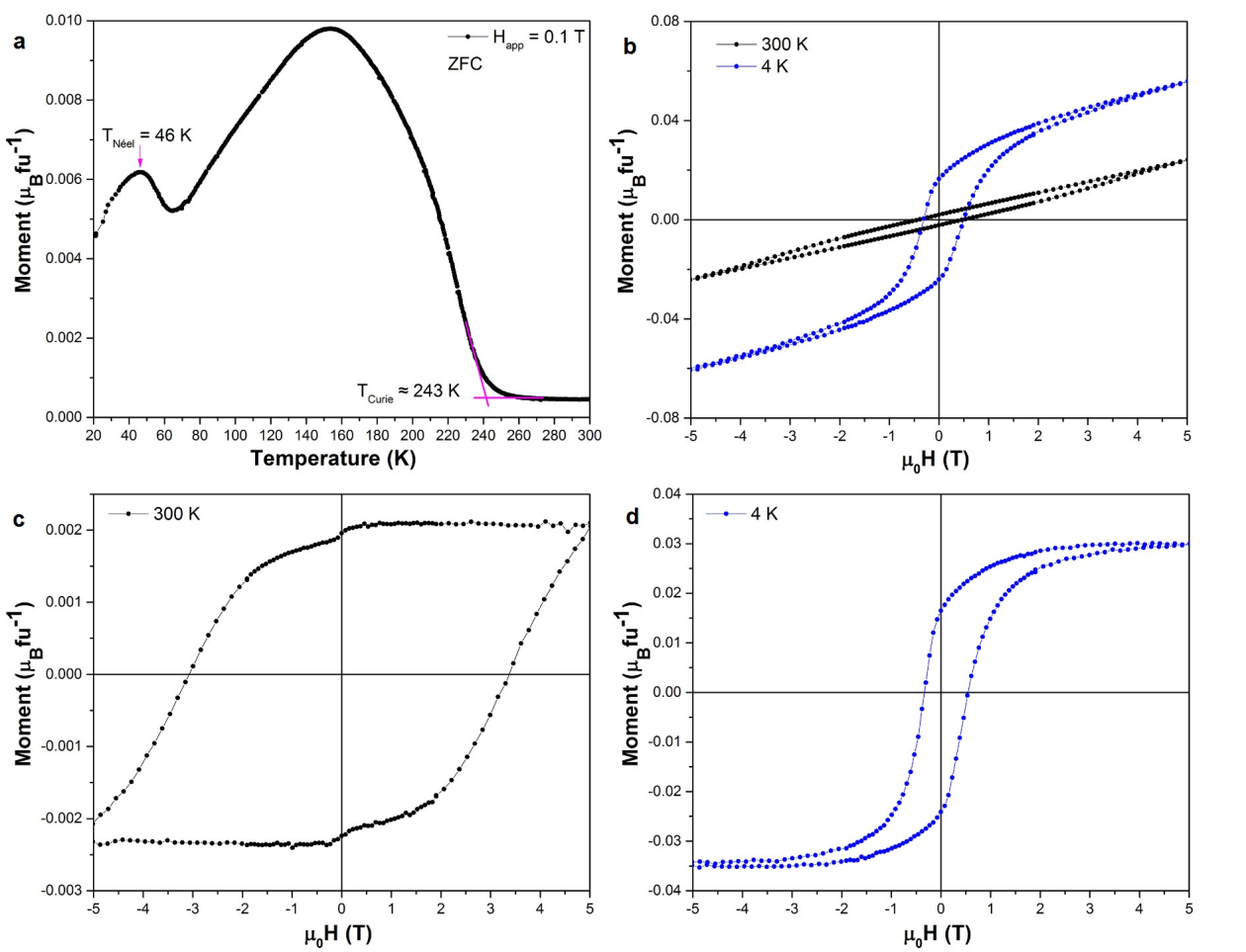}
\mycaption{(\textbf{a}) Zero field cooled magnetization curve of $\text{Mn}_{2}\text{PtV}$ as
a function of temperature and (\textbf{b}) magnetization curves at 300~K and
4~K (\textbf{c}-\textbf{d}) after correction for the paramagnetic slope.}
\label{fig:art109:Mn2PtV}
\efig

\fig
\includegraphics[width=\linewidth]{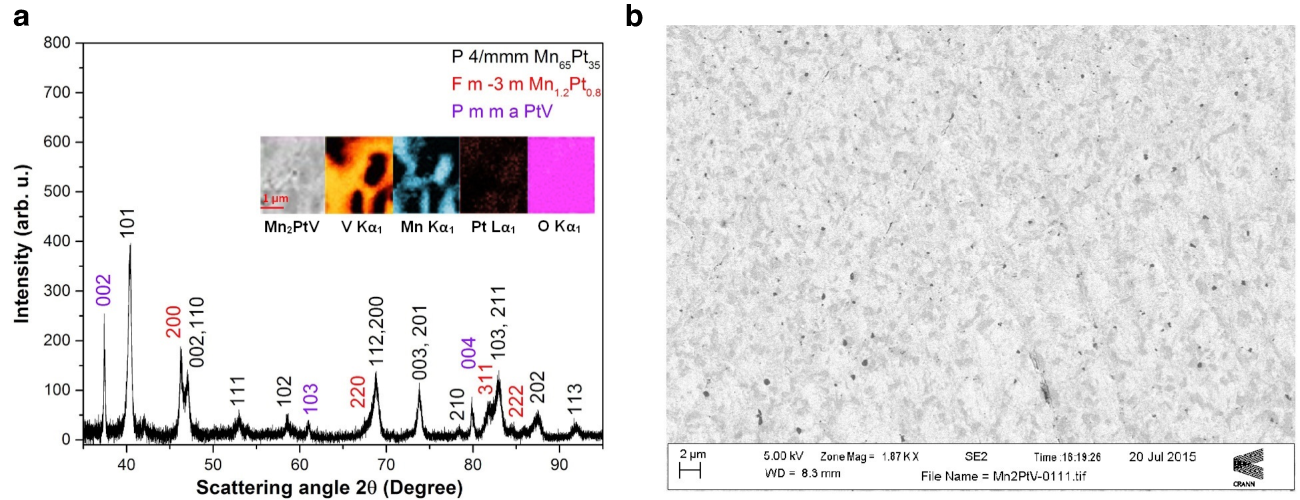}
\mycaption{(\textbf{a}) XRD pattern of $\text{Mn}_{2}\text{PtV}$ with main $\text{AuCu}$
structure and (\textbf{b}) SEM image of $\text{Mn}_{2}\text{PtV}$.}
\label{fig:art109:XRD_Mn2PtV}
\efig

Mn$_{2}$PtV is unstable and decomposes into three main phases: tetragonal Mn$_{65}$Pt$_{35}$
(AuCu structure, $P4/mmm$), cubic Mn$_{1.2}$Pt$_{0.8}$ ($Fm\overline{3}m$) and orthorhombic PtV (AuCd structure
$Pmma$). Mn$_{65}$Pt$_{35}$ and Mn$_{1.2}$Pt$_{0.8}$ are ferromagnetic with Curie temperature $\sim$~250~K
and 540~K respectively~\cite{A.KjekshusR.Mollerud1966,Kren1968}. Our data reveals an unidentified transition at
$\sim$46~K and a peak at $\sim$243~K, which we associate with Mn$_{65}$Pt$_{35}$ (Figure \ref{fig:art109:Mn2PtV}).

The room temperature magnetization curve is dominated by the ferromagnetic Mn$_{1.2}$Pt$_{0.8}$, with
paramagnetic contributions from Mn$_3$O$_4$ and Mn$_{65}$Pt$_{35}$. After correction for the paramagnetic
slope due to Mn$_{65}$Pt$_{35}$, we obtained a loop 3~T coercivity and low magnetization, 0.002
$\mu_{\text{B}}\ \text{f.u.}$ The hysteresis curve at 4~K is dominated by Mn$_{65}$Pt$_{35}$ with 0.35~T coercivity.

Mn$_{65}$Pt$_{35}$ has a tetragonal structure with lattice parameter $a=273$~pm and $c=386$~pm whereas
Mn$_{1.2}$Pt$_{0.8}$ is cubic with lattice parameter $a=390$~pm and PtV is orthorhombic with lattice parameters
$a=446$~pm, $b=266$~pm and $c=480$~pm, as is shown in Figure~\ref{fig:art109:XRD_Mn2PtV}. SEM images prove the
decomposition of Mn-Pt and Pt-V phases (Figure \ref{fig:art109:XRD_Mn2PtV}). Generally vanadium and manganese
do not co-exist in the same area.

According to point and line spectrum; dark grey parts and small black points belongs to Pt-V, light grey and white
areas indicates Mn-rich Mn-Pt and big black regions indicates Pt rich Pt-V phases. In Figure~\ref{fig:art109:XRD_Mn2PtV}
dark regions indicates the elemental rich part and light shade color regions indicates elemental poor phases.

\clearpage

\subsection{Distribution of \texorpdfstring{$T_\mathrm{S}$}{TS} for the intermetallic Heuslers}
We report here histograms for the distribution of the entropic temperatures of the 8776 intermetallic Heuslers
presenting negative enthalpy of formation and for the 248 found stable after the convex hull diagram analysis.

\vspace{1cm}

\figsec
\includegraphics[width=0.65\linewidth]{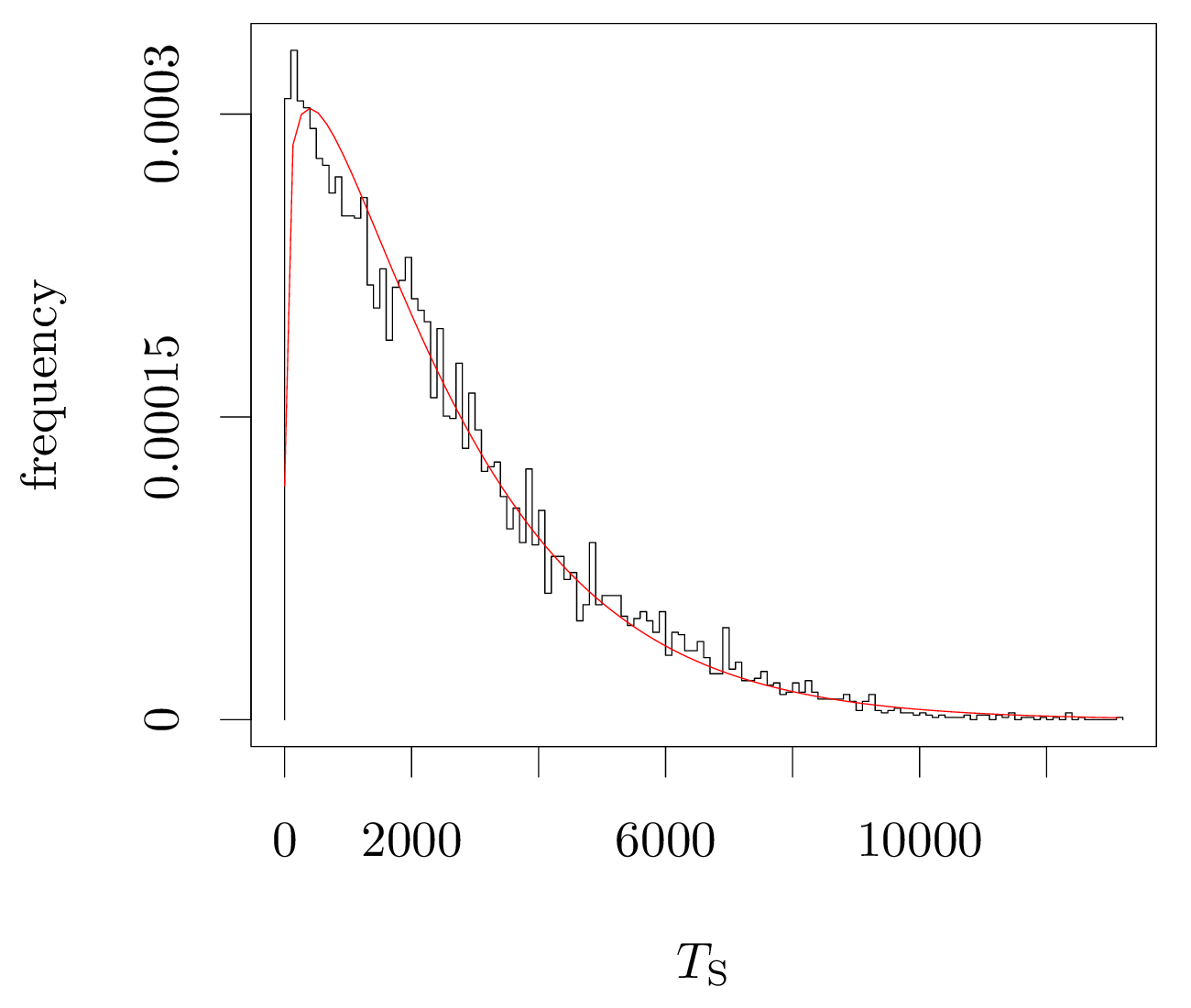}
\mycaption[Histogram of the entropic temperature, $T_\mathrm{S}$, for all the 8776 intermetallic Heuslers displaying
negative enthalpy of formation ($H_{\mathrm{f}}<0$ and $T_\mathrm{S}>0$).]
{The continuous red line is our best fit to a two-parameter Weibull distribution with a shape of 1.13
and a scale of 2585.63.}
\label{fig:art109:histogram_full}
\efig

\figsec
\includegraphics[width=1.0\linewidth]{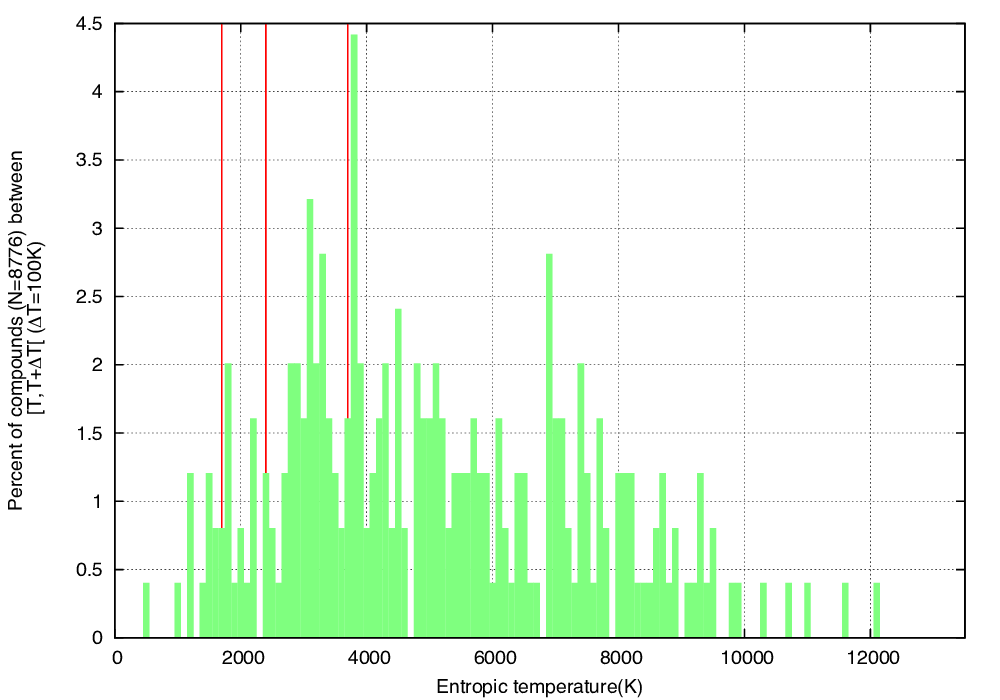}
\mycaption[Histogram of the entropic temperature, $T_\mathrm{S}$, for all the 248 intermetallic Heuslers estimated stable
after the construction of the convex hull diagrams for the ternary phase.]
{The red lines indicate three compounds present in the \ICSD\ database.}
\efig

\clearpage

\subsection{Tetragonal distortion for \texorpdfstring{Mn$_2$}{Mn2}PtPd} \label{subsec:art109:tet_disorder_Mn2PtPd}

The total energy of Mn$_2$PtPd is calculated for different $c/a$ ratio (and constant volume) for both the
ferromagnetic and antiferromagnetic state. Note that, while in the ferromagnetic configuration the energy
minimum is found for the cubic solution, in the antiferromagnetic case (lower in energy) this is found for
$c/a=1.3$, in agreement with the experimental data.

\vspace{1cm}

\figsec
\includegraphics[width=0.5\linewidth]{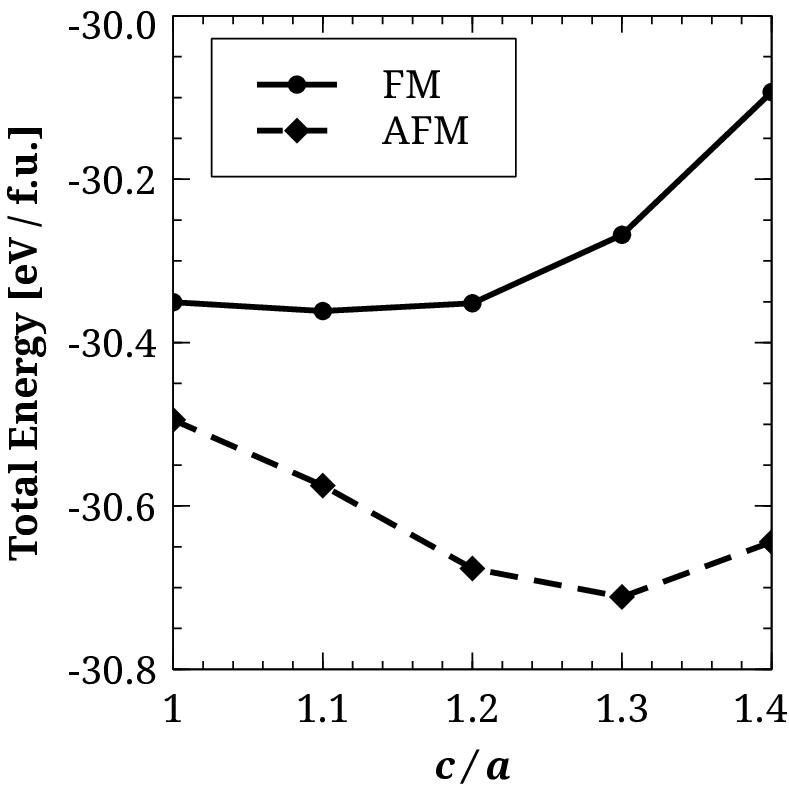}
\mycaption{Total energy as a function of the $c/a$ ratio for Mn$_2$PtPd calculated with \GGA-\DFT.}
\efig

\clearpage

\subsection{List of all stable intermetallic Heuslers}

\tabsec
\mycaption{Summary table of magnetic Heuslers (1/8).}
\tabvspace
\begin{tabular}{l|r|r|r}
material & volume (\AA$^3$) & $H_{\mathrm{f}}$ (eV) & $T_\mathrm{S}$~\K\  \\
\hline
Zn$_2$AgAu &	64.64	&	-0.15	&	1723	\\
Pd$_2$AgCd &	68.13	&	-0.27	&	2958	\\
Ag$_2$CdSc	&	77.7808	&	-0.248922	&	2778.26	\\
Ag$_2$CdY	&	85.8372	&	-0.301022	&	3359.76	\\
Ag$_2$CdZr	&	77.9156	&	-0.098514	&	1099.53	\\
Hg$_2$AgLa	&	99.102	&	-0.392808	&	4384.2	\\
Pd$_2$AgHg	&	69.3436	&	-0.146835	&	1638.85	\\
Hg$_2$AgSc	&	81.9652	&	-0.256216	&	2859.68	\\
Sc$_2$AgHg	&	84.1676	&	-0.364257	&	4065.54	\\
Hg$_2$AgY	&	89.648	&	-0.349733	&	3903.43	\\
Sc$_2$AgOs	&	72.1864	&	-0.376927	&	4206.95	\\
Sc$_2$AgRu	&	72.4904	&	-0.44129	&	4925.31	\\
Y$_2$AgRu	&	87.0664	&	-0.346082	&	3862.68	\\
Au$_2$CdLa	&	94.5472	&	-0.66943	&	7471.63	\\
Pd$_2$AuCd	&	68.5296	&	-0.301286	&	3362.7	\\
Au$_2$CdY	&	85.436	&	-0.674423	&	7527.36	\\
Au$_2$CdZr	&	78.2676	&	-0.457602	&	5107.38	\\
Cu$_2$AuPd	&	57.166	&	-0.115899	&	1293.57	\\
Au$_2$CuZn	&	62.4144	&	-0.142872	&	1594.62	\\
Au$_2$HfZn	&	71.7456	&	-0.438785	&	4897.35	\\
Au$_2$HgLa	&	94.9036	&	-0.627046	&	6998.57	\\
Pd$_2$AuHg	&	69.7384	&	-0.162896	&	1818.12	\\
Zn$_2$AuRh	&	59.324	&	-0.312353	&	3486.23	\\
Sc$_2$AuRu	&	71.926	&	-0.675774	&	7542.43	\\
Au$_2$TiZn	&	66.8216	&	-0.352571	&	3935.11	\\
Au$_2$ZnZr	&	73.2548	&	-0.467891	&	5222.21	\\
Cu$_2$CdZr	&	65.8668	&	-0.155451	&	1735.01	\\
Rh$_2$CdHf	&	67.1976	&	-0.68254	&	7617.94	\\
Hg$_2$CdLa	&	103.328	&	-0.460008	&	5134.23	\\
Hg$_2$CdSc	&	86.474	&	-0.265346	&	2961.58	\\
Hg$_2$CdY	&	94.0524	&	-0.381128	&	4253.84	\\
\end{tabular}
\label{fig:art109:magnetic_heusler_1}
\etab

\tabsec
\mycaption{Summary table of magnetic Heuslers continued (2/8).}
\tabvspace
\begin{tabular}{l|r|r|r}
material & volume (\AA$^3$) & $H_{\mathrm{f}}$ (eV) & $T_\mathrm{S}$~\K\  \\
\hline
Pd$_2$CdSc	&	70.7684	&	-0.725422	&	8096.56	\\
Pd$_2$CdY	&	78.308	&	-0.731543	&	8164.88	\\
Pd$_2$CdZr	&	72.2028	&	-0.58712	&	6552.94	\\
Rh$_2$CdSc	&	67.0008	&	-0.622274	&	6945.31	\\
Rh$_2$CdZr	&	68.4788	&	-0.627501	&	7003.65	\\
Hf$_2$CoRe	&	66.8792	&	-0.412526	&	4604.27	\\
Co$_2$HfSc	&	61.3956	&	-0.38894	&	4341.02	\\
Hf$_2$CoTc	&	66.1292	&	-0.493898	&	5512.48	\\
Co$_2$HfZn	&	53.9212	&	-0.326005	&	3638.6	\\
Sc$_2$CoIr	&	64.4424	&	-0.71918	&	8026.89	\\
Ti$_2$CoIr	&	56.8924	&	-0.622184	&	6944.3	\\
Ti$_2$CoMn	&	52.0108	&	-0.382265	&	4266.53	\\
Ti$_2$CoRe	&	56.7704	&	-0.444075	&	4956.4	\\
Sc$_2$CoRu	&	63.6324	&	-0.467309	&	5215.72	\\
Ti$_2$CoTc	&	56.0352	&	-0.510928	&	5702.56	\\
Zr$_2$CoTc	&	68.3008	&	-0.359379	&	4011.09	\\
Co$_2$TiZn	&	48.8244	&	-0.350328	&	3910.07	\\
Co$_2$ZnZr	&	55.166	&	-0.268346	&	2995.06	\\
V$_2$CrFe	&	47.8092	&	-0.167619	&	1870.82	\\
Ti$_2$CrIr	&	57.4292	&	-0.551684	&	6157.44	\\
V$_2$CrMn	&	48.2312	&	-0.193973	&	2164.97	\\
Nb$_2$CrOs	&	62.176	&	-0.200243	&	2234.95	\\
Ta$_2$CrOs	&	62.2812	&	-0.311877	&	3480.92	\\
V$_2$CrOs	&	52.3748	&	-0.302942	&	3381.19	\\
V$_2$CrRe	&	53.0104	&	-0.258046	&	2880.1	\\
Ta$_2$CrRu	&	61.7168	&	-0.280556	&	3131.34	\\
V$_2$CrRu	&	51.9164	&	-0.25086	&	2799.9	\\
Hf$_2$CuRe	&	69.632	&	-0.296279	&	3306.82	\\
Hf$_2$CuTc	&	69.12	&	-0.339081	&	3784.54	\\
Cu$_2$HfZn	&	58.7964	&	-0.19888	&	2219.74	\\
\end{tabular}
\label{fig:art109:magnetic_heusler_2}
\etab

\tabsec
\mycaption{Summary table of magnetic Heuslers continued (3/8).}
\tabvspace
\begin{tabular}{l|r|r|r}
material & volume (\AA$^3$) & $H_{\mathrm{f}}$ (eV) & $T_\mathrm{S}$~\K\  \\
\hline
Sc$_2$CuIr	&	68.0976	&	-0.699208	&	7803.98	\\
Sc$_2$CuOs	&	67.3716	&	-0.408716	&	4561.75	\\
Zr$_2$CuOs	&	71.076	&	-0.345336	&	3854.35	\\
Pd$_2$CuZn	&	56.2556	&	-0.403379	&	4502.19	\\
Sc$_2$CuPt	&	70.4452	&	-0.801364	&	8944.16	\\
Rh$_2$CuTa	&	58.3456	&	-0.455697	&	5086.11	\\
Sc$_2$CuRu	&	67.33	&	-0.46985	&	5244.08	\\
Y$_2$CuRu	&	81.6212	&	-0.318052	&	3549.83	\\
Zr$_2$CuTc	&	71.3516	&	-0.26889	&	3001.13	\\
Cu$_2$TiZn	&	53.8756	&	-0.169069	&	1887	\\
Cu$_2$ZnZr	&	60.172	&	-0.223658	&	2496.28	\\
Hf$_2$FeOs	&	65.9256	&	-0.524889	&	5858.38	\\
Ti$_2$FeMn	&	51.8856	&	-0.336061	&	3750.83	\\
Ti$_2$FeOs	&	55.8712	&	-0.568209	&	6341.88	\\
Hf$_2$IrMn	&	66.5552	&	-0.641543	&	7160.37	\\
Hf$_2$IrMo	&	70.62	&	-0.605585	&	6759.04	\\
Hf$_2$IrRe	&	69.8952	&	-0.743454	&	8297.82	\\
Hf$_2$IrTc	&	69.3832	&	-0.854328	&	9535.3	\\
Ir$_2$HfZn	&	63.0396	&	-0.732469	&	8175.21	\\
Hf$_2$MoRh	&	70.6316	&	-0.529099	&	5905.36	\\
Tc$_2$HfMo	&	64.7628	&	-0.293247	&	3272.98	\\
Tc$_2$HfNb	&	67.0276	&	-0.447695	&	4996.8	\\
Ni$_2$HfZn	&	55.6964	&	-0.431443	&	4815.41	\\
Hf$_2$OsRu	&	68.636	&	-0.769146	&	8584.58	\\
Os$_2$HfSc	&	67.5148	&	-0.560224	&	6252.76	\\
Hf$_2$OsTc	&	69.0408	&	-0.626591	&	6993.49	\\
Hf$_2$PdRe	&	71.6136	&	-0.559388	&	6243.42	\\
Hf$_2$PdTc	&	71.1996	&	-0.620052	&	6920.51	\\
Pd$_2$HfZn	&	65.598	&	-0.675223	&	7536.29	\\
Hf$_2$ReRh	&	69.9132	&	-0.700075	&	7813.66	\\
Hf$_2$ReZn	&	71.8108	&	-0.299023	&	3337.45	\\
\end{tabular}
\label{fig:art109:magnetic_heusler_3}
\etab

\tabsec
\mycaption{Summary table of magnetic Heuslers continued (4/8).}
\tabvspace
\begin{tabular}{l|r|r|r}
material & volume (\AA$^3$) & $H_{\mathrm{f}}$ (eV) & $T_\mathrm{S}$~\K\  \\
\hline
Hf$_2$RhTc	&	69.3144	&	-0.78918	&	8808.18	\\
Rh$_2$HfZn	&	61.8484	&	-0.857463	&	9570.3	\\
Ru$_2$HfSc	&	66.8988	&	-0.728346	&	8129.19	\\
Hf$_2$RuTc	&	68.6544	&	-0.669049	&	7467.38	\\
Tc$_2$HfTa	&	66.898	&	-0.509943	&	5691.57	\\
Tc$_2$HfW	&	64.88	&	-0.346078	&	3862.64	\\
Ti$_2$IrMn	&	56.4032	&	-0.694021	&	7746.08	\\
Ti$_2$IrMo	&	61.0552	&	-0.626827	&	6996.13	\\
Sc$_2$IrNi	&	65.8112	&	-0.801053	&	8940.69	\\
Sc$_2$IrPd	&	69.9072	&	-0.991899	&	11070.8	\\
Y$_2$IrPd	&	83.818	&	-0.881856	&	9842.55	\\
Ti$_2$IrRe	&	60.1416	&	-0.756525	&	8443.71	\\
Sc$_2$IrRh	&	67.5008	&	-1.04502	&	11663.7	\\
Y$_2$IrRh	&	81.2688	&	-0.841385	&	9390.84	\\
Sc$_2$IrRu	&	66.5136	&	-0.830031	&	9264.12	\\
Sc$_2$IrZn	&	70.6392	&	-0.724073	&	8081.51	\\
Ti$_2$IrTc	&	59.6952	&	-0.840269	&	9378.39	\\
Zr$_2$IrTc	&	71.522	&	-0.693835	&	7744.01	\\
Ir$_2$TiZn	&	57.6528	&	-0.695974	&	7767.89	\\
Ir$_2$ZnZr	&	64.3208	&	-0.634848	&	7085.65	\\
Mn$_2$NbTi	&	54.8624	&	-0.227403	&	2538.09	\\
Ti$_2$MnNi	&	53.3032	&	-0.342964	&	3827.88	\\
Ti$_2$MnOs	&	56.2816	&	-0.502285	&	5606.09	\\
Ti$_2$MnRh	&	56.0512	&	-0.577568	&	6446.34	\\
Mn$_2$TaTi	&	54.9728	&	-0.27885	&	3112.3	\\
Mn$_2$TiV	&	49.6572	&	-0.274813	&	3067.23	\\
Mn$_2$TiW	&	52.8752	&	-0.237692	&	2652.92	\\
Nb$_2$MoOs	&	65.8108	&	-0.281237	&	3138.94	\\
Nb$_2$MoRe	&	66.5	&	-0.250455	&	2795.37	\\
Nb$_2$MoRu	&	65.5172	&	-0.256633	&	2864.32	\\
Mo$_2$NbTa	&	67.5352	&	-0.166161	&	1854.55	\\
\end{tabular}
\label{fig:art109:magnetic_heusler_4}
\etab

\tabsec
\mycaption{Summary table of magnetic Heuslers continued (5/8).}
\tabvspace
\begin{tabular}{l|r|r|r}
material & volume (\AA$^3$) & $H_{\mathrm{f}}$ (eV) & $T_\mathrm{S}$~\K\  \\
\hline
Nb$_2$MoTc	&	66.0904	&	-0.252979	&	2823.54	\\
Mo$_2$NbW	&	65.644	&	-0.113515	&	1266.96	\\
Ti$_2$MoNi	&	58.4756	&	-0.291743	&	3256.19	\\
Ta$_2$MoOs	&	65.8048	&	-0.393033	&	4386.71	\\
V$_2$MoOs	&	56.3172	&	-0.312104	&	3483.45	\\
Ti$_2$MoPd	&	62.484	&	-0.393388	&	4390.67	\\
Ti$_2$MoPt	&	62.1896	&	-0.647475	&	7226.58	\\
Ta$_2$MoRe	&	66.5052	&	-0.341358	&	3809.96	\\
Re$_2$MoTi	&	61.2988	&	-0.294221	&	3283.85	\\
V$_2$MoRe	&	56.986	&	-0.280895	&	3135.12	\\
Ti$_2$MoRh	&	60.9012	&	-0.515086	&	5748.97	\\
Ta$_2$MoRu	&	65.4864	&	-0.37169	&	4148.5	\\
V$_2$MoRu	&	56.0544	&	-0.266878	&	2978.68	\\
Ta$_2$MoTc	&	66.1236	&	-0.348888	&	3894	\\
Mo$_2$TaW	&	65.598	&	-0.138309	&	1543.69	\\
Tc$_2$MoTi	&	60.4636	&	-0.348792	&	3892.93	\\
Mo$_2$TiW	&	63.4916	&	-0.13852	&	1546.05	\\
Mo$_2$VW	&	61.3504	&	-0.111196	&	1241.07	\\
Os$_2$NbSc	&	65.272	&	-0.455798	&	5087.24	\\
Ta$_2$NbOs	&	67.6176	&	-0.32288	&	3603.72	\\
Nb$_2$OsW	&	66.2136	&	-0.199311	&	2224.54	\\
Re$_2$NbTa	&	65.8404	&	-0.370123	&	4131.01	\\
Nb$_2$ReTc	&	65.302	&	-0.339447	&	3788.63	\\
Re$_2$NbTi	&	63.3304	&	-0.398599	&	4448.84	\\
Rh$_2$NbZn	&	60.0404	&	-0.492704	&	5499.15	\\
Ru$_2$NbSc	&	64.652	&	-0.549806	&	6136.48	\\
Ta$_2$NbRu	&	67.4116	&	-0.270899	&	3023.55	\\
Ru$_2$NbZn	&	59.4144	&	-0.275852	&	3078.83	\\
Tc$_2$NbTa	&	64.8712	&	-0.435369	&	4859.23	\\
Tc$_2$NbTi	&	62.4432	&	-0.468812	&	5232.5	\\
Tc$_2$NbZr	&	67.9524	&	-0.372639	&	4159.09	\\
Sc$_2$NiOs	&	65.1596	&	-0.499591	&	5576.03	\\
\end{tabular}
\label{fig:art109:magnetic_heusler_5}
\etab

\tabsec
\mycaption{Summary table of magnetic Heuslers continued (6/8).}
\tabvspace
\begin{tabular}{l|r|r|r}
material & volume (\AA$^3$) & $H_{\mathrm{f}}$ (eV) & $T_\mathrm{S}$~\K\  \\
\hline
Sc$_2$NiPt	&	68.0772	&	-0.888355	&	9915.08	\\
Ti$_2$NiRe	&	57.7528	&	-0.434031	&	4844.29	\\
Zn$_2$NiRh	&	52.1512	&	-0.344857	&	3849.01	\\
Sc$_2$NiRu	&	64.7916	&	-0.579752	&	6470.71	\\
Ti$_2$NiTc	&	57.1568	&	-0.486273	&	5427.38	\\
Ni$_2$TiZn	&	50.568	&	-0.405025	&	4520.55	\\
Sc$_2$OsPd	&	68.9692	&	-0.693907	&	7744.82	\\
Sc$_2$OsPt	&	68.3896	&	-0.8455	&	9436.77	\\
Ta$_2$OsRe	&	65.1812	&	-0.351313	&	3921.07	\\
Ti$_2$OsRu	&	59.0072	&	-0.744346	&	8307.77	\\
Zr$_2$OsRu	&	70.7628	&	-0.593618	&	6625.47	\\
Os$_2$ScTa	&	65.0652	&	-0.533786	&	5957.68	\\
Sc$_2$OsZn	&	69.8296	&	-0.439858	&	4909.33	\\
Os$_2$ScZr	&	68.6728	&	-0.476955	&	5323.38	\\
Ta$_2$OsTc	&	64.6524	&	-0.405699	&	4528.07	\\
Os$_2$TaTi	&	62.2512	&	-0.496833	&	5545.25	\\
Ta$_2$OsW	&	66.2384	&	-0.299962	&	3347.93	\\
Ti$_2$OsTc	&	59.5392	&	-0.632543	&	7059.92	\\
V$_2$OsTc	&	55.1728	&	-0.345037	&	3851.02	\\
Zr$_2$OsTc	&	71.0824	&	-0.476841	&	5322.1	\\
Sc$_2$PdPt	&	72.2524	&	-1.08971	&	12162.5	\\
Zn$_2$PdRh	&	56.2356	&	-0.51684	&	5768.54	\\
Sc$_2$PdRu	&	68.874	&	-0.78368	&	8746.79	\\
Pd$_2$ScZn	&	64.8108	&	-0.783946	&	8749.75	\\
Ti$_2$PdTc	&	61.282	&	-0.542858	&	6058.93	\\
Zr$_2$PdTc	&	73.412	&	-0.522887	&	5836.03	\\
Pd$_2$TiZn	&	60.4704	&	-0.57928	&	6465.44	\\
Pd$_2$ZnZr	&	67.0388	&	-0.641322	&	7157.91	\\
Zn$_2$PtRh	&	56.5432	&	-0.518725	&	5789.58	\\
Sc$_2$PtRu	&	68.1764	&	-0.962407	&	10741.6	\\
Pt$_2$ScZn	&	65.1192	&	-0.926966	&	10346	\\
\end{tabular}
\label{fig:art109:magnetic_heusler_6}
\etab

\tabsec
\mycaption{Summary table of magnetic Heuslers continued (7/8).}
\tabvspace
\begin{tabular}{l|r|r|r}
material & volume (\AA$^3$) & $H_{\mathrm{f}}$ (eV) & $T_\mathrm{S}$~\K\  \\
\hline
Sc$_2$PtZn	&	73.0916	&	-0.836692	&	9338.47	\\
Zn$_2$PtSc	&	62.8012	&	-0.667899	&	7454.54	\\
Ti$_2$PtTc	&	61.1772	&	-0.743086	&	8293.71	\\
Ti$_2$ReRh	&	60.04	&	-0.667826	&	7453.72	\\
Ta$_2$ReRu	&	64.9088	&	-0.376608	&	4203.39	\\
Ta$_2$ReTc	&	65.2984	&	-0.465143	&	5191.55	\\
Re$_2$TaTi	&	63.47	&	-0.457933	&	5111.07	\\
Ta$_2$ReW	&	66.8768	&	-0.278984	&	3113.79	\\
Re$_2$TiV	&	58.2544	&	-0.402985	&	4497.78	\\
Re$_2$TiW	&	61.6096	&	-0.350499	&	3911.97	\\
Ti$_2$ReZn	&	61.3876	&	-0.316406	&	3531.47	\\
Sc$_2$RhRu	&	66.6712	&	-0.818639	&	9136.97	\\
Rh$_2$ScZn	&	60.8824	&	-0.779193	&	8696.71	\\
Rh$_2$TaZn	&	59.9676	&	-0.548351	&	6120.24	\\
Ti$_2$RhTc	&	59.5088	&	-0.741616	&	8277.31	\\
Zr$_2$RhTc	&	71.5328	&	-0.64339	&	7180.99	\\
Rh$_2$TiZn	&	56.792	&	-0.783097	&	8740.29	\\
Rh$_2$VZn	&	55.0032	&	-0.416055	&	4643.66	\\
Rh$_2$ZnZr	&	63.148	&	-0.778072	&	8684.2	\\
Ru$_2$ScTa	&	64.4184	&	-0.625766	&	6984.29	\\
Ru$_2$ScTi	&	62.122	&	-0.656051	&	7322.31	\\
Ru$_2$ScV	&	59.5772	&	-0.460194	&	5136.31	\\
Sc$_2$RuZn	&	70.06	&	-0.491623	&	5487.1	\\
Ru$_2$ScZr	&	68.1104	&	-0.649445	&	7248.57	\\
Ta$_2$RuTc	&	64.312	&	-0.412004	&	4598.45	\\
Ru$_2$TaTi	&	61.542	&	-0.554291	&	6186.54	\\
Ta$_2$RuW	&	66.0004	&	-0.285381	&	3185.19	\\
Ru$_2$TaY&	70.2656	&	-0.340037	&	3795.22	\\
Ru$_2$TaZn	&	59.4956	&	-0.344438	&	3844.33	\\
Ti$_2$RuTc	&	59.1864	&	-0.643868	&	7186.32	\\
V$_2$RuTc	&	54.8572	&	-0.320533	&	3577.52	\\
\end{tabular}
\label{fig:art109:magnetic_heusler_7}
\etab

\tabsec
\mycaption{Summary table of magnetic Heuslers continued (8/8).}
\tabvspace
\begin{tabular}{l|r|r|r}
material & volume (\AA$^3$) & $H_{\mathrm{f}}$ (eV) & $T_\mathrm{S}$~\K\  \\
\hline
Zr$_2$RuTc	&	70.7428	&	-0.524685	&	5856.11	\\
Ru$_2$VZn	&	54.4836	&	-0.218631	&	2440.18	\\
Ru$_2$WZn	&	57.7132	&	-0.126584	&	1412.82	\\
Tc$_2$TaTi	&	62.5436	&	-0.530531	&	5921.35	\\
Tc$_2$TaZr	&	67.8028	&	-0.431941	&	4820.97	\\
Tc$_2$TiV	&	57.4784	&	-0.450416	&	5027.17	\\
Tc$_2$TiW	&	60.7084	&	-0.403323	&	4501.56	\\
Ti$_2$TcZn	&	61.0412	&	-0.346197	&	3863.97	\\
Tc$_2$WZr	&	65.7664	&	-0.258379	&	2883.81	\\
\end{tabular}
\label{fig:art109:magnetic_heusler_8}
\etab

\subsection{Conclusion}

In conclusion we have demonstrated a new systematic pathway to the discovery of novel magnetic materials. We have
created an extensive library of Heusler compounds including about 250,000 structures. For the sub-class of intermetallic
alloys we have been able to establish the materials stability against decomposition of 20 novel magnetic HAs,
belonging to Co$_2YZ$, Mn$_2YZ$ and $X_2$Mn$Z$ classes. A simple machine learning method, correlating calculated
microscopic electronic structure quantities with macroscopic measured properties, has been used to predict the magnetic
$T_\mathrm{C}$ of such compounds. The method has been put to the test with the experimental synthesis of four
compounds and validated by the growth of two. In particular we have discovered a new high-performance ferromagnet,
Co$_2$MnTi and a tetragonally distorted antiferromagnet, Mn$_2$PtPd. Our method offers a new high-throughput tool
for the discovery of new magnets, which can now be applied to other structural families, opening new possibilities for
designing materials for energy, data storage and spintronics applications.
\clearpage
\chapter{Conclusion}
Modeling approaches promise a direct and systematic path to materials discovery.
To justify their application, these methods need to bridge several gaps:
\textbf{i.} prediction of synthesizability (as property prediction/optimization becomes irrelevant if the material cannot form),
\textbf{ii.} treatment of more ``real-world'' phenomena (\vs\ the ideal systems modeled \abinitiospace),
and
\textbf{iii.} identification of structure-property relationships (harnessing the information for practical design rules).
Recent progress has been driven by data-centric approaches~\cite{curtarolo:art13}
facilitated by large, programmatically-accessible materials databases.

Frameworks like \AFLOW~\citeAFLOW\ have characterized millions of compounds
without the need for laborious human intervention~\citeAFLOWLIB.
Combinatorial exploration of various structure prototypes offers a means for sampling
candidate stable structures~\cite{curtarolo:art130,aflowANRL}.
The gamut of extractable features derives from electronic, magnetic, chemical, crystallographic, thermomechanical,
and thermodynamic characterizations --- each warranting
robust algorithms
that scale with the panoply of structures in the database.
For example, convenient definitions for the primitive cell representation~\cite{aflowPAPER} and
high-symmetry Brillouin Zone path~\cite{aflowBZ} have not only standardized electronic structure calculations,
but also optimized their computation.
Moreover, careful treatment of spatial tolerance and proper validation schemes have finally
enabled accurate and autonomous determination of the
complete symmetry profile of crystals~\cite{curtarolo:art134}.
Elasticity~\cite{curtarolo:art115} and phonon~\cite{aflowPAPER,curtarolo:art114,curtarolo:art119,curtarolo:art125}
calculations are incredibly sensitive to the quality of the symmetry analysis.
The scheme resolves experimentally-validated space groups and
accommodates even the most skewed unit cells, meeting the demand for high-throughput thermomechanical characterizations.

The development of the \AFLOWorg\ repository has motivated both
broad-scale thermodynamic formability modeling and adoption of \ML\ algorithms.
Ensembles of ordered phases are successfully employed to
\textbf{i.} construct phase diagrams forecasting stability~\cite{curtarolo:art146}
and
\textbf{ii.} formulate descriptors and models to predict the formation/properties of disordered materials~\cite{curtarolo:art110}.
These methods go beyond standard modeling approaches, leveraging
several \abinitio\ calculations in each analysis
and encouraging
the continued expansion of these large materials databases.

As the proliferation of high-throughput approaches
increases the wealth of data in the field, the gap between accumulated-information and derived-knowledge widens.
The divergence must be addressed autonomously, reciprocating the pace of data generation.
\ML\ models
are constructed for rapid predictions and exposing subtle/hidden trends
that would have otherwise evaded human detection/understanding.
Useful examples include models
predicting electronic and thermomechanical properties
from basic features of the structure and composition, \ie, not requiring additional calculations
or experiments, affording easy integration into virtually any materials design workflow~\cite{curtarolo:art124}.

\ML\ models are also employed to identify meaningful correlations among materials/properties,
leading to enhanced understanding of fundamental physical mechanisms.
For many phenomena, the connection between the arrangement of elements into solid compounds
and the observed macroscopic behavior is still largely unknown, as with
high-temperature superconductors.
These materials are particularly difficult to address within automated \abinitio\ frameworks because
the underlying \DFT\ theory fails to capture the strong interactions and correlations
responsible for the effect~\cite{DFT}.
However, as demonstrated by the materials cartography approach~\cite{curtarolo:art94},
other similarities between materials, such as the electronic density of states and band structure,
can be exploited to reveal interesting candidates.
Alternatively, \DFT\ data can be avoided altogether.
Instead, models have been constructed leveraging empirical
information retrieved from the SuperCon database~\cite{SuperCon} for more than 12,000 materials~\cite{curtarolo:art137}.
Distinct driving mechanisms are resolved by comparing important features
of a general model, trained on all data,
with that of family-specific models, trained on
low-$T_{\mathrm{c}}$, cuprate, and iron-based superconductors, respectively.

Structure-property relationships have also been resolved
in perovskites ($ABX_{3}$ where $X$ = F and O) for high-temperature thermoelectric applications~\cite{curtarolo:art120}.
The thermal conductivity of fluorides is strongly influenced by substitutions of the $B$ site,
while in oxides the same is true for the $A$ site --- presenting a useful engineering opportunity.
For example, to mitigate costs in device production, substitutions in the less influential site
can be expected not to affect the thermoelectric performance.

Finally, thermodynamic descriptors and regression analyses among classes of ground-state compounds
contributed to the screening of 36,540 Heusler compounds for new magnetic systems~\cite{curtarolo:art109}.
An attempt to synthesize four candidates yielded two novel materials.
Of these, Co$_{2}$MnTi promises to be a high-performance ferromagnet with $T_{\mathrm{C}}=938$~K,
as predicted by the Slater-Pauling curve --- illustrating the predictive power of data-driven approaches.
These methods will accelerate the path to synthesis and, ultimately,
transform the practice of traditional materials discovery to one of rational and autonomous materials design.
\clearpage

\newcommand{\Ozolins}{Ozoli\c{n}\v{s}}

\biography
Corey Oses is a PhD Candidate at Duke University in the Department of Mechanical Engineering and Materials Science
and a National Science Foundation Graduate Fellow.
He received his Bachelors of Science in Applied and Engineering Physics from Cornell University in 2013.
His research interests include design of data-driven thermodynamic descriptors for magnetic and
disordered materials and development of autonomous frameworks for chemical and crystallographic materials properties.
He has coauthored eighteen manuscripts --- fifteen of which are already published in
journals such as
Nature Communications~\cite{curtarolo:art124}, Science Advances~\cite{curtarolo:art109}, Physical Review X~\cite{curtarolo:art120},
Chemistry of Materials~\cite{curtarolo:art94,curtarolo:art110}, and
MRS Bulletin~\cite{curtarolo:art142} --- and three book chapters.
Several of these publications have been selected for Editors' Choice
awards~\cite{curtarolo:art94,curtarolo:art104,aflux,curtarolo:art136},
and highlighted in the
press~\cite{molecular_tornado_duke_press_2015,materials_fingerprints_mrs_press_2015,computers_new_magnets_duke_press_2017,plmf_unc_press_2017,plmf_mrs_press_2017}.
These are the fruits of many productive collaborations, including ones with
UNC-Chapel Hill~\cite{curtarolo:art94,curtarolo:art124,curtarolo:art136},
Trinity College-Dublin~\cite{curtarolo:art109,curtarolo:art146},
UC San Diego~\cite{curtarolo:art110},
CEA-Grenoble in France~\cite{curtarolo:art120,curtarolo:art136},
Tel Aviv University~\cite{curtarolo:art130},
University of Maryland~\cite{curtarolo:art137,curtarolo:art146},
and
the US Naval Academy~\cite{curtarolo:art146}.
His up-to-date biography can be found at {\sf coreyoses.com}.

\end{document}